\documentclass[11pt,a4paper,oneside]{scrartcl}
\usepackage[english]{babel}
\usepackage[margin=3.0cm]{geometry}
\usepackage[pdftex]{graphicx}
\usepackage[font=smaller]{subfig}
\usepackage{mathtools} 
\usepackage{color}
\usepackage{placeins}
\usepackage{epstopdf}
\usepackage[amssymb,thinqspace,cdot]{SIunits}
\usepackage{amsfonts, amsmath, amsthm, bm}
\usepackage[colorlinks=true]{hyperref}
\usepackage[font=small,format=hang,labelfont=bf,textfont=up]{caption}
\usepackage[all,defaultlines=3]{nowidow}
\usepackage{pdflscape}
\usepackage{graphbox}
\usepackage{enumitem}
\usepackage[export]{adjustbox}

\usepackage{tikz}
\usepackage{xcolor}
\usepackage{tabularx}
\usepackage{rotating}
\usetikzlibrary{arrows}
\usetikzlibrary{intersections}
\usetikzlibrary{calc}
\usetikzlibrary{arrows.meta,
                chains,
                positioning,
                shapes.geometric,
                shapes.misc,
                patterns
                }

\usepackage{arydshln} 

\definecolor{cornflowerblue}{RGB}{178, 178, 240}
\definecolor{coldpurple}{RGB}{57, 57, 200}

\definecolor{myContractedBlue}{rgb}{0.3,0.3,1.0}

\definecolor{myHighlightedRed}{rgb}{1.0,0.2,0.2}
\definecolor{myHighlightedGreen}{rgb}{0.0,0.95,0.0}

\definecolor{myUnfocussedGray}{rgb}{0.5,0.5,0.5}

\tikzset{
    vertDotLab/.style={circle,shape=circle,inner sep=1pt,fill,name=#1,label={$#1$}},
    vertDot/.style={   circle,shape=circle,inner sep=1pt,fill,name=#1},
    clip even odd rule/.code={\pgfseteorule}, 
    invclip/.style={
        clip,insert path=
            [clip even odd rule]{
                [reset cm](-\maxdimen,-\maxdimen)rectangle(\maxdimen,\maxdimen)
            }
    }
}
\tikzstyle{unfocussedEdge} = [myUnfocussedGray, densely dashed]

\renewcommand{\[}{\!\left[}
\renewcommand{\]}{\right]}
\renewcommand{\(}{\!\left(}
\renewcommand{\)}{\right)}
\makeatletter
\newcommand{\vast}{\bBigg@{4}}
\newcommand{\Vast}{\bBigg@{5}}
\makeatother

\newcommand{\norm}[1]{\left\Vert{#1}\right\Vert}

\renewcommand{\phi}{\ensuremath{\varphi}}
\newcommand{\eps}{\ensuremath{\varepsilon}}

\newcommand{\Ecal}{\ensuremath{{\mathcal{E}}}}

\newcommand{\Ncal}{\ensuremath{{\mathcal{N}}}}
\newcommand{\Acal}{\ensuremath{{\mathcal{A}}}}
\newcommand{\weps}{\ensuremath{{w_\eps}}}
\newcommand{\ceps}{\ensuremath{{c_\eps}}}
\newcommand{\wcontr}{\ensuremath{{w^\mathrm{c}}}}
\newcommand{\multiMergeCorr}{\ensuremath{{m_{\ge 2}}}}

\newcommand{\penalAngle}{\ensuremath{{\Pi^\angle}}}
\newcommand{\penalAngleFunc}{\ensuremath{{\Pi_\delta}}}
\newcommand{\muAngle}{\ensuremath{{\mu^\angle}}}

\newcommand{\penalLen}{\ensuremath{{\Pi^{\norm{\!\!\cdot\!\!}}}}}
\newcommand{\muLen}{\ensuremath{{\mu^{\norm{\!\!\cdot\!\!}}}}}

\newcommand{\nb}{\ensuremath{_\mathrm{nb}}}
\newcommand{\acc}{\ensuremath{_\mathrm{acc}}}

\numberwithin{equation}{section}

\begin{document}

\author{Roald Frederickx and Philip Dutr\'e\\KU Leuven, Department of Computer Science\footnote{Celestijnenlaan 200A, 3001 Heverlee, Belgium}}
\title{\LARGE Stackless Ray-Object Intersections Using \\ Approximate Minimum Weight Triangulations:\\ Results in 2D That Outperform Roped KD-Trees\\ (And Massively Outperform BVHs)}
\date{November 2021}
\maketitle

\newcommand{\PE}{Pont-l'Ev\^eque}
\newcommand{\topk}{Topkapı}

\begin{abstract}
\small{
Computing ray-object intersections is a key operation of ray tracers.
Two well-known data structures to accelerate this computation are the kd-tree (which partitions space) and the Bounding Volume Hierarchy (BVH, which partitions the primitives).
A third type of structure is a Constrained Convex Space Partitioning (CCSP), which --- like the kd-tree --- partitions space, but it does this in such a way that the geometric primitives exactly overlap with the boundaries of its cells.
As a consequence, it is robust against ill-fitting cells that plague methods with axis-aligned cells (kd-tree, BVH) and it permits an efficient, stackless traversal.

Within the computer graphics community, such CCSPs have received some attention in both 2D and 3D, but their construction methods were never directly aimed at minimizing their traversal cost --- even having fundamentally opposing goals for Delaunay-type methods.
Instead, for an isotropic and translation-invariant ray distribution the traversal cost is minimized when minimizing the \emph{weight}: the total boundary size of all cells in the structure.

We study the two-dimensional case using triangulations as CCSPs and explicitly minimize their total edge length using a simulated annealing process that allows for topological changes and varying vertex count.
Standard Delaunay-based triangulation techniques have total edge lengths ranging from $10\%$ higher to twice as high as our optimized triangulations for a variety of scenes, with a similar difference in traversal cost when using the triangulations for ray tracing.
Compared to a idealised roped kd-tree with stackless traversal, our triangulations require less traversal steps for all scenes that we tested and they are robust against the kd-tree's pathological behaviour when geometry becomes misaligned with the world axes.
Moreover, the stackless traversal of our triangulations strongly outperforms a BVH, which always requires a top-down descent in the hierarchy.
In fact, we show several scenes where the number of traversal operations for our triangulations \emph{decreases} as the number of geometric primitives $N$ increases, in contrast to the increasing $\log N$ behaviour of a BVH.
}
\end{abstract}

\newpage

\tableofcontents

\section{Introduction}

\subsection{Accelerating Ray-Object Intersections}

Efficiently finding the closest intersection of a ray with the geometry of a scene is a fundamental operation within computer graphics.
Several acceleration techniques are known that reduce the average case complexity of such a ray-object intersection to $O(\log n)$ for typical scenes with $n$ geometric primitives (e.g.\ triangles in 3D).
Each of these techniques have their own, specific acceleration data structures and traversal methods thereof.

The technique most commonly used in contemporary production renderers is the Bounding Volume Hierarchy (BVH), which hierarchically partitions the \emph{primitives} of the scene \cite{hyperion,arnold,renderman,manuka}. 
A well-known complementary technique is the kd-tree, which hierarchically partitions the \emph{space} of the scene into axis-aligned cells and stores pointers in the leaf cells to all objects that overlap the cell \cite{pbrt}.

\subsection{Hierarchies with Stackless Traversal}
For a kd-tree, if the leaf-cell containing the starting point of the ray is already known (as is the case for recursive rays in a path tracer) or its search can be amortized (e.g.\ all rays of a pinhole camera share the same origin), then the initial top-down descent of the hierarchy can be avoided and the kd-tree can be traversed iteratively in a stackless manner by enhancing the leaf nodes with ropes: links to adjacent neighbour-cells for each boundary plane of the original leaf cell \cite{havran1998ropeTrees}.
Due to the convex nature of axis aligned cells, the closest geometric intersection with the ray can be found by simply walking along the cells one by one in the order that they get pierced by the ray.
When a leaf-cell then finds one or more intersections that lie \emph{within}\footnote{
Because objects can overlap multiple cells, a closest intersection between ray and object can potentially lie in a cell $c'$ that is `behind' the currently visited cell, and one cannot be sure that this is indeed the closest intersection overall until all cells up to and including $c'$ have been traversed. Nonetheless, for `flat' geometric primitives (such as a triangle or quad in 3D) a ray can only intersect such a primitive in one point (or infinitely many points in the degenerate case, e.g.\ where a ray lies in the same plane as a triangle that it intersects in 3D). For such `flat' primitives, this multiple-intersection nuance is avoided.}
the volume of the cell, the closest of those intersections will be the closest intersection of the ray with any object.
As an aside, a BVH can also be adapted for a stackless traversal at the expense of visiting (but not intersecting) some nodes more than once \cite{stacklessBVH}.

For efficient implementations on hardware with limited high-bandwidth on-chip memory --- such as GPUs --- having a stackless traversal becomes crucial.
Moreover, regardless of underlying hardware, the cost that can be saved by avoiding the initial descent that would otherwise be necessary in a hierarchical structure becomes substantial when expansive scenes are rendered of production level complexity --- of which potentially only a part is visible or relevant for the light transport calculations.

Nonetheless, due to the strict axis alignedness of the spatial partitioning in a kd-tree, the tightness of the leaf cells can be poor, leading either to large leaf cells with many associated primitives which all need to be checked, or small leaf cells where neighbouring leaf cells share many `overlapping' associated geometric primitives which lead to redundant intersection tests on the same geometric primitives\footnote{Redundant computation of the actual primitive intersection can be partly avoided by techniques such as mailboxing, but this merely lowers the overall constant of the time complexity: one still has to loop over all pointers in the leaf node even if a full intersection test is not needed for each primitive. Alternatively, geometric primitives can be clipped to the splitting planes so that each piece fits in their cell, which has the trade-off of increasing memory usage.}.
There are spatial partition techniques which break free from strict axis aligned cells and have more freedom to tightly bound the actual geometry, such as k-DOP based partitions \cite{kammaje2007study,budge2008accelerated}, but which cause the cell traversal complexity to increase and still suffer from the underlying non-tightness problem that is fundamentally due to the disconnect between the cell boundaries and the geometric primitives.

\subsection{Constrained Convex Space Partitionings}

In the natural limit of perfect cell tightness, the geometrical primitives should exactly overlap with the cell boundaries. 
The cells schould furthermore still be convex to keep the efficient iterative traversal method where one simply walks through the cells in the order that they get pierced.
These two conditions lead to a general Constrained Convex Space Partitioning (CCSP) \cite{aresCDT,maria17architecturalCCSP}, where `constrained' denotes that the scene's geometric primitives (e.g. triangles, quads) exactly lie on the cell boundaries.
Note that there may also be intermediate cell boundaries which do not coincide with a geometric primitive, but each geometric primitive (e.g.~a triangle or quad in 3D) should at least coincide with one (or possibly the union of several, non-overlapping) cell boundary(ies) that fully cover it. Table \ref{tabComparisonOfAccelStruct} gives an overview and comparison of CCSPs with a more classical (roped) kd-tree and BVH.

\begin{table}[htb]
\caption{Comparison of different ray-object acceleration data structures. The roped kd-tree and BVH are discussed in more detail in Sec.\ \ref{secOtherMethodsForComparison}.
\label{tabComparisonOfAccelStruct}}
\small{
{\renewcommand{\arraystretch}{1.8}
\begin{tabular}{@{\!\!}c@{}|c@{}c@{}c@{\!}}
& CCSP (e.g.\ triangulation)
& Roped KD-tree
& BVH
\\\hline
Partition & Partitions space & Partitions space & Partitions objects
\\
Duplication
& No duplication
& {\renewcommand{\arraystretch}{1.0}\begin{tabular}{c}1 object in many cells \\ `ownership overlap'\end{tabular}}
& {\renewcommand{\arraystretch}{1.0}\begin{tabular}{c}1 position in many cells \\ `spatial overlap'\end{tabular}}
\\[4mm]
{\renewcommand{\arraystretch}{1.0}
    \begin{tabular}{c}
    Intersection\\
    cost terms
    \end{tabular}
}
&
\# cells traversed
&
{\renewcommand{\arraystretch}{1.0}
    \begin{tabular}{l}
    \# nodes visited\\
    \# rope search steps\\
    \# primitives tested
    \end{tabular}
}
&
{\renewcommand{\arraystretch}{1.0}
    \begin{tabular}{l}
    \# nodes visited ($\ge \log N$)\\
    \# primitives tested
    \end{tabular}
}
\end{tabular}
}
}
\end{table}

\begin{figure}[htb]
\newcommand{\geoSteinerScale}{0.35}
\begin{center}
    \subfloat[Without Steiner vertex]
    {\adjustbox{margin=5mm 2mm}{
    \scalebox{1.0}[0.8]{
    \includegraphics[scale=\geoSteinerScale]{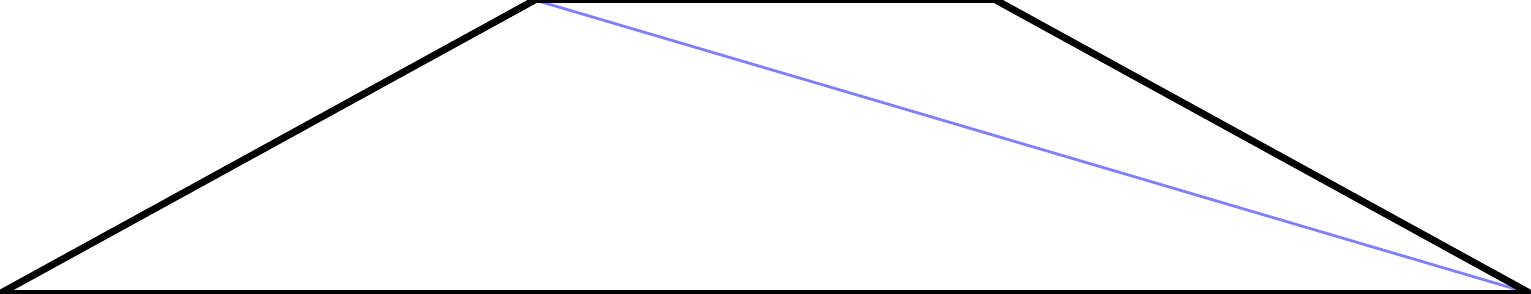}}}}
    \hspace{10mm}
    \subfloat[With `constrained' Steiner vertex on bottom edge]
    {\adjustbox{margin=5mm 2mm}{
    \scalebox{1.0}[0.8]{
    \includegraphics[scale=\geoSteinerScale]{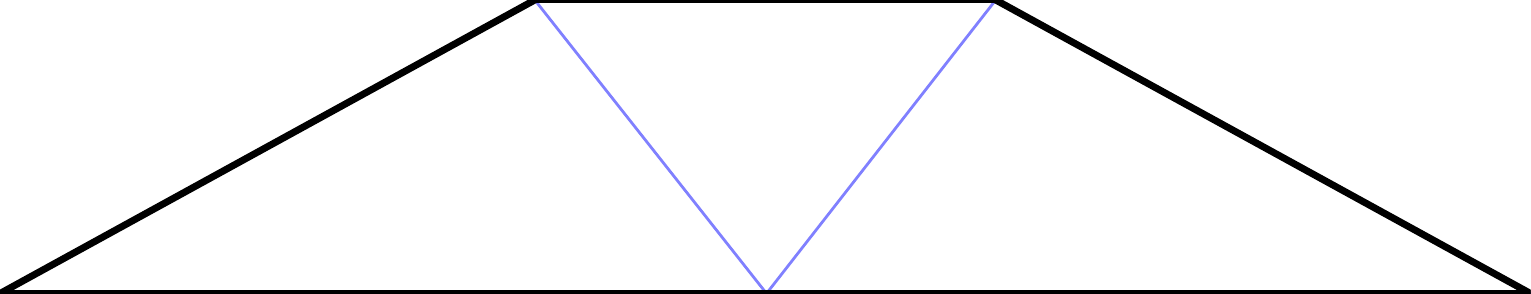}}}}
\end{center}
\caption{
    Example of a triangulation (blue) of a simple geometric object (black) where the introduction of a Steiner vertex that is `constrained' to lie on an edge of the original geometry leads to a CCSP (in this 2D case a triangulation) with less total weight (total edge length). In this 2D example, the `constrained' Steiner vertex only has one remaining positional degree of freedom: its distance along the bottom edge of the original geometry.
    \label{figGeoSteinerExample}}
\end{figure}

\begin{figure}[htb]
\newcommand{\freeSteinerScale}{0.35}
\begin{center}
    \subfloat[Without Steiner vertex]
    {\adjustbox{margin=4mm 2mm}{\includegraphics[scale=\freeSteinerScale]{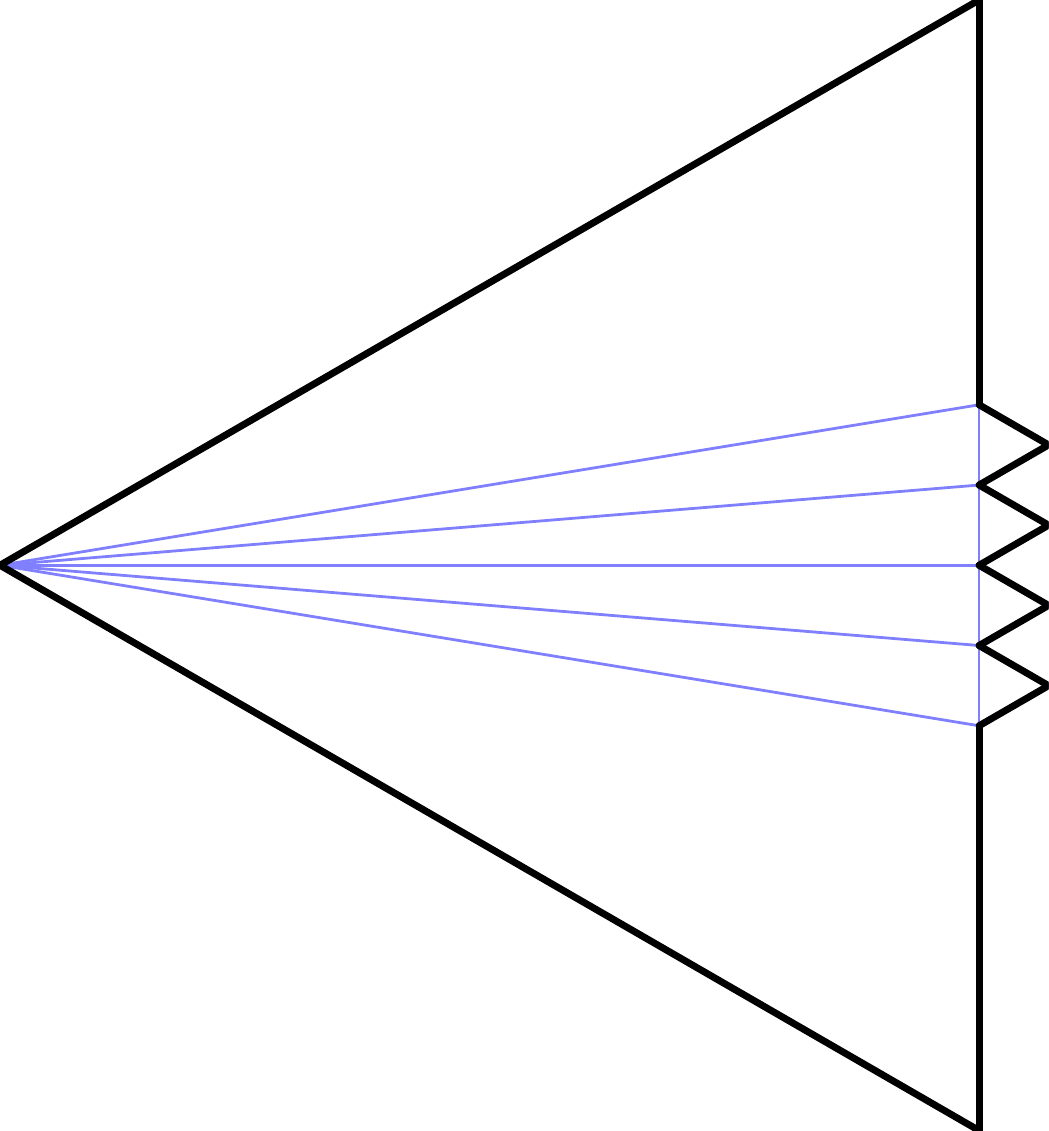}}}
    \hfill
    \subfloat[With `free' Steiner vertex \label{figFreeSteinerExampleMinWeight}]
    {\adjustbox{margin=4mm 2mm}{\includegraphics[scale=\freeSteinerScale]{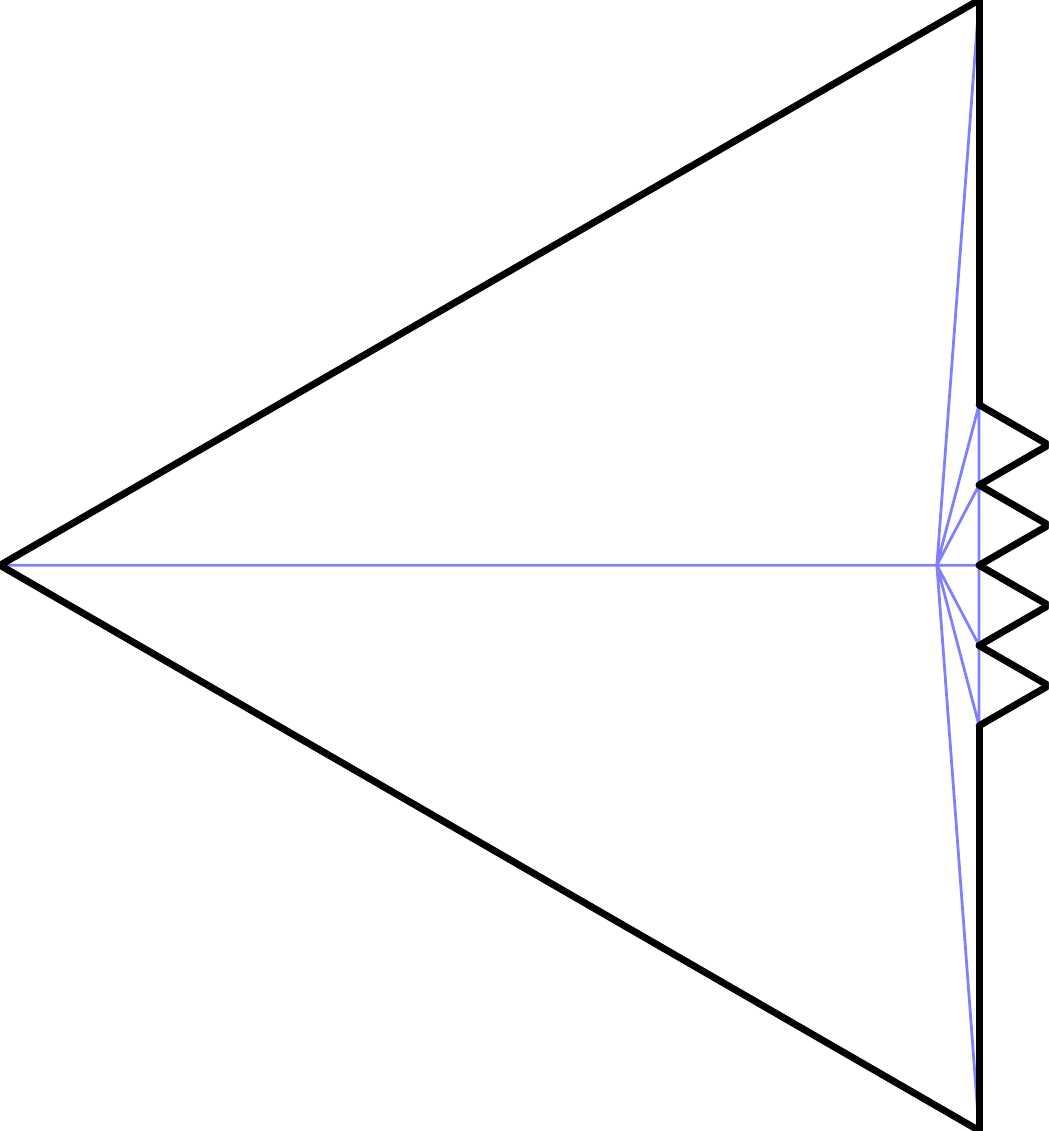}}}
    \hfill
    \subfloat[With angles $\ge 25\degree$ \label{figFreeSteinerExampleQuality}]
    {\adjustbox{margin=4mm 2mm}{\includegraphics[scale=\freeSteinerScale]{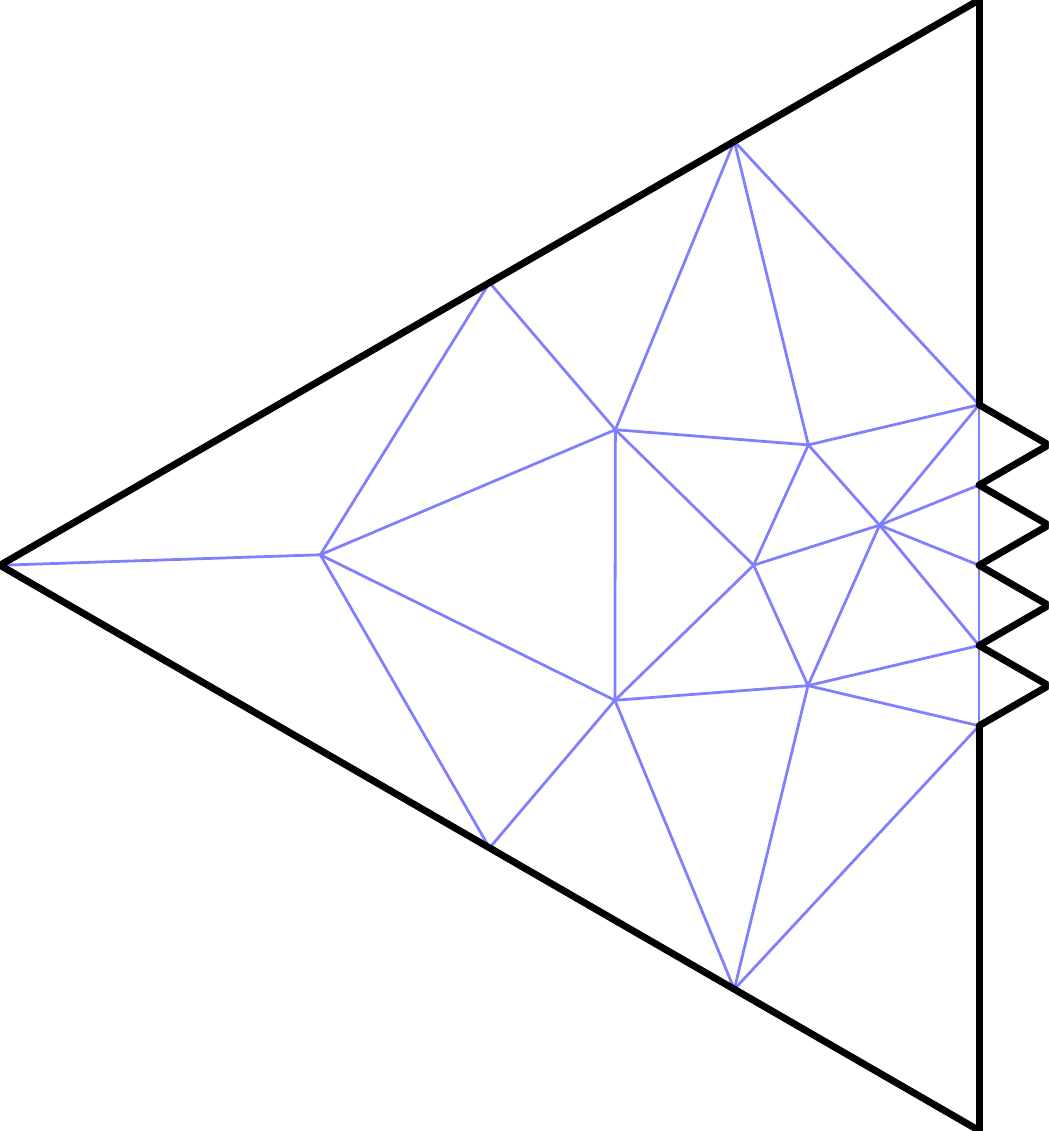}}}
\end{center}
\caption{
    Example where the introduction of a `free' Steiner vertex leads to a triangulation with lower total edge length from (a) $\to$ (b), and an example of `free' and `constrained' Steiner vertices to satisfy a lower bound on the minimum angle in (c). In this 2D example, the position of the `free' Steiner vertex in (b) has the full two degrees of freedom.
    Note that moving the Steiner vertex in (b) closer to the right decreases the overall edge length, but it cannot be moved \emph{completely} to the right (i.e.\ to coincide with the closest vertex of the original geometry) as this would lead to topological problems with `squashed', zero-area triangles which we want to avoid (discussed in Sec.~\ref{secMainDiscussionAgainstBadConditioning}).
    Also note that in (b) no `constrained' Steiner vertex that lies on the original geometry can produce a triangulation with less total edge length than the triangulation with this `free' Steiner vertex --- often `constrained' Steiner vertices suffice to produce an optimal triangulation, but some cases truly require a `free' Steiner vertex.
    Lastly, the triangulation in (c) is an example of a triangulation that places a lower bound of $25\degree$ on the angles of the triangles, which will be discussed in Sec.~\ref{secOptimalityOfCCSP}. Note that, as a result, the total edge length has increased greatly.
    \label{figFreeSteinerExample}}
\end{figure}

Due to the overlap with primitives and cell boundaries, the data structure for such a CCSP should thus certainly contain the vertices of the scene's geometry, but there can be additional vertices --- called Steiner vertices --- as well.
This can happen in two ways: 
(1) an original geometric primitive can be subdivided into multiple fragments (each a boundary of their own cell in the structure) by the introduction of one or more Steiner vertices that are then `constrained' to lie on the original geometric primitive, or
(2) several internal cell boundaries in the structure which do not coincide with a geometric primitive can meet at a `free' Steiner vertex. Examples of both cases in a 2D setting with a triangulation as CCSP are shown in Figs.~\ref{figGeoSteinerExample} and \ref{figFreeSteinerExample}, respectively.

Such Constrained Convex Space Partitionings have been studied only sparsely within the domain of computer graphics. Lagae and Dutr\'e \cite{aresCDT} used a Constrained Delaunay Tetrahedralisation (CDT) for accelerating ray-object intersections of triangle meshes in 3D, where the triangles of the mesh correspond to faces of the tetrahedral cells. This was later improved by Maria et al. to increase the numerical robustness, which allowed for implementation on GPUs \cite{maria17robustCDTopGPU}.
Maria et al.\ also introduced a dedicated CCSP for architectural scenes \cite{maria17architecturalCCSP}, which leverages the `2.5D' nature of architectural scenes by representing them as a polygonal 2D floor plan with extruded vertical walls, possibly with simple openings such as doors or windows. The 2D floor plan is then decomposed into convex polygonal cells \cite{fernandez2008practical}, with an added heuristic to avoid generating many large cells near detailed concave areas (e.g.\ round pillars represented by a regular $n$-gon).

\subsection{Optimality of a CCSP \label{secOptimalityOfCCSP}}

The underlying methods used to construct the convex space partitioning for both the CDT of Lagae and Dutr\'e \cite{aresCDT} and the architectural CCSP of Maria et al.\ \cite{maria17architecturalCCSP} are, however, not inherently optimized to produce a CCSP that has low cost for ray-object intersections.

The goal of the underlying polygonal CCSP algorithm employed by Maria et al.\ \cite{maria17architecturalCCSP}, for instance, is to achieve a partition in as few convex polygons as possible (and without introducing extra Steiner vertices), which was originally designed for application in `location problems' \cite{fernandez2008practical}.
As a consequence of minimizing the number of cells, the resulting convex polygons can have a large number of edges, leading to a high traversal cost of a single cell when finding the edge where the ray exits. This traversal cost is linear in the number of edges of the cell, and whereas a more fine subdivision (e.g.\ into `cheap-to-traverse' triangular cells) would lead to more individual cells being traversed to cover the same distance, the total number of edges that is checked is potentially reduced by more quickly cutting away distant edges.

For the case of a CDT \cite{aresCDT}, on the other hand, the cost of traversing a single tetrahedron is a constant time operation as there are only four possible exit-faces to check (three if the entrance face is discarded).
The expected cost of traversing a ray through an entire tetrahedralisation is then determined by the average number of tetrahedra that are visited until an exit face is found that corresponds to a triangle of the original geometry.
To minimize this average number of traversed tetrahedra, assuming an isotropic and translation invariant ray distribution, the tetrahedralisation should minimize its \emph{weight}: the total surface area of the faces of its tetrahedral cells (in a 2D world: a triangulation should minimize the total length of the edges of its triangles) \cite{aronovFortune97}.
For intuition: the underlying reasoning is rather analogous to the well-known Surface Area Heuristic for BVHs or kd-trees.

In contrast, the refinement process of the Delaunay tetrahedralisation employed by Lagae and Dutr\'e \cite{aresCDT} is optimized to yield `quality' or `well shaped' tetrahedra, meaning `close to regular' tetrahedra without skinny or long `slivers' \cite{si2006refinement, si2015tetgen}.
Such properties are desirable when using the tetrahedralisation for simulation purposes (e.g. in finite element methods) and imposing a minimum quality can help to convert an initial pathological concentration of long, thin tetrahedra with large surface areas to more regular tetrahedra (by adding extra Steiner vertices) and thereby decreasing the total weight. However, requesting too high a quality leads to a strong increase in the number of (near perfectly regular) tetrahedra and correspondingly an increase in the total weight of the tetrahedralisation \cite{si2015tetgen}. A similar behaviour could be seen in Fig.~\ref{figFreeSteinerExampleQuality} in 2D.
As such, the goal of a quality tetrahedralisation is fundamentally incompatible with the goal of a minimum weight tetrahedralisation.

If, for simplicity, we focus on CCSPs where the cells have a low, fixed number of boundaries --- specifically a tetrahedralisation in 3D and a triangulation in 2D --- then the traversal cost per cell is constant and finding an optimal CCSP reduces to finding a minimum weight CCSP. Exact solutions to such minimization problems are known to be inherently difficult. For instance, the related problem of finding the minimum weight triangulation of a point set was shown to be NP-hard \cite{mulzer2008MWTisNPHard}. For finding an \emph{approximate} minimum weight constrained Steiner triangulation, Aronov and Fortune \cite{aronovFortune97} developed an octree-based method that runs in $O(n^6)$ time and returns a provably constant-factor approximation to the minimum weight tetrahedralisation, although the value of this constant factor is very large for practical purposes.
Cheng and Dey \cite{cheng99approxMinWgtSteinerTri} improved on this method and obtain a constant-factor approximation algorithm that runs in $O(n^3\log n)$ time.
In both cases, the large value of the constant factor makes these results primarily interesting from a theoretical standpoint, rather than directly applicable in practice.

\subsection{Goal and Overview of this Text}
In this work, we constrain ourselves to the two-dimensional case of finding (approximate) minimum-weight triangulations.
This 2D case can be seen as an initial exploration before going to a more traditional 3D context,
but the 2D version also has direct merit in itself and it can be used for `2.5D' architectural scenes \cite{maria17architecturalCCSP}.

The goals of this work are to:
\begin{itemize}
\item Examine the structure of (approximately) minimum weight triangulations;
\item Compare various levels of optimization to approximate these minimum weight triangulation with regards to the compute versus optimality trade-off;
\item Compare the ray tracing performance of our triangulations to classical structures (BVH and roped kd-tree).
\end{itemize}

Our strategy to obtain a minimum weight triangulation will be to start from a sufficiently fine initial triangulation $\tau$, and optimize
(1) any edge connections that can be `flipped' directly, and
(2) the positional degrees of freedom associated with the Steiner vertices
with regards to an objective function $f$. This objective function measures the total edge length, but (1) it allows for topology changes by (fuzzily) contracting edges that are shorter than a contraction length scale $\eps$ and (2) it penalises some configurations that have bad (numerical) conditioning.
In Sec.~\ref{secApproxMWT} we piece together the objective function $f$, followed by a discussion of the optimization process in Sec.~\ref{secOptimOfObjFunc} and its results in Sec.~\ref{secOptimResults}. The resulting (approximate) minimum weight triangulations are then used for ray tracing in Secs.~\ref{secRTwithTheTriangs} and~\ref{secRTresults}.

\section{Approximate Minimum Weight Triangulations\label{secApproxMWT}}

In this section we construct the objective function $f$ that allows us to optimize a given triangulation with respect to the total edge length. This objective function supports topology changes through fuzzy contraction of short edges and it includes some safeguards to avoid bad numerical conditioning.

\subsection{A Restriction on Conditioning \label{secMainDiscussionAgainstBadConditioning}}
Before we dive into the actual objective function that we will try to minimize, we first make a small detour to discuss a potential problem with bad (numerical) conditioning that can arise in truly minimum weight triangulations.

In Fig.~\ref{figFreeSteinerExampleMinWeight}, we already saw an example where minimizing the total edge length of a triangulation (i.e.\ moving the Steiner vertex all the way to the right) would lead to a pathological topology with several overlapping zero-area triangles that are all squished to a flat line. These cases could be allowed in principle, although great care should then be taken when building and traversing such a structure. 

To start, computing the orientation (the sign of the signed area) of a very thin (near-zero-area) triangle is badly conditioned. We use such signed areas to ensure that the topology of a triangulation stays valid during the optimization. Errors in computing this orientation can lead to non-planar and inconsistent triangulations.

Furthermore, infinitesimally thin triangles can also be problematic for ray traversal.
Indeed, when there are several overlapping edges from different overlapping `squished' triangles, trying to traverse these triangles in the order that they are pierced by a ray can easily lead to loops: all edges are pierced at the exact same point and their traversal order is thus ill-defined.

\begin{figure}
\begin{center}
\begin{tikzpicture}[scale=1.5,
main node/.style={black,draw,minimum width=1.5cm,minimum height=1.2cm, align=center, rounded corners=2mm}]

\tikzstyle{every node}=[font=\small]

\draw[-Stealth,ultra thick] (0,0.3)--(5,0.3) node[right] {Lower total edge length};
\draw[-Stealth,ultra thick] (5,0)--(0,0) node[left]  {Better conditioning};

\coordinate (MinWgtOnAx) at (5,-0.1);
\coordinate (OursOnAx) at (4.7,-0.1);
\coordinate (CDTSmallOnAx) at (2,-0.1);
\coordinate (CDTLargeOnAx) at (0.5,-0.1);

\node[main node] (MinWgt) at (6,-1.2) {Minimum Weight\\Triangulation};
\node[main node] (Ours) at (4.3,-1.2) {Ours};

\node[main node] (CDTSmall) at (2,-1.2) {Refined CDT\\Small Min.~Angle};
\node[main node] (CDTLarge) at (-0.5,-1.2) {Refined CDT\\Large Min.~Angle};

\draw[dashed] (MinWgt) to [in=270,out=90] (MinWgtOnAx) -- +(0,0.55);
\draw[dashed] (Ours) to [in=270,out=90] (OursOnAx) -- +(0,0.55);
\draw[dashed] (CDTSmall) to [in=270,out=90] (CDTSmallOnAx) -- +(0,0.55);
\draw[dashed] (CDTLarge) to [in=270,out=90] (CDTLargeOnAx) -- +(0,0.55);

\end{tikzpicture}
\end{center}
\caption{Schematic representation of the trade-off between total edge length and conditioning. A triangulation with truly minimal edge length often leads to conditioning problems as we saw in Fig.~\ref{figFreeSteinerExampleMinWeight}. Our approximate minimum weight triangulations trade a slight increase in total edge length for improved conditioning. In contrast, the primary goal of refined Constrained Delaunay Triangulations (CDTs) is to achieve very good conditioning, at the expense of greatly increased total edge length.
    \label{figEdgeLenVsConditioning}}
\end{figure}
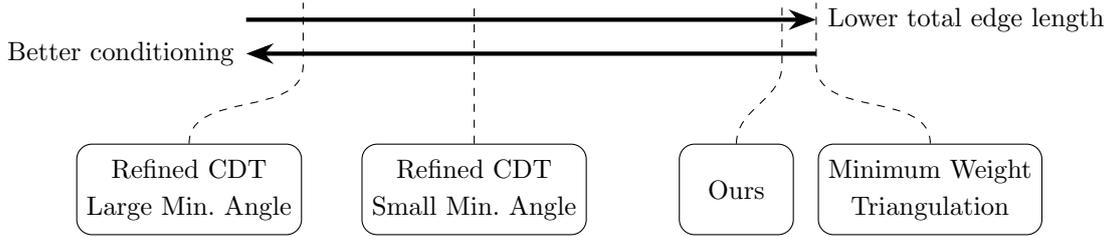

Given that numerical robustness of ray traversal in triangulations without pathological squished triangles is already critical and non-trivial \cite{maria17robustCDTopGPU}, we opt to avoid squished triangles by penalizing badly conditioned triangles.
For cases such as in Fig.~\ref{figFreeSteinerExampleMinWeight}, this means that we deliberately trade off a bit of extra edge length in return for ease of traversal, as represented schematically in Fig.~\ref{figEdgeLenVsConditioning}.
The details will be given in Sec.~\ref{secPenalisingBadConditioningInPractice}.

\subsection{Objective Function for minimization with Fuzzy Topology \label{secIncludingTopologyChanges}}

A straightforward optimization of a given triangulation could use a general purpose optimizer to directly minimize the total edge length as a function of the positional degrees of freedom of the Steiner vertices.
A gradient-descent based optimizer can quickly find such a (local) minimum, but this minimum is at best the minimum weight for all triangulations \emph{with the same topology}.

In order to incorporate topology changes into the search space, we modify the objective function to allow edges to be dynamically `contracted' to a single point if the removal of that edge would lead to a lower overall weight, which we explain in this section. If one then starts with a sufficiently subdivided triangulation $\tau$ of the original geometry (as will be explained in Sec.~\ref{secSubdividing})
then there should be sufficient topological freedom (in the added Steiner vertices) such that the global minimum of $f$ for $\eps \to 0$ is expected to approach the minimum weight possible for any constrained triangulation of the original geometry that satisfies our conditioning requirements. The (approximate) minimum weight triangulation can then be obtained by
contracting\footnote{An alternative approach, which directly changes the topology during the optimization by explicitly adding or removing vertices instead of using a fuzzy edge contraction, could be inspired by the physical analog of the grand canonical ensemble, where the number of particles is allowed to vary. This could have the advantage of reducing the typical dimensionality by only introducing extra degrees of freedom when they are `necessary', thereby speeding up the optimization iterations. We currently opted for the conceptually simpler choice of a fixed number of degrees of freedom (with fuzzy contracion) for simplicity.}
all edges smaller than $\eps$ in the triangulation that minimizes $f$, or a slightly more advanced process in that vein as discussed in Sec.~\ref{secContractionOfOptimizedTopology}.

\subsubsection{Contraction Through Edge Weighting \label{secEdgeWeighting}}
The goal of our objective function with fuzzy contraction is to only count the edge lengths in the triangulation that would remain after all edges with lengths below the contraction scale $\eps$ have been contracted to a single vertex.
For a triangle that has a `fuzzily contracted' edge that is shorter than $\eps$, its two other edges effectively overlap and merge into a single edge and they should thus only count as one edge in the weight --- or, equivalently, their own length should be downweighted by a factor 1/2. Therefore, in order to enable such contraction of edges, we define a `contractible' weight $\wcontr$ of a triangulation $\tau$ as a weighted sum of edge lengths
\begin{equation}
\label{eqwcontr}
\wcontr(\tau; \eps)
=
    \sum_{e \in \Ecal(\tau)} \weps(e) \norm{e}
    ,
\end{equation}
where $\eps$ determines characteristic length at which edges become contracted. The edge weight $\weps$ is given by
\begin{equation}
\label{eqweps}
    \weps(e)
    =
    \multiMergeCorr\(
    \prod_{e' \in \Ncal(e)}
    \ceps(e')
    \)
    ,
\end{equation}
where $\multiMergeCorr$ is a correction function for multiply-merged edges that will be discussed below and the product runs over the other edges of the (one or two) triangle(s) that also have $e$ as edge. The `contraction weight' $\ceps(e')$ interpolates between 1 for $\norm{e'} \gg \eps$ and 1/2 for $\norm{e'} \ll \eps$ if $e'$ is a contractible edge, or it is fixed to 1 if the edge $e'$ is incontractible (more on that in Sec.~\ref{secContractibleEdges}).
Many functions fit the asymptotic requirements of $\ceps$ --- and the specific behaviour can be quite important when using a local, gradient-based optimizer --- but for the optimization strategy that we use, a simple linear ramp on $[0,\eps]$ works sufficiently well:
\begin{equation}
\label{eqCeps}
\ceps(e) = 
    \begin{cases}
        1/2 + \min(1, \norm{e}\!/\eps)/2
        & \text{if $e$ is contractible}\\
        1
        & \text{otherwise.}
    \end{cases}
\end{equation}

\begin{figure}
\begin{center}
\newcommand{\CtopL}{(-1, 1)}
\newcommand{\CtopR}{( 1, 1)}
\newcommand{\CbotL}{(-1,-1)}
\newcommand{\CbotR}{( 1,-1)}

\newcommand{\Ccentr}{(-0.05,0.1)}
\newcommand{\CtwoTop}{( 0.05,  0.2)}
\newcommand{\CtwoBot}{(-0.15, -0.0)}

\newcommand{\CthreeTop}{( 0.15,  0.2)}
\newcommand{\CthreeBot}{(-0.15, -0.1)}
\newcommand{\CthreeXtra}{(-0.15, 0.2)}

\newcommand{\drawWContrSituation}{
\begin{tikzpicture}[scale=1.4]
    \node[vertDot=vtopL] at \CtopL {};
    \node[vertDot=vtopR] at \CtopR {};
    \node[vertDot=vbotL] at \CbotL {};
    \node[vertDot=vbotR] at \CbotR {};

    \node[vertDot=vcentr] at \Ccentr {};

    \draw (vtopL) -- (vtopR) -- (vbotR) -- (vbotL) -- (vtopL);

    \draw (vtopL) -- (vcentr);
    \draw (vtopR) -- (vcentr);
    \draw (vbotL) -- (vcentr);
    \draw (vbotR) -- (vcentr);
\end{tikzpicture}
}

\newcommand{\drawWContrOneEdgeCalculation}{
\begin{tikzpicture}[scale=1.4]
    \node[vertDot=vtopL] at \CtopL {};
    \node[vertDot=vtopR,myUnfocussedGray] at \CtopR {};
    \node[vertDot=vbotL] at \CbotL {};
    \node[vertDot=vbotR,myUnfocussedGray] at \CbotR {};

    \node[vertDot=vtwoTop] at \CtwoTop {};
    \node[vertDot=vtwoBot] at \CtwoBot {};

    \draw[myHighlightedGreen,ultra thick] (vtopL) -- (vtwoBot);

    \draw[unfocussedEdge] (vtopL) -- (vtopR) -- (vbotR) -- (vbotL);
    \draw[ultra thick] (vbotL) -- (vtopL);

    \draw[ultra thick] (vtopL) -- (vtwoTop);
    \draw[ultra thick] (vtwoBot) -- (vbotL);

    \draw[unfocussedEdge] (vtopR) -- (vtwoTop) -- (vbotR);
    \draw[unfocussedEdge] (vtwoBot) -- (vbotR);

    \draw[myContractedBlue,ultra thick] (vtwoTop) -- (vtwoBot);
\end{tikzpicture}
}

\newcommand{\drawWContrSimpleEdgeMerge}{
\begin{tikzpicture}[scale=1.4]
    \node[vertDot=vtopL] at \CtopL {};
    \node[vertDot=vtopR] at \CtopR {};
    \node[vertDot=vbotL] at \CbotL {};
    \node[vertDot=vbotR] at \CbotR {};

    \node[vertDot=vtwoTop] at \CtwoTop {};
    \node[vertDot=vtwoBot] at \CtwoBot {};

    \draw[myHighlightedGreen,ultra thick] (vtwoTop) -- (vtopL) -- (vtwoBot);

    \draw (vtopL) -- (vtopR) -- (vbotR) -- (vbotL) -- (vtopL);

    \draw (vtwoBot) -- (vbotL);

    \draw (vtwoTop) -- (vbotR);
    \draw (vtwoBot) -- (vbotL);

    \draw (vtopR) -- (vtwoTop) -- (vbotR);
    \draw (vbotL) -- (vtwoBot) -- (vbotR);

    \draw[myContractedBlue,ultra thick] (vtwoTop) -- (vtwoBot);
\end{tikzpicture}
}

\newcommand{\drawWContrMultiEdgeMerge}{
\begin{tikzpicture}[scale=1.4]
    \node[vertDot=vtopL] at \CtopL {};
    \node[vertDot=vtopR] at \CtopR {};
    \node[vertDot=vbotL] at \CbotL {};
    \node[vertDot=vbotR] at \CbotR {};

    \node[vertDot=vthreeTop] at \CthreeTop {};
    \node[vertDot=vthreeBot] at \CthreeBot {};
    \node[vertDot=vthreeXtra] at \CthreeXtra {};

    \draw[myHighlightedGreen,ultra thick] (vthreeTop) -- (vtopL) -- (vthreeBot);

    \draw (vtopL) -- (vtopR) -- (vbotR) -- (vbotL) -- (vtopL);

    \draw (vthreeTop) -- (vbotR);

    \draw (vtopR) -- (vthreeTop) -- (vbotR);
    \draw (vbotL) -- (vthreeBot) -- (vbotR);

    \draw[myContractedBlue,ultra thick] (vthreeTop) -- (vthreeBot) -- (vthreeXtra) -- (vthreeTop);
    \draw[myHighlightedRed,ultra thick](vtopL) -- (vthreeXtra);
\end{tikzpicture}
}

\newcommand{\drawWContrMultiEdgeMergeCalculation}{
\begin{tikzpicture}[scale=1.4]
    \node[vertDot=vtopL] at \CtopL {};
    \node[vertDot=vtopR,myUnfocussedGray] at \CtopR {};
    \node[vertDot=vbotL,myUnfocussedGray] at \CbotL {};
    \node[vertDot=vbotR,myUnfocussedGray] at \CbotR {};

    \node[vertDot=vthreeTop] at \CthreeTop {};
    \node[vertDot=vthreeBot] at \CthreeBot {};
    \node[vertDot=vthreeXtra] at \CthreeXtra {};

    \draw[ultra thick] (vthreeTop) -- (vtopL) -- (vthreeBot);

    \draw[unfocussedEdge] (vtopL) -- (vtopR) -- (vbotR) -- (vbotL) -- (vtopL);

    \draw[unfocussedEdge] (vthreeTop) -- (vbotR);

    \draw[unfocussedEdge] (vtopR) -- (vthreeTop) -- (vbotR);
    \draw[unfocussedEdge] (vbotL) -- (vthreeBot) -- (vbotR);

    \draw[myContractedBlue,ultra thick] (vthreeBot) -- (vthreeXtra) -- (vthreeTop);
    \draw[unfocussedEdge] (vthreeTop) -- (vthreeBot);
    \draw[myHighlightedRed,ultra thick](vtopL) -- (vthreeXtra);
\end{tikzpicture}
}
\newcommand{\multiContrEdgeMargin}[1]{\adjustbox{margin=7mm 0mm}{#1}}
\begin{tabular}{m{3cm}c}
\subfloat[Contracted state]{\drawWContrSituation}
&
\begin{tabular}{c}
\begin{tabular}{cc}
\subfloat[One contracted edge \label{figWContrSimpleEdgeMerge}]
{\multiContrEdgeMargin{\drawWContrSimpleEdgeMerge}}
&
\subfloat[Calculation detail\label{figWContrOneEdgeCalculation}]
{\multiContrEdgeMargin{\drawWContrOneEdgeCalculation}}
\end{tabular}
\\
\\[-3mm]\hdashline
\begin{tabular}{cc}
\subfloat[Three contracted edges \label{figWContrMultiEdgeMerge}]
{\multiContrEdgeMargin{\drawWContrMultiEdgeMerge}}
&
\subfloat[Calculation detail \label{figWContrMultiEdgeCalculation}]
{\multiContrEdgeMargin{\drawWContrMultiEdgeMergeCalculation}}
\end{tabular}
\end{tabular}
\end{tabular}
\caption{An example of how the fuzzy contraction calculates the total edge length of the contracted state. The same contracted state (a) can be represented by the triangulation (b) with a single contracted edge (blue), or a triangulation with three contracted edges (d). For the case in (b) of a single contracted edge, the two edges highlighted in green each receive a contraction weight $\wcontr$ of 1/2. The calculation for the bottom green edge $e$ is shown in (c): of its four `neighbouring' edges $\Ncal(e)$, three are uncontracted (black) and yield a factor 1, and one of them is contracted (blue) and yields a factor $1/2$ in Eq.~\eqref{eqweps}, giving a weight $\wcontr$ of 1/2 as per Eq.~\eqref{eqwcontr}. The top green edge of (b) follows a similar calculation, as do the two green edges in the case of three contracted edges in (d). The red edge in (d), however, should then no longer contribute anything to the weighted edge length, as it would get merged with the two green edges which have already fully accounted for the resulting length with their edge weight $\wcontr$ of 1/2. Ignoring $\multiMergeCorr$, a calculation of $\weps$ for this red edge would yield a result of 1/4 due to its two contracted `neighbouring' edges as shown in (e), and this indeed yields an edge weight of zero due to the correction function $\multiMergeCorr$ within $\wcontr$. Note that in this triangulation, all three contracted edges themselves also receive an edge weight of zero, which is consistent with the fact that they will get merged to a single vertex in the contracted state.}
\end{center}
\end{figure}
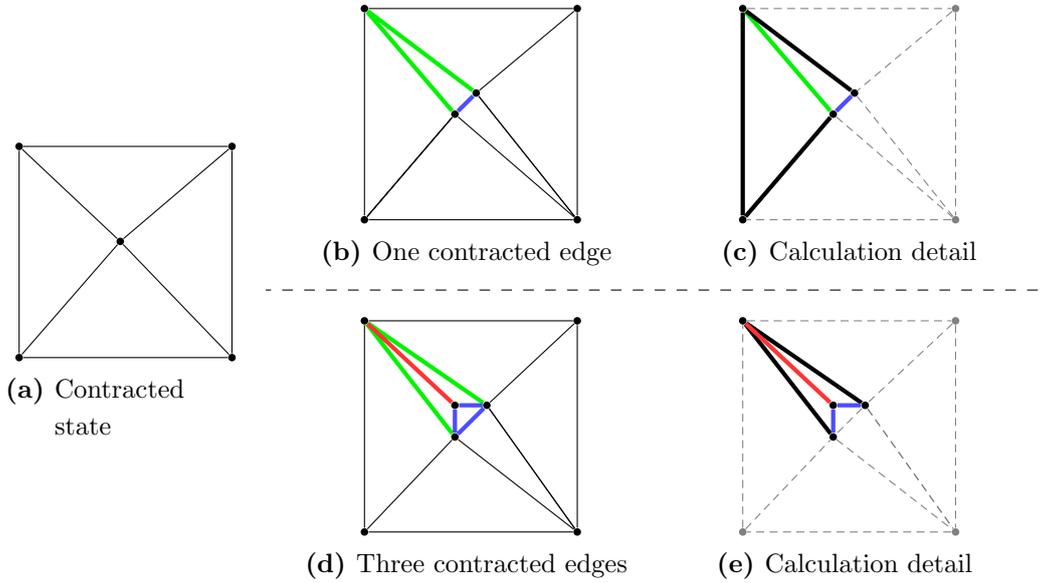

If the correction function $\multiMergeCorr$ were simply the identity function, then the above strategy would already work when edges are contracted `in isolation', as illustrated in Figs.~\ref{figWContrSimpleEdgeMerge} and \ref{figWContrOneEdgeCalculation}. In this case, the two edges that get merged together indeed have a weight $\wcontr$ of 1/2. However, when multiple edges are contracted to a single vertex such that more than two edges would get merged into a single edge, we would still have an overcounting of the edge lengths as illustrated in Fig.~\ref{figWContrMultiEdgeMerge}. This problem can be addressed by discarding the inner `multiply merged' edge(s) marked in red in Fig.~\ref{figWContrMultiEdgeMerge}. Such `multiply merged' edges $e$ would get a weight of 1/4 or below: 1/4, 1/8 or 1/16 if there are two, three or four contracted neighbouring edges in $\Ncal(e)$ (again, assuming that $\multiMergeCorr$ is simply the identitity function). The actual correction function $\multiMergeCorr$ makes sure that such `multiply contracted' edges are not counted: it is zero on $[0,1/4]$, a simple linear ramp up to one on $[1/4,1/2]$ and stays one on $[1/2,1]$, i.e.
\begin{equation}
\label{eqMultiMergeCorr}
\multiMergeCorr(x) =
    \begin{cases}
        0 & \text{if $x < 1/4$} \\
        4x - 1 & \text{if $1/4 \leq x \leq 1/2$} \\
        1 & \text{if $x > 1/2$}
        .
    \end{cases}
\end{equation}
This simple choice ensures that $\wcontr$ behaves as it should for `multiply contracted' contracted edges, and it does so without introducing discontinuities. An example is shown in Fig.~\ref{figWContrMultiEdgeCalculation}.

\subsubsection{Contractible Edges \label{secContractibleEdges}}
Not all edges of a constrained triangulation are contractible, as was already hinted at in Eq.~\eqref{eqCeps}: edges don't get downweighted according to our fuzzy contraction if an edge that is shorter than $\eps$ cannot actually be contracted for the final trianculation\footnote{This introduces discontinuities in the objective function $f$ as the `contracatbility' of an edge can change dynamically based on the configuration of the neighbouring vertices as we will see. For gradient-based optimizers this discontinuity should be smoothed out, but as our optimization procedure does not use gradient information we have simply kept it as-is and have not noticed any problems in this regard.}.
Contraction can be prohibited due to constraints imposed by the fixed vertices and edges of the original geometry or in general because the planar topology of the triangulation would be violated otherwise. We discuss the various cases that we detect in what follows.

\paragraph{Direct Restrictions due to Original Geometry}
Trivially, edges that have both endpoints on a fixed vertex of the original geometry cannot be contracted as their vertices are locked in position. Furthermore, edges that have either an original vertex and a constrained Steiner vertex or two constrained Steiner vertices cannot be contracted if those vertices are not part of the \emph{same} original edge.

\paragraph{Conditioning Restrictions due to Original Geometry Constraints}
There are also indirect ways in which the constraints of vertices on original geometry can prohibit the contraction of edges if this would guarantee a badly conditioned triangle. Indeed, the contraction of an edge with at least one original or constrained Steiner vertex leads to a decrease in degrees of freedom (DoF) of the other vertex as it gets `snapped' to the most-constrained vertex, for example free\,(2D) + constrained\,(1D) $\to$ constrained\,(1D), or constrained\,(1D) + original\,(0D) $\to$ original\,(0D). If such an edge were to get contracted to a single vertex, the triangles that contained the initial vertex with the higher DoF may now be forced into a badly conditioned configuration if that vertex gets replaced by the vertex with the lower DoF:
\begin{itemize}[noitemsep,topsep=2pt,parsep=2pt,partopsep=0pt]
    \item
If a resulting triangle has only original vertices, then such bad conditioning can directly be detected by checking the degree of collinearity between the vertices.
    \item
If a resulting triangle has two original vertices and one constrained Steiner vertex, we heuristically use a collinearity check for the Steiner vertex at both endpoints and the middle of its original edge and forbid contraction if all positions yield bad conditioning.
    \item
If a resulting triangle has one original vertex and two constrained Steiner vertices, then there are conditioning problems if both original edges of the constrained vertices are themselves collinear and they are also collinear with the original vertex.
    \item
If all vertices of a resulting triangle are constrained Steiner vertices, then bad conditioning occurs if all their original edges are collinear.
    \item
Lastly, if at least one vertex of a resulting triangle is free\,(2D), then there are always enough DoF to avoid bad conditioning\footnote{A free vertex can still be \emph{effectively} constrained by nearby original geometry in a region that leads to bad conditioning, but such cases are harder to detect and this did not cause significant problems in practice.}.
\end{itemize}
All tests of (near-)collinearity between three points are performed by comparing the area of the triangle with those points as vertices to the area of a square with the same perimeter as this triangle. If this ratio drops below a small, critical value, then the points are deemed (near-)collinear.

\begin{figure}
\begin{center}
\begin{tikzpicture}[scale=1.8]
\tikzstyle{every node}=[font=\small]
    \node[vertDot,label=$F$] (top) at (-0.8,1.1) {};
    \node[vertDot,label={[anchor=north,yshift=-2]$C_0$}] (left) at (-1.3,0) {};
    \node[vertDot,label={[anchor=north,yshift=-2]$C_N$}] (right) at (4.0,0) {};
    \node[vertDot,label={[anchor=north,yshift=-2]$C_1$}] (lmid1) at ($(left)!0.1!(right)$) {};
    \node[vertDot,label={[anchor=north,yshift=-2]$C_2$}] (lmid2) at ($(left)!0.17!(right)$) {};
    \node[vertDot,label={[anchor=north,yshift=-2]$C_m$}] (lmidi) at ($(left)!0.31!(right)$) {};

    \node[vertDot,label={[anchor=north,yshift=-2]$C_{m+1}$}] (rmidip1) at ($(left)!0.76!(right)$) {};
    \node[vertDot,label={[anchor=north,yshift=-2]$C_{m+2}$}] (rmidip2) at ($(left)!0.87!(right)$) {};

    \draw[ultra thick,myContractedBlue] (left) -- (lmid1);
    \draw[ultra thick,myContractedBlue] (lmid1) -- (lmid2);
    \draw[ultra thick,myContractedBlue,dotted] (lmid2) -- (lmidi);
    \draw[ultra thick] (lmidi) -- (rmidip1);
    \draw[ultra thick,myContractedBlue] (rmidip1) -- (rmidip2);
    \draw[ultra thick,myContractedBlue,dotted] (rmidip2) -- (right);

    \draw[ultra thick,myContractedBlue] (left) -- (top);

    \draw (top) -- (lmid1);
    \draw (top) -- (lmid2);
    \draw (top) -- (lmidi);
    \draw (top) -- (rmidip1);
    \draw (top) -- (rmidip2);
    \draw (top) -- (right);
\end{tikzpicture}
\end{center}
\caption{
    An example of flat triangles when contracting to original edges. The segment $C_0C_N$ is part of a single edge of the original geometry (or several original edges that are all collinear), hence all vertices here are constrained to stay on the line $C_0C_N$ and the endpoint $C_N$ is chosen such that this is the longest uninterrupted straight line possible (which implies that $C_N$ is a fixed vertex of the original geometry). The edge of interest is $FC_0$, consisting of the free (2D) vertex $F$ and the constrained (1D or 0D) vertex $C_1$. If this edge were to be contracted, the less-constrained vertex $F$ would get `snapped' to $C_0$, and all triangles that had $F$ as a vertex now effectively get $C_0$ instead. In particular, the triangles $C_1FC_N$, $C_2FC_N$, etc.\ now become the zero-area triangles $C_1C_0C_N$, $C_2C_0C_N$, creating `trapped' vertices $C_1$, $C_2$, etc. Such contraction is only valid if these trapped vertices can themselves be removed by contraction, either by contracting to $C_0$ or to $C_N$. In other words, there can only be one edge $C_mC_{m+1}$ that is longer than $\eps$, and all other edges $C_iC_{i+1}$ should be contractible and shorter than $\eps$.
\label{figAllowContractionForGeomFreeCombo}}
\end{figure}
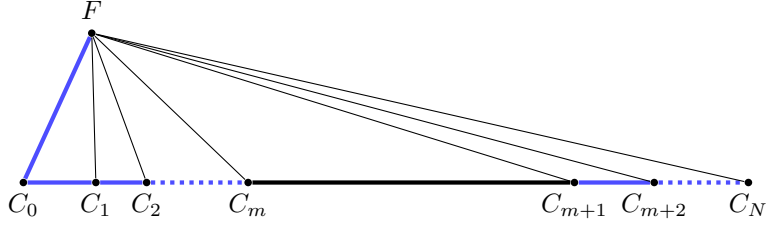

\paragraph{Flat Triangles When Contracting to Original Edges}
When contracting an edge composed of one free vertex and either a constrained vertex or possibly a vertex from the original geometry, there is a possibility of creating zero-area, flat triangles if the free vertex is connected to the original geometry through other edges. This leads to `trapped' vertices when `pulling' the free vertex onto the original geometry as in Fig.~\ref{figAllowContractionForGeomFreeCombo}. We flag the edge as incontractible unless all of those `trapped' vertices are themselves already contracted to the Steiner or original-geometry vertex that the edge gets contracted to, or to the original-geometry vertex at the end of the original edge ($C_0$ and $C_N$ respectively in Fig.~\ref{figAllowContractionForGeomFreeCombo}).

\begin{figure}
\begin{center}
\begin{tikzpicture}[scale=1.8]
\tikzstyle{every node}=[font=\small]
    \node[vertDot,label={[anchor=south,yshift=-2]$V_1$}] (top) at (-0.8,1.1) {};
    \node[vertDot,label={[anchor=north,yshift=-1]$V_2$}] (left) at (-1.3,0) {};
    \node[vertDot,label={[anchor=north,yshift=-1]$H$}] (right) at (2.0,0) {};
    \node[vertDot,label={[anchor=north,yshift=-1]$V_3$}] (mid) at (1.0,0.27) {};

    \draw[ultra thick,myContractedBlue] (left) -- (top);
    \draw[ultra thick,myContractedBlue] (mid) -- (right);

    \draw (top) -- (mid);
    \draw (left) -- (mid);
    \draw (top) -- (right);
    \draw (left) -- (right);
\end{tikzpicture}
\end{center}
\caption{
    An example of a possible topology violation from a 1-hop connection. When determining the contractibility of the edge $V_1V_2$ within the triangle $V_1V_2V_3$, we look for a vertex $H$ that is an indirect `1-hop connection' to $V_1$ and $V_2$, i.e.\ $H$ is connected to the edge $V_1V_2$ through two edges ($V_1H$ and $HV_2$) that are not themselves part of our triangle $V_1V_2V_3$. If $V_1$ and $V_2$ were to be contracted to a single point $P$, then the two remaining triangles (originally $V_1HV_3$ and $V_2HV_3$) would both overlap as `$PHV_3$' in an illegal topological state --- or, viewed differently, one of them would have their original orientation (the sign of its signed area) flipped which again violates the topology.
    However, if $V_3$ is itself already fuzzily contracted to $H$ ($\norm{V_3H}<\eps$), then we \emph{can} view $V_1V_2$ as contractible as the entire drawing would simply represent a single edge $PH$.
    Lastly, note that there can be additional small (fuzzily contracted) triangles in-between $V_3$ and $H$, which is why we simply use the Euclidian distance between $V_3$ and $H$ as a criterion instead of looking for a direct edge $V_3H$ that is shorter than $\eps$.
\label{figPlanarityViolation1HopConnection}}
\end{figure}
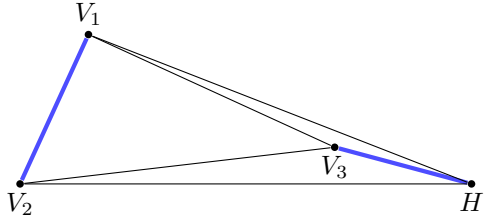

\paragraph{Topology Violation From 1-Hop Connections}
If the two vertices of an edge are connected `indirectly' through a nontrivial 1-hop connection (a connection along an intermediate vertex that is not itself a vertex of the triangles that share the edge), then contraction of that edge will lead to topology violations as in Fig.~\ref{figPlanarityViolation1HopConnection}. We flag the edge as incontractible unless all such nontrivial 1-hop connection vertices are within contraction distance $\eps$ of a trivial 1-hop connection vertex (the one/two vertices of the one/two triangle(s) that share our edge which are themselves not vertices of our edge).

\subsubsection{Penalizing Bad Conditioning \label{secPenalisingBadConditioningInPractice}}
As mentioned in Sec.~\ref{secMainDiscussionAgainstBadConditioning}, we want to stay away from badly conditioned, thin triangles with near-zero area. A natural way to avoid such triangles is to add a penalty term to the objective function for triangles with very sharp edges.
However, this would conflict with the edge weighting of Sec.~\ref{secEdgeWeighting} which encourages short, `contracted' edges and thus leads to long, thin triangles. After performing the actual contraction of the short edges, these thin `badly conditioned' triangles become a single edge and thus pose no problem for a ray traversal algorithm. These thin `contracted' triangles should thus be excluded from the conditioning penalty.

A more `indirect' way to avoid (most) badly conditioned triangles without competing with the contraction goals of edge weighting is not to directly penalize sharp angles, but to penalize obtuse angles that are nearly $180\degree$.
Triangles with an angle that is bigger than $180\degree-\alpha$ are indeed guaranteed to have two angles that are sharper than $\alpha$, as the sum of all three angles should equal $180\degree$. 
Nonetheless, a triangle can still have a single sharp angle (for example because one of its edges is shorter than $\eps$ and thus effectively contracted) without being penalized.

In concrete terms, we introduce an angle penalty term $\penalAngle$ into the objective function of the form
\begin{equation}
\label{eqPenalAngle}
    \penalAngle(\tau; \delta) = 
    \!\!\!\!
    \sum_{\{e_1, e_2\} \in \Acal(\tau)}
    \!\!\!\!
    \penalAngleFunc(\langle e_1,e_2 \rangle)
,
\end{equation}
where $\Acal(\tau)$ is the set of `angle pairs' (each triangle contributes three unordered combinations of two edges) of the triangulation $\tau$, $\langle e_1,e_2 \rangle$ is the angle's cosine or dot product between the direction vectors of the edges $e_1$ and $e_2$ (taken with the directions pointing away from the shared vertex) 
and $\penalAngleFunc$ is given by
\begin{equation}
\penalAngleFunc(x)
=
    \frac{\max(0,\, -1+\delta - x)}{\delta}
    ,
\end{equation}
which is a simple linear ramp starting at 0 for an angle cosine of $(-1+\delta)$ and ending at 1 for a cosine of $-1$ (i.e.\ an angle of $180\degree$). We use $\delta=\cos(179\degree)+1\approx 1.5\times10^{-4}$, which enables the penalty term for angles over $179\degree$\footnote{Note that angles that are formed by edges of the original geometry are fixed and cannot be adjusted, so the sum in Eq.~\eqref{eqPenalAngle} can optionally exclude these angles as their contribution to the total penalty is a constant --- although hopefully none of those original angles will have a nonzero penalty term as this means an inherent bad conditioning of the triangulation that cannot be avoided.}.

In case we are `polishing' a triangulation with fixed topology (i.e.\ without edge weighting for contracted edges, more on that in Sec.~\ref{secSuccessiveContractionAndPolishing}), we use a modified $\penalAngleFunc$ that \emph{does} directly penalise sharp edges as there is no longer a conflict with the edge weighting. In that case we use a linear ramp from 0 to 1 for angle cosines $1-\delta$ to 1 (corresponding to an angle of 0).

Lastly, for numerical stability reasons during the optimization (e.g.\ when checking the topology), we add a penalty term $\penalLen$ that avoids edge lengths that become \emph{too} small (much smaller than the contraction length $\eps$). This penalty takes the form
\begin{equation}
\label{eqPenalLen}
\penalLen(\tau; \eps_0)
=
    \sum_{e \in \Ecal(\tau)} \frac{\max(0,\, \eps_0 - \norm{e})}{\eps_0}
    ,
\end{equation}
with $\eps_0 \ll \eps$. We use double precision numbers and take $\eps_0 = 10^{-10}$ with the triangulation normalized to the unit square.

\subsubsection{The Final Objective Function}
The final objective function $f$ that we want to optimize is then the weight of the triangulation with fuzzy contraction $\wcontr$ [Eq.~\eqref{eqwcontr}] combined with the angle and edge length penalties $\penalAngle$ [Eq.~\eqref{eqPenalAngle}] and $\penalLen$ [Eq.~\eqref{eqPenalLen}], respectively
\begin{equation}
\label{eqObjFunc}
f(\tau; \eps, \delta, \eps_0)
=
    \wcontr(\tau; \eps)
    + \muAngle \penalAngle(\tau; \delta)
    + \muLen  \penalLen(\tau; \eps_0)
,
\end{equation}
which is parametrized by the contraction length $\eps$, the angle cosine penalty threshold $\delta$ and the minimum edge length penalty threshold $\eps_0$. The weights $\muAngle$ and $\muLen$ control the relative impact of the angle and minimum edge length penalties and we use $\muAngle=1$ and $\muLen=10$ for a triangulation that is normalized to the unit square.

\section{Optimization of the Objective Function \label{secOptimOfObjFunc}}
The minimization of the objective function [Eq.~\eqref{eqObjFunc}] is not a trivial task --- certainly not as the contraction length $\eps$ approaches zero, which causes the valleys corresponding to states with contracted edges to become very narrow. One could use a gradient descent method, but this would quickly get stuck in one of the many local minima.
Instead we use simulated annealing to approximate the global minimum, after which the fuzzy topology gets baked in by actually contracting the fuzzily contracted edges, as we discuss below.

\subsection{Simulated Annealing \label{secSimulatedAnnealing}}
Simulated annealing is a stochastic optimization technique that aims to find (an approximation of) the global minimum of a target function by interpreting it the energy of a physical system that undergoes gradual cooling (`annealing').
For such a system, the probability density of observing some state $x$ is given by the Boltzmann distribution (also called the Gibbs distribution for a non-finite state space)\footnote{We set the Boltzmann constant $k_\mathrm{B}=1$ for simplicity here.} \cite{numericalRecipes}
\begin{equation}
\label{eqBoltzmann}
p(x) \propto e^{-E(x)/T}
\end{equation}
for temperature $T$ and energy $E(x)$ of state $x$.

The sampling of the distribution given in Eq.~\eqref{eqBoltzmann} is typically performed indirectly through a Markov Chain method which starts at some initial state and samples new (perturbed) states in such a way that the asymptotic distribution of all sampled states approaches that of Eq.~\eqref{eqBoltzmann}.

Initially, the temperature is set high (w.r.t.\ typical variations in energy) such that all states have roughly equal probability according to Eq.~\eqref{eqBoltzmann}. Then, the temperature is gradually decreased such that the system favours states with lower energy. For a sufficiently slow cooling scheme such that the Markov Chain is always approximately in equilibrium with the target distribution of Eq.~\eqref{eqBoltzmann}, the final (low-temperature) state will be a good approximation to the global minimum of $E$.

The Markov Chain transition probability for state $x$ to some new state $y$ is typically given by the Metropolis-Hastings update rule
\begin{equation}
\label{eqMarkovTransitionProb}
    p(y \mid x) = g(y \mid x) A(x \to y)
    + \delta(y - x) R
    ,
\end{equation}
where $g(y \mid x)$ is some easily-sampled proposal density for generating a new tentative state $y$ given the current state $x$, $A(x \to y)$ is an acceptance probability for that proposed state and $R$ is the fraction of proposed states that are rejected according to $A$ (which leave the state unchanged, hence the Dirac delta distribution). The acceptance probability $A$ is chosen to make the resulting sampled states satisfy the distribution of Eq.~\eqref{eqBoltzmann}
\begin{equation}
\label{eqAcceptanceProbAsymmetric}
    A(x \to y) = \min\(1,\, \frac{g(x \mid y)}{g(y \mid x)} e^{[E(x) - E(y)]/T} \)
    .
\end{equation}

For symmetrical proposal densities where $g(x \mid y) = g(y \mid x)$ (i.e.\ you are `equally likely' to propose a jump from $x$ to $y$ as the reversed jump from $y$ back to $x$), the acceptance probability simplifies to
\begin{equation}
\label{eqAcceptanceProbSymmetricEnergy}
    A(x \to y) = \min\(1,\, e^{[E(x) - E(y)]/T} \)
    .
\end{equation}
The method for sampling proposal states that we discuss in Sec.~\ref{secPerturbations} will be symmetric by design to use this simplified version, which has the benefit that no explicit formulation or evaluation of the transition probability $g(y\mid x)$ is needed.

In our case where we want to optimize triangulations, the state $x$ becomes the triangulation $\tau$ and the energy $E$ becomes the objective function $f$ given by Eq.~\eqref{eqObjFunc}. The acceptance probability for a new triangulation $\tau'$ that was proposed from triangulation $\tau$ thus reads
\begin{equation}
\label{eqAcceptanceProbSymmetricTriang}
    A(\tau \to \tau') = \min\(1,\, \exp\[\frac{f(\tau; \eps,\delta,\eps_0) - f(\tau'; \eps,\delta,\eps_0)}{T}\] \)
    .
\end{equation}
Note that our temperature $T$ thus has the same units as our objective function $f$, which measures the total edge length (including fuzzy contraction and two penalty terms) --- the temperature can thus be interpreted as the length scale of typical variations in the total edge length that are thermally accessible to the system.

\subsection{Perturbations \label{secPerturbations}}
The heart of a Markov chain method (as used in simulated annealing) lies in its proposal distribution $g$ [Eq.~\eqref{eqMarkovTransitionProb}]. This proposal distribution can be conditioned on the current state, and is typically constructed by applying some perturbation to this current state (note that this makes the next sample correlated with the previous one, but if it were easy to directly sample from the target distribution of Eq.~\eqref{eqBoltzmann} as independent samples, then there would be no need for a Markov chain).
The proposed states should be chosen for optimal state space exploration of the Markov chain (i.e.\ make large steps with minimal correlation to the current state, ideally guided by the target distribution of Eq.~\eqref{eqBoltzmann}), but too extreme a perturbation and the proposed state can end up in an unfavorable region with high objective function and low acceptance probability $A$.
To increase the robustness and efficiency of the exploration, we use an ensemble of different perturbation strategies (discussed below).

\subsubsection{Several Remarks Before Listing The Different Perturbations}
Before we list the various perturbation strategies, we give some general remarks that are applicable to all of them

\paragraph{Efficient Updates of the Objective Function}
Our perturbation strategies typically only change a (small) subset of the vertices or edges, which enables an efficient sampling of the Markov Chain. Indeed, the acceptance probability of Eq.~\eqref{eqAcceptanceProbSymmetricTriang} only depends on the difference in objective function $f$, and all three components of $f$ ($\wcontr$, $\muAngle$ and $\muLen$) are themselves simple sums whose terms only depend on a few vertices or edges each. The difference in $f$ can thus be found efficiently by only computing the difference in those terms that depend on the vertices or edges that were actually changed (as the remaining contribution stays constant and cancels out)\footnote{For perturbation strategies that only change a fixed number of vertices, this means that sampling the next state in the Markov Chain can be done in constant time. Of course, when only updating a fixed number of vertices at a time, it will take $O(N)$ such steps with $N$ the total number of vertices in order to have a final state that is properly decorrelated with the initial state.}. A full evaluation of $f$ is never needed.

\paragraph{Adaptive Aggressiveness}
Each perturbation strategy has a parameter $\lambda$ which sets the aggressiveness of the perturbation (for instance by determining how far vertices will be moved, or how many vertices will be moved).
This parameter gets controlled during the simulation to keep the acceptance rate of a perturbation strategy near the ideal 23\% for high-dimensional problems \cite{optimalAccRateMCMC}, which is the sweet spot between sufficiently high acceptance rate to make progress, but still sufficiently aggressive perturbations to avoid high autocorrelation in the samples.
Note that such on-line modifications of the perturbation's proposal distribution gives slight modifications of the resulting sampling compared to the true target distribution of Eq.~\eqref{eqBoltzmann}, but this is typically not an issue for simulated annealing as we are simply looking to minimize our objective function and not obtain a true sampling of the exact distribution.
Also note that we will use the same symbol $\lambda$ for all perturbation strategies below for notational simplicity, even though each perturbation strategy has its own distinct parameter value that is controlled independently.

\paragraph{Constrained Vertices}
The various perturbation strategies will often output a 2D displacement vector to update the position of the movable (i.e.\ Steiner) vertices. For `free' Steiner vertices with the full two degrees of freedom, this displacement is applied directly. On the other hand, when the vertex in question is a constrained vertex (i.e.\ a vertex that lays on an edge of the original geometry), the displacement vector is projected onto the original edge to ensure that the constrained vertices are only moved along their single degree of freedom and never moved away from their edge. All update rules $p_i \to p_i + \delta$ for some position $p_i$ and displacement $\delta$ that are written below should be interpreted as being wrapped in this constraint-preserving projection.

\paragraph{Topological Constraints}
As a last remark before we discuss the different strategies, note that it can happen that a perturbation leads to a triangulation that no longer has planar topology. In such cases, we automatically reject the perturbation such that our state is always a valid triangulation. This is equivalent to adding an extra penalty term to the objective function which effectively becomes infinite for topology-violating triangulations and is zero otherwise.

\subsubsection{Directly Perturbing Vertex Position \label{secPertSingleVert}}
The first and simplest perturbation strategy amounts to adding a form of noise to one or more vertex positions. The position $p_i$ of the $i$th vertex is then updated as
\begin{equation}
p_i \to p_i + \lambda \xi_i
,
\end{equation}
where $\xi_i$ is a sample of some 2D distribution with zero mean and $\lambda$ is an overall scale factor that is used to control the resulting acceptance rate of the perturbation.
There is quite some freedom in choosing the distribution $\xi_i$, but typically heavy tailed distributions lead to a more efficient exploration of the search space than more localised distributions such as a Gaussian \cite{heavyTailedProposalIsBetter}.

Our perturbation distribution $\xi_i = \eta_i \xi$ is given by a 2D heavy tailed base distribution $\xi$ that gets scaled by a per-vertex factor $\eta_i$. The base distribution $\xi$ is generated by first sampling a point in the unit disk and then modulating its radius by a factor sampled from a heavy tailed Pareto distribution with pdf $\alpha/x^{\alpha+1}$ and $\alpha=1/2$ for $x>1$.
To make the perturbation size well-matched to the typical length scales in the local neighbourhood of a vertex, we use an additional per-vertex scale factor $\eta_i$ equal to the square root of the combined surface area of all triangles that share this vertex.

We have two variants of this simple vertex perturbation strategy that act on different numbers of vertices. In the first variant, we only perturb a single vertex at a time. This allows for a large perturbation scale $\lambda$ due to the limited impact on the overall triangulation, but it can fail to provide sensible perturbations with reasonable acceptance probabilities when vertex positions become highly correlated (for example when several are contracted into a tight cluster). A second variant therefore perturbs a fixed fraction (2\% in our case) of the movable vertices in one go. This demands a smaller overall scale $\lambda$, but it has the potential to move several correlated vertices more coherently. (Note: for the case of a tight cluster of contracted vertices, we discuss a more specialised perturbation in Sec.~\ref{secGroupPert})

\subsubsection{Perturbing a Group of Contracted Vertices \label{secGroupPert}}
The previous strategy of independently perturbing individual vertices tends to become inefficient when many vertices have been contracted into tight clusters, which requires small perturbation sizes $\lambda$. In such cases, it makes more sense to move the entire contracted cluster coherently like the single contracted vertex that it represents.

We therefore have a perturbation strategy that select a single `central' vertex with position $p_c$ and updates all vertices $p_i$ that are within a distance $d$ with the same perturbation $\lambda \xi_c$
\begin{equation}
    p_i \to p_i + \lambda \xi_c
    \text{\quad if \quad}
    \norm{p_i - p_c} \le d
    .
\end{equation}
Here, $\xi_c = \eta_c \xi$ is again composed of the same heavy tailed distribution $\xi$ as in the previous perturbation strategy
and the per-vertex factor $\eta_c$ now equals the square root of the combined surface area of all triangles that share \emph{any} vertex $p_i$ that is within a distance $d$ of $p_c$. The exact value of $d$ is not that critical and we choose it stochastically for each perturbation from a uniform distribution between $\eps/2$ and $5\eps$, with $\eps$ the characteristic contraction length of Section \ref{secIncludingTopologyChanges}.

Similar to the previous perturbation strategy, we have two variants, where the first one acts on a single vertex group, and the second one perturbs a fixed set of groups (2\% of the number of movable vertices in our case).

Note that to facilitate efficient range queries to find all affected $p_i$, we use a simple uniform 2D grid that holds the vertex positions and which is incrementally updated whenever vertices are moved. The grid enables finding a neighbour in amortized constant time.

\subsubsection{Resampling a Vertex Within its Neighbouring Triangles \label{secPertResampleInNeighb}}

\newcommand{\drawResampWithinTris}[1] {
\pgfmathsetmacro\hiddenEdgesLen{0.4}
\begin{tikzpicture}[scale=1.2]
    \coordinate (v1) at (-0.2,-0.13);
    \coordinate (v2) at (-0.19,0.3);
    \coordinate (v3) at (0.15,1.5);
    \coordinate (v4) at (2.2,1.6);
    \coordinate (v5) at (2.5,-1.2);
    \coordinate (v6) at (0.9,-1.7);
    \coordinate (v7) at (0.1,-0.9);
    \coordinate (v) at (0,0);
    
    \def\tris{(v)
    -- (v1) -- (v2) -- (v)
    -- (v2) -- (v3) -- (v)
    -- (v3) -- (v4) -- (v)
    -- (v4) -- (v5) -- (v)
    -- (v5) -- (v6) -- (v)
    -- (v6) -- (v7) -- (v)
    -- (v7) -- (v1) -- cycle}

    \def\thecirc{(v) circle [radius=1.1]}

    \ifnum#1=0
        \draw[fill=lightgray,densely dashed] \thecirc;
        \begin{scope}
        \begin{pgfinterruptboundingbox}
        \clip[invclip] \tris;
        \end{pgfinterruptboundingbox}
        \fill[pattern=north west lines, pattern color=darkgray!80] \thecirc;
        \end{scope}
    \else
        \fill[fill=lightgray] \tris;
    \fi

    \draw \tris;
    \foreach \i in {1,...,7} {\node[vertDot] at (v\i) {};}
    \node[vertDot,label={[anchor=west,xshift=2.3]$v$}] at (v) {};

    \draw[densely dotted] (v1) -- ++(178:\hiddenEdgesLen);

    \draw[densely dotted] (v2) -- ++(190:\hiddenEdgesLen);
    \draw[densely dotted] (v2) -- ++(125:\hiddenEdgesLen);

    \draw[densely dotted] (v3) -- ++(220:\hiddenEdgesLen);
    \draw[densely dotted] (v3) -- ++(55:\hiddenEdgesLen);
    \draw[densely dotted] (v3) -- ++(125:\hiddenEdgesLen);
    \draw[densely dotted] (v3) -- ++(105:\hiddenEdgesLen);

    \draw[densely dotted] (v4) -- ++(110:\hiddenEdgesLen);
    \draw[densely dotted] (v4) -- ++(-10:\hiddenEdgesLen);
    \draw[densely dotted] (v4) -- ++(-40:\hiddenEdgesLen);

    \draw[densely dotted] (v5) -- ++(30:\hiddenEdgesLen);
    \draw[densely dotted] (v5) -- ++(-50:\hiddenEdgesLen);

    \draw[densely dotted] (v7) -- ++(155:\hiddenEdgesLen);
    \draw[densely dotted] (v7) -- ++(-120:\hiddenEdgesLen);

    \draw[densely dotted] (v6) -- ++(-140:\hiddenEdgesLen);
    \draw[densely dotted] (v6) -- ++(-110:\hiddenEdgesLen);
    \draw[densely dotted] (v6) -- ++(-30:\hiddenEdgesLen);

\end{tikzpicture}
}

\begin{figure}
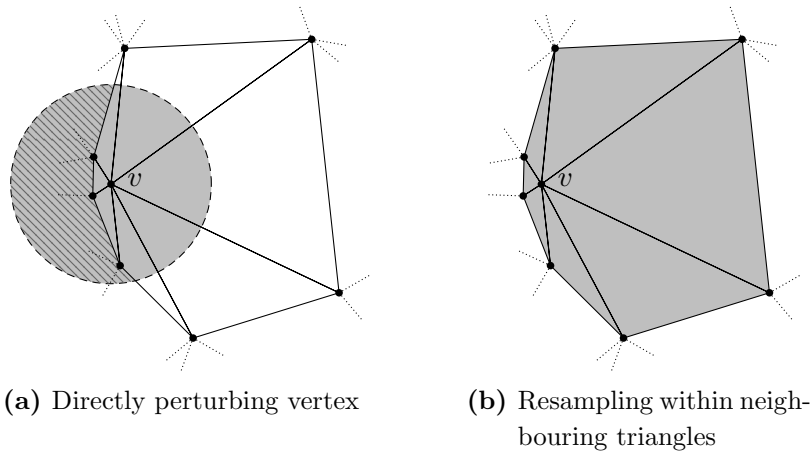

\begin{center}
    \subfloat[Directly perturbing vertex \label{subfigDirectPerturb}]
    {\drawResampWithinTris{0}}
    \hspace{10mm}
    \subfloat[Resampling within neighbouring triangles \label{subfigResampWithinTris}]
    {\drawResampWithinTris{1}}
\end{center}
\caption{
    A case where directly perturbing a vertex in a small (locally-scaled) neighbourhood can still lead to many rejected perturbations because the neighbouring triangles have inhomogeneous sizes.
    If $v$ were to be perturbed into the hatched region shown on the left in (a), the result would violate the topology and be rejected.
    If $v$ were instead moved within the triangles that share $v$ (gray region in (b)), then the topology is guaranteed to stay correct.
    Of course, the perturbation itself can be quite drastic for large neighbouring triangles and thus can still be rejected by the Metropolis-Hastings criterion.
    \label{figResampWithinTris}}
\end{figure}

The previous two perturbation strategies used a per-vertex scaling factor based on the size of the neighbouring triangles to adapt the perturbation size to the typical local length scales. One could alternatively directly resample a vertex's position somewhere within its neighbouring triangles, which also results in a locally adaptive sampling and which can better handle neighbouring triangles with strongly varying sizes, as illustrated in Fig.~\ref{figResampWithinTris}. As a bonus, when the neighbouring triangles of a vertex form a convex region, then the topology is guaranteed to remain valid when that vertex is moved within this region.

Our implementation simply resamples the position of a vertex within its neighbouring triangles with uniform surface area measure. The acceptance rate is controlled by resampling $\lambda$ vertices at once.

One small caveat here is that the behaviour of large and small (nearly-contracted) triangles during optimization is different. For example, if several triangles are contracted together in a cluster, then resampling a vertex within neighbouring triangles that are part of this cluster is very likely to be accepted. Therefore we use two variants of our strategy (each with their own $\lambda$ parameter) depending on whether the surface area of the neighbouring triangles is above or below a threshold (equal to $10\eps^2$ in our case).

\subsubsection{Swapping a Contracted Edge \label{secSwapContrEdge}}

\newcommand{\Ctop}{(-0.1, 1.1)}
\newcommand{\Cbot}{(0,-0.9)}
\newcommand{\Clef}{(-1.5,0)}
\newcommand{\Crig}{(1.5,0.3)}

\newcommand{\swapContrEdgeOne} {
\pgfmathsetmacro\vInteriorX{-1.2}
\pgfmathsetmacro\vInteriorY{-0.08}
\pgfmathsetmacro\vLabAngle{-1}
\begin{tikzpicture}[scale=1.4]
    \filldraw[fill=lightgray,densely dashed] \Ctop circle [radius=0.45];

    \node[vertDot=vtop,label={[yshift=-1]$v\nb$}] at \Ctop {};
    \node[vertDot=vbot] at \Cbot {};
    \node[vertDot=vlef] at \Clef {};
    \node[vertDot=vrig] at \Crig {};
    \node[vertDot=vint,label={[anchor=west,yshift=-4.0,xshift=5]$v$}] at (\vInteriorX,\vInteriorY) {};

    \draw (vtop) -- (vlef) -- (vbot) -- (vrig) -- (vtop);
    \draw (vtop) -- (vint);
    \draw[myContractedBlue,ultra thick] (vlef) -- (vint);
    \draw (vbot) -- (vint);
    \draw (vrig) -- (vint);
\end{tikzpicture}
}

\newcommand{\swapContrEdgeTwo} {
\pgfmathsetmacro\vInteriorX{-0.25}
\pgfmathsetmacro\vInteriorY{0.8}
\pgfmathsetmacro\vLabAngle{-95}
\begin{tikzpicture}[scale=1.4]
    \node[vertDot=vtop,label={[yshift=-1]$v\nb$}] at \Ctop {};
    \node[vertDot=vbot] at \Cbot {};
    \node[vertDot=vlef] at \Clef {};
    \node[vertDot=vrig] at \Crig {};
    \node[vertDot=vint,label={[anchor=north,yshift=-2.0,xshift=5.6]$v$}] at (\vInteriorX,\vInteriorY) {};

    \draw (vtop) -- (vlef) -- (vbot) -- (vrig) -- (vtop);
    \draw (vtop) -- (vint);
    \draw (vlef) -- (vint);
    \draw (vbot) -- (vint);
    \draw (vrig) -- (vint);
\end{tikzpicture}
}

\begin{figure}
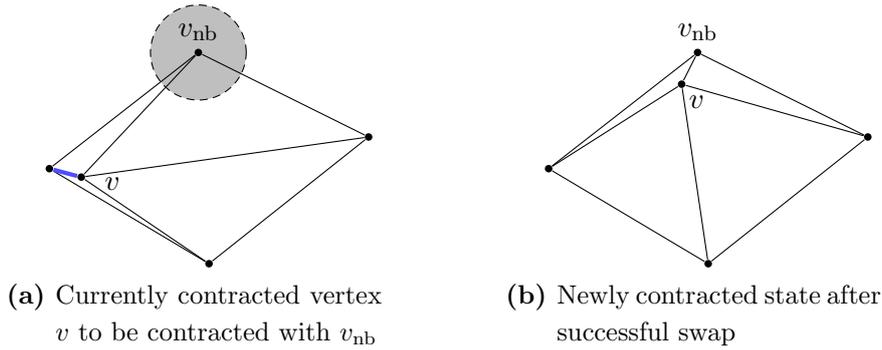

\begin{center}
    \subfloat[Currently contracted vertex $v$ to be contracted with $v\nb$]
    {\swapContrEdgeOne}
    \hspace{15mm}
    \subfloat[Newly contracted state after successful swap]
    {\swapContrEdgeTwo}
\end{center}
\caption{Example of swapping a contracted edge. The vertex $v$ is currently contracted (blue edge in (a) is shorter than the length threshold $l$) and a neighbour $v\nb$ has been sampled as a contraction target. The vertex $v$ is then resampled within a radius $l$ around $v\nb$ to produce a new contracted state (b).
\label{figSwapContrEdge}}
\end{figure}

In a low-temperature regime with many contracted vertex clusters, the Markov Chain dynamics tend to exhibit a critical slowing down where all vertices in a contracted vertex cluster remain glued together instead of `exchanging' vertices with neighbouring vertices (or vertex clusters).
Indeed, the small perturbations of Section \ref{secPertSingleVert} require several steps to first `uncontract', `move towards a new neighbour' and then `recontract'.
The increased edge length associated with the intermediate uncontracted configuration acts as a barrier and effectively prohibits such paths.
On the other hand, resampling vertices directly within the neighbouring triangles as per Sec.~\ref{secPertResampleInNeighb} can potentially skip such intermediate `uncontracted' configurations by immediately sampling a new position within a distance $\eps$ of a neighbouring vertex (that isn't part of the same cluster). However, as $\eps$ decreases the probability of sampling such a `contracted' position decreases as $\eps^2$ and becomes highly unlikely.

We therefore use a dedicated perturbation type that directly moves a vertex that is already contracted such that it becomes contracted with a different neighbouring vertex. This way, there is no edge length penalty associated with an `uncontracted' intermediary configuration.
Contracted vertices can thus more easily `hop' along the triangulation (within the constraints imposed by the topology)

The process is illustrated in Fig.~\ref{figSwapContrEdge}. Concretely, we sample a length threshold $l$ that is on the order of $\eps$ (we choose $l$ uniformly on $0.2\eps\ldots2\eps$) and sample a vertex $v$ that is part of a contracted edge (an edge whose length is below the threshold $l$). For this vertex $v$, we uniformly sample one of its neighbouring vertices $v\nb$ and sample a new location for $v$ uniformly in a disk with radius $l$ centered around $v\nb$. 
This simple strategy can certainly be improved, for example by taking into account topological constraints instead of simply sampling in a disk, or by only selecting neighbours that are themselves further away than $l$, but the current strategy is easy to implement, trivially symmetrical
and we found it to be sufficiently effective.

The acceptance rate is controlled by performing the perturbation for $\lambda$ vertices at once.

\subsubsection{Contracting to a Neighbour \label{secContrToNeighb}}
Similarly motivated by barriers due to intermediary uncontracted configurations as in the previous perturbation method, we use a perturbation strategy that directly tries to contract a given vertex to one of its neighbours (or, symmetrically, tries to uncontract a contracted vertex). In this context, a contracted vertex is a vertex that is part of at least one edge with length below $\eps$.

In practice, we use a rejection step on top of the technique of Sec.~\ref{secPertResampleInNeighb} to only allow transitions from a contracted to an uncontracted vertex, and vice versa. In all other cases (contracted to new contracted position, or uncontracted to new uncontracted position) we reject the attempt and keep the vertex at its original position. This explicit rejection in the proposed configuration 
focusses the perturbation on actual contraction changes (and thus effective topology changes) and avoids rejections by the Metropolis-Hastings criterion for changes that did not alter the effective topology (e.g.\ when resampling an uncontracted vertex to a new uncontracted position).

Note that --- unlike traditional rejection sampling --- we only `try once' and hence the resulting pdf for the vertex position in the case of acceptance is $1/A\nb$ with $A\nb$ the surface area of the neighbouring triangles, regardless of whether we went from contracted to uncontracted or the other way around (in the case of rejection the pdf is a Dirac delta on the original position, weighted with a factor $A\acc/A\nb$ where $A\acc$ is the surface area of the regions that would have yielded acceptance had we sampled the new vertex position there).
This is in contrast with traditional rejection sampling which keeps trying until a sample is accepted, thereby renormalizing the effective pdf to be constrained to the valid region.

The acceptance rate of the perturbation is controlled by attempting to (un)contract $\lambda$ vertices at once.

\subsubsection{Explicit Topology Perturbation with Edge Flipping \label{secEdgeFlipping}}

\newcommand{\edgeFlipOne} {
\begin{tikzpicture}[scale=1.4]
    \node[vertDot=vtop] at \Ctop {};
    \node[vertDot=vbot] at \Cbot {};
    \node[vertDot=vlef] at \Clef {};
    \node[vertDot=vrig] at \Crig {};

    \draw (vtop) -- (vlef) -- (vbot) -- (vrig) -- (vtop);
    \draw (vlef) -- (vrig);
\end{tikzpicture}
}

\newcommand{\edgeFlipTwo} {
\begin{tikzpicture}[scale=1.4]
    \node[vertDot=vtop] at \Ctop {};
    \node[vertDot=vbot] at \Cbot {};
    \node[vertDot=vlef] at \Clef {};
    \node[vertDot=vrig] at \Crig {};

    \draw (vtop) -- (vlef) -- (vbot) -- (vrig) -- (vtop);
    \draw (vtop) -- (vbot);
\end{tikzpicture}
}

\newcommand{\edgeFlipNotStrictConvex} {
\begin{tikzpicture}[scale=1.4]
    \node[vertDot=vtop] at (0.2,1.1) {};
    \node[vertDot=vlef] at (-1.5,0) {};
    \node[vertDot=vrig] at (1.5,0) {};
    \node[vertDot=vmid] at (0,0) {};

    \draw (vlef) -- (vtop) -- (vrig);
    \draw[ultra thick] (vlef) -- (vmid) -- (vrig);
    \draw[ultra thick,myHighlightedGreen] (vmid) -- (vtop);
\end{tikzpicture}
}

\begin{figure}
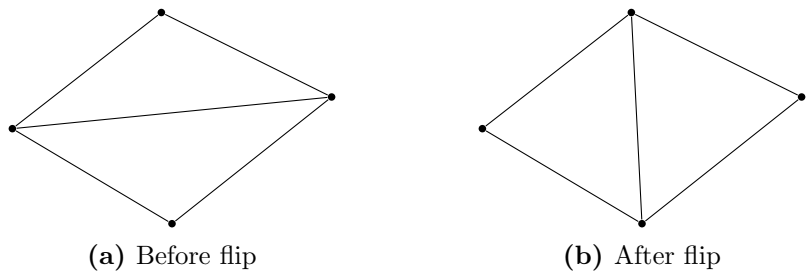

\begin{center}
    \subfloat[Before flip]
    {\edgeFlipOne}
    \hspace{15mm}
    \subfloat[After flip]
    {\edgeFlipTwo}
\end{center}
\caption{Example of directly modifying the topology by flipping an edge.
\label{figEdgeFlip}}
\end{figure}

All perturbation strategies that we discussed so far only act on vertex positions, hence they can only change the \emph{effective} topology (the topology obtained when merging each cluster of contracted vertices to a single vertex). We use one final perturbation strategy that changes the \emph{actual} topology of the triangulation by flipping the shared edge of two neighbouring triangles as in figure \ref{figEdgeFlip}.
These explicit topology changes can be thought of as perturbations in an additional discrete configuration space that is added to the continuous state space of vertex positions on which the previous perturbation strategies operate.
Alternatively, if only the space of vertex positions is regarded as the actual state space, then these topology changes can be viewed as changes in the energy function (or Hamiltonian) of the system, similar to Hamiltonian Replica Exchange methods \cite{sugita2000multidimensional, fukunishi2002hamiltonian}.

\begin{figure}
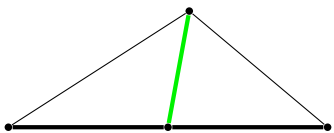

\begin{center}
    \edgeFlipNotStrictConvex
\end{center}
\caption{An example where flipping the green edge would give an invalid state with a zero-area squished triangle. The bottom two edges (thick black) are part of the original geometry and thus their direction is constrained. The 4-gon formed by the two triangles is not \emph{strictly} convex and the edge flipping perturbation will not be attempted.
\label{figEdgeFlipNotStrictConvex}}
\end{figure}

In order to keep the topology valid, edge flips are only allowed if the 4-gon formed by the two neighbouring triangles is strictly convex. This assures that the flipped edge is still internal to the triangles. The demand for \emph{strict} convexity comes into play when one of the vertices of the edge-to-be-flipped is part of the original geometry, for example when it is a constrained vertex that subdivides an original edge and those sub-edges are part of the two neighbouring triangles of the edge-to-be-flipped, as illustrated in Fig.~\ref{figEdgeFlipNotStrictConvex}. In that case, flipping the edge would force a zero-area triangle with all edges overlapping the edge of the original geometry. 

The acceptance rate of the perturbation is controlled by attempting to flip $\lambda$ edges at once.

\subsection{Details of the Annealing Process \label{secDetailsOfAnnealing}}

The idea behind the simulated annealing process is to start a Markov Chain in a regime where all states are roughly equally probable and then slowly tweak the probability distribution such that states with low energy (low objective function) become ever more probable.
In practice, the probability distribution is typically taken to be the Boltzmann distribution of Eq.~\eqref{eqBoltzmann}, which gets tweaked by starting from a high simulated temperature $T$ and gradually lowering $T$ to favour low-energy states. The actual global minimum of the energy $E$ is then found in the limit $T \to 0$ (for $T$ decreasing sufficiently slowly).

\subsubsection{Annealing both $T$ and $\eps$}

When sampling states acconding tho the Boltzmann distribution of Eq.~\eqref{eqBoltzmann}, the high temperature case is usually easy to sample (approximately), as all states have roughly equal probability. The low temperature case, on the other hand, is difficult to (approximately) sample directly, because the (local) minima of the energy $E$ will be surrounded by steep barriers and the location of these minima is not known a priori --- indeed, finding the minimum is precisely the goal of running simulated annealing.

In the case of our objective function with fuzzy topology, we have a parameter that behaves somewhat similarly to temperature: the contraction length $\eps$. For values of $\eps$ much larger than the typical edge length, most edges are effectively `contracted' with quasi-constant contraction weight [Eq.~\eqref{eqCeps}] $\ceps \approx 1/2$. On the other hand, as $\eps \to 0$, the transition of $\ceps$ from 1/2 ($\norm{e}\ll\eps$) to 1 ($\norm{e}\ge\eps$) happens very quickly and leads to steep valleys and barriers in the objective function. Furthermore, analogous to temperature, the result that we are after manifests itself in the limit $\eps \to 0$, where the fuzzy contraction behaves as a true pointlike contraction.

Given the similarity between $T$ and $\eps$, it seems advantageous to start the annealing process (high $T$) at a value of $\eps$ that is larger than the target value for the low-$T$ endpoint, or to have access to larger $\eps$ values during the simulation.
There are several ways to achieve this goal.
One possibility is to simply decrease the contraction length $\eps$ in step with the decrease of temperature during annealing.
More sophisticated methods can borrow from parallel tempering \cite{earl2005parallel} or hamiltonian exchange methods \cite{fukunishi2002hamiltonian} where multiple chains can be run in parallel and states can be exchanged between chains. One could, for instance, run multiple chains at varying $\eps$ values but equal (and decreasing) $T$, or fix $\eps$ to the target value and run multiple chains with varying $T$, or run a `2D grid' of chains with both varying $\eps$ and $T$ values.

Of all these possibilities, we found that the simplest method (simply decreasing $\eps$ alongside $T$ in a traditional simulated annealing approach) gave the best results when compared to the computational overhead of the more elaborate methods. This is what we use throughout this work.

\subsubsection{Further Implementation Details}

The actual cooling schedule that we used follows a simple exponential decrease in both $T$ and $\eps$. The initial value for $T$ is chosen such that there is a high degree of mobility in the vertices, and its value can be inspired by the typical length scales of the largest features in the scene. We use triangulations that are normalized to the unit square and choose an initial $T$ value of 0.02 for coarser scenes with longer edges and the slightly lower vale of 0.01 for the more detailled scenes such as the biggest real-world scenes. The initial contraction length $\eps$ can similarly be inspired by the typical length scales of the scene, and we choose it around 0.05 for scenes with large open spaces between the geometry (such that vertices in such openings are sure to `find' each other and can become contracted if that is beneficial) up to 0.01 for highly detailed scenes where primitives are mostly close together with few large openings. The final values for $T$ and $\eps$ can be chosen based on the size of the smallest features in the geometry. Certainly we would like $\eps$ to be smaller that the finest detail in the geometry so we can resolve and differentiate between small scale detail of the geometry and deliberate fuzzy contraction --- which should then happen at an even smaller scale. We use a final $\eps$ value as $10^{-4}$, which can resolve the finest detail in the overwelming majority of scenes. Similarly, the final temperature is chosen as $10^{-4}$, or $10^{-5}$ for the most detailed real-world floorplan scenes.

As discussed in Sec.~\ref{secPerturbations}, when computing the acceptance probability for a perturbed state, we only compute the terms of the objective function that have actually changed. For the perturbations that move a group of vertices at once (Sec.~\ref{secGroupPert}), we use a uniform grid of vertex positions for efficient neighbour queries that is incrementally updated whenever vertices are moved.

Furthermore, the different perturbation strategies get chosen with a weight such that each strategy receives an equal computation time. This way, `cheap' perturbation strategies can run for many more iterations than `expensive' ones (and thus sample the chain more effectively) with comparatively minor impact on the total computational time.

Lastly, we parallelize the sampling of the Markov chain over all available cores (ranging from 4 to 6 on current consumer hardware) following the adaptive resampling method of Lou and Reinitz \cite{lou2016parallel}.

\subsection{Contraction of Optimized Fuzzy Topology \label{secContractionOfOptimizedTopology}}
After having optimized the triangulation with our objective function, we have a triangulation with fuzzy topology where edges with lengths near or below the contraction length $\eps$ effectively represent single vertices. We now want to turn this fuzzy topology into a true topology, by contracting those short edges.

\subsubsection{Contracting an Edge}
There are several cases to consider when contracting an edge, depending on the degrees of freedom of its vertices.
For an edge where both vertices are free (2 DoF) or both are constrained on the same original-geometry edge (1 DoF), the resulting contracted vertex has the same (nonzero) degrees of freedom and can simply be positioned at the middle of the edge.
If one vertex is a vertex of the original geometry (0 DoF), then this becomes the resulting vertex after contraction (edges with two original geometry vertices cannot be contracted, cf.\ Sec.~\ref{secContractibleEdges}).
Similarly, an edge with a constrained (1 DoF) and a free (2 DoF) vertex gets contracted to its constrained vertex.

In case a simple contraction (e.g.\ to the middle of an edge) leads to a topology violation (see also the discussion in Sec.~\ref{secContractibleEdges}), the remaining degrees of freedom of the contracted vertex can sometimes be used to avoid this. If the contracted vertex is a free vertex, we try to move it to several positions along the original uncontracted edge to see if one of those positions leads to a valid topology. Similarly, when the contracted vertex is a constrained vertex, we try placing it at several positions between the two neighbouring vertices along its original-geometry edge. If no suitable position was found, the edge is flagged as (temporarily) incontractible and we move on to the next (contractible) edge that is shorter than $\eps$.
Occasionally, we re-try to contract the edges that were flagged as incontractible, as their topological restrictions typically get less problematic when neighbouring vertices get contracted: think, for example, of a tight fuzzily-contracted vertex cluster versus a single vertex.

\subsubsection{Successive Contraction and Polishing \label{secSuccessiveContractionAndPolishing}}
One could simply contract all edges with lengths below $\eps$ in one go in order to turn the fuzzy topology into a true topology, but this often leaves some possible optimization on the table.
Instead, we use an iterative method depicted in Fig.~\ref{figOptimAndContrDiagram} which we now discuss. 

\tikzstyle{block}    = [rectangle, draw, text width=5.7em, text centered, rounded corners, minimum height=3.5em]
\tikzstyle{fatblock} = [rectangle, draw, text width=6.5em, text centered, rounded corners, minimum height=3.5em]
\tikzstyle{line} = [draw, -Stealth]

\newcommand{\drawOptimAndContrDiagram}{
\begin{tikzpicture}[node distance=0.5cm and 0.9cm, auto]
    \node (initTriang) {$\tau_0$};
    \node [block, right=of initTriang] (mainOptim) {Main\\Optimization};
    \node [block, right=of mainOptim] (contr1) {Contract\\\& Polish};
    \node [block, right=of contr1] (contr2) {Contract\\\& Polish};
    \node [right=of contr2] (dots) {\ldots};

    \node [block, below=of contr1] (flip1)  {Flip \&\\Polish};
    \node [block, below=of contr2] (flip2)  {Flip \&\\Polish};
    \node [below=of flip1] (triang1) {$\tau_1$};
    \node [below=of flip2] (triang2) {$\tau_2$};

    \path [line] (initTriang) -- (mainOptim);
    \path [line] (mainOptim) -- (contr1);
    \path [line] (flip1) -- +(1.62,0) |- node [midway,above] {\ $\eps\uparrow$} (contr2);
    \path [line] (flip2) -- +(1.62,0) |- node [midway,above] {\ $\eps\uparrow$} (dots);

    \path [line] (contr1) -- (flip1);
    \path [line] (contr2) -- (flip2);

    \path [line] (flip1) -- (triang1);
    \path [line] (flip2) -- (triang2);
\end{tikzpicture}
}

\usetikzlibrary{fit,positioning}
\usetikzlibrary{decorations.pathreplacing}

\newcommand{\drawFullPipelineDiagram}{
\begin{tikzpicture}[node distance=0.5cm and 0.9cm, auto]

    \node (initTriang) {$\tau_0$};
    \node [block, right=of initTriang] (mainOptim) {Main\\Optimization};
    \node [block, right=of mainOptim] (contr1) {Contract\\\& Polish};
    \node [block, right=of contr1] (contr2) {Contract\\\& Polish};
    \node [right=of contr2] (topdots) {\ldots};

    \node [block, below=of contr1] (flip1)  {Flip \&\\Polish};
    \node [block, below=of contr2] (flip2)  {Flip \&\\Polish};
    \node [below=of flip1] (triang1) {$\tau_1$};
    \node [below=of flip2] (triang2) {$\tau_2$};
    \node (triangdots) at (triang2 -| topdots) {$\ldots$};

    \path [line] (initTriang) -- (mainOptim);
    \path [line] (mainOptim) -- (contr1);
    \path [line] (flip1) -- +(1.62,0) |- node [midway,above] {\ $\eps\uparrow$} (contr2);
    \path [line] (flip2) -- +(1.62,0) |- node [midway,above] {\ $\eps\uparrow$} (dots);

    \path [line] (contr1) -- (flip1);
    \path [line] (contr2) -- (flip2);

    \path [line] (flip1) -- (triang1);
    \path [line] (flip2) -- (triang2);

    \draw [decorate,decoration={brace,amplitude=10pt,mirror}]
        ($(triang1.west)+(0,-0.2)$) -- ($(triangdots.east)+(0,-0.2)$)
        node[pos=0.5,below=10pt] (triangmin) {$\tau_\mathrm{min}$};

    \node (innerOptimisation) [
        fit=(initTriang) (mainOptim) (contr1) (contr2) (topdots) (triang1) (triang2) (triangmin),
        yshift=1mm, inner ysep=2mm,inner xsep=1mm,
        draw, rounded corners,
        label={[anchor=south west]south west:Optimization \& Contraction}
        ] {};

    \node [fatblock, above=of initTriang, yshift=8.5mm] (initCDT) {Optimally\\Refined CDT};

    \node [fatblock, right=of initCDT] (subdiv) {Subdivide};

    \node [above=of initCDT] (inputGeom) {Input geometry};

    \path [line] ($(initCDT.south)+(-0.09,0)$) -- ($(initTriang.north)+(-0.09,0)$);
    \path [line, dashed] (subdiv.south) -- +(0,-0.30) -| ($(initTriang.north)+(0.09,0)$);

    \path [line] (inputGeom) -- (initCDT);

    \node [below=of triangmin] (outputTriang) {Output triangulation};
    \path [line] (triangmin) -- (outputTriang);

    \path [line, dashed] (outputTriang) |- +(4,0) |- (subdiv);
\end{tikzpicture}
}

\begin{figure}[htb]
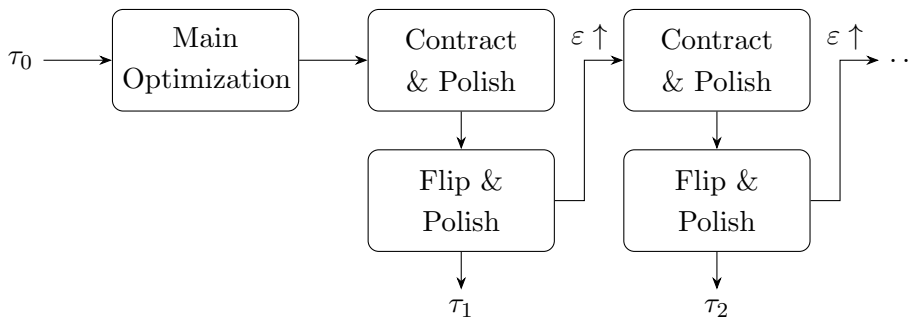

\begin{center}
\drawOptimAndContrDiagram
\end{center}
    \caption{
Optimizing and contracting an initial triangulation $\tau_0$. After the main optimization with fuzzy contraction length scale $\eps$, edges that are shorter than $\eps$ get contracted and the resulting triangulation gets polished.
A greedy edge flipping step tries to further improve the triangulation, which gets alternated with a polishing step without fuzzy contraction to give a triangulation $\tau_1$. The contraction length $\eps$ is then increased, the triangulation is polished and the same overall steps are repeated to obtain triangulations $\tau_2$, $\tau_3$, etc.\ until the resulting total edge length starts to increase. The returned triangulation is the triangulation $\tau_i$ with the smallest total edge length.
\label{figOptimAndContrDiagram}
}
\end{figure}

From the resulting triangulation from the main optimization, a `Contract \& Polish' step first contracts all edges smaller than $\eps$ in increasing order of edge length.
The contracted triangulation is then `polished' by a short simulated annealing run that already starts at a low temperature, after which edges smaller than $\eps$ are contracted again should they have re-appeared. This process of polishing and contracting can be repeated until no significant further progress is made or no edges shorter than $\eps$ show up.

The contracted triangulation may further be improved by greedily flipping (cf.\ Sec.~\ref{secEdgeFlipping}) shared edges of adjacent triangles if the flipped edge is shorter\footnote{Ideally, the edge flip perturbations of the previous polishing step from `Contract \& Polish' should have already optimized those edges such that no such edges remain. However, one extra sweep over all edges is not costly and can catch some remaining edges, although such a greedy flipping approach by itself can typically only reach a local optimum \cite{bern1992mesh}.}. This deterministic flipping step is alternated by a polishing step until no further progress is made. The polishing step has fuzzy contraction disabled by forcing $\wcontr=1$ and the contraction-based perturbation types (Sec.\ \ref{secGroupPert}, \ref{secSwapContrEdge} and \ref{secContrToNeighb}) are not used.

The total edge length of the resulting triangulation is computed and the triangulation is saved if it is the best so far. If this edge length is larger than some multiple of the best length seen so far (we use a factor of $1.1$) we stop and return the best triangulation, otherwise we increase $\eps$ (say, by 30\%) and start the process of polishing and contracting again.

Due to the contraction steps in the `Contract \& Polish' iterations, each iteration deals with less degrees of freedom and is thus faster to compute and more easy to optimize in the subsequent polishing step due to the lower dimensionality. This is one of the reasons for using such an elaborate, iterated contraction scheme that performs several polishing optimizations. One could leave out the polishing and allocate the freed-up computational resources to the main simulated annealing run at the start followed by a simple contraction of all edges shorter than $\eps$, but this would require disproportionally more computational resources to get a similar result due to the dimensionality.

\subsection{Iterated Optimization with Intermediate Subdivision: The Full Strategy\label{secIteratedoptimization}}

In the previous sections, we have seen the simulated annealing process that we use (Secs.~\ref{secSimulatedAnnealing} and \ref{secDetailsOfAnnealing}), its perturbation strategies (Sec.~\ref{secPerturbations}) and the contraction from fuzzy topology to a final triangulation (Sec.~\ref{secContractionOfOptimizedTopology}).
We will now discuss how we obtain an initial triangulation for this optimization process, and how we iterate it with intermediate subdivision to increase the topological degrees of freedom and obtain a better triangulation.

\subsubsection{Obtaining an Initial Triangulation \label{secInitialTriangulation}}
In order to optimize a triangulation according to our objective function, we need some kind of triangulation as a starting point. This section describes how we obtain such an initial triangulation.

One standard way to triangulate a set of \emph{points} in the plane is by using a Delaunay triangulation, which has the property that no point lies inside the circumcircle of any triangle \cite{toth2017handbook}. Such triangulations maximize the minimum angle of their triangles, which leads to maximally `regular' triangles.
Delaunay-type triangulations can also be lifted from points to `constrained' triangulations with predefined \emph{edges} as input, which fit our purpose.

More specifically, we start from a quality constrained Delaunay triangulation (CDT) \cite{shewchuk2002delaunay}, where the `quality' refers to the regularity of the triangles, achieved by refining the triangulation by introducing new vertices until a lower bound on the minimum angle of the triangles is met. Additionally, an upper bound on triangle area can be enforced. 
We use the Triangle utility by Shewchuk \cite{shewchuk1996triangle} and use the minimum-angle lower bound and triangle area upper bound combination that gives the lowest weight triangulation (i.e.\ the triangulation with the shortest combined length of all edges).
This starting point can be viewed as a good proxy for the lowest-weight triangulation that can be achieved when sticking to Delaunay-type triangulations.

\subsubsection{Subdividing to Increase Topological Degrees of Freedom \label{secSubdividing}}
Because our objective function [Eq.~\eqref{eqObjFunc}] allows for a fuzzy contraction of edges, it is useful to subdivide an initial triangulation to allow for more topological freedom during the optimization. We use a simple subdivision strategy that turns each triangle into four triangles by splitting each edge into two pieces as shown in Fig.~\ref{figSubdiv}.

\newcommand{\drawSubdiv}[1]{
\pgfmathsetmacro\hiddenEdgesLen{0.4}
\begin{tikzpicture}[scale=1.2]
    \coordinate (v1) at (-0.2,1.1);
    \coordinate (v2) at (-1.3,0);
    \coordinate (v3) at (1,0.2);

    \draw (v1) -- (v2) -- (v3) -- cycle;

    \foreach \i in {1,...,3} {\node[vertDot] at (v\i) {};}

    \draw[densely dotted] (v1) -- ++(158:\hiddenEdgesLen);
    \draw[densely dotted] (v1) -- ++(48:\hiddenEdgesLen);

    \draw[densely dotted] (v2) -- ++(-60:\hiddenEdgesLen);
    \draw[densely dotted] (v2) -- ++(-150:\hiddenEdgesLen);
    \draw[densely dotted] (v2) -- ++(108:\hiddenEdgesLen);

    \draw[densely dotted] (v3) -- ++(240:\hiddenEdgesLen);
    \draw[densely dotted] (v3) -- ++(75:\hiddenEdgesLen);
    \draw[densely dotted] (v3) -- ++(12:\hiddenEdgesLen);

    \ifnum#1=1
        \coordinate (v12) at ($(v1)!0.5!(v2)$);
        \coordinate (v23) at ($(v2)!0.5!(v3)$);
        \coordinate (v13) at ($(v1)!0.5!(v3)$);

        \node[vertDot] at (v12) {};
        \node[vertDot] at (v23) {};
        \node[vertDot] at (v13) {};

        \draw (v12)--(v23)--(v13)--cycle;

        \draw[densely dotted] (v12) -- ++(110:\hiddenEdgesLen);
        \draw[densely dotted] (v12) -- ++(170:\hiddenEdgesLen);

        \draw[densely dotted] (v23) -- ++(-62:\hiddenEdgesLen);
        \draw[densely dotted] (v23) -- ++(-120:\hiddenEdgesLen);

        \draw[densely dotted] (v13) -- ++(80:\hiddenEdgesLen);
        \draw[densely dotted] (v13) -- ++(25:\hiddenEdgesLen);
    \fi

\end{tikzpicture}
}

\begin{figure}
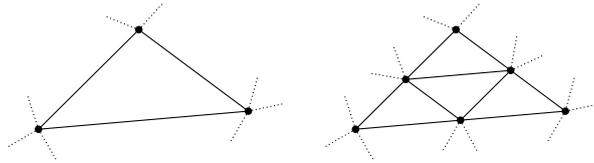

\begin{center}
    \drawSubdiv{0}
    \drawSubdiv{1}
\end{center}
\caption{
    A simple subdivision strategy that splits every triangle into four subtriangles. The additional vertices give extra topological degrees of freedom, especially when combined with our fuzzy contraction.
    \label{figSubdiv}
}
\end{figure}

This subdivision step can optionally be repeated to yield even more adaptability of the topology through fuzzy contraction, but the corresponding steep increase in vertex count makes it more difficult for the simulated Markov Chains to reach equilibrium on the resulting high-dimensional space and thus hampers the optimization. Instead, an iterative approach is more practical.

\subsubsection{The Full Pipeline}
\begin{figure}[htb]
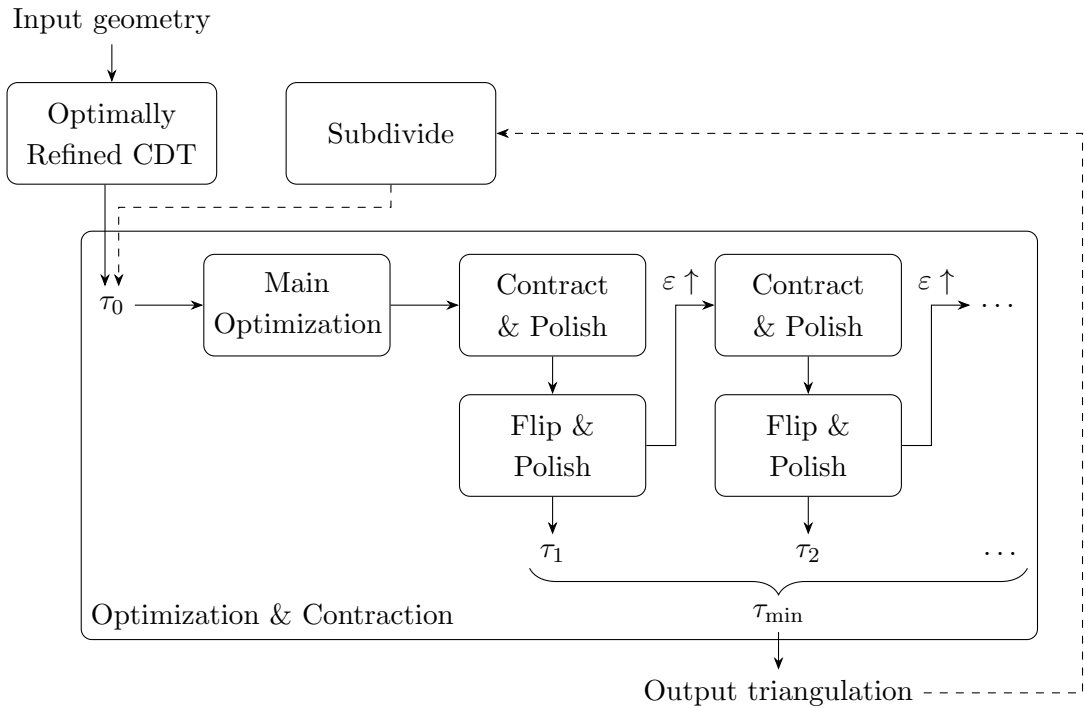

\begin{center}
\drawFullPipelineDiagram
\end{center}
    \caption{
Overview of the full pipeline from input geometry to approximately minimal weight output triangulation. The dashed line represents the optional iterative path with intermediate subdivisions to increase the topological degrees of freedom. The `Optimization \& Contraction' block is identical to Fig.~\ref{figOptimAndContrDiagram}.
\label{figFullPipelineDiagram}
}
\end{figure}

The full, iterative approach to obtain an approximate minimum-weight triangulation from a given scene (i.e.\ a collection of line segments) is sketched in Fig.~\ref{figFullPipelineDiagram}.

In a first step, the optimally refined CDT is run through the optimization and contraction procedure without any prior subdivision. The goal is to immediately cull spurious vertices that may have appeared due to the conflicting goals of a refined CDT versus a minimum weight triangulation, and already try to achieve a reasonably good triangulation with the topological degrees of freedom that are available. Afterwards, the resulting optimized triangulation can optionally be subdivided and run through the full optimization process again to further improve areas that could use more vertices than were available in the initial optimally refined CDT.

Apart from specially handcrafted scenes, we found that one such intermediate subdivision followed by a second full optimization yielded results that were very close to what could be achieved with this method for typical scenes. A second subdivision followed by a third full optimization usually yielded highly diminishing returns, as it seems that at that point the limitations is not so much that there are insufficient degrees of freedom to represent a more optimal topology, but instead that the high dimensionality hampers the optimization process.

\FloatBarrier

\section{Optimization Results \label{secOptimResults}}

This section discusses the results of our pipeline for generating approximately minimum weight triangulations, i.e.\ triangulations that minimize their total edge length.

\subsection{Illustrative Example: the Trouble with Maximising Minimal Angle
\label{secIllustrativeExample}}
Let us start by looking at an illustrative example in Fig.~\ref{figMinAngleContrOneTwoLines} to see why a Delaunay-type refinement (which maximises the minimal angle) can lead to triangulations with many spurious edges and thus suboptimal total edge length.
Directly building a CDT using only the original vertices (top row) yields many narrow triangles and results in a large total edge length. When the CDT is refined to increase the minimal angle (middle row), those long triangles get split and the resulting total edge length decreases. However, when there are two long parallel segments in the original geometry (middle row, right column), the two long and slender triangles in between these segments also get split due to their sharp angles and the total edge length increases markedly. The bottom row shows our optimized triangulation as an approximation of the ideal, minimum-weight triangulation. This optimal triangulation shows no notable structural differences between the one and two line case, as expected. 

Our optimized triangulations in Fig.~\ref{figMinAngleContrOneTwoLines} are the result of the full pipeline of Sec.~\ref{secOptimOfObjFunc} as visualized in Fig.~\ref{figFullPipelineDiagram}, which include topology changes through fuzzy contraction. In Fig.~\ref{figMinAngleContrPolishOfOneLine} we show results for the scene with a single top line when starting from the optimal refined CDT, and only optimizing the vertex positions but keeping the topology fixed (left) or optimizing the vertex positions and allowing some topology changes through edge flipping (right). These simpler optimization strategies are less computationally expensive than a full optimization over (fuzzily) contacted states, but compared to the fully optimized triangulation (Fig.~\ref{figMinAngleContrOneTwoLines}, bottom left) they leave some further optimization potential on the table. 

\newcommand{\exampleWithCrop}[1]{
    \includegraphics[width=0.31\textwidth,align=c]{#1.pdf}
    &
    \begin{tabular}{@{}c@{}}
    \includegraphics[viewport={11.3cm 7.2cm 13.1cm 8.6cm},clip,width=0.13\textwidth,align=c]{#1_thin.pdf}
    \\[8.5mm]
    \includegraphics[viewport={11.3cm 4.1cm 13.1cm 5.5cm},clip,width=0.13\textwidth,align=c]{#1_thin.pdf}
    \end{tabular}
}

\begin{figure}
\begin{center}
\begin{tabular}{c@{\ } c@{\,}c @{\ \ } c@{\,}c}
& Single top line & & Double top line & \\
\rotatebox[origin=c]{90}{\small Unrefined CDT}
&
    \exampleWithCrop{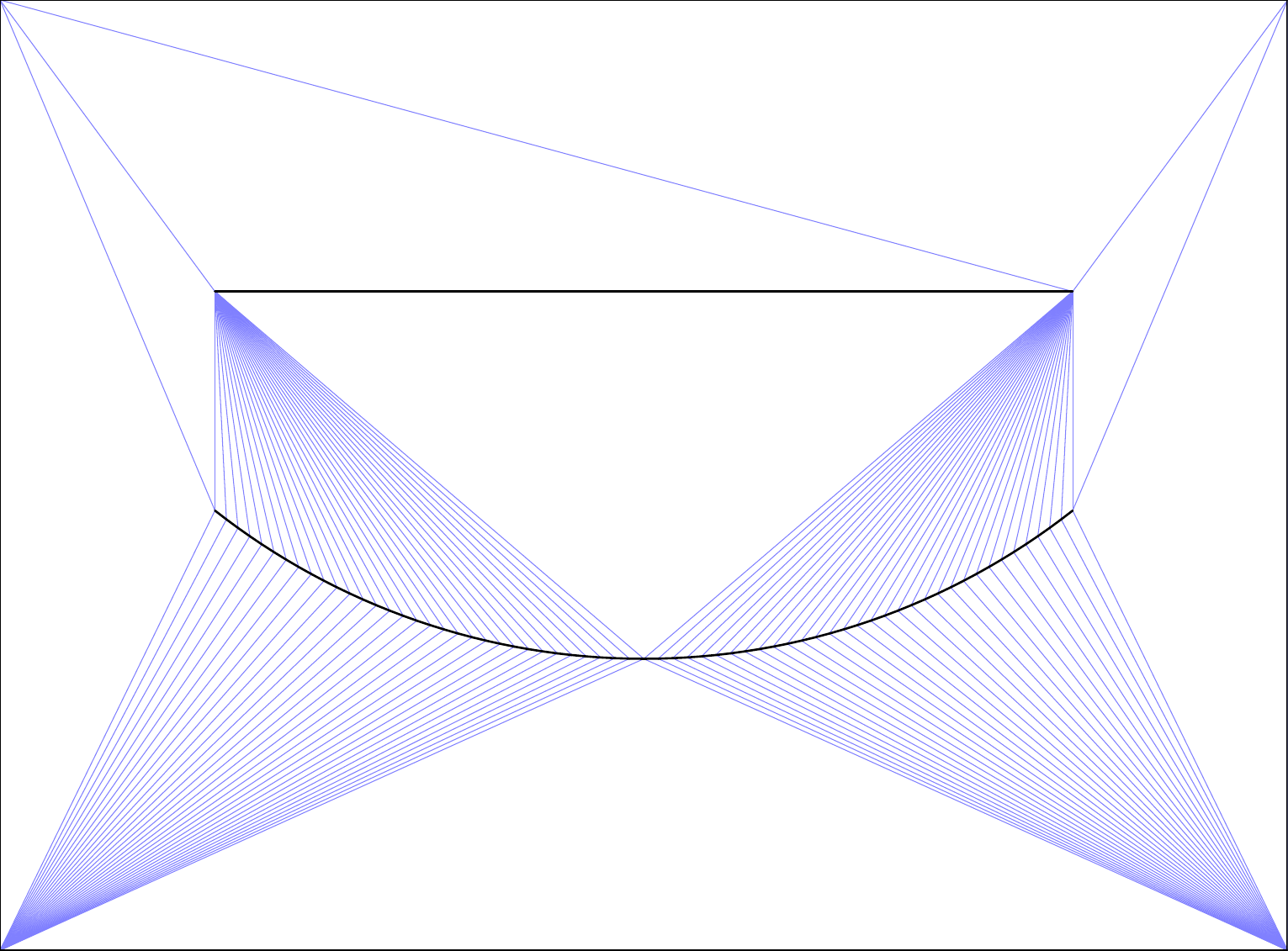}
    &
    \exampleWithCrop{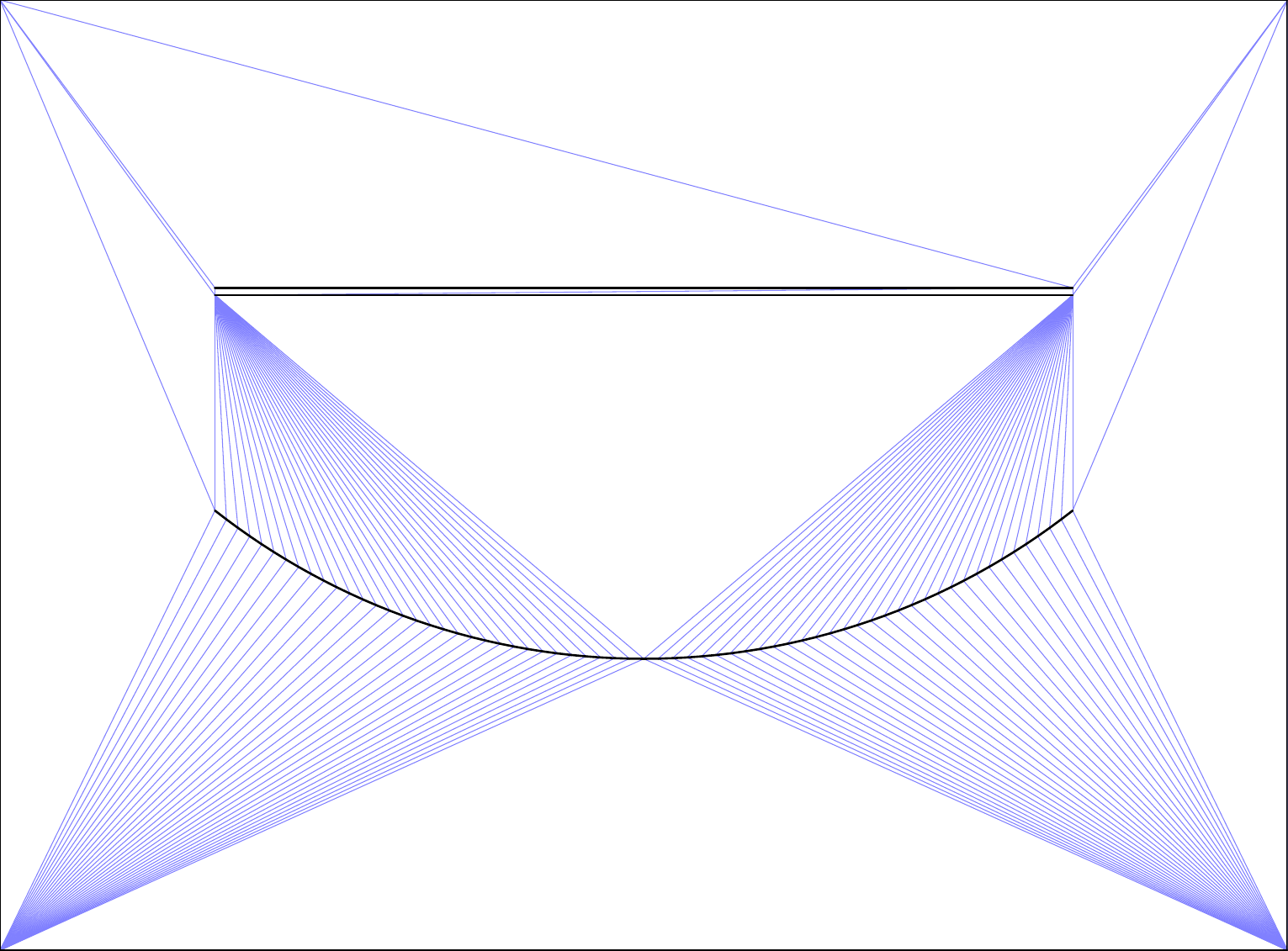}
    \\[-0.5mm]
    & \small 55.5 & & \small 57.3 \\[1mm]
\rotatebox[origin=c]{90}{\small Optim.\ refined CDT}
&
    \exampleWithCrop{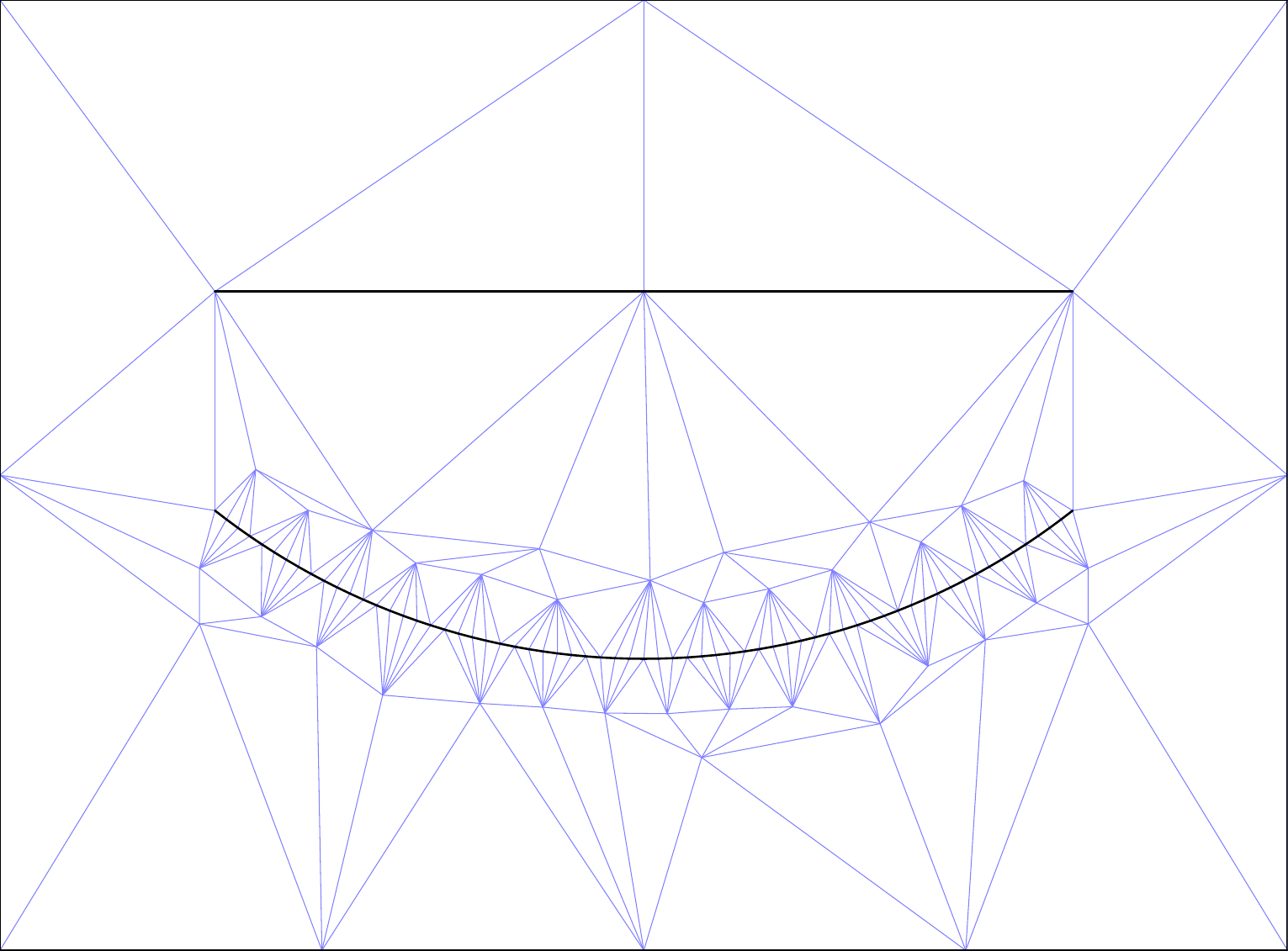}
    &
    \exampleWithCrop{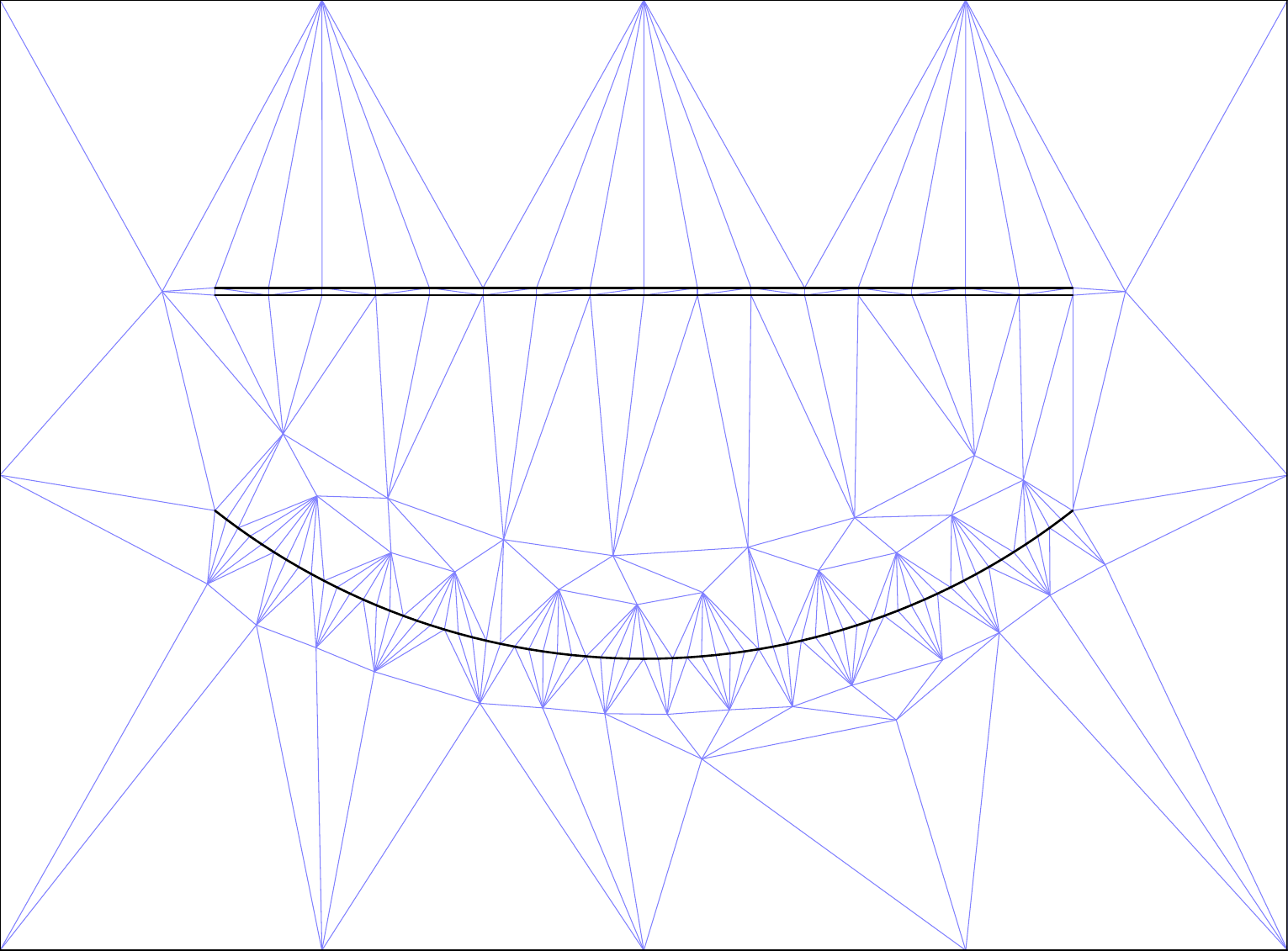}
    \\[-0.5mm]
    & \small 24.9 & & \small 33.2 \\[1mm]
\rotatebox[origin=c]{90}{\small Optimized (ours)}
&
    \exampleWithCrop{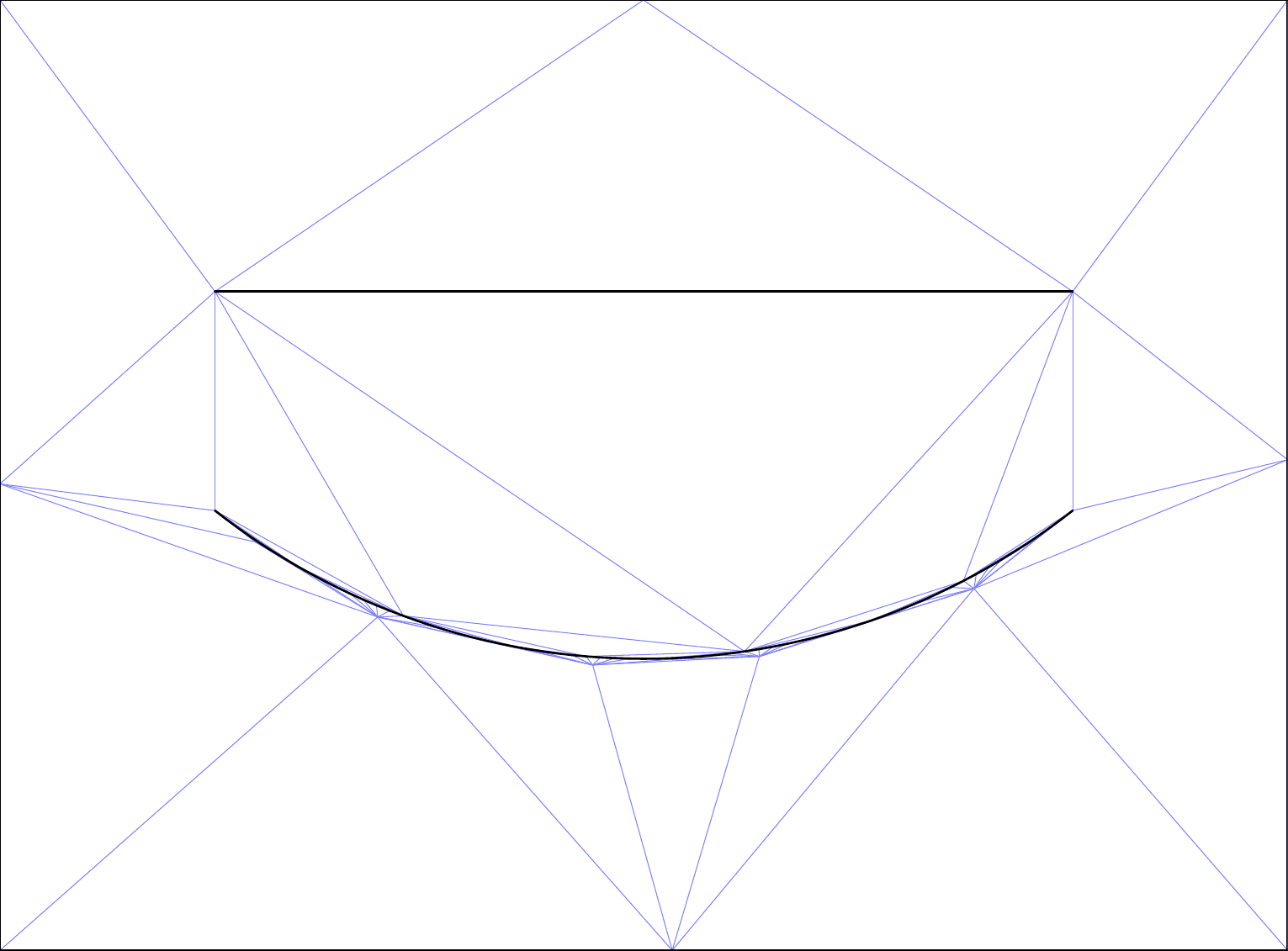}
    &
    \exampleWithCrop{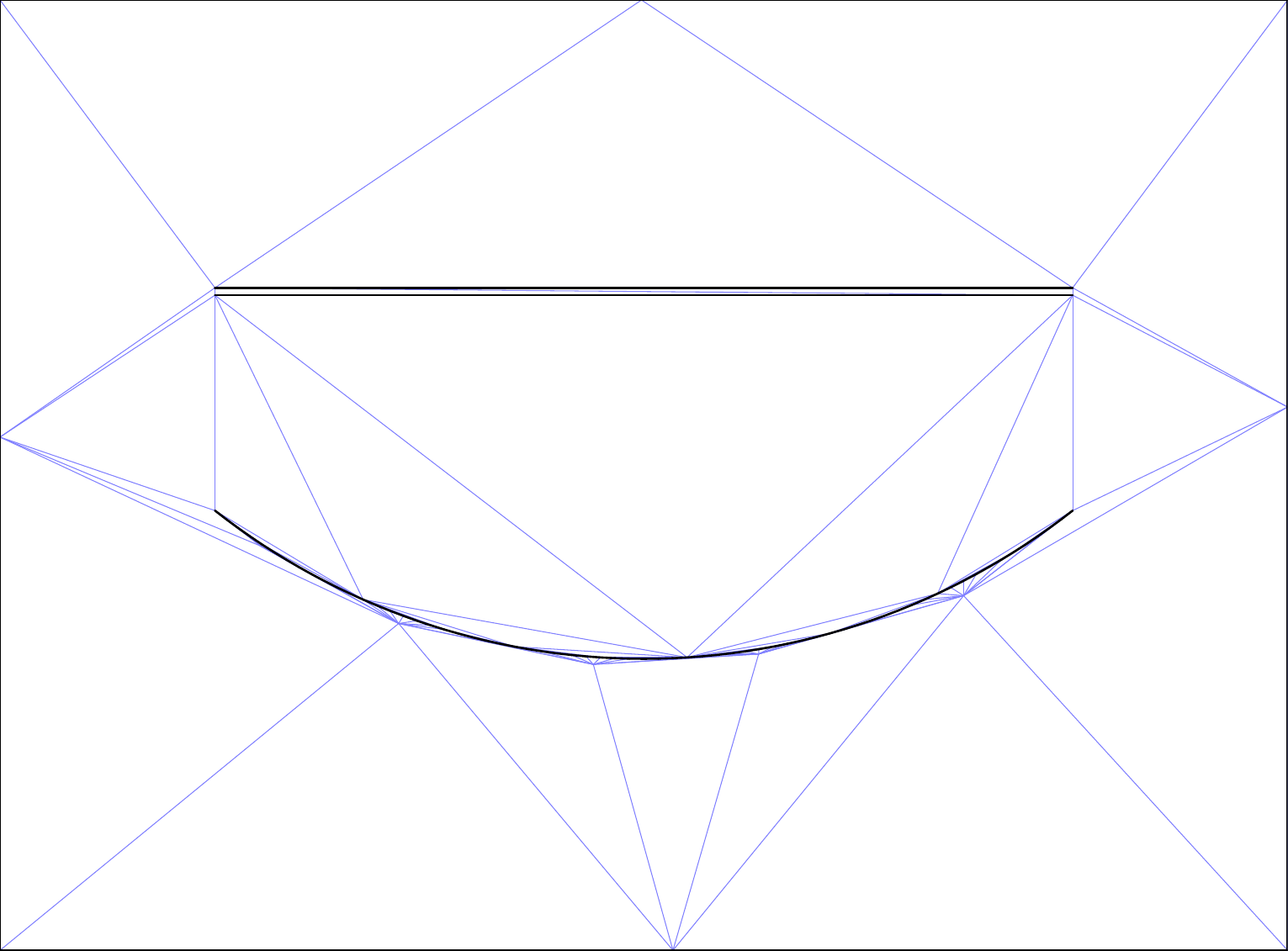}
    \\[-0.5mm]
    & \small 17.5 & & \small 19.3 \\
\end{tabular}
\end{center}
\caption{
Triangulations (blue) of a synthetic scene (black) containing a 64-segment curved line at the bottom and one long horizontal line segment (left column) or two closely spaced long horizontal line segments (right column). 
To the right of each scene, close-ups are shown of the rightmost part of the horizontal line(s) (top) and curved segment (bottom).
The middle row shows refined CDTs with optimal lower bound on angle and upper bound on triangle surface area to yield the smallest total edge length. The bound on the minimal angle leads to a better triangulation of the curved segment, but it leads to spurious edges for the two closely spaced horizontal line segments in the right column, as the two initial triangles between these lines have sharp angles and thus also get subdivided. 
The bottom row shows our approximately optimal triangulation. Here, the top line segments are not needlessly subdivided and the triangles around the curved geometry are tightly fitting.
Total edge lengths are given below the scenes.
\label{figMinAngleContrOneTwoLines}
}
\end{figure}

\begin{figure}
\begin{center}
\begin{tabular}{c@{\quad}c}
Polish fixed topology & Polish with edge flip perturbation \\
    \includegraphics[width=0.42\textwidth,align=c]{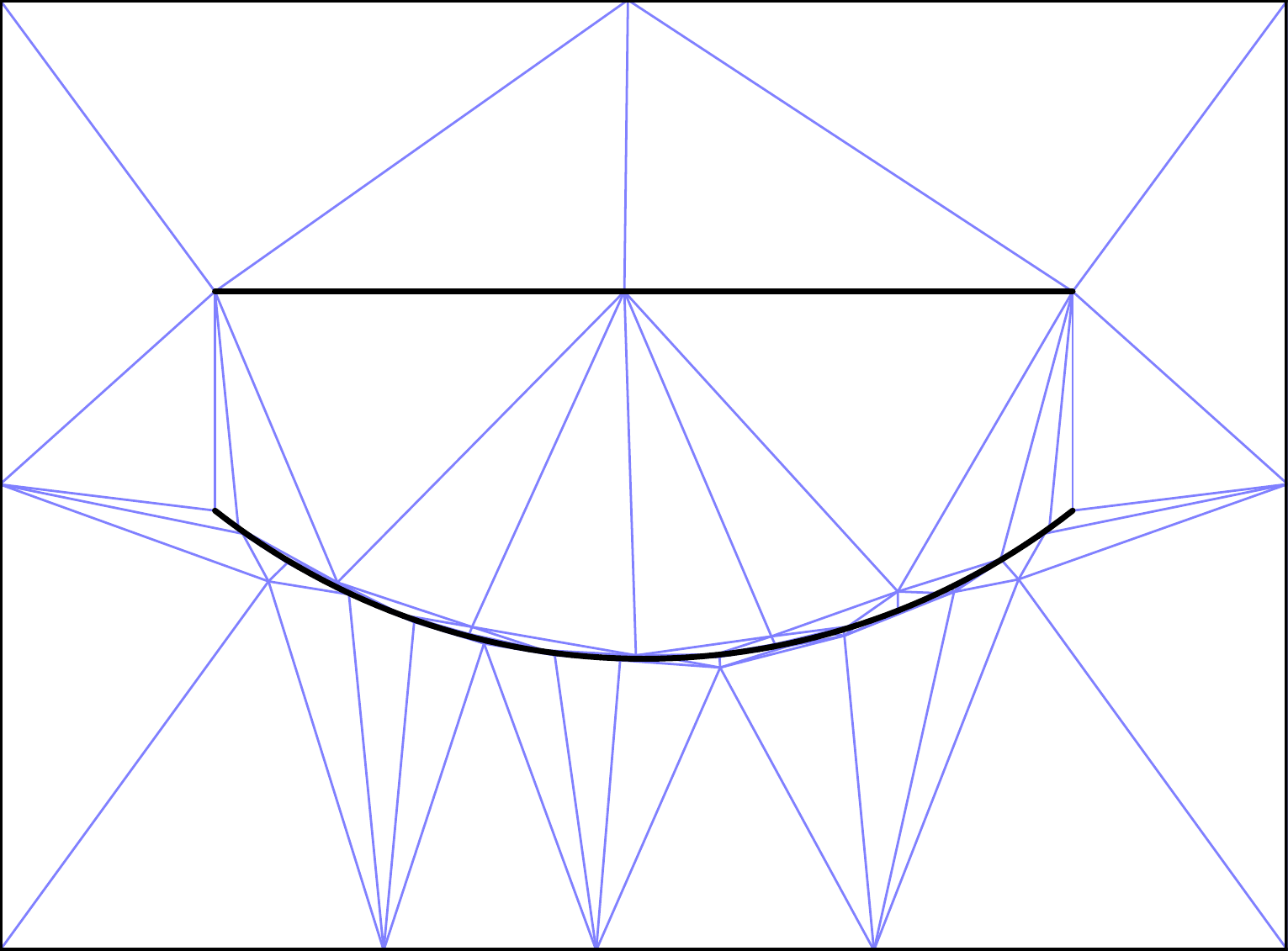}
    &
    \includegraphics[width=0.42\textwidth,align=c]{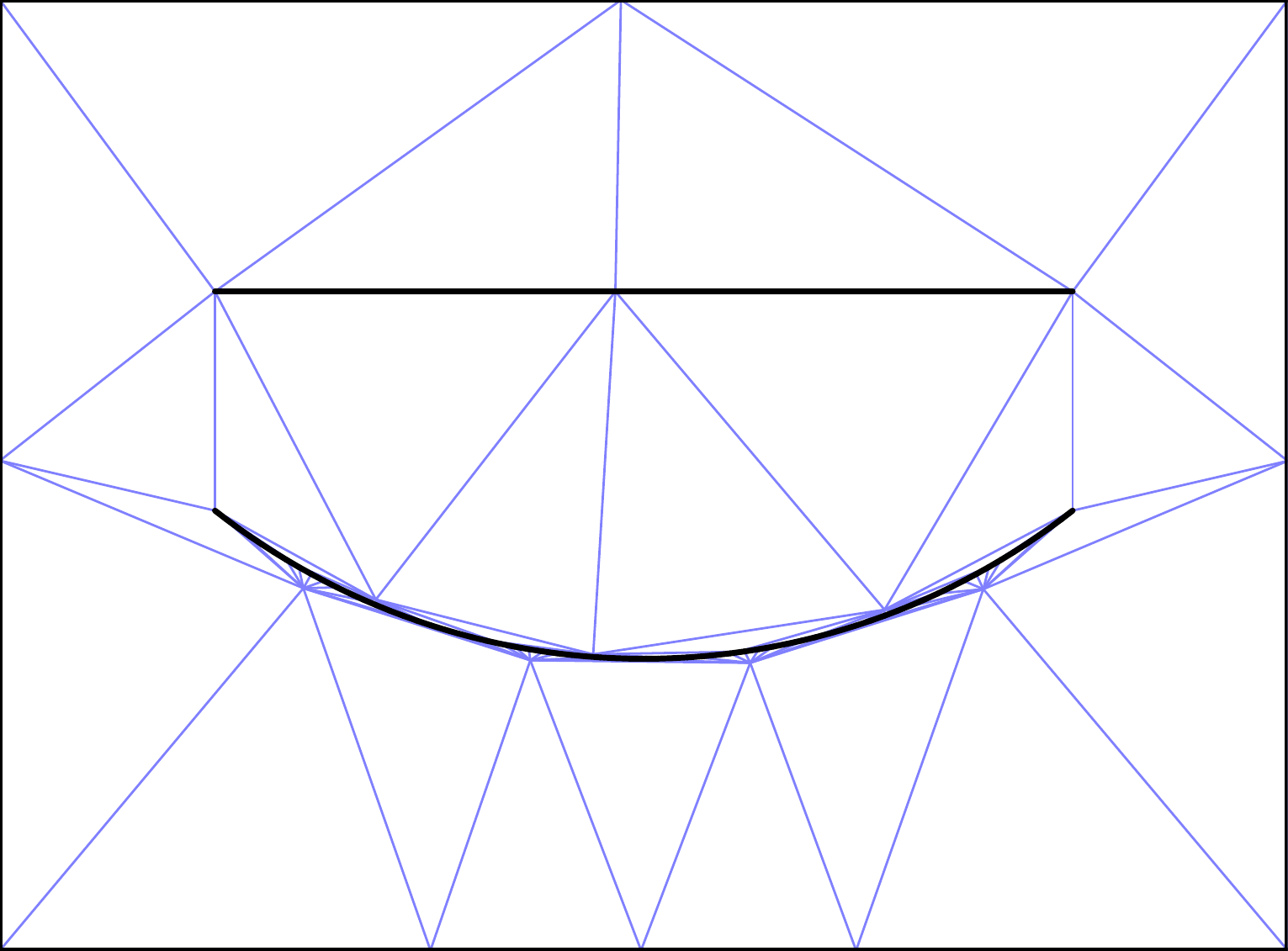}
    \\[-0.5mm]
    \small 20.2 & \small 18.4 \\[1mm]
\end{tabular}
\end{center}
\caption{
Optimized triangulations for the `Single top line' case starting from the optimally refined CDT (middle row, left in Fig.~\ref{figMinAngleContrOneTwoLines}) without fuzzy contraction. The left image shows the result of a direct edge length minimization with fixed topology which causes the triangles to closely hug the curved geometry, yielding a total edge length of 20.2. The right image additionally uses the edge flip perturbation to enable \emph{some} topology changes while keeping the number of vertices constant and has edge length 18.4. Compared to the optimized result at the bottom left of Fig.~\ref{figMinAngleContrOneTwoLines} with edge length 17.5,
these results illustrate the additional benefit of (fuzzy) contraction during the optimization. Note that the difference would be even more dramatic for the scene with two top lines as it has many spurious vertices.
\label{figMinAngleContrPolishOfOneLine}
}
\end{figure}

\FloatBarrier

\subsection{Synthetic Scenes: Lines \label{secOptimLines}}

To further explore the behaviour of CDTs and optimized triangulations, we will look at a simple family of synthetic scenes shown in Fig.~\ref{figLinesPolySC1}. The scenes are parametrised in the number of lines, $N$, and the preferred orientation of the line segments.
The line length gets scaled as $1/\sqrt{N}$ to keep a constant overall surface area density when considering small perturbations in the line's orientation, irrespective of $N$.
Furthermore, the basic length of each line segment can be scaled up or down: Fig.~\ref{figLinesPolySC3} shows scenes where each segment is scaled by a length factor of $3$ (clamped to a maximum of 0.95 for low $N$ so that everything fits inside the unit square), and Fig.~\ref{figLinesPolySC0p1} shows scenes where every line is scaled by a factor of $0.1$.

\begin{landscape}
\begin{figure}
\newcommand{\linePlotScale}{1.0}
\begin{center}
\begin{tabular}{c@{\ }c@{\ }c@{\ }c}
& \small Length factor 0.1
& \small Length factor 1
& \small Length factor 3
\\
\rotatebox[origin=c]{90}{\small Relative to unrefined CDT} &
    \includegraphics[scale=\linePlotScale,align=c]{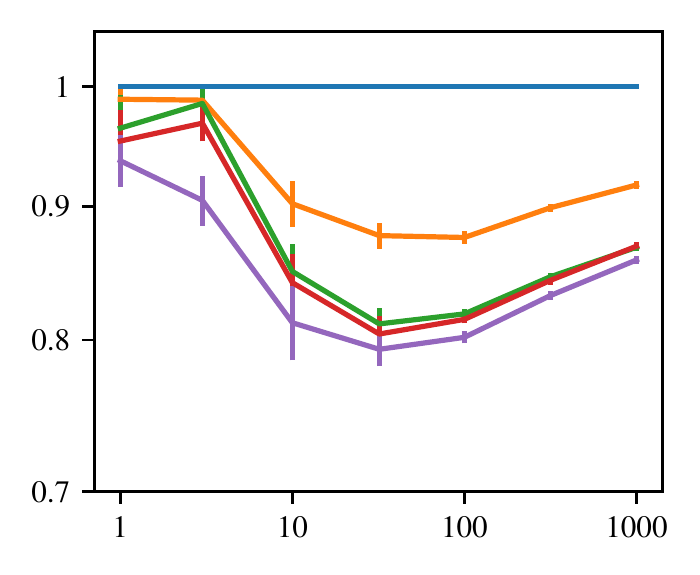} &
    \includegraphics[scale=\linePlotScale,align=c]{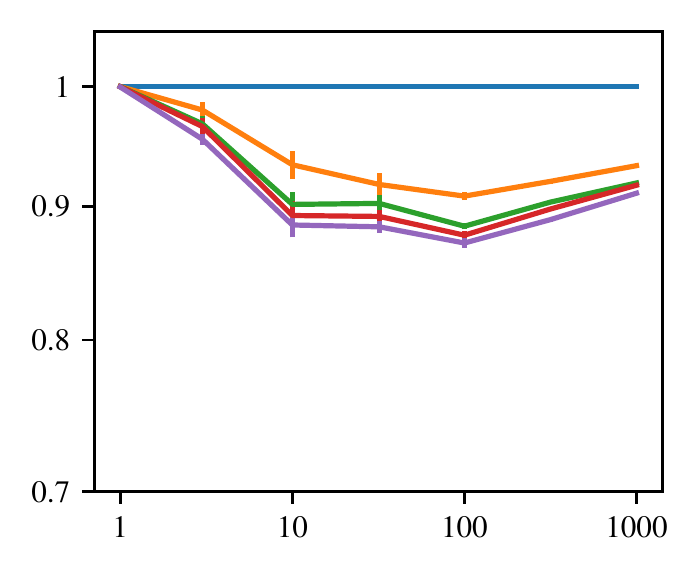} &
    \includegraphics[scale=\linePlotScale,align=c]{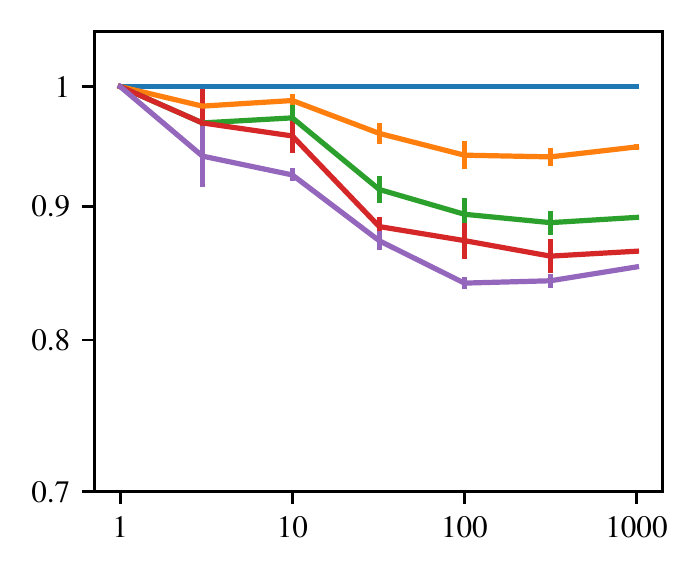}
\\[-3mm]
\rotatebox[origin=c]{90}{\small Relative to optim.~refined CDT} &
    \includegraphics[scale=\linePlotScale,align=c]{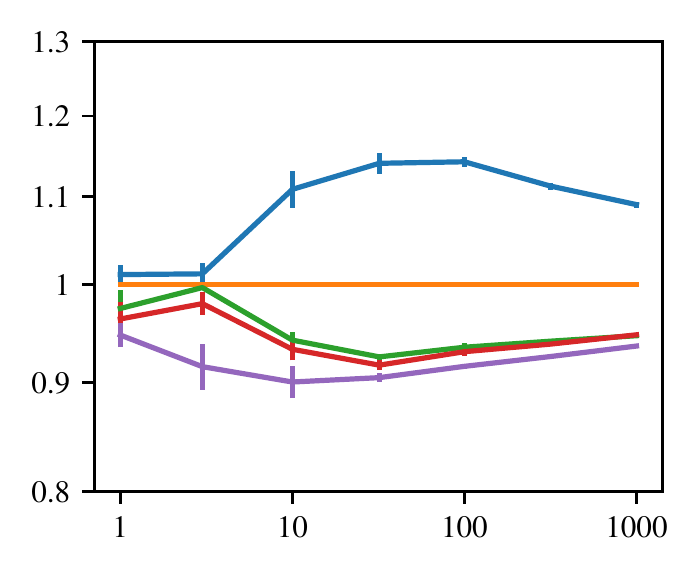} &
    \includegraphics[scale=\linePlotScale,align=c]{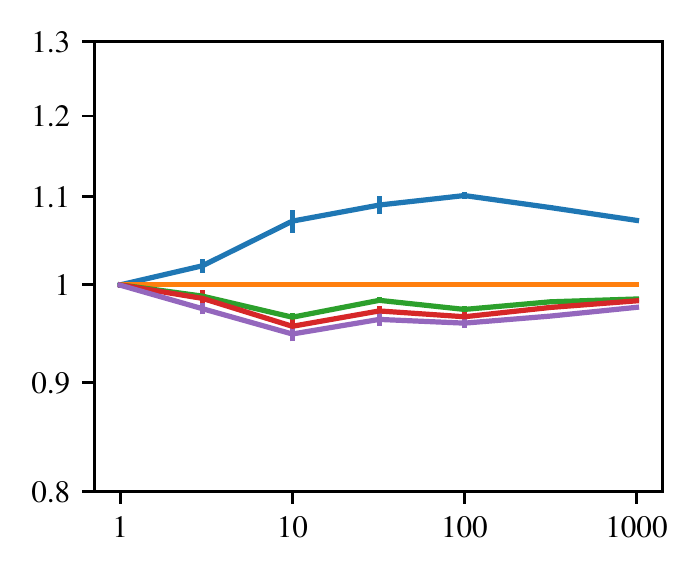} &
    \includegraphics[scale=\linePlotScale,align=c]{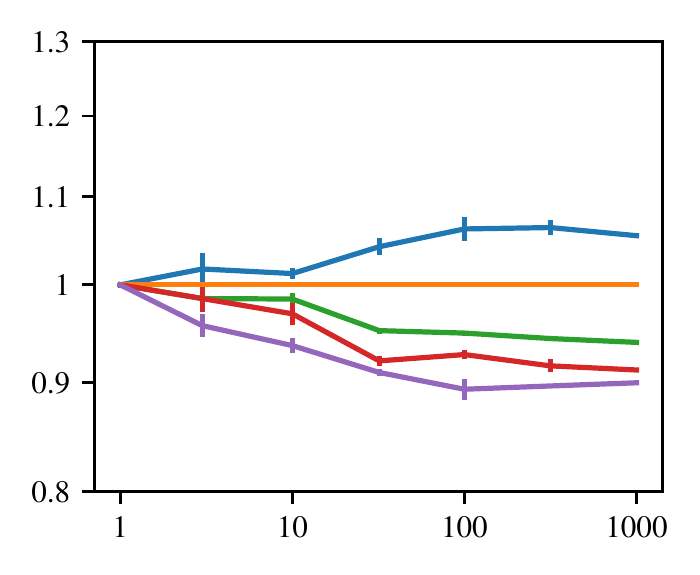}
\\[-5mm] 
& \small ~~~~$N$
& \small ~~~~$N$
& \small ~~~~$N$
\\
& \multicolumn{3}{c}{\includegraphics[align=c]{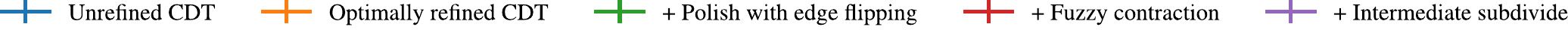}}
\end{tabular}
\end{center}
\caption{
    Relative total edge lengths for triangulations of the line scene with uniform orientation and varying length factor for successively powerful optimization strategies. The top plots show the total edge lengths relative to the total edge length of an unrefined CDT, the bottom plot shows the total edge lengths relative to that of the optimally refined CDT. Data points and their error bars represent the average and its standard error over five independently generated scenes.
}
\label{figPlotLines}
\end{figure}
\end{landscape}

The total edge lengths of triangulations for these line scenes with uniform orientation with various degrees of optimization are shown in Fig.~\ref{figPlotLines}, which we will now discuss. The results for lines with different preferred orientation (vertical or diagonal) are largely similar and are not shown here (their relevance will become clear in Sec.~\ref{secRTResultsLines}). The interested reader can find visualisations of the optimized triangulations in Appendix \ref{appTriangsKDBVH}.

From the plots in Fig.~\ref{figPlotLines}, we observe that as $N \to 1$, the difference between the various levels of optimization shrinks (or even disappears completely). In the simple limit of only one line segment, the most straightforward connection to vertices of the `world' bounding square often already yields an optimal triangulation.

As $N$ increases, the freedom in connecting the vertices (and potentially including extra Steiner vertices) gives more opportunity to make unfavorable choices. The gap between a simple CDT and the various optimized triangulations (and therein between the simple and more powerful optimization strategies themselves) becomes larger accordingly.

\begin{figure}
\newcommand{\linePlotScale}{1.0}
\begin{center}
\begin{tabular}{c@{\ }c@{\ }c}
& \small Unrefined CDT
& \small Optimized (ours)
\\
\rotatebox[origin=c]{90}{\small Total edge length} &
    \includegraphics[scale=\linePlotScale,align=c]{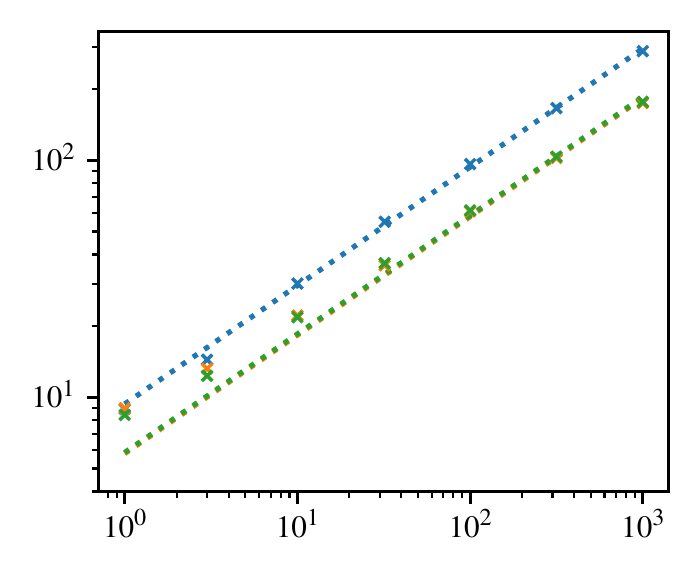} &
    \includegraphics[scale=\linePlotScale,align=c]{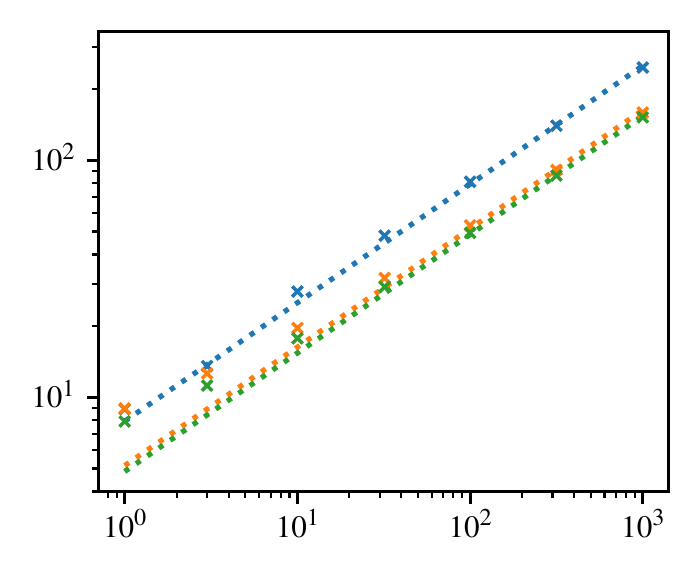}
\\[-5mm] 
& \small ~~~~~$N$
& \small ~~~~~$N$
\\[1mm]
& \multicolumn{2}{c}{\includegraphics[align=c]{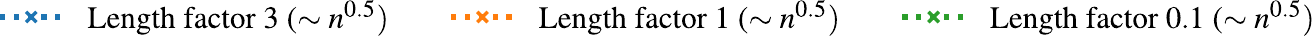}}
\end{tabular}
\end{center}
\caption{Total edge lengths for triangulations of the line scene with uniform orientation for the unrefined CDTs (left) and our fully optimized triangulation (right). Error bars are smaller that symbol sizes. For large $N$, the total edge length roughly starts to scale as $\sqrt{N}$ (as shown by the dotted lines), similar to the minimum weight triangulation of a uniform point set \cite{golin1996limit,lingas1986greedy}.
\label{figPlotLinesAbsolute}}
\end{figure}

For very large $N$, the shrinking line segments become point-like. The differences between the various optimized triangulations now gets compressed again again as there are less unfavorable choices to make regarding potential constrained Steiner vertices on the short line segments themselves as they become nearly point-like. We also compare the total edge length of our triangulations for these short line segments with the theoretical $\sqrt{N}$ scaling for minimum weight triangulations of uniform \emph{point sets} without Steiner points \cite{golin1996limit,lingas1986greedy} in Fig.~\ref{figPlotLinesAbsolute} and find reasonable similarity.

\newcommand{\linesScale}{0.35}
\begin{figure}
\begin{center}
\begin{tabular}{c@{\ }c@{\ }c@{\ }c}
& \multicolumn{3}{c}{\textsc{Length factor 1}}\\
& uniform orientation & preferentially vertical & preferentially diagonal
\\
\rotatebox[origin=c]{90}{\small 1 line} &
    \includegraphics[scale=\linesScale,align=c]{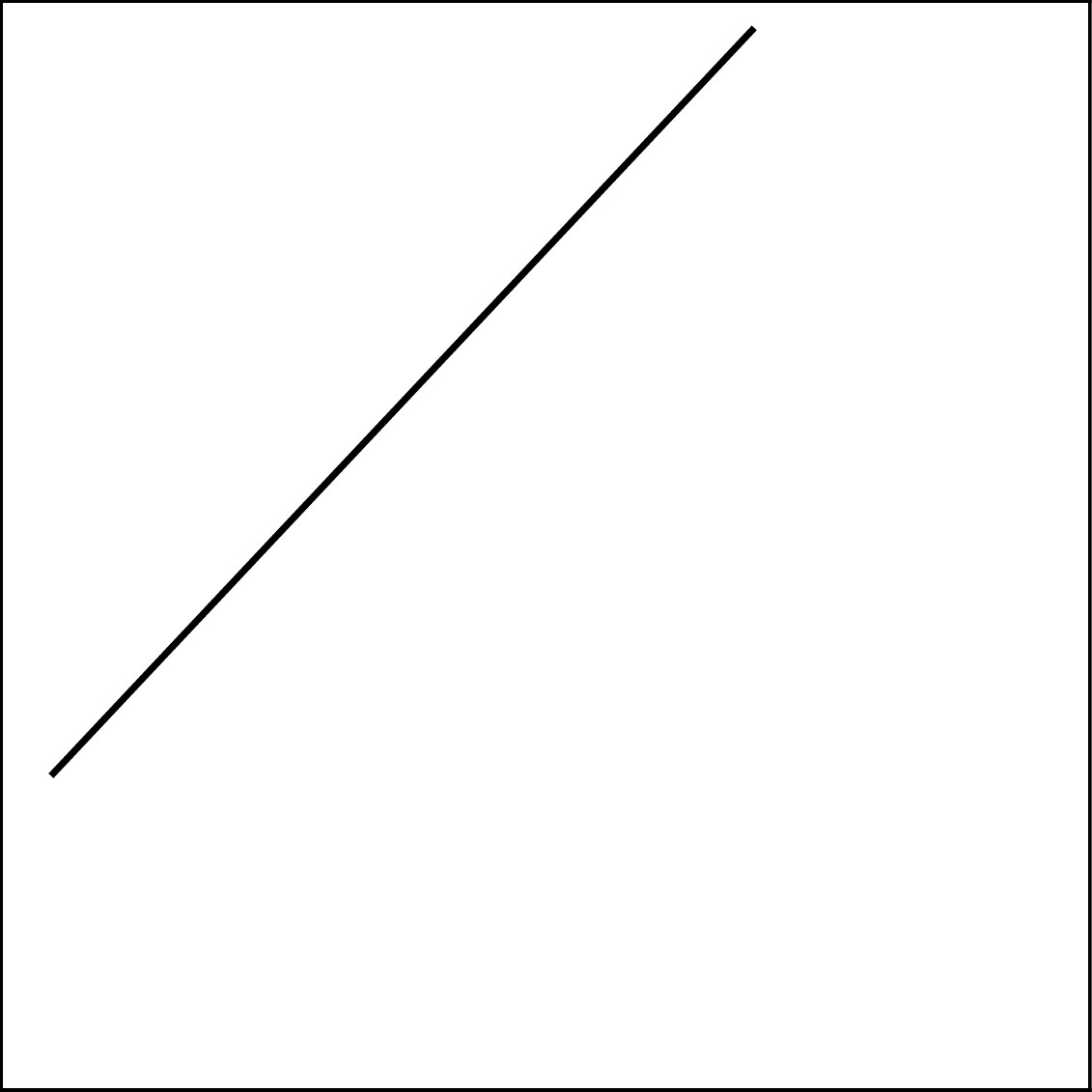}
    &
    \includegraphics[scale=\linesScale,align=c]{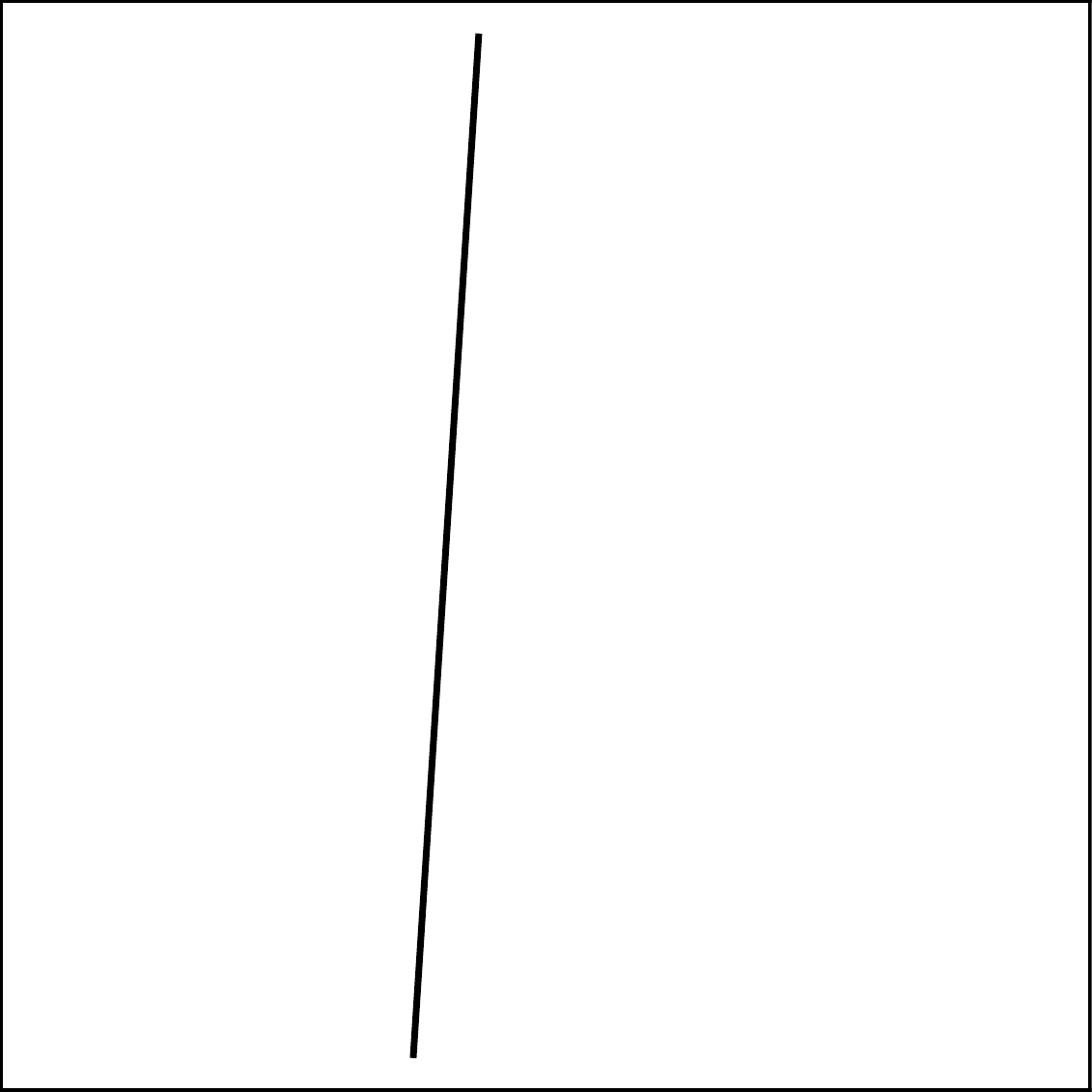}
    &
    \includegraphics[scale=\linesScale,align=c]{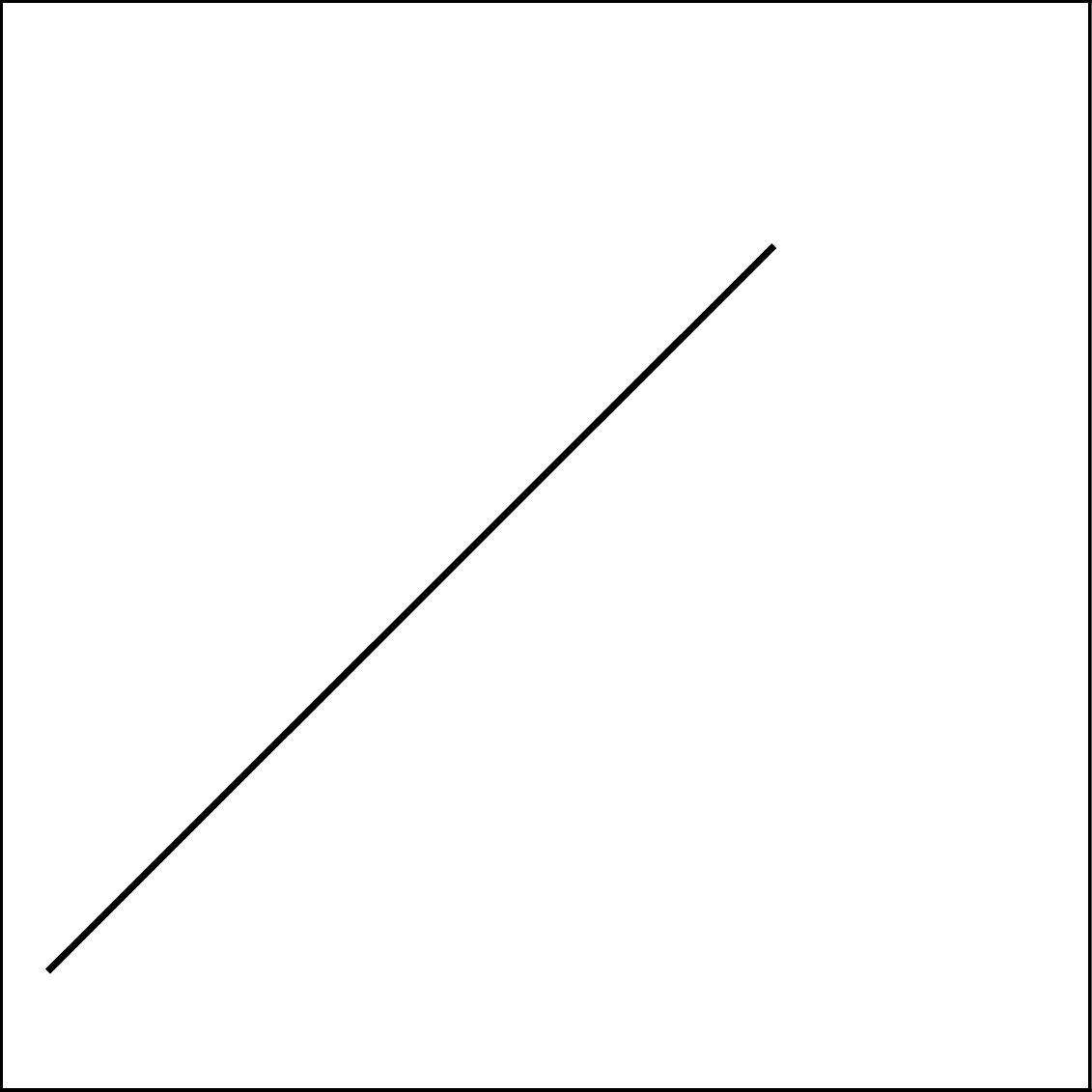}
\\
\rotatebox[origin=c]{90}{\small 10 lines} &
    \includegraphics[scale=\linesScale,align=c]{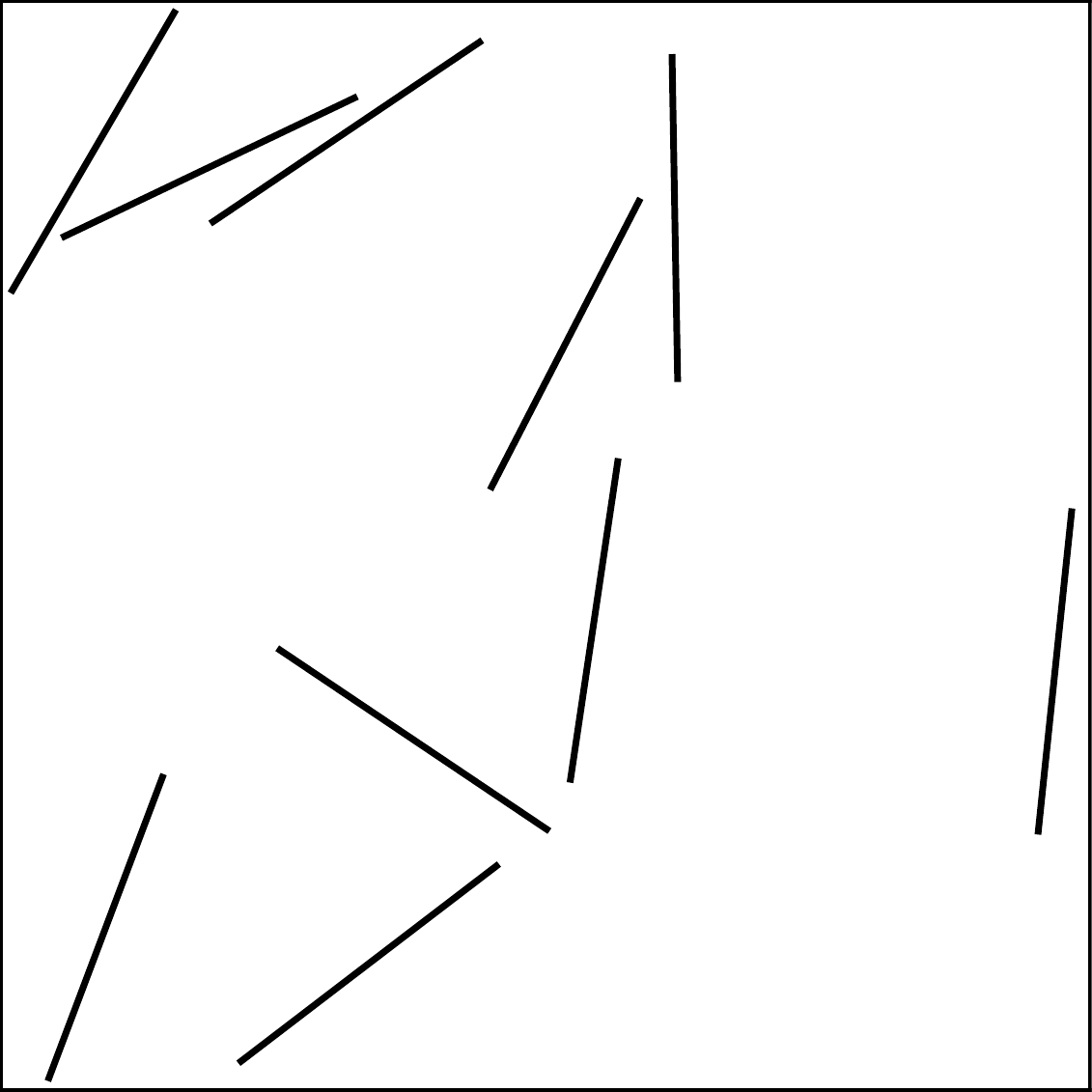}
    &
    \includegraphics[scale=\linesScale,align=c]{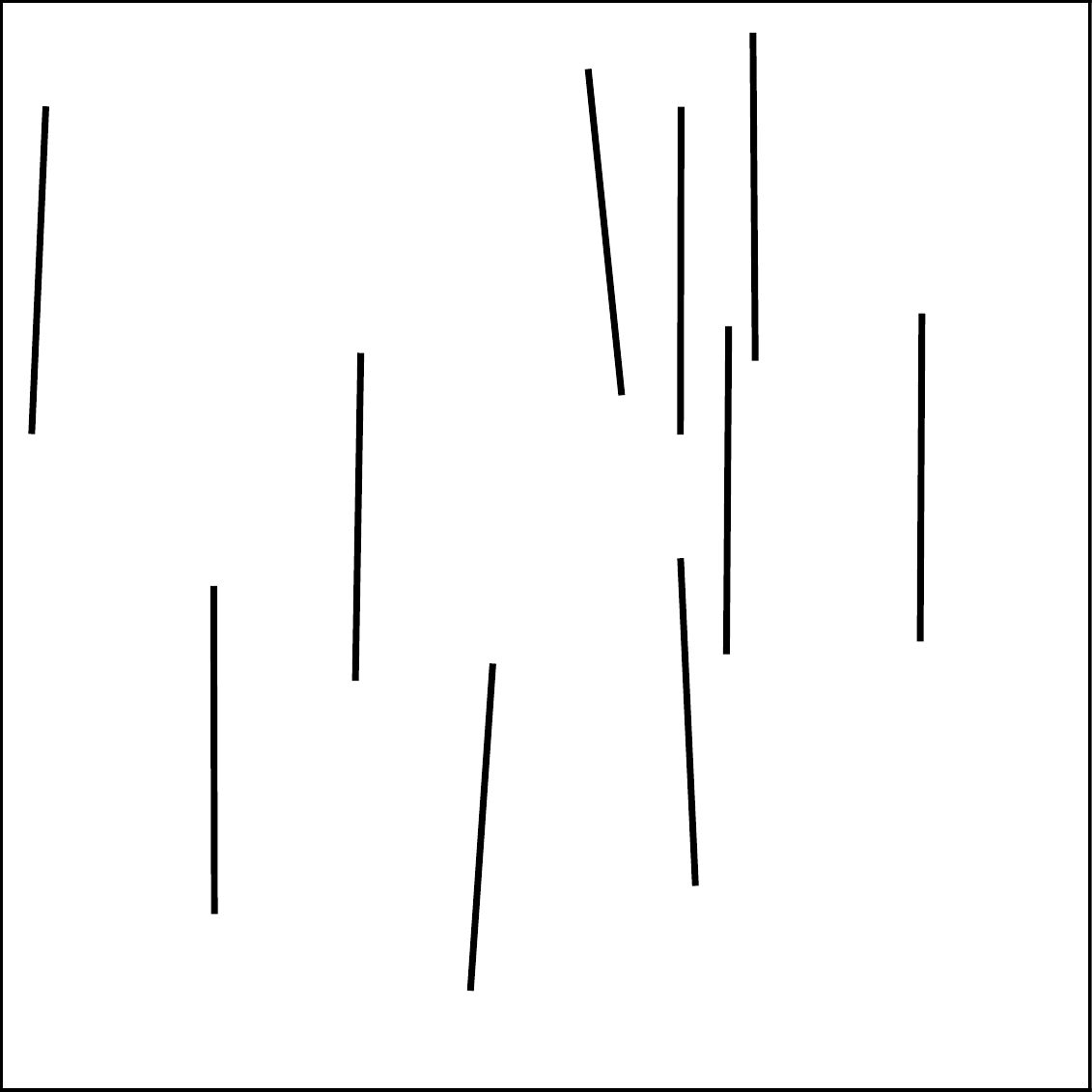}
    &
    \includegraphics[scale=\linesScale,align=c]{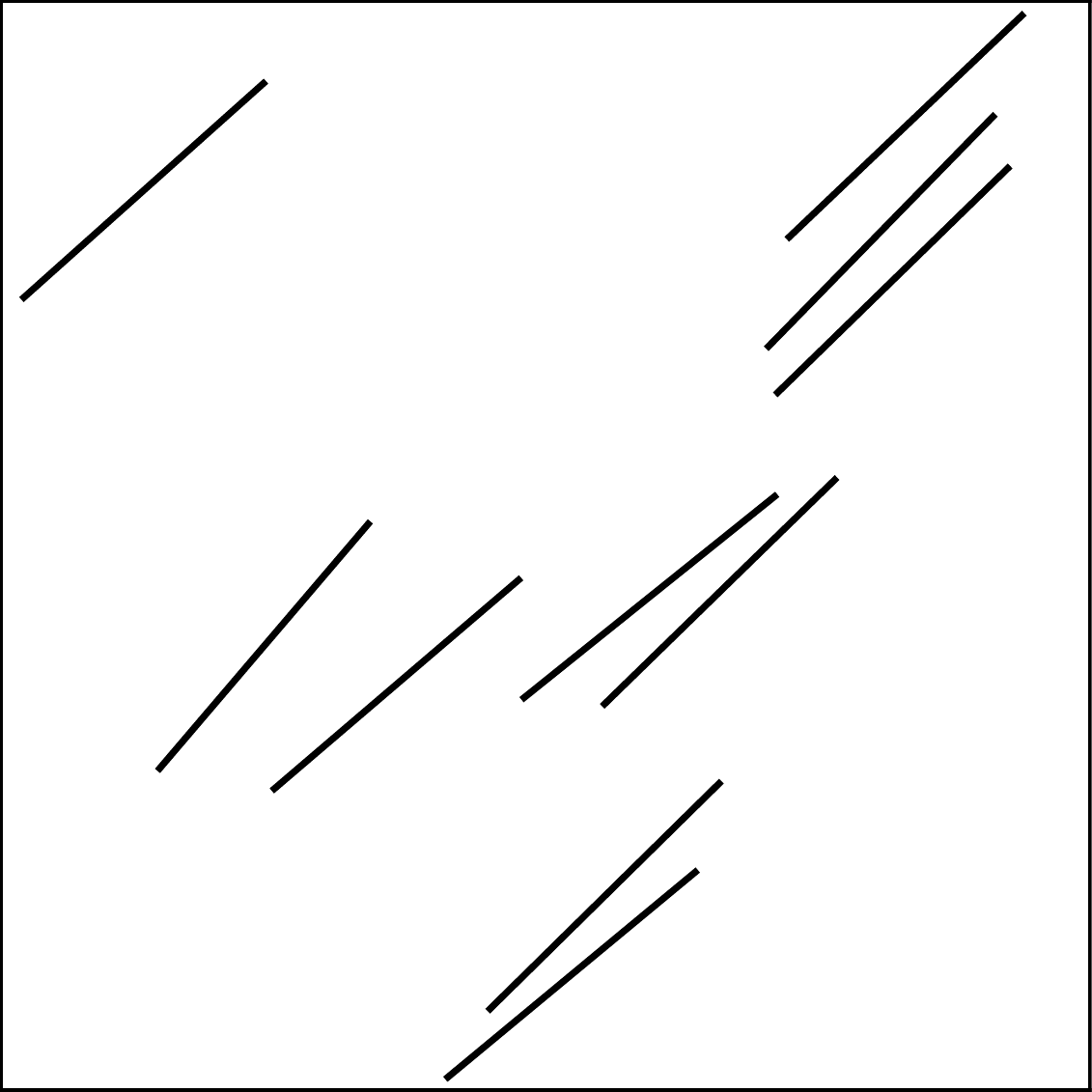}
\\
\rotatebox[origin=c]{90}{\small 100 lines} &
    \includegraphics[scale=\linesScale,align=c]{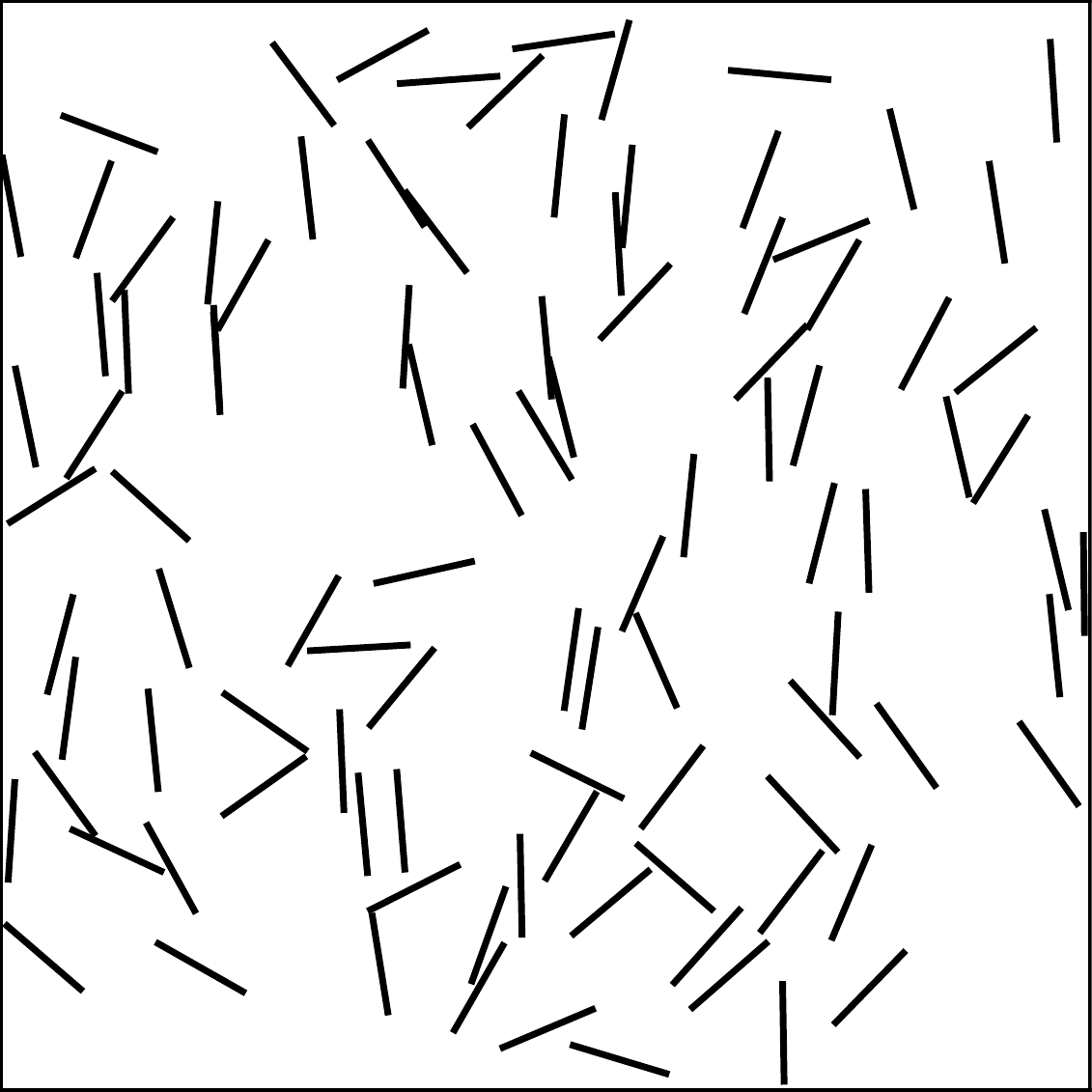}
    &
    \includegraphics[scale=\linesScale,align=c]{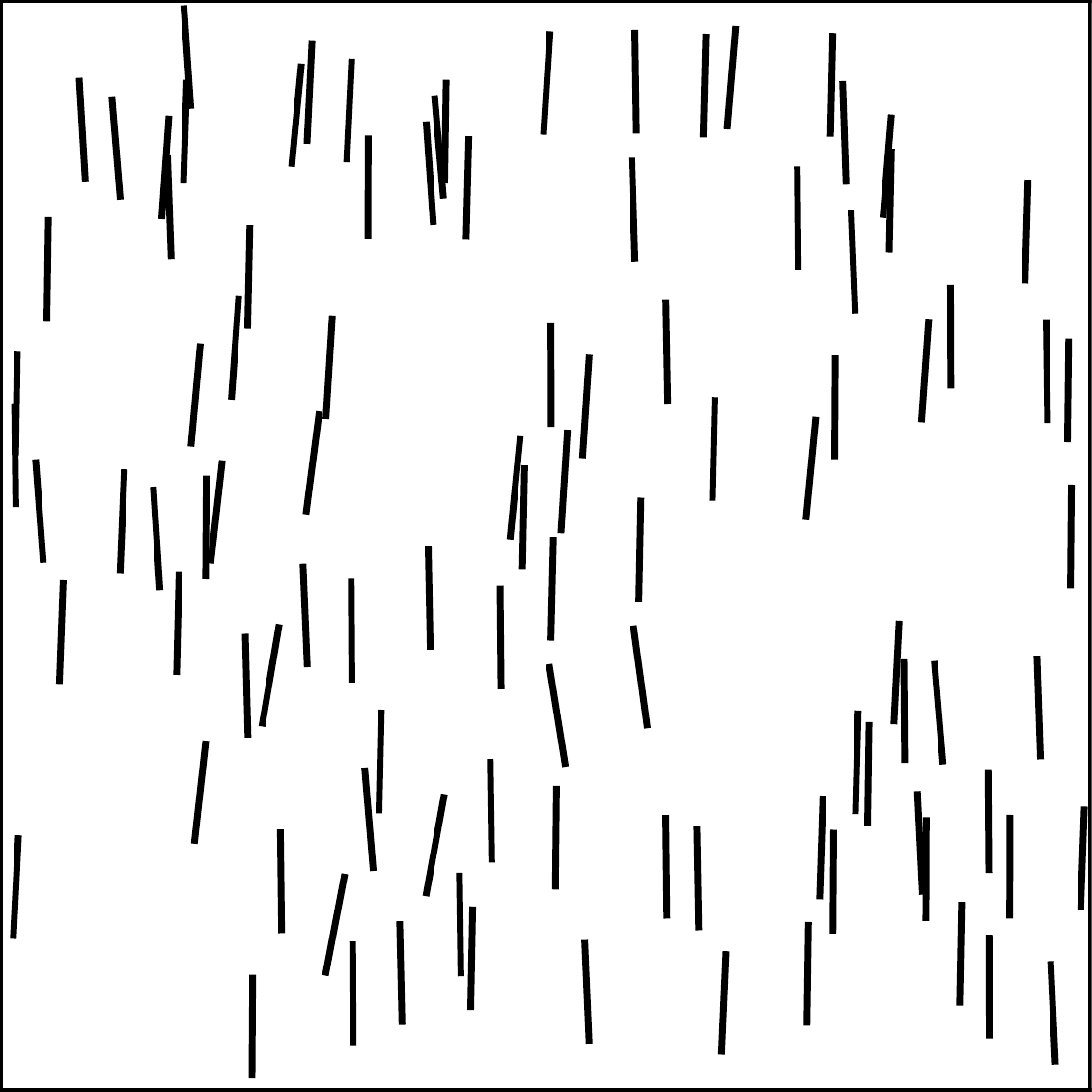}
    &
    \includegraphics[scale=\linesScale,align=c]{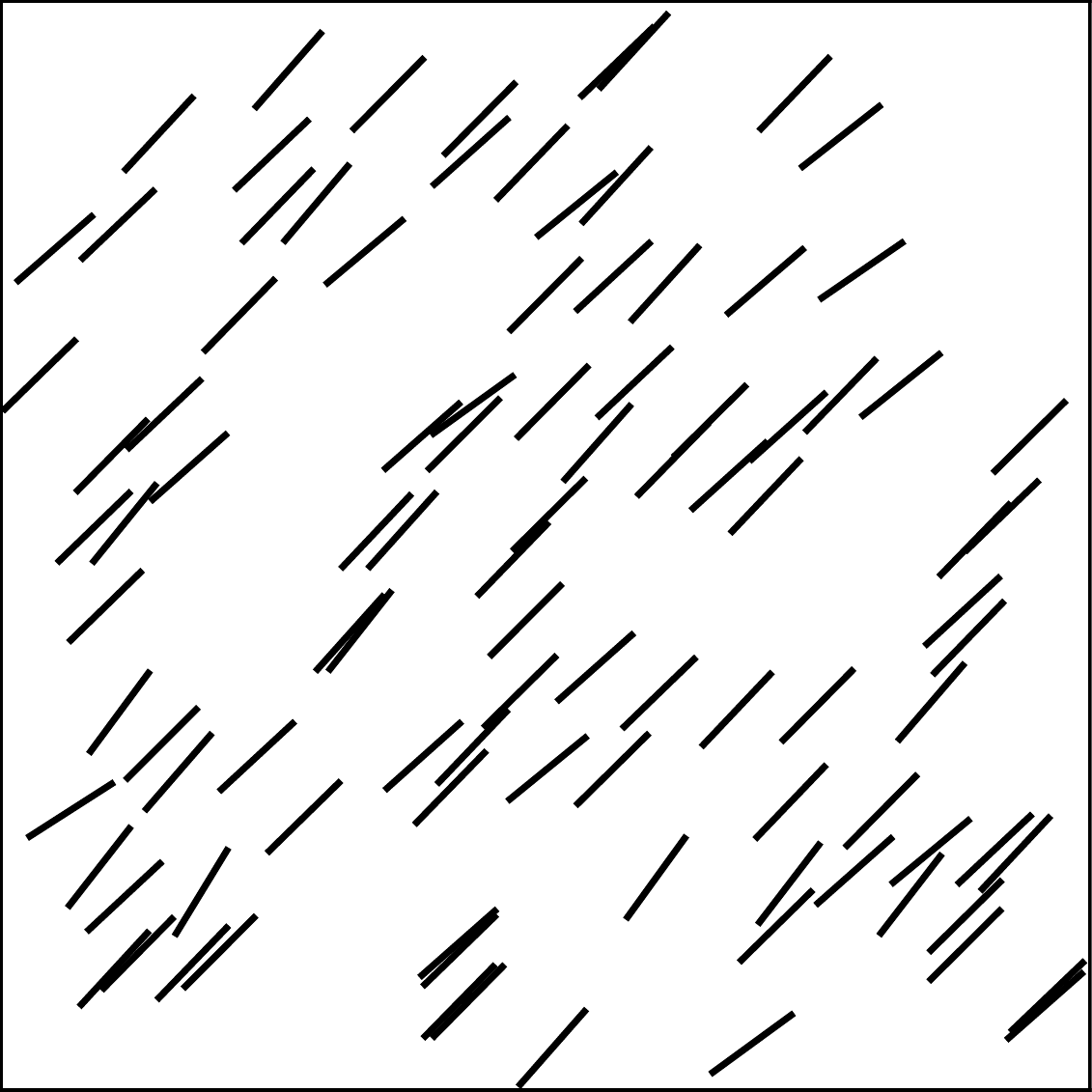}
\\
\rotatebox[origin=c]{90}{\small 1000 lines} &
    \includegraphics[scale=\linesScale,align=c]{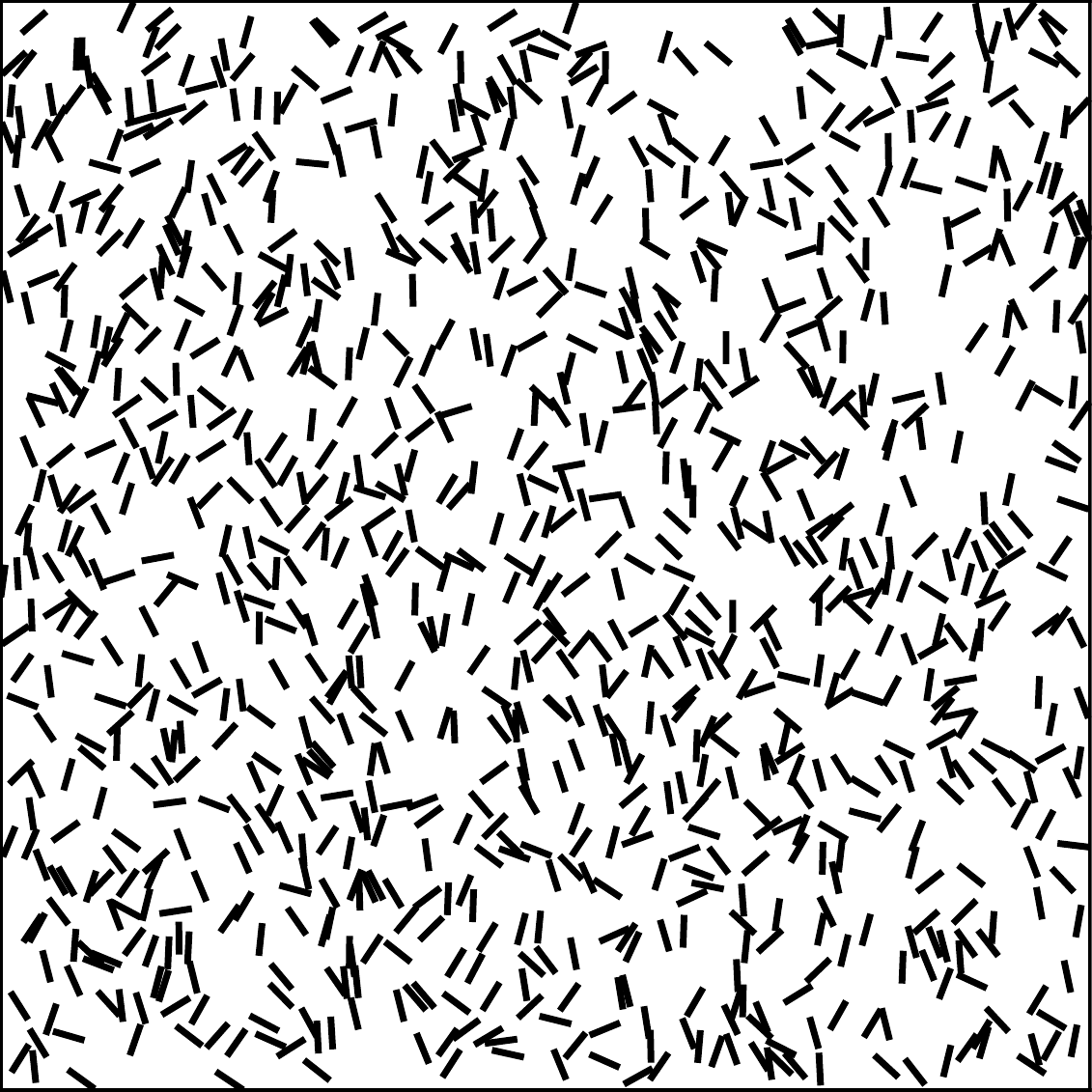}
    &
    \includegraphics[scale=\linesScale,align=c]{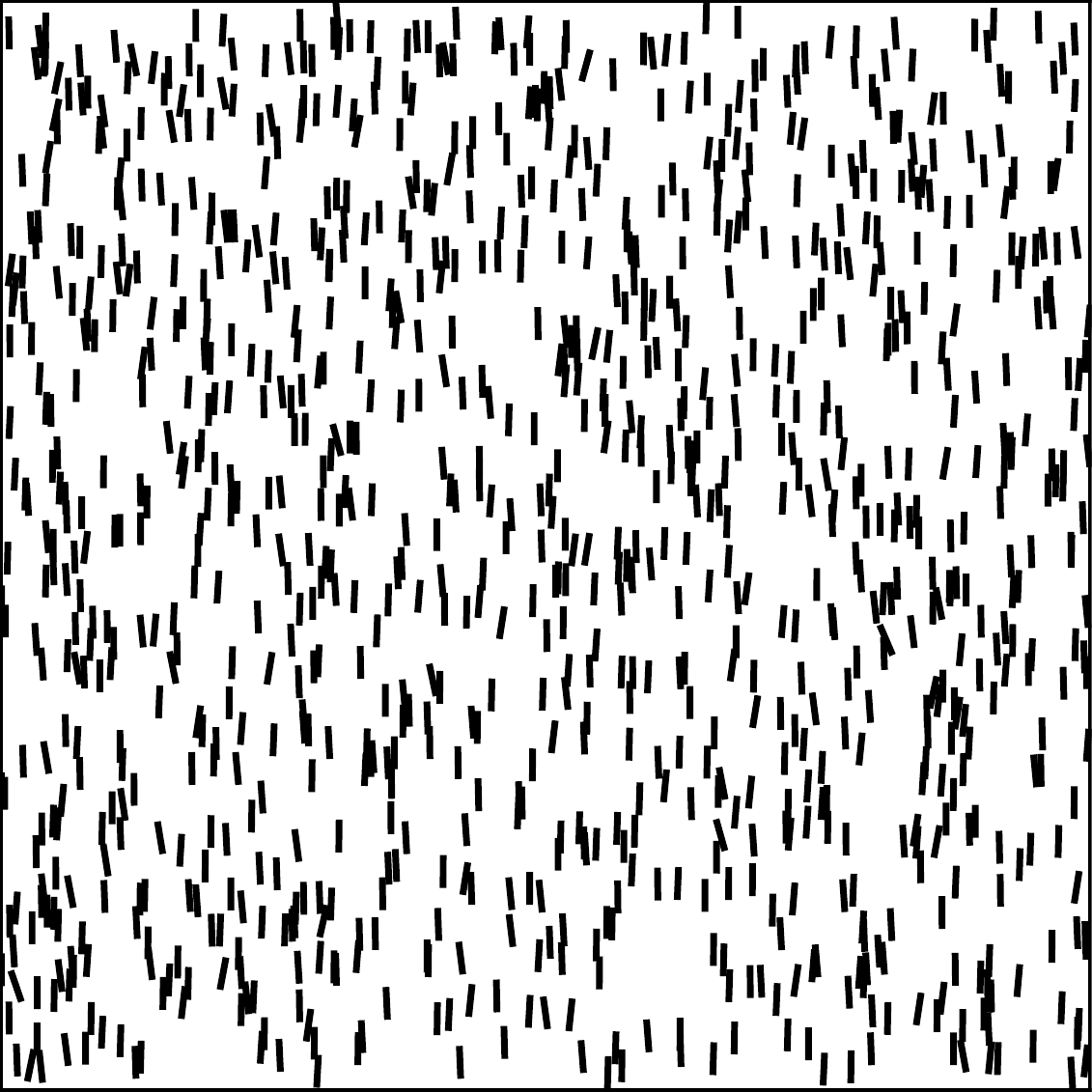}
    &
    \includegraphics[scale=\linesScale,align=c]{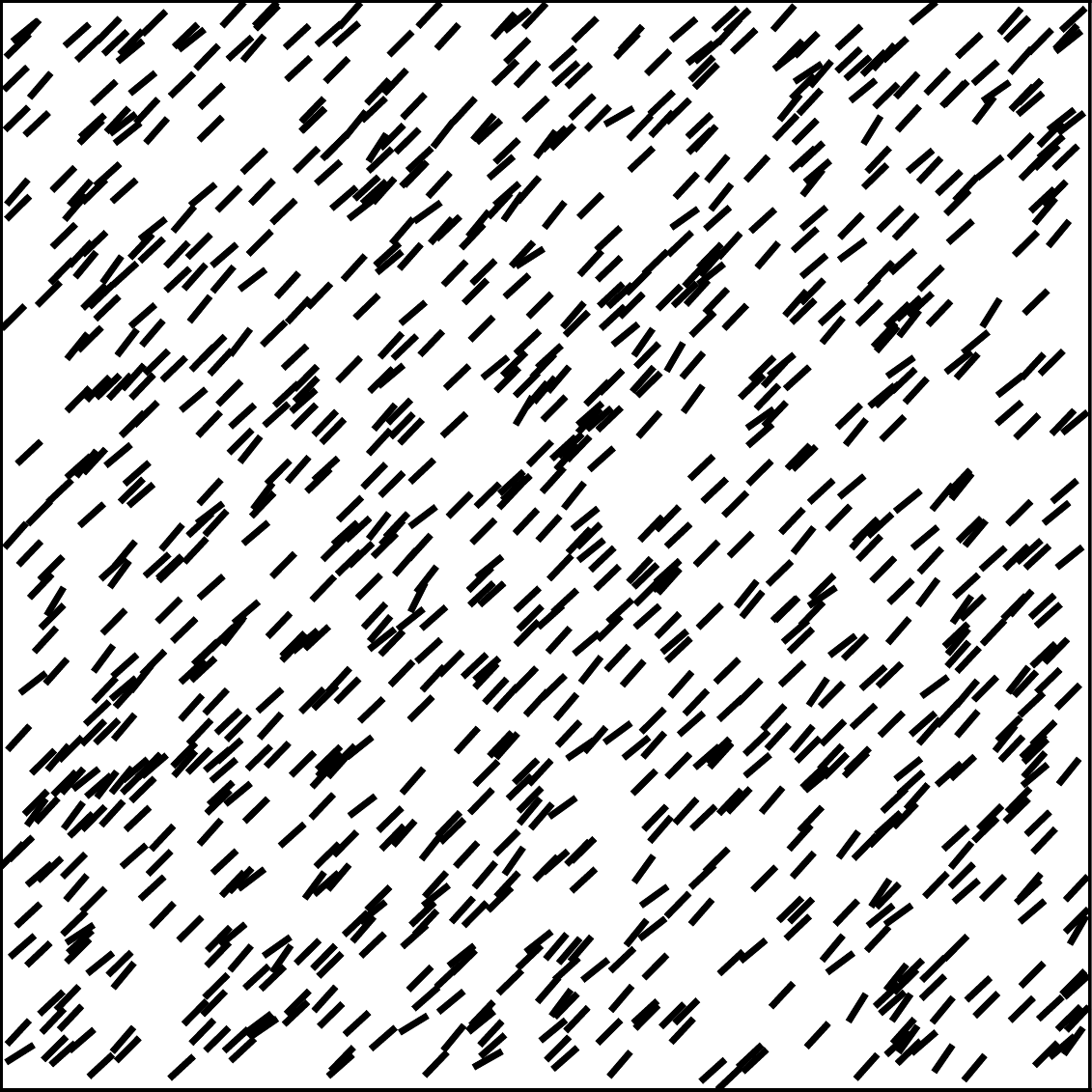}
\end{tabular}
\end{center}
\caption{Examples of the `lines' scene for different orientations and number of lines $N$ in the unit square. The line length for $N=1$ is 0.95 and gets scaled as $1/\sqrt{N}$ to keep a constant overall surface area density when considering small perturbations in the line's orientation, irrespective of $N$.
}
\label{figLinesPolySC1}
\end{figure}

\begin{figure}
\begin{center}
\begin{tabular}{c@{\ }c@{\ }c@{\ }c}
& \multicolumn{3}{c}{\textsc{Length factor 3}}\\
& uniform orientation & preferentially vertical & preferentially diagonal
\\
\rotatebox[origin=c]{90}{\small 1 line} &
    \includegraphics[scale=\linesScale,align=c]{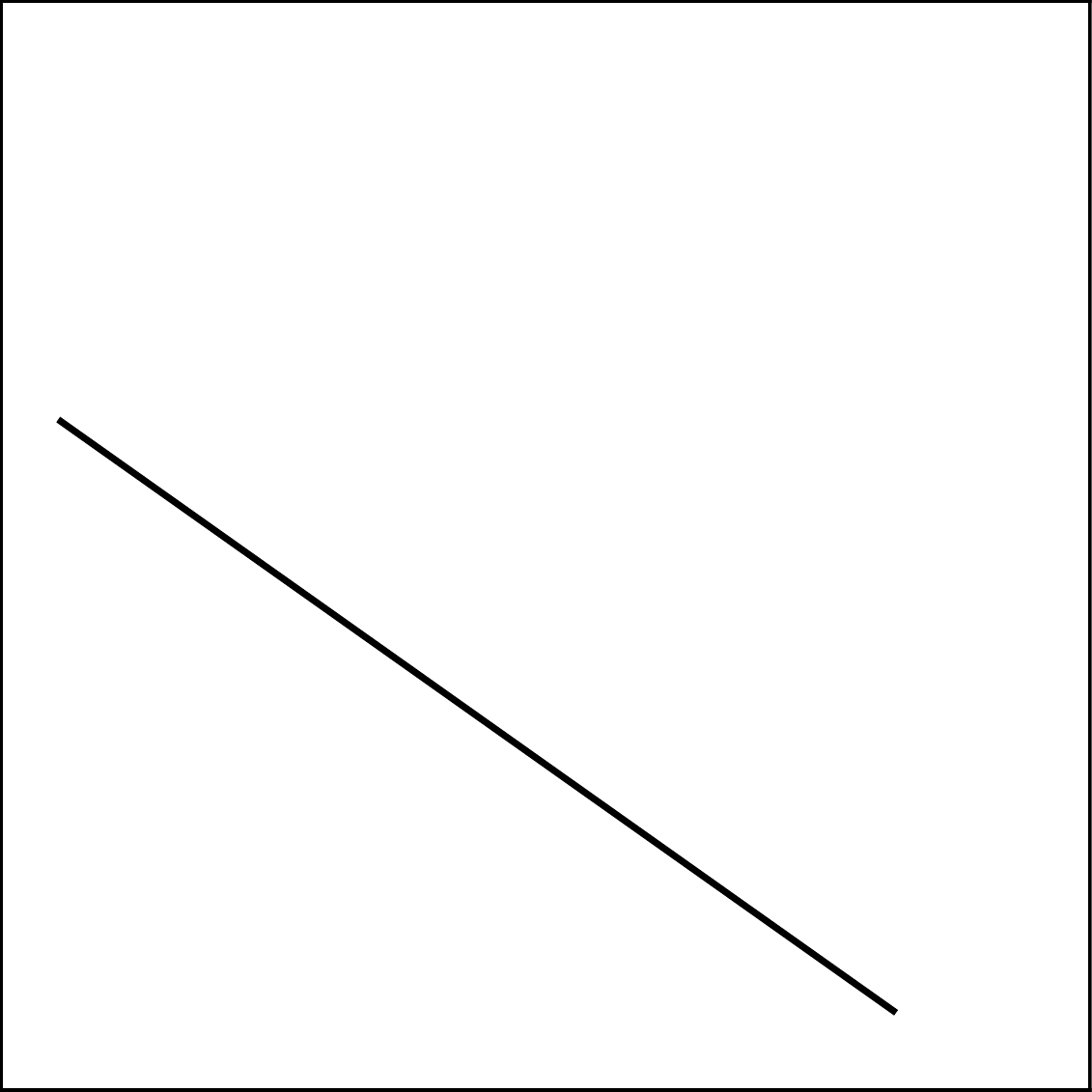}
    &
    \includegraphics[scale=\linesScale,align=c]{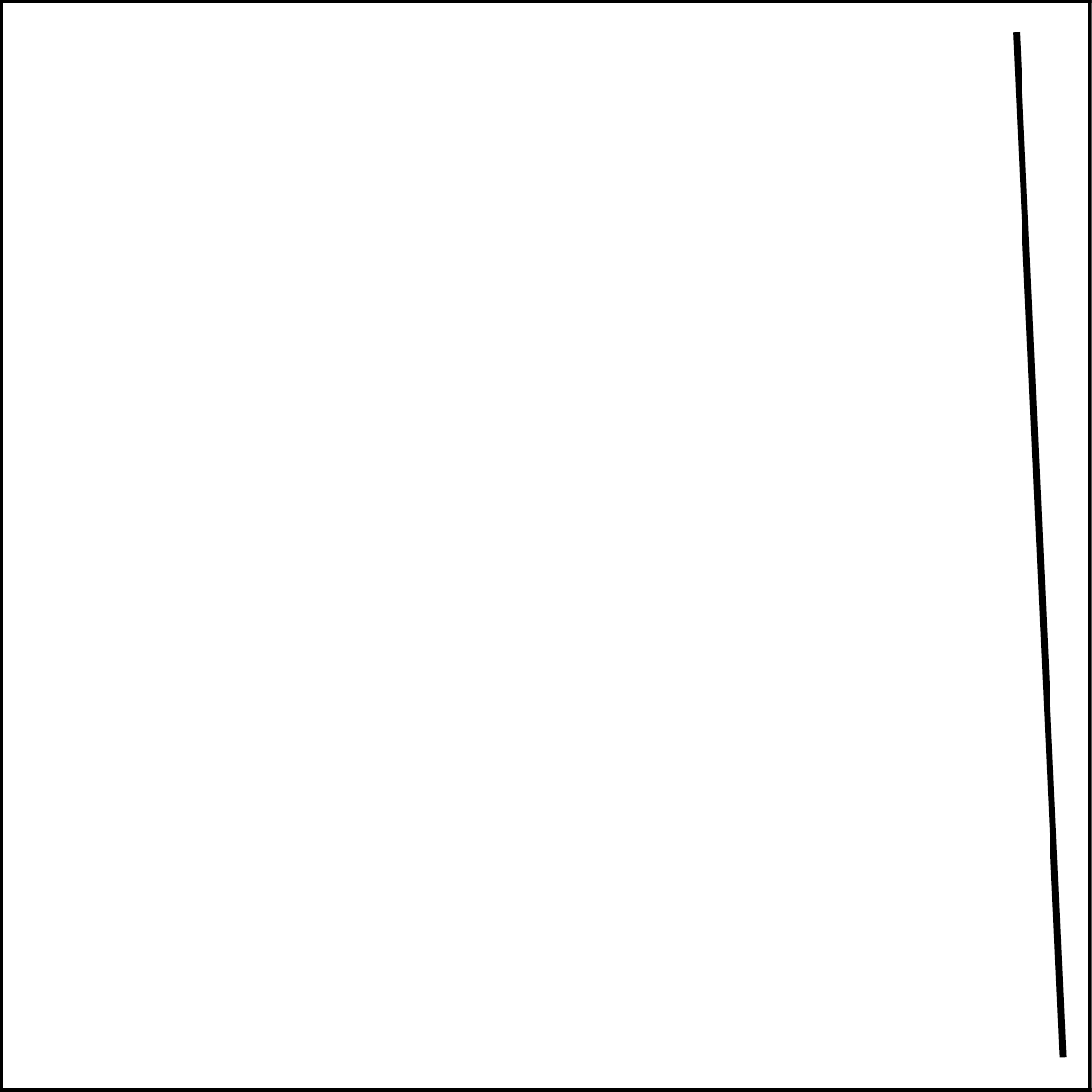}
    &
    \includegraphics[scale=\linesScale,align=c]{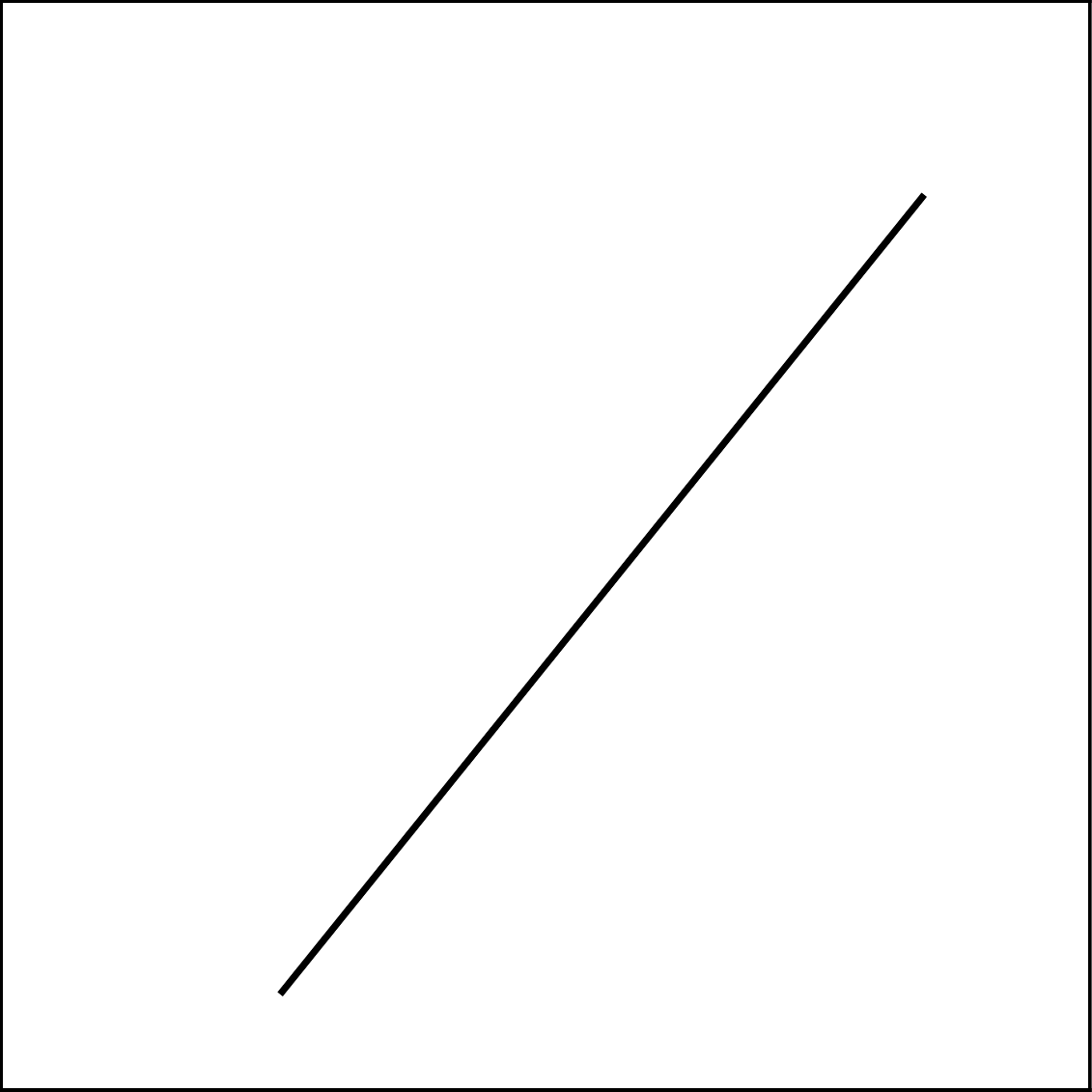}
\\
\rotatebox[origin=c]{90}{\small 10 lines} &
    \includegraphics[scale=\linesScale,align=c]{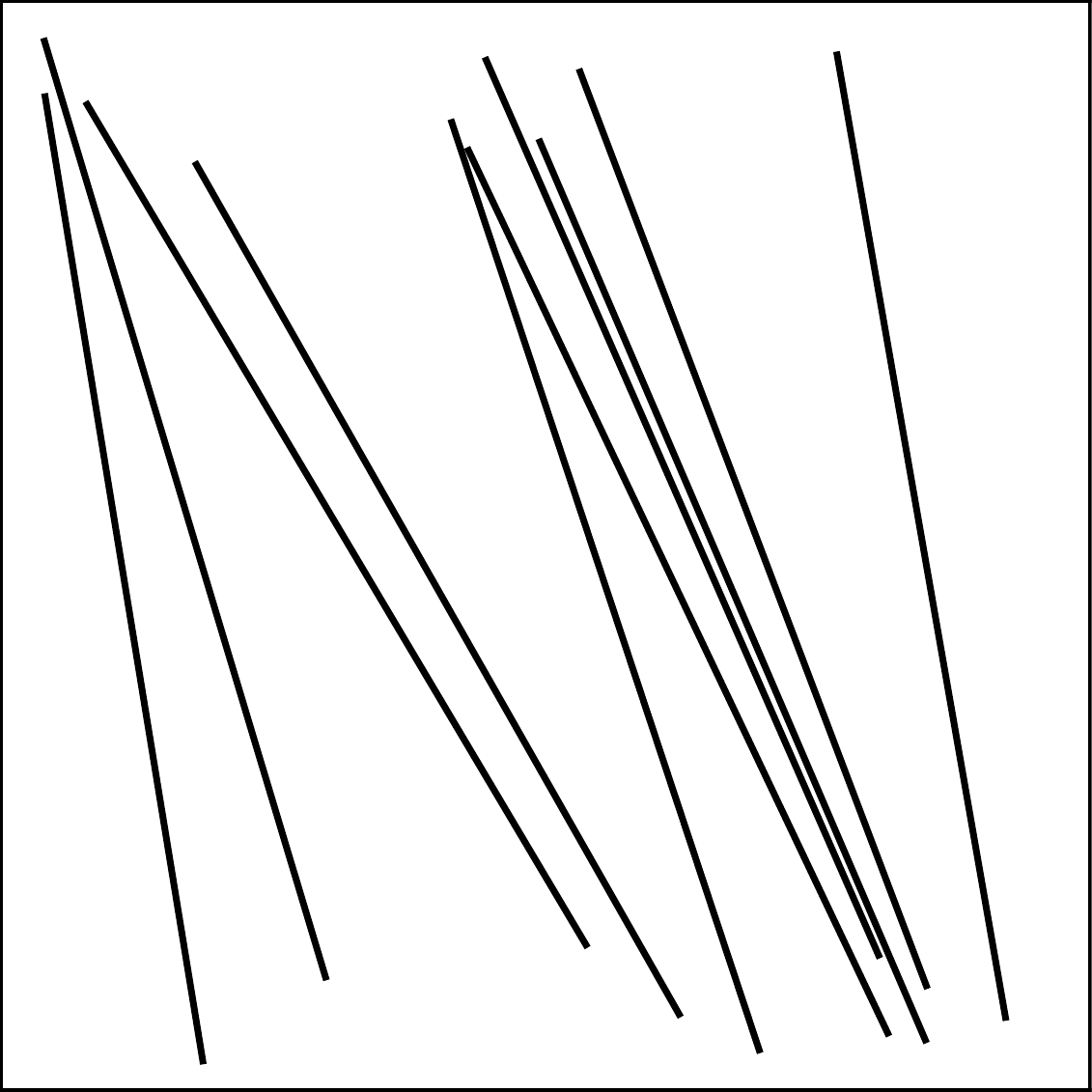}
    &
    \includegraphics[scale=\linesScale,align=c]{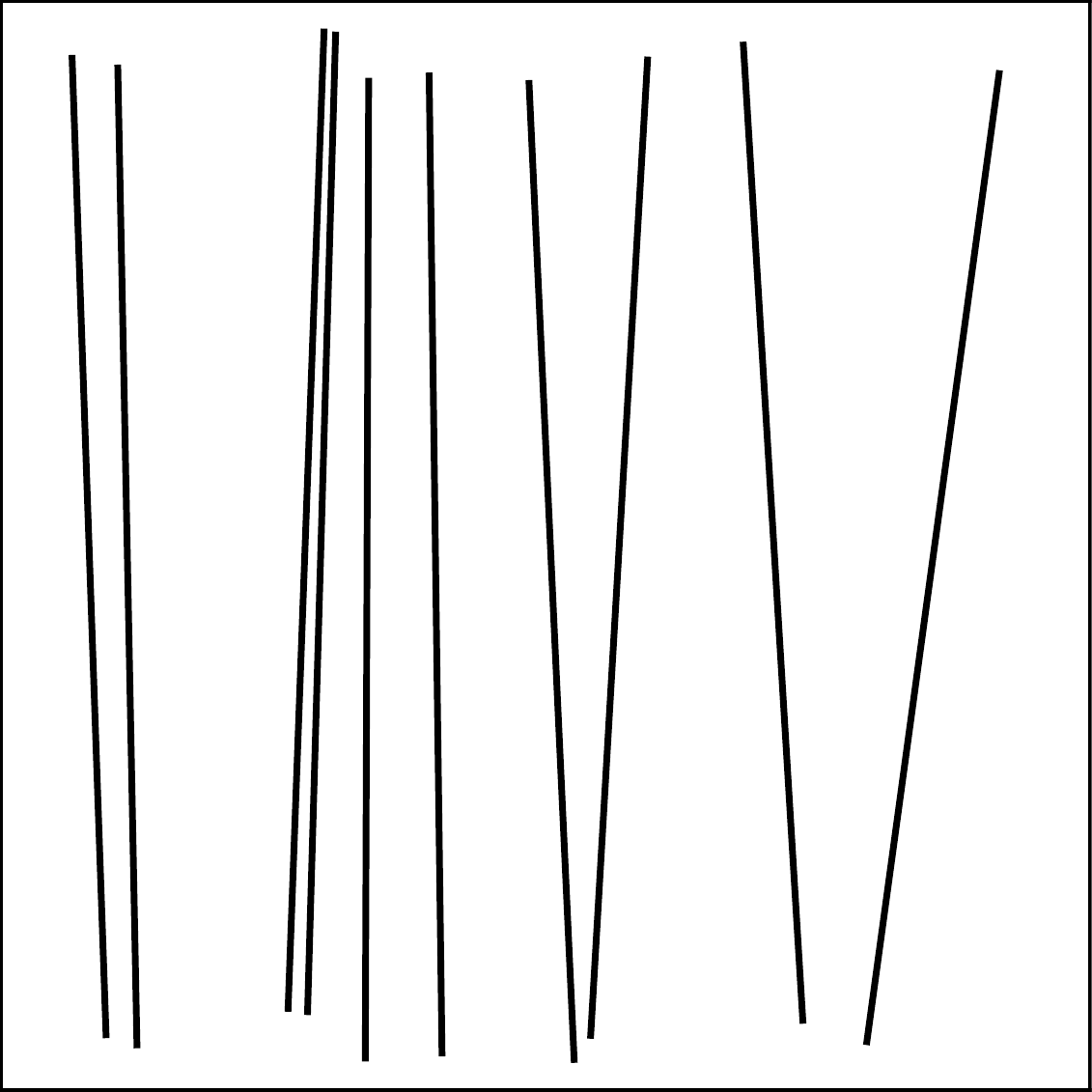}
    &
    \includegraphics[scale=\linesScale,align=c]{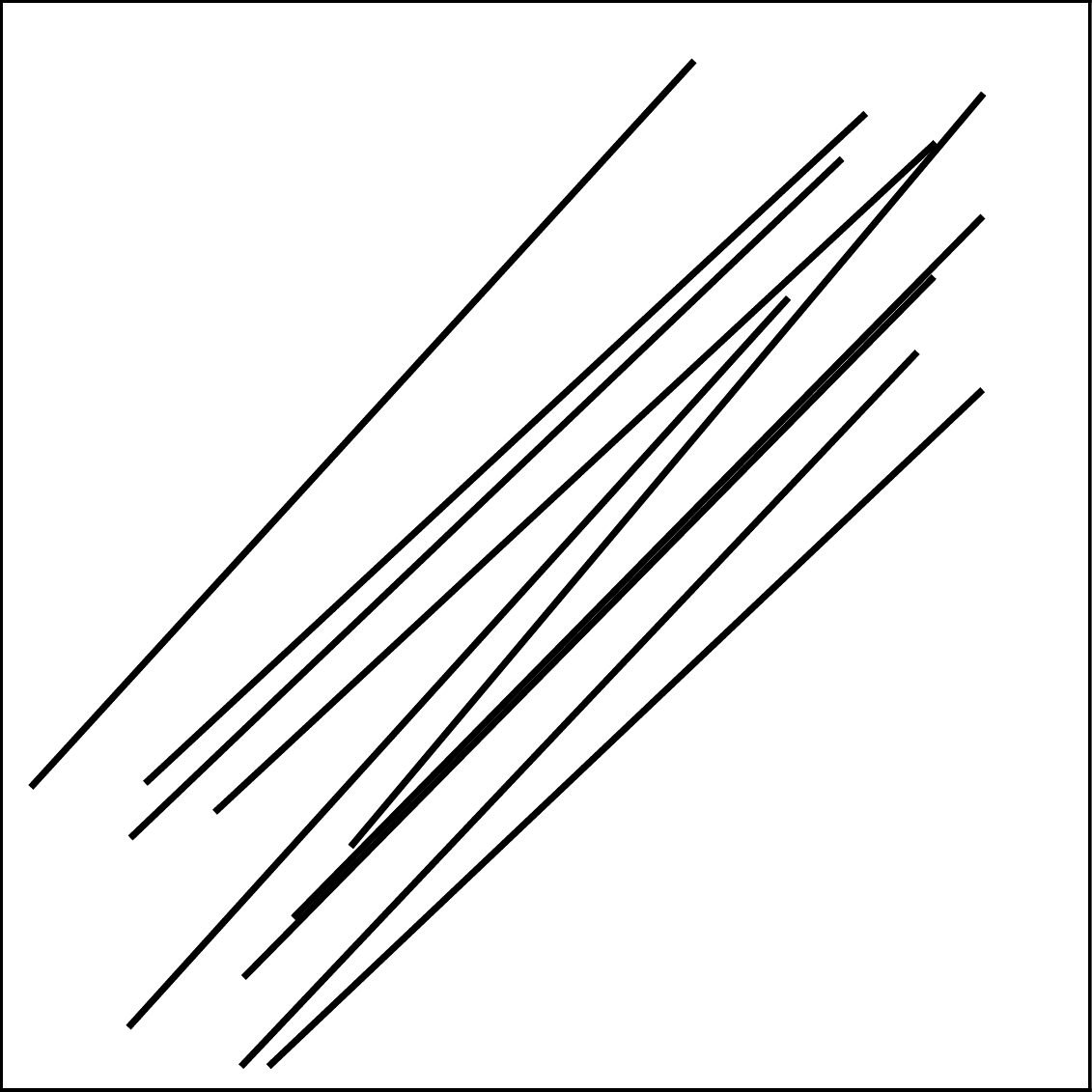}
\\
\rotatebox[origin=c]{90}{\small 100 lines} &
    \includegraphics[scale=\linesScale,align=c]{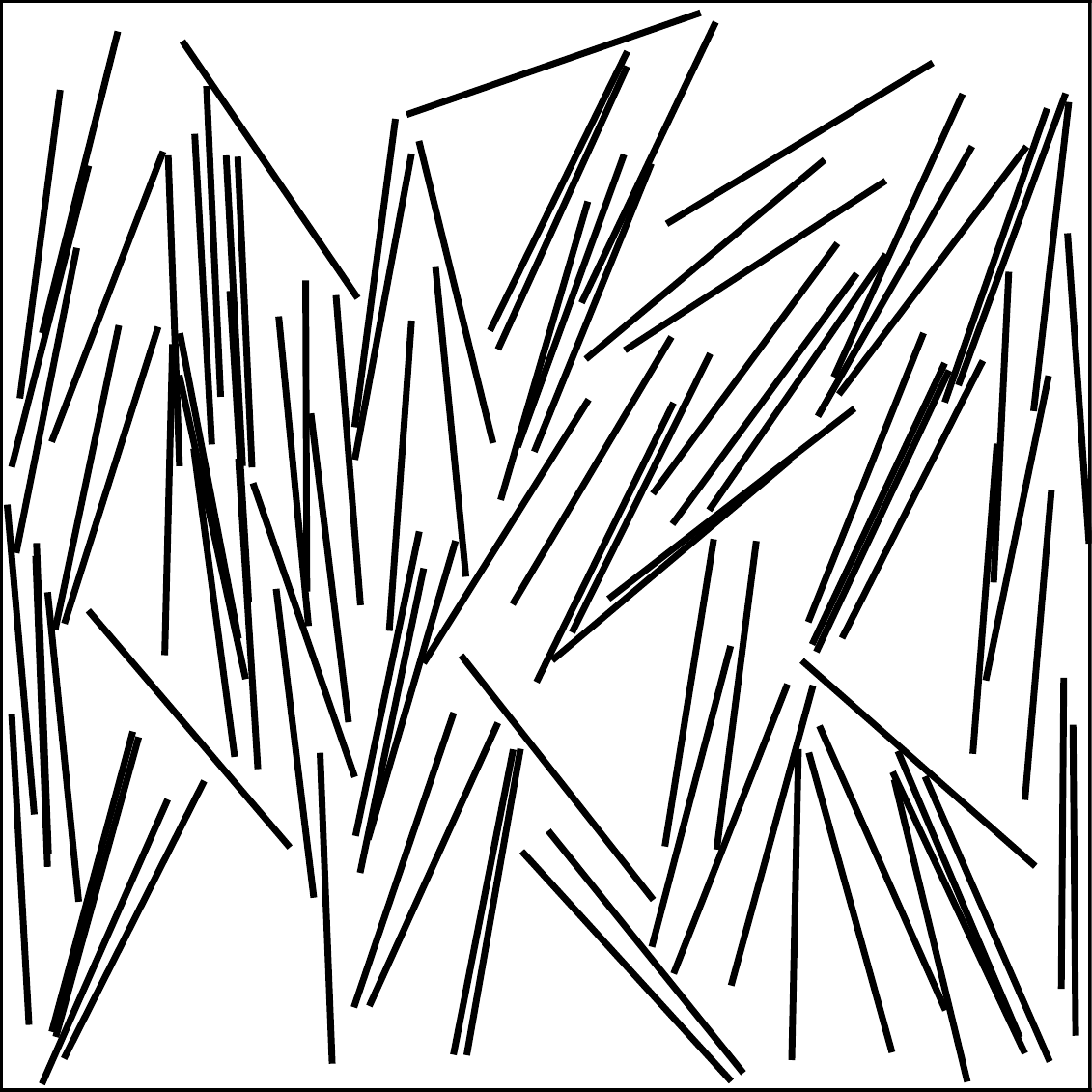}
    &
    \includegraphics[scale=\linesScale,align=c]{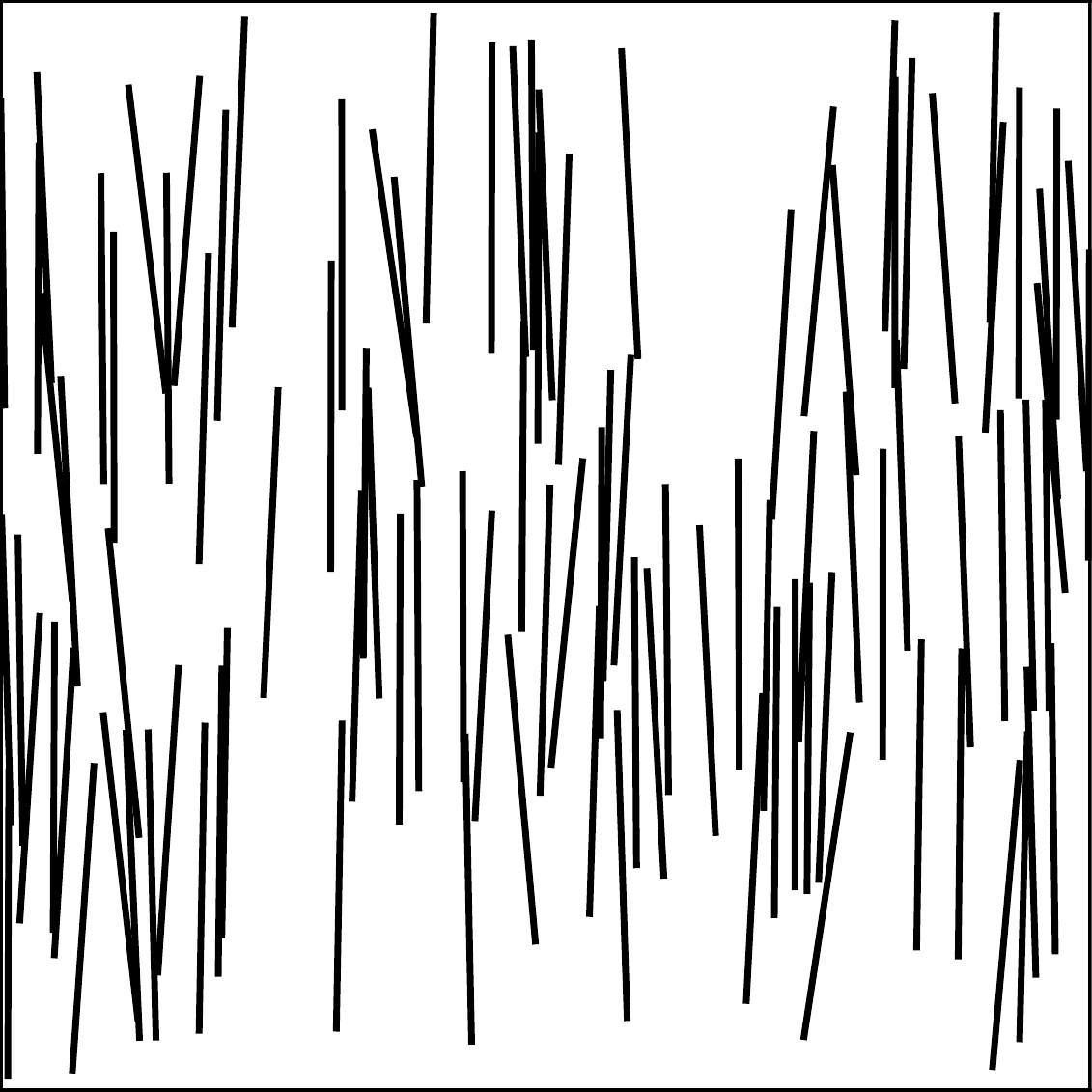}
    &
    \includegraphics[scale=\linesScale,align=c]{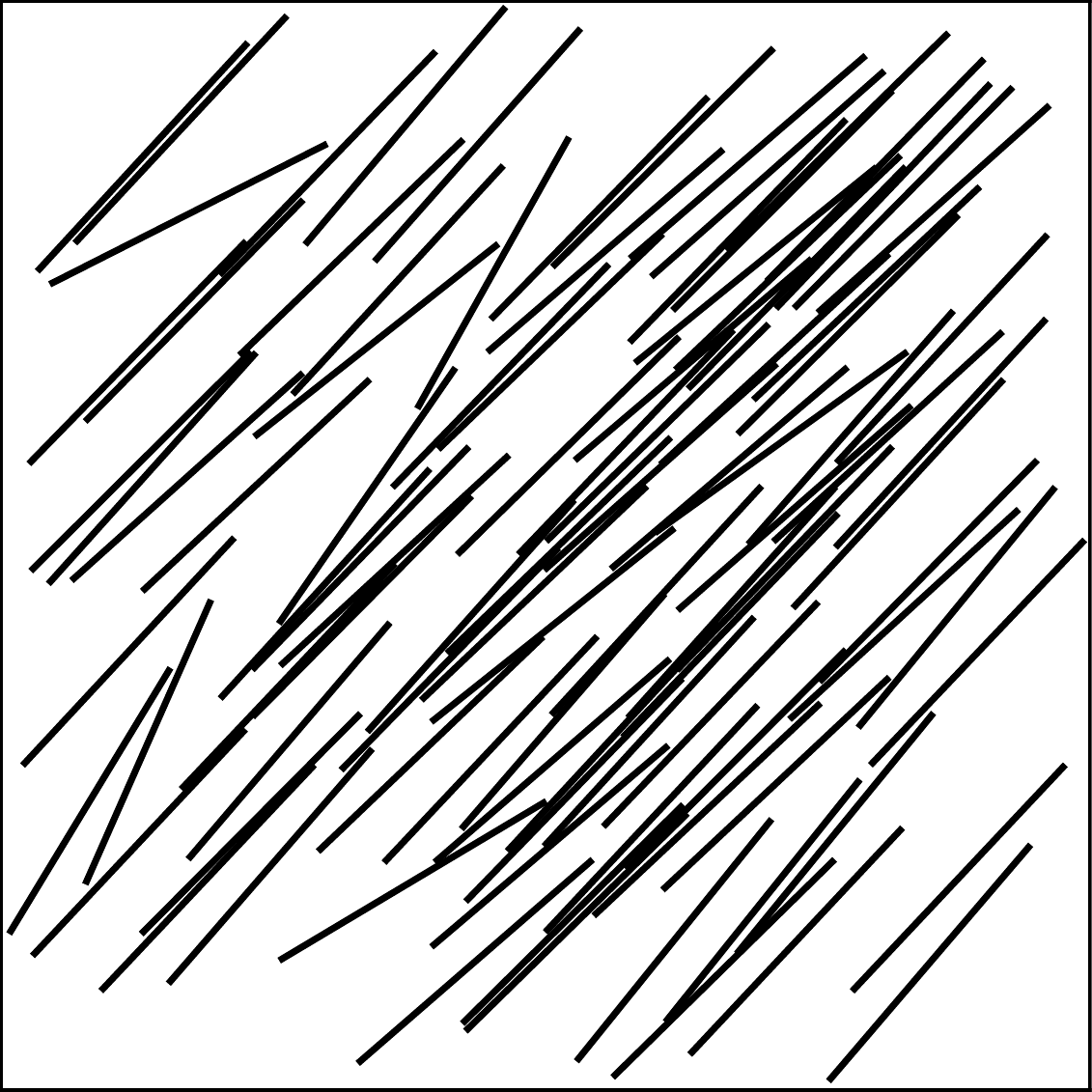}
\\
\rotatebox[origin=c]{90}{\small 1000 lines} &
    \includegraphics[scale=\linesScale,align=c]{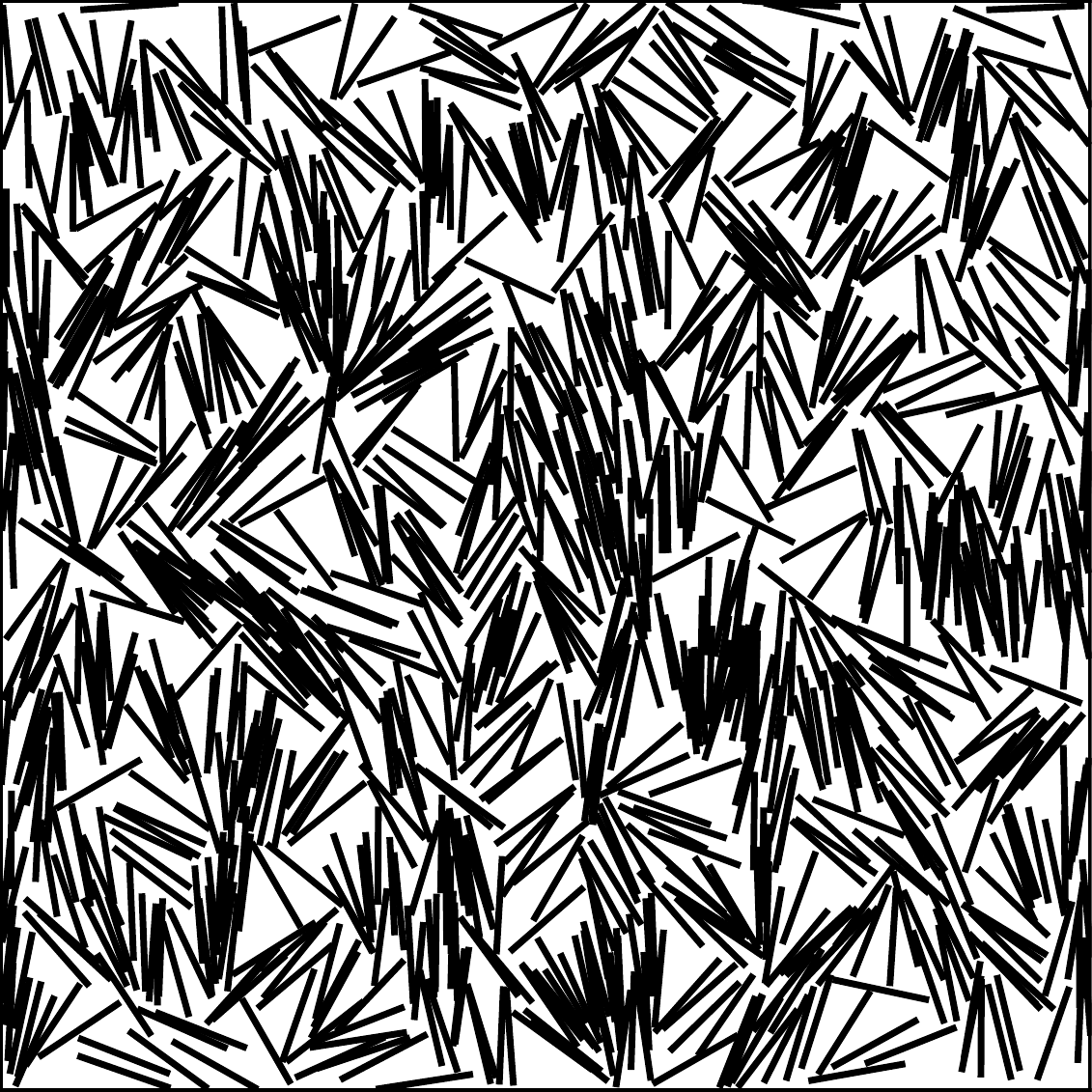}
    &
    \includegraphics[scale=\linesScale,align=c]{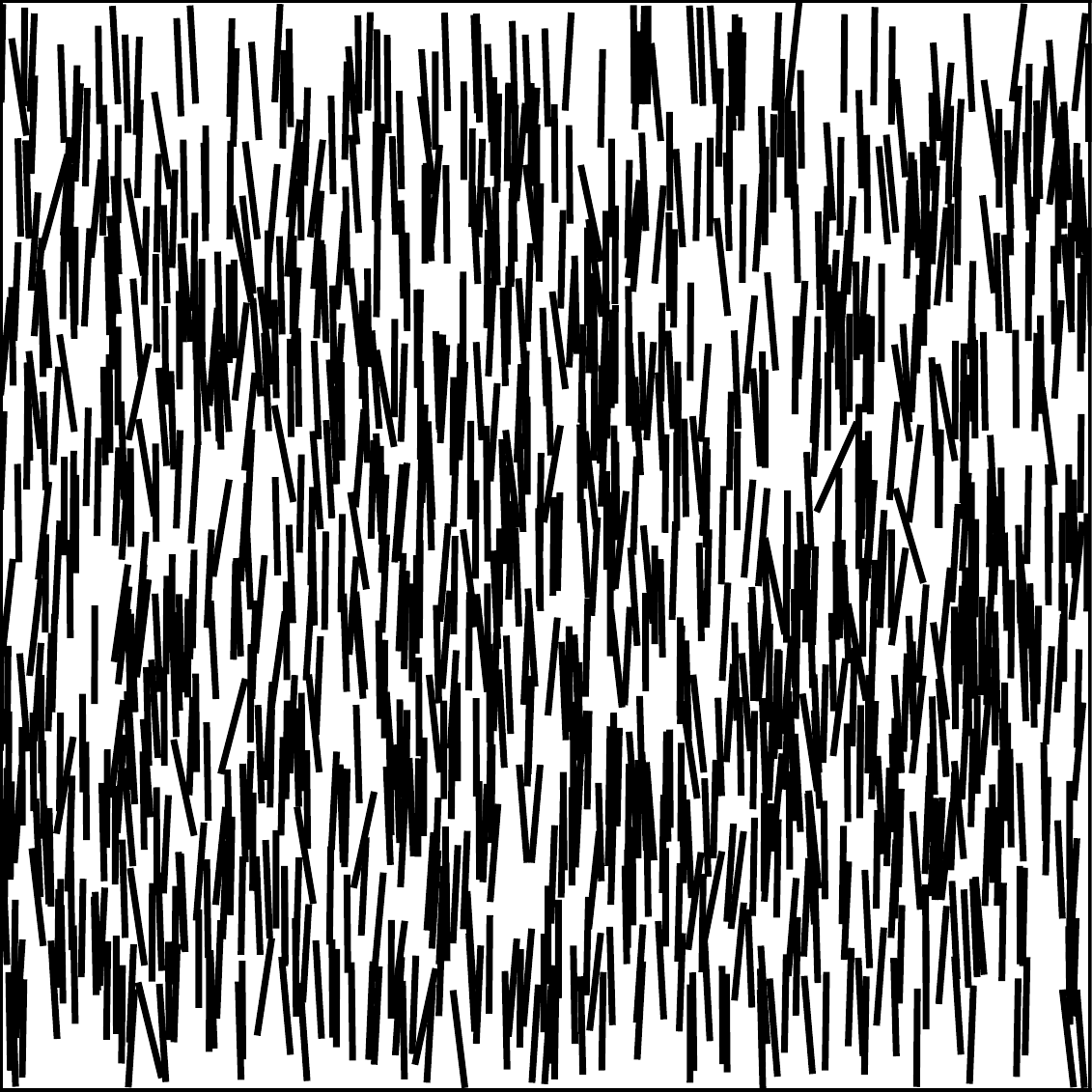}
    &
    \includegraphics[scale=\linesScale,align=c]{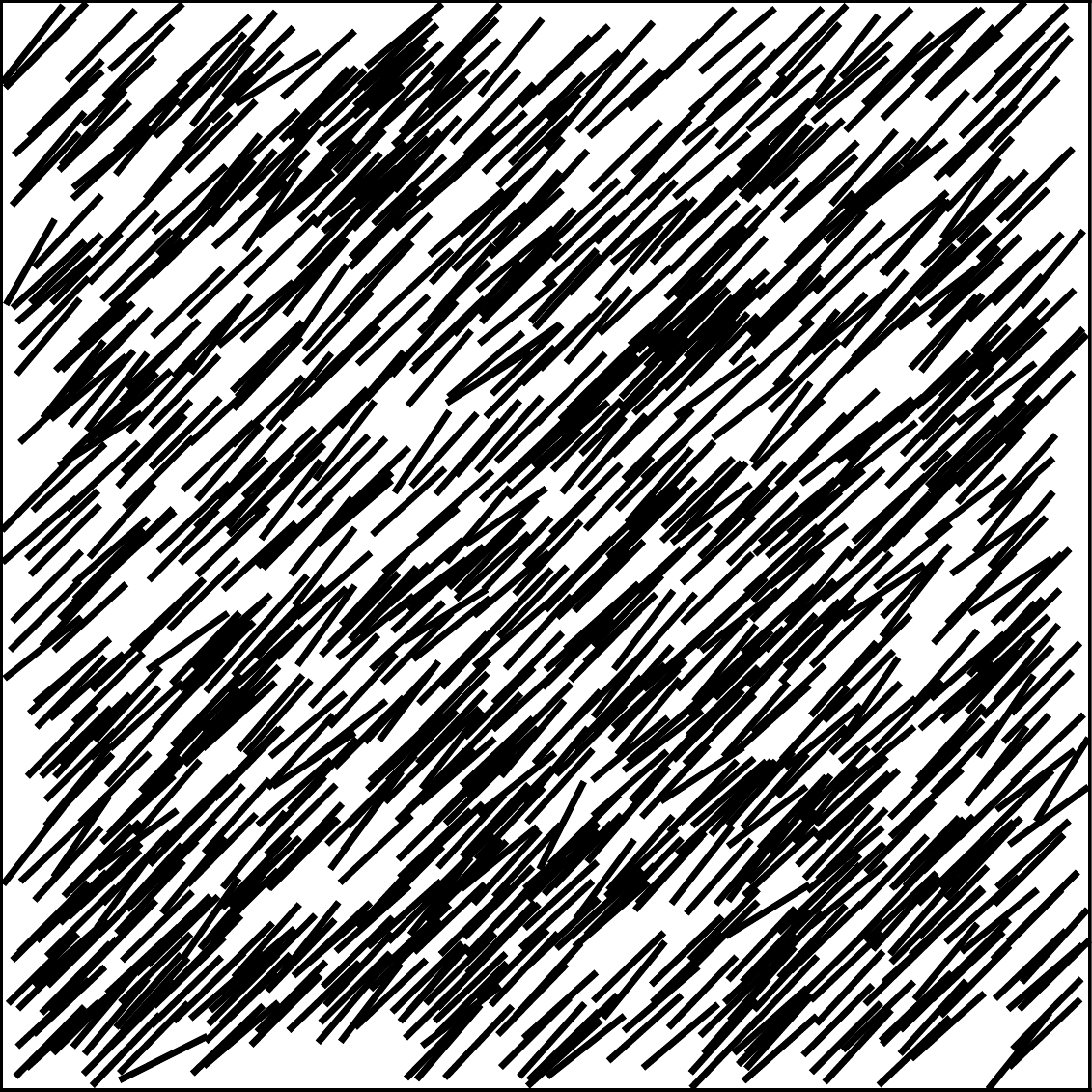}
\end{tabular}
\end{center}
\caption{Similar scenes as those in Fig.~\ref{figLinesPolySC1}, except that the line lengths are increased by a factor of 3 (and clamped to a maximum of 0.95 for low $N$). This tighter packing leads to local correlations for the `uniform orientation' case (left).
\\ 
}
\label{figLinesPolySC3}
\end{figure}

\begin{figure}
\begin{center}
\begin{tabular}{c@{\ }c@{\ }c@{\ }c}
& \multicolumn{3}{c}{\textsc{Length factor 0.1}}\\
& uniform orientation & preferentially vertical & preferentially diagonal
\\
\rotatebox[origin=c]{90}{\small 1 line} &
    \includegraphics[scale=\linesScale,align=c]{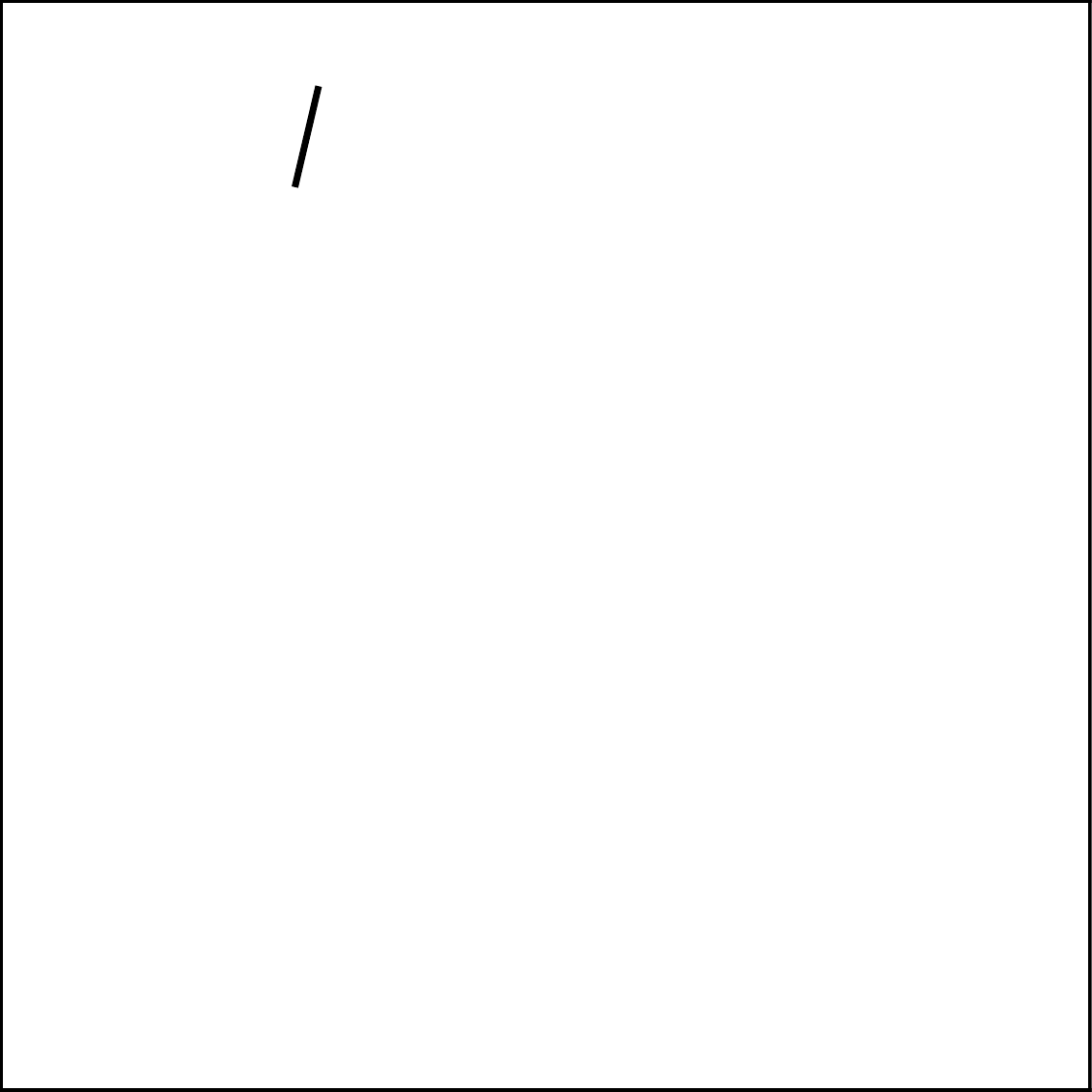}
    &
    \includegraphics[scale=\linesScale,align=c]{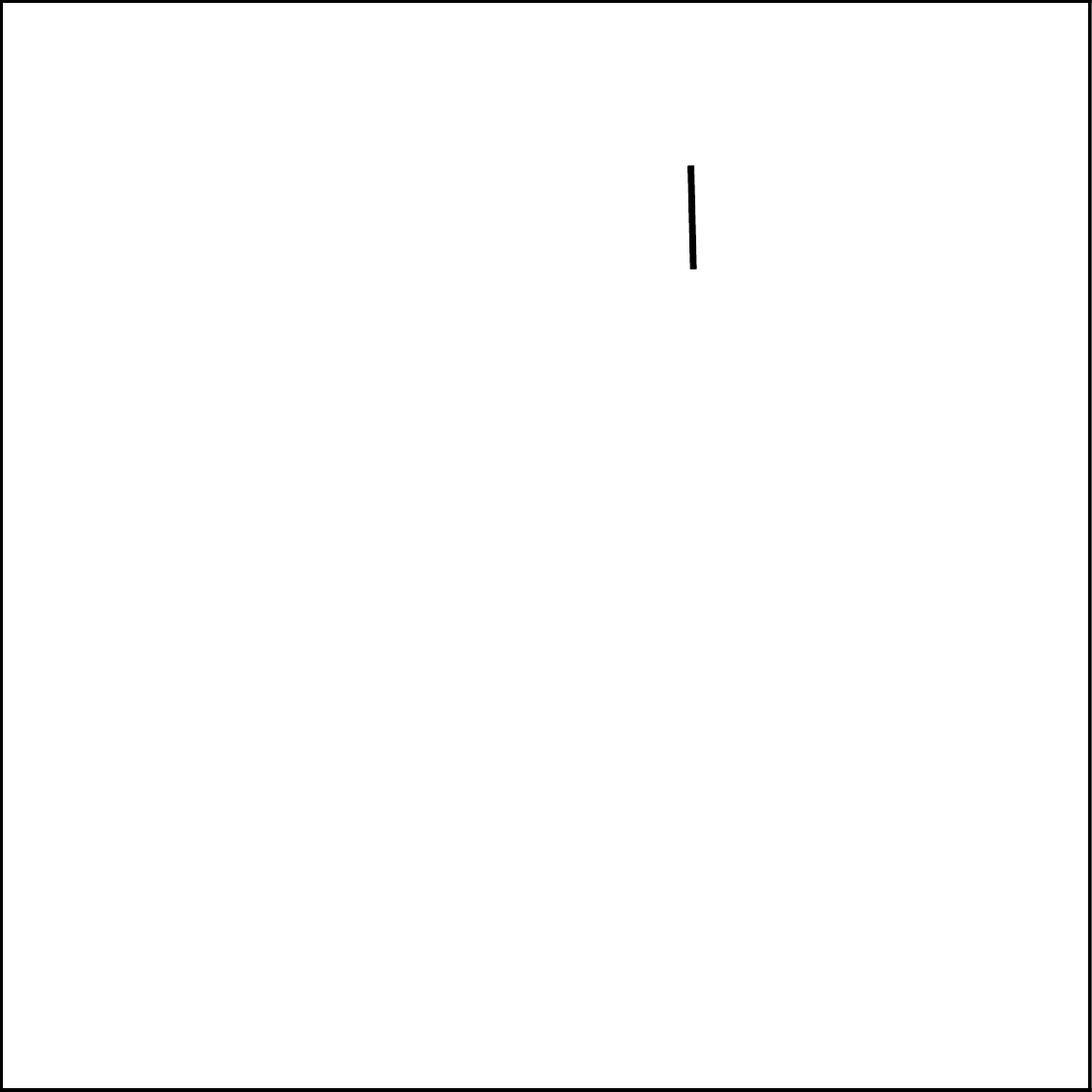}
    &
    \includegraphics[scale=\linesScale,align=c]{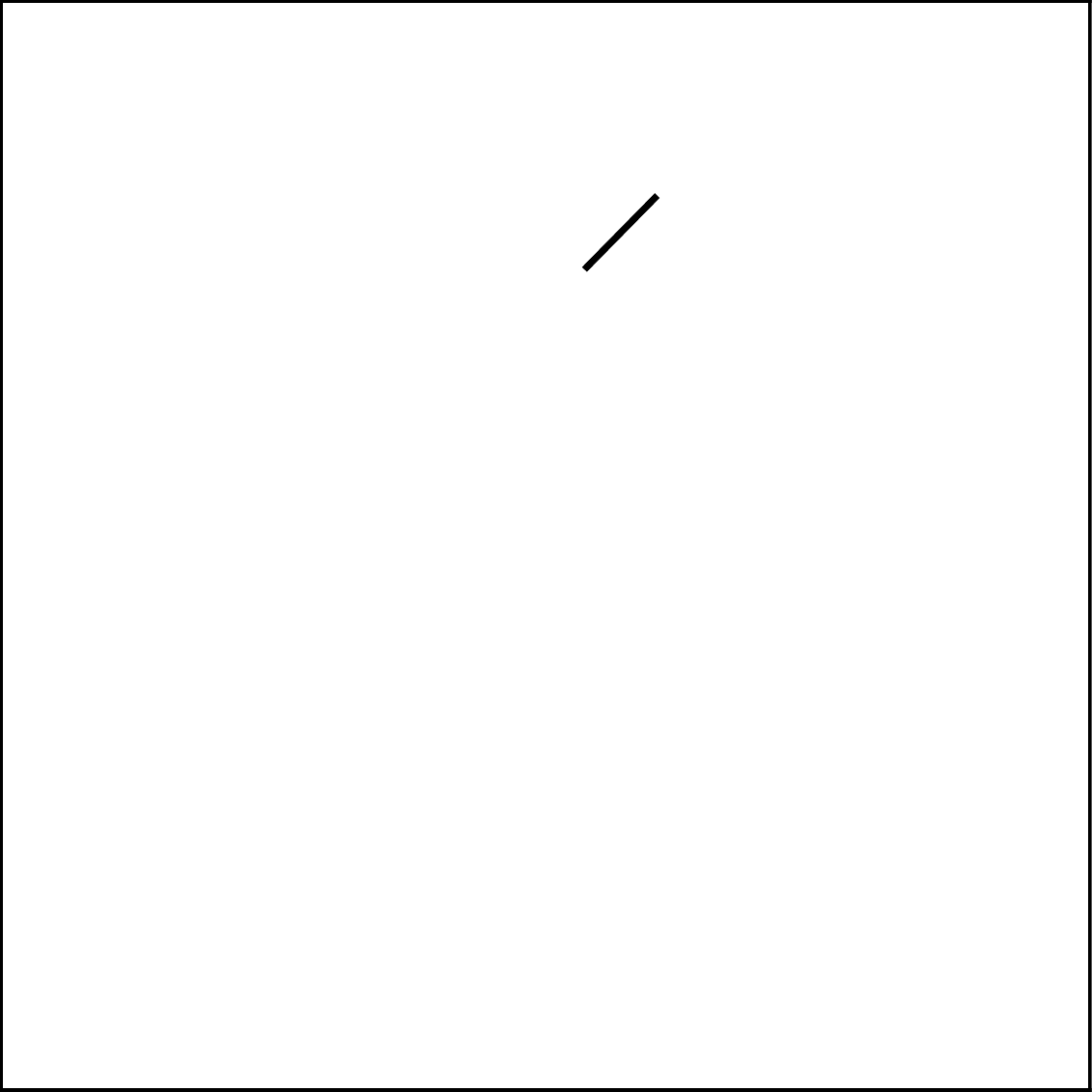}
\\
\rotatebox[origin=c]{90}{\small 10 lines} &
    \includegraphics[scale=\linesScale,align=c]{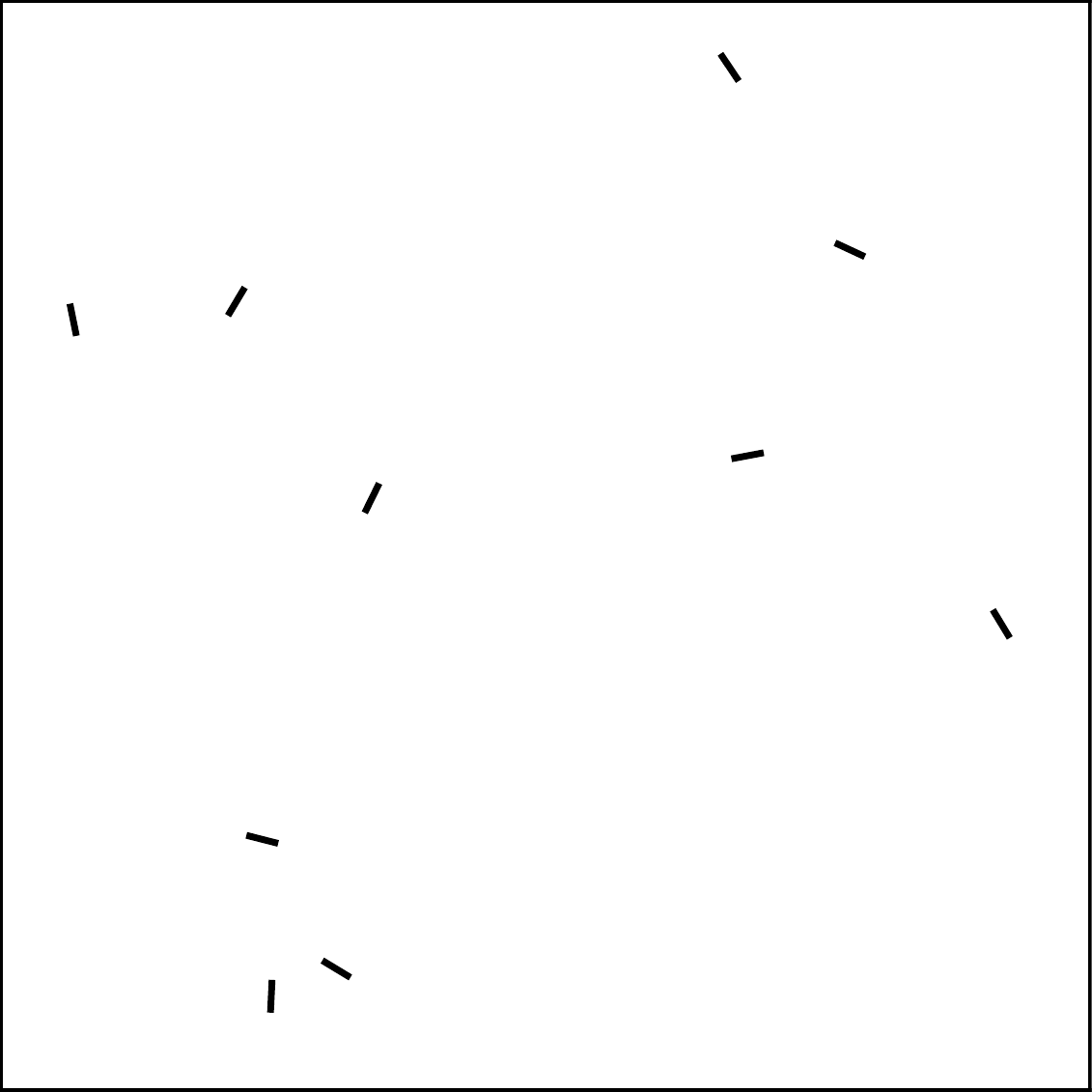}
    &
    \includegraphics[scale=\linesScale,align=c]{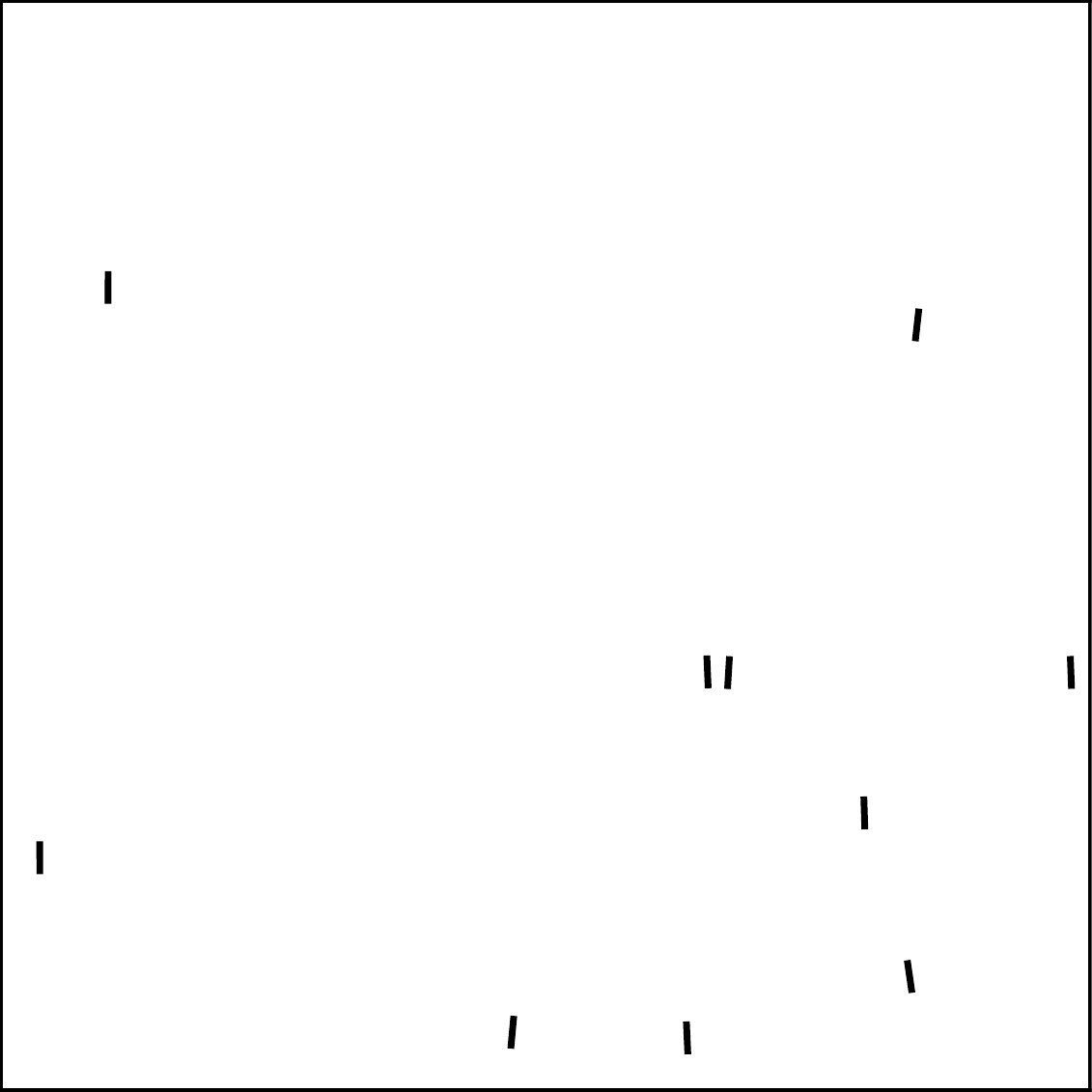}
    &
    \includegraphics[scale=\linesScale,align=c]{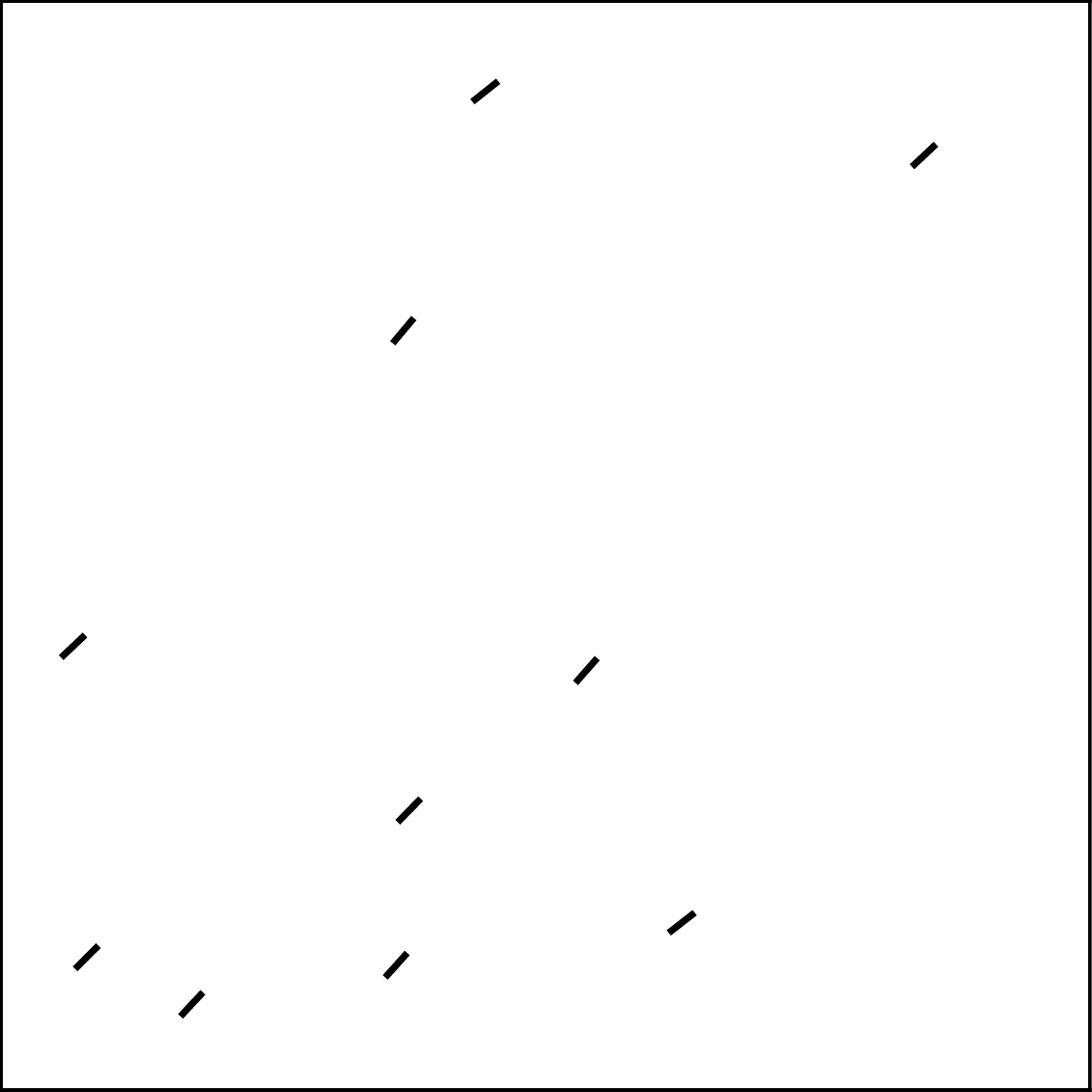}
\\
\rotatebox[origin=c]{90}{\small 100 lines} &
    \includegraphics[scale=\linesScale,align=c]{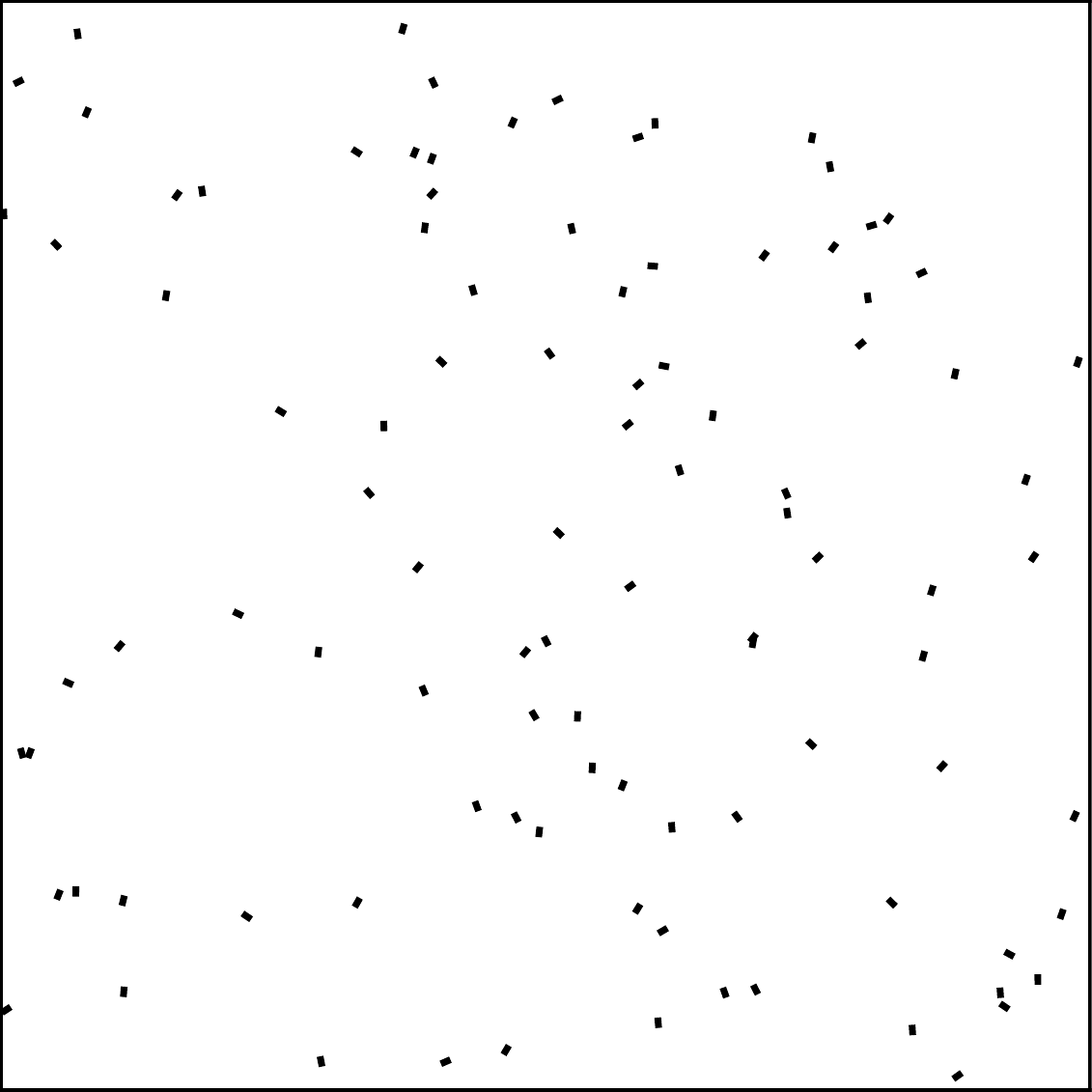}
    &
    \includegraphics[scale=\linesScale,align=c]{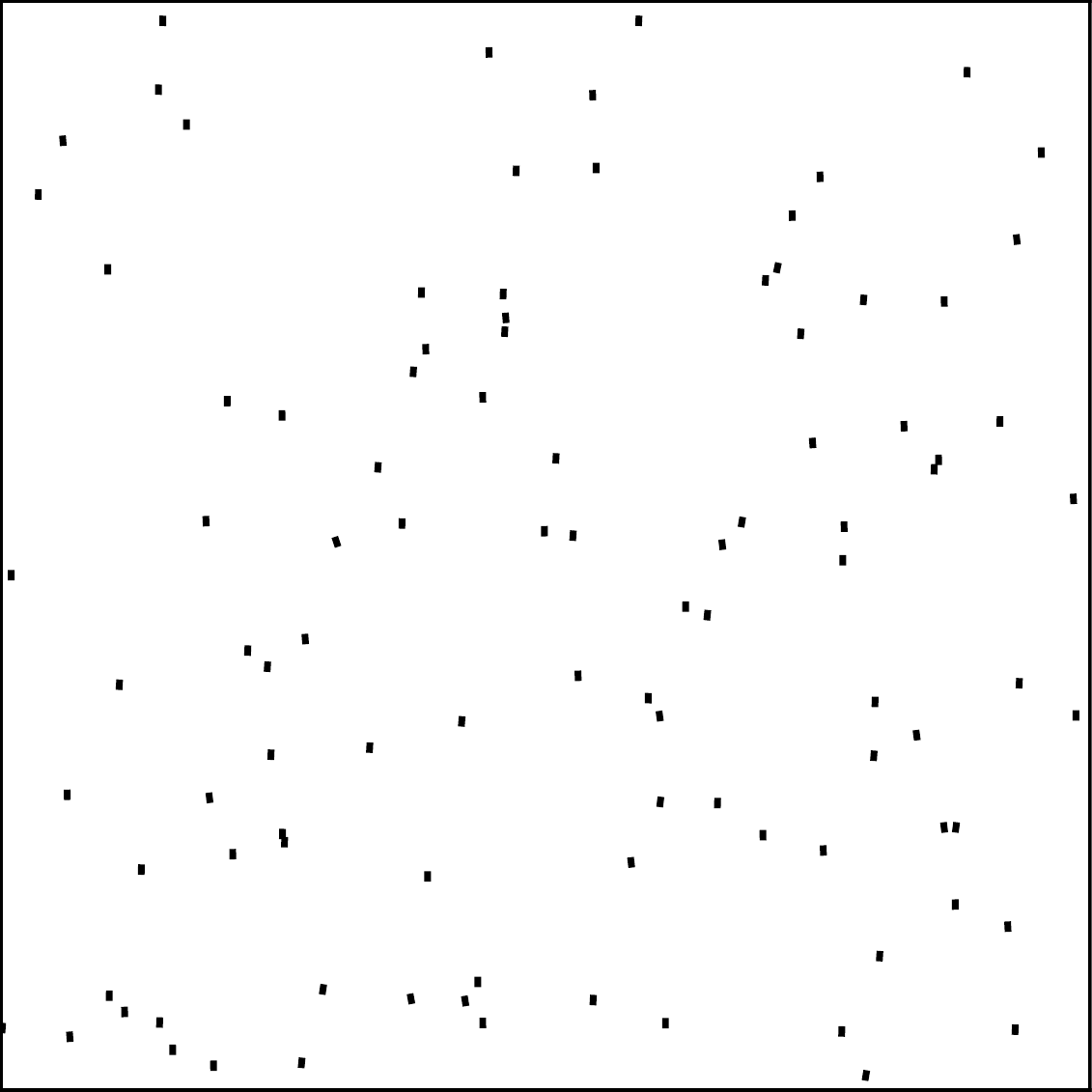}
    &
    \includegraphics[scale=\linesScale,align=c]{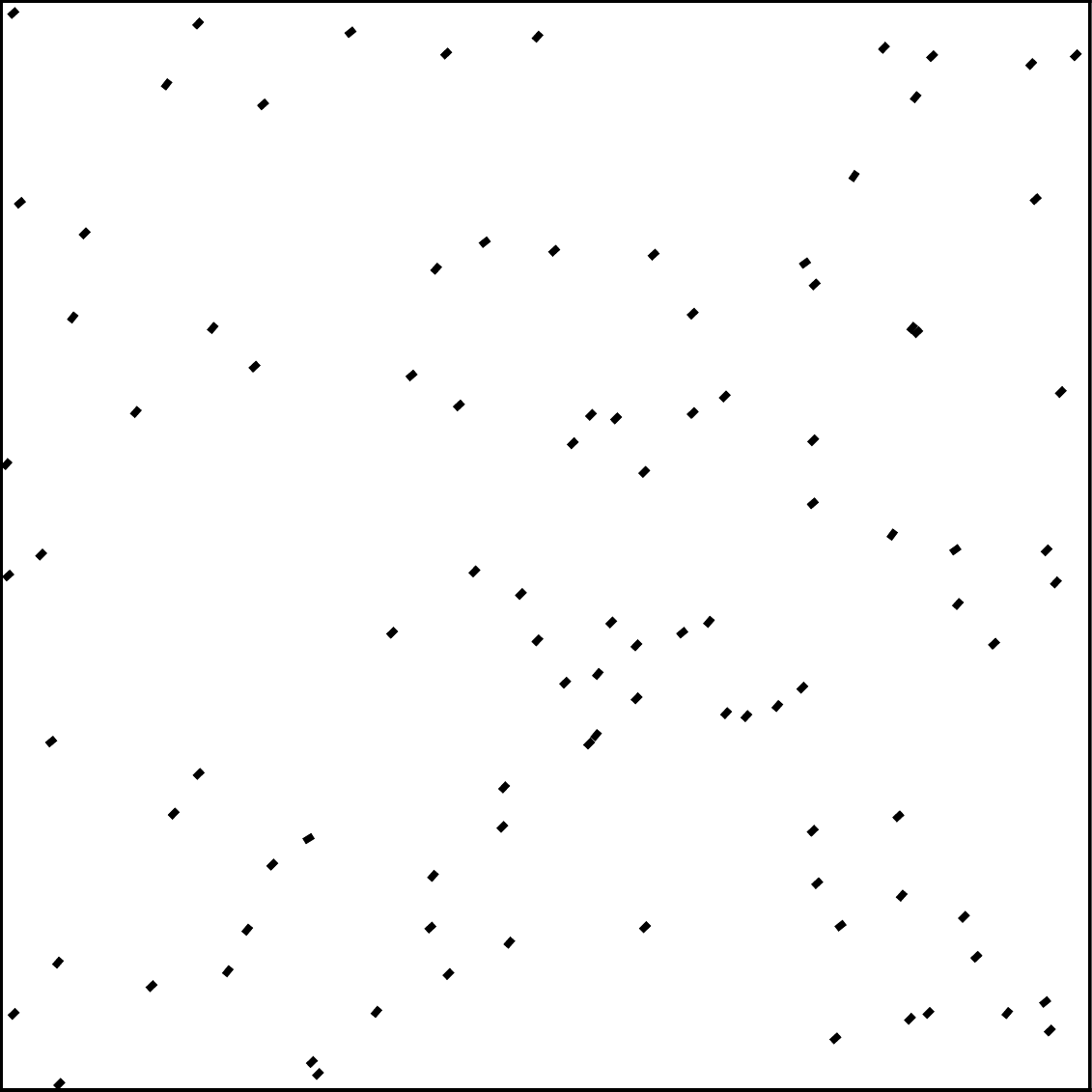}
\\
\rotatebox[origin=c]{90}{\small 1000 lines} &
    \includegraphics[scale=\linesScale,align=c]{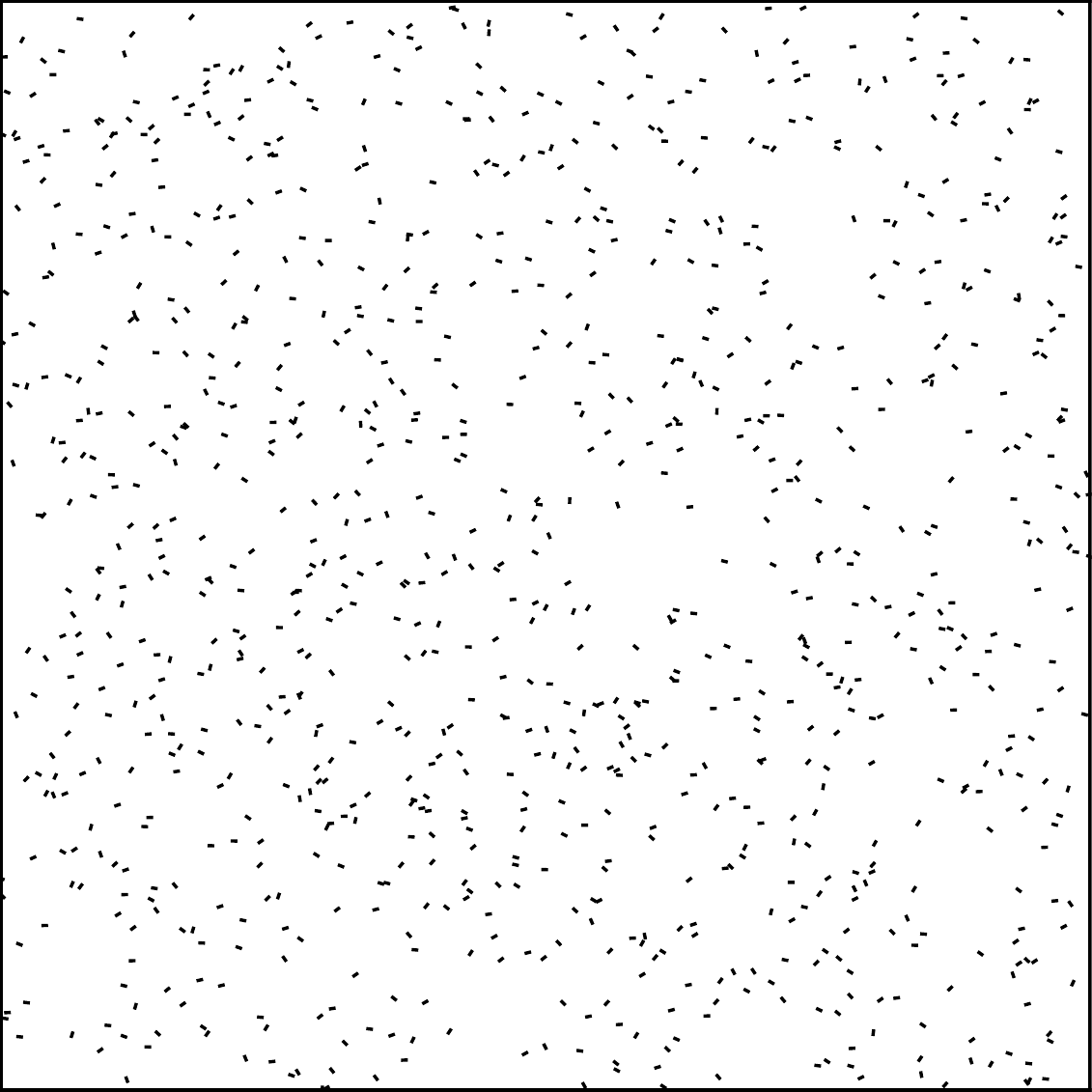}
    &
    \includegraphics[scale=\linesScale,align=c]{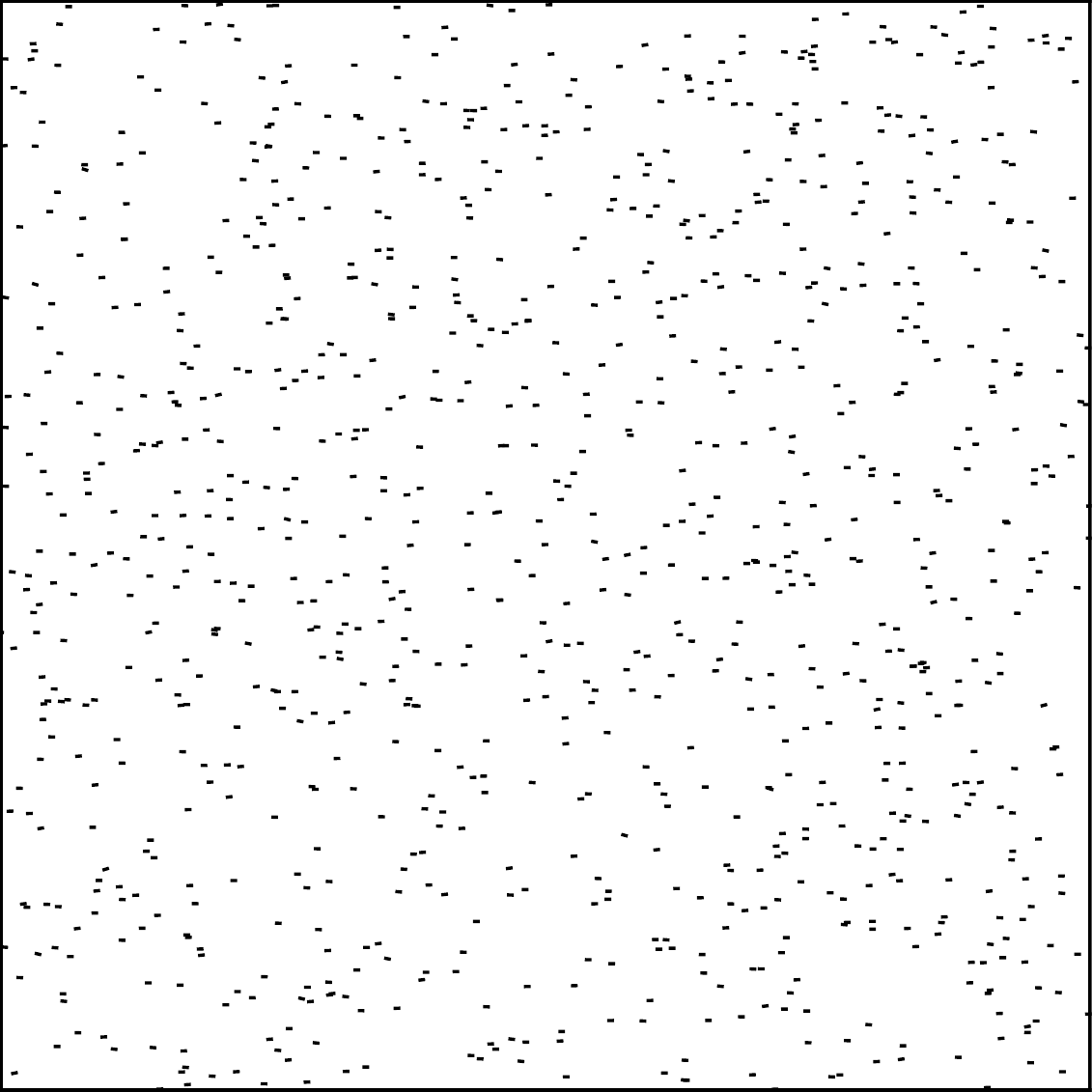}
    &
    \includegraphics[scale=\linesScale,align=c]{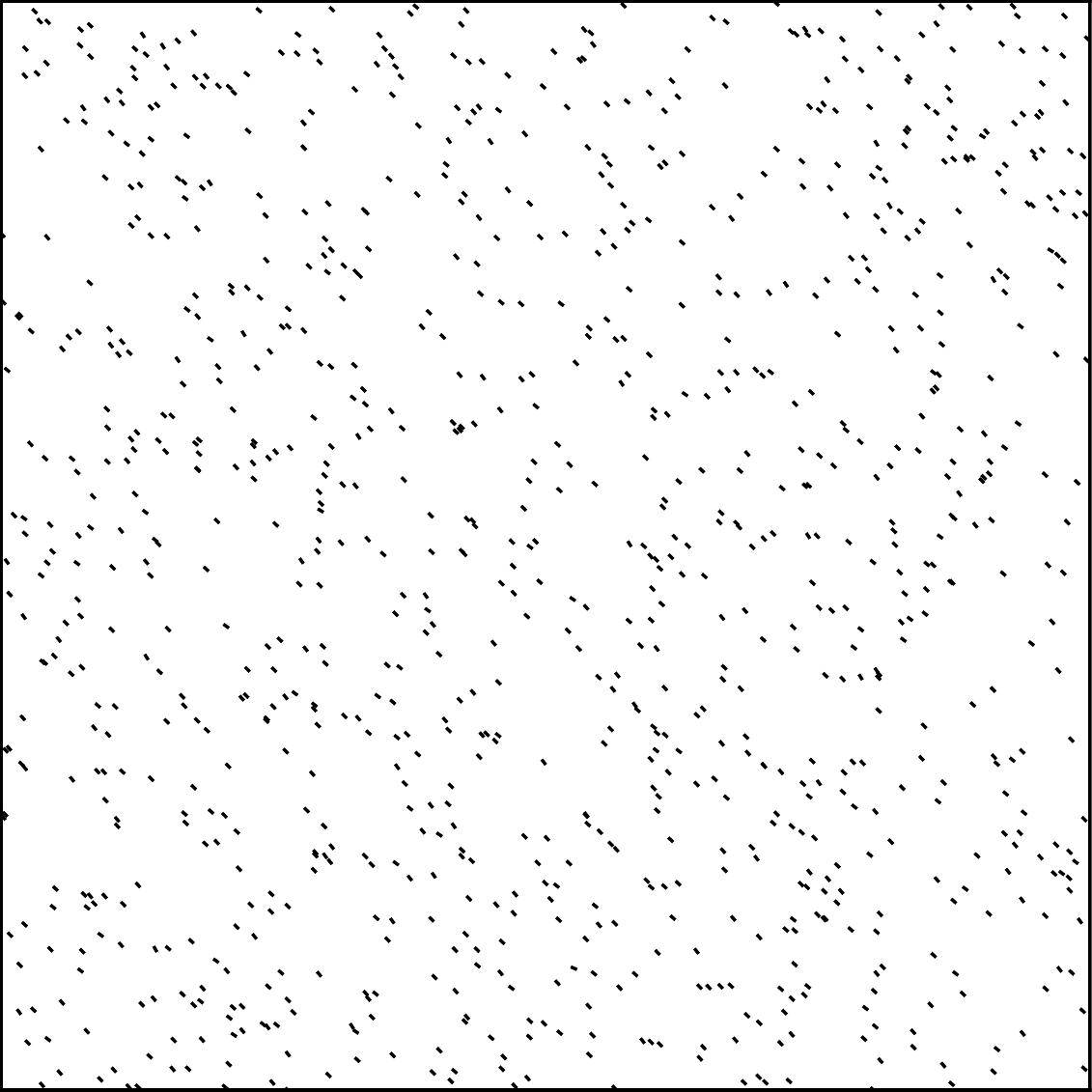}
\end{tabular}
\end{center}
\caption{Similar scenes as those in Fig.~\ref{figLinesPolySC1}, except that the line lengths are shortened with a factor of 0.1.
The line segments become point-like and, their triangulation behaves similarly to triangulations of uniform point sets, as we will see.
\\ 
}
\label{figLinesPolySC0p1}
\end{figure}

\usetikzlibrary{calc,spy}
\newcommand{\linesExampleHeight}{4.5cm} 
\tikzstyle{closeup} = [
  opacity=1.0,
  height=2.1cm,
  width=2.1cm,
]
\tikzstyle{largewindow} = [line width=0.50mm] 
\tikzstyle{smallwindow} = [line width=0.50mm]
\newcommand{\linesSpyImage}[1]{
\begin{tikzpicture}[x=6cm, y=6cm, spy using outlines={every spy on node/.append style={smallwindow}}]
\node[anchor=south,inner sep=0,outer sep=0] (theFig) at (0,0) {
    \includegraphics[height=\linesExampleHeight]{#1}
};
\spy [closeup,magnification=2.5,green] on ($(theFig)+(-0.00,+0.29)$) 
    in node[largewindow,anchor=north west,green] at ($(theFig.south west) + (0.25mm,-0.03)$); 
\spy [closeup,magnification=2.5,red] on ($(theFig)+(+0.1,+0.13)$) 
    in node[largewindow,anchor=north east,red] at ($(theFig.south east) + (-0.25mm,-0.03)$);
\end{tikzpicture}
}
\begin{figure}
\begin{center}
\begin{tabular}{c@{\ \,}c@{\ \,}c}
{\small Unrefined CDT} &
{\small Optim.\ refined CDT} &
{\small Optimized (ours)}
\\
    \linesSpyImage{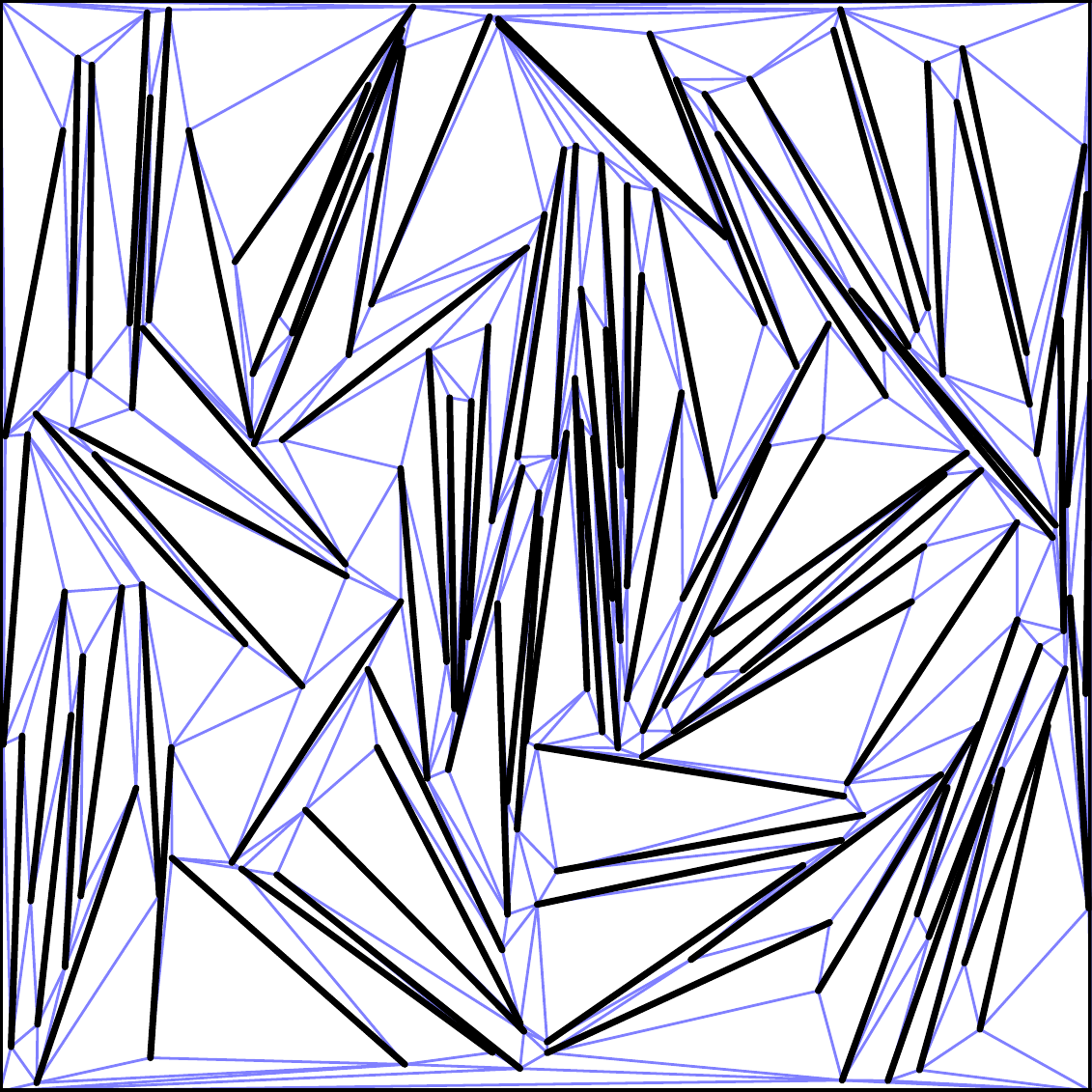}

    &
    \linesSpyImage{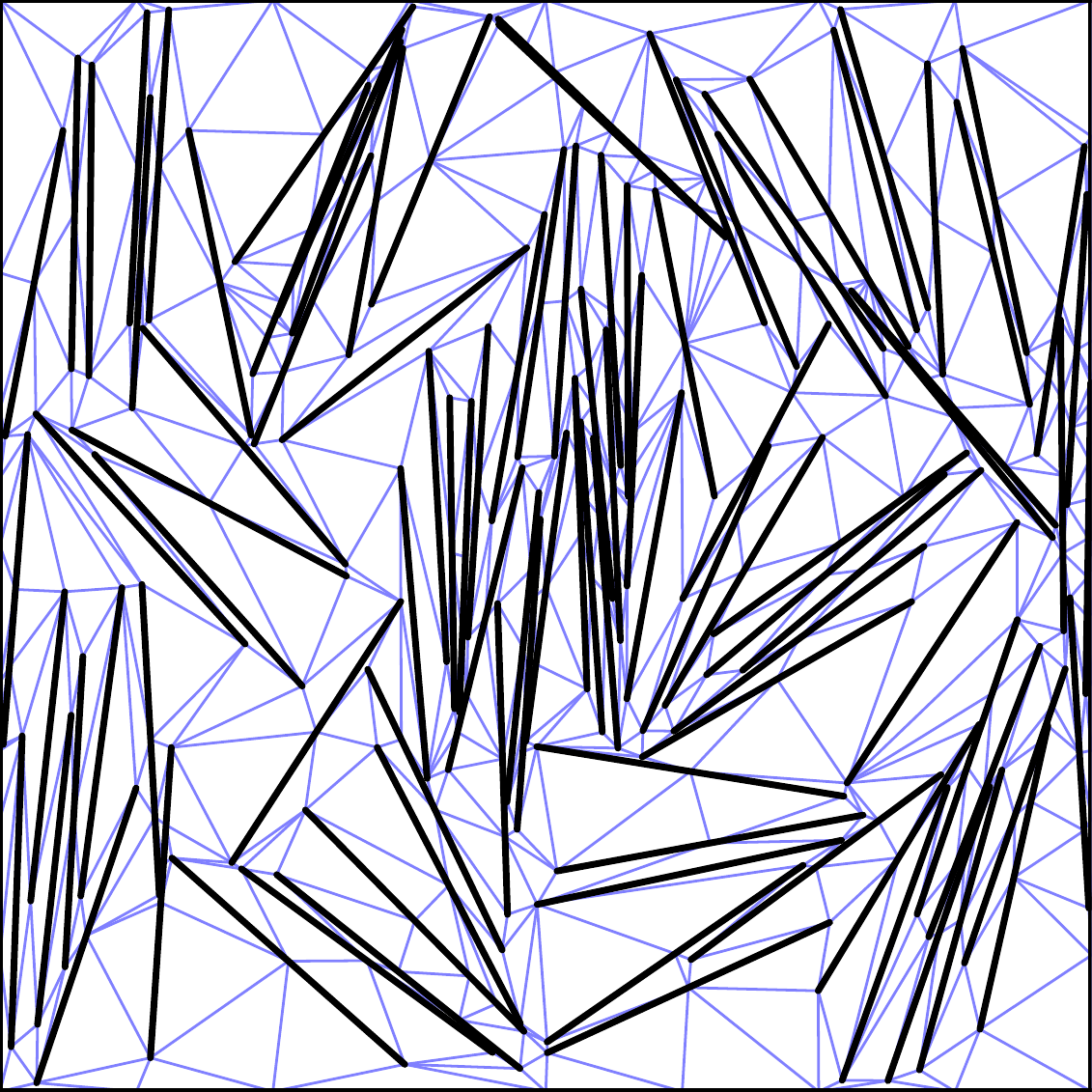}
    &
    \linesSpyImage{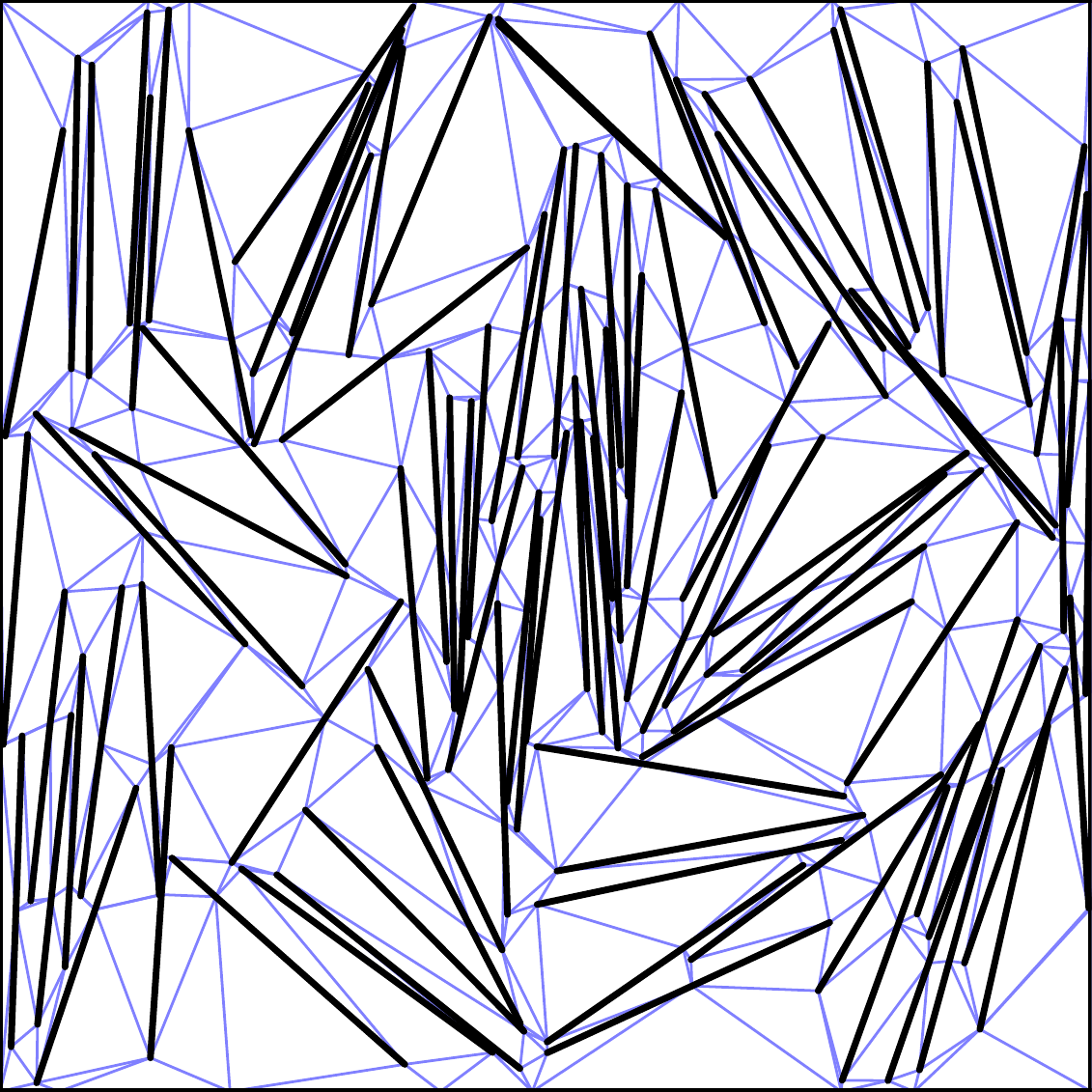}
\\
\small 94.4 & \small 88.2 & \small 80.7
\end{tabular}
\end{center}
\caption{
    An example of the `lines' scene ($N=100$, long uniformly oriented lines with length factor 3 as in Fig.~\ref{figLinesPolySC3}) that shows the trouble with maximising the minimal angle.
    Compared to the unrefined CDT (left), the optimal refined CDT (middle) avoids the long triangles around the borders of the scene as well as within the scene (green close-up), but it leads to an oversplitting of triangles elsewhere (red close-up). Our optimized triangulation (right) improves on the best of both cases. The total edge lengths are given below the scenes.
}
\end{figure}

\FloatBarrier

\subsection{Synthetic Scenes: Grass \label{secOptimGrass}}

To examine the behaviour of (refined) CDTs in comparison to our approximately optimal minimum weight triangulations more generally, we have generated the `grass' scenes shown in Figs.~\ref{figGrassPoly} resembling two-dimensional `grass' where $N$ (non-intersecting) leaves are sampled uniformly along the horizontal dimension. The leaves are enclosed in a rectangle that is slightly larger than a tightly fitting bounding box and the scene is normalized so that its longest side has unit length.

\begin{figure}
\newcommand{\grassPlotScale}{1.0}
\begin{center}
\begin{tabular}{c@{\ }c@{\ }c}
& \small 3 segments/leaf 
& \small 10 segments/leaf
\\
\rotatebox[origin=c]{90}{\small Relative to unrefined CDT} &
    \includegraphics[scale=\grassPlotScale,align=c]{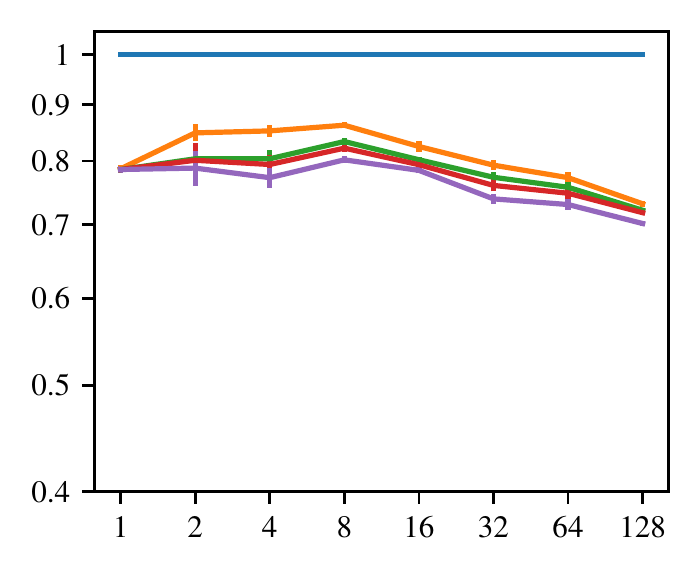} &
    \includegraphics[scale=\grassPlotScale,align=c]{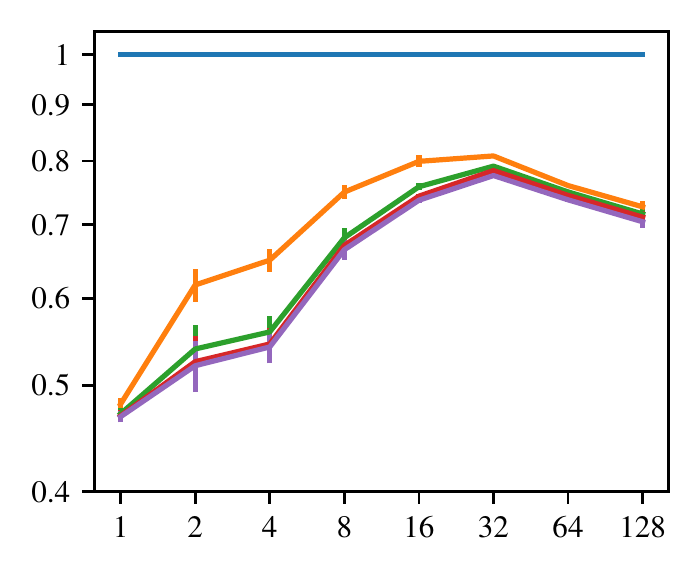}
\\[-3mm]
\rotatebox[origin=c]{90}{\small Relative to optim.~refined CDT} &
    \includegraphics[scale=\grassPlotScale,align=c]{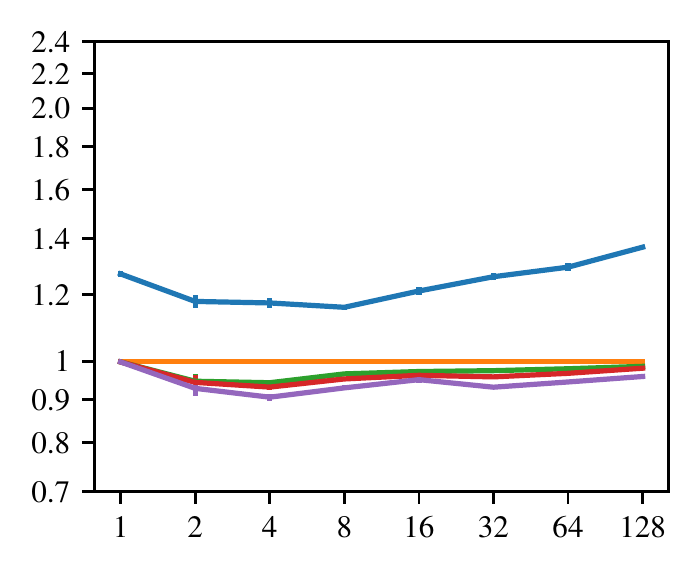} &
    \includegraphics[scale=\grassPlotScale,align=c]{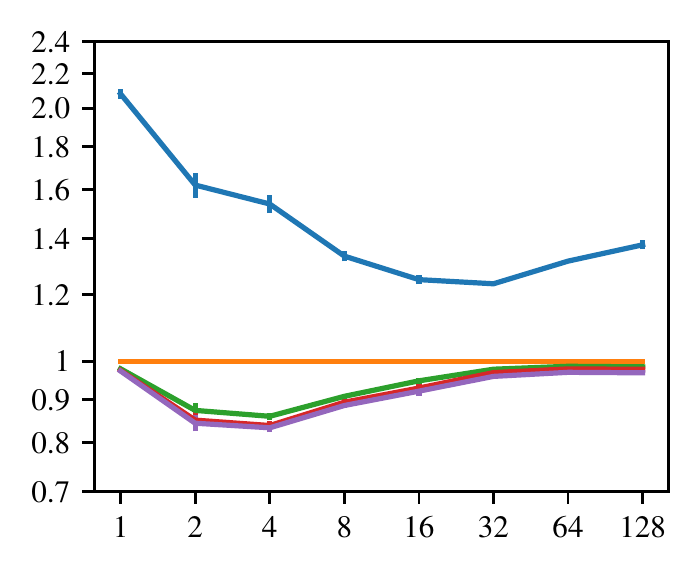}
\\[-4mm] 
& \small ~~~~~$N$
& \small ~~~~~$N$
\\[1mm]
& \multicolumn{2}{c}{\includegraphics[align=c]{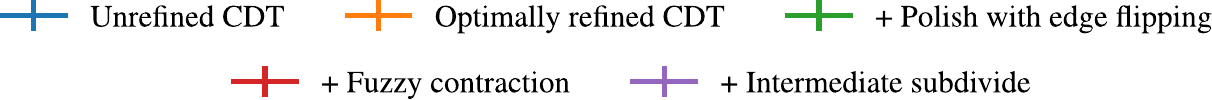}}
\end{tabular}
\end{center}
\caption{
    Relative total edge lengths for triangulations of the grass scene for successively powerful optimization strategies. The top plots show the total edge lengths relative to the total edge length of an unrefined CDT, the bottom plot shows the total edge lengths relative to that of the optimally refined CDT.
    }
\label{figGrassPlots}
\end{figure}

The relative impact of successively more powerful optimization strategies on the total edge length of triangulations for these grass scenes is shown in Fig.~\ref{figGrassPlots}. Comparing the left and right plots shows that the different techniques move further apart for the more finely modelled leaves (10 segments/leaf). Overall, the improved edge length over an unrefined CDT for the 10 segments/leaf (right top plot) dives down to 50\%, whereas for the coarser leaves with only 3 segments (left top plot), the best improvement is around 70\% of the unrefined CDT. The differences relative to the optimally refined CDT (bottom plots) are also more pronounced for the 10 segments/leaf scenes than the 3 segments/leaf scenes, although the additional gain from our full optimization is most prominent around $N = 2 \sim 8$, and becomes negligible for $N \to 1$ and for very high $N$. We discuss and explain this behaviour by examining Figs.~\ref{figGrassNOneTriangs}, \ref{figGrassNFourTriangs} and \ref{figGrassNOneTwoEightTriangs}. 

Figure \ref{figGrassNOneTriangs} shows a generated scene for a single ($N = 1$) leaf of grass with 10 segments. The unrefined CDT has many long edges that connect vertices of the leaf to one of the edges of the scene's bounding box. This situation gets improved when refining the CDT by introducing Steiner vertices on the sides of the bounding box. The resulting optimally refined CDT is already very close to our fully optimized result due to the simplicity of this scene. For $N=1$ scenes with 3 segments/leaf instead of 10, a similar observation holds, although the gap between the edge length of the suboptimal unrefined CDT and the optimal triangulation is less wide due to the fewer segments and thus relatively fewer `bad connections' to corners of the bounding box in the unrefined CDT.

For less trivial scenes with $N \ne 1$, a more interesting dynamic appears. Figure \ref{figGrassNFourTriangs} shows examples of scenes with $N=4$ leaves. Here, the regions between the outermost leaves and the edges of the bounding box behave similarly as in the $N=1$ case of Fig.~\ref{figGrassNOneTriangs}, but there are now also spaces between two leaves that need to be triangulated. When these `voids' are wide relative to the segment length, then a straightforward direct connection between neighbouring leaves leads to a highly suboptimal triangulation. The optimally refined CDT can alleviate this problem somewhat by introducing Steiner vertices, but when the goal is to minimize the total edge length, then this still leaves room for improvement as can be seen in our optimized triangulations. This observation explains the dip around $N=4$ in the bottom plots of Fig.~\ref{figGrassPlots}.

As the number of leaves goes up, the gap between the optimally refined CDT and a fully optimized triangulation starts to narrow, as can be seen in Fig.~\ref{figGrassNOneTwoEightTriangs} for $N=128$. As the space between leaves gets smaller then the segment length, a straightforward connection of neighbouring segment vertices becomes the optimal strategy that cannot be improved. The fully optimized triangulations can only improve (1) voids between leaves that are larger than the segment length and (2) the connections between the leaves and the scene's bounding box, which contribute relatively less to the total edge length as $N$ goes up.

\newcommand{\grassScale}{0.35}
\begin{figure}
\begin{center}
\begin{tabular}{c@{\ }c@{\ }c}
& 3 segments/leaf & 10 segments/leaf
\\
\rotatebox[origin=c]{90}{\small 2 leaves} &
    \includegraphics[scale=\grassScale,align=c]{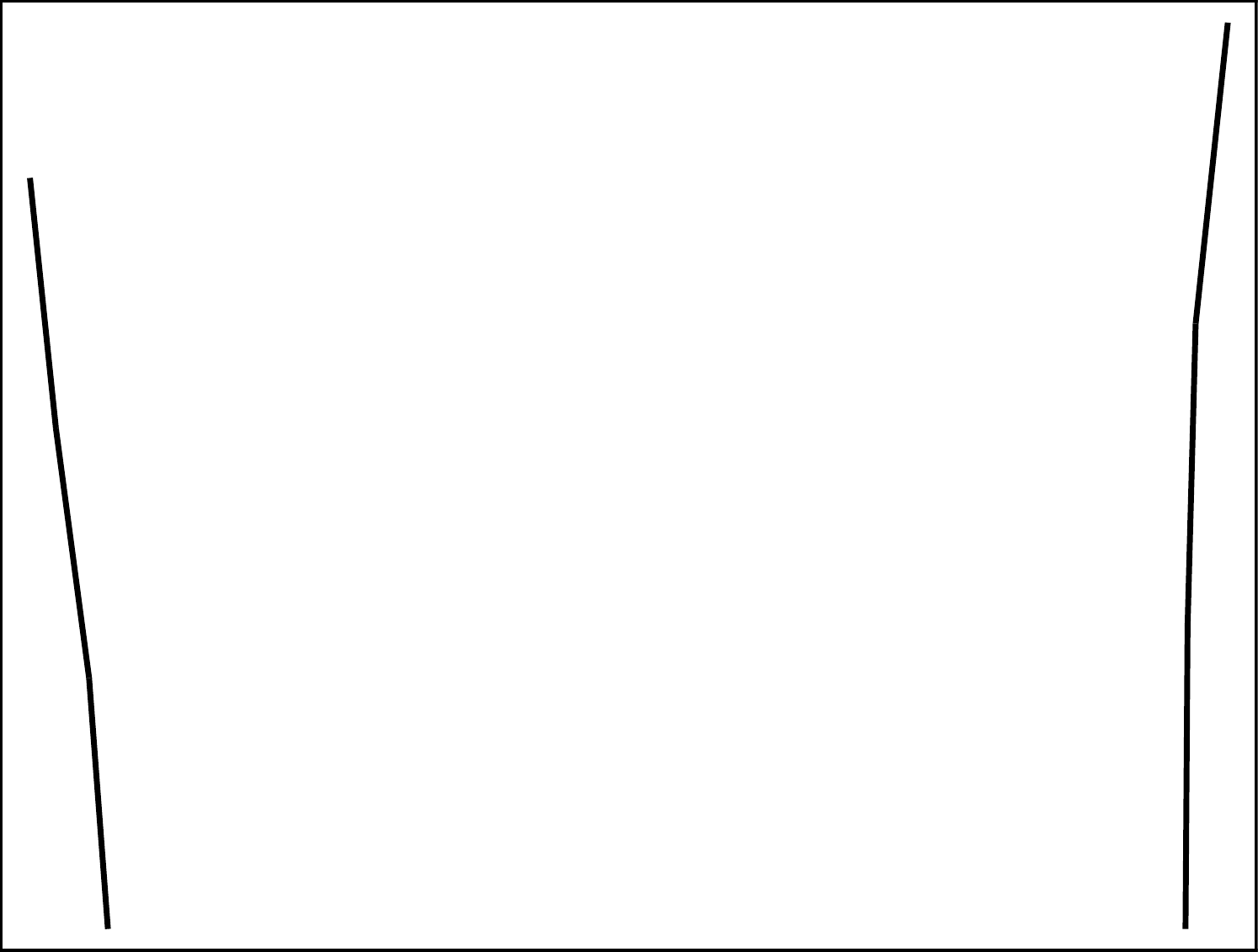}
    &
    \includegraphics[scale=\grassScale,align=c]{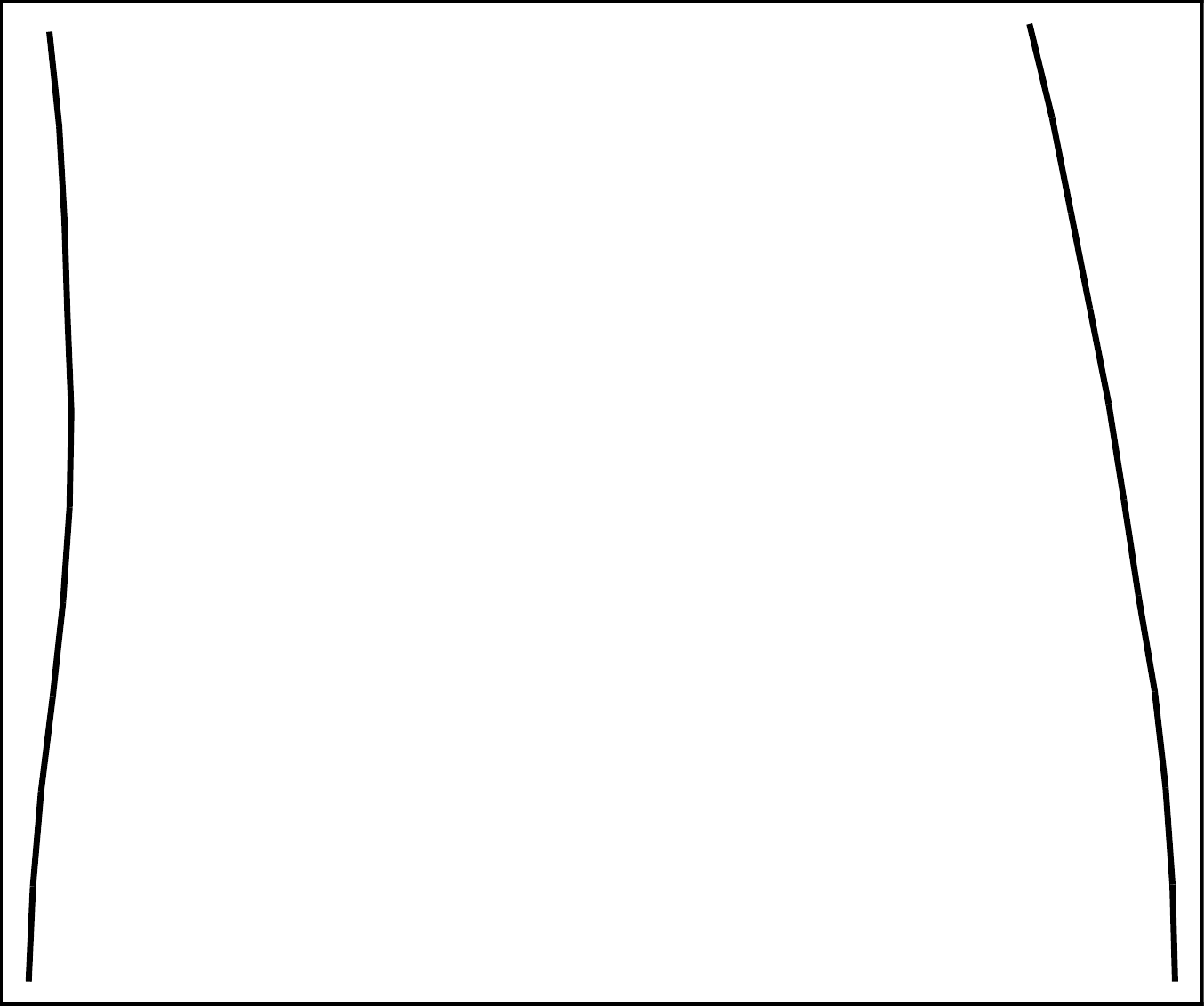}
\\
\rotatebox[origin=c]{90}{\small 8 leaves} &
    \includegraphics[scale=\grassScale,align=c]{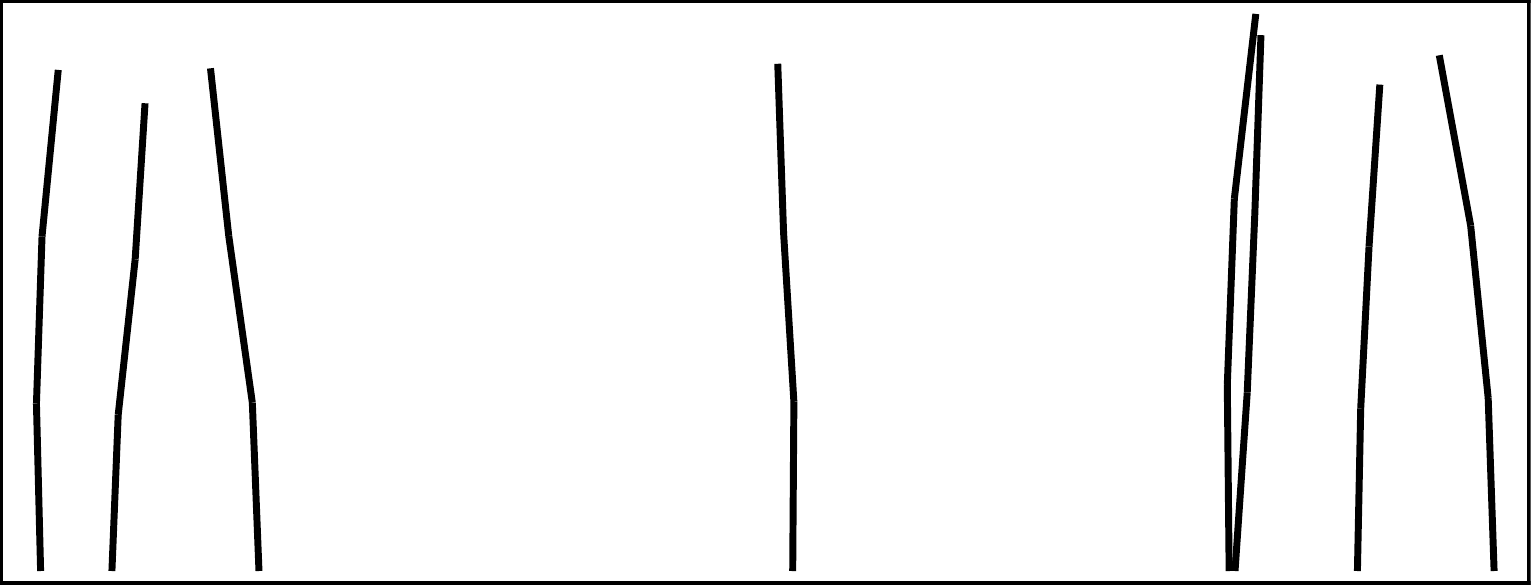}
    &
    \includegraphics[scale=\grassScale,align=c]{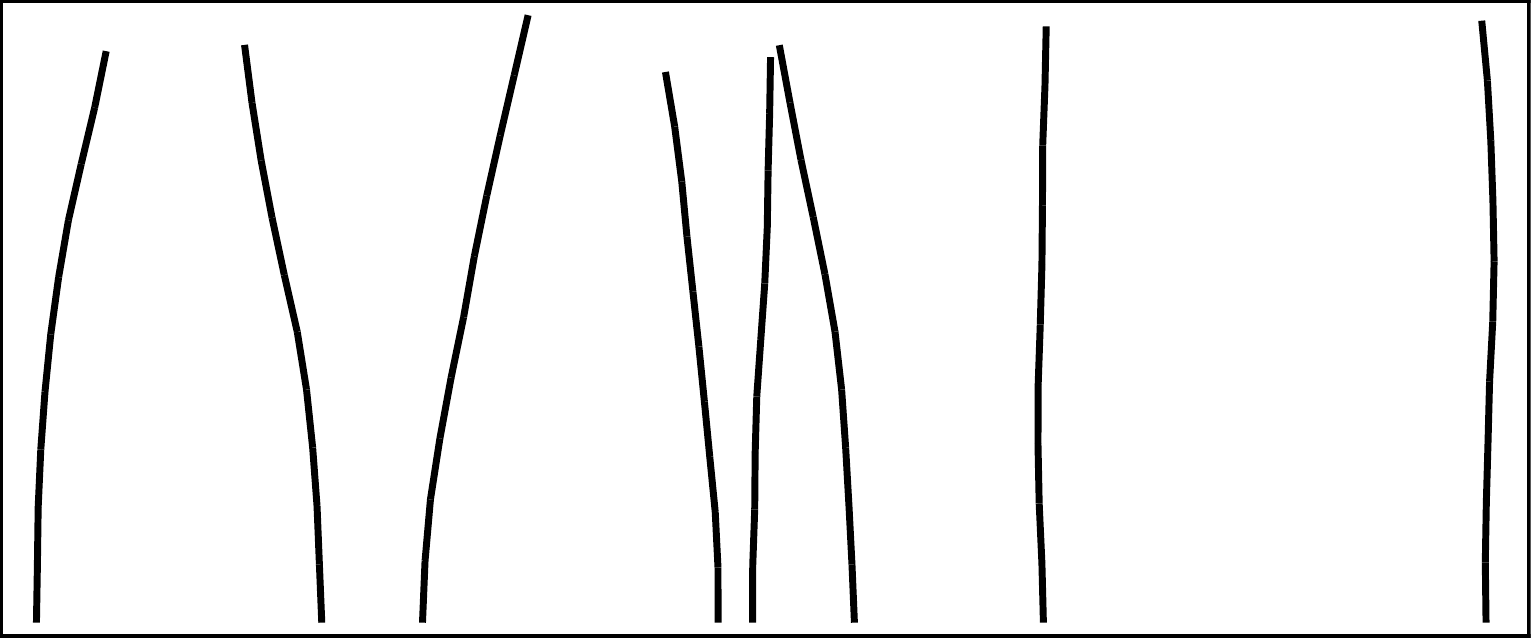}
\\
\rotatebox[origin=c]{90}{\small 32 leaves} &
    \includegraphics[scale=\grassScale,align=c]{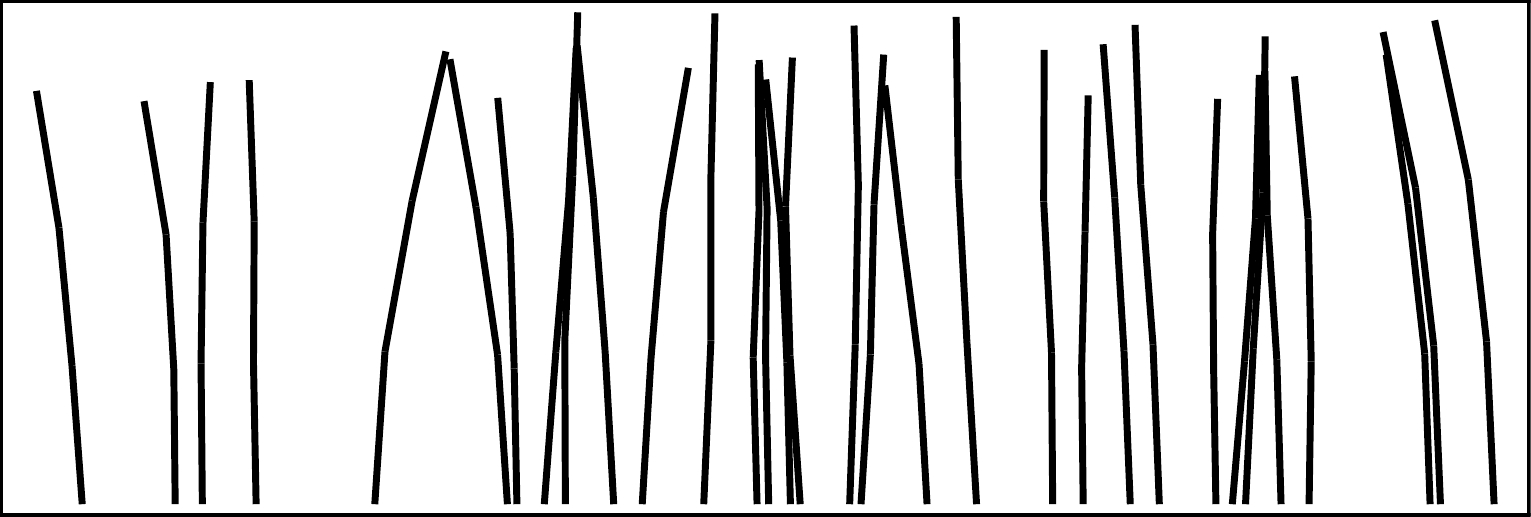}
    &
    \includegraphics[scale=\grassScale,align=c]{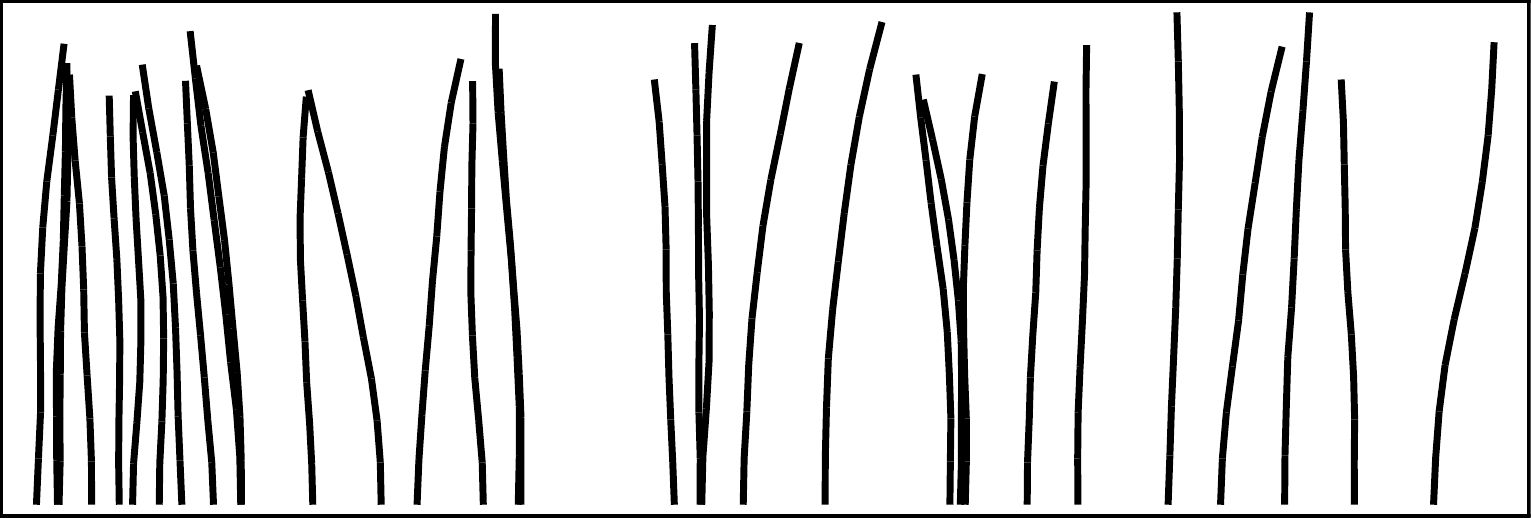}
\\
\rotatebox[origin=c]{90}{\small 128 leaves} &
    \includegraphics[scale=\grassScale,align=c]{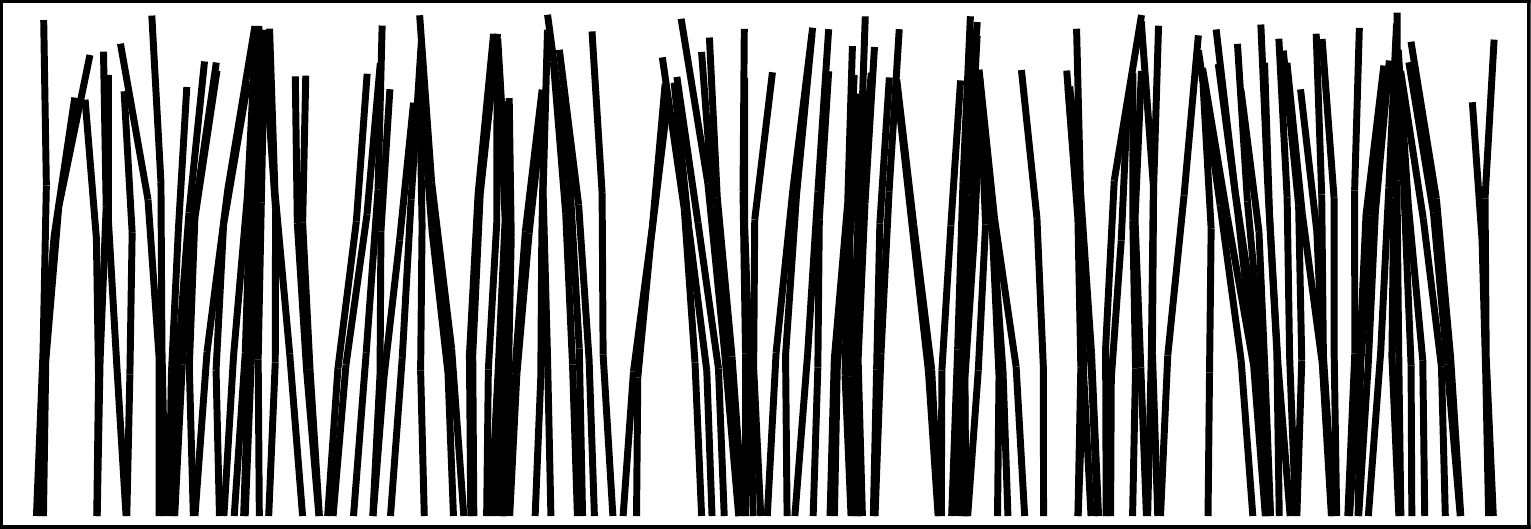}
    &
    \includegraphics[scale=\grassScale,align=c]{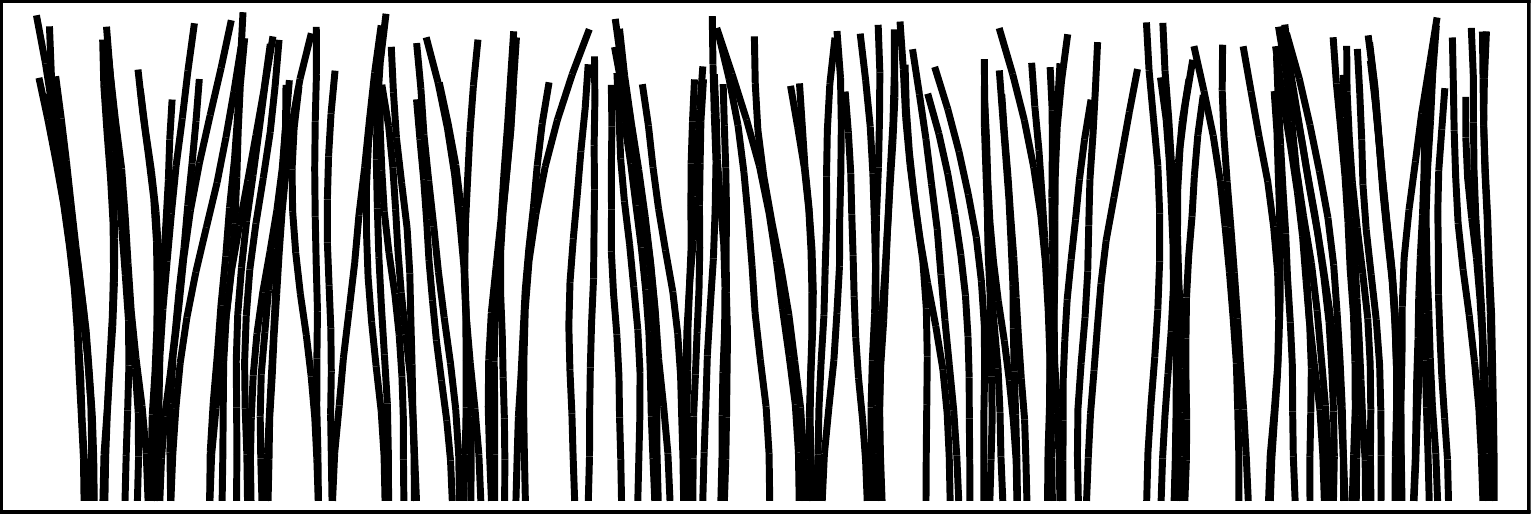}
\end{tabular}
\end{center}
\caption{Examples of the `grass' scene with varying number of leaves and segments per leaf.
\label{figGrassPoly}}
\end{figure}

\newcommand{\grassNOneExampleHeight}{3.8cm}
\newcommand{\grassNOneArrowSpace}{2.5mm}
\begin{figure}
\begin{center}
\begin{tabular}{ccccccc}
\mbox{\hspace{-1cm}\small Unref.\ CDT         \hspace{-1cm}}&&
\mbox{\hspace{-1cm}\small Optim.\ ref.\ CDT       \hspace{-1cm}}&&
\mbox{\hspace{-1cm}\small + Polish           \hspace{-1cm}}&&
\mbox{\hspace{-1cm}\small \begin{tabular}{@{}c@{}}+ Intermed.\ subdiv.\\\& fuzzy contr.\end{tabular}\hspace{-1cm}}
\\[1mm]
    \includegraphics[height=\grassNOneExampleHeight,align=c]{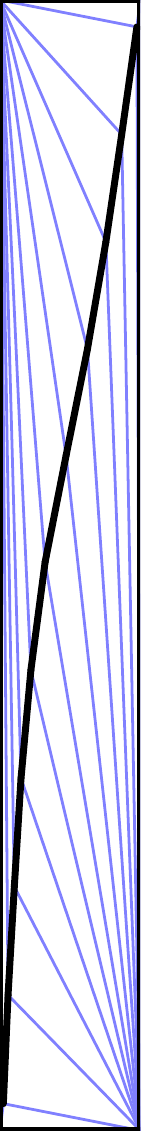}
    & \hspace{\grassNOneArrowSpace} $\to$ \hspace{\grassNOneArrowSpace} &
    \includegraphics[height=\grassNOneExampleHeight,align=c]{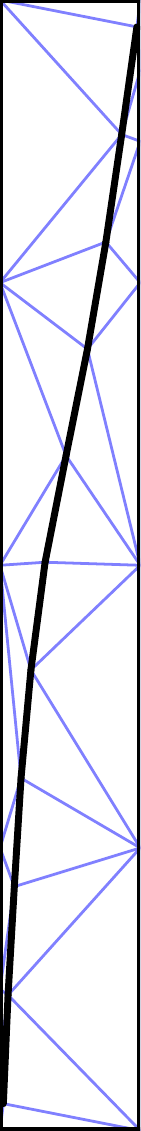}
    & \hspace{\grassNOneArrowSpace} $\to$ \hspace{\grassNOneArrowSpace} &
    \includegraphics[height=\grassNOneExampleHeight,align=c]{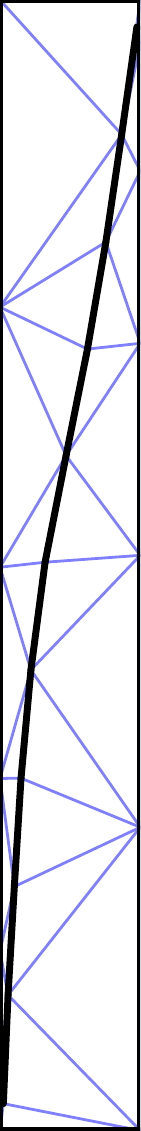}
    & \hspace{\grassNOneArrowSpace} $\to$ \hspace{\grassNOneArrowSpace} &
    \includegraphics[height=\grassNOneExampleHeight,align=c]{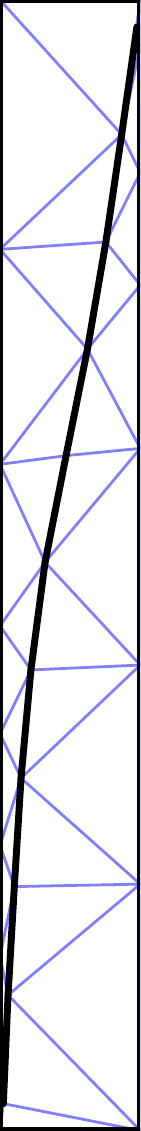}
\\[-0.03cm]
\small 13.79 &&
\small 6.56&&
\small 6.41&&
\small 6.38
\end{tabular}
\end{center}
\caption{
    An example of the successive optimization strategies for a scene with only a single leaf of grass ($N=1$) with 10 segments. Total edge lengths are given below the triangulations.
    The unrefined CDT (left) behaves especially bad due to the connections to the bounding box, which can only happen in corner points of the bounding box as the unrefined case does not create Steiner vertices.
    The optimally refined CDT subdivides the long slender triangles by introducing Steiner vertices on the bounding box, thereby halving the total edge length and already bringing it very close to the optimum. A further polishing step can flip a handful of edges and optimizes vertex positions for a 2\,\% decrease in edge length. Finally, after a subdivision and full optimization with fuzzy contraction, we can further improve the edge length slightly (0.4\,\%) by keeping three extra Steiner vertices on the bounding box.  
    The simplicity of the scene (low degrees of freedom for triangulations) and the ideal placement of Steiner vertices that only occur on the bounding box are reasons for the limited improvement of the fully optimized triangulation compared to the optimally refined CDT.
    \label{figGrassNOneTriangs}
}
\end{figure}

\newcommand{\grassExampleHeight}{3.2cm}
\begin{figure}
\begin{center}
\begin{tabular}{c@{\ }c@{\quad}c}
& 3 segments/leaf & 10 segments/leaf
\\
\rotatebox[origin=c]{90}{\small Unrefined CDT} &
    \includegraphics[height=\grassExampleHeight,align=c]{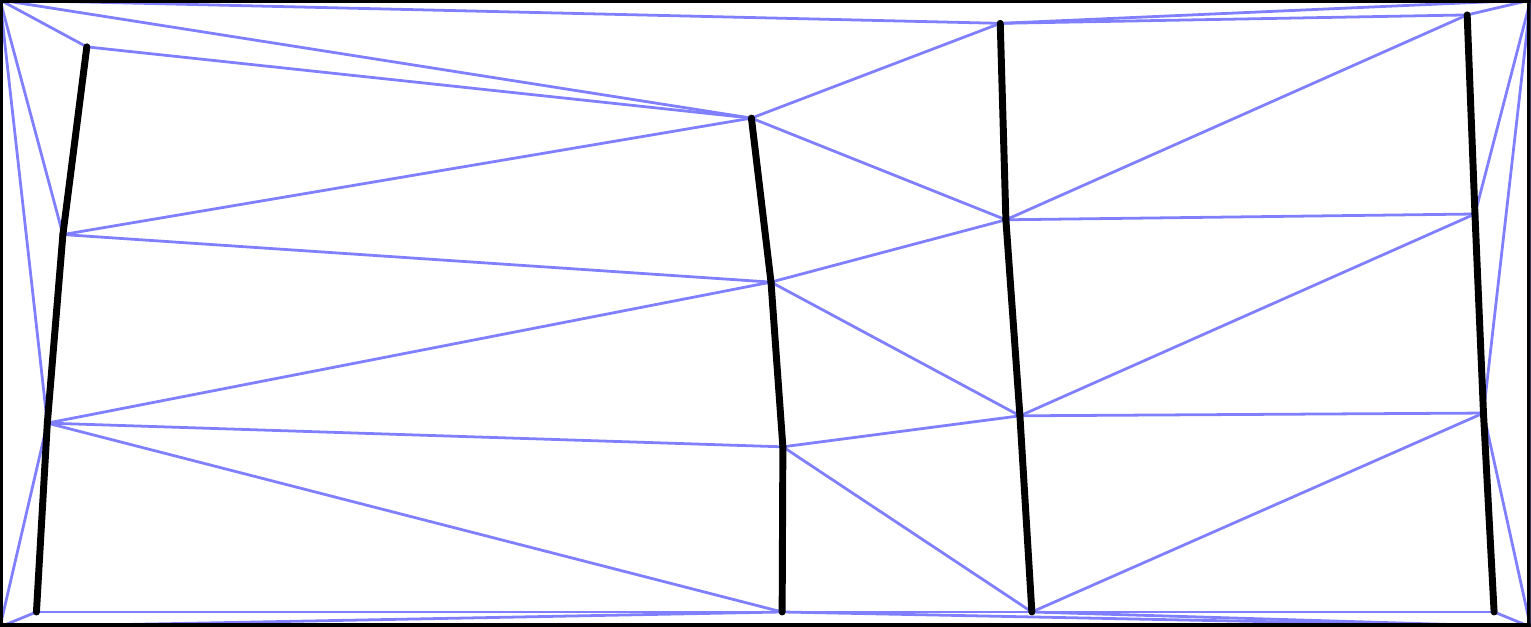}
    &
    \includegraphics[height=\grassExampleHeight,align=c]{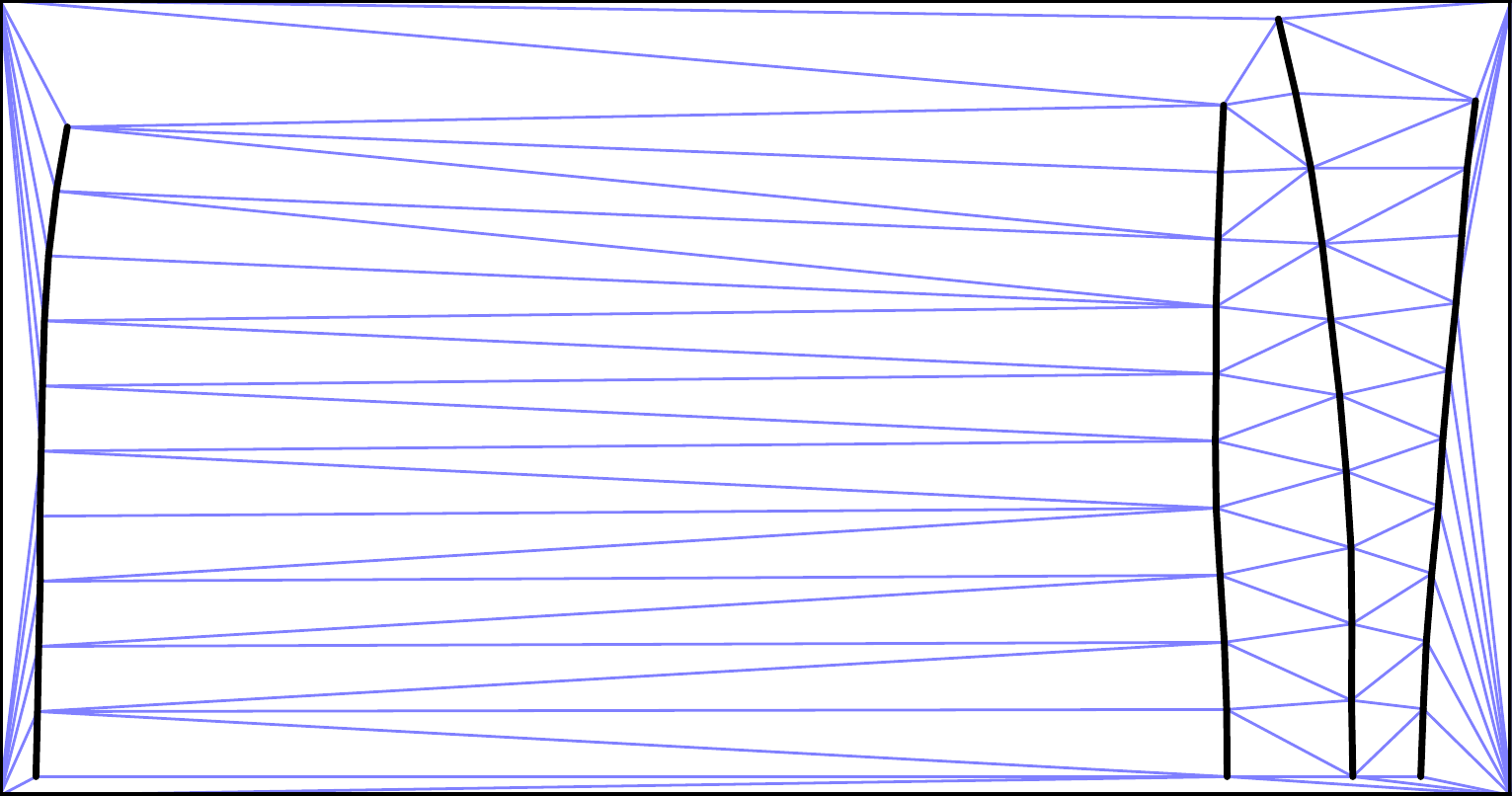}
\\[-0.2mm]
& \small 15.1 & \small 31.4
\\[1mm]
\rotatebox[origin=c]{90}{\small Optim.\ refined CDT} &
    \includegraphics[height=\grassExampleHeight,align=c]{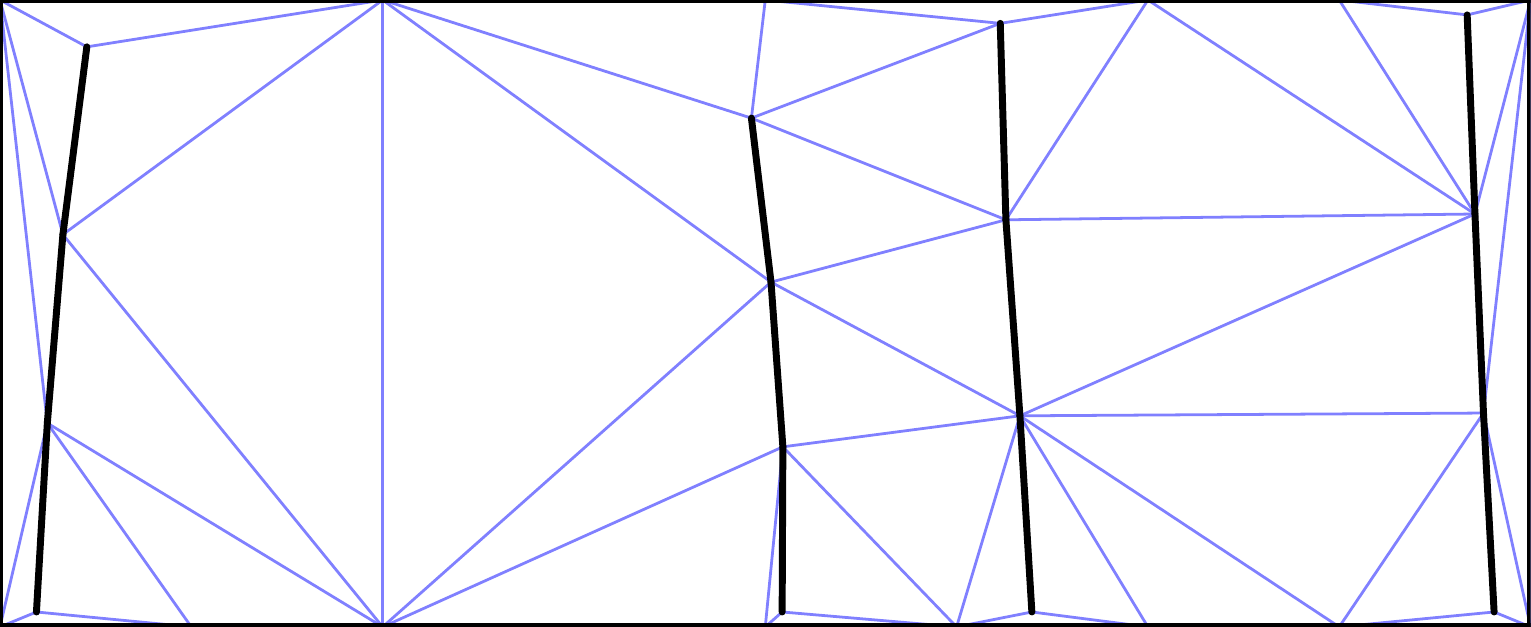}
    &
    \includegraphics[height=\grassExampleHeight,align=c]{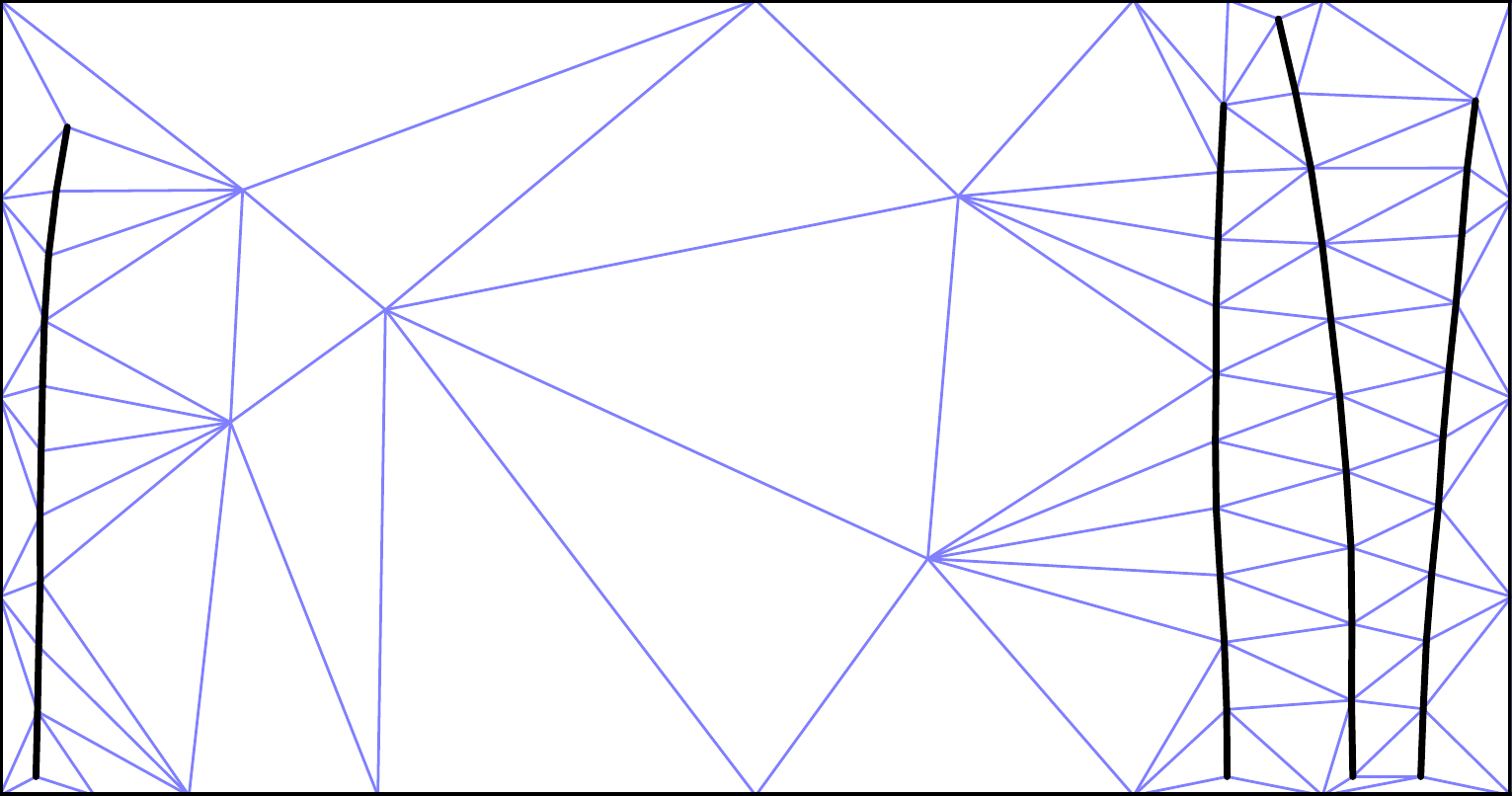}
\\[-0.2mm]
& \small 12.6 & \small 18.7
\\[1mm]
\rotatebox[origin=c]{90}{\small Optimized (ours)} &
    \includegraphics[height=\grassExampleHeight,align=c]{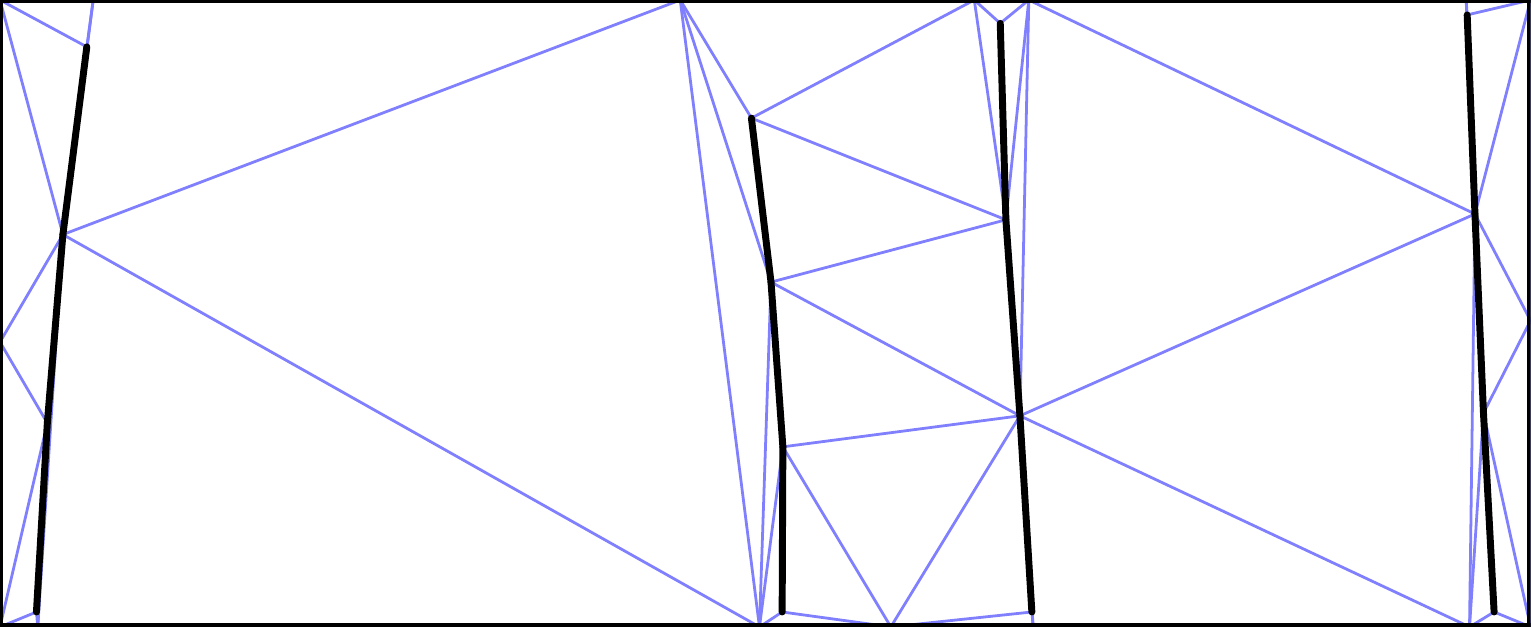}
    &
    \includegraphics[height=\grassExampleHeight,align=c]{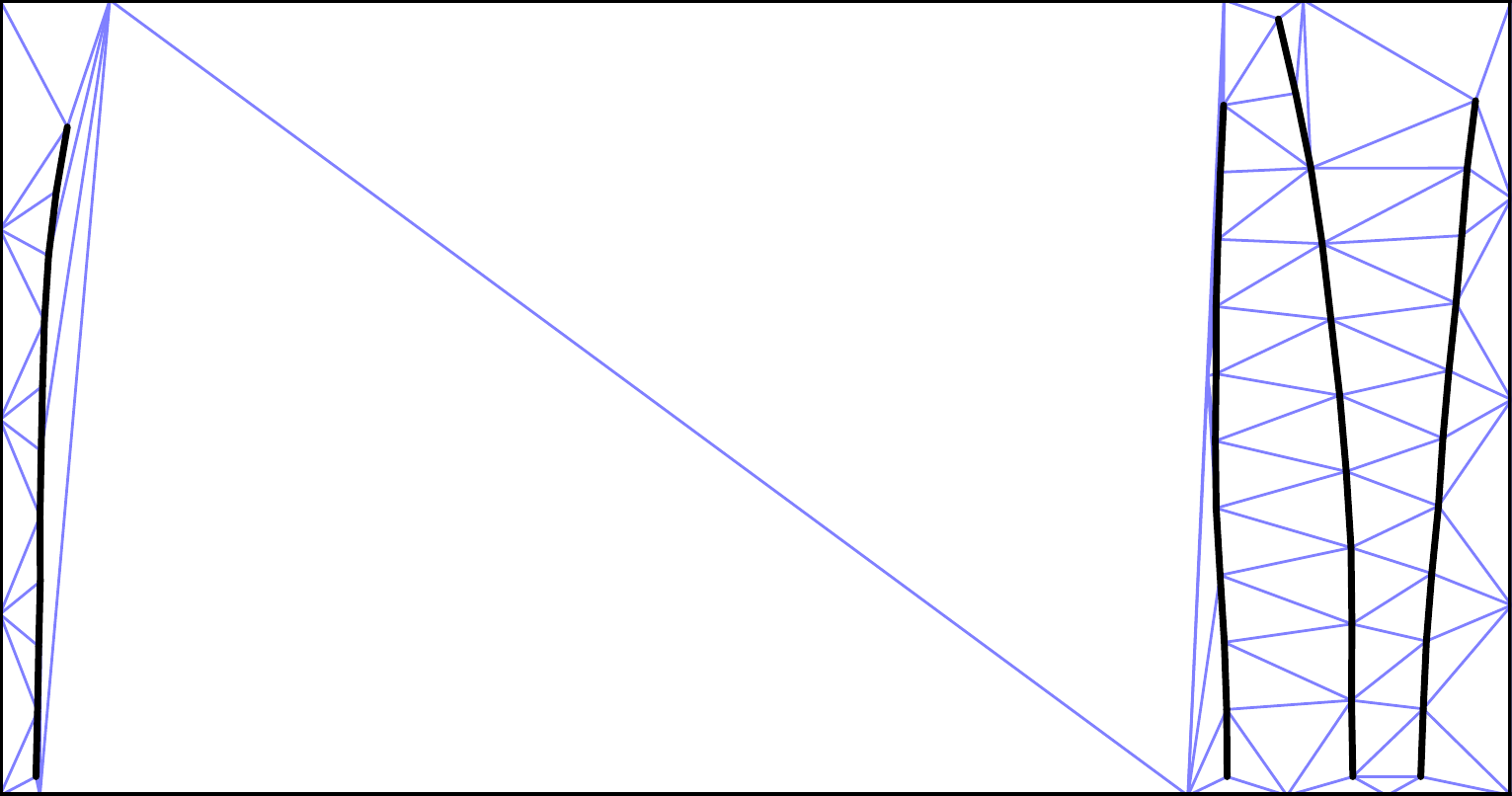} 
\\[-0.2mm]
& \small 11.5 & \small 15.3
\end{tabular}
\end{center}
\caption{Two examples of the `grass' scene with four leaves ($N=4$, original geometry in black). Large empty regions between (clusters of) leaves get filled with spurious edges in the optimal refined CDT (top). Our optimized result (bottom) merges these edges and improves the total edge length (given below the images).
\label{figGrassNFourTriangs}
}
\end{figure}

\newcommand{\grassOneTwoEightHeight}{2.9cm}
\begin{landscape}
\begin{figure}
\begin{center}
\begin{tabular}{c@{\ }c@{\ \ }c}
& 3 segments/leaf & 10 segments/leaf \\
\rotatebox[origin=c]{90}{\small Unrefined CDT}
    &
    \includegraphics[height=\grassOneTwoEightHeight,align=c]{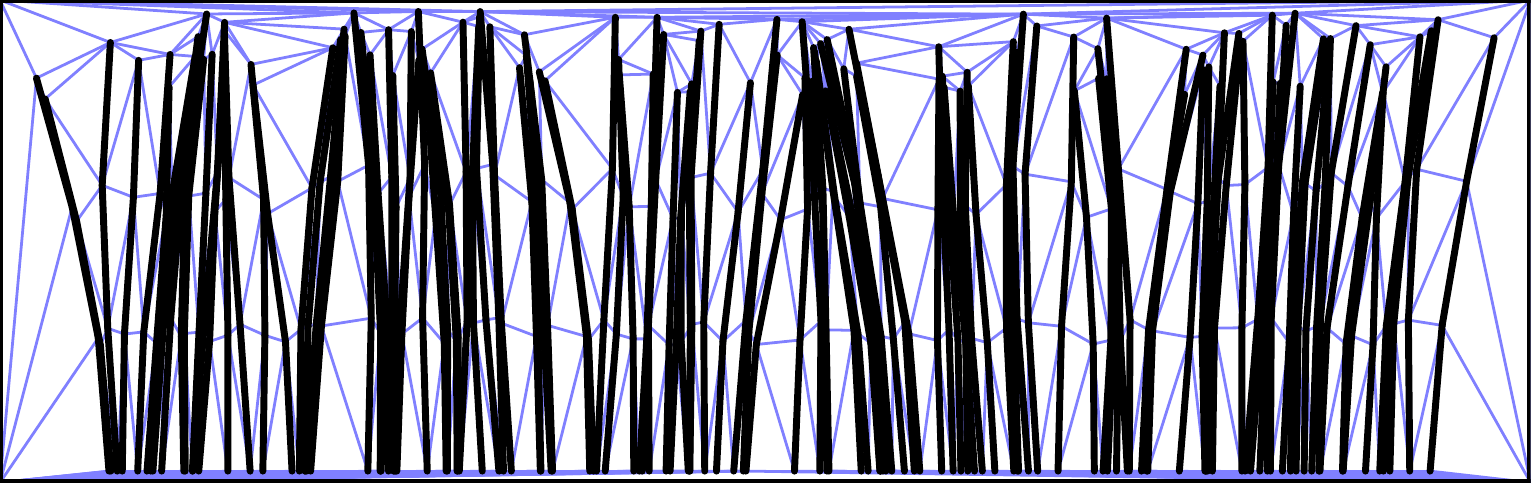}
    &
    \includegraphics[height=\grassOneTwoEightHeight,align=c]{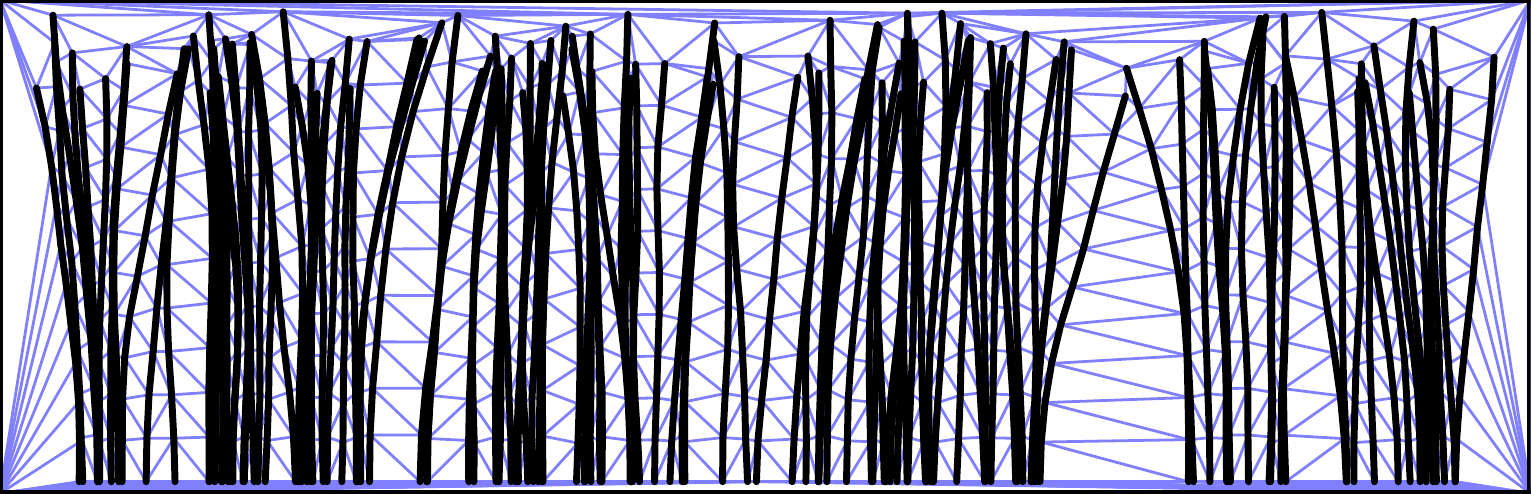}
\\[-0.2mm]
& \small 120.1 & \small 129.1
\\[0.5mm]
\rotatebox[origin=c]{90}{\small Optim.\ refined CDT}
    &
    \includegraphics[height=\grassOneTwoEightHeight,align=c]{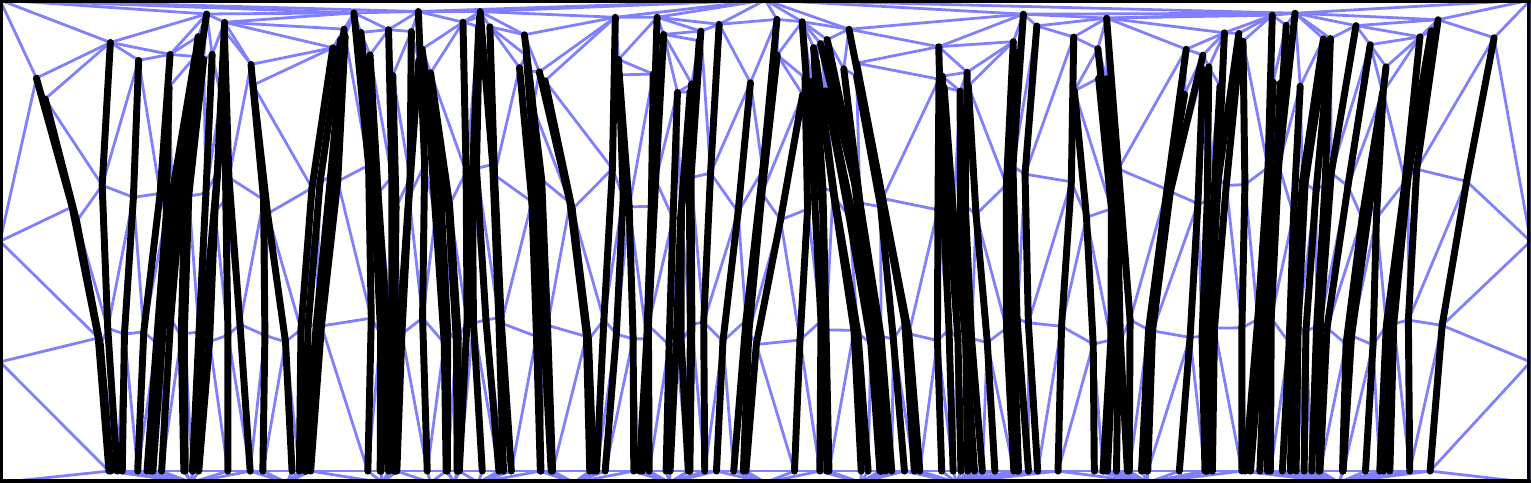}
    &
    \includegraphics[height=\grassOneTwoEightHeight,align=c]{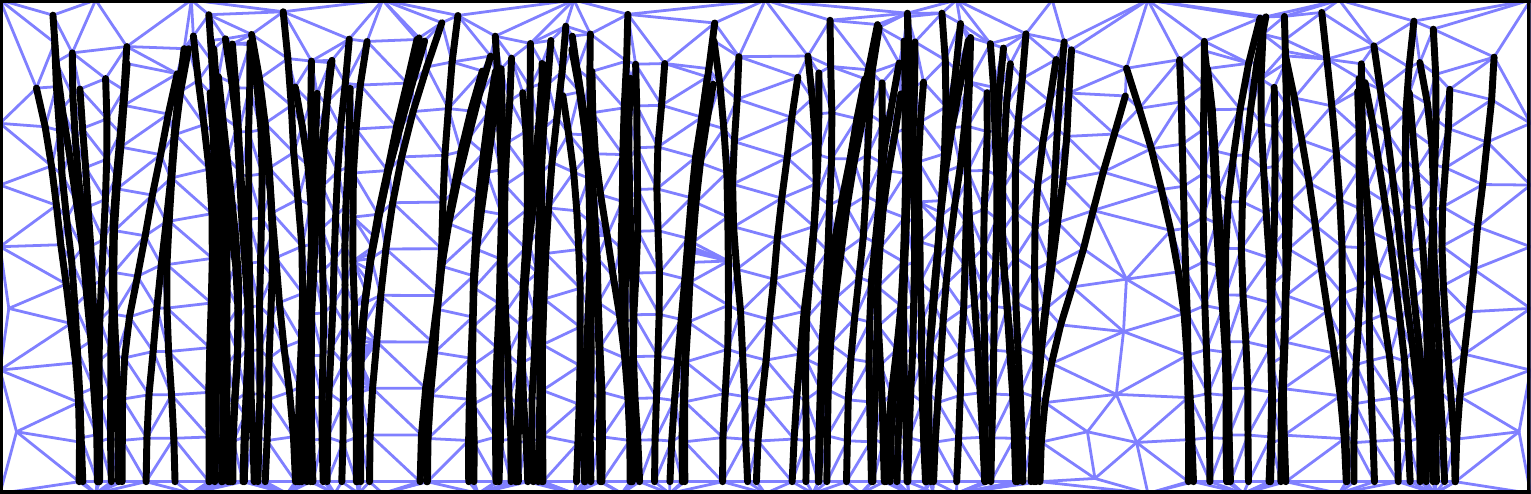}
\\[-0.2mm]
& \small 86.5 & \small 94.3
\\[0.5mm]
\rotatebox[origin=c]{90}{\small Optimized (ours)}
    &
    \includegraphics[height=\grassOneTwoEightHeight,align=c]{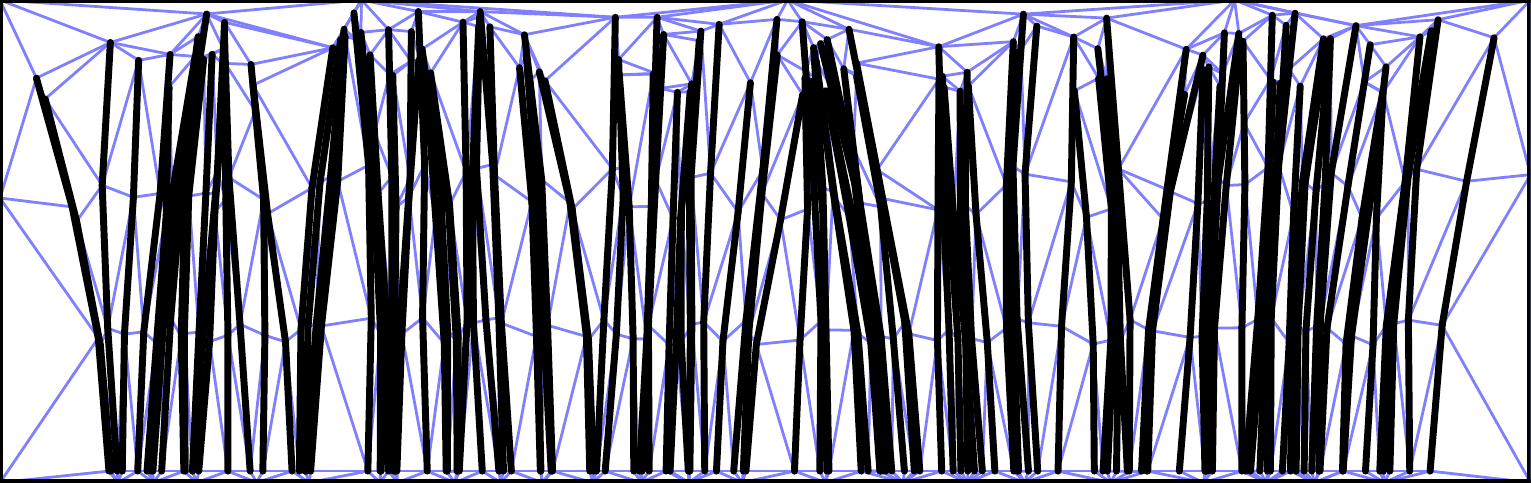}
    &
    \includegraphics[height=\grassOneTwoEightHeight,align=c]{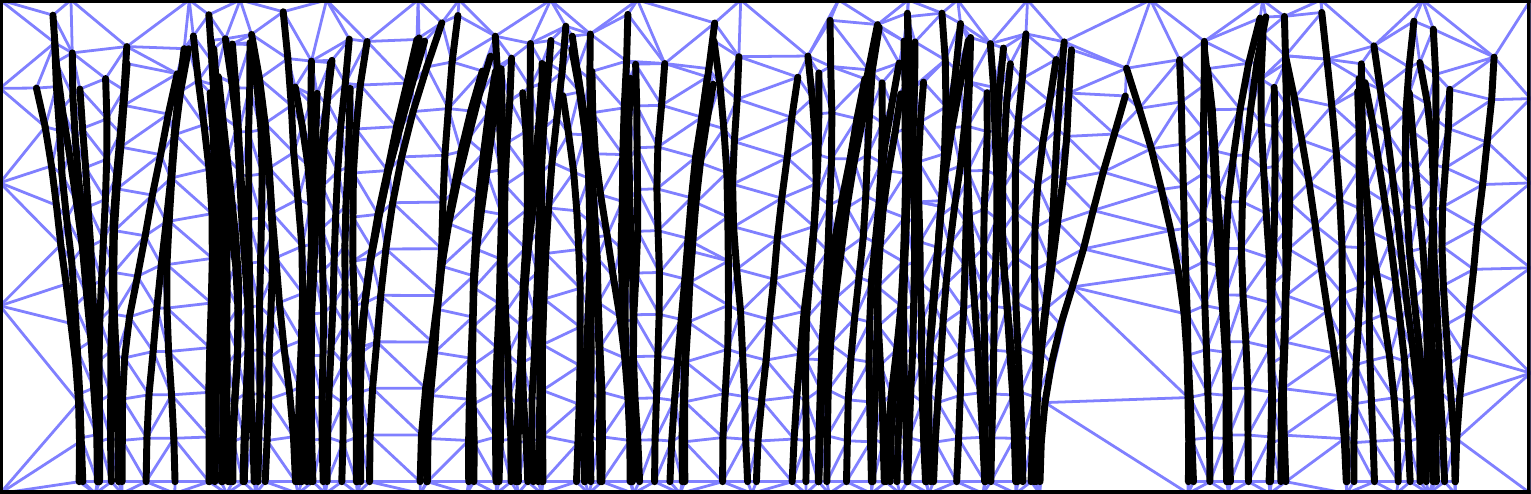}
\\[-0.2mm]
& \small 83.2 & \small 91.3
\end{tabular}
\end{center}
\caption{
    Two examples of the `grass' scene with $N=128$ leaves.
    The scene with 3 segments/leaf (left) shows no large gaps (relative to the segment length) between neighbouring leaves, and the interior part of the fully optimized triangulation is almost identical to even the trivial unrefined CDT: there is no better way to connect those vertices than through straightforward, direct connections. The only difference between the varying levels of optimized triangulations occurs for edges near the bounding box (e.g.\ more Steiner vertices on the bottom of the bounding box from our subdivision). For the scene with 10 segments/leaf (right), a similar story can be told, apart from the presence of one larger void between the leaves, which our optimization with fuzzy contraction can improve. Near the middle of the scene, the CDT based methods also show an occurrence of spurious edges due to thin triangles between two nearly-overlapping strands in close proximity.
\label{figGrassNOneTwoEightTriangs}
}
\end{figure}
\end{landscape}

\subsection{Synthetic Scenes: Hair \label{secOptimHair}}
In the previous `grass' scenes, all segments of the geometry were roughly vertical and we found that our fully optimized triangulations could especially improve scenes that have a combination of fine geometry (many segments/leaf) with large voids in-between. We complement this with the `hair' scenes of Fig.~\ref{figHairPoly}, which vary the orientation of the segments and contain a central empty region.

\begin{figure}
\newcommand{\hairPlotScale}{1.0}
\begin{center}
\begin{tabular}{c@{\ }c@{\ }c}
& \small ~~~ 5 segments/strand 
& \small ~~~ 20 segments/strand
\\
\rotatebox[origin=c]{90}{\small Relative to unrefined CDT} &
    \includegraphics[scale=\hairPlotScale,align=c]{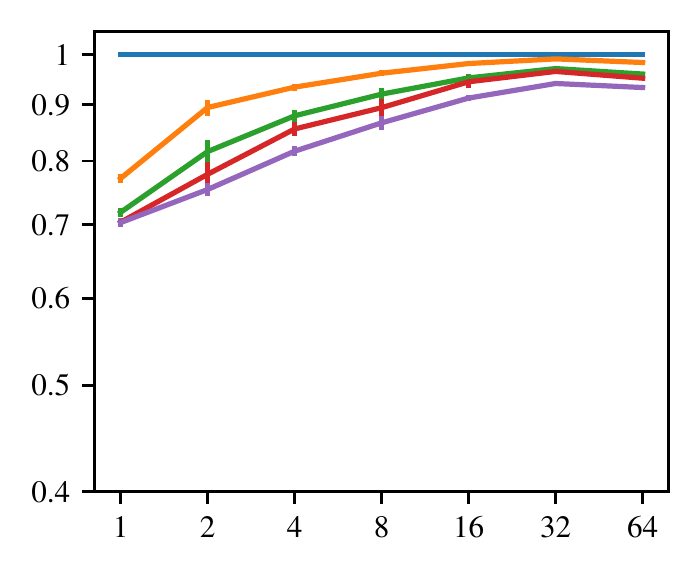} &
    \includegraphics[scale=\hairPlotScale,align=c]{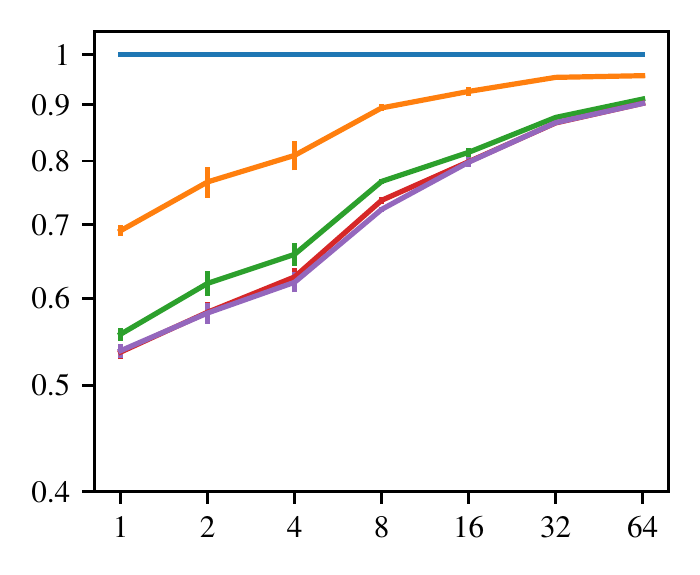}
\\[-3mm]
\rotatebox[origin=c]{90}{\small Relative to optim.~refined CDT} &
    \includegraphics[scale=\hairPlotScale,align=c]{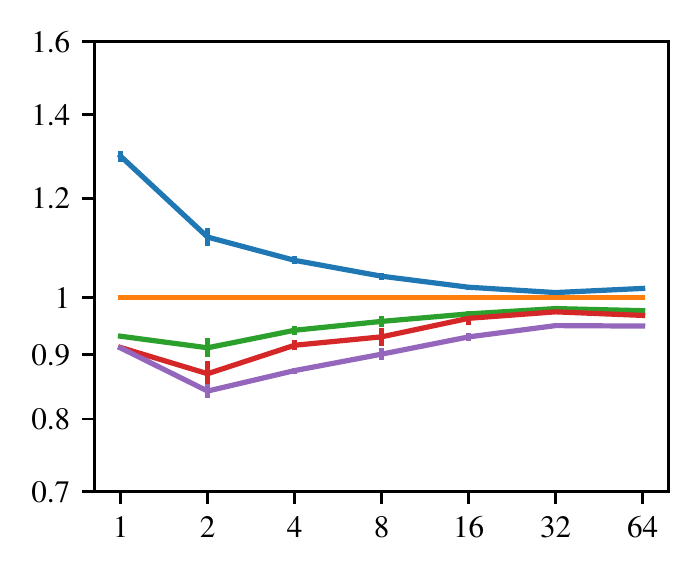} &
    \includegraphics[scale=\hairPlotScale,align=c]{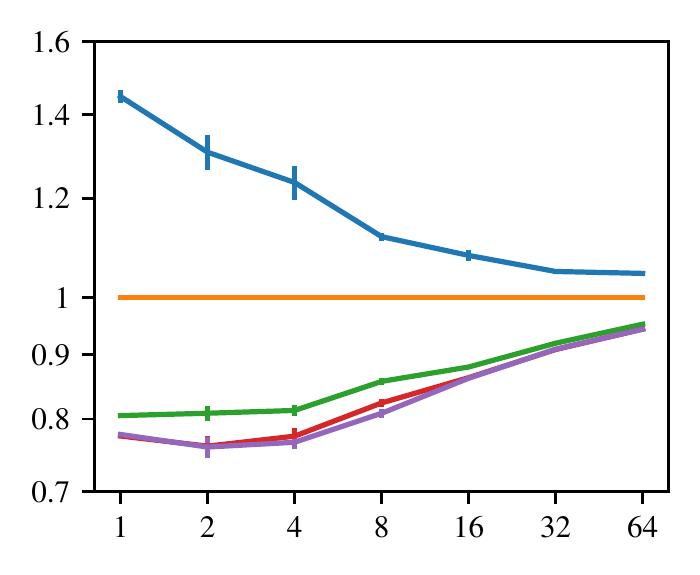}
\\[-4mm] 
& \small ~~~~~$N$
& \small ~~~~~$N$
\\[1mm]
& \multicolumn{2}{c}{\includegraphics[align=c]{plots/gen_l_legend_MANUALLY_FIXED_textNowHardcodedGlyphs_cropAndMoreCleanup_2lines.pdf}}
\end{tabular}
\end{center}
\caption{
    Relative total edge lengths for triangulations of the hair scene for successively powerful optimization strategies. The top plots show the total edge lengths relative to the total edge length of an unrefined CDT, the bottom plot shows the total edge lengths relative to that of the optimally refined CDT.
\label{figHairPlots}
}
\end{figure}

Figure \ref{figHairPlots} shows the relative impact of successively more powerful optimization strategies on the total edge length of triangulations for the hair scenes.
Similar to the grass scenes (Fig.~\ref{figGrassPlots}), the difference between the optimization strategies grows larger as the geometry gets more detailed, i.e.\ as the number of segments per strand increase. The underlying principle is similar: open spaces that are large relative to the segment size of the geometry are not triangulated optimally with CDT-based methods, and these suboptimal regions become more abundant as the segment size becomes smaller. 

In terms of the number of strands $N$, the difference between the strategies now shrinks monotonically as $N$ increases.
Figures \ref{figHairNEightTriangs} and \ref{figHairNThreeTwoTriangs} show samples of the resulting triangulations for $N=8$ and $N=32$ respectively.
The part of the triangulation \emph{in between} the hair strands is nearly optimal for all strategies, even the simplest one.
The main differences between triangulations of different optimization strategies are found in the area \emph{around} the actual hair geometry.
The edge length contribution of the triangulation in this area grows smaller as the number of strands $N$ increases, when taken relative to the edge length of the actual hair strands themselves and the edges of the triangulation between the strands. This explains the diminishing relative difference between the techniques as $N$ increases.

Another observation is that the difference between the pipeline without fuzzy contraction (red curve), and the iterated pipeline with an intermediate subdivision (purple curve) is markedly smaller for the detailed scenes of 20 segments/strand than for those of 5 segments/strand. Indeed, the high number of vertices in the detailed scenes already provide sufficient topological freedom to optimize the triangulation and benefit less from an additional subdivision step than the coarser scenes.

Moreover, for these detailed scenes with 20 segments/strand, a simple polishing step with edge flipping (but without fuzzy contraction) starting from the optimally refined CDT already ends up close to our more advanced strategies. An explanation can again be found in the central empty area, which can already be `cleaned up' with a simple polishing step with edge flipping.

\newcommand{\hairScale}{0.455}
\begin{figure}
\begin{center}
\begin{tabular}{c@{\ }c@{\ }c}
& 5 segments/strand & 20 segments/strand
\\
\rotatebox[origin=c]{90}{\small 2 strands} &
    \includegraphics[scale=\hairScale,align=c]{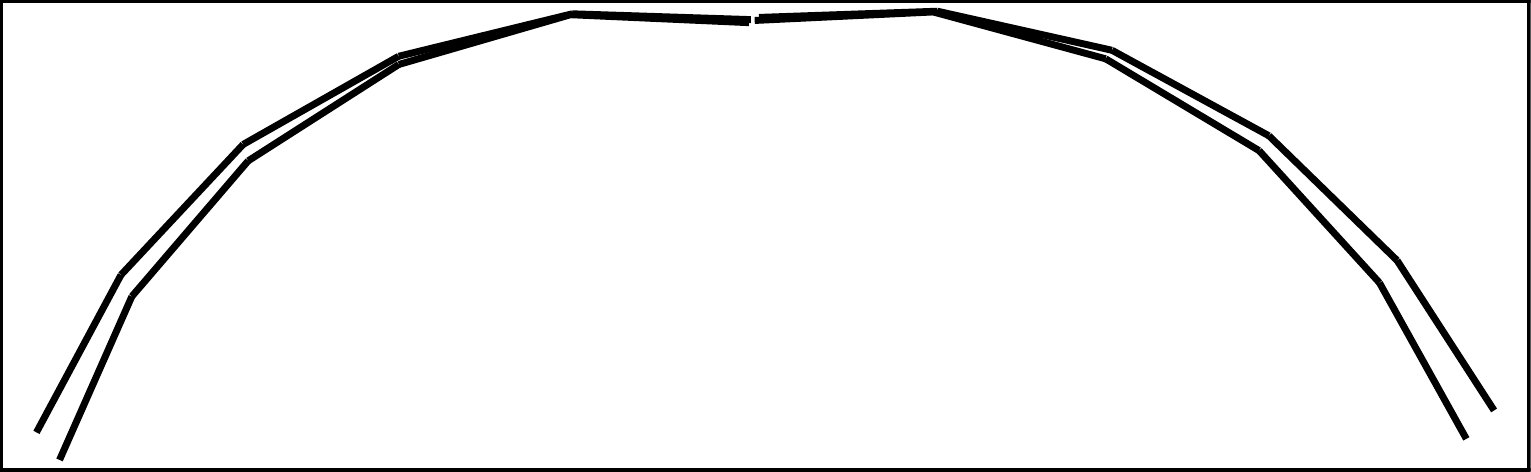}
    &
    \includegraphics[scale=\hairScale,align=c]{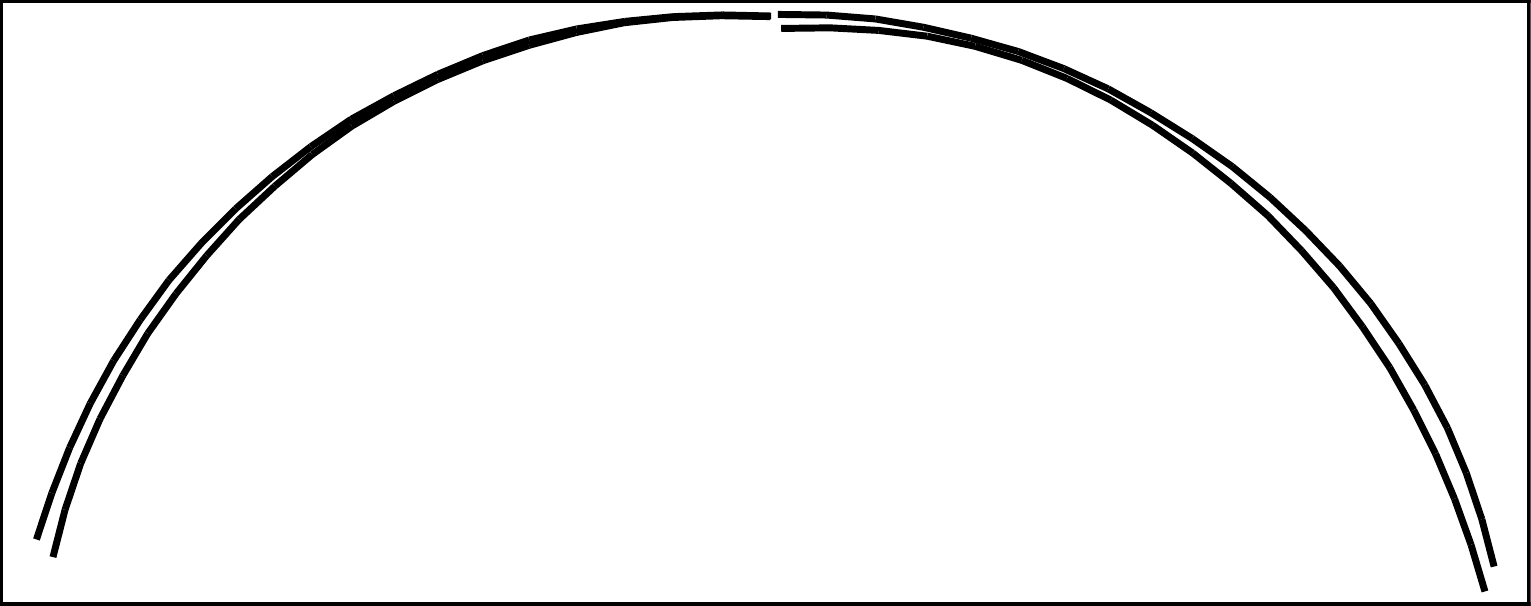}
    \\
\rotatebox[origin=c]{90}{\small 8 strands} &
    \includegraphics[scale=\hairScale,align=c]{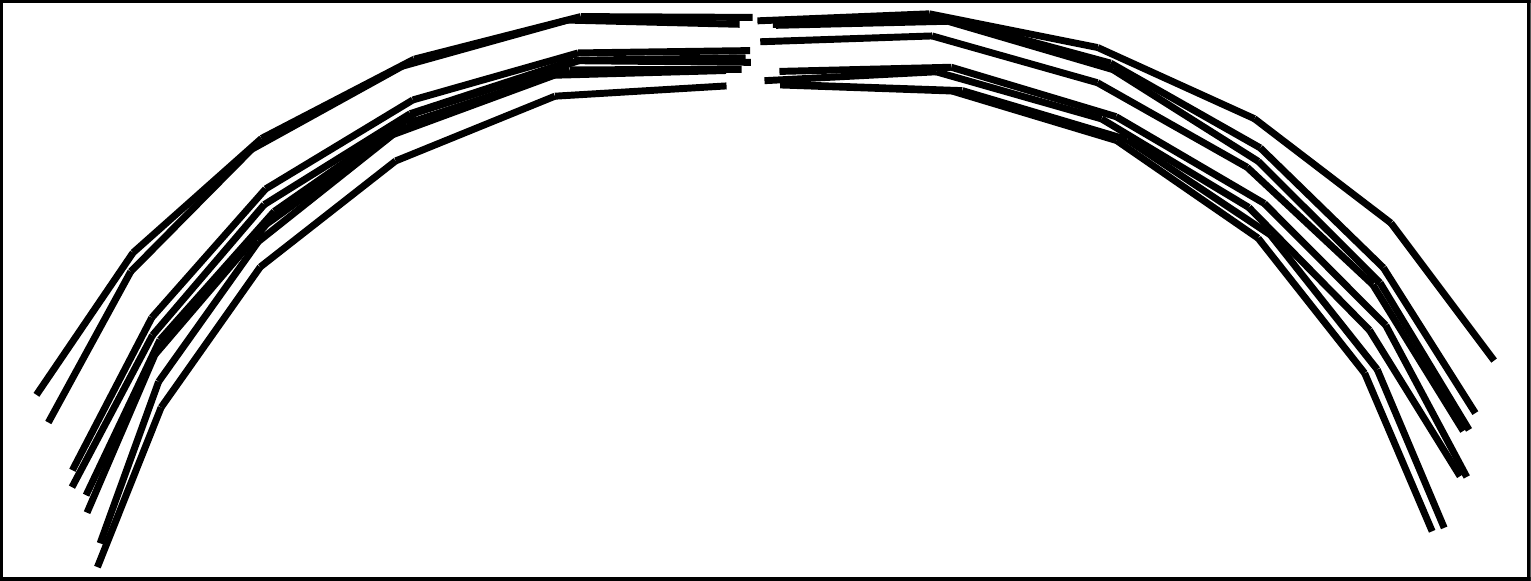}
    &
    \includegraphics[scale=\hairScale,align=c]{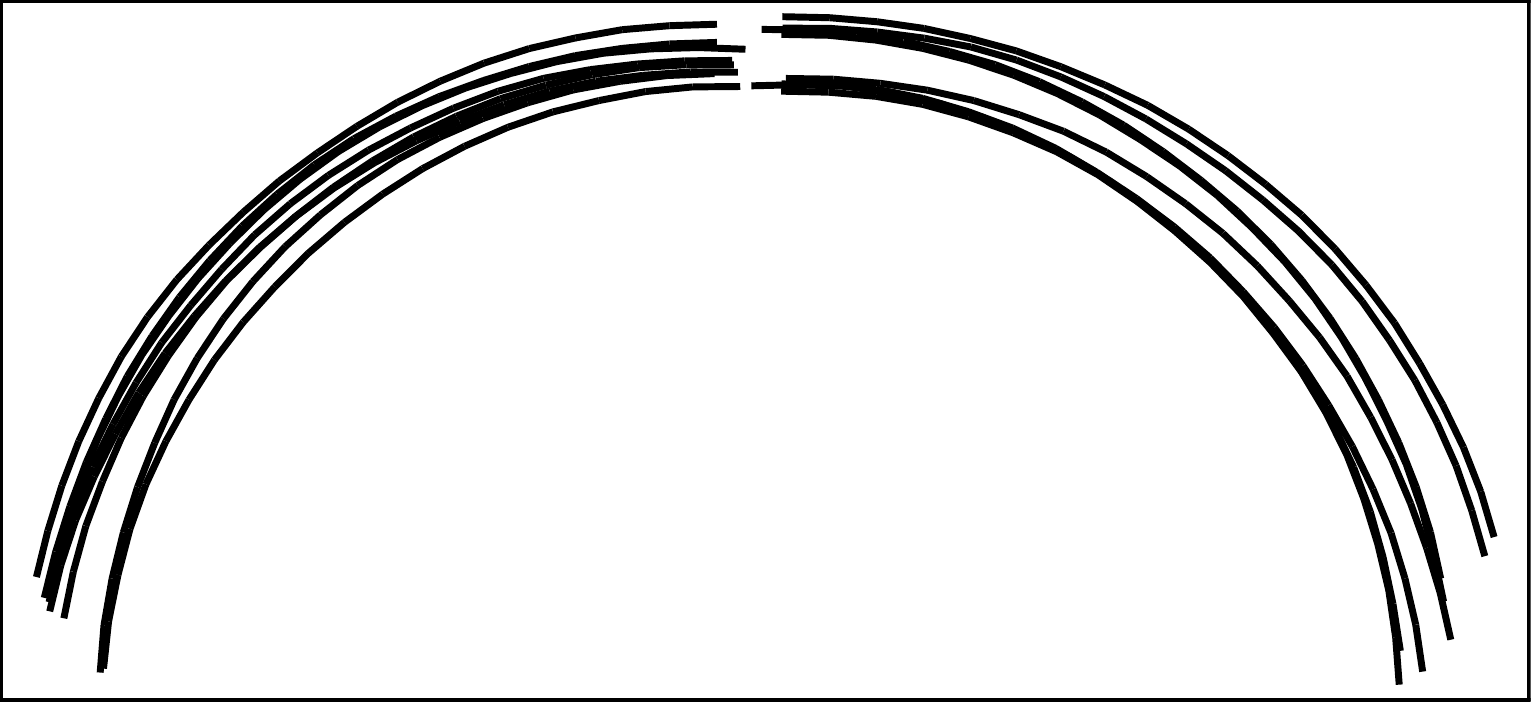}
    \\
\rotatebox[origin=c]{90}{\small 32 strands} &
    \includegraphics[scale=\hairScale,align=c]{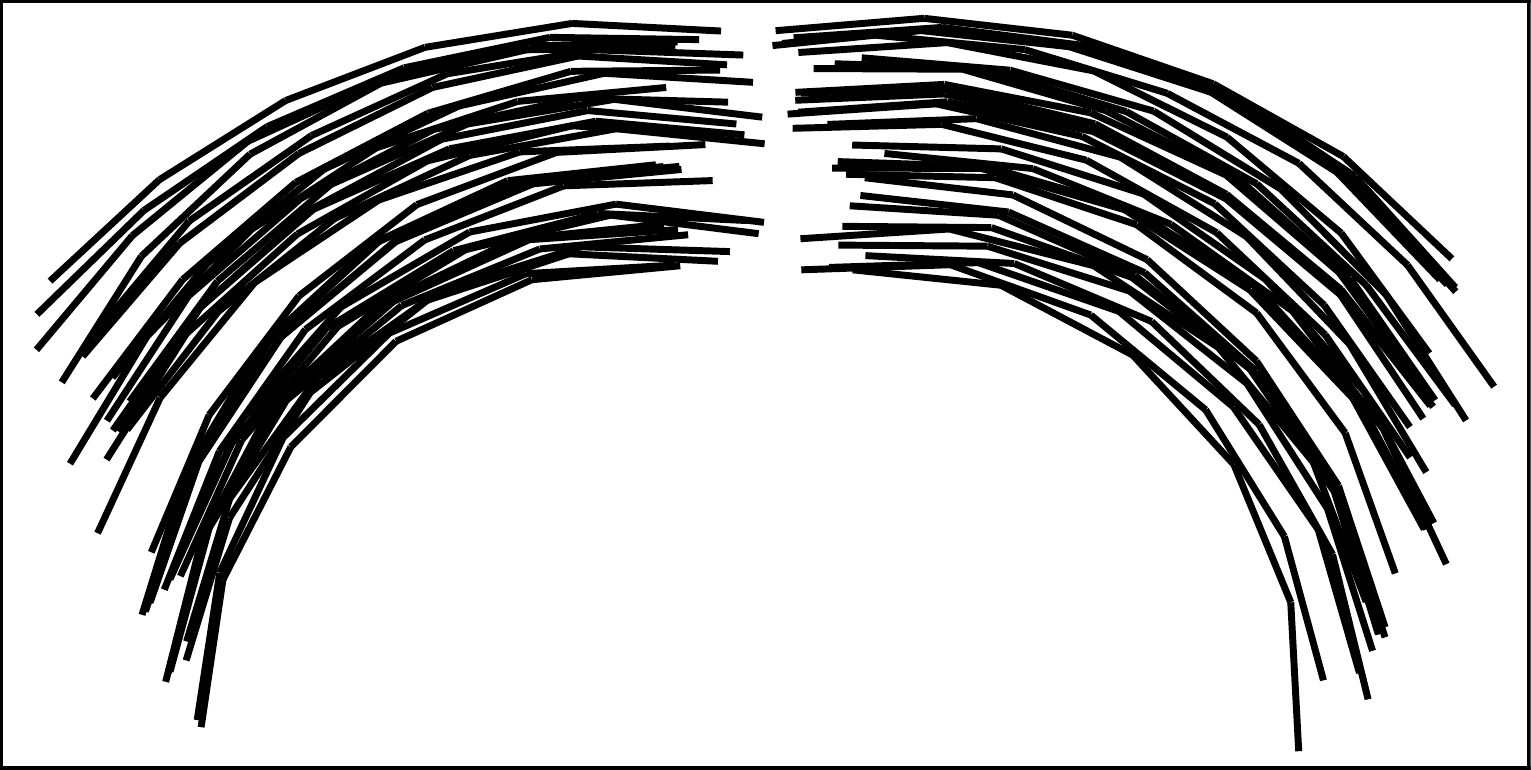}
    &
    \includegraphics[scale=\hairScale,align=c]{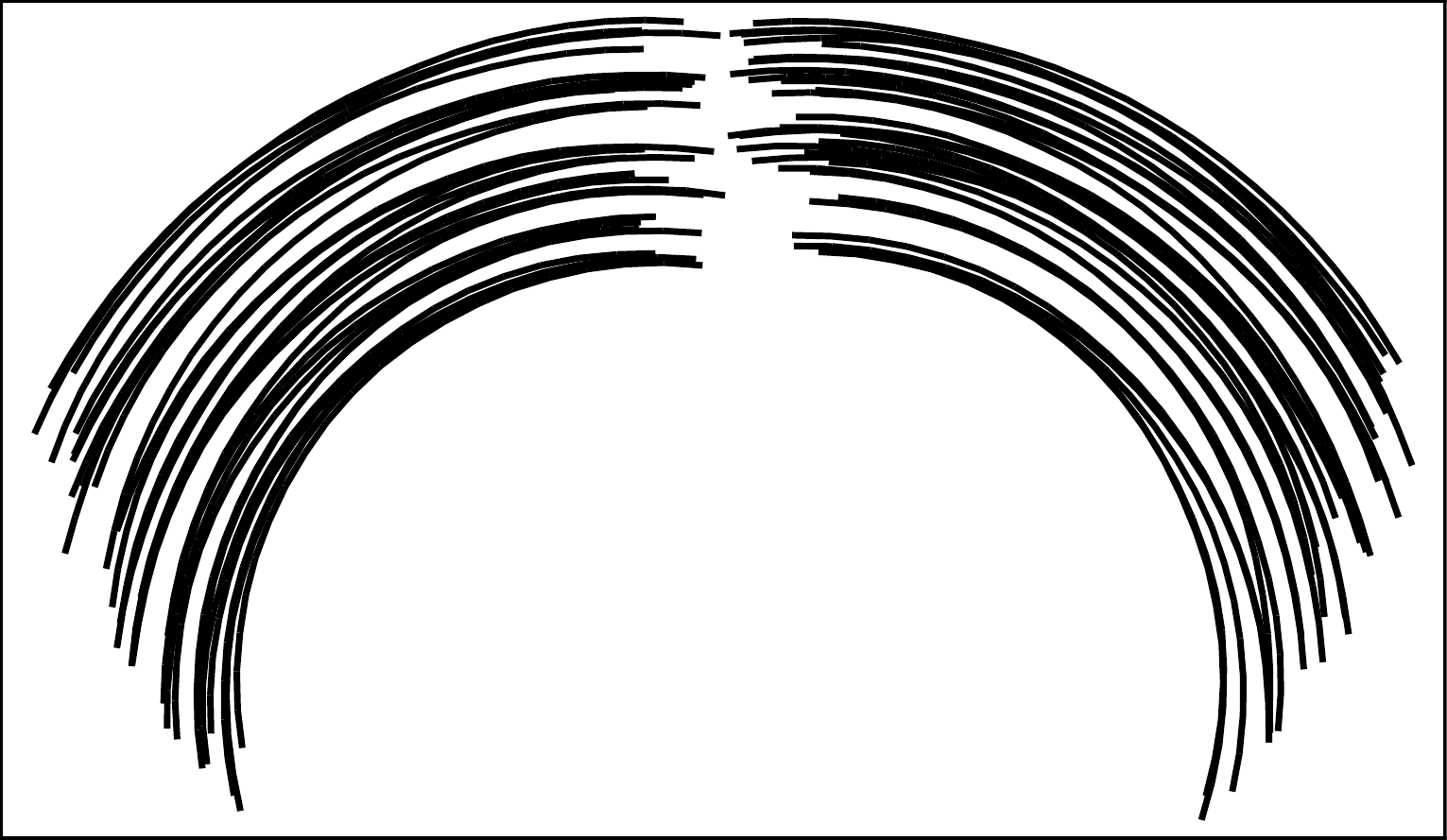}
    \\
\rotatebox[origin=c]{90}{\small 64 strands} &
    \includegraphics[scale=\hairScale,align=c]{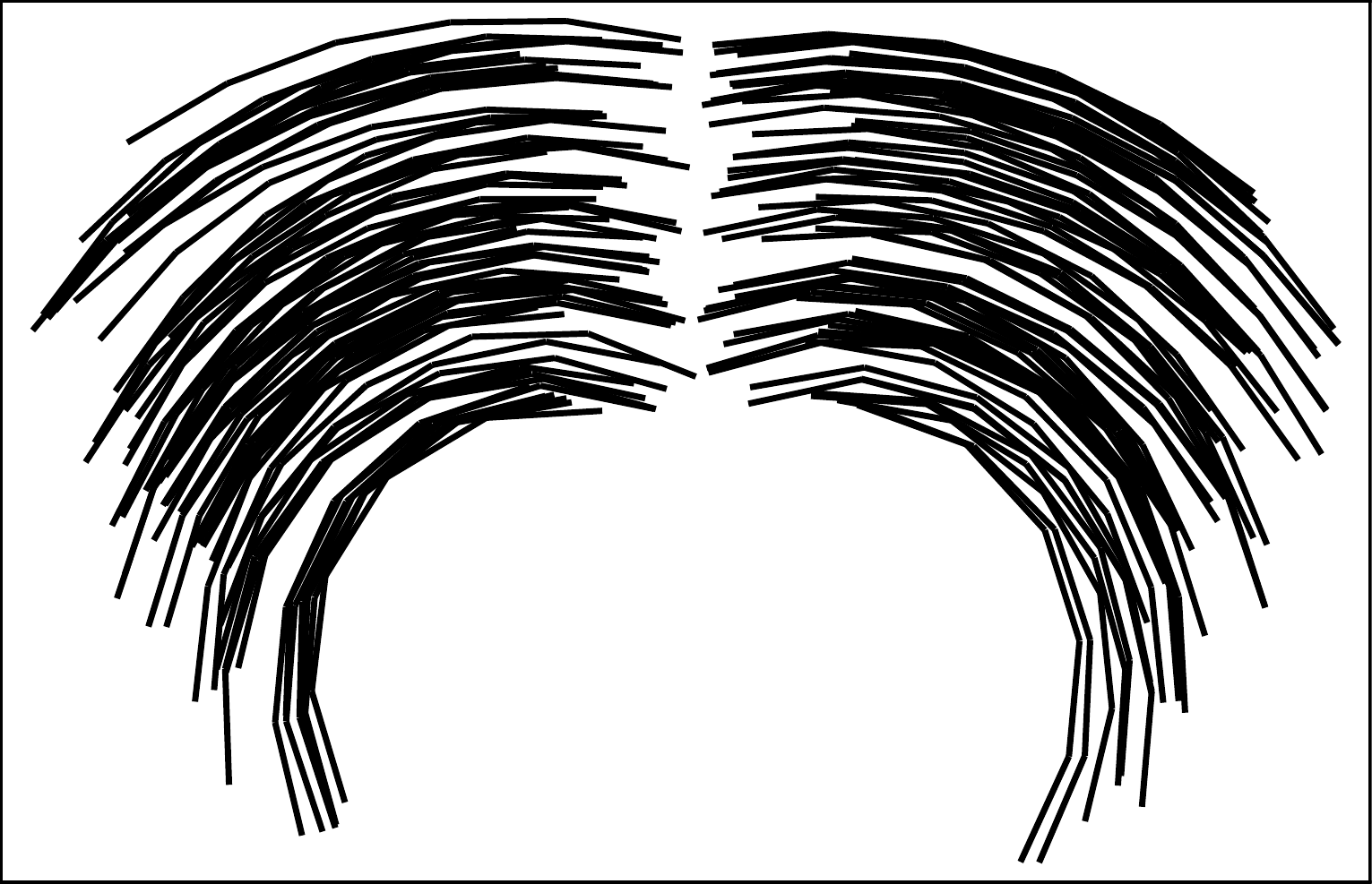}
    &
    \includegraphics[scale=\hairScale,align=c]{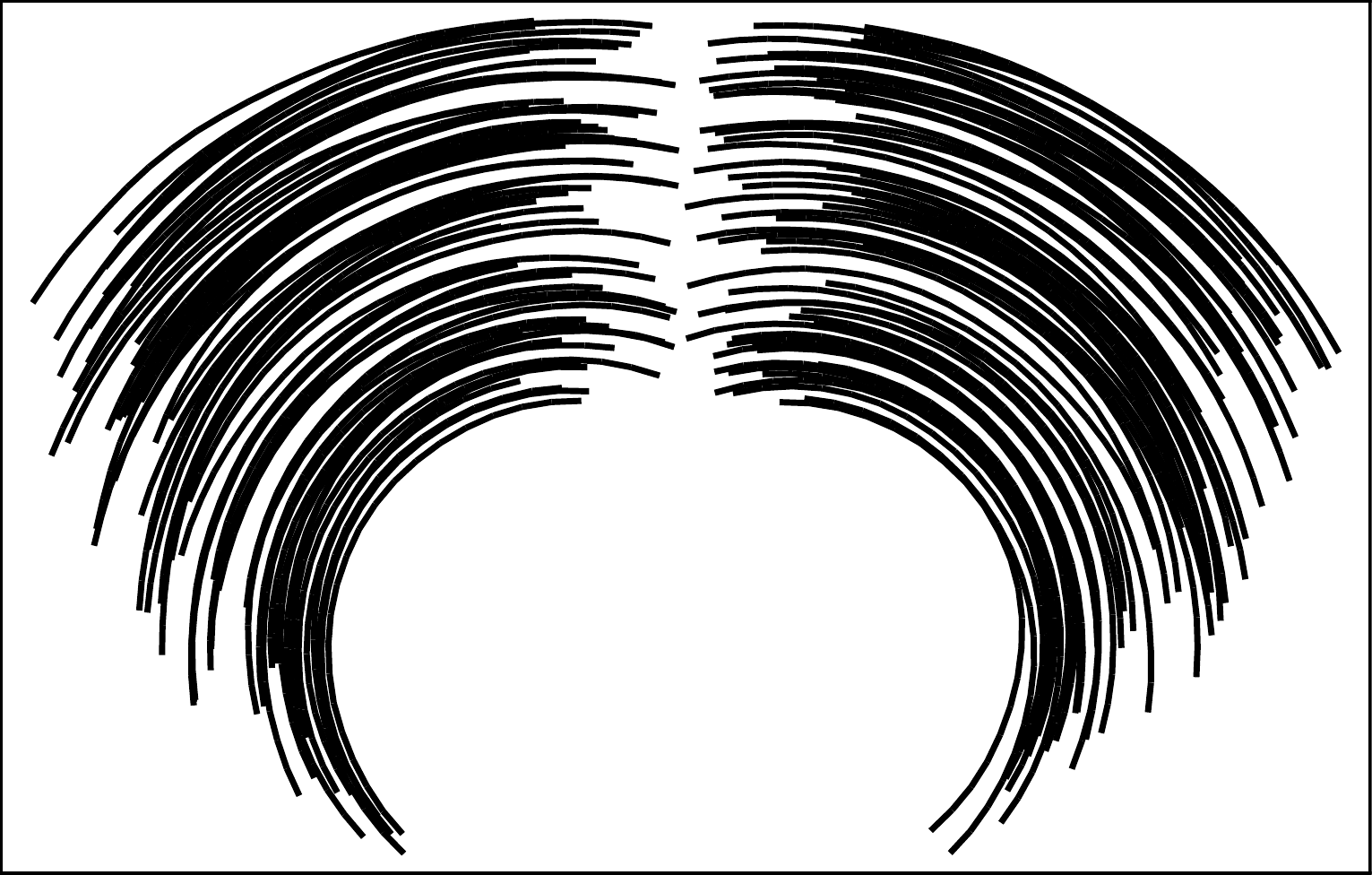}
\end{tabular}
\end{center}
\caption{Examples of the `hair' scene with varying number of strands (per side) and segments per strand.}
\label{figHairPoly}
\end{figure}

\newcommand{\hairExampleHeightEight}{2.73cm}
\begin{figure}
\begin{center}
\begin{tabular}{c@{\ }c@{\quad}c}
& 5 segments/strand & 20 segments/strand
\\
\rotatebox[origin=c]{90}{\small Unrefined CDT} &
    \includegraphics[height=\hairExampleHeightEight,align=c]{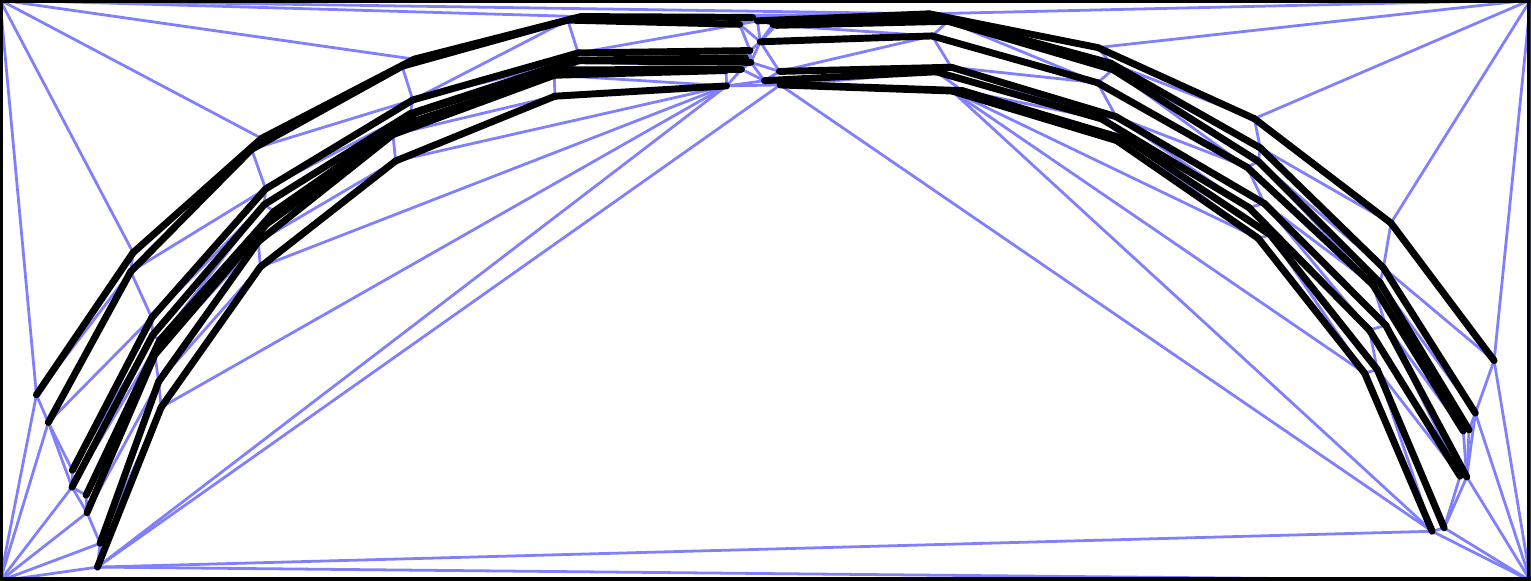}
    &
    \includegraphics[height=\hairExampleHeightEight,align=c]{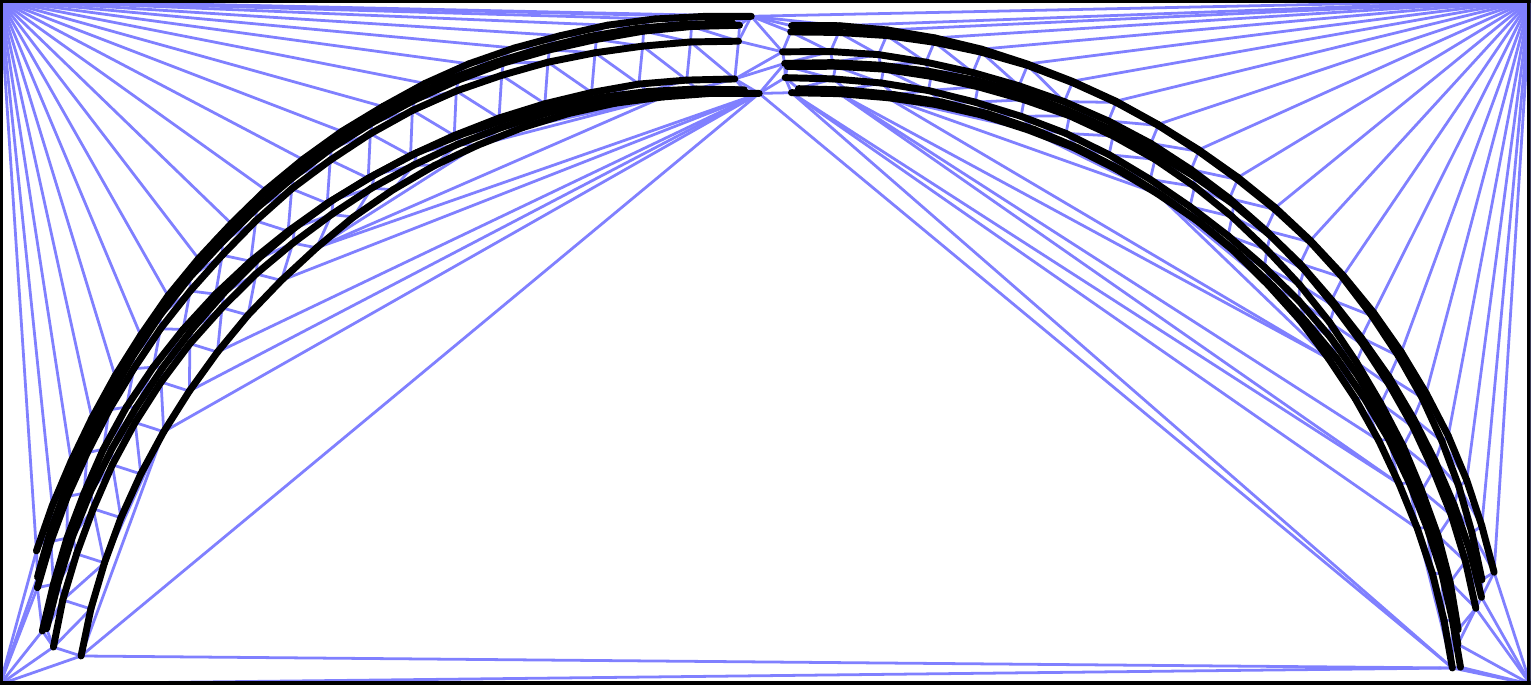}
\\[-0.2mm]
& \small 30.4 & \small 47.7
\\[1mm]
\rotatebox[origin=c]{90}{\mbox{\hspace{-1cm}\small Optim.\ refined CDT \hspace{-1cm}}}&
    \includegraphics[height=\hairExampleHeightEight,align=c]{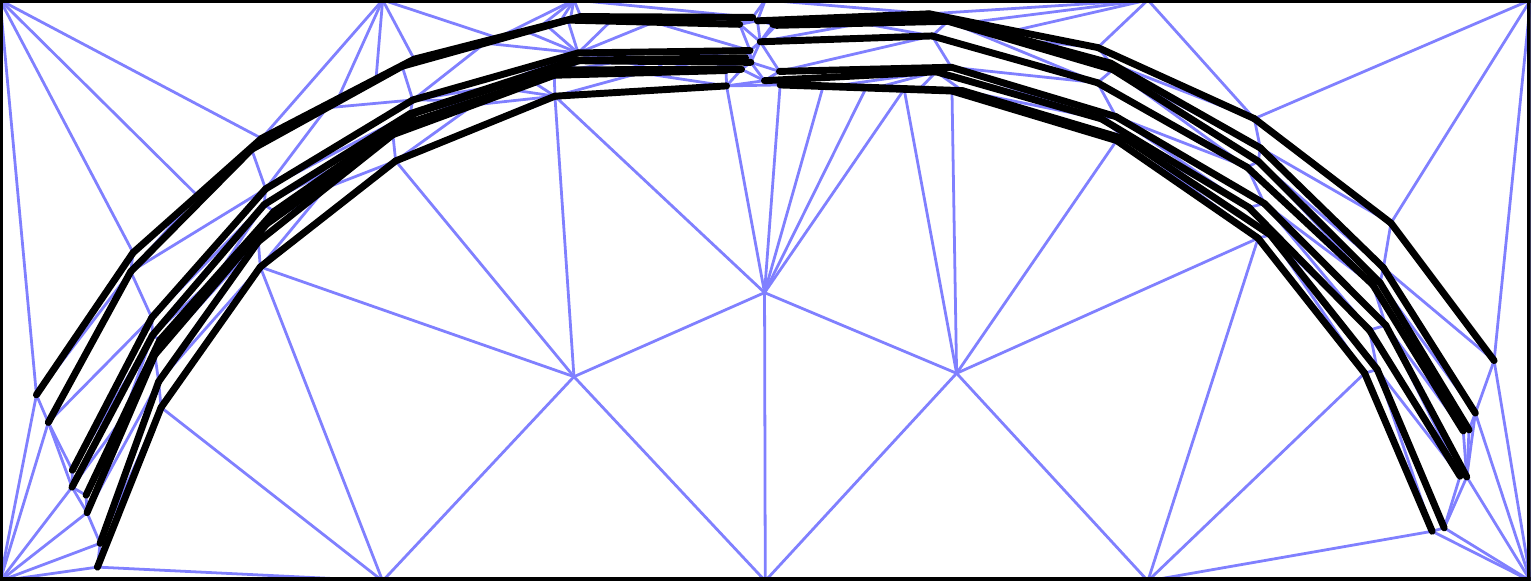}
    &
    \includegraphics[height=\hairExampleHeightEight,align=c]{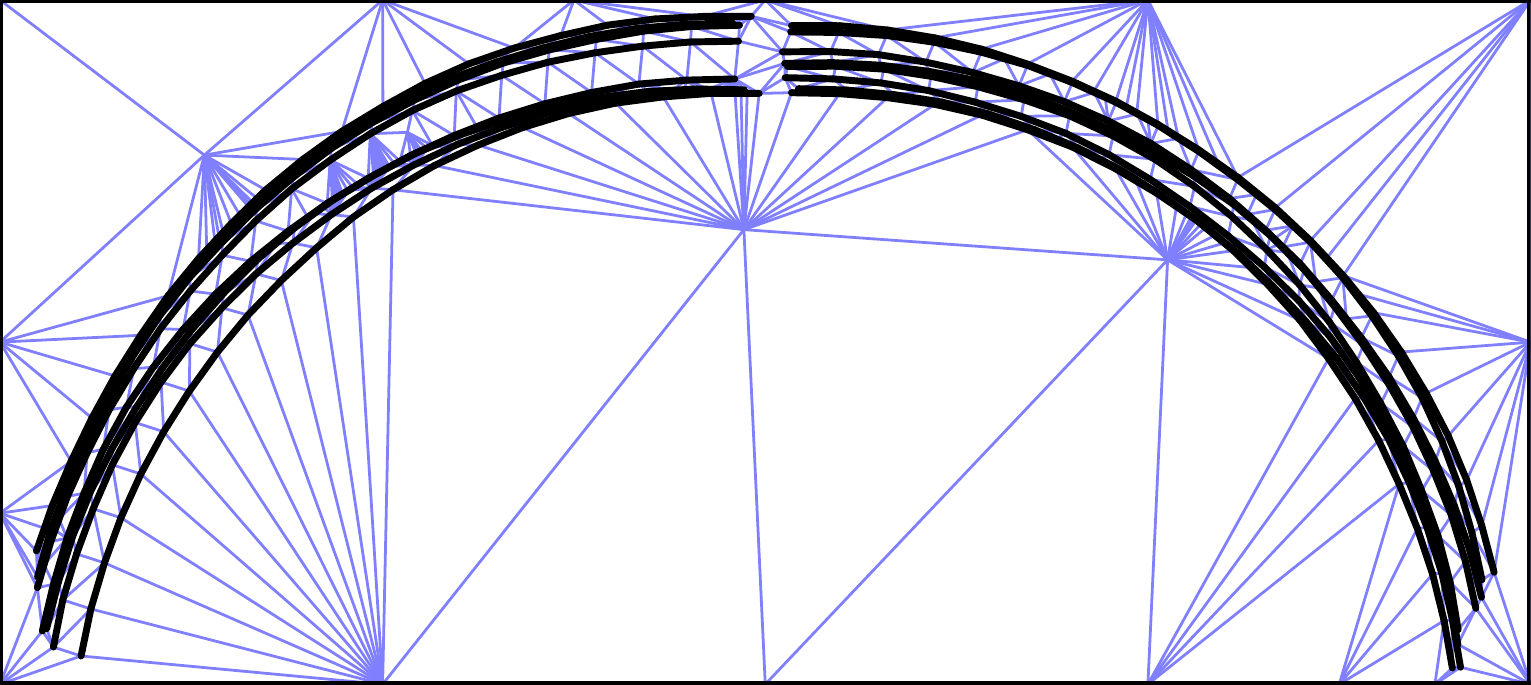}
\\[-0.2mm]
& \small 29.6 & \small 42.0
\\[1mm]
\rotatebox[origin=c]{90}{\small \begin{tabular}{c}Optimized\\[-0.5mm](no subdiv.)\end{tabular}} &
    \includegraphics[height=\hairExampleHeightEight,align=c]{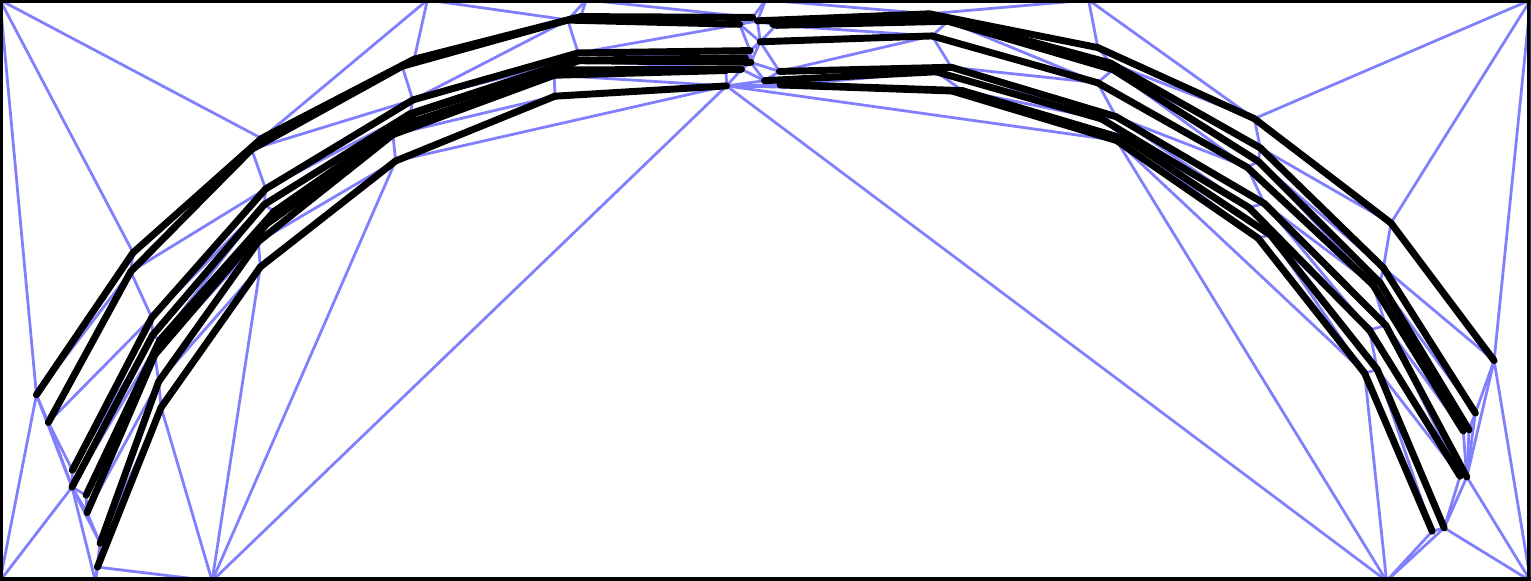}
    &
    \includegraphics[height=\hairExampleHeightEight,align=c]{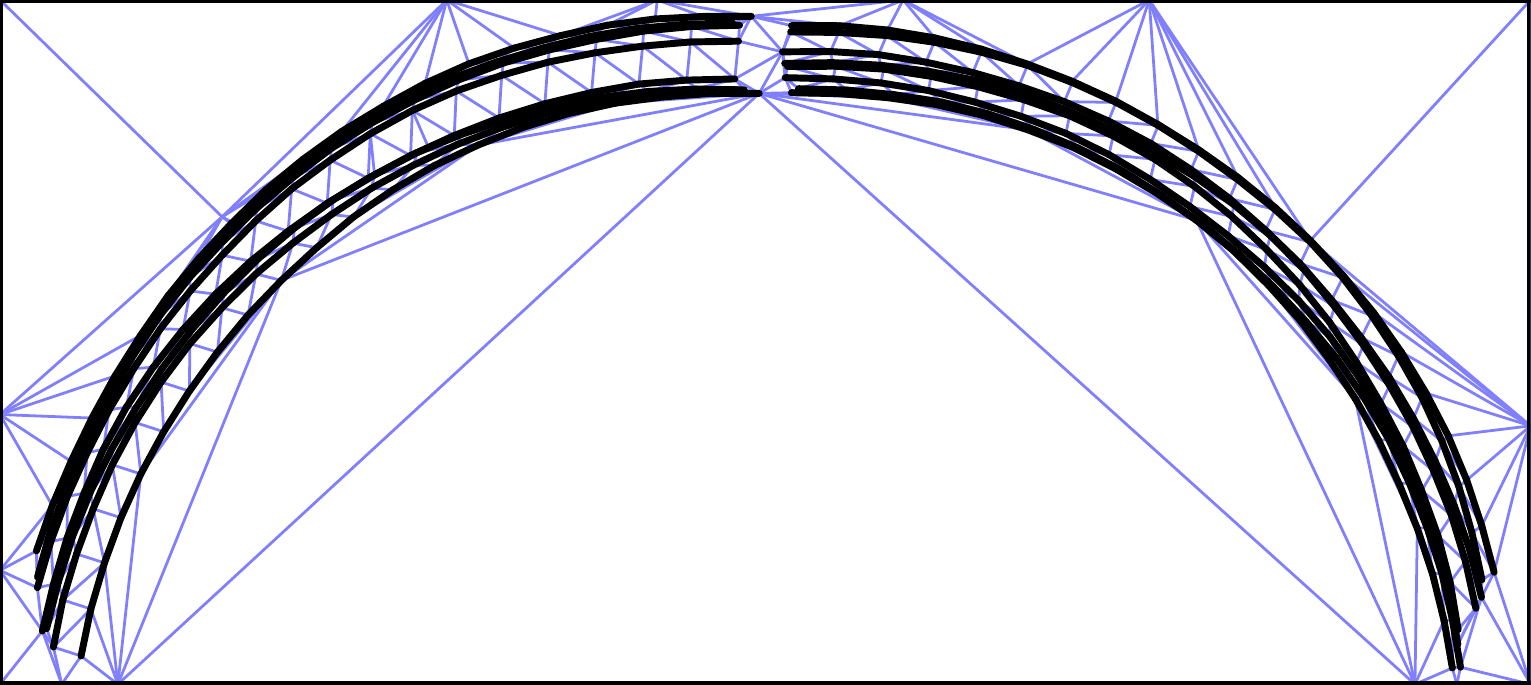}
\\[-0.2mm]
& \small 26.5 & \small 34.5
\\[1mm]
\rotatebox[origin=c]{90}{\small \begin{tabular}{c}Optimized\\[-0.5mm](with subdiv.)\end{tabular}} &
    \includegraphics[height=\hairExampleHeightEight,align=c]{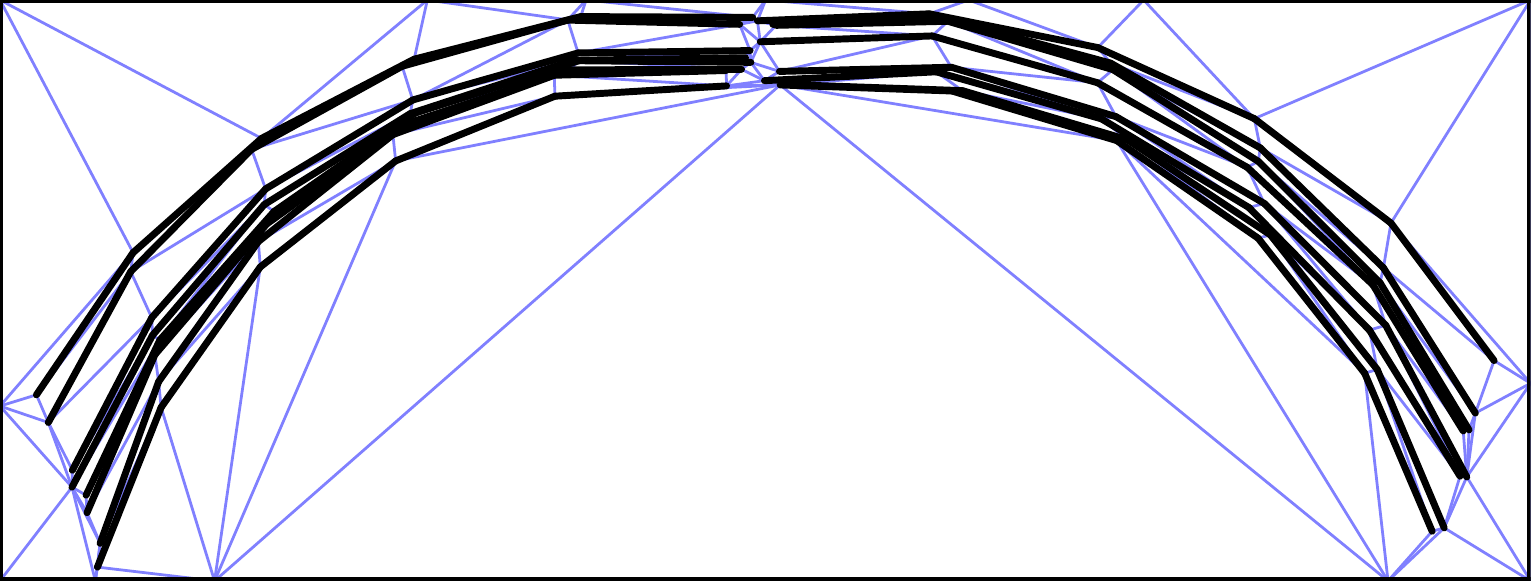}
    &
    \includegraphics[height=\hairExampleHeightEight,align=c]{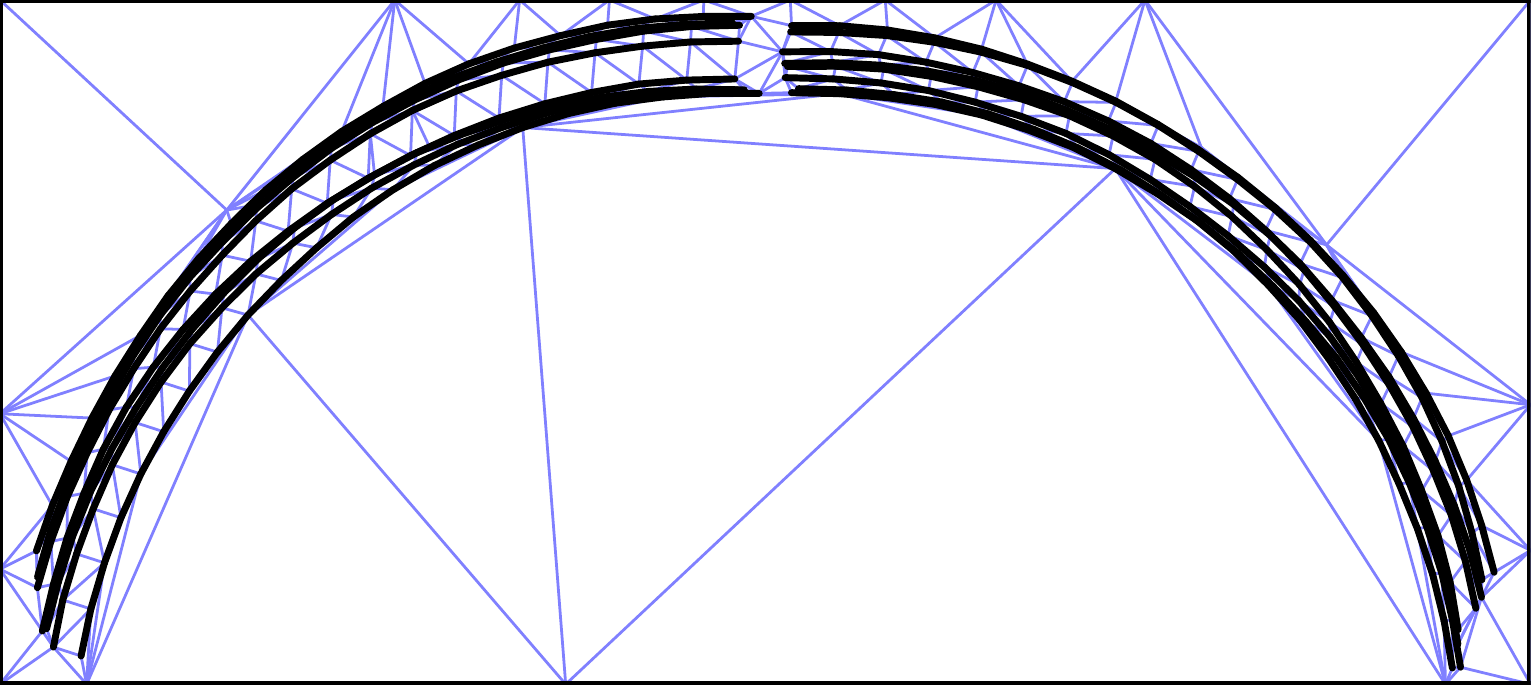}
\\[-0.2mm]
& \small 26.1 & \small 34.1
\end{tabular}
\end{center}
\caption{
    Two examples of the `hair' scene with eight strands per side ($N=8$, original geometry in black). 
    The CDT-based triangulations handle the region between neighbouring strands almost perfectly, but again shows problems near the bounding box and in the central empty space. Our optimization with fuzzy contraction removes the spurious edges, and a second iteration with an intermediate subdivision further improves the triangulation by retaining several Steiner vertices on the bounding box as well as  on the interface between the lowest strands and the central empty space (this is most clearly seen on 20 segments/strand image in the bottom right).
\label{figHairNEightTriangs}
}
\end{figure}

\newcommand{\hairExampleHeightMany}{3.74cm}
\begin{figure}
\begin{center}
\begin{tabular}{c@{\ }c@{\quad}c}
& 5 segments/strand & 20 segments/strand
\\
\rotatebox[origin=c]{90}{\small Unrefined CDT} &
    \includegraphics[height=\hairExampleHeightMany,align=c]{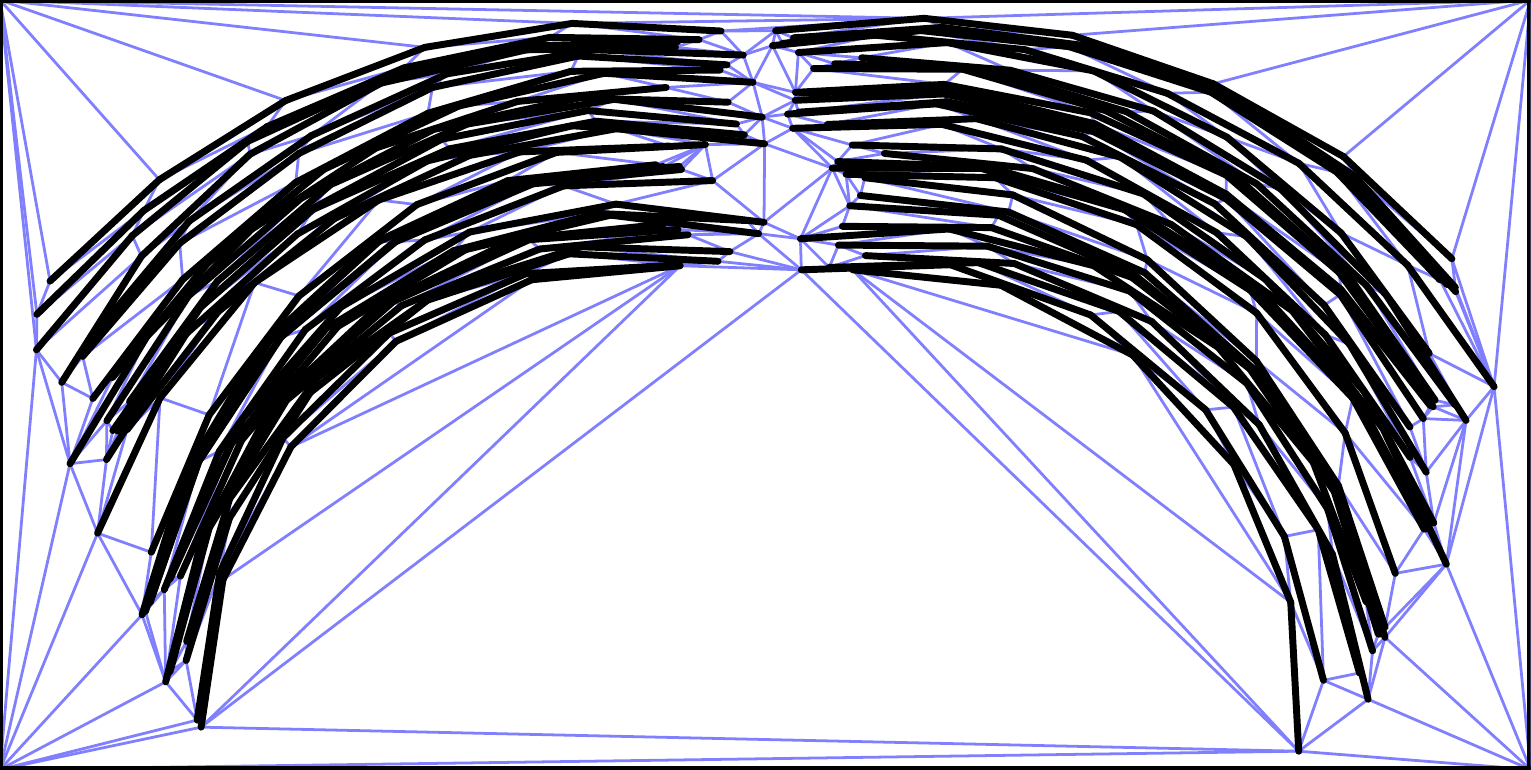}
    &
    \includegraphics[height=\hairExampleHeightMany,align=c]{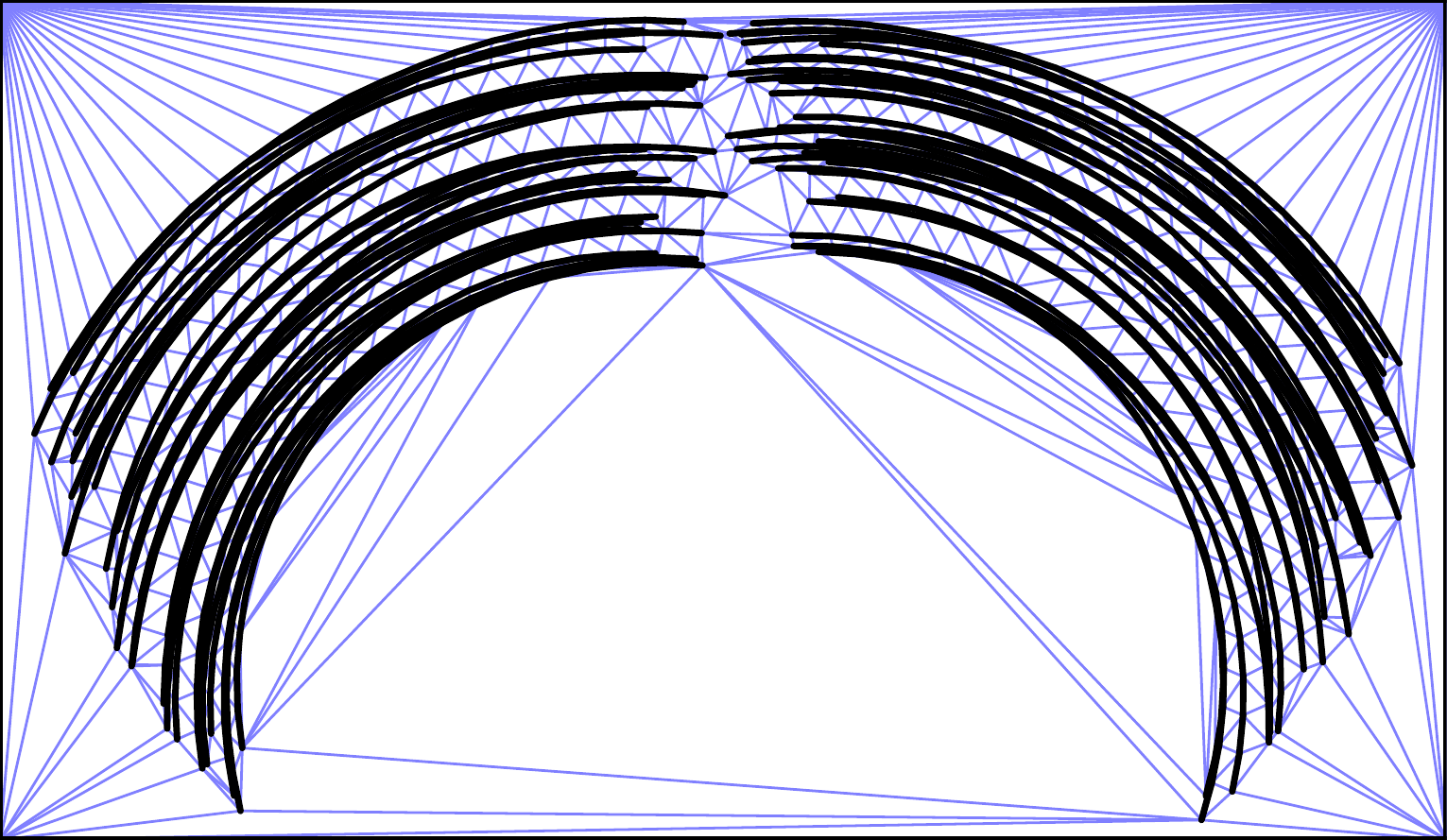}
\\[-0.2mm]
& \small 80.9 & \small 103.8
\\[1mm]
\rotatebox[origin=c]{90}{\mbox{\hspace{-1cm}\small Optim.\ refined CDT \hspace{-1cm}}}&
    \includegraphics[height=\hairExampleHeightMany,align=c]{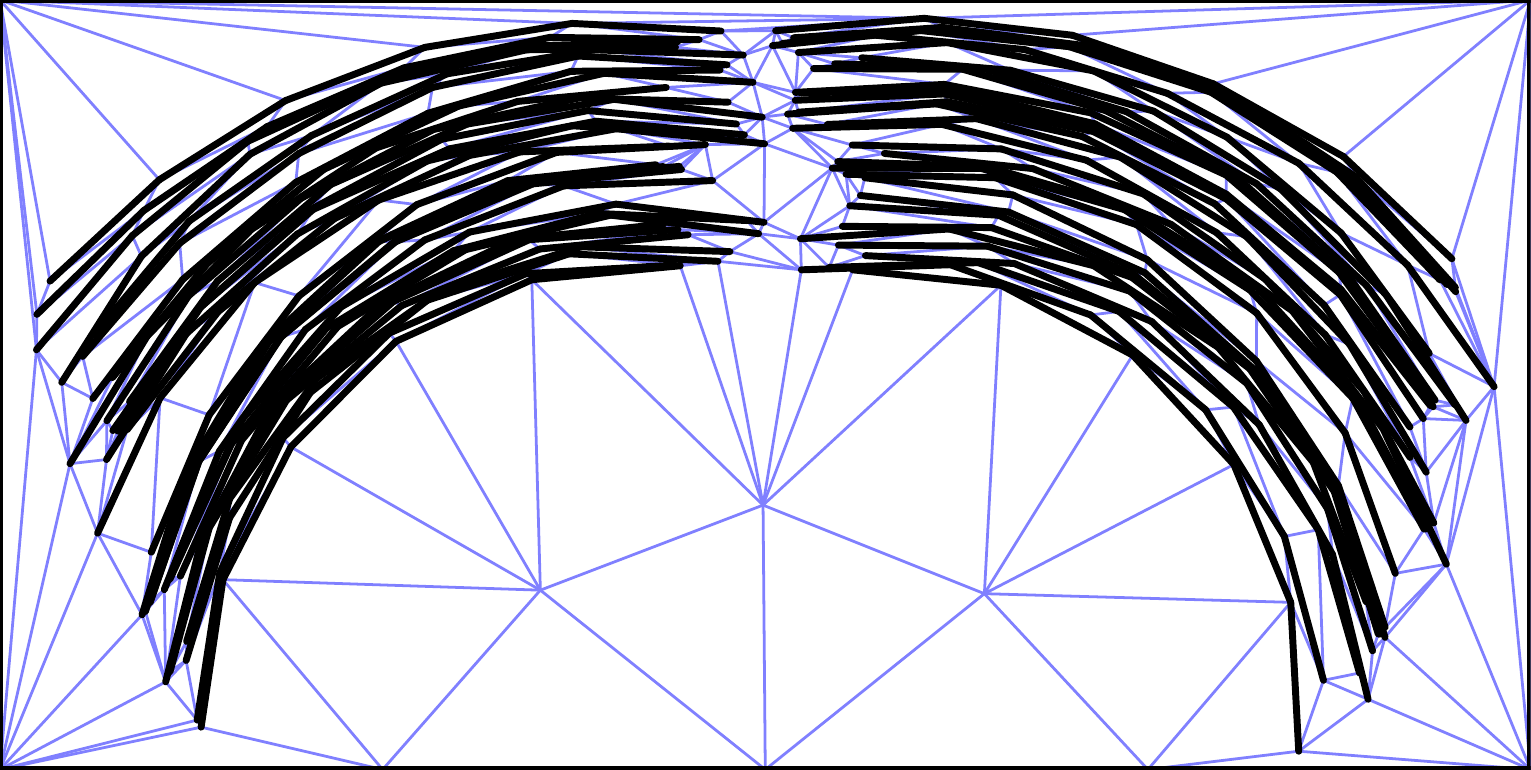}
    &
    \includegraphics[height=\hairExampleHeightMany,align=c]{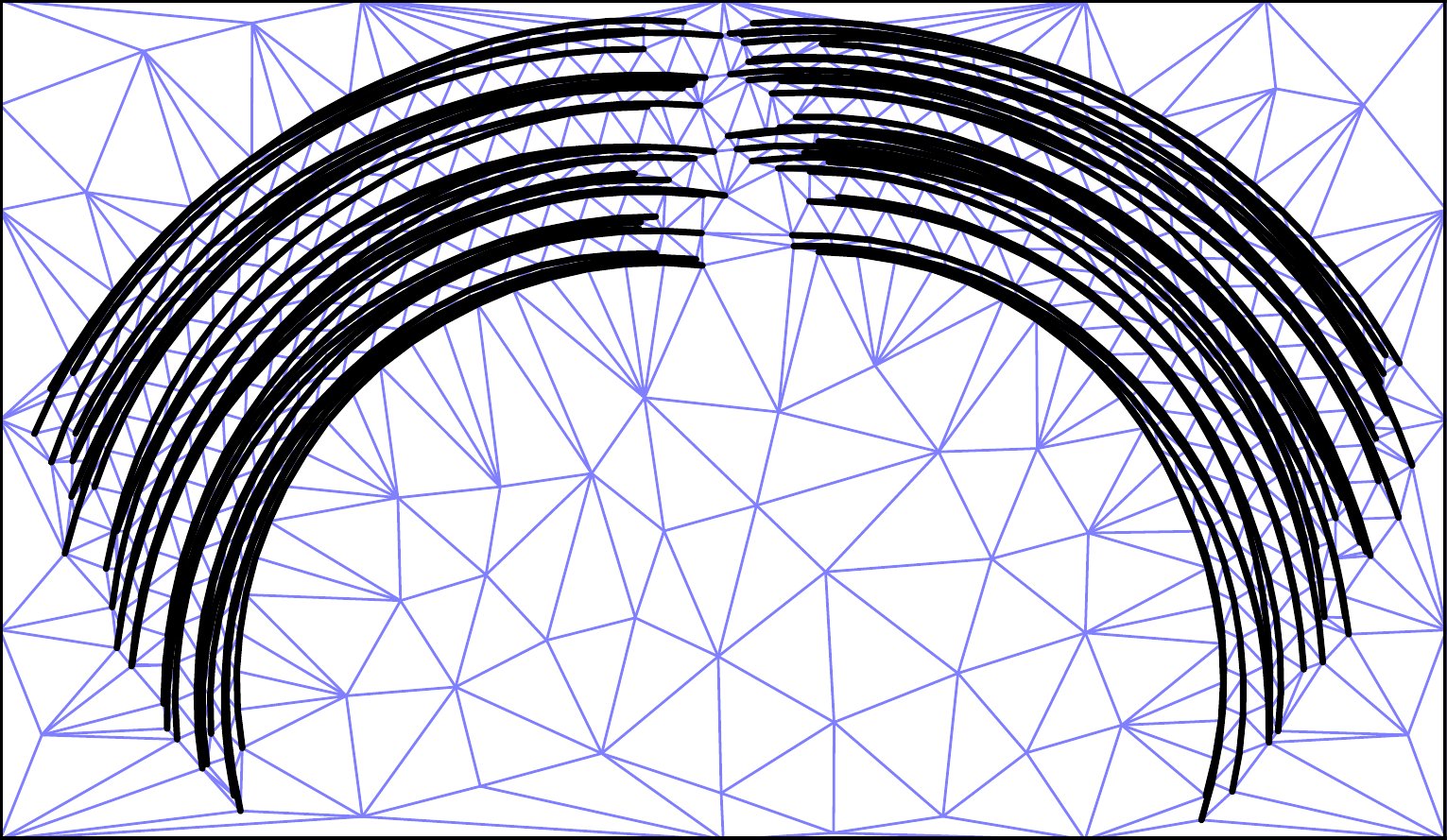}
\\[-0.2mm]
& \small 80.2 & \small 100.6
\\[1mm]
\rotatebox[origin=c]{90}{\small Optimized (ours)} &
    \includegraphics[height=\hairExampleHeightMany,align=c]{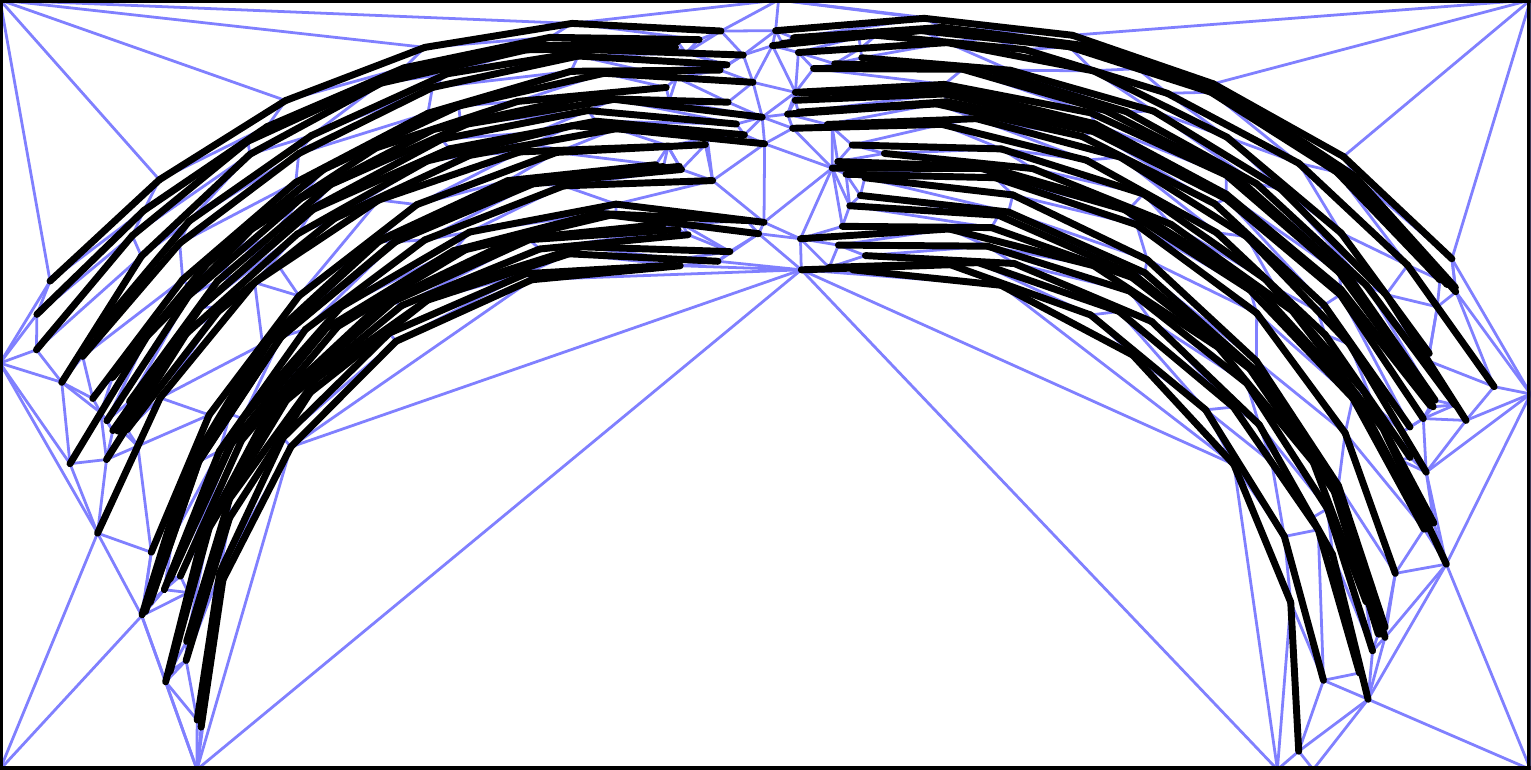}
    &
    \includegraphics[height=\hairExampleHeightMany,align=c]{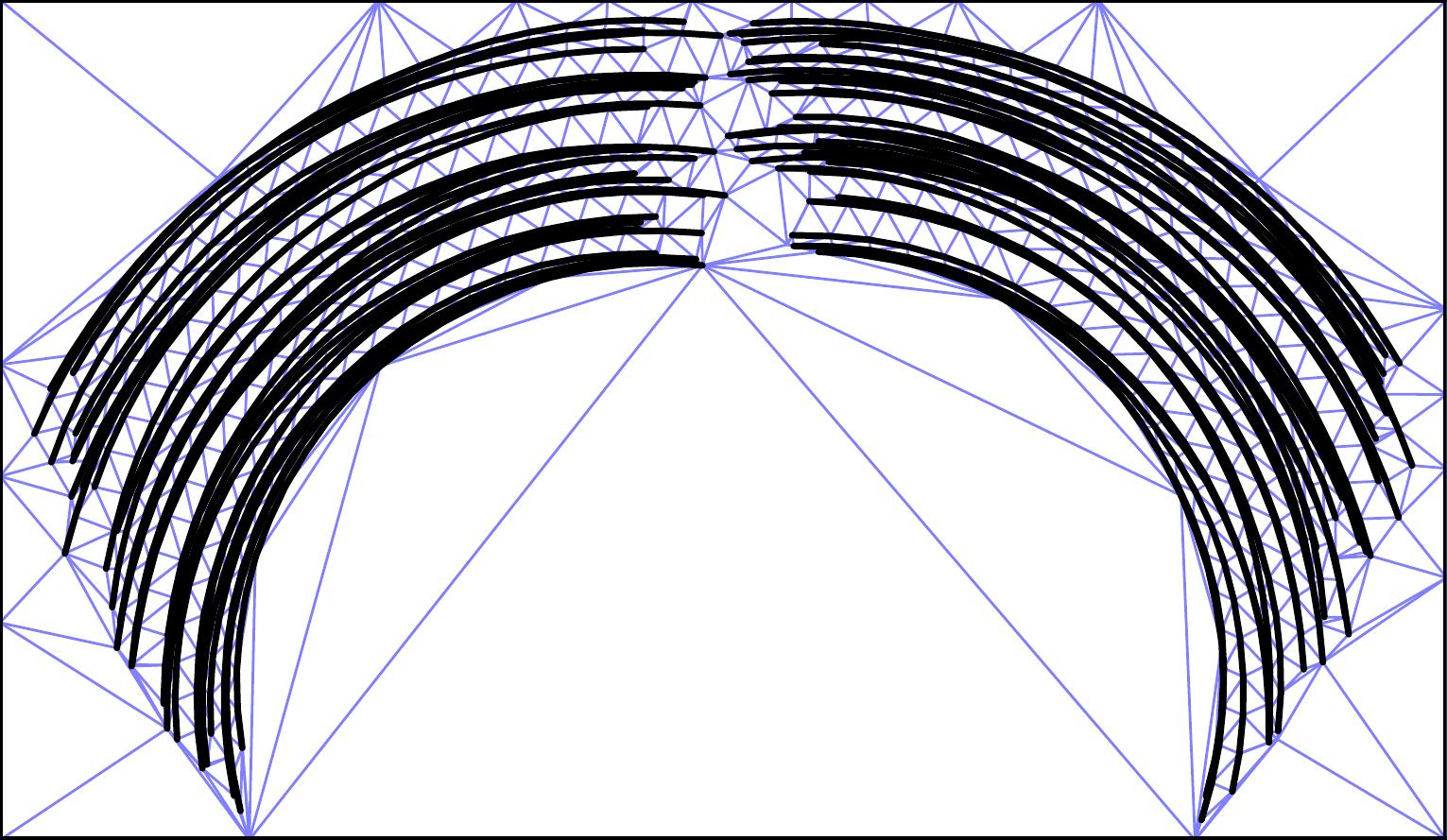}
\\[-0.2mm]
& \small 76.6 & \small 90.5
\end{tabular}
\end{center}
\caption{
    Two examples of the `hair' scene with $N=32$ strands per side. 
    We observe similar behaviour to the $N=8$ case of Fig.~\ref{figHairNEightTriangs}. The bottom row shows our fully optimized triangulation (including an intermediate subdivide). The relative gains in edge length by our optimization for these scenes are less than for those in Fig.~\ref{figHairNEightTriangs} because as $N$ increases, the total edge length gets dominated by (1) the original hair segments themselves and (2) the part of the triangulation between those hair segments, which is already nearly optimal in the CDT based methods.
\label{figHairNThreeTwoTriangs}
}
\end{figure}

\FloatBarrier

\subsection{Real-World Floor Plans \label{secOptimFloorPlans}}

The synthetic scenes that we used so far have been rather homogeneous in their distribution of geometry in both position and segment length.
As a final heterogeneous `real world' test set, we selected six public domain vectorised (SVG) images depicting floor plans of varying complexity. We discretised B\'ezier curves into straight line segments and cleaned some mesh issues by merging vertices that were intended to be overlapping but were not exactly identical numerically (note that the input geometry should be well-behaved in order to have a triangulation, the topology should be flat and free from intersecting segments).

The selected scenes and their triangulations with various levels of optimization  are shown in Figs.~\ref{figFloorPlanTriangsSmall} and \ref{figFloorPlanTriangsLarge}. 
For scale: the bounding box of each scene has unit length along the largest dimension.

We see the same issues pop up in the CDT-based triangulations that we saw before (cfr.\ the illustrative example in Sec.~\ref{secIllustrativeExample}), being the trade-off between keeping many inefficient long slender triangles that fill a large empty region versus needlessly subdividing slender triangles that are constrained to their slender shape by the input geometry.
The region around (segmented) curved geometry is especially prone to spurious slender triangles, as can be seen in the \PE\ and White House scenes (Fig.~\ref{figFloorPlanTriangsSmall}, middle and bottom)
and in lesser extent for the Seville scene (Fig.~\ref{figFloorPlanTriangsLarge}, middle).
These triangles can typically be optimized by a relatively cheap polishing step with edge flipping, and a full iteration with an intermediate subdivide followed by fuzzy contraction can often further improve these regions by retaining additional Steiner vertices, albeit at a higher computational cost.

\begin{landscape}
\begin{figure}
\newcommand{\lscapeTabWidth}{5.3cm}
\begin{center}
\begin{tabular}{c@{\ }c@{\ }c@{\ }c@{\ }c}
& Optimally refined CDT & + Polish with edge flipping & + Fuzzy contraction & + Intermediate subdivide
\\
\rotatebox[origin=c]{90}{\small Louvre} &
    \includegraphics[width=\lscapeTabWidth,align=c]{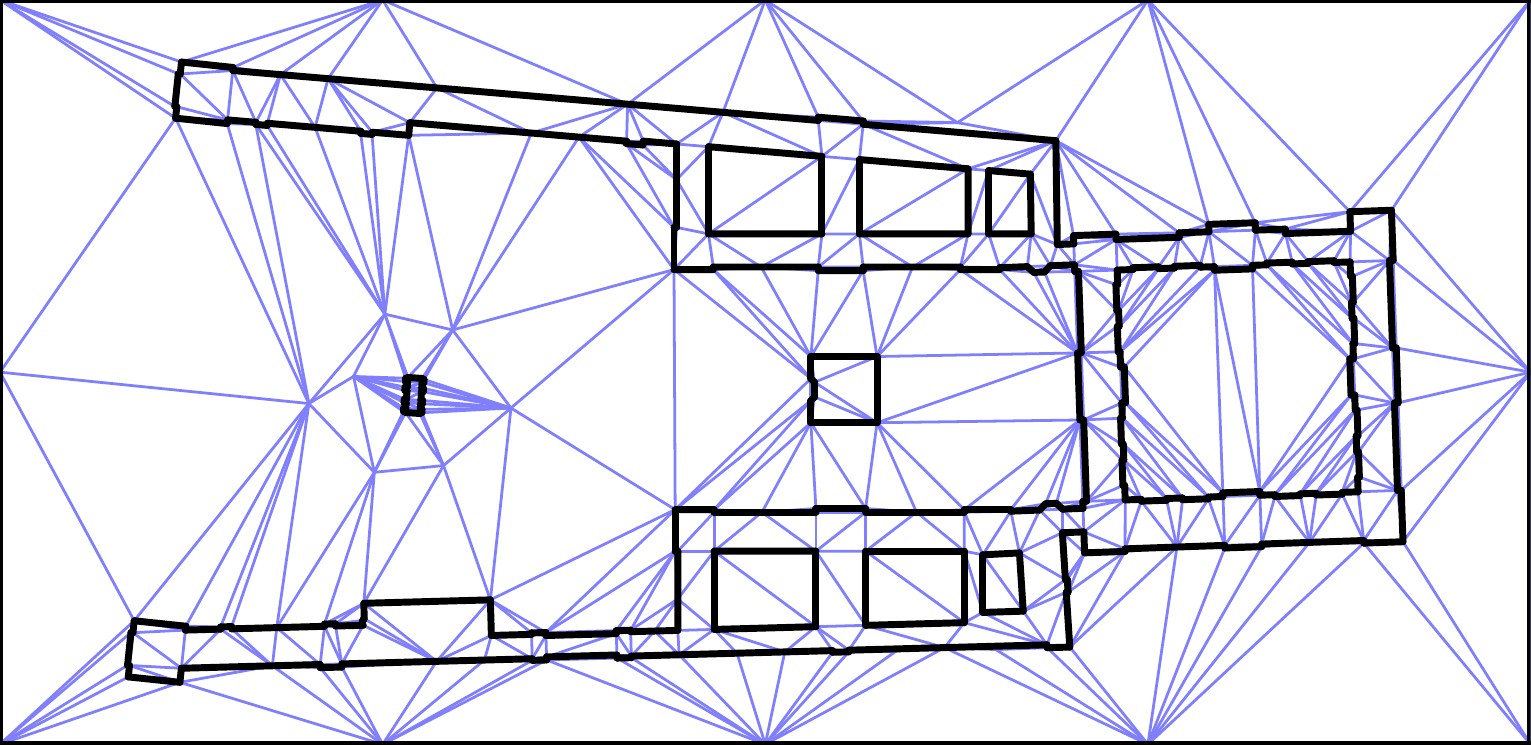}
    &
    \includegraphics[width=\lscapeTabWidth,align=c]{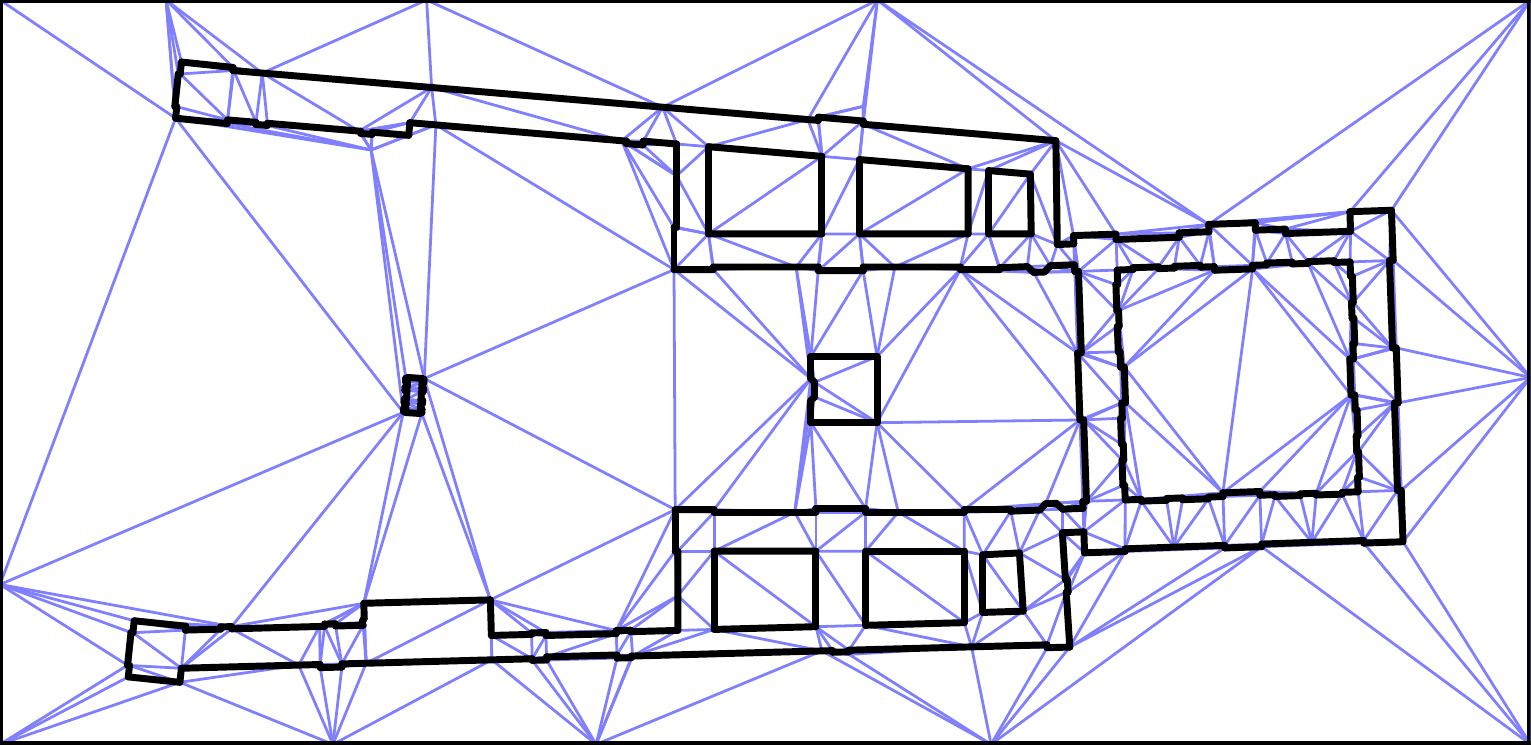}
    &
    \includegraphics[width=\lscapeTabWidth,align=c]{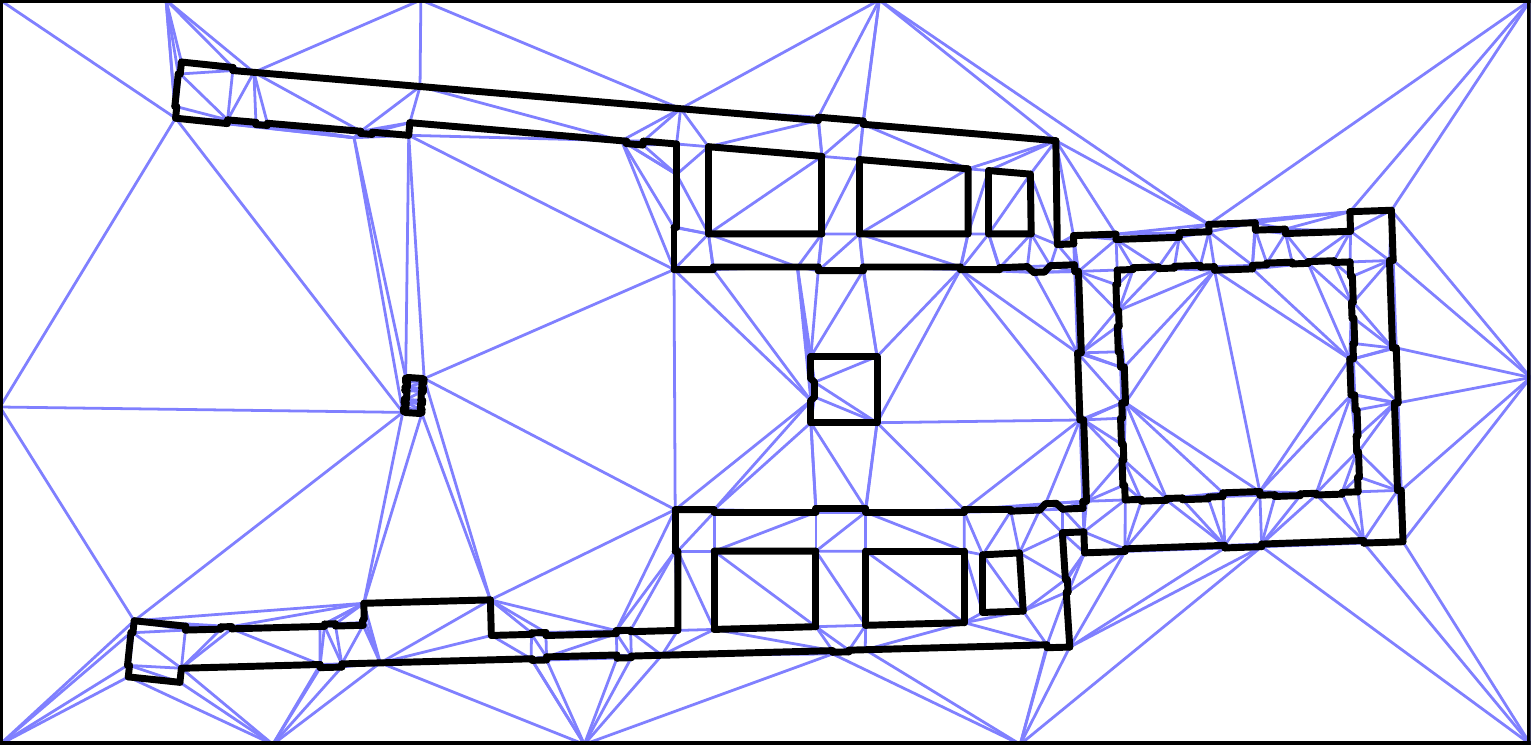}
    &
    \includegraphics[width=\lscapeTabWidth,align=c]{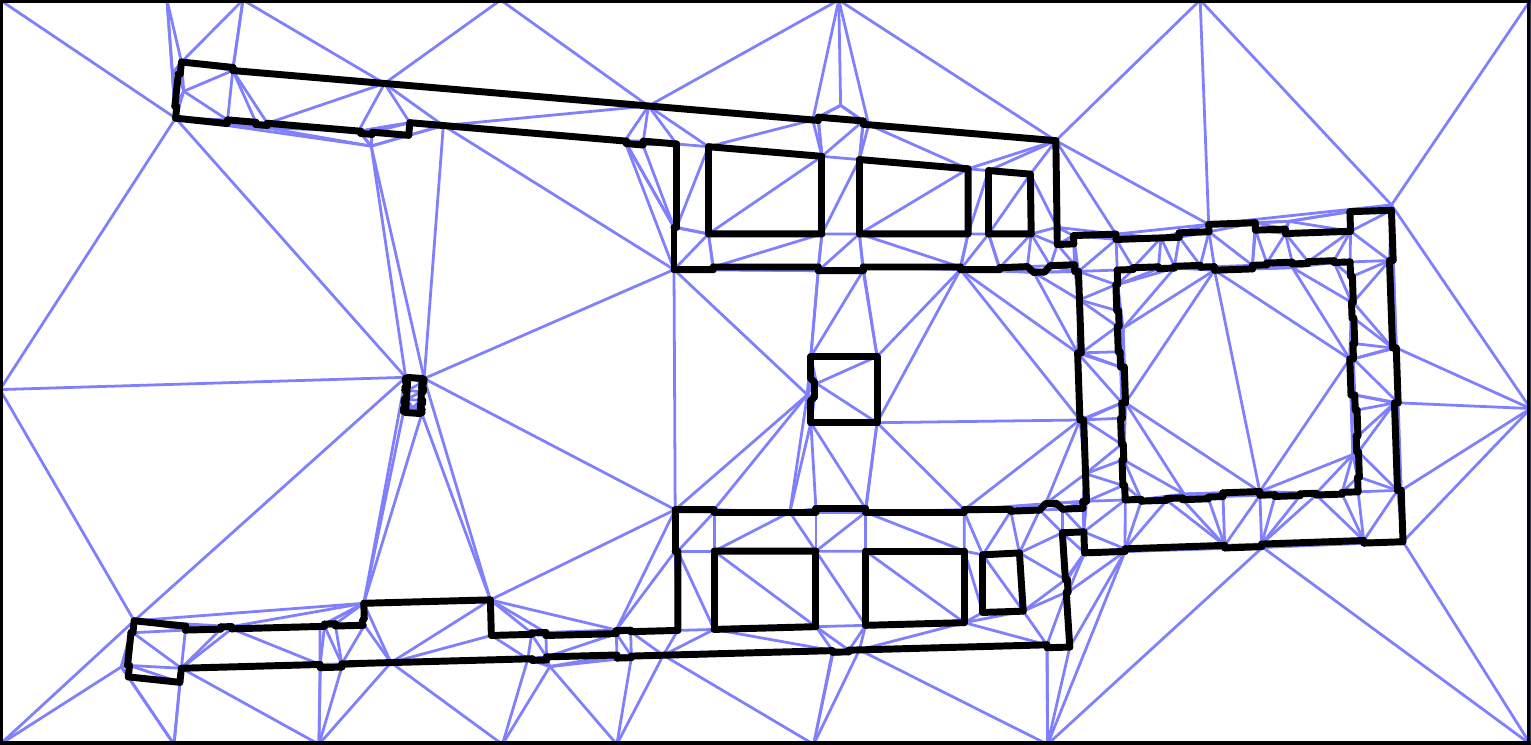}
    \\[-0.2mm]
    & \small 41.8 (0.88, 1.00) & \small 37.1 (0.78, 0.89) & \small 36.9 (0.78, 0.88) & \small 36.2 (0.76, 0.87) \\[0.5mm]
\rotatebox[origin=c]{90}{\small Pont-l'Ev\^eque} &
    \includegraphics[width=\lscapeTabWidth,align=c]{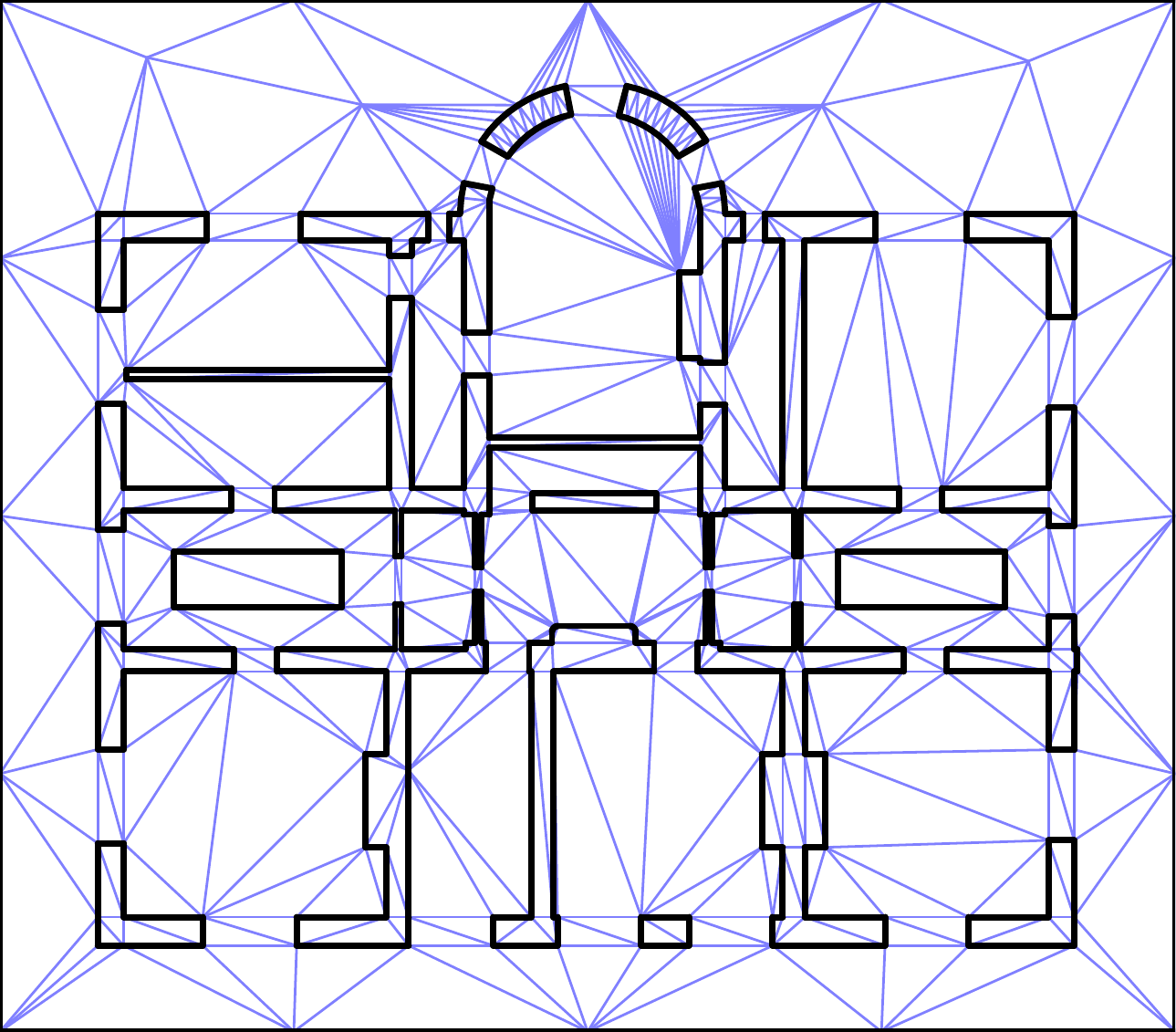}
    &
    \includegraphics[width=\lscapeTabWidth,align=c]{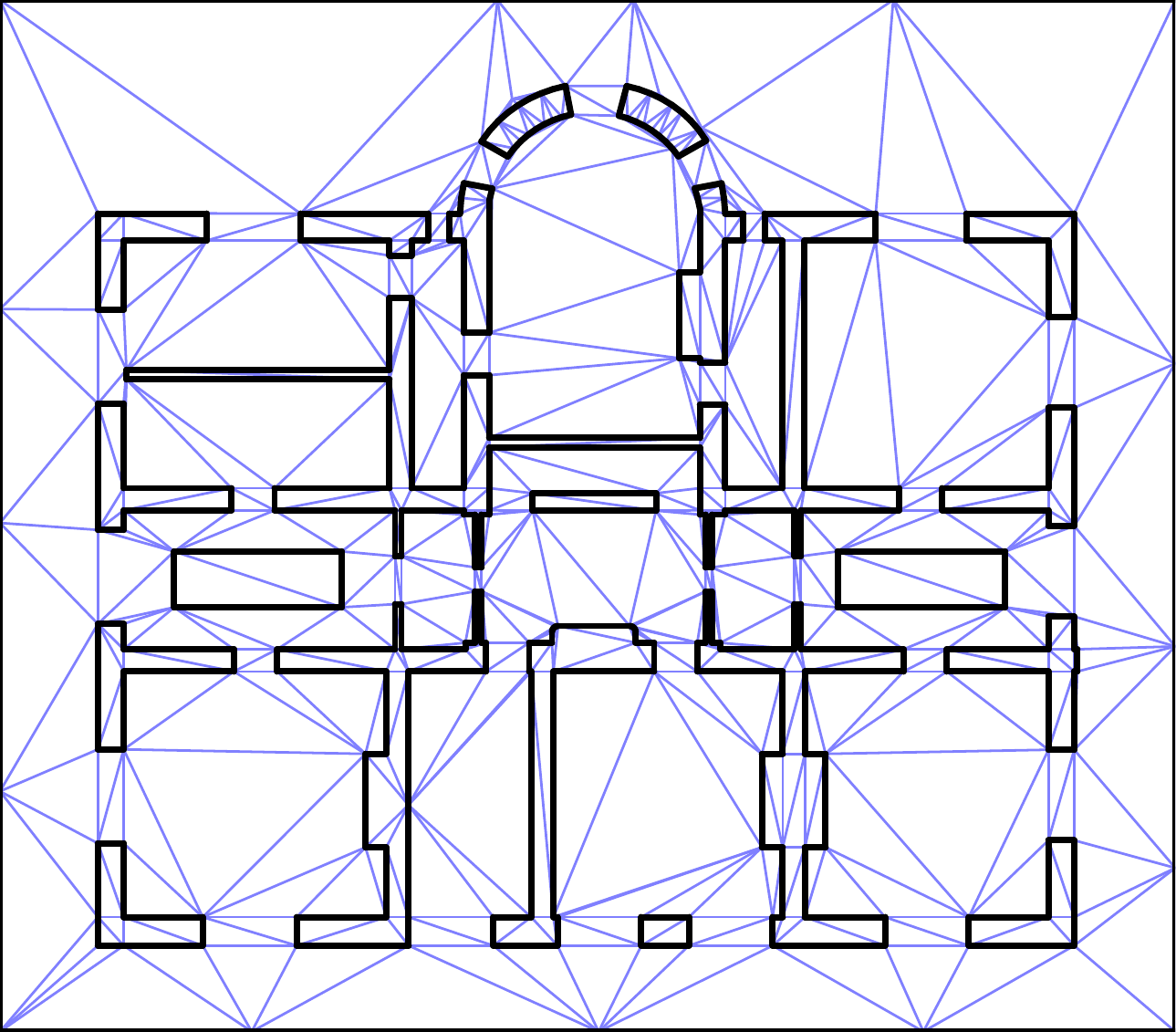}
    &
    \includegraphics[width=\lscapeTabWidth,align=c]{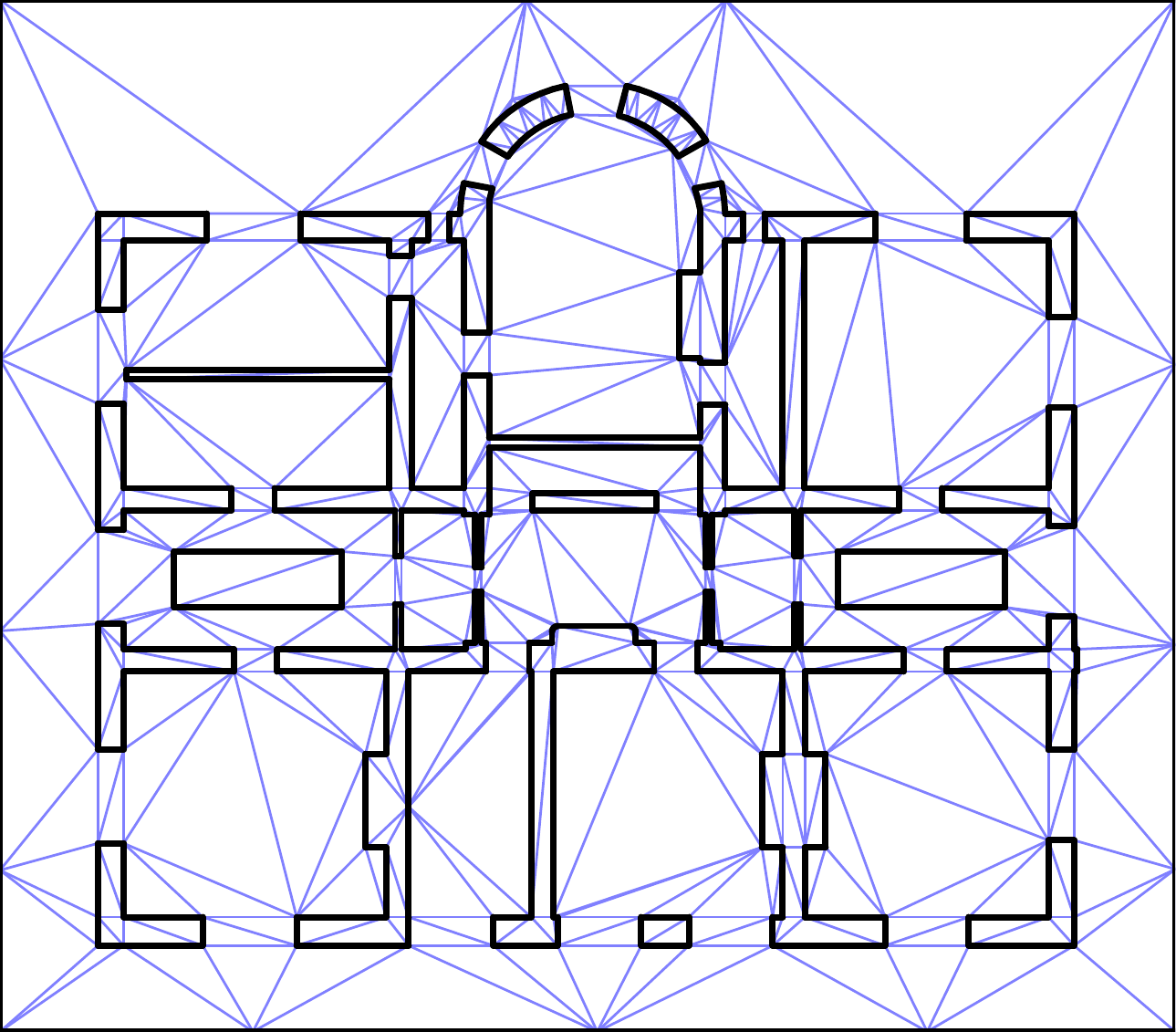}
    &
    \includegraphics[width=\lscapeTabWidth,align=c]{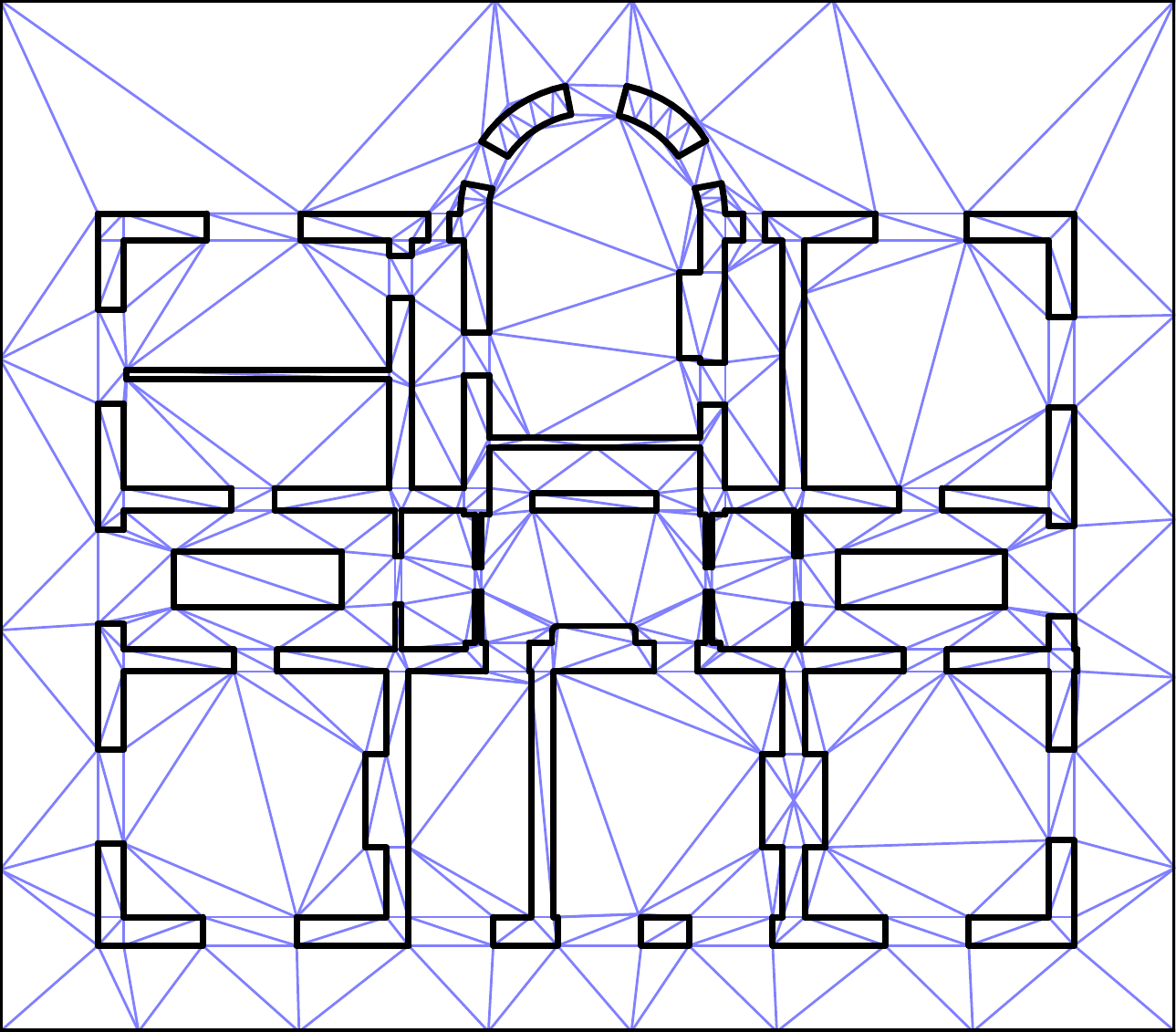}
    \\[-0.2mm]
    & \small 57.5 (0.86, 1.00) & \small 53.7 (0.81, 0.93) & \small 53.6 (0.80, 0.93) & \small 53.0 (0.79, 0.92) \\[0.5mm]
\rotatebox[origin=c]{90}{\small White House} &
    \includegraphics[width=\lscapeTabWidth,align=c]{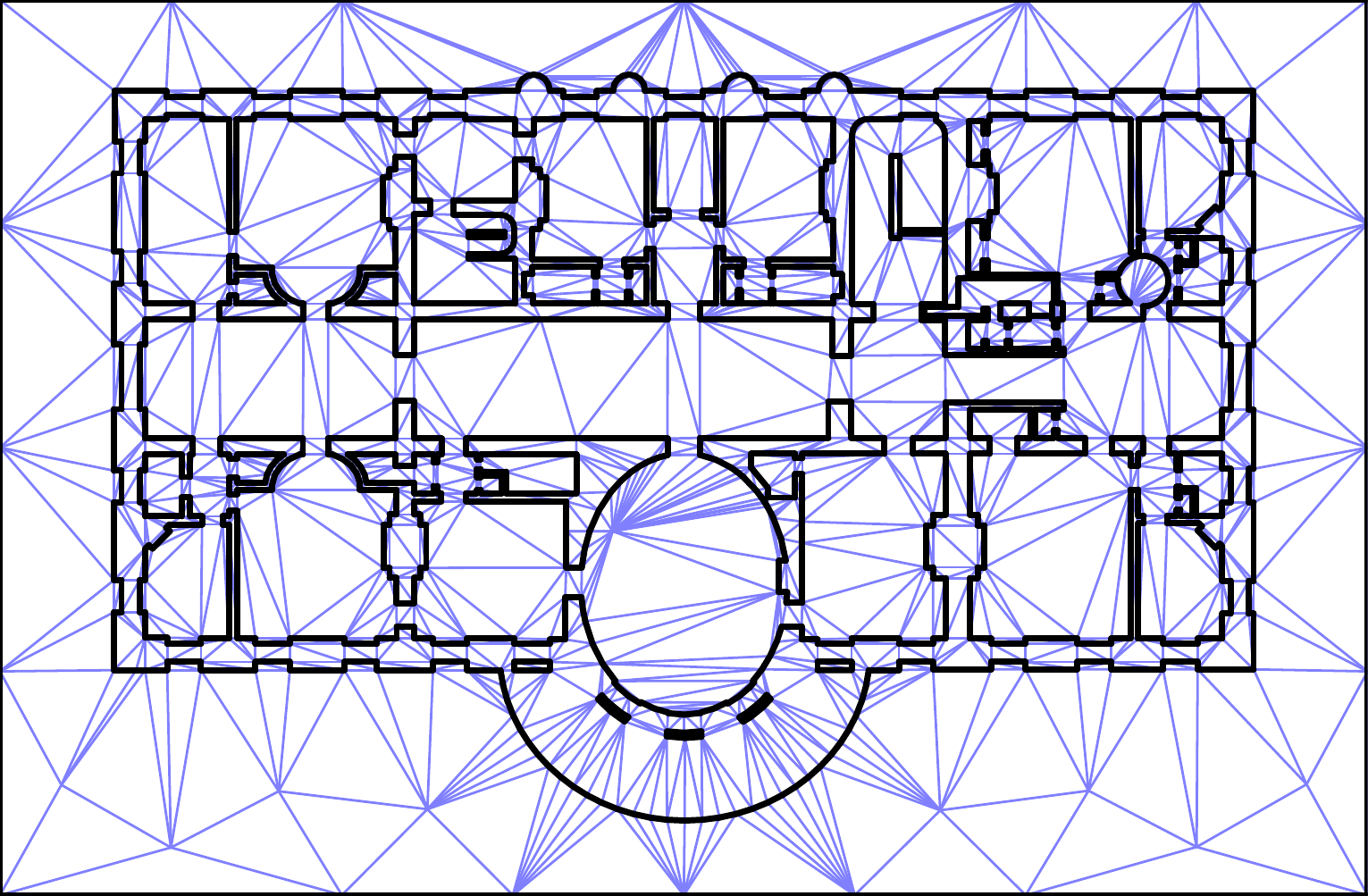}
    &
    \includegraphics[width=\lscapeTabWidth,align=c]{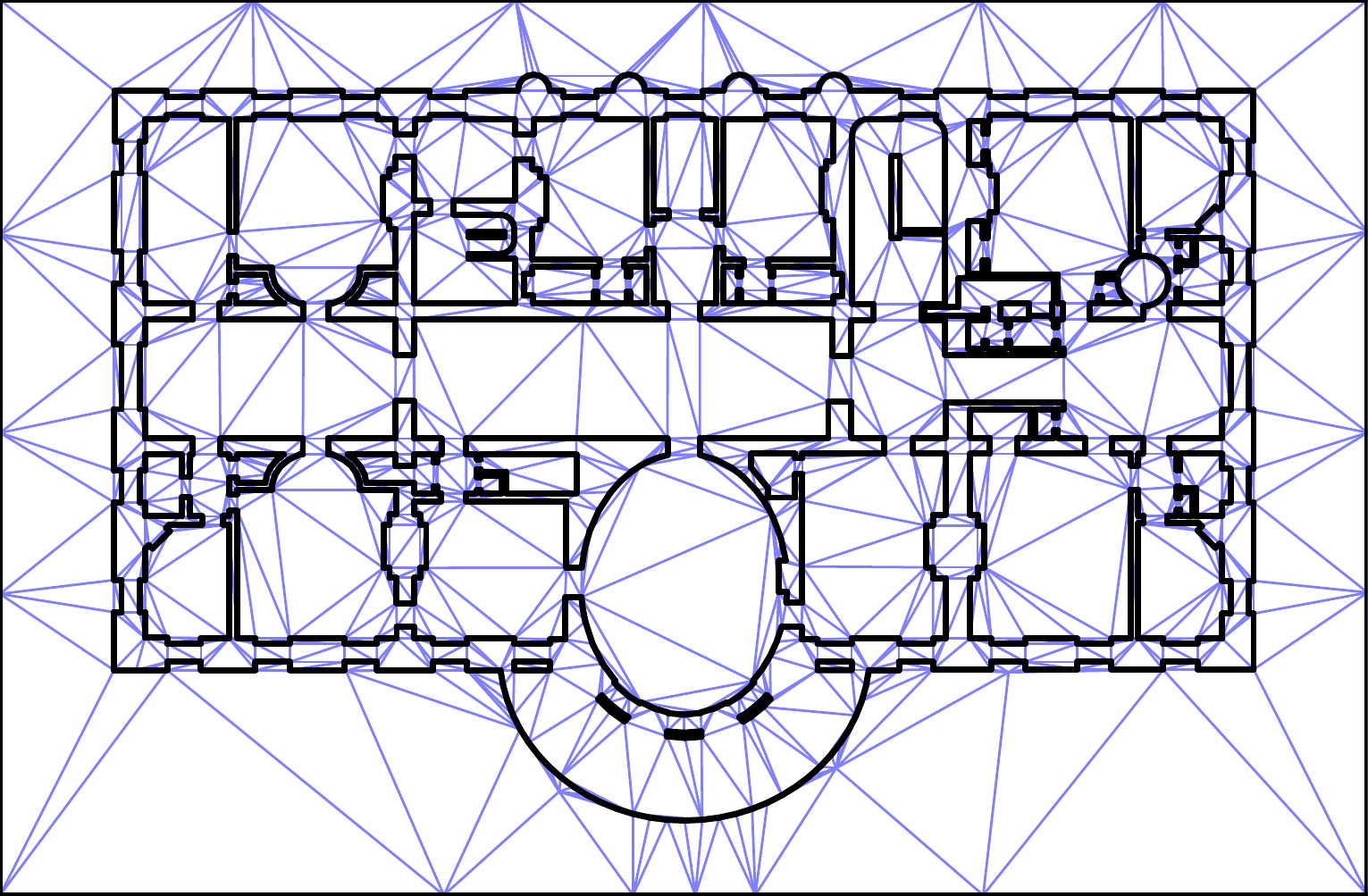}
    &
    \includegraphics[width=\lscapeTabWidth,align=c]{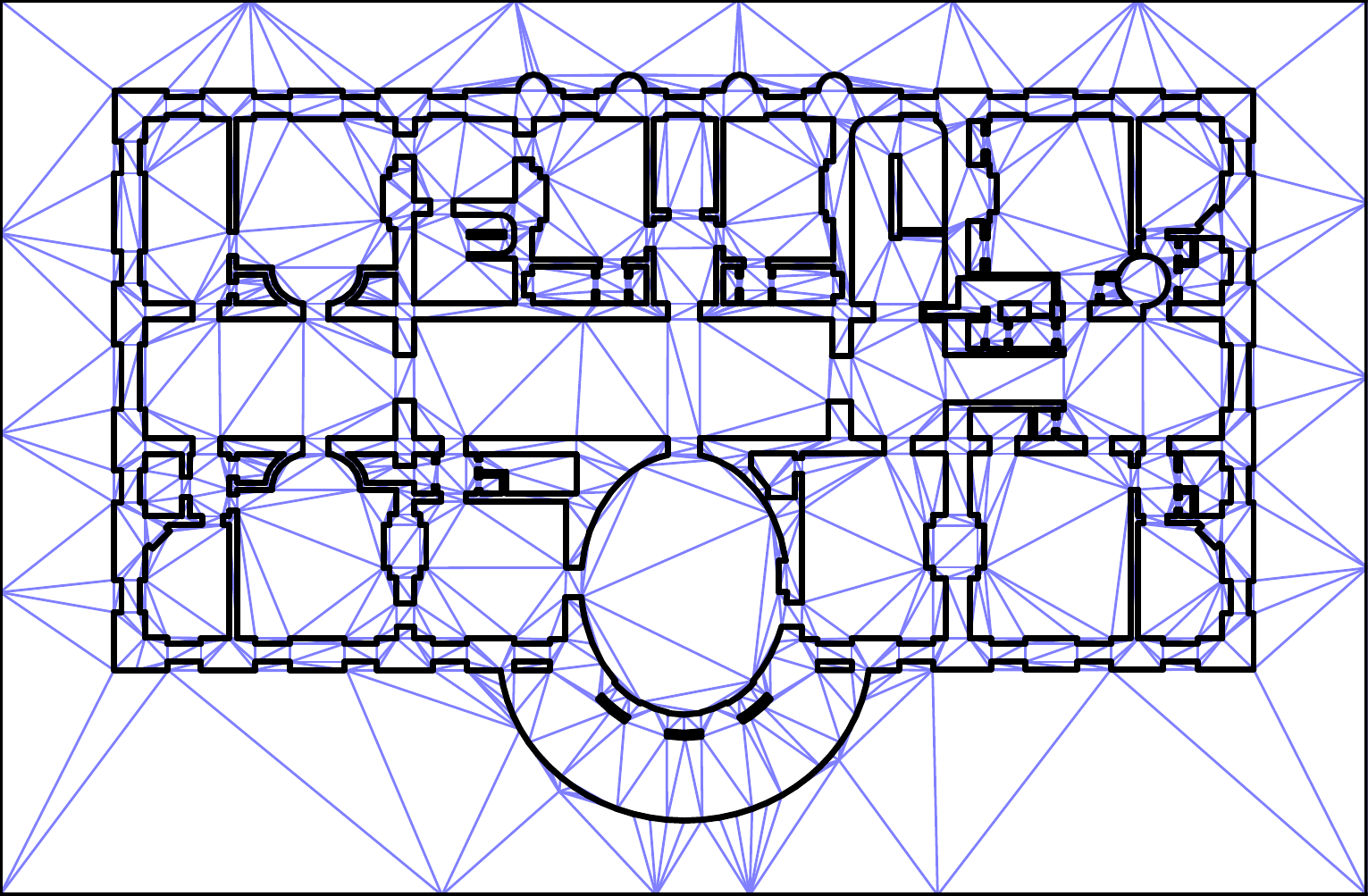}
    &
    \includegraphics[width=\lscapeTabWidth,align=c]{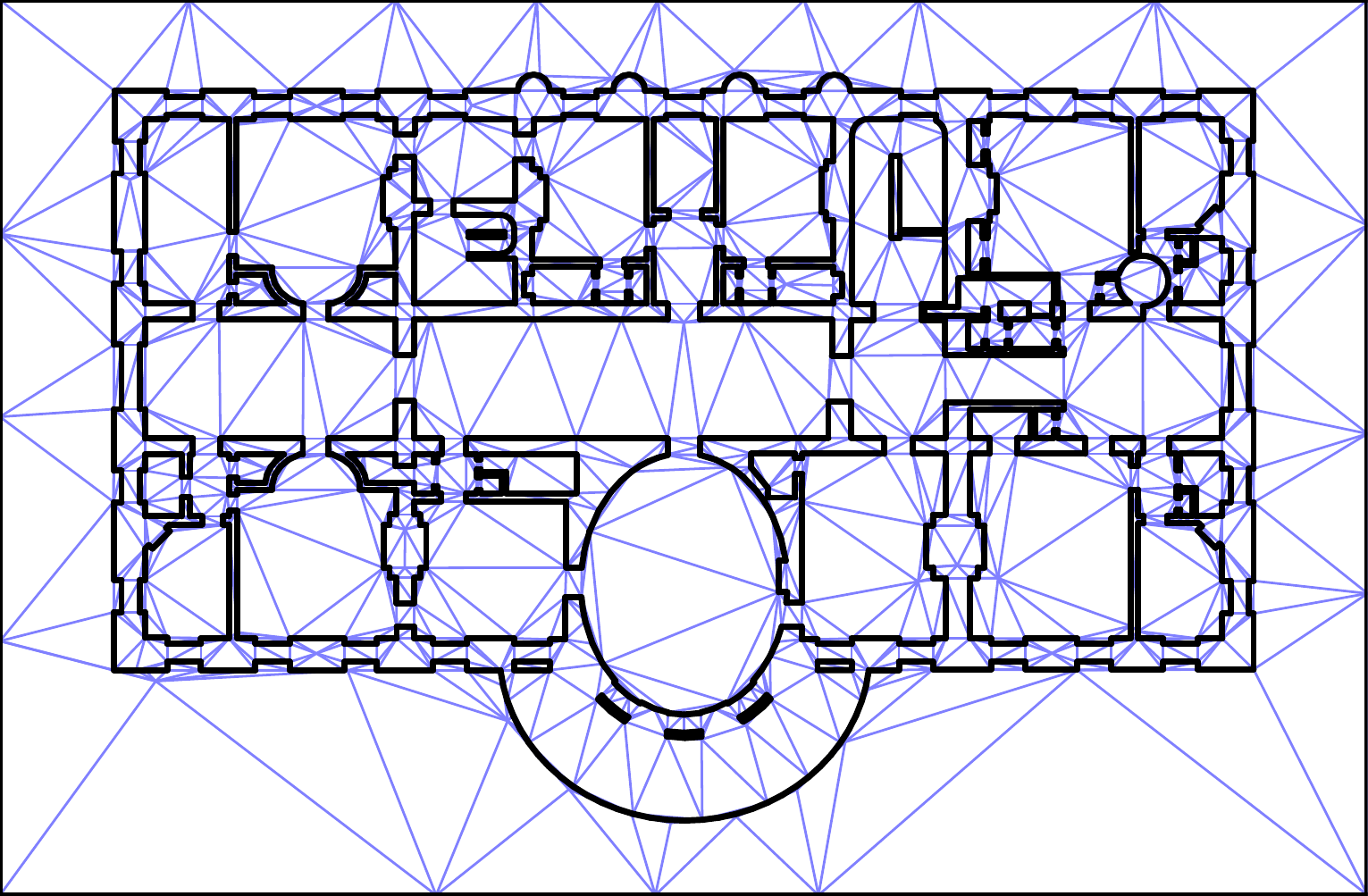}
    \\[-0.2mm]
    & \small 87.4 (0.88, 1.00) & \small 80.3 (0.81, 0.92) & \small 80.1 (0.81, 0.92) & \small 78.6 (0.79, 0.90)
\end{tabular}
\vspace{-0.2cm}
\end{center}
\caption{From top to bottom: triangulations of floor plans of the Louvre (326 segments, combined length 8.91), the prison of Pont-l'Ev\^eque (290 segments, combined length 16.0) and the White House (1152 segments, combined length 19.2).
The total edge lengths of the triangulations are given below each image, followed by the relative lengths compared to an unrefined CDT and the optimally refined CDT in parentheses.
The edge lengths for the unrefined CDT (not shown) are: Louvre:\! 47.4, Pont l'Ev\^eque:\! 66.7, White House:\! 99.3.
\label{figFloorPlanTriangsSmall}
}
\end{figure}

\begin{figure}
\newcommand{\lscapeTabWidth}{5.1cm}
\vspace{-1.2cm}
\begin{center}
\begin{tabular}{c@{\ }c@{\ }c@{\ }c@{\ }c}
& Optimally refined CDT & + Polish with edge flipping & + Fuzzy contraction & + Intermediate subdivide
\\
\rotatebox[origin=c]{90}{\small Topkapı} &
    \includegraphics[angle=270,origin=c,width=\lscapeTabWidth,align=c]{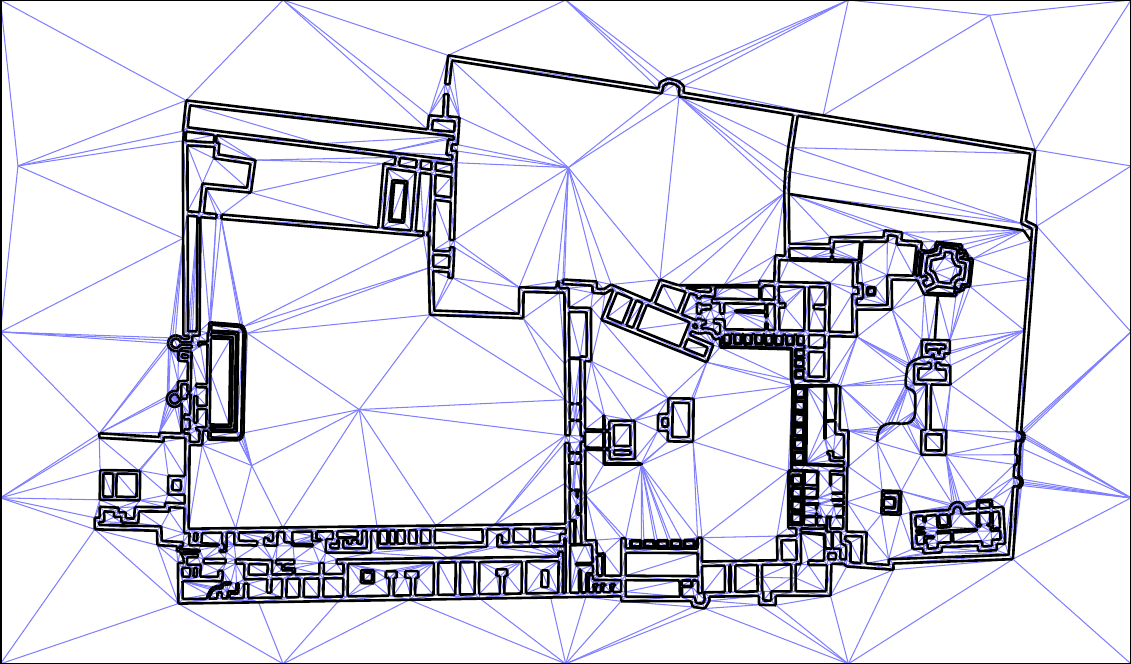}
    &
    \includegraphics[angle=270,origin=c,width=\lscapeTabWidth,align=c]{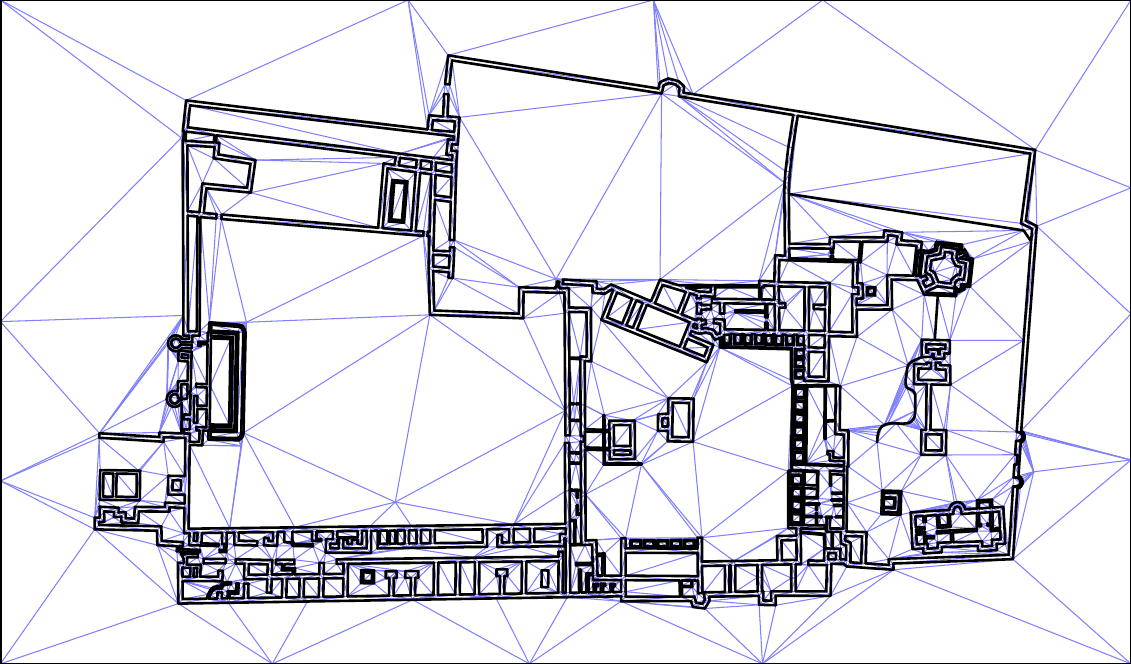}
    &
    \includegraphics[angle=270,origin=c,width=\lscapeTabWidth,align=c]{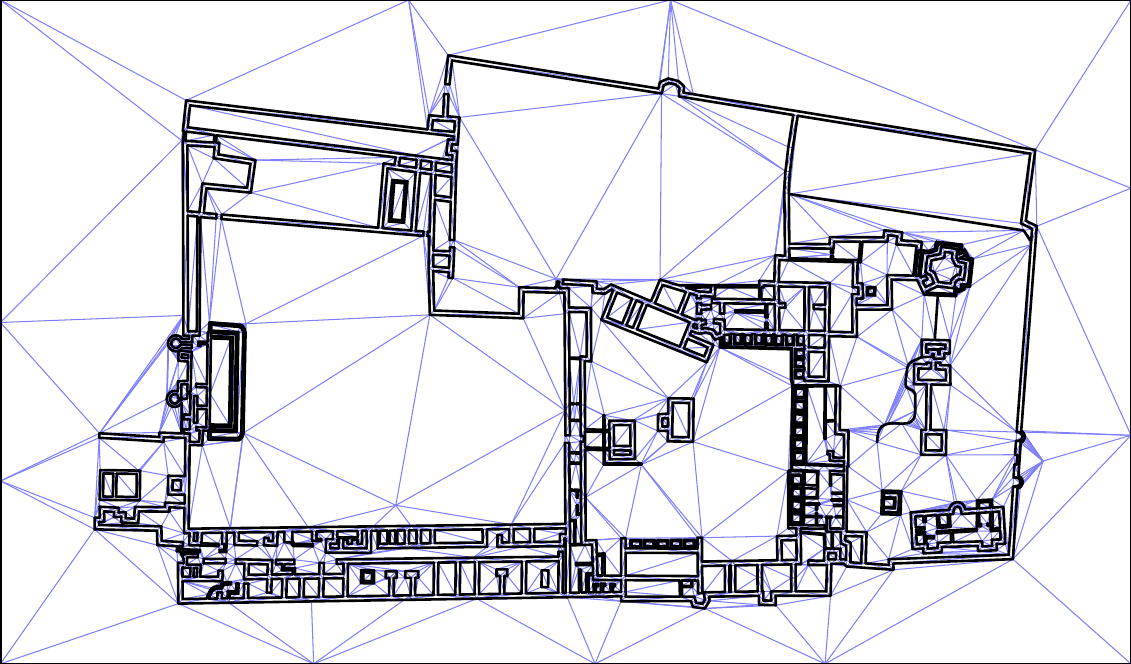}
    &
    \includegraphics[angle=270,origin=c,width=\lscapeTabWidth,align=c]{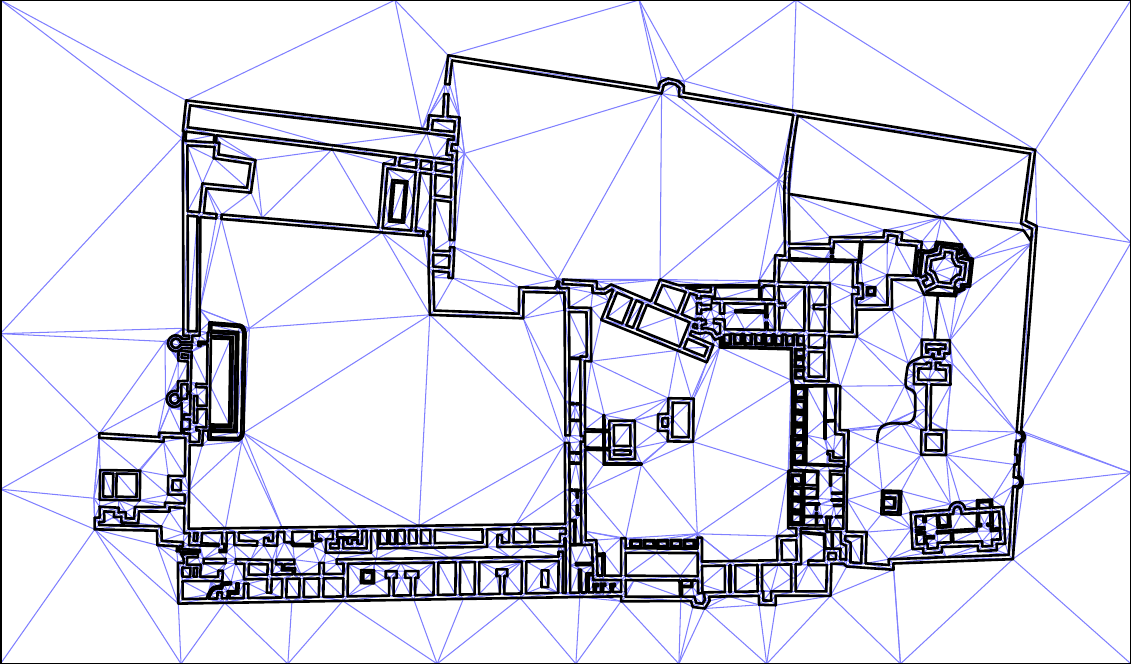}
    \\[-0.2mm]
    & \small 93.0 (0.89, 1.00) & \small 86.6 (0.83, 0.93) & \small 86.4 (0.83, 0.93) & \small 83.2 (0.80, 0.89) \\[0.5mm]
\rotatebox[origin=c]{90}{\small Seville} &
    \includegraphics[width=\lscapeTabWidth,align=c]{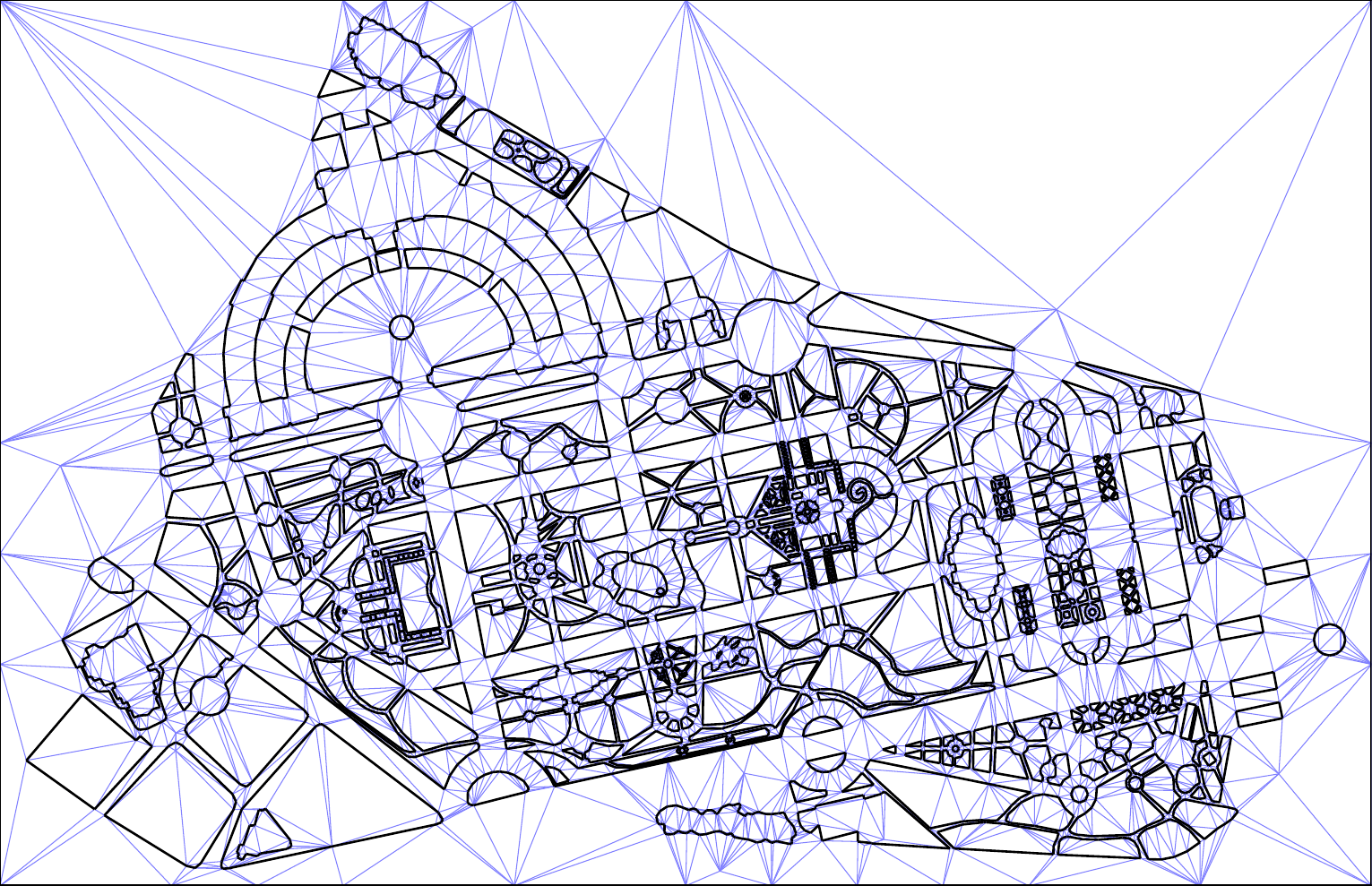}
    &
    \includegraphics[width=\lscapeTabWidth,align=c]{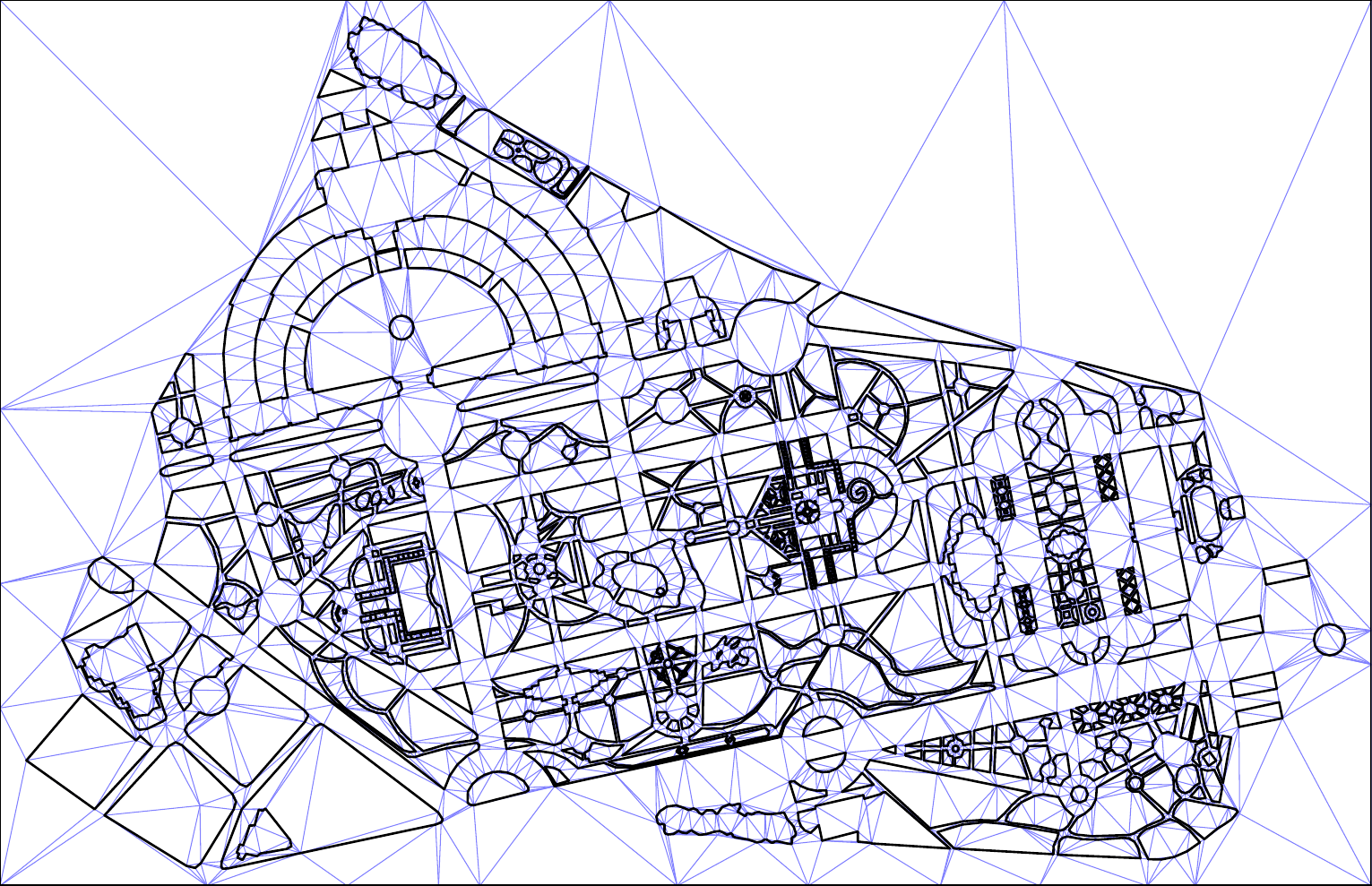}
    &
    \includegraphics[width=\lscapeTabWidth,align=c]{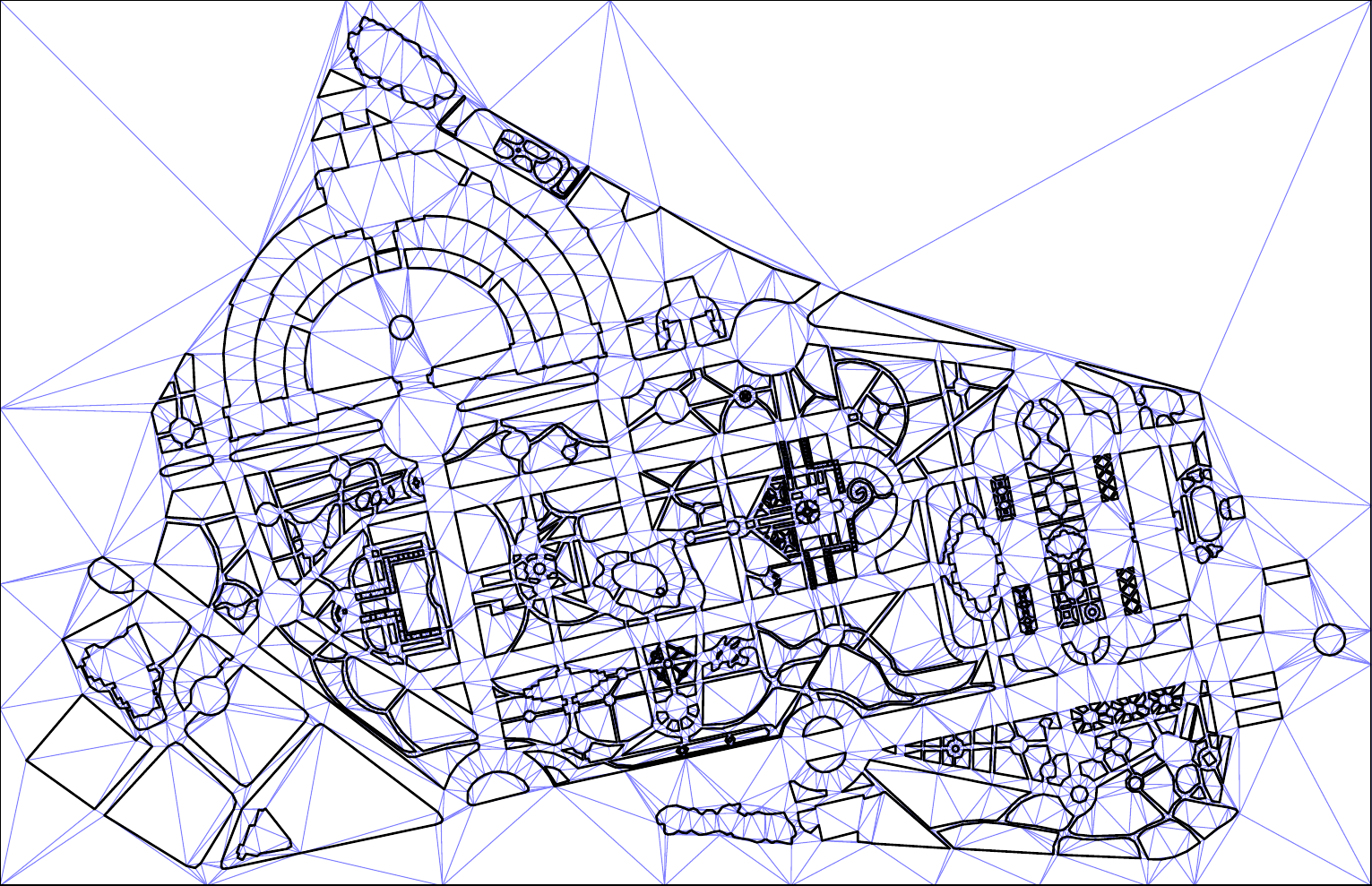}
    &
    \includegraphics[width=\lscapeTabWidth,align=c]{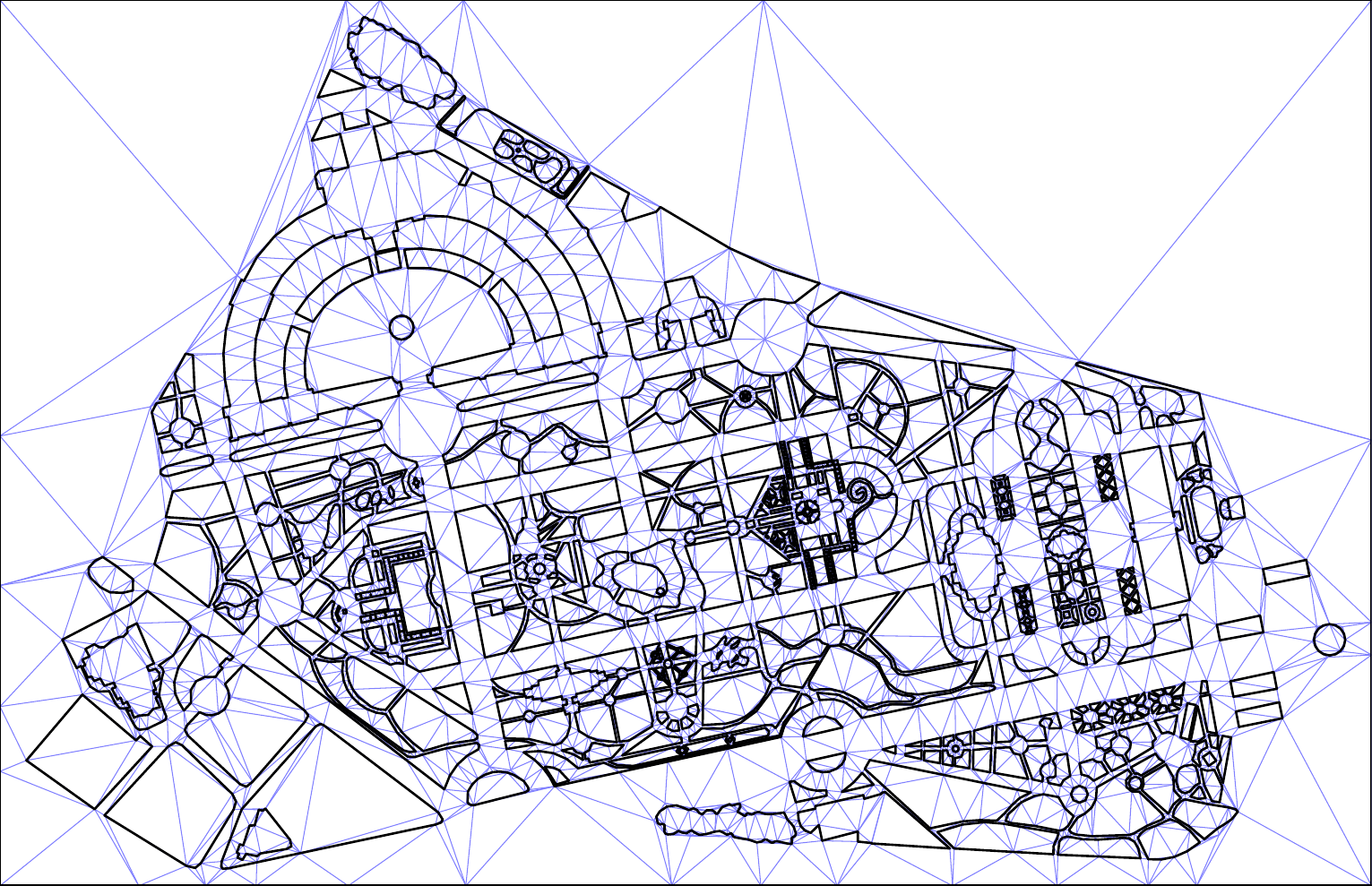} 
    \\[-0.2mm]
    & \small 146.6 (0.92, 1.00) & \small 139.0 (0.78, 0.95) & \small 138.2 (0.86, 0.94) & \small 136.5 (0.85, 0.93)\\[0.5mm]
\rotatebox[origin=c]{90}{\small Thessaloniki} &
    \includegraphics[width=\lscapeTabWidth,align=c]{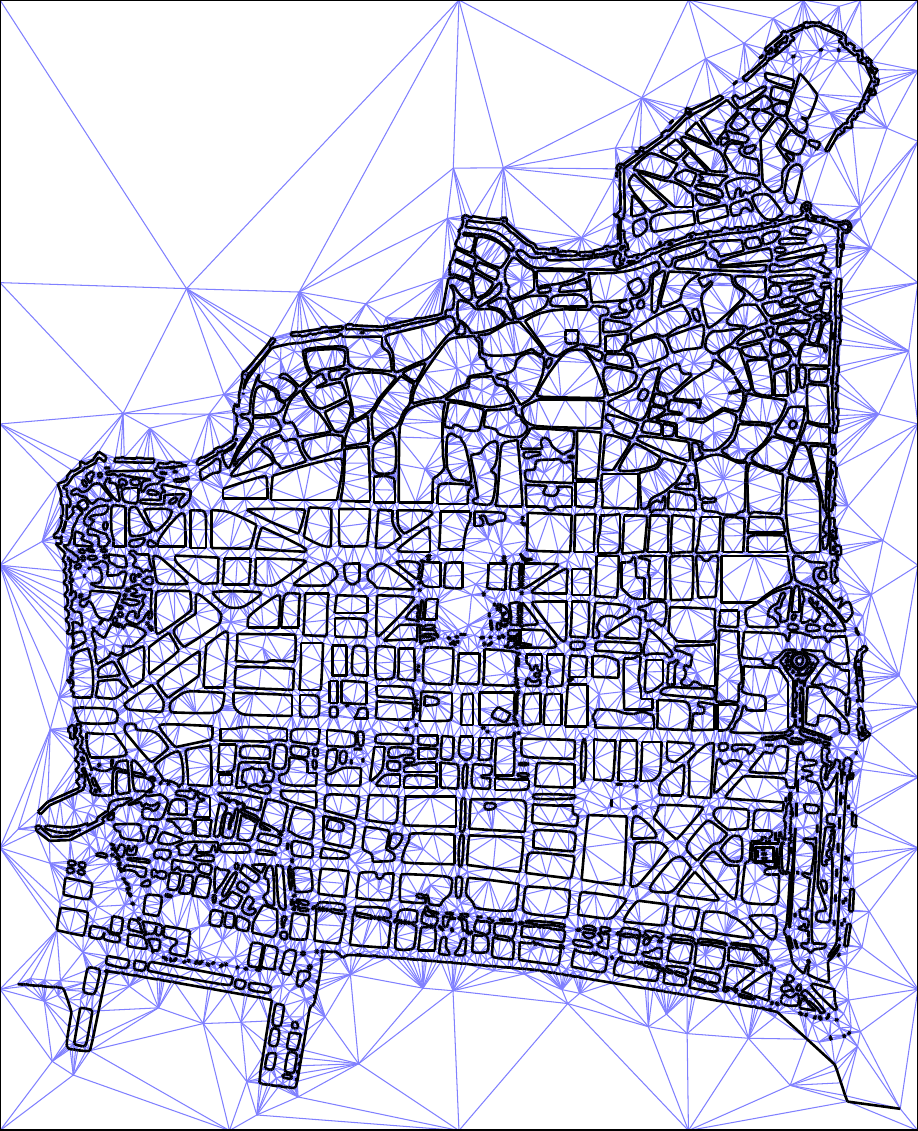}
    &
    \includegraphics[width=\lscapeTabWidth,align=c]{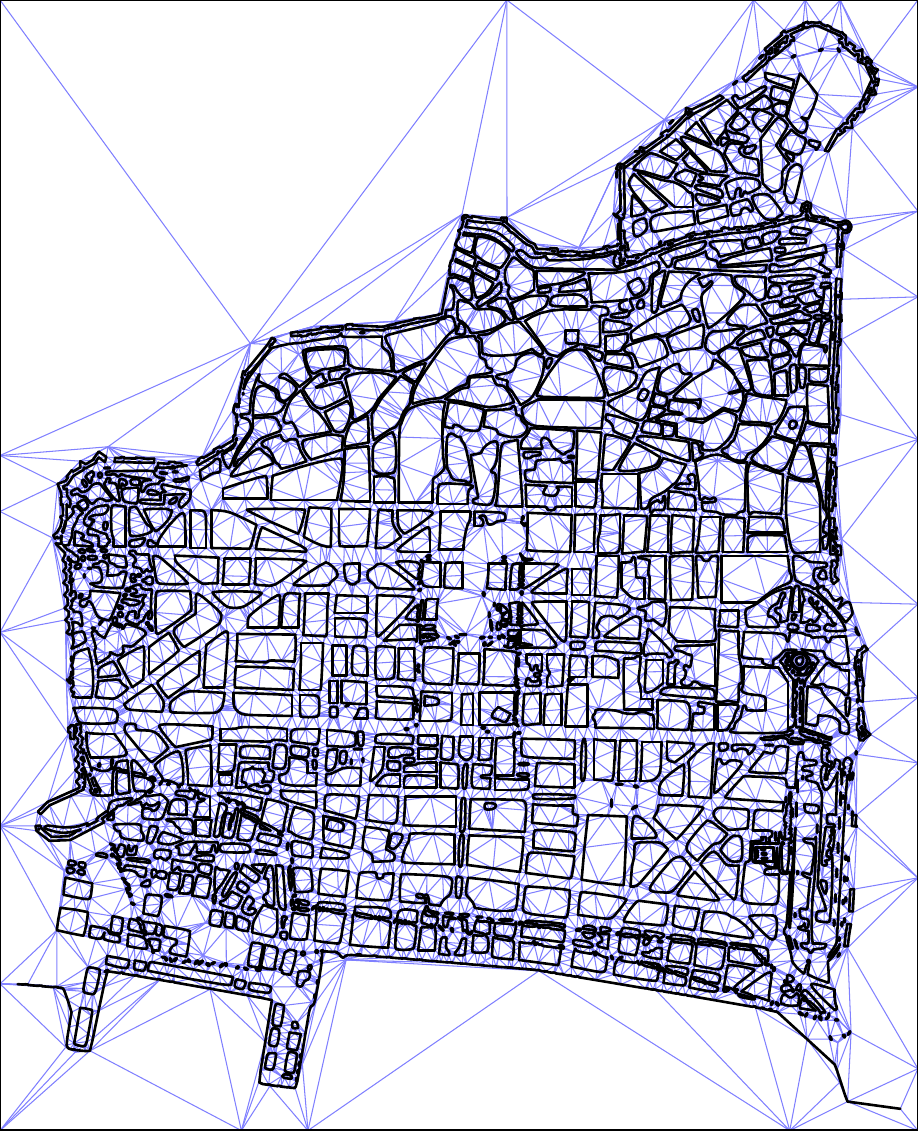}
    &
    \includegraphics[width=\lscapeTabWidth,align=c]{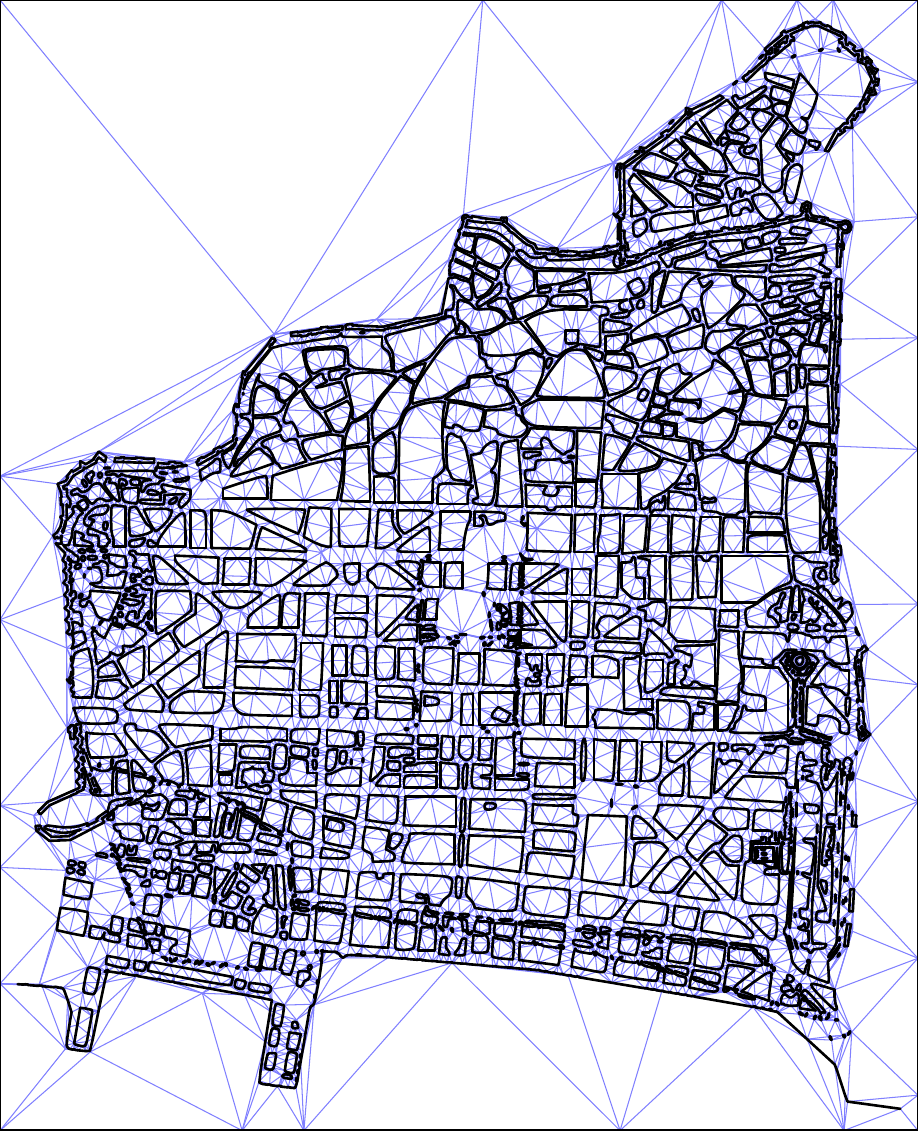}
    &
    \begin{tabular}{c}
    (\small no further improvement\\\small in reasonable time)
    \end{tabular}
    \\[-0.2mm]
    & \small 308.0 (0.94, 1.00) & \small 286.3 (0.88, 0.93) & \small 282.4 (0.86, 0.92) & 
\end{tabular}
\vspace{-0.3cm}
\end{center}
\caption{From top to bottom: triangulations of floor plans of the Topkapı palace (1915 segments, combined length 29.1), the Mar\'ia Luisa Park in Seville (4634 segments, combined length 39.3) and the historic centre of Thessaloniki (16\,812 segments, combined length 63.9).
The total edge lengths of the triangulations are given below each image, followed by the relative lengths compared to an unrefined CDT and the optimally refined CDT in parentheses.
The total edge lengths of the triangulations are given below each image. The edge lengths for the unrefined CDT (not shown) are: Topkapı:\! 104.1, Seville:\! 160.1, Thessaloniki:\! 327.0.
\label{figFloorPlanTriangsLarge}
}
\end{figure}
\end{landscape}

\subsection{Computational Cost}
We end the section on optimization results with a discussion of the computational cost of the various levels of optimization of the triangulations. All computations were performed on a small cluster of contemporary consumer CPUs with four or six cores each, with the different scenes spread over the nodes of the cluster but each scene's optimization running only on a single CPU.

Starting with the least optimized triangulations, we have the CDT, which can be built and refined in $N \log N$ time for $N$ input segments \cite{shewchuk1996triangle}, which for the largest scenes was still on a timescale on the order of seconds. For a maximally fair comparison, we searched for the combination of refinement parameters (minimum angle and maximum area) that yielded the optimally refined CDT with the lowest total edge length. For our present purpose, we simply used a brute force search of the parameter space which could be accomplished on a reasonable timeframe.
Of course, the iterative nature of refinement process of a CDT might be directly modified to facilitate this search, for example by only accepting a newly added Steiner vertex if this decreases the total edge length (or less greedy: if the total edge length decreases after having added several new vertices successively, with backtracking otherwise). We leave such modifications of the CDT refinement process for future work.

The next step up in computational cost was to polish this optimally refined CDT with our simulated annealing process, but with only edge flipping as topology-changing operation without any fuzzy contraction. The simulated annealing process, by its very nature, has no determinate runtime but a time-optimality trade-off. We ran this polishing step with as a rough goal to have the resulting total edge length optimal to one decimal, which took on the order of minutes for the simplest scenes to hours for the most complex scene.

Including fuzzy contraction in the optimization, together with the contraction pipeline of Fig.~\ref{figOptimAndContrDiagram}, further increased the computational cost --- although it should again be viewed through the lens of a time-optimality trade-off. 
As we want to examine the structure and behaviour of a (close-to-)truly minimum weight triangulation, we erred on the side of `optimality' here, with
optimization times ranging from around an hour for the simplest scenes, to roughly a day for the most complex scenes.

Lastly, the full pipeline of Fig.~\ref{figFullPipelineDiagram} with an intermediate subdivision took significantly longer to optimize due to the increased dimensionality of the state space. Optimization times in this case ranged from a day to a week, with improvements to the most complex scenes (e.g.\ the Thessaloniki floor plan) becoming constrained by available computational time.

Of course, such large computational costs are typically not feasible in practice when using (approximately) minimal weight triangulations for example in the context of ray tracing. However, the goal of this work is predominantly to examine the structure and behaviour of such close-to-minimal weight triangulations (in which case we want to approximate the true minimum as close as possible) and use that optimal lower bound for comparison and evaluation of the less powerful --- but computationally less expensive --- refinement or optimization techniques.

\section{Ray Tracing with the Triangulations \label{secRTwithTheTriangs}}

Now that we have triangulations at various levels of optimization from the previous section, it is time to put them to use! This section describes how they are used as an acceleration data structure for ray tracing in 2D `flatland', and the other methods (BVH and roped kd-trees with mailboxing) which we use for comparison.

\subsection{Ray Traversal}
If a ray starts within a known triangle $T$ of the triangulation, then the ray can be traced through the scene by calculating which edge $E$ of $T$ gets intersected by the ray. If this edge $E$ is part of the original geometry, then we have found the closest intersection and can stop the traversal. In the other case where $E$ is not part of the original geometry but simply an internal edge in the triangulation, the traversal continues by looking up which other triangle $T'$ shares this edge $E$ with the original triangle $T$, and repeating the same traversal process for this new triangle $T'$. If no other triangle shares the edge $E$, then we are at the exterior boundary of the triangulation and can conclude that the ray did not intersect any geometry.

Note that care should be taken to have numerically stable and consistent computations of the intersected boundaries and neighbouring cells when performing a traversal of a constrained convex space partitioning (such as a triangulation in 2D). Failure to do so may result in problems during traversal, such as getting stuck in a loop (e.g.\ going back and forth between two cells) or not finding any exiting boundary for a cell that is pierced by a ray \cite{maria17robustCDTopGPU}.

\subsection{Finding the Starting Triangle \label{secFindingStartingPoint}}
To follow the traversal algorithm and trace rays using the triangulation, the triangle that contains the starting point (a.k.a.\ origin) of the ray should be known. Often this triangle is already at hand, for instance when generating a secondary ray starting from where the previous ray intersected: the final triangle in the traversal of the original ray is the starting triangle for traversal of the secondary ray. In other cases, the cost of finding the initial triangle can be amortised over many rays, for instance when tracing rays from a pinhole camera which all share the same origin. In such cases, even a brute force search for the triangle that contains the camera's origin might be acceptable.

When the starting triangle for a ray is not known and when its search cannot be amortised over many rays, there are still several options to efficiently find it. One option would be to build a space partitioning (e.g. a regular grid or a kd-tree) and store a list in each cell of all triangles that overlap that cell. The cell containing the origin $o$ of ray can then be queried, and the associated triangles that overlap that cell can be searched to find the one containing the ray's origin.

Another option that avoids an extra data structure is to directly use the triangulation itself (or in general the constrained convex space partition): if \emph{any} arbitrary point $p$ is known with it's associated triangle (convex cell), then one can simply follow the traversal algorithm to trace a ray from $p$ to the origin $o$ of the ray that we actually want to trace. 
Ideally, the point $p$ with known associated convex cell should be `close' to $o$ (close in terms of number of number of cells traversed along the way). A straightforward generalisation would be a set of points (with their associated convex cells) on the centers of a regular grid, such that for any point $o$, the regular grid can be queried in constant time to find a nearby point $p$ with known convex cell from which traversal towards $o$ can be started.

Lastly, if the convex space partition is embedded within a more traditional ray traversal data structure
to handle some particularly challenging geometry (e.g.\ a specialised structure for dense grass within a classical BVH or kd-tree that holds the rest of the scene), then many traversal queries can be expected for a ray that starts on the outside boundary of the structure, for which a separate optimization can be made. Often the natural choice for such an outside boundary is an axis aligned (bounding) box so that it can be easily integrated within a BVH or kd-tree. For such axis aligned boundaries, the part of the convex space partition that lies on the boundary is itself a convex space partition (CSP) in one dimension less. For 3D geometry within an axis aligned bounding box with a tetrahedralisation as CSP, the faces of the tetrahedra that make up the sides of the bounding box are themselves a 2D triangulation such as the ones examined here. For an external ray that intersects the CSP's bounding box, the search for the starting tetrahedron can efficiently be preformed as a 2D traversal within the triangulation of the boundary plane (with any of the above methods to select a known starting point) instead of a full 3D traversal.

\subsection{Other Methods for Comparison \label{secOtherMethodsForComparison}}
We compare the ray traversal performance of the triangulations from Sec.~\ref{secOptimResults} with a BVH and (roped) kd-tree as classical acceleration data structures for the same scenes. This section briefly describes how those structures were obtained and what was measured.

\subsubsection{BVH \label{secBVH}}
A Bounding Volume Hierarchy (BVH) is built on the line segments of the original geometry using the 2D version of the standard Surface Area Heuristic (SAH) \cite{pbrt}, effectively becoming a `perimeter length' heuristic.
The BVH uses a greedy top-down construction based on an exhaustive sweeping partitioning by sorting the objects according to the projection of their centroids on the splitting axis in the order $o_1, o_2, \ldots, o_n$ and testing all partitions $\(\{o_1, o_2,\ldots, o_k\},\{o_{k+1}, o_{k+2}, \ldots, o_n\}\)$ for $k\in\{1,2,\ldots, n-1\}$ as possible children. Both the $x$ and $y$ axes are tested as splitting axis and from all candidates the split with lowest SAH cost is used. Splitting stops when the (2D) SAH metric predicts a cost for the split node that is lower than the cost of simply intersecting all objects in the node.

For simplicity, when counting the number of traversal operations in our results, we simply add the counts for each operation --- intersecting an Axis Aligned Bounding Box (AABB) and intersecting a geometric primitive --- without any reweighting based on their actual computational cost (which depends on implementation and hardware details). Nonetheless, for a maximally fair comparison, we build multiple BVHs for each scene where we vary the parameter within the SAH that denotes the relative cost of an AABB versus geometric primitive intersection (and thus influences the stopping criterium) and use the BVH that gives the lowest overall measured number of traversal operations (i.e.\ with the number of AABB and geometric primive intersection tests simply added without weighting) for the identical set of rays that will be used in the final measurement.

\subsubsection{Roped KD-Tree \label{secKD}}
Whereas a BVH does not allow for stackless traversal like our triangulations, we also compare the triangulations to a roped kd-tree, which \emph{does} support stackless traversal.

We use a standard greedy top-down construction of a (roped) kd-tree over the line segments of the original geometry using the 2D version of the SAH. Due to the piecewise linearity of the SAH cost function for kd-trees, the optimal splitting position has to lie on a vertex of the original geometry \cite{wald2006kdnlogn}. We exhaustively test all such splitting positions for both $x$ and $y$ splitting axes and use the candidate with the lowest SAH cost.
To enable a stackless traversal without initial top-down descent, the leaf nodes are enhanced with ropes: links to adjacent neighbour-nodes for each boundary segment of the original leaf node.

For a maximally fair comparison, we count the traversal cost of our roped kd-tree assuming the best case scenario. We therefore emulate an ideal rope tree \cite{havran1998ropeTrees} by counting $\log n$ steps to traverse a boundary that has $n$ ropes. We furthermore emulate ideal (hashed) mailboxing \cite{tao2008kdMailboxing, wald2001interactive} by only counting each uniquely intersected object once in case of duplicate intersection tests. Lastly, for each scene we build multiple kd-trees with varying traversal cost estimate parameter in the SAH metric (and thus varying stopping criteria and leaf node sizes) and select the kd-tree that gave the lowest overall measured traversal cost, i.e.\ the lowest combined count of (1) the number of visited nodes in the kd-tree, (2) the total number of rope links visited when searching for the next kd-nodes, assuming optimal $\log n$ behaviour, and (3) the number of geometric primitives for which an intersection test was performed, discarding duplicate tests. This optimal kd-tree is selected based on the traversal cost for the entire set of rays that will be used in the final measurement.

\subsection{Experimental Setup}
To measure the performance of our approximate minimum weight triangulations and the conventional BVH and kd-tree, we shoot rays from a uniform and isotropic distribution within the scene and count the number of traversal operations that were performed for each method when finding the closest intersection.
For the triangulation, there is only one type of operation that is performed during traversal: determining which outgoing edge of the current triangle is pierced by the ray, which can be part of the original geometry (in which case the traversal is complete) or it can be an internal edge after which traversal continues int the adjoining triangle.
For the BVH, there are two types of operations: there are (1) intersection tests on the Axis Aligned Bounding Boxes (AABB) of the nodes that are visited and (2) there are intersection tests on the actual geometry in leaf nodes.
Lastly, the roped kd-tree has three types of operations: nodes of the hierarchy can be (1) \emph{visited} through downwards traversal of the hierarchy by checking which child node(s) is (are) intersected or by sidewards or upwards traversal of the hierarchy by (2) \emph{selecting} and following the appropriate rope, and (3) intersection tests on the \emph{geometry} associated with leaf nodes.
For the stackless methods (triangulations and roped kd-tree) we always assume that the node that contains the ray's origin is known a priori (cf.\ Sec.~\ref{secFindingStartingPoint}).
As discussed in Secs.~\ref{secBVH} and \ref{secKD}, we count the number of operations for the `comparison' methods (BVH and roped kd-tree) in the most optimiztic way for a maximally fair comparison.

Although there may be different computational costs associated with the different traversal operations
(e.g.\ intersecting an AABB may be cheaper than intersecting an arbitrary line segment), we are principally interested in the behaviour or (asymptotic) scaling of the \emph{number} of these operations and therefore simply compare the different methods by an unweighted sum of all traversal operations that are performed by those methods, regardless of constant multiplicative factors.

\FloatBarrier
\section{Ray Tracing Results \label{secRTresults}}
This section describes the results of using the triangulations that were obtained in Sec.~\ref{secOptimResults} for ray tracing as per Sec.~\ref{secRTwithTheTriangs}.

\subsection{Synthetic Scenes: Lines \label{secRTResultsLines}}
\begin{figure}
\newcommand{\lineRaytracePlotScale}{1.0}
\newcommand{\lineRandomRaysScale}{0.415}
\begin{center}
\textsc{Length factor 0.1 -- Uniform orientation}
\begin{tabular}{c@{}c@{\ }c}
\rotatebox[origin=c]{90}{\small ~~ Number of traversal operations} &
    \includegraphics[scale=\lineRaytracePlotScale,align=c]{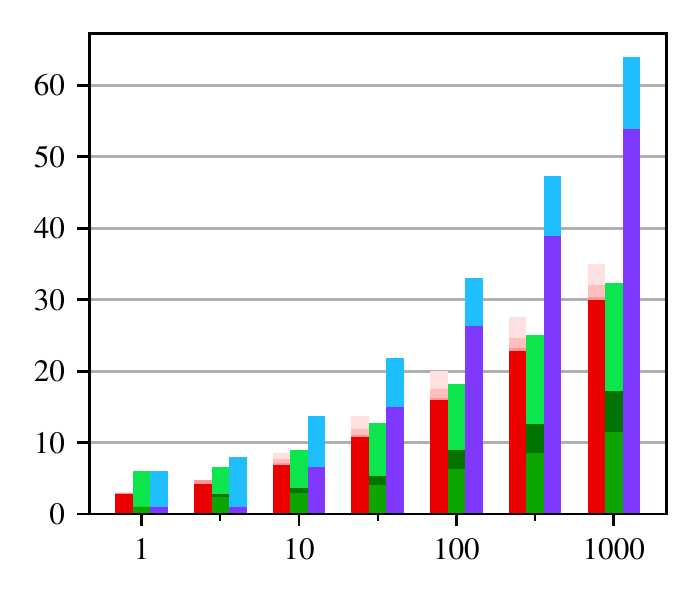}
    &
    \begin{tabular}{c}
    \includegraphics[scale=\lineRaytracePlotScale,align=c]{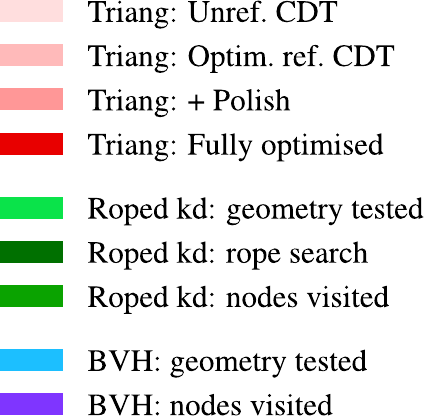}
    \\ \vspace{-1mm} 
    \end{tabular}
\\[-5mm] 
& \small ~~~~$N$
\\[2mm]
\end{tabular}
\begin{tabular}{c@{\ }c@{\ }c}
    \includegraphics[scale=\lineRandomRaysScale]{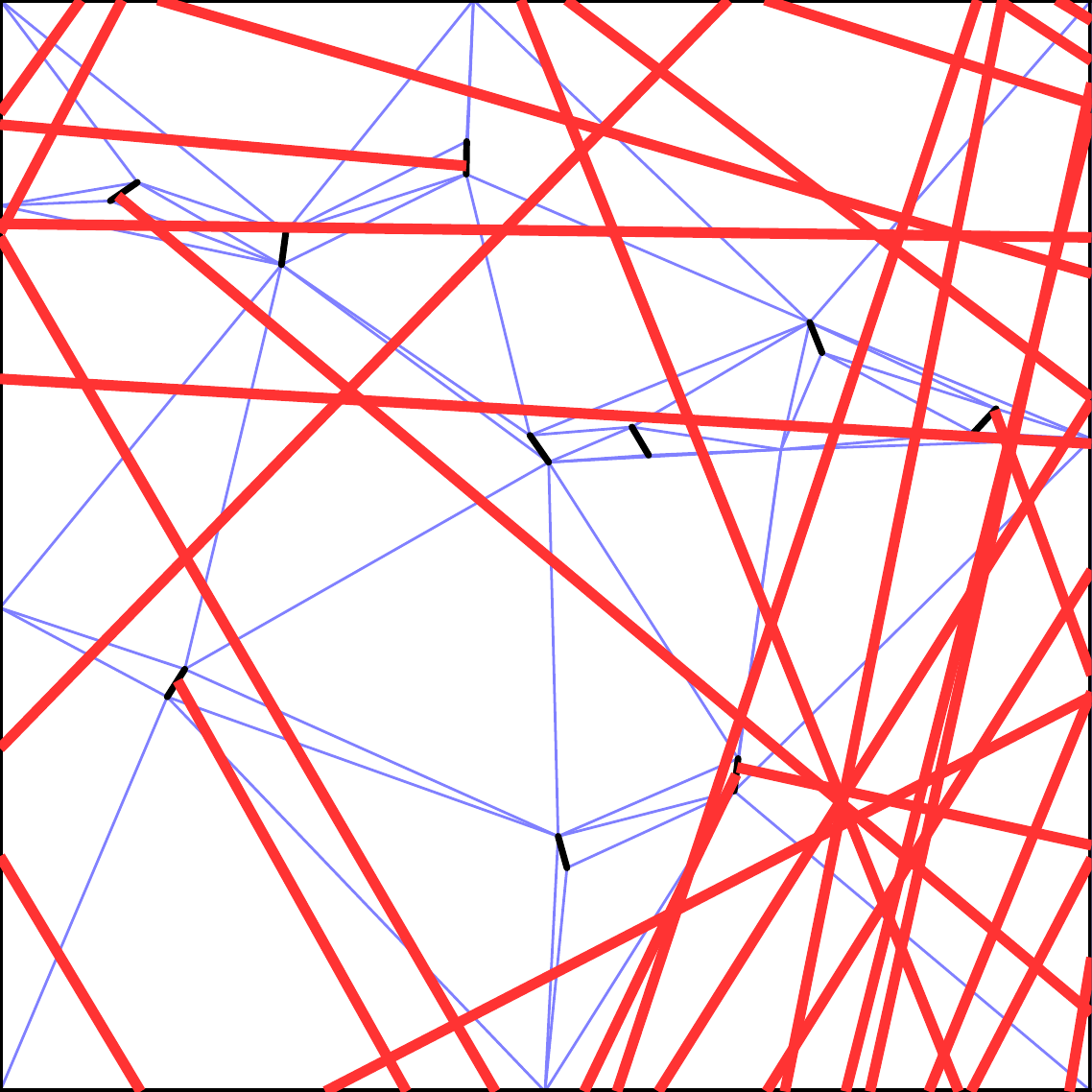}
    &
    \includegraphics[scale=\lineRandomRaysScale]{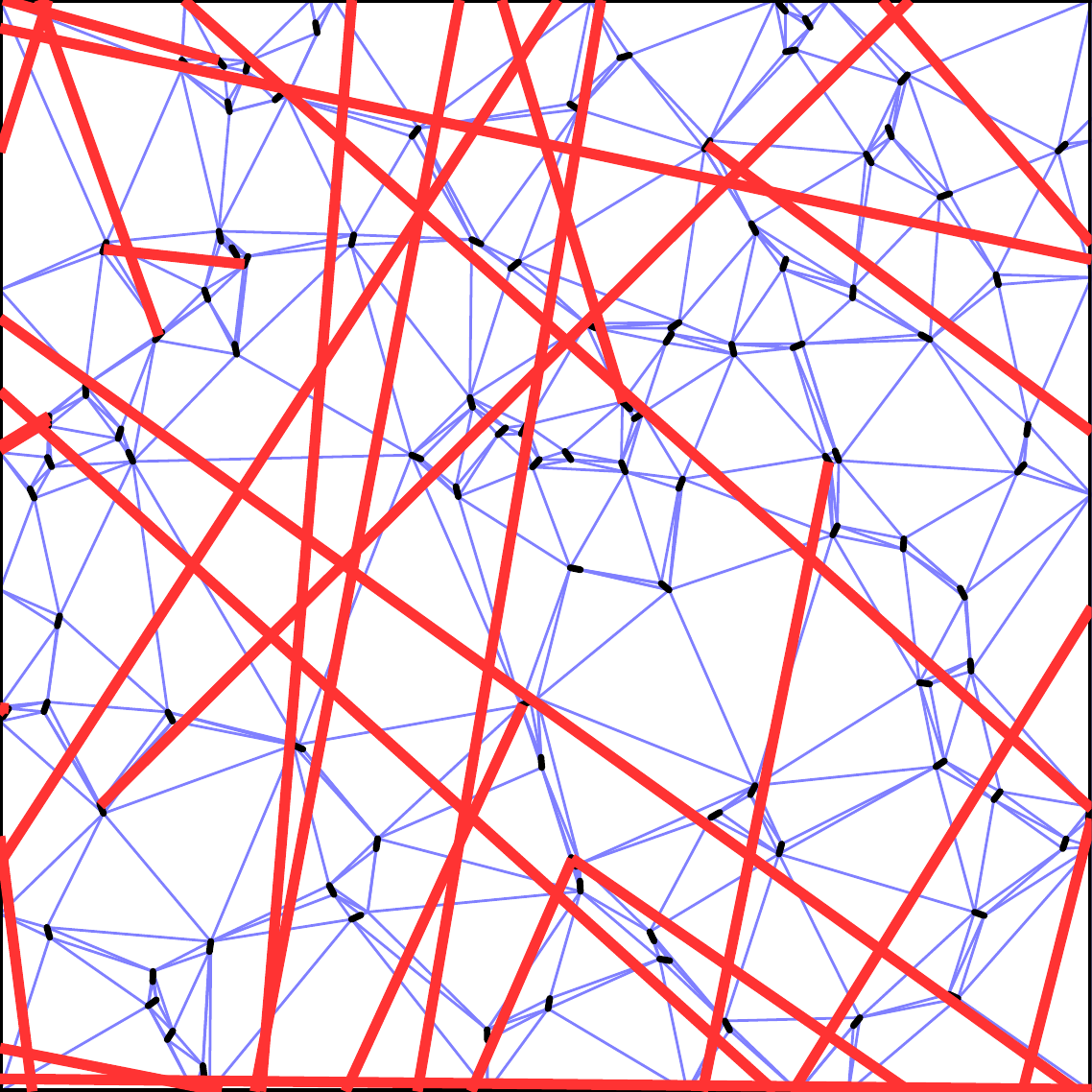}
    &
    \includegraphics[scale=\lineRandomRaysScale]{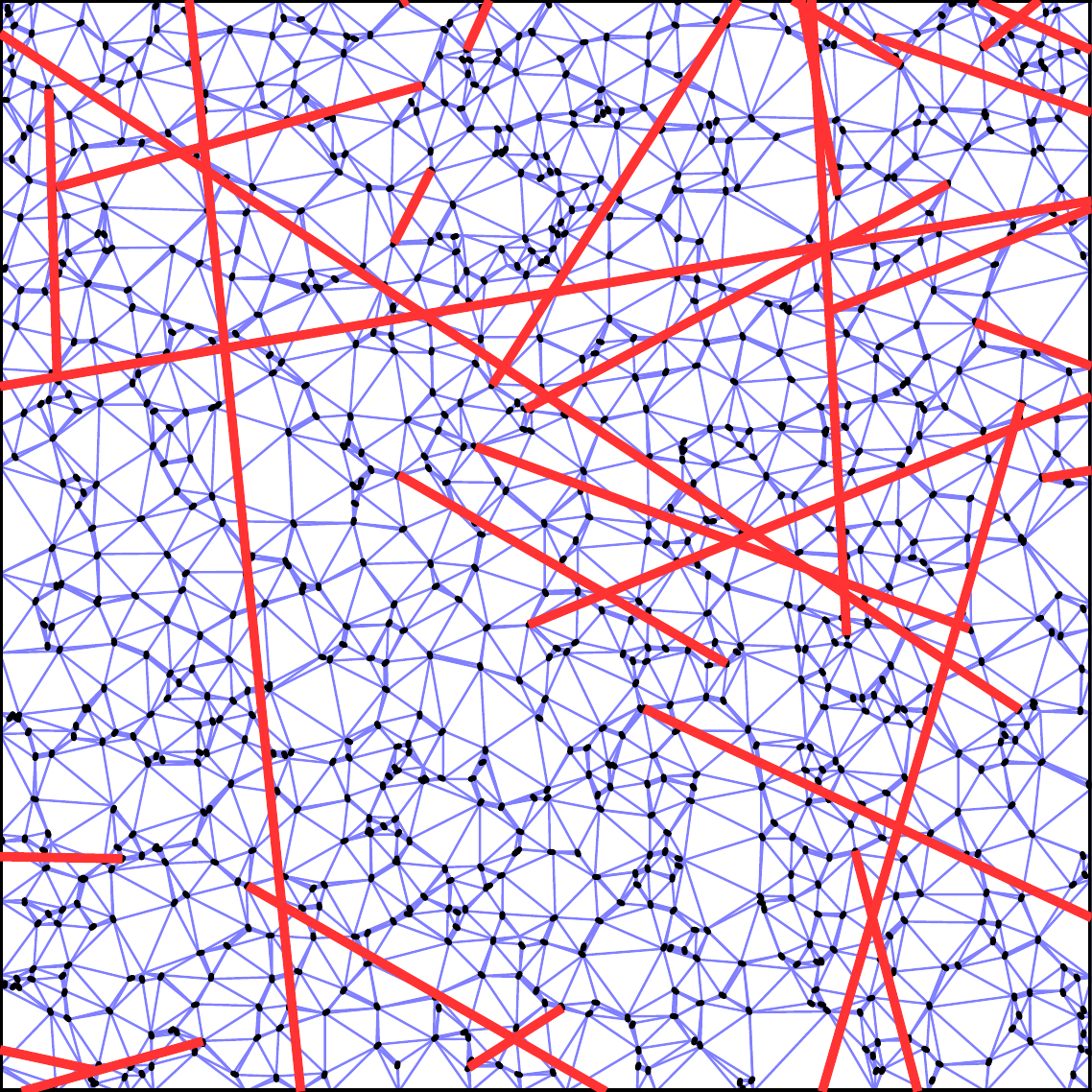}
    \\[-1mm]
    \small $N=10$ &
    \small $N=100$ &
    \small $N=1000$
\end{tabular}
\end{center}
\caption{
    Comparison of the lines scene with length factor 0.1 and uniform orientation (results for mostly vertical and diagonal lines are quasi-identical). Top graph: the number of triangles traversed for the triangulations with various levels of optimization are overlayed (e.g.\ the result for the best triangulation is given by only the dark red bar, and the result for the simplest triangulation is the height of the lightest red bar), whereas the results for the kd-tree and BVH are shown split into their components and the total cost for each data structure is thus given by the top of the uppermost sub-segment.
    Bottom: visualisation of the geometry (black) with the optimized triangulation (blue) and sampled rays (red) for three values of $N$.
    For these sparse scenes with very short line segments, rays travel over large distances --- even for high $N$ --- before intersecting geometry and the total number of operations for the triangulation and roped kd-tree are similar (although the triangulation shows more operations for low $N$, it gets on par with and even slightly outperforms the roped kd-tree for the highest values of $N$)
}
\label{figPlotLinesRaytraceSC0p1}
\end{figure}

The results for the short line scenes of Sec.~\ref{secOptimLines} (length factor 0.1) are shown in Fig.~\ref{figPlotLinesRaytraceSC0p1}. For this `sparse' scene of very short line segments, the performance of all three methods (triangulation, roped kd-tree, BVH) --- despite their fundamentally different approaches --- is very similar.

The benefit of the stackless methods with known starting cell (triangulation and roped kd-tree) over the BVH can nonetheless still be seen, although the difference is rather small. An explanation can be found in the sparsity of the scene due to the short line segments: rays are expected to travel long distances before intersecting geometry --- or even traverse the entire scene without finding an intersection. Consequently, many cells of the acceleration data structure will be visited, and the relative advantage of knowing the first cell a priori is not that big.

\begin{landscape}
\begin{figure}
\newcommand{\lineRaytracePlotScale}{1.0}
\vspace{-10mm}
\begin{center}
\begin{tabular}{c@{\ \ }c@{\!\!\!\!}c@{\ }c@{\ }c}
&
& \small\textsc{Mostly vertical}
& \small\textsc{Uniform orientation}
& \small\textsc{Mostly diagonal}
\\
\rotatebox[origin=c]{90}{\small\textsc{Length factor 1}} &
\rotatebox[origin=c]{90}{\small ~~ Number of traversal operations} &
    \includegraphics[scale=\lineRaytracePlotScale,align=c]{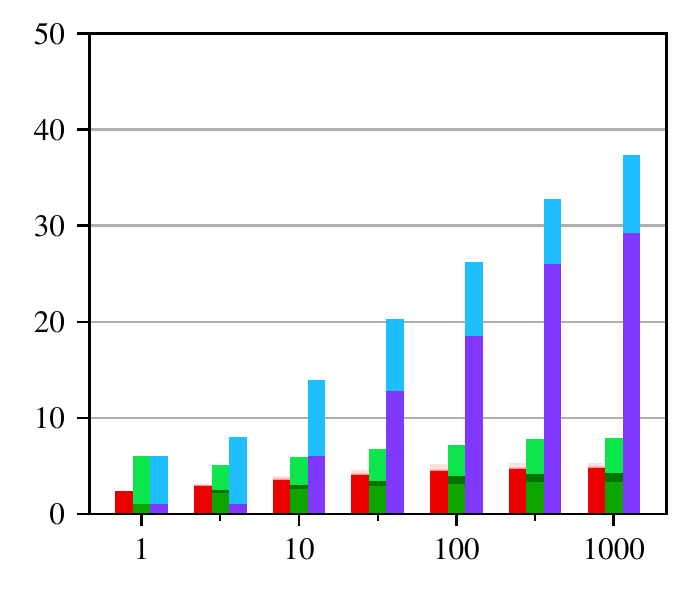} &
    \includegraphics[scale=\lineRaytracePlotScale,align=c]{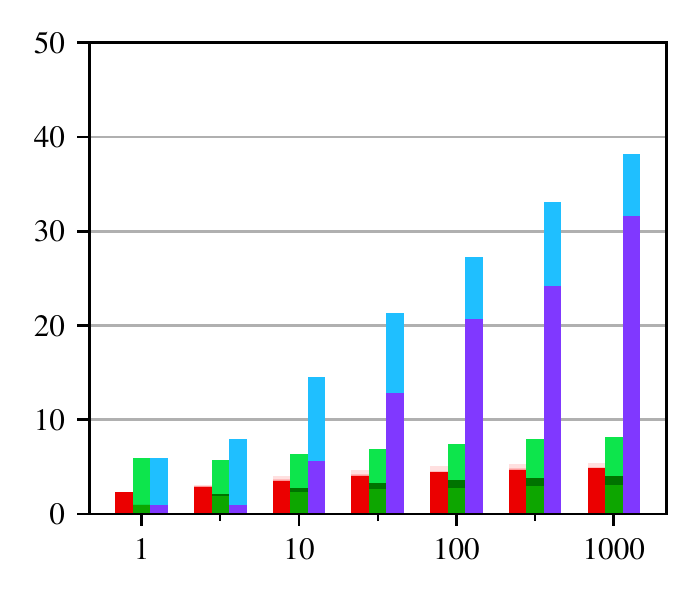} &
    \includegraphics[scale=\lineRaytracePlotScale,align=c]{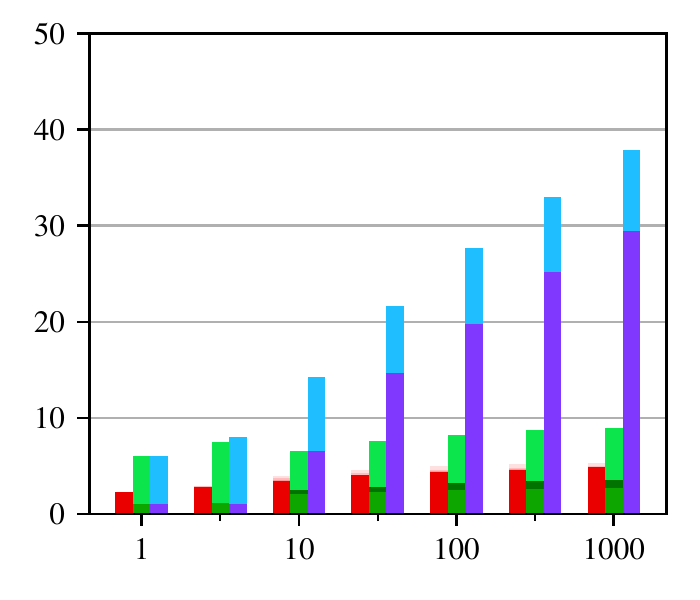}
\\[-4mm]
\rotatebox[origin=c]{90}{\small\textsc{Length factor 3}} &
\rotatebox[origin=c]{90}{\small ~~ Number of traversal operations} &
    \includegraphics[scale=\lineRaytracePlotScale,align=c]{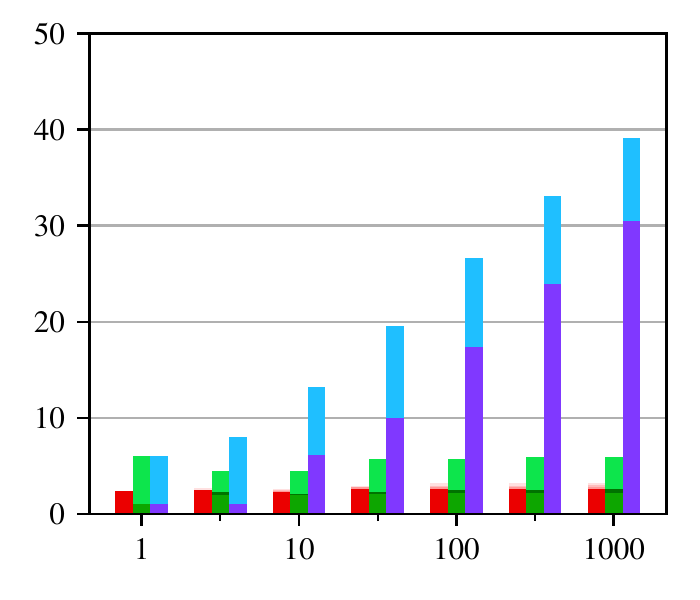} &
    \includegraphics[scale=\lineRaytracePlotScale,align=c]{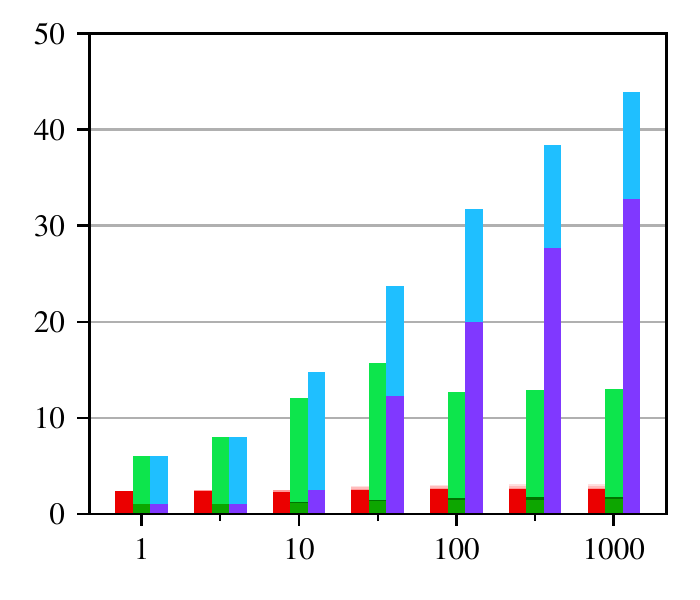} &
    \includegraphics[scale=\lineRaytracePlotScale,align=c]{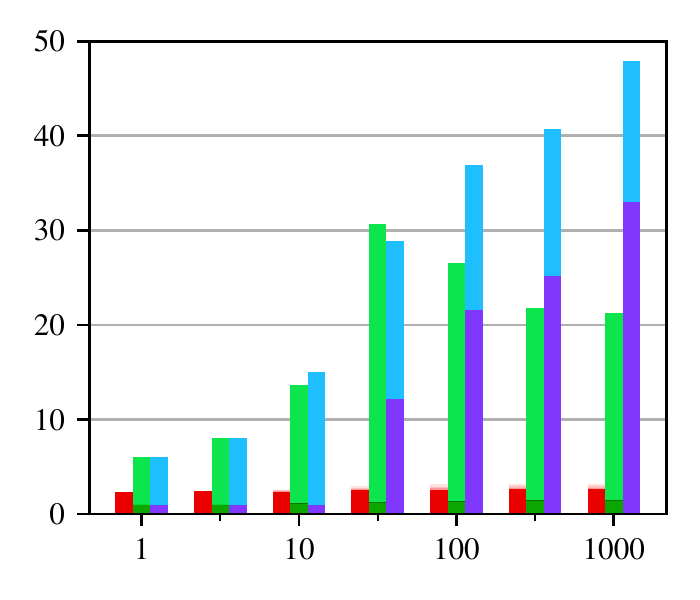}
\\[-5mm] 
& \small ~~~~$N$
& \small ~~~~$N$
& \small ~~~~$N$
\\[-5mm]
\end{tabular}
\end{center}
\caption{
    Traversal costs for the lines scenes with length factors 1 (top) and 3 (bottom) for different preferred orientations (columns). The legend is identical to the one in Fig.~\ref{figPlotLinesRaytraceSC0p1}.
    The behaviour of the BVHs (blue) asymptotically approaches $\log N$ (straight line on this graph with logarithmic $N$-axis) due to the top-down descent in the hierarchy. 
    The triangulation (red) and roped kd-tree (green) immediately start their (stackless) traversal from the known cell that contains the ray's origin and --- in these dense scenes with relatively long line segments --- only have to traverse a handful of cells before an intersection with geometry is found, although the kd-tree is highly sensitive to the preferred orientation of the line segments.
    In the maximally problematic case of long diagonal lines (bottom right plot), the traversal of the kd-tree is linear in $N$ up to $N \approx 30$ (note that this looks like an `exponential' curve with the logarithmic $N$-axis) as the best kd-tree here is simply one leaf node with all geometry. Even for $N\ge100$, when the kd-tree is no longer a single leaf node, the few leaves still contain many line segments which shows itself in the high number of geometry tests (light green).
    See Fig.~\ref{figPlotLinesRaytraceSC1and3NoBvh} for a zoomed-in comparison of the triangulations and kd-trees.
   }
\label{figPlotLinesRaytraceSC1and3}
\end{figure}
\end{landscape}

\begin{landscape}
\begin{figure}
\newcommand{\lineRaytracePlotScale}{1.0}
\vspace{-10mm}
\begin{center}
\begin{tabular}{c@{\ \ }c@{\!\!\!\!}c@{\ }c@{\ }c}
&
& \small\textsc{Mostly vertical}
& \small\textsc{Uniform orientation}
& \small\textsc{Mostly diagonal}
\\
\rotatebox[origin=c]{90}{\small\textsc{Length factor 1}} &
\rotatebox[origin=c]{90}{\small ~~ Number of traversal operations} &
    \includegraphics[scale=\lineRaytracePlotScale,align=c]{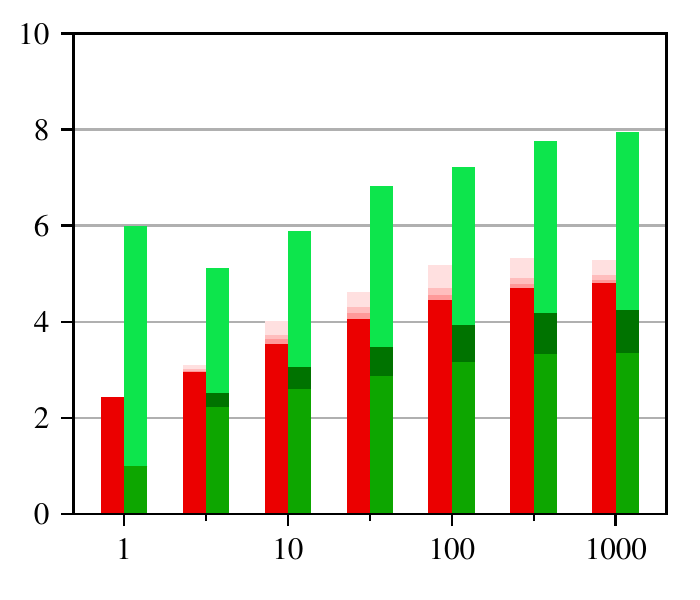} &
    \includegraphics[scale=\lineRaytracePlotScale,align=c]{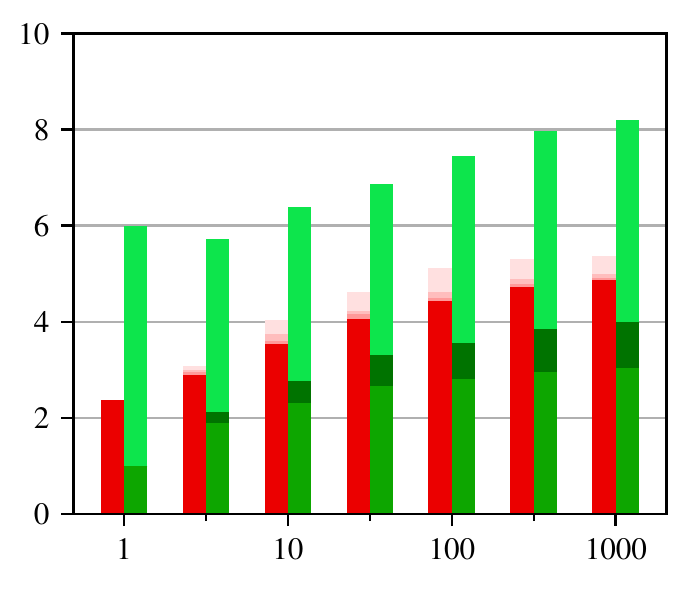} &
    \includegraphics[scale=\lineRaytracePlotScale,align=c]{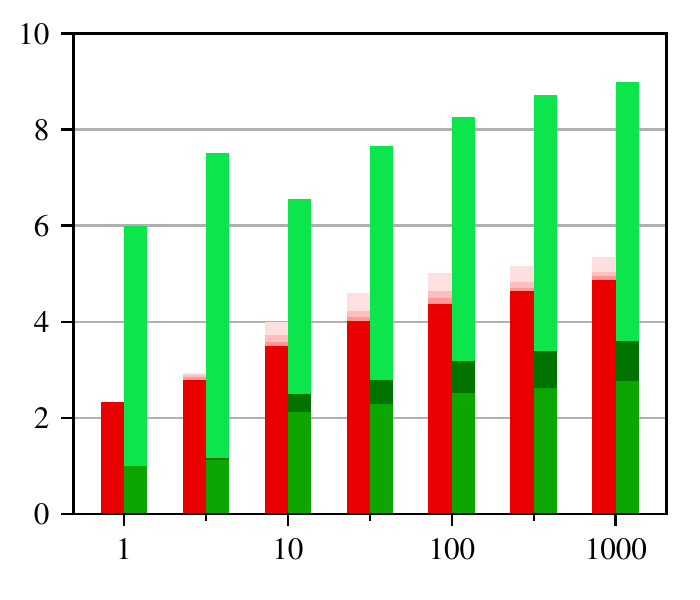}
\\[-4mm]
\rotatebox[origin=c]{90}{\small\textsc{Length factor 3}} &
\rotatebox[origin=c]{90}{\small ~~ Number of traversal operations} &
    \includegraphics[scale=\lineRaytracePlotScale,align=c]{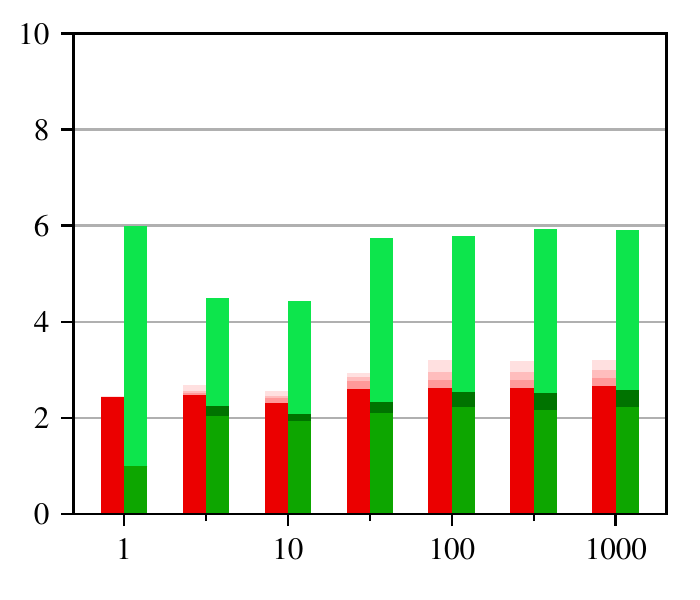} &
    \includegraphics[scale=\lineRaytracePlotScale,align=c]{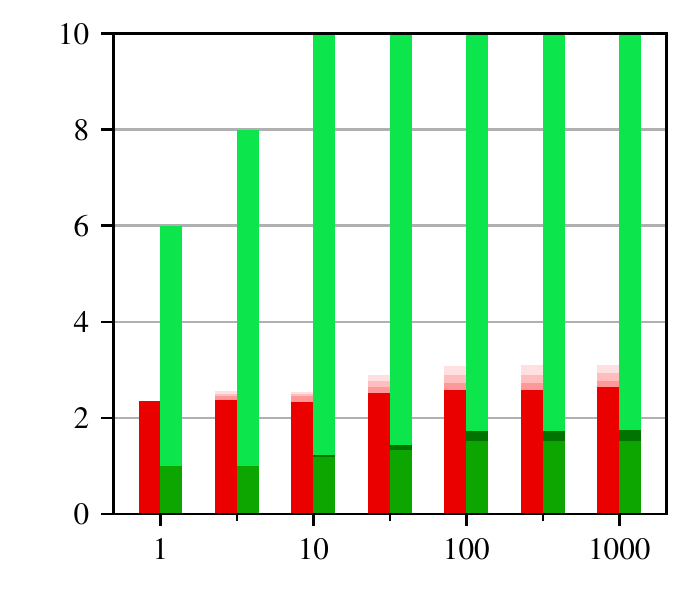} &
    \includegraphics[scale=\lineRaytracePlotScale,align=c]{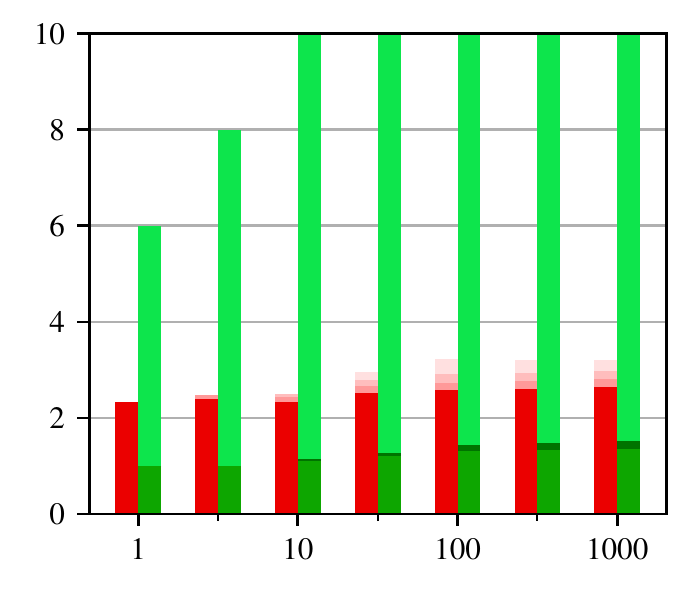}
\\[-5mm] 
& \small ~~~~$N$
& \small ~~~~$N$
& \small ~~~~$N$
\\[-5mm]
\end{tabular}
\end{center}
\caption{
    Identical to Fig.~\ref{figPlotLinesRaytraceSC1and3} but with the BVH results omitted for clarity.
    The asymptotic behaviour for the triangulations (red) and roped kd-tree (green) becomes sub-$\log N$ and even approaches a constant for the triangulations in the very dense scenes with length factor 3 (bottom), regardless of the preferred orientation of the line segments (column). 
    The behaviour of the kd-trees, on the other hand, are strongly dependent on the preferred orientation of the line segments. For the mostly vertial segments, the spatial partitioning consisting of axis aligned cells matches well with the actual geometry yielding small and tight fitting cells with low amounts of geometric duplication in leaf nodes. For lines that are preferentially diagonal, the case is completely opposite with large, ill-fitting cells and many (possibly duplicated) geometry pointers in the leaves, leading to more geometry tested (light green) and slightly less nodes visited (medium green).
}
\label{figPlotLinesRaytraceSC1and3NoBvh}
\end{figure}
\end{landscape}

The results for the scenes with longer lines are shown in Fig.~\ref{figPlotLinesRaytraceSC1and3} (with zoomed-in detail in Fig.~\ref{figPlotLinesRaytraceSC1and3NoBvh}). Here, the advantage of the stackless methods over the BVH becomes readily apparent. Whereas the BVH shows the typical $\log N$ behaviour for descent in a hierarchical structure, the stackless methods show sub-$\log N$ scaling or even evolution to a \emph{constant} traversal cost independent of $N$ for the triangulations of dense scenes with length factor 3. Indeed, the high density of those scenes results in rays that only travel a short distance before intersecting geometry, and thus only a few cells in the acceleration data structure need to be visited (visualisations of the scenes, the data structures and several representative rays can be found in Appendix \ref{appTriangsKDBVH}).

The remaining difference between the stackless methods for these dense scenes lies in the `tightness' of their cells. By construction, triangulations have cells that perfectly fit the geometry because the geometric primitves \emph{are} edges of the cells.
Somewhat similarly, in the scenes with the `mostly vertical' orientation, the geometric primitives can be tightly bounded by axis aligned bounding boxes such that cells of the roped kd-tree fit the geometry reasonably well. For this vertically oriented geometry, the behaviour of the roped kd-tree follows that of the triangulations, although with a slightly higher overall factor.
In contrast, axis aligned bounding boxes are a poor fit in the `mostly diagonal' scenes, or the correlated line clusters in the `uniform orientation' scenes with length factor 3. Here, the low number of nodes that are visited by the stackless kd-tree is overshadowed by the many geometric primitive intersection tests in the ill-fitting leaf nodes --- in some cases even surpassing the total number of traversal operations of a BVH.
The roped kd-tree is thus highly sensitive to `ill-fitting' bounding boxes in dense scenes which can cause pathological behaviour, whereas the triangulations automatically have perfectly fitting cells and remain well-behaved.

In summary, we find that triangulations are on par with BVHs and roped kd-trees for the sparse scenes, and they handle the dense scenes efficiently (approaching a \emph{constant} traversal complexity for increasing $N$) due to their stackless traversal, which is robust to the orientation of the underlying geometry (in contrast to the roped kd-tree).

\FloatBarrier
\subsection{Synthetic Scenes: Grass and Hair}
Transitioning from the idealised `lines' scenes to the somewhat more real-world inspired `grass' and `hair' scenes, we find a similar story. The results for these scenes are shown in Fig.~\ref{figRaytracePlotsGH} and visualisations of the data structures are again given in Appendix \ref{appTriangsKDBVH} for the interested reader.

The grass scenes bear resemblance to the `mostly vertical' line scenes with long line segments, and the results indeed show similarities: the BVH has the expected $\log N$ behaviour, the roped kd-tree is sub-logarithmic and the triangulation performs the least number of traversal steps --- it even shows an asymptotically \emph{decreasing} number of operations for increasing $N$. This counterintuitive observation can be explained by the increased density of the scene as $N$ grows, which shortens the expected traversal length of a ray --- and thus also the expected number of triangles that are traversed in the triangulation as many rays simply travel from one leaf to the directly neighbouring leaf.

The geometry in the hair scenes is somewhat similar to the correlated line clusters in the `lines' scene with long line segments and uniform orientation, albeit with large empty parts of the scene that contain no geometry. The ill-fitting problem of axis aligned bounding boxes again leads to large leaves in the kd-tree with excessive geometry intersection tests, especially for the 5~segments/strand scenes where the roped kd-tree nearly performs as bad as a BVH. On the other hand, the number of traversal operations for the triangulations is monotonically \emph{decreasing} in terms of $N$ due to the density and short expected traversal length of the rays.

\begin{figure}
\newcommand{\thisScale}{1.0}
\begin{center}
\begin{tabular}{c@{\ }c@{\ }c}
&\multicolumn{2}{c}{\textsc{Grass scenes}}\\[-2mm]
& \small 3 segments/leaf 
& \small 10 segments/leaf
\\
\rotatebox[origin=c]{90}{\small ~~ Number of traversal operations} &
    \includegraphics[scale=\thisScale,align=c]{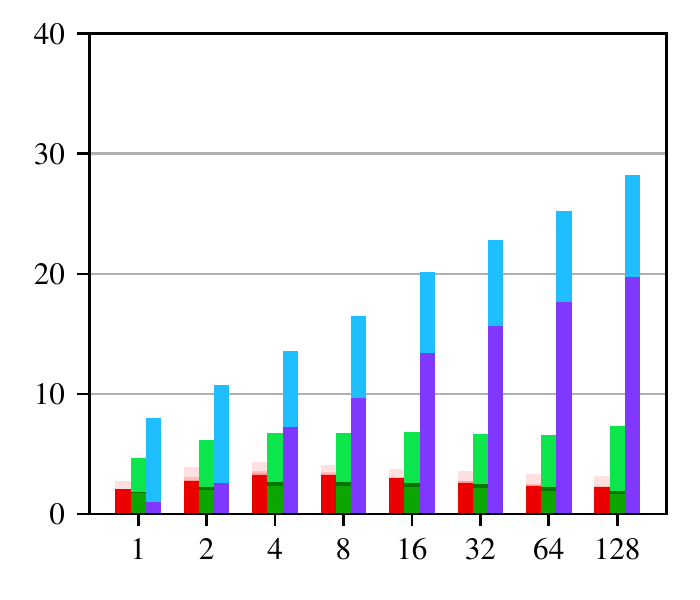}&
    \includegraphics[scale=\thisScale,align=c]{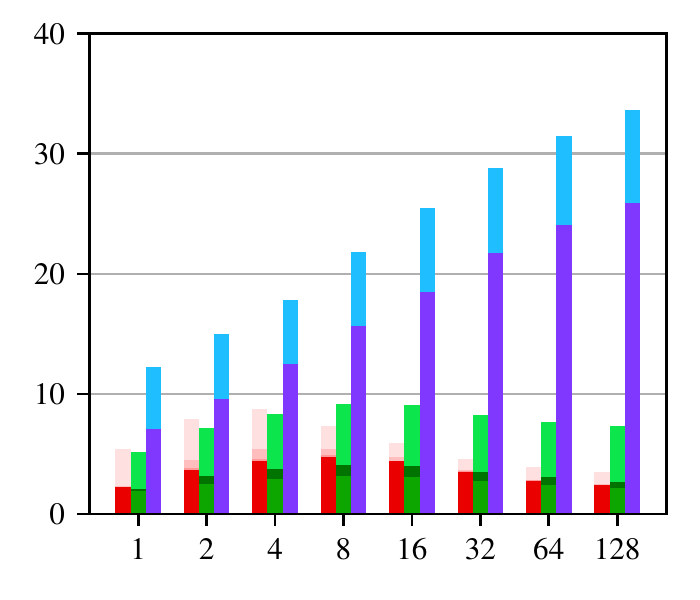}
\\[-4mm] 
& \small ~~~~~$N$
& \small ~~~~~$N$
\\[1mm]
&\multicolumn{2}{c}{\textsc{Hair scenes}}\\[-2mm]
& \small 5 segments/strand 
& \small 20 segments/strand
\\
\rotatebox[origin=c]{90}{\small ~~ Number of traversal operations} &
    \includegraphics[scale=\thisScale,align=c]{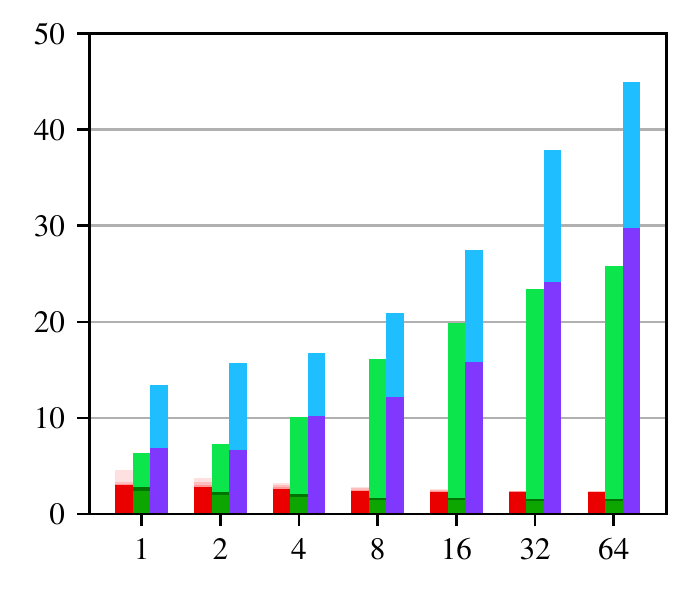}&
    \includegraphics[scale=\thisScale,align=c]{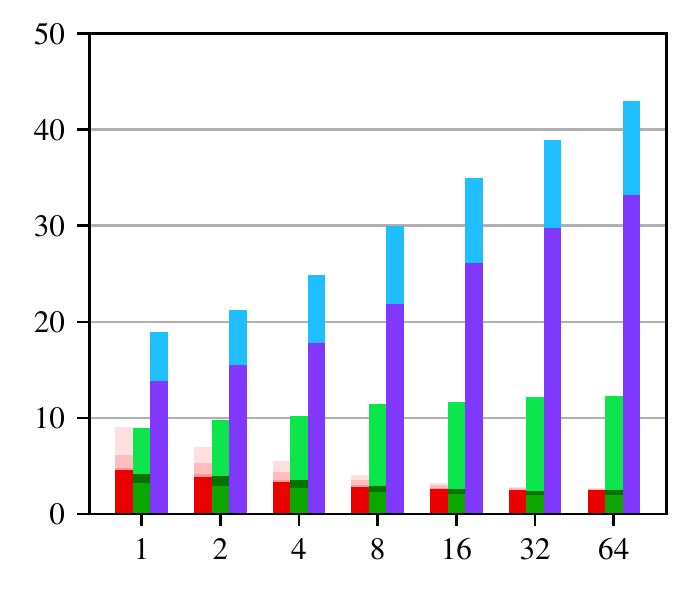}
\\[-4mm] 
& \small ~~~~~$N$
& \small ~~~~~$N$
\\[2mm]
& \multicolumn{2}{c}{\includegraphics[align=c]{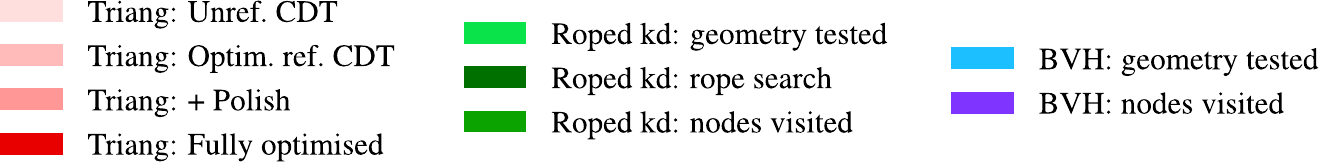}\vspace{-2mm}}
\end{tabular}
\end{center}
\caption{
    Traversal costs for the grass (top) and hair (bottom) scenes. The BVH (blue) shows the expected $\log N$ behaviour, whereas the stackless methods (triangulation and roped kd-tree) tend towards sub-logarithmic behaviour.
    The triangulations (red) even show an asymptotically \emph{decreasing} number of traversal operations as $N$ increases, due to the higher density of high-$N$ scenes and thus shorter expected traversal lengths. The roped kd-tree (green) shows somewhat similar behaviour for the grass scenes (top), albeit with a higher overall factor. In the hair scenes (bottom), however, the kd-tree does tend towards a decreasing number of visited nodes (bottom green segment) as $N$ increases, but this is overshadowed by an increasing number of geometry tests (top green segment) due to ill-fitting cells in the regions where the strands are diagonal with respect to the world axes. The triangulations are immune to this `ill-fitting' problem because the boundaries of their cells can be the geometric primitives themselves.
\label{figRaytracePlotsGH}
}
\end{figure}

\subsection{Real-World Floor Plans \label{secOptimFloorPlansRaytraceResults}}

So far we have seen that triangulations perform favourably in dense scenes where rays are expected to find an intersection in a very short distance.
We have even seen cases where the number of traversal operations \emph{decreased} as more geometry was added to the scene.
However, the synthetic scenes thus far have had a very high density for large $N$, which may not be a particularly realistic or typical case and one could argue that it gives the triangulations an unfair advantage as this is precisely where they perform best.  
A comparison of real-world scenes is thus in order.

The traversal costs for the different techniques on the real-world floor plans of Sec.~\ref{secOptimFloorPlans} are shown in Fig.~\ref{figRaytracePlotsRealWorld}, and the structures are visualised in Appendix \ref{appTriangsKDBVH}. We find that for every scene the triangulations perform the least number of traversal operations, followed by the roped kd-tree within a factor of two, and with the BVH trailing (far) behind.

\begin{figure}
\begin{center}
\begin{tabular}{c@{\,}c}
\rotatebox[origin=c]{90}{\small ~~ Number of traversal operations}
&
    \includegraphics[scale=1.0,align=c]{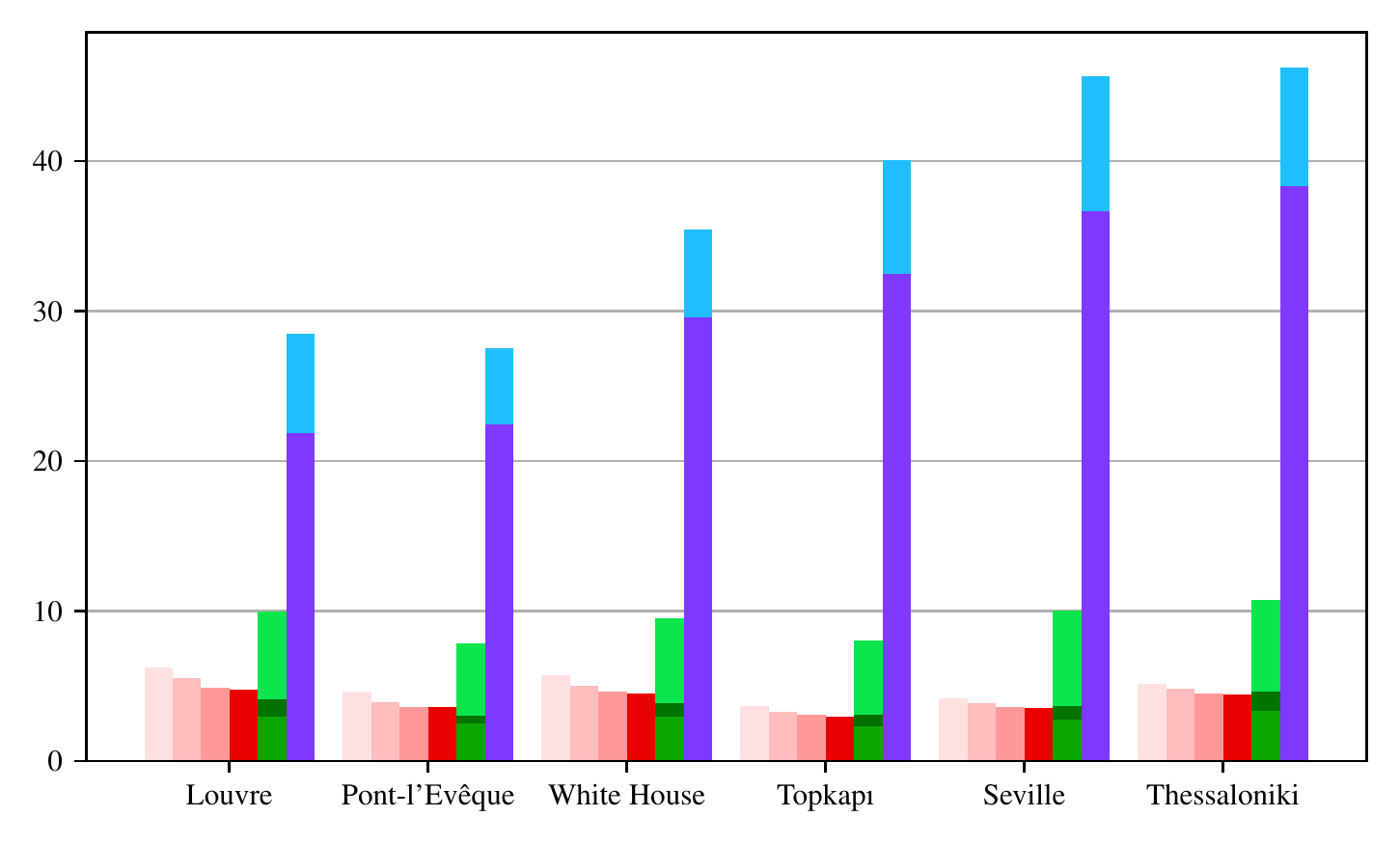}
    \\
    &
    \adjustbox{padding=3mm}{\includegraphics[scale=1.0,align=c]{plots/raytracePlotLegend_manueel_cleaned_horiz.pdf}}
\end{tabular}
    \caption{
        Traversal costs for the real-world floorplans. The triangulations require less operations than a roped kd-tree for each scene, and both techniques significantly outperform a BVH.
        \label{figRaytracePlotsRealWorld}
        }
\end{center}
\end{figure}

\FloatBarrier
\section{Discussion}

\subsection{Conclusions}

We have presented a full pipeline for generating an approximate minimum weight triangulation for a given set of input edges.
The process starts from an initial triangulation (e.g.\ a CDT) which gets optimized with respect to an objective function that minimizes the total edge length with safeguards against bad conditioning and with the ability to (fuzzily) contract edges to a point if such topological changes would improve the total edge length.
The optimization of this objective function happens through simulated annealing with several custom perturbation strategies to facilitate efficient exploration of the topological configuration space.
After optimization, the fuzzy topology is then baked in through a process of contraction and polishing. 
The resulting triangulation can be used as-is, or if desired it can be subdivided and used as a new initial triangulation for the entire process once again, in the hopes that the added topological degrees of freedom lead to a further improvement in the final total edge length.

Apart from our full pipeline, a simpler subset can be used to optimize or `polish' an existing triangulation with less computational effort, for instance by disabling fuzzy contraction.
Computation times for the full pipeline can reach up to the order of a week on current consumer hardware for our most detailed scenes when trying to approximate the true minimum.
A simpler `polishing' step takes around the order of hours for such scenes.
The resulting improvement in total edge length over the optimally refined CDT is less than what we found for the full pipeline, but for some scenes the majority of the improvement potential is already extracted by such a simple polishing step. This was especially visible in the detailed hair scenes, for which this simple strategy already handles the large empty central area quite well.

The structure of our resulting minimum weight triangulations can be markedly different than what is obtained from traditional (constrained) Delaunay-type triangulation construction and refinement methods due to their fundamentally incompatible goals: minimizing the total edge length versus guaranteeing well-behaved triangles with no sharp edges.
For specific scenes, the additional edge length in such Delaunay triangulations over our minimum can represent a substantial factor as we saw in Sec.~\ref{secIllustrativeExample}.
For more `typical' scenes we find differences of around $10$ to $20$ percent in edge length between the optimally refined CDT, and our minimum weight triangulations. The total edge lengths of a simple \emph{unrefined} CDT (i.e.\ without any Steiner vertices) ranged between an additional 10 to 20 percent for the homogeneous lines scenes, up to twice our minimum for more inhomogeneous scenes such as the hair scenes.

When used for ray tracing, we find a similar variation in number of traversal operations when comparing triangulations of various levels of optimization. A factor up to two can be discerned between an unrefined CDT and our optimized triangulations (e.g.\ in the detailed grass and hair scenes with low $N$), but this gap lessens for the majority of other typical scenes roughly ranging around an additional 10 to 20 percent.

Compared to traditional ray acceleration data structures such as a BVH and a (roped) kd-tree, we find that the triangulations require less traversal operations across the board.
Firstly, by virtue of their stackless traversal they can avoid the initial $\log N$ descent in a hierarchical data structure, which immediately gives them an edge over a BVH --- indeed, we even observed cases where the number of traversal operations \emph{decreased} for increasing number of primitives $N$.
Moreover, because triangulations directly incorporate the geometric primitives as boundaries for their cells, their traversal cost is robust against the specific orientation and position of the geometry in the coordinate system --- in contrast to the axis-aligned cells in a BVH or kd-tree which show pathological behaviour when the input geometry becomes maximally misaligned with the world axes.

\subsection{Limitations and Future Work}
The most obvious limitation of the current work is that it is limited to a two-dimensional setting. Although a 2D (or `2.5D') case has merit of itself \cite{maria17architecturalCCSP}, it is not not immediately obvious that our results --- in particular $N$-dependence --- scale up to 3D. 

Another inherent limitation of triangulations, or CCSPs in general, is that the input geometry should be well-behaved and consistent. Self-intersections, for instance, are not allowed as they would lead to topological inconsistencies. A solution could be to split or clip the geometry to resolve self-intersections in a preprocessing step, or inspiration could be taken from contemporary methods for robustly triangulating (or tetrahedralizing) arbitrary geometry \cite{tetMeshingInTheWild,fastTetMeshingInTheWild,triWild}. A related point of attention is the importance of numerical stability during traversal of these structures \cite{maria17robustCDTopGPU}.

On the topic of numerical stability: we have come across a few cases where our indirect restriction on badly conditioned triangles when fuzzy contraction is enabled was insufficient to avoid a highly sharp triangle. Due to the conflicting goal of fuzzy contraction, this conditioning restriction is only applied indirectly by avoiding angles of nearly 180$\degree$ instead of directly avoiding angles close to $0\degree$. As the former is a necessary but not sufficient requirement for the latter, some near-zero angles can occasionally slip through in the odd case.

In comparing triangulations with classical acceleration data structures, we have simply counted the different traversal operations without weighting them with their actual cost as we are mostly interested by their overall (asymptotic) behaviour or scaling. However, in practice their actual costs can differ.
For instance, remember that the node traversal cost in the SAH was chosen per scene such that the total (unweighted) sum of all operations was minimized for the roped kd-tree (and again separately for the BVH). One might argue that the computational cost of a geometry test in the kd-tree (i.e.\ a ray -- line segment intersection) should be weighted at roughly half the cost of traversing a triangle in the triangulation (i.e.\ finding which of the two\footnote{During traversal, one can assume that the incoming edge of the current triangle is known and thus only two possible outgoing edges remain --- unless we are at the very first triangle of the traversal, in which case all three edges are candidates.}
possible line segments is the outgoing edge of the current triangle).
Moreover, the operations for descent in the kd-tree and ascent using the rope tree are even cheaper due to their axis aligned nature.
If the traversal cost in the SAH were chosen to optimize a \emph{weighted} sum of the traversal operations based on their actual computational cost, then the balance between the number of geometry tests and node visits (and rope searches) might be altered, so as to decrease the (weighted) sum of all operations and effectively scoring more favourable relative to the triangulation.
In our current treatment, we ignored these (implementation- and hardware-dependent) subtleties and focused on the overall (asymptotic) behaviour and scaling of the total (unweighted) number of operations.

Our optimization procedure is based on a general-purpose simulated annealing approach (albeit with custom perturbation strategies targeted at the specific problem domain), started from any triangulation and aimed at finding an optimal triangulation with as little assumptions as possible about how such a minimum weight triangulation should be structured. Of course, it might be beneficial to modify an existing procedure, such as the refinement process of a CDT, to directly aim towards a minimum weight triangulation.
This can either be by outputting a triangulation that is already `good enough', or by outputting a triangulation that is already well-suited for our polishing step or full pipeline (e.g.\ with extra Steiner vertices in regions that are expected to be less-than-ideal, and few Steiner vertices in regions where it is reasonably assumed that the local configuration is already near-optimal).

\section{Acknowledgements}
This work was funded by the FWO (Fonds Wetenschappelijk Onderzoek -- Vlaanderen), grant 
G099617N.
The authors would like to thank Jonathan Shewchuk for open sourcing the Triangle utility and the various Wikimedia contributors for releasing the images that we used for the real-world floor plans in the public domain.

\appendix

\newpage

\FloatBarrier
\section{Triangulations, KD-Trees and BVHs of Selected Scenes \label{appTriangsKDBVH}}
This appendix contains visualisations of the triangulations, kd-trees and (the leaf nodes of) BVHs of a subset of generated scenes (lines, grass, hair) and of the real-world scenes. The original geometry is shown in black, the acceleration data structure in blue and several randomly sampled rays in red.

For the lines scene, we show the scenes with $N=10$, 100 and 1000, as well as $N=32$ (the logarithmic midpoint between $N=10$ and $N=100$) as this lies in the region where there is a transition in behaviour for the kd-trees and BVHs in the case of long, non-axis aligned lines as discussed in Sec.~\ref{secRTResultsLines}.

\newcommand{\linesAccelCompareSize}{height=42mm}
\newcommand{\grassAccelCompareHeight}{scale=0.55}
\newcommand{\grassAccelCompareHeightFirst}{scale=0.35}
\newcommand{\hairAccelCompareHeight}{scale=0.49}

\newcommand{\accelStructComparisonA}[7]{
    \begin{landscape}
    \begin{figure}[htb]
    \begin{center}
    \begin{tabular}{c@{\ }c@{\ }c@{\ }c@{\ }c}
    & \multicolumn{4}{c}{\textsc{#3}} \\[1mm]
    \rotatebox[origin=c]{90}{\small Optimized triangulation} &
        \expandafter\includegraphics\expandafter[#1,align=c,frame]{#4} &
        \expandafter\includegraphics\expandafter[#2,align=c,frame]{#5} &
        \expandafter\includegraphics\expandafter[#2,align=c,frame]{#6} &
        \expandafter\includegraphics\expandafter[#2,align=c,frame]{#7}
        \\[23mm]
}
\newcommand{\accelStructComparisonB}[6]{
    \rotatebox[origin=c]{90}{\small Roped kd-tree} &
        \expandafter\includegraphics\expandafter[#1,align=c,frame]{#3} &
        \expandafter\includegraphics\expandafter[#2,align=c,frame]{#4} &
        \expandafter\includegraphics\expandafter[#2,align=c,frame]{#5} &
        \expandafter\includegraphics\expandafter[#2,align=c,frame]{#6}
        \\[23mm]
}
\newcommand{\accelStructComparisonC}[6]{
    \rotatebox[origin=c]{90}{\small BVH leaves} &
        \expandafter\includegraphics\expandafter[#1,align=c,frame]{#3} &
        \expandafter\includegraphics\expandafter[#2,align=c,frame]{#4} &
        \expandafter\includegraphics\expandafter[#2,align=c,frame]{#5} &
        \expandafter\includegraphics\expandafter[#2,align=c,frame]{#6}
        \\[1mm]
}
\newcommand{\accelStructComparisonDlines}{
        &
        \small $N=10$ &
        \small $N=32$ &
        \small $N=100$ &
        \small $N=1000$
    \end{tabular}
    \end{center}
    \end{figure}
    \end{landscape}
}
\newcommand{\accelStructComparisonDhair}{
        &
        \small $N=2$ &
        \small $N=8$ &
        \small $N=32$ &
        \small $N=64$
    \end{tabular}
    \end{center}
    \end{figure}
    \end{landscape}
}
\newcommand{\accelStructComparisonDgrass}{
        &
        \small $N=2$ &
        \small $N=8$ &
        \small $N=32$ &
        \small $N=128$
    \end{tabular}
    \end{center}
    \end{figure}
    \end{landscape}
}

\accelStructComparisonA{\linesAccelCompareSize}{\linesAccelCompareSize}
    {Length factor 0.1 -- Vertical orientation}
    {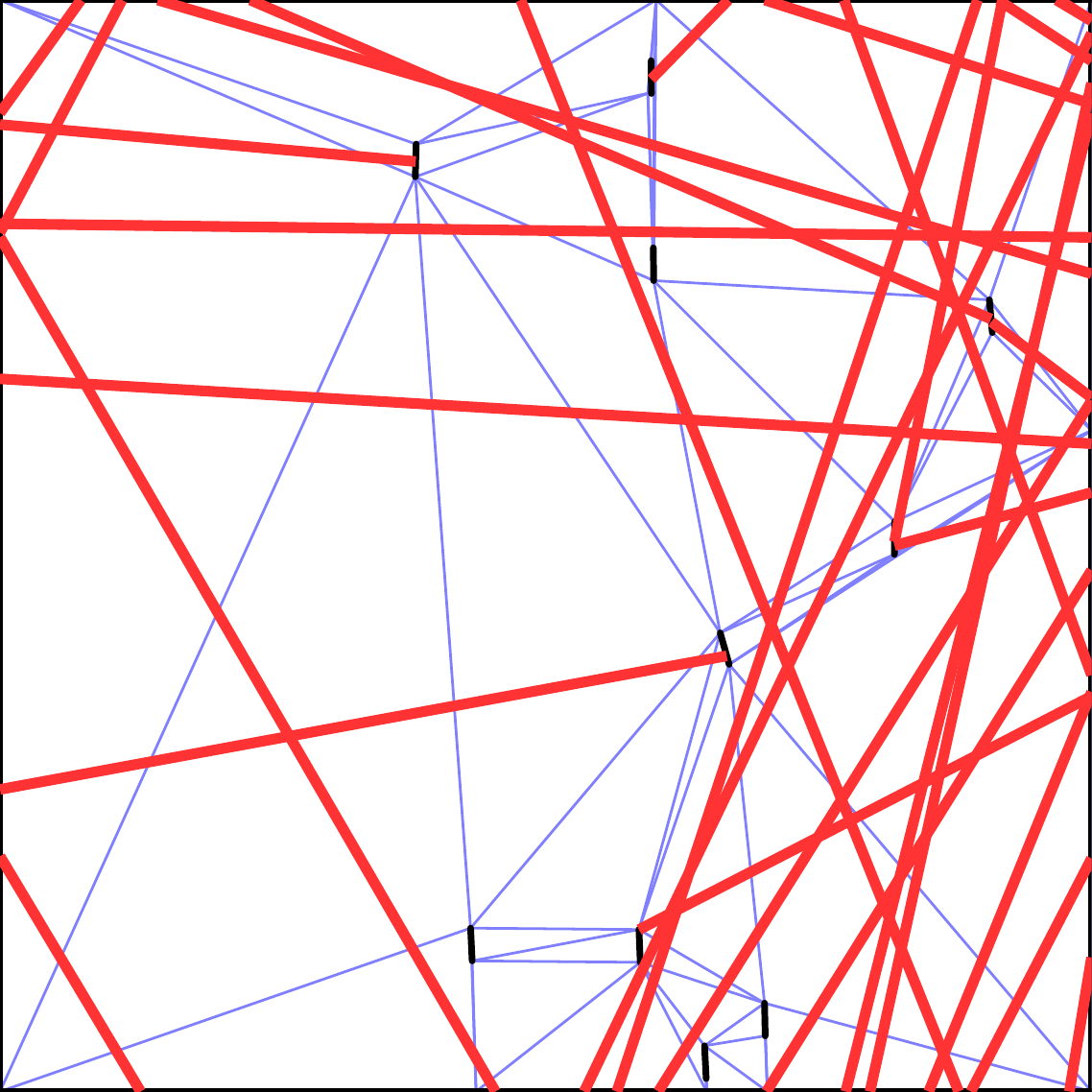}
    {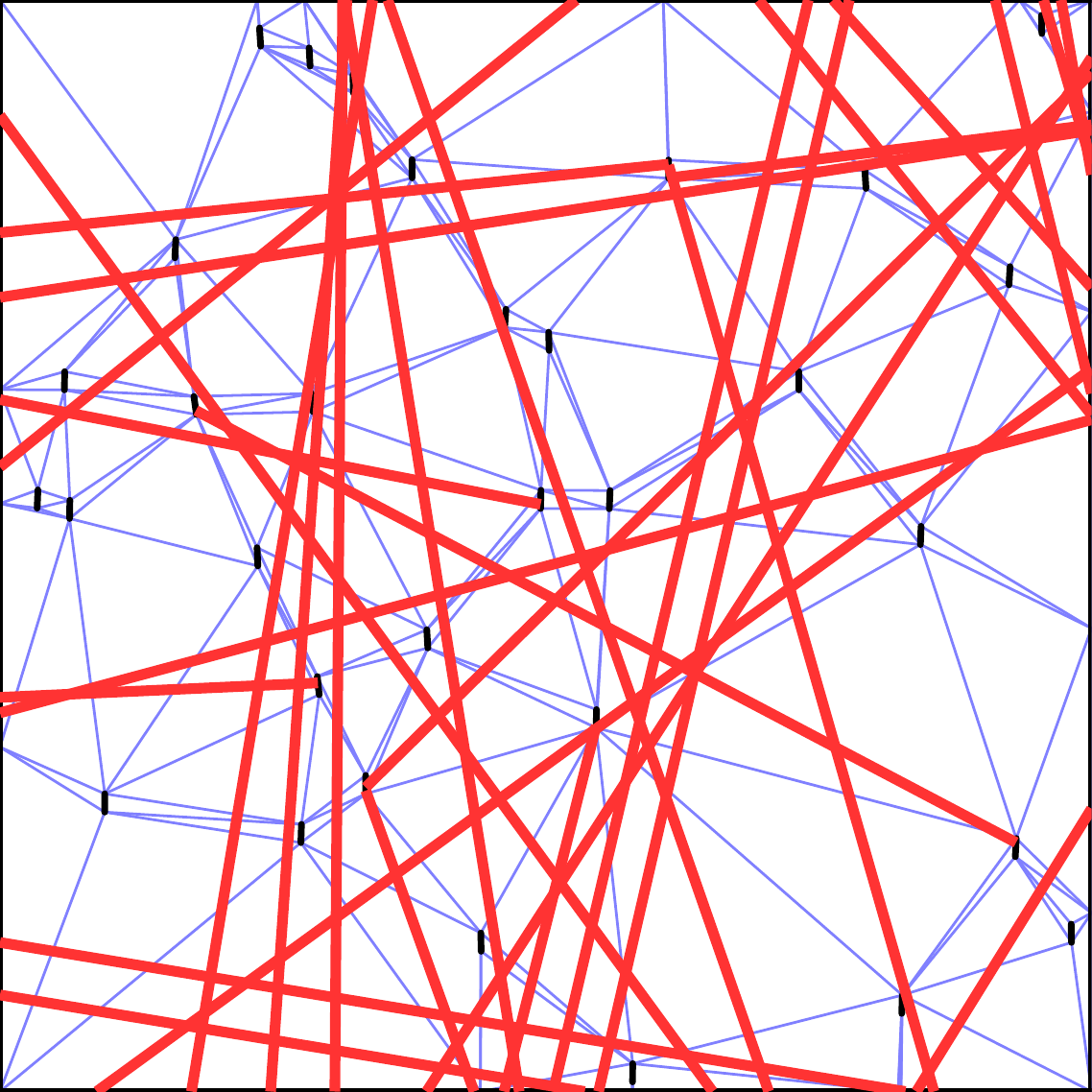}
    {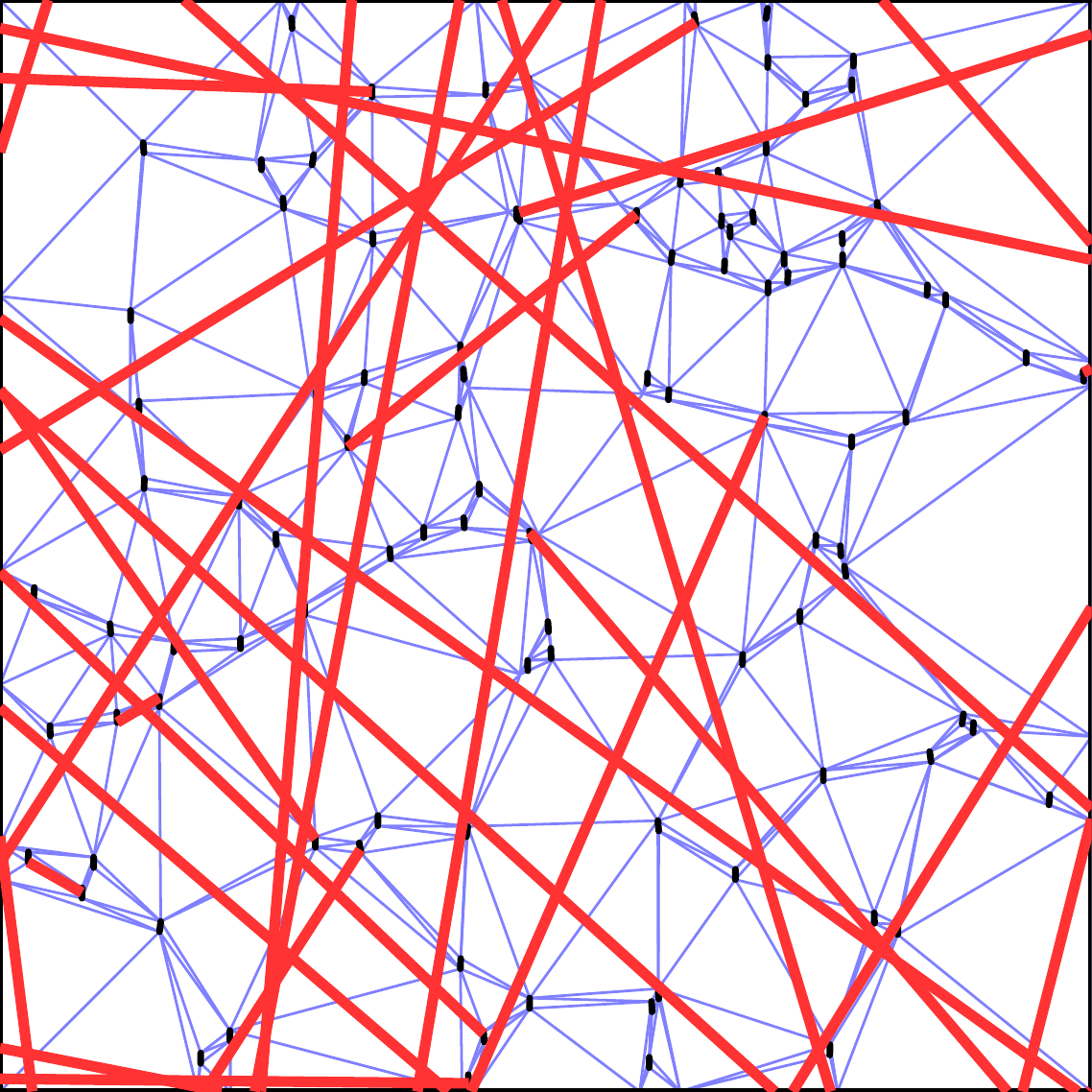}
    {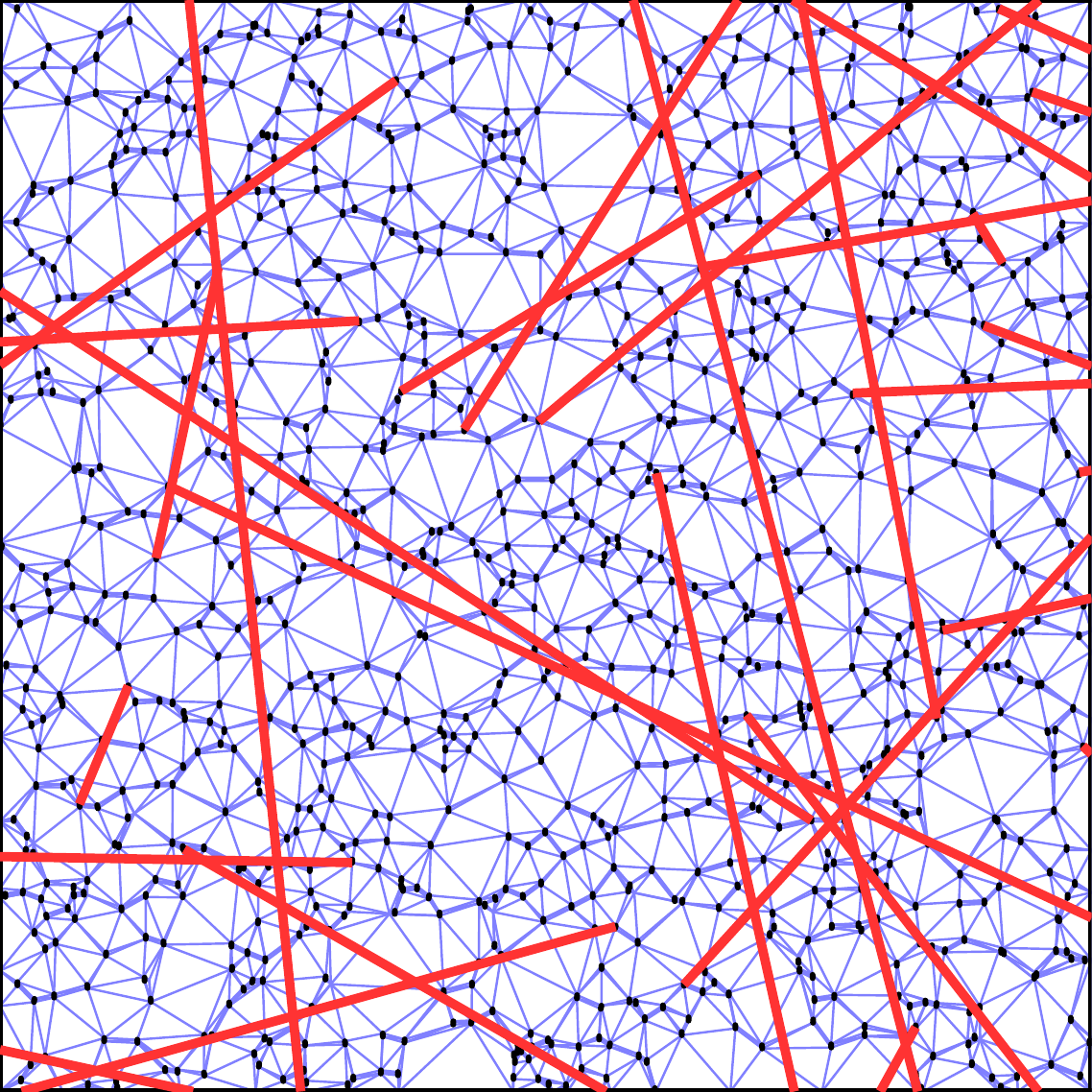}
\accelStructComparisonB{\linesAccelCompareSize}{\linesAccelCompareSize}
    {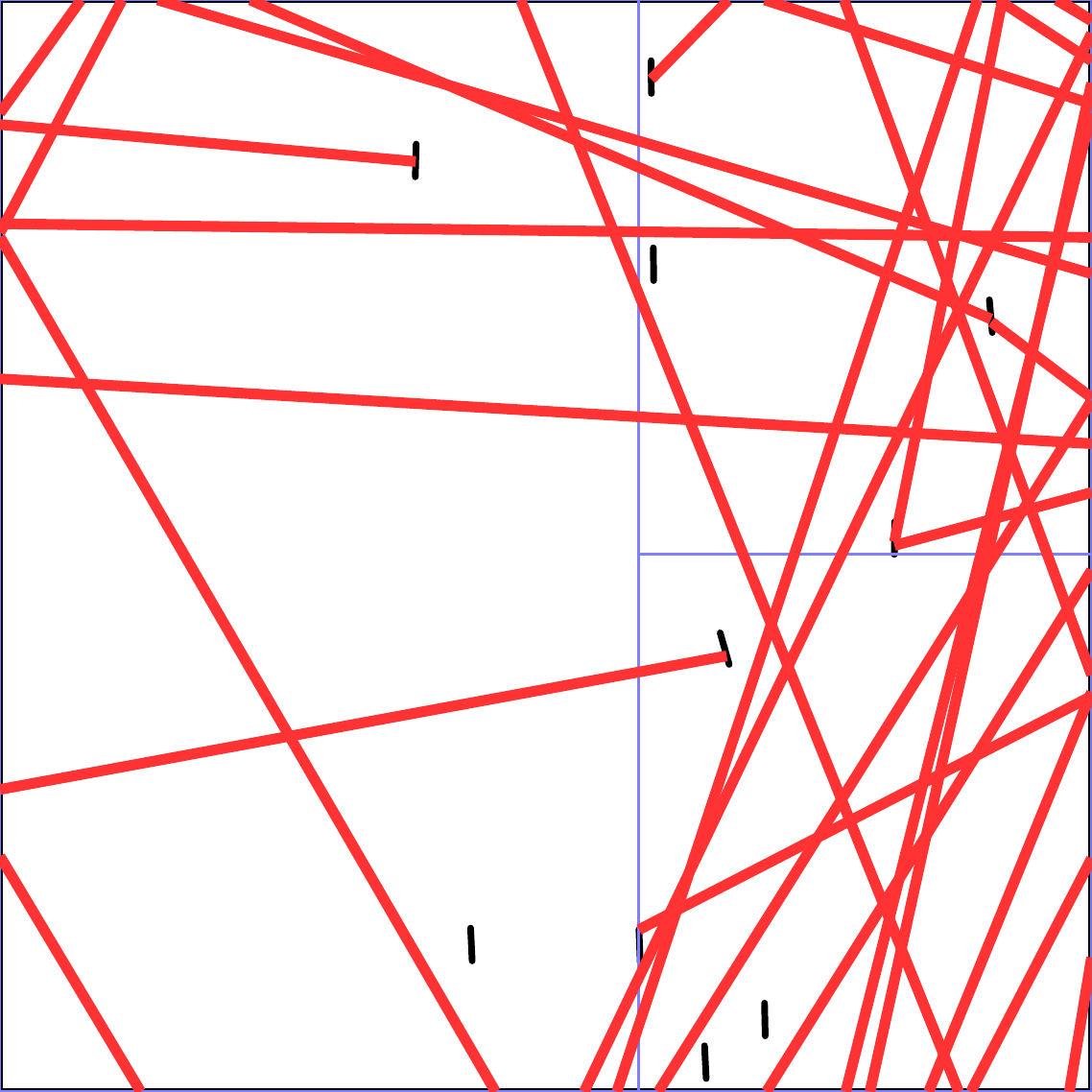}
    {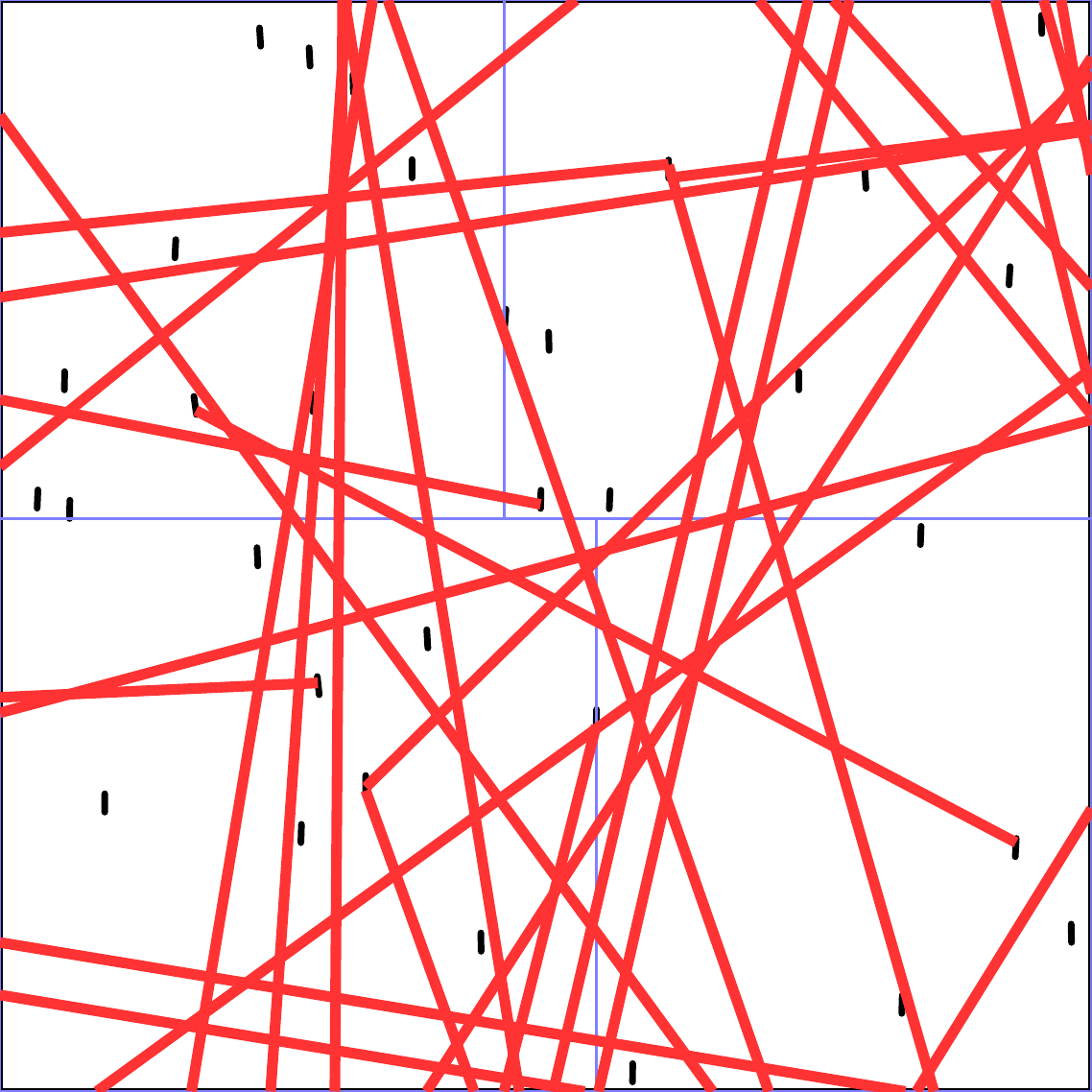}
    {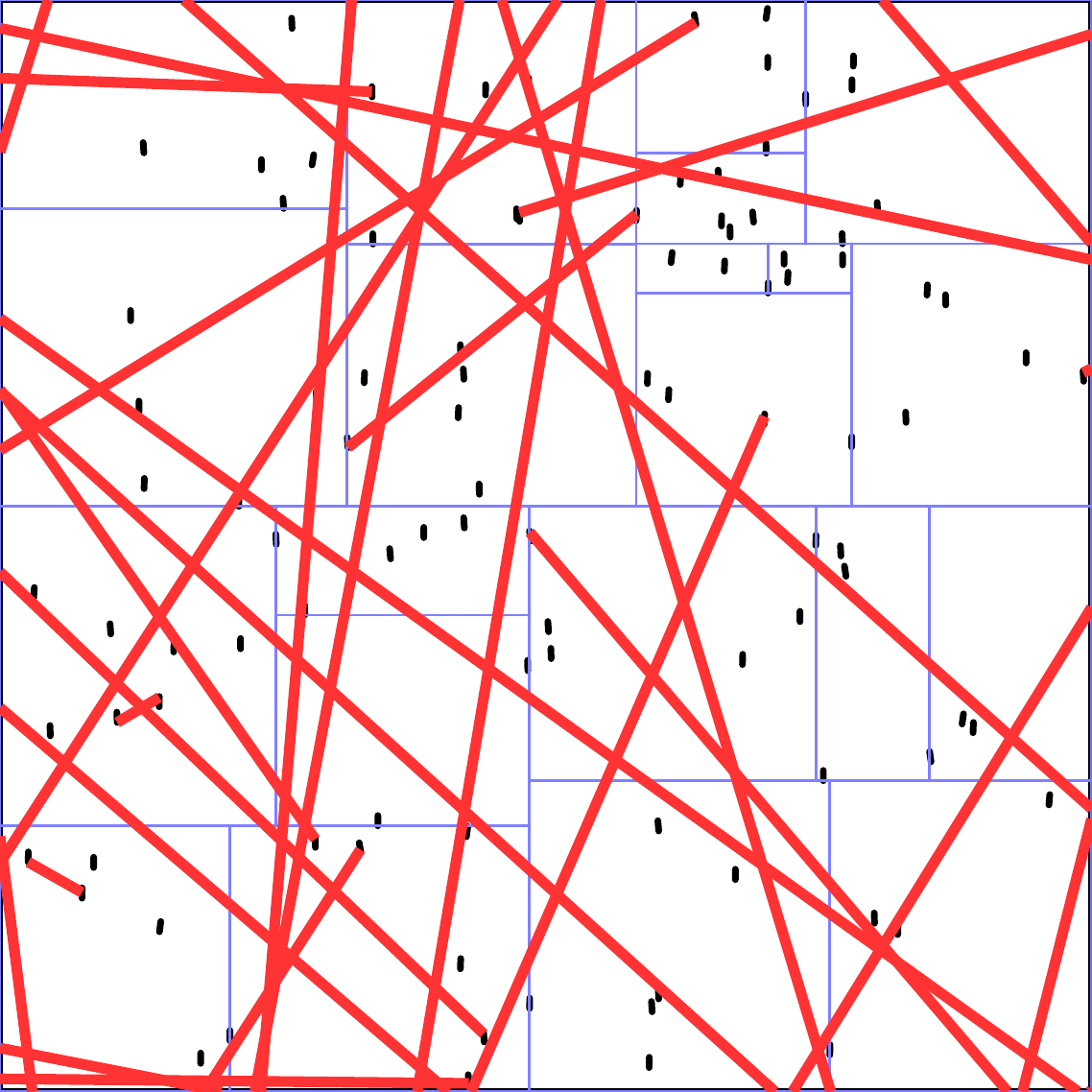}
    {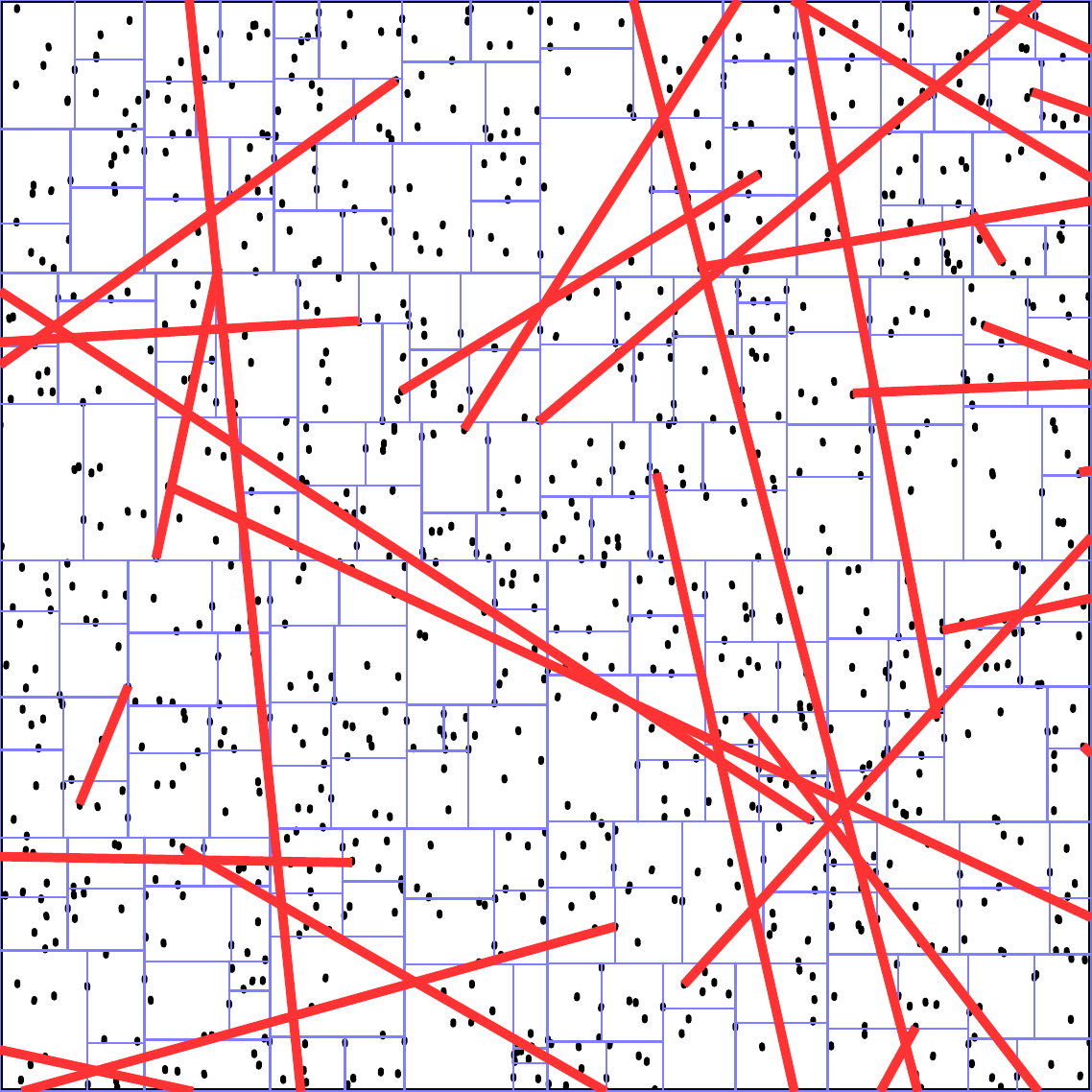}
\accelStructComparisonC{\linesAccelCompareSize}{\linesAccelCompareSize}
    {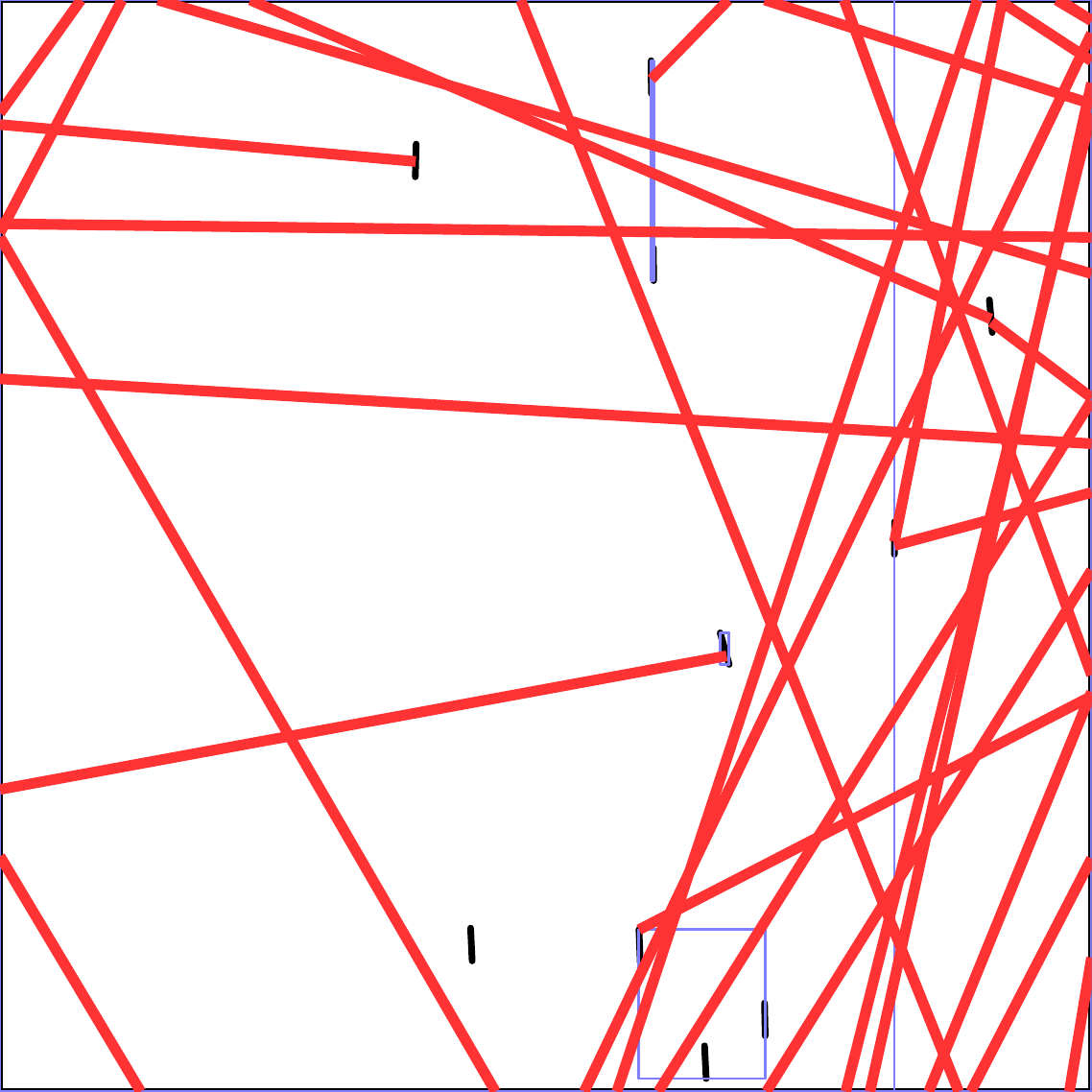}
    {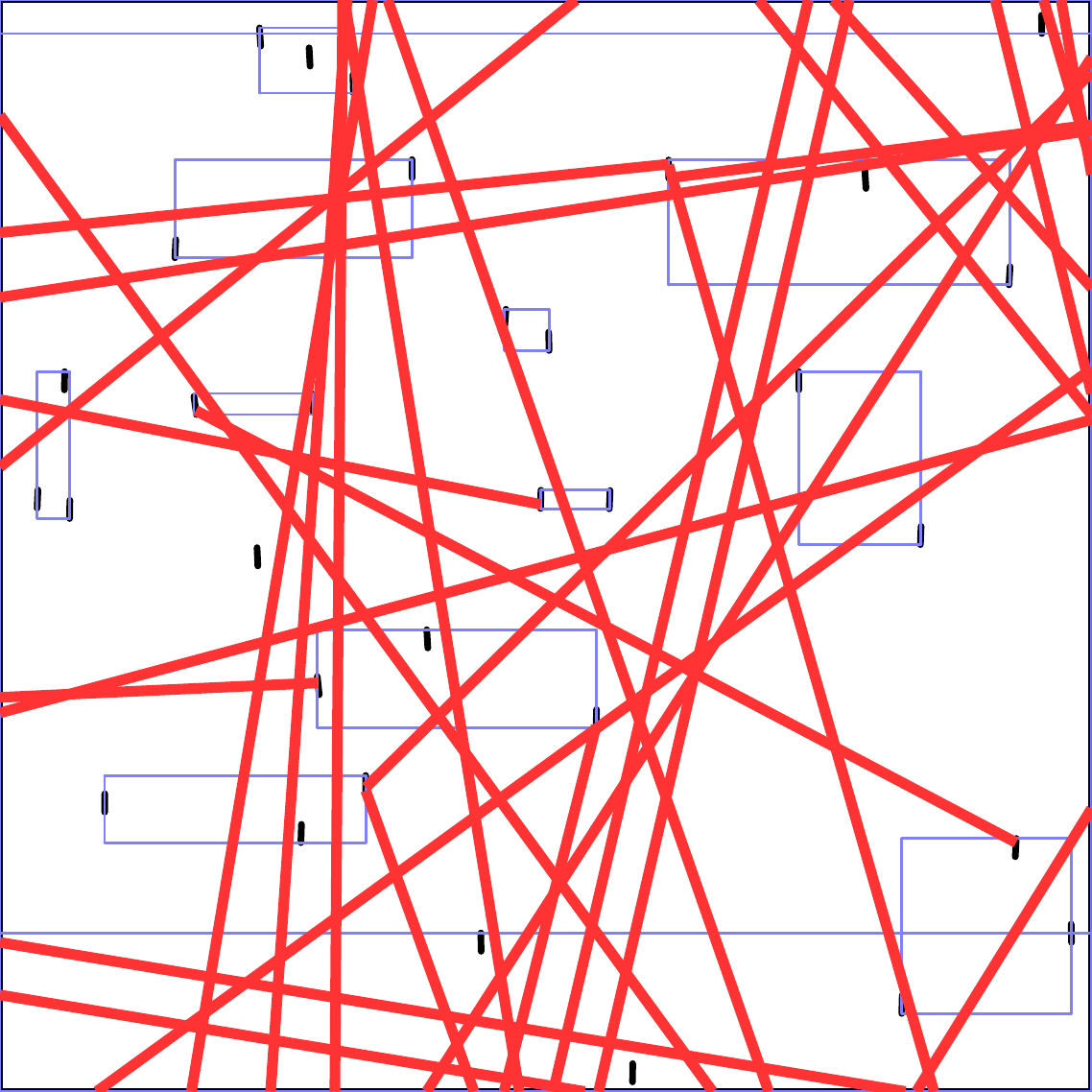}
    {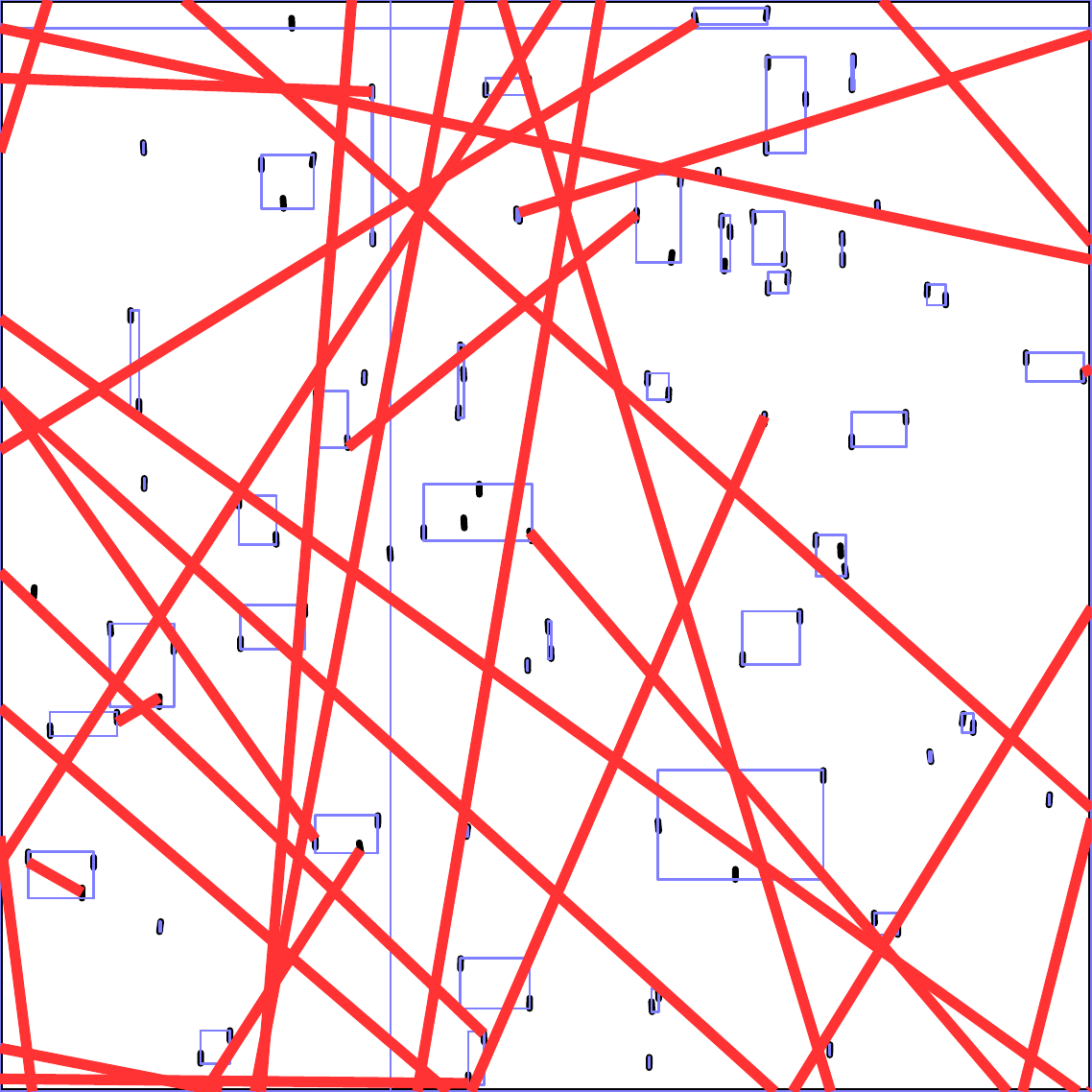}
    {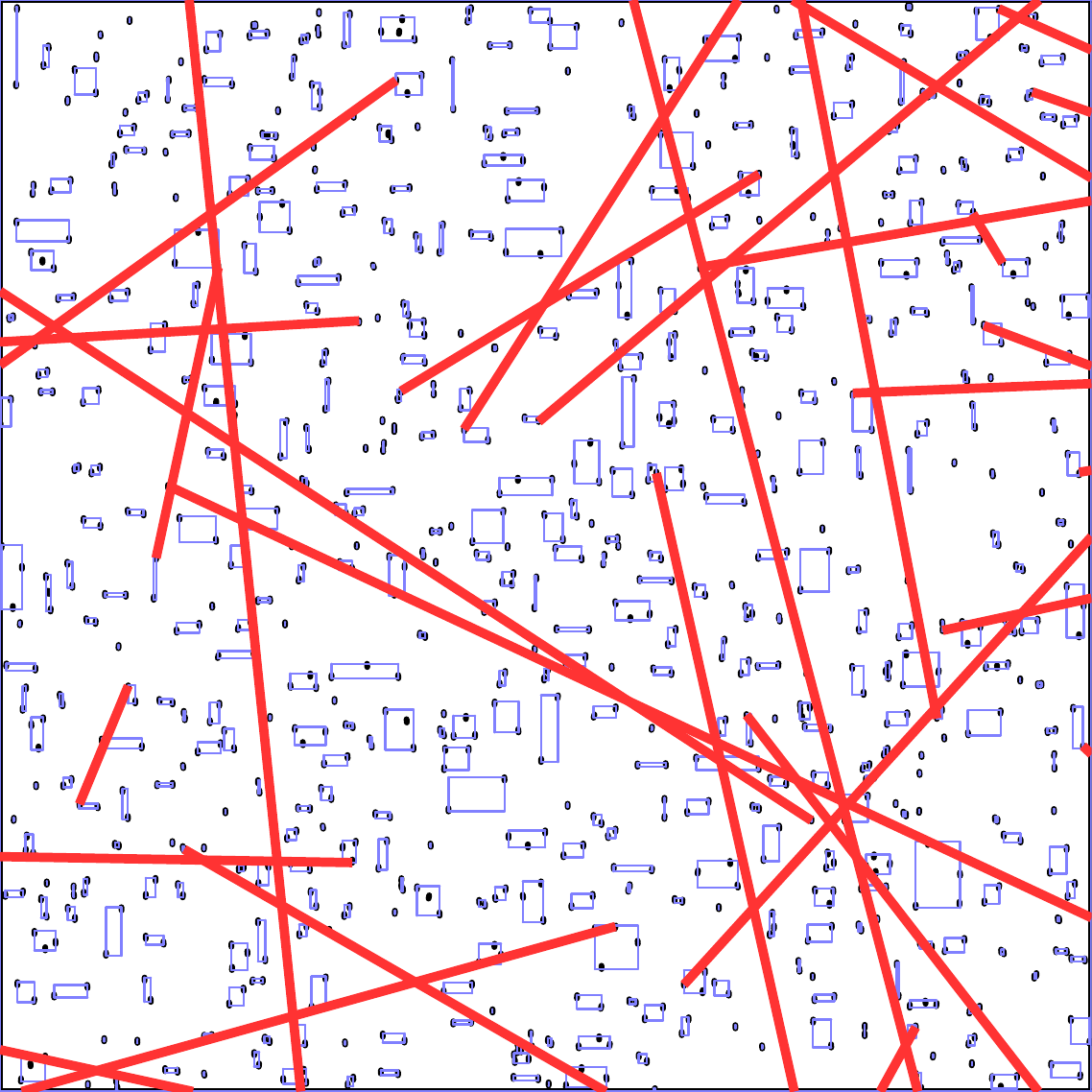}
\accelStructComparisonDlines

\accelStructComparisonA{\linesAccelCompareSize}{\linesAccelCompareSize}
    {Length factor 0.1 -- Uniform orientation}
    {plots/gen_l_sc-0.1_oa0_sh0.05_n0010_v1_randomRays.pdf}
    {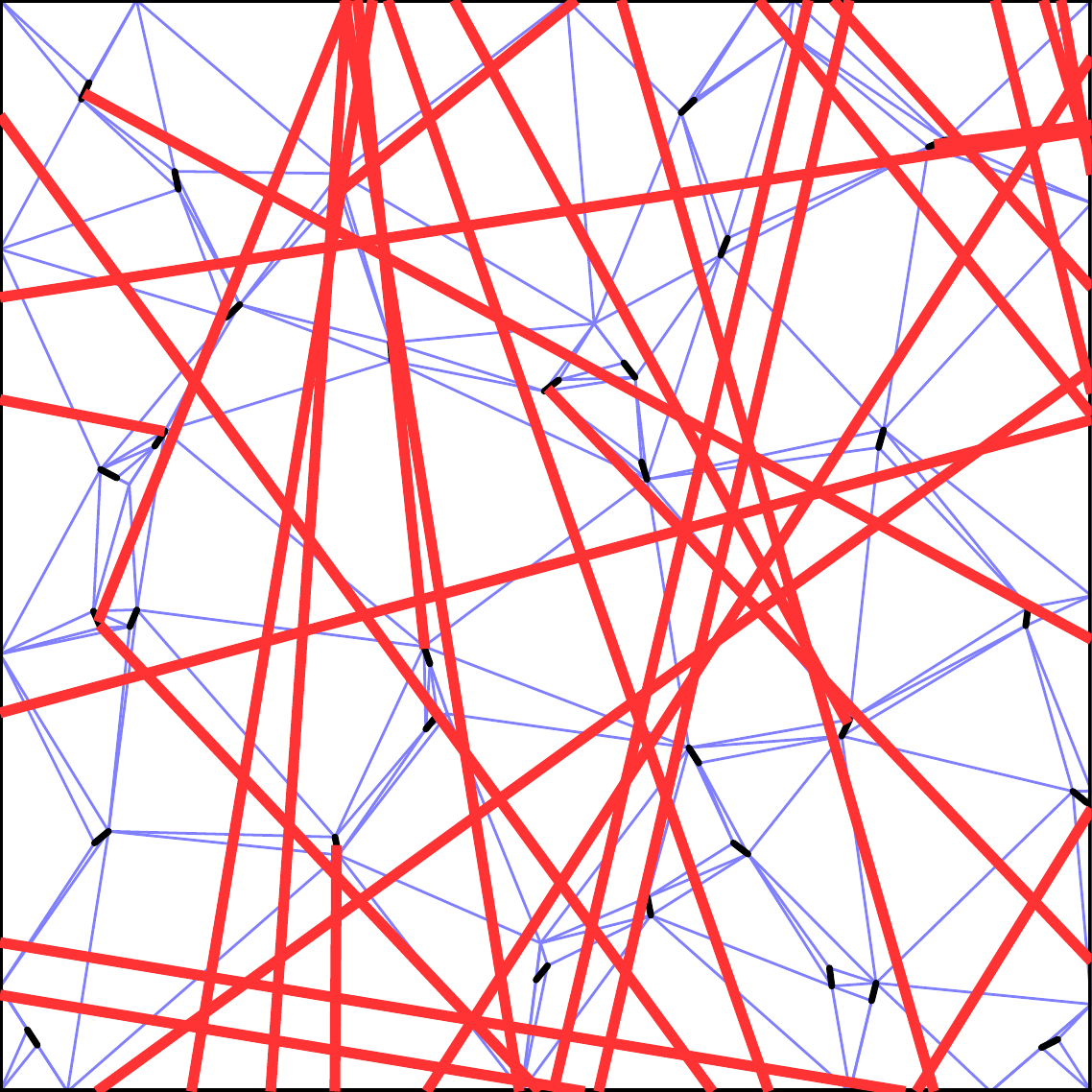}
    {plots/gen_l_sc-0.1_oa0_sh0.05_n0100_v1_randomRays.pdf}
    {plots/gen_l_sc-0.1_oa0_sh0.05_n1000_v1_randomRays.pdf}
\accelStructComparisonB{\linesAccelCompareSize}{\linesAccelCompareSize}
    {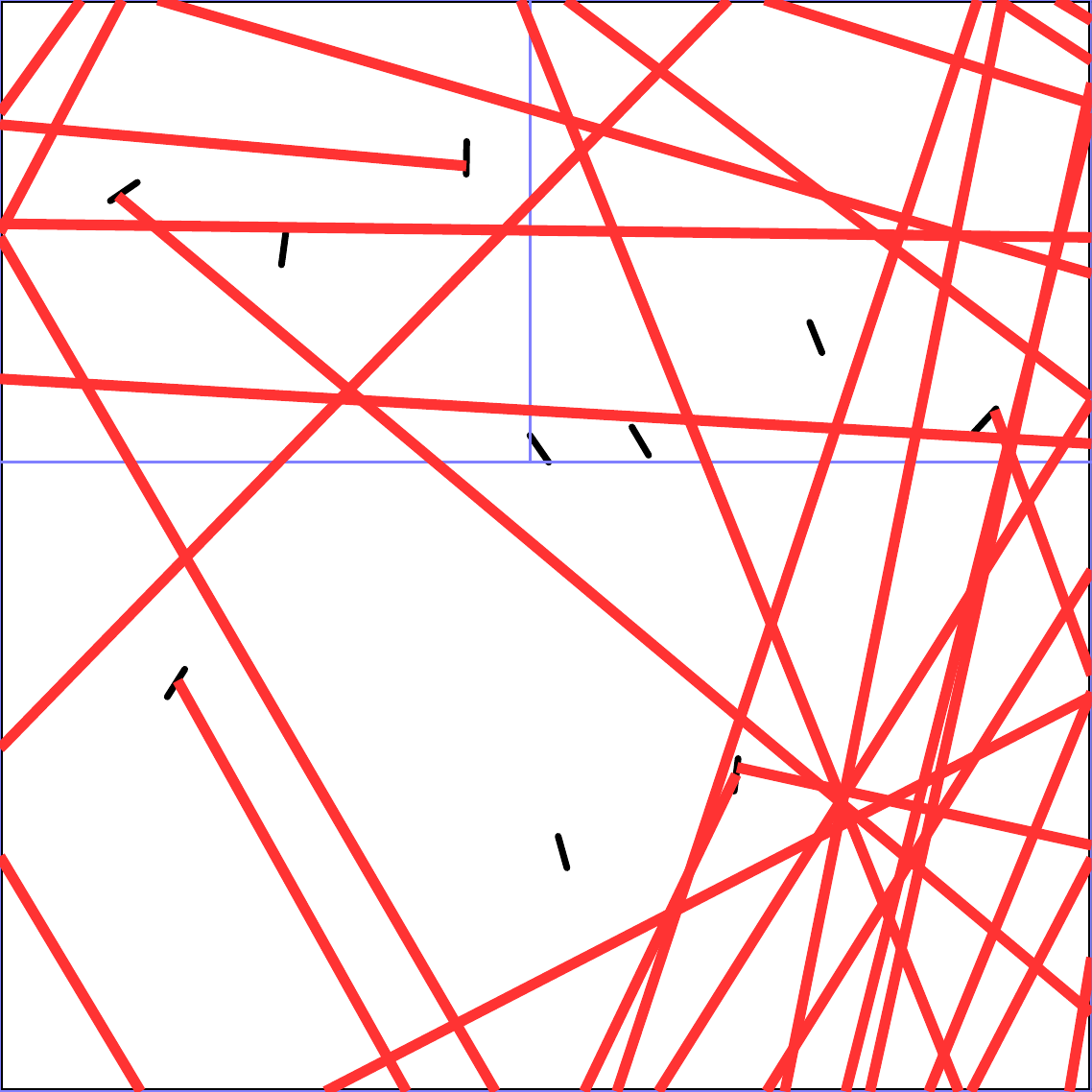}
    {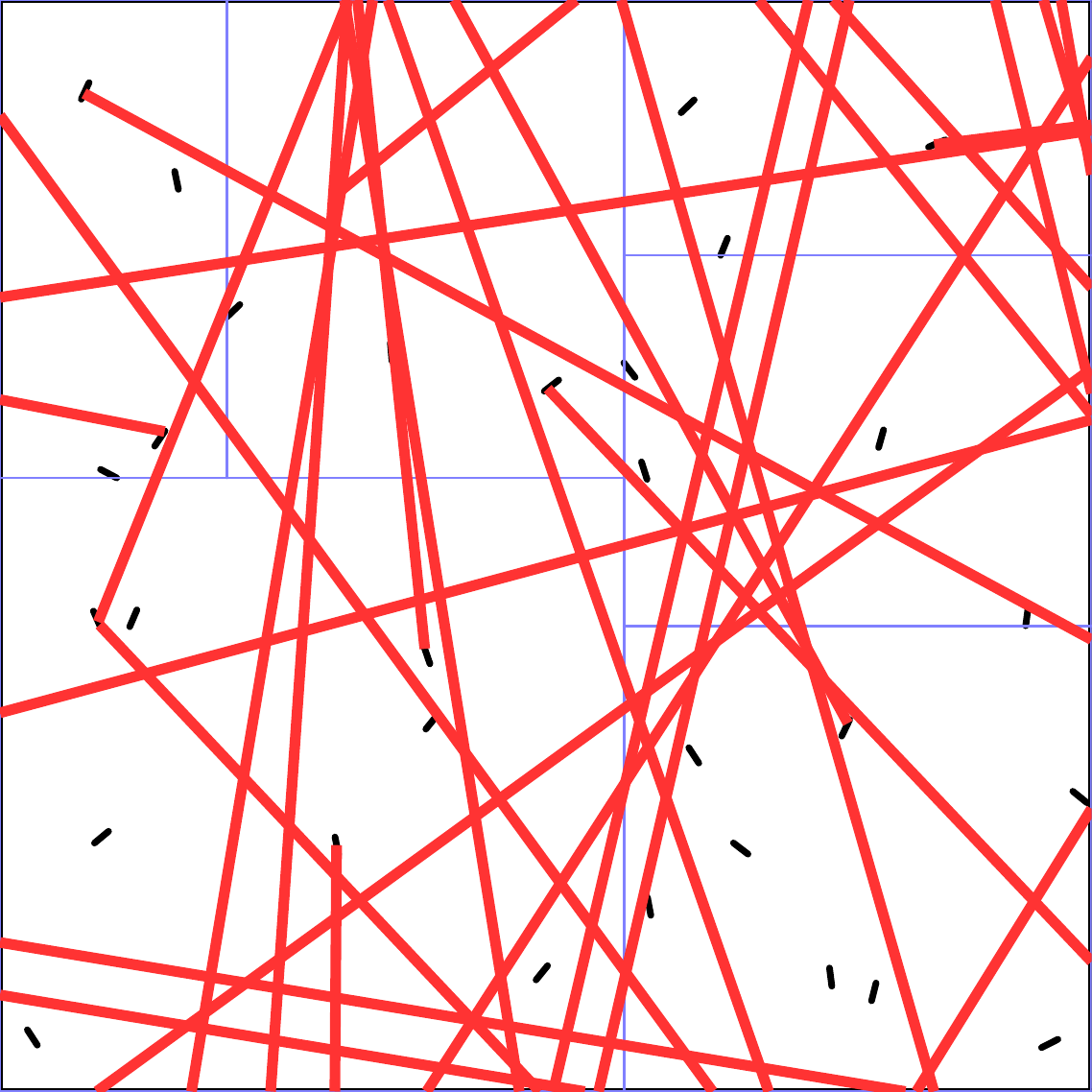}
    {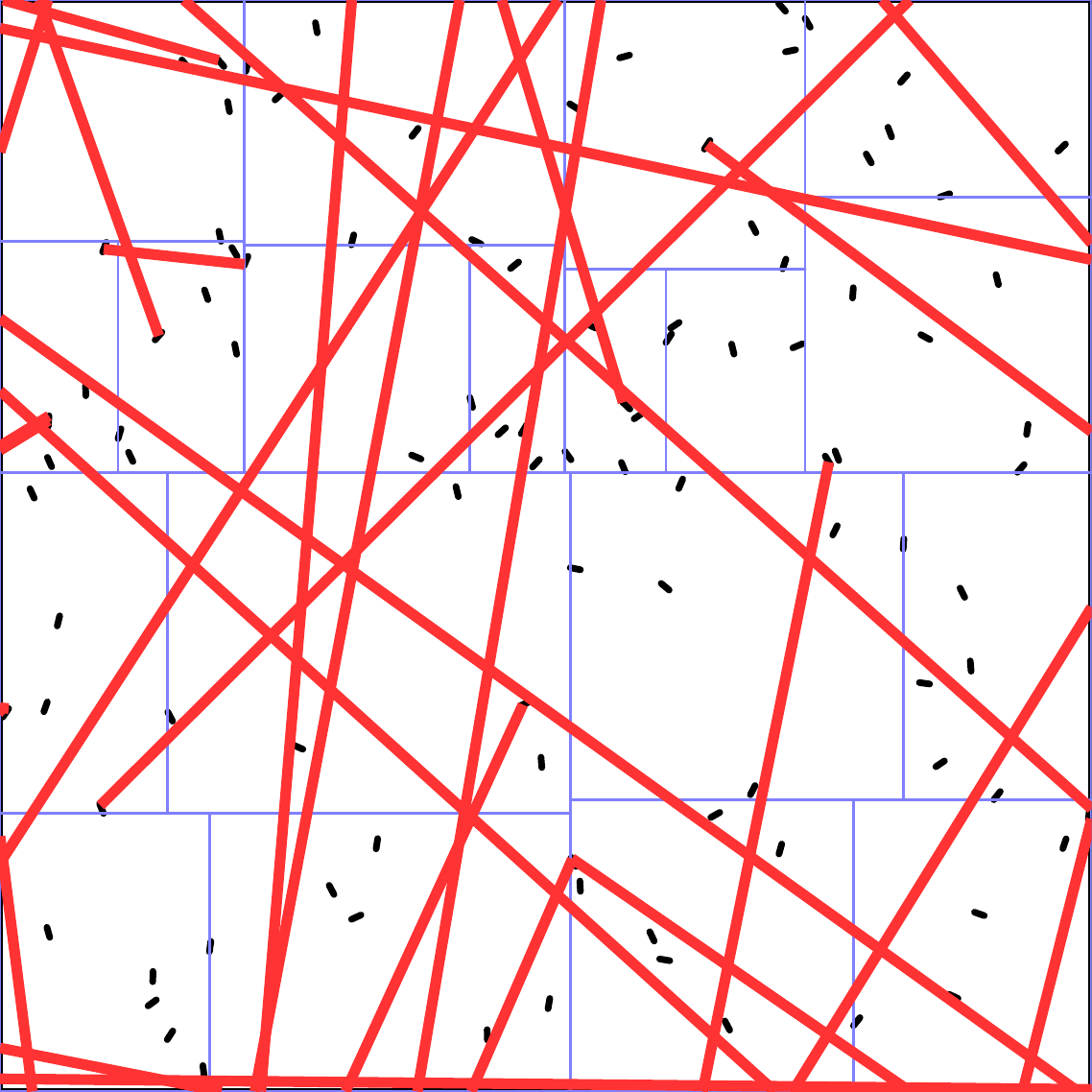}
    {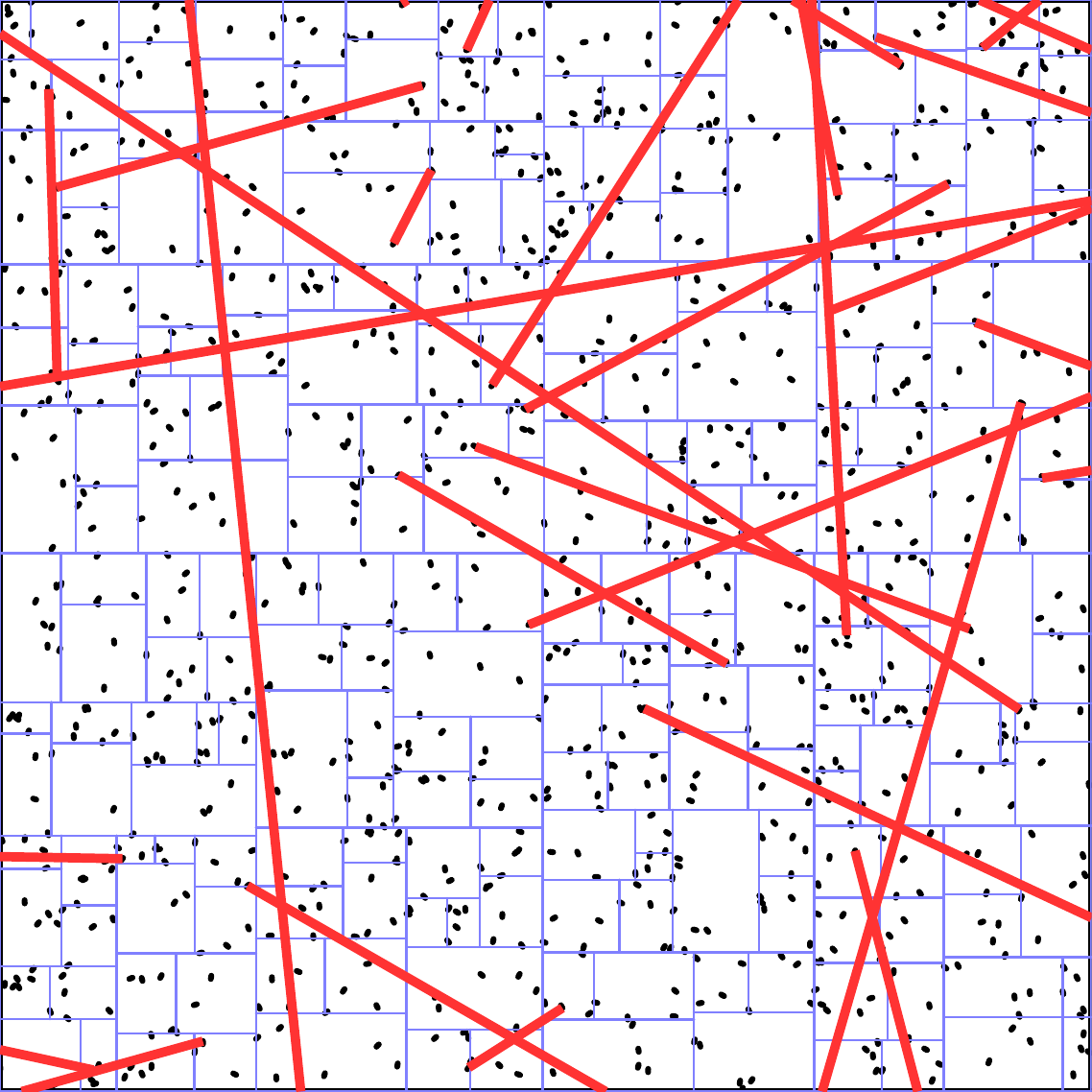}
\accelStructComparisonC{\linesAccelCompareSize}{\linesAccelCompareSize}
    {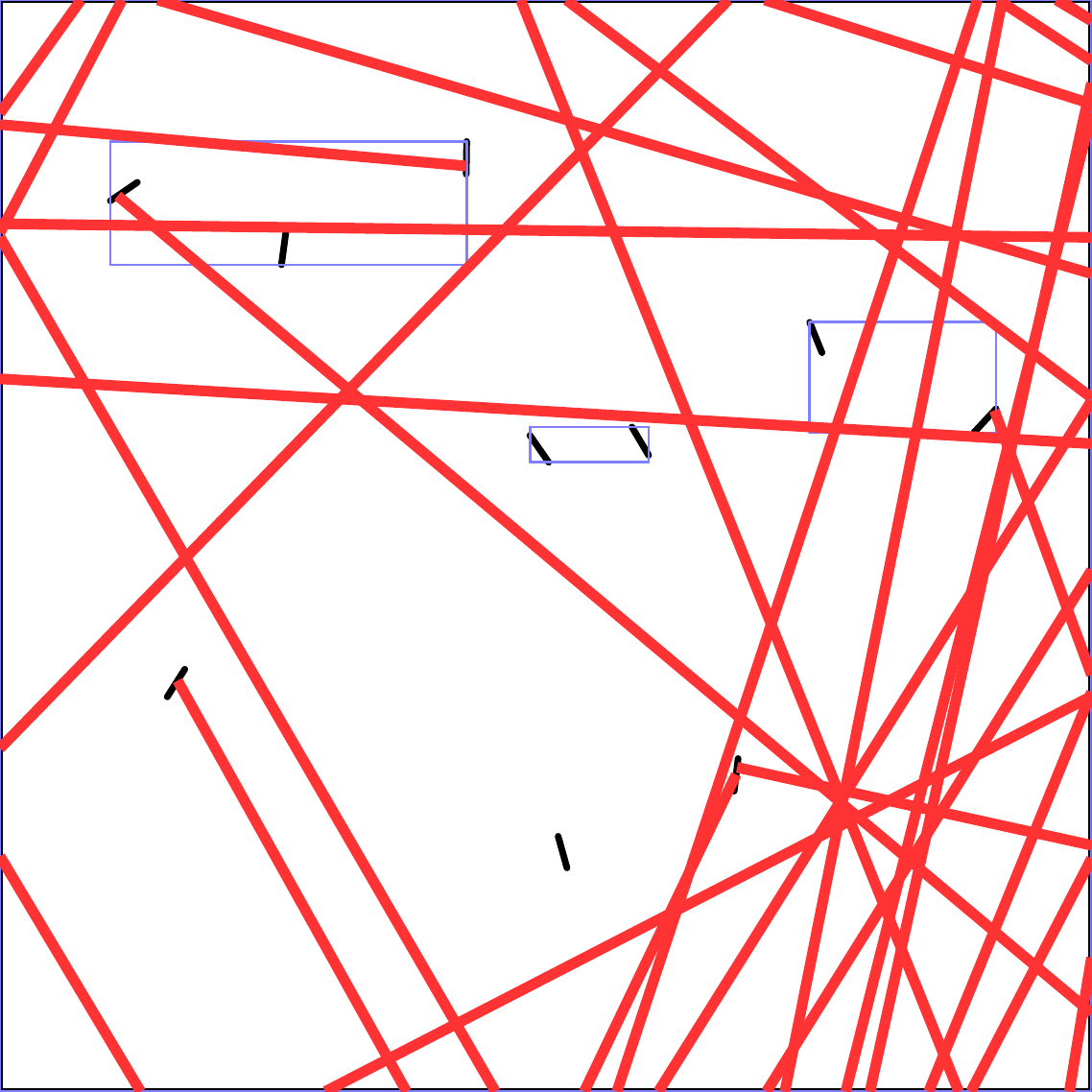}
    {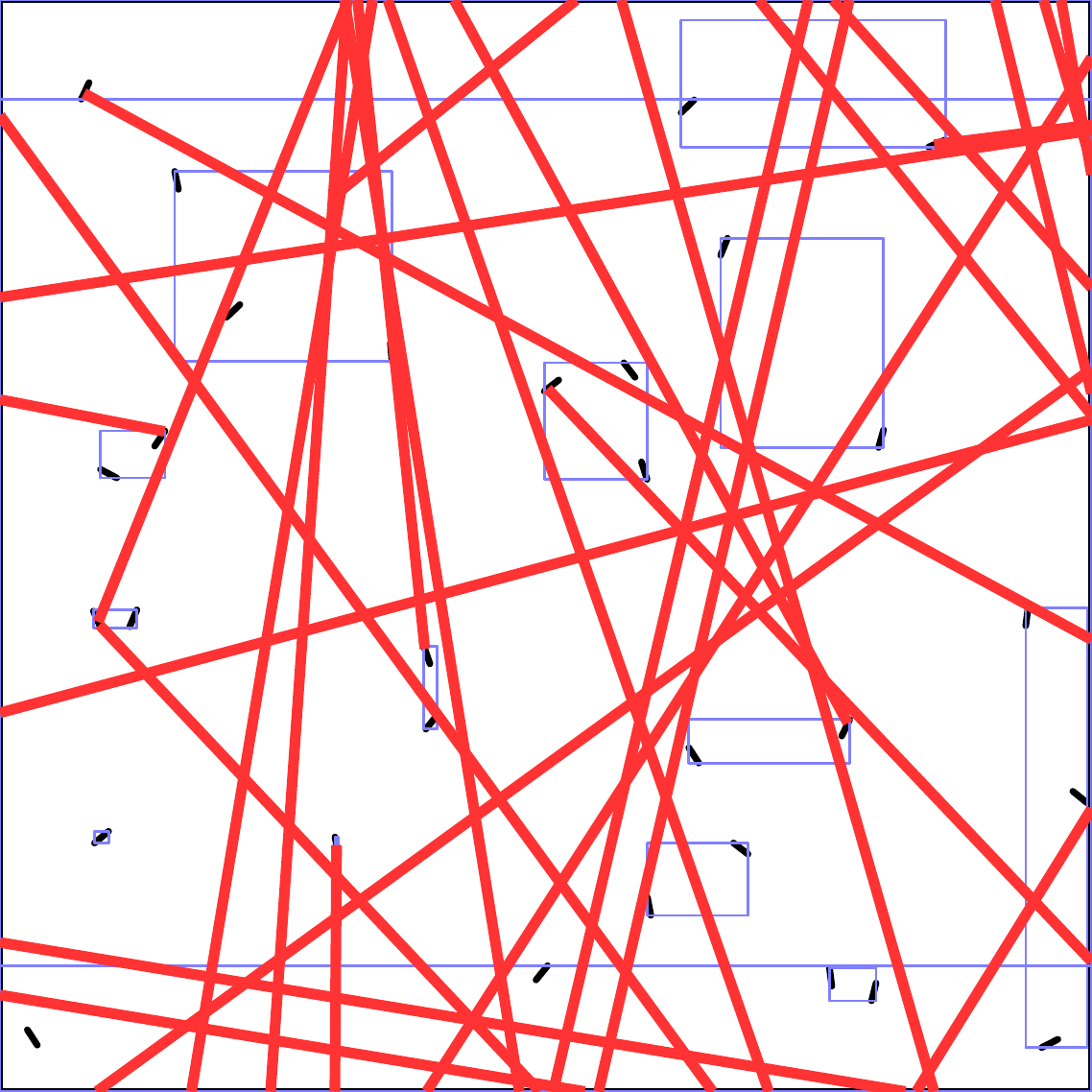}
    {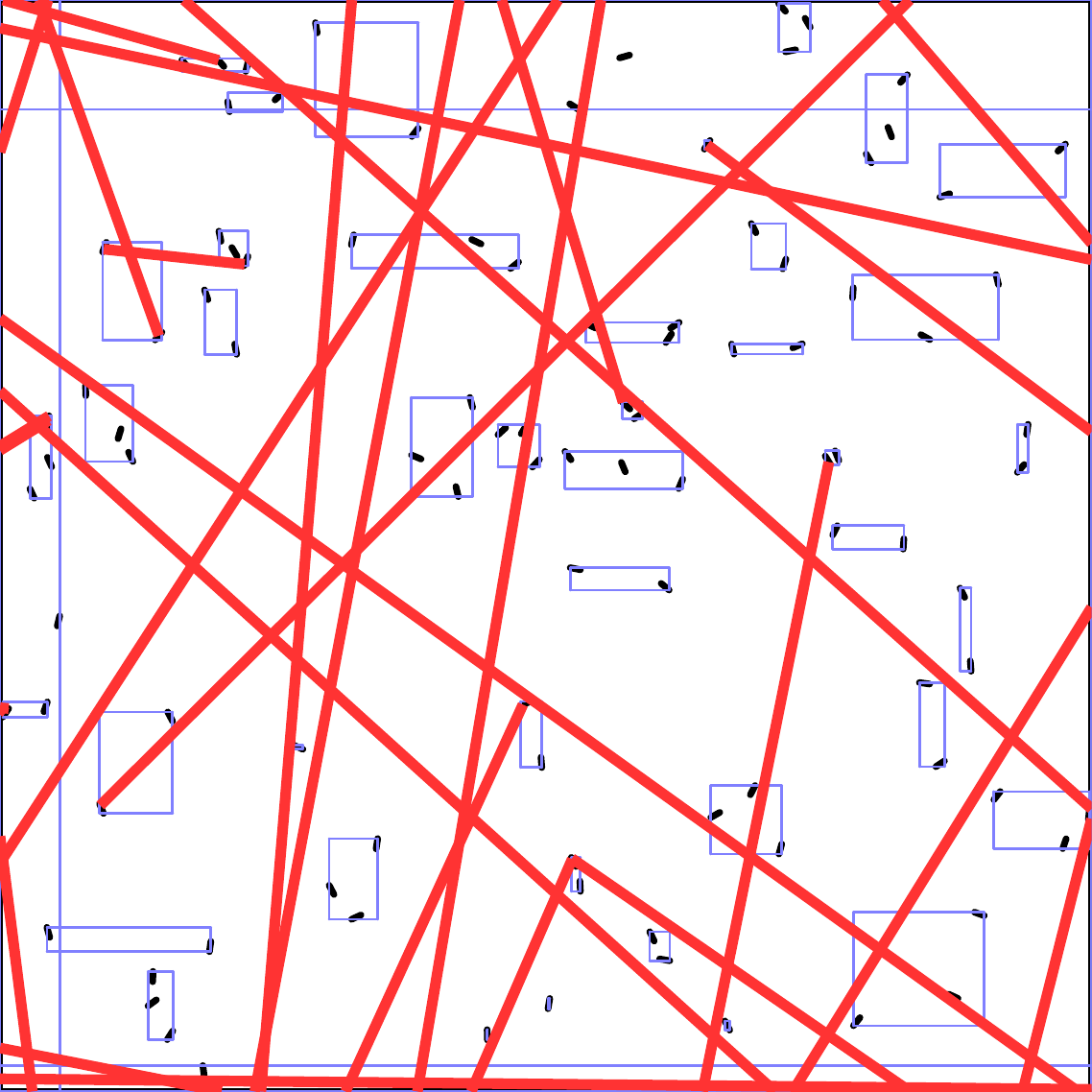}
    {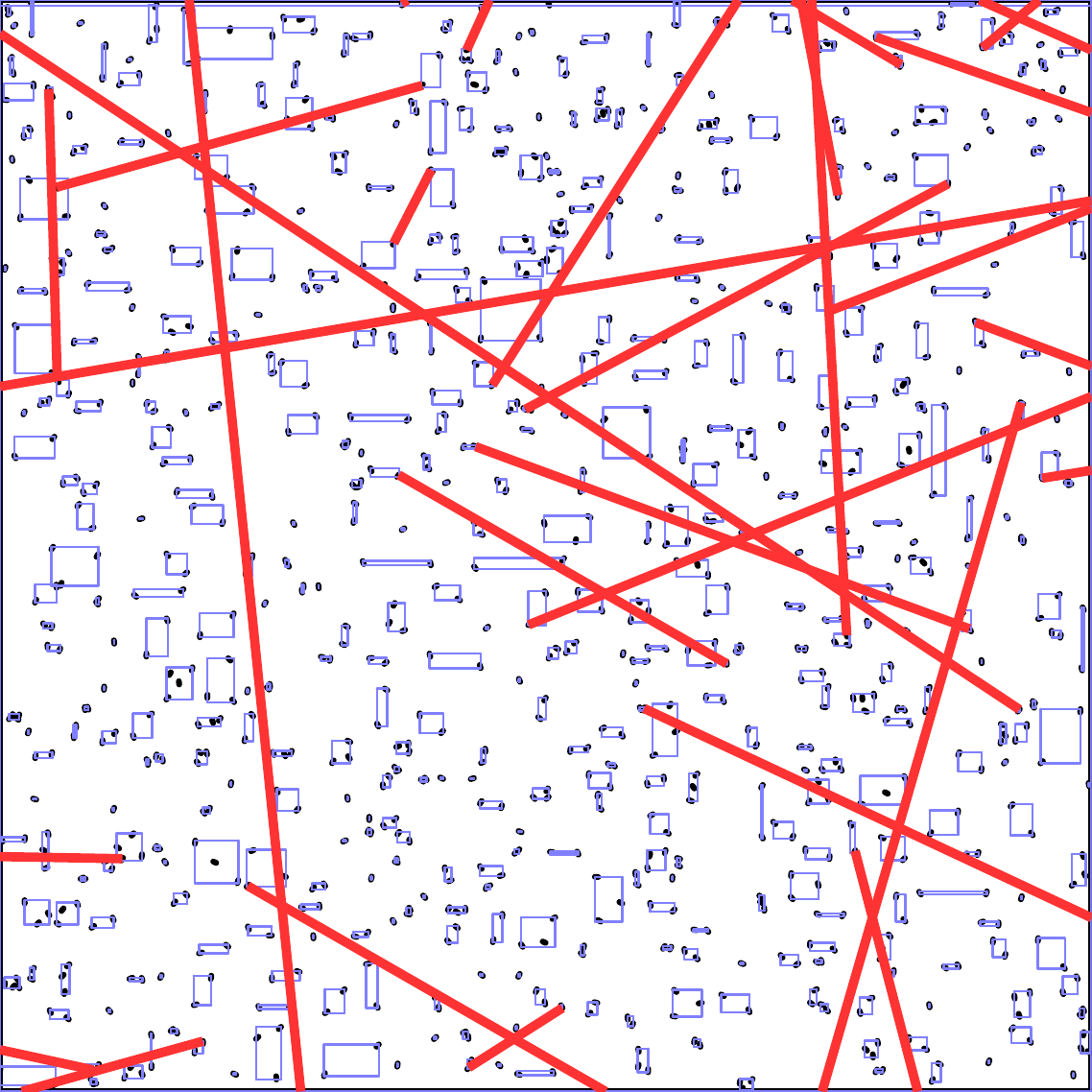}
\accelStructComparisonDlines

\accelStructComparisonA{\linesAccelCompareSize}{\linesAccelCompareSize}
    {Length factor 0.1 -- Diagonal orientation}
    {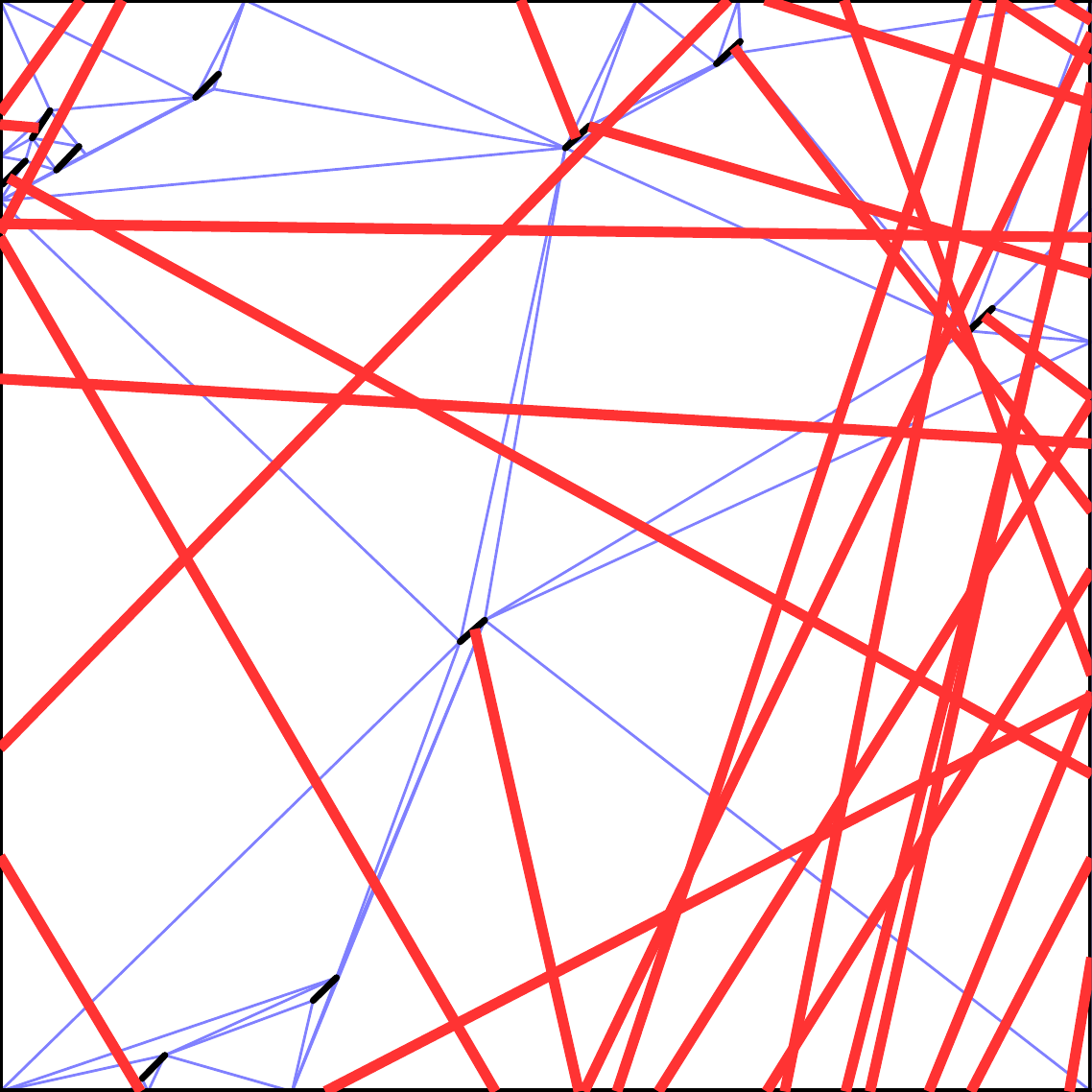}
    {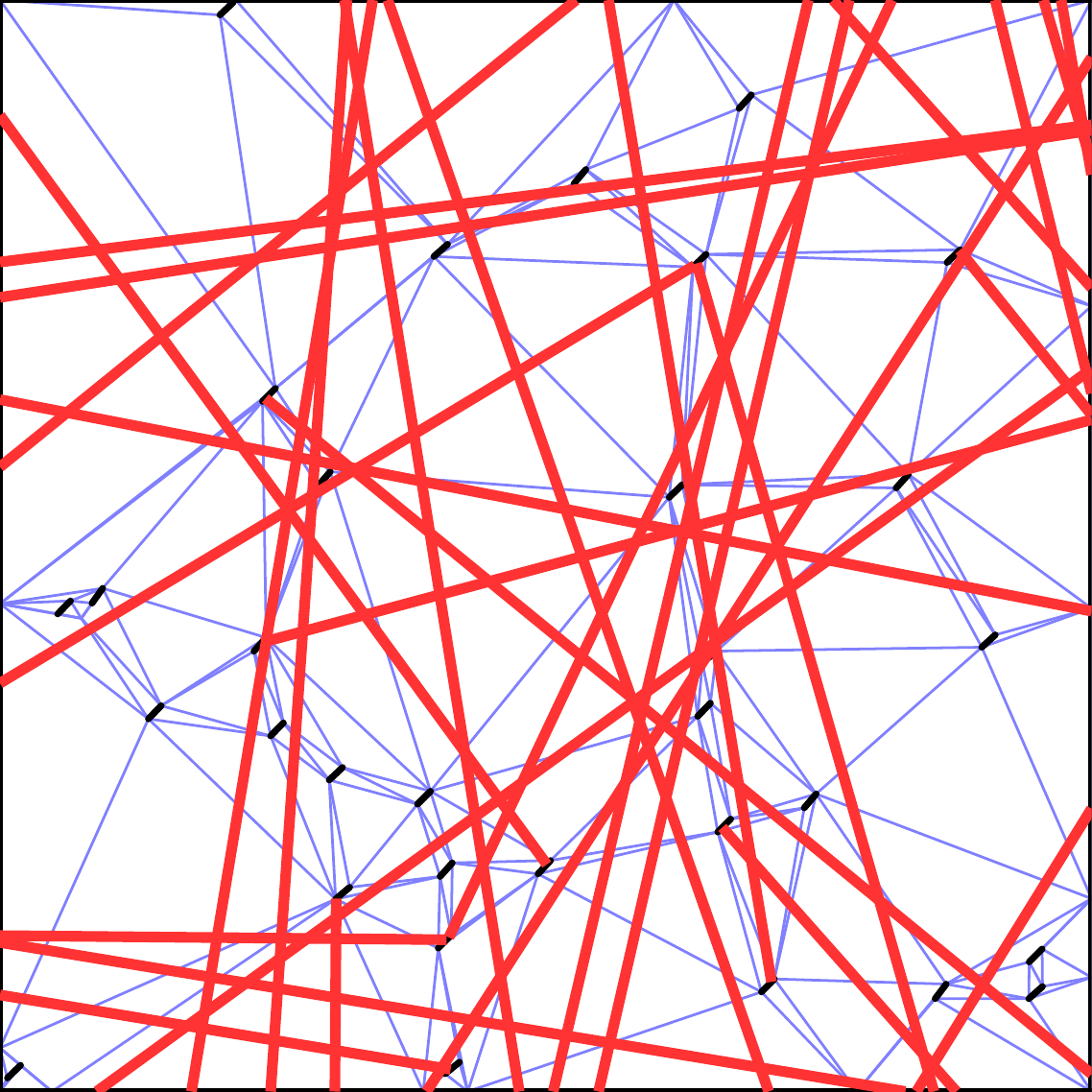}
    {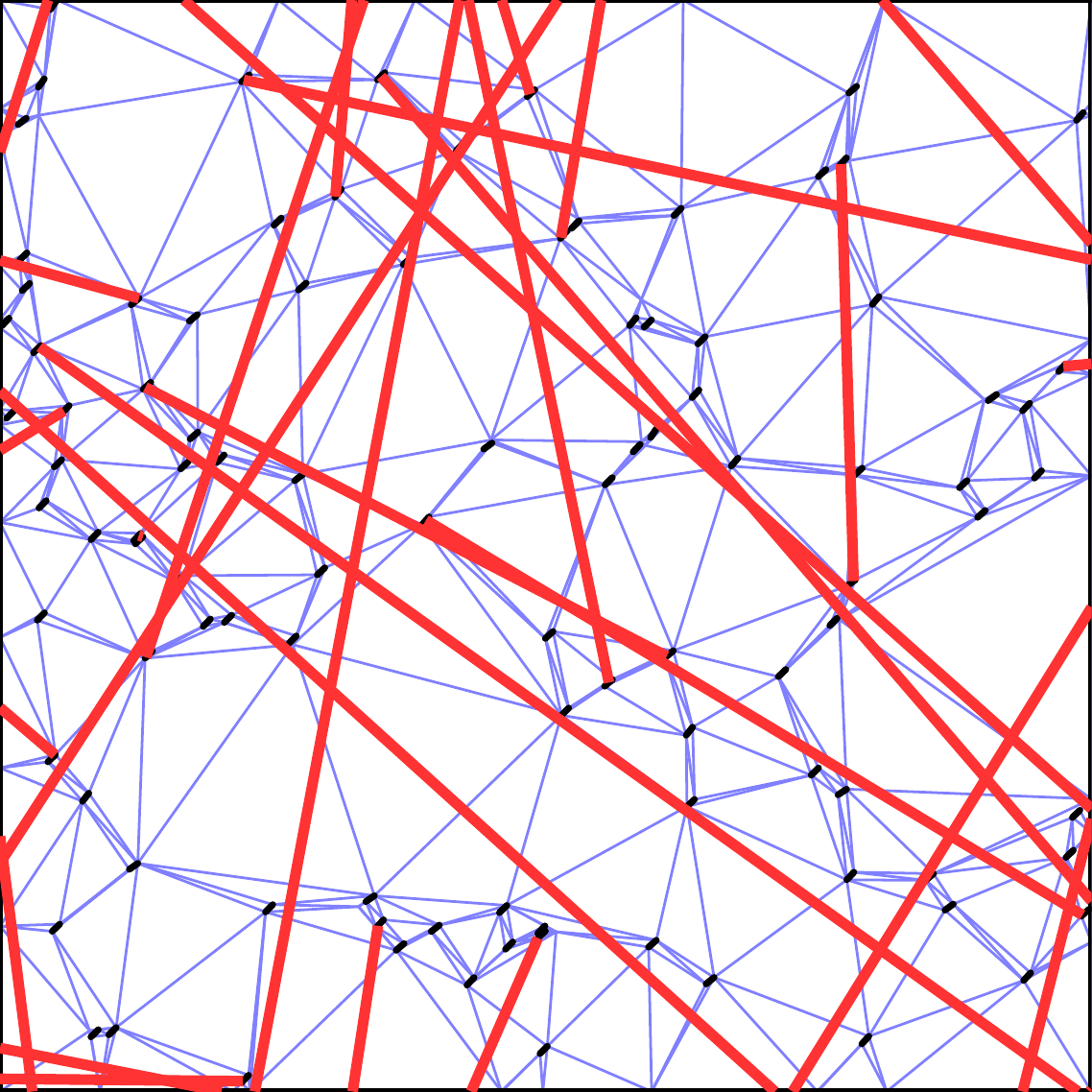}
    {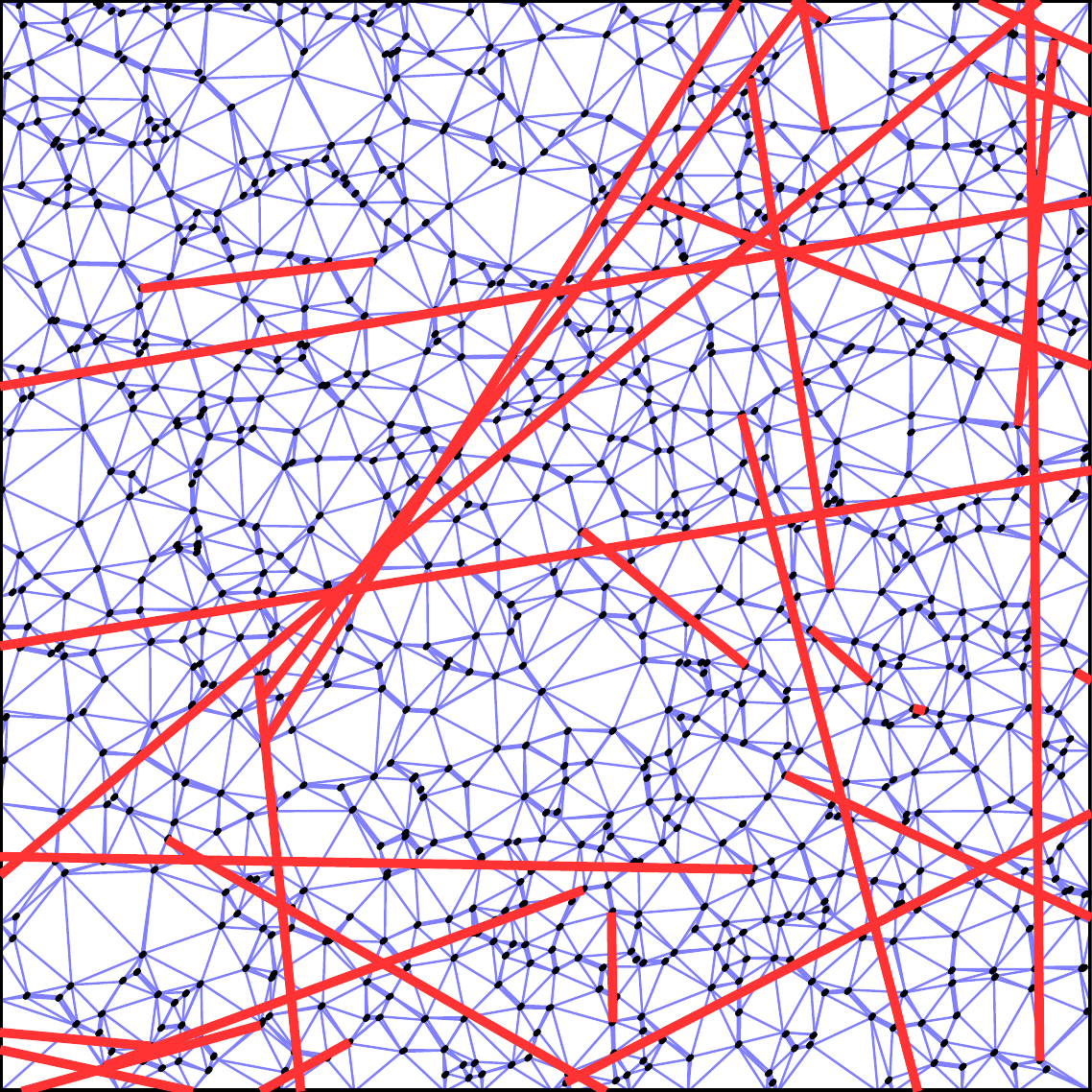}
\accelStructComparisonB{\linesAccelCompareSize}{\linesAccelCompareSize}
    {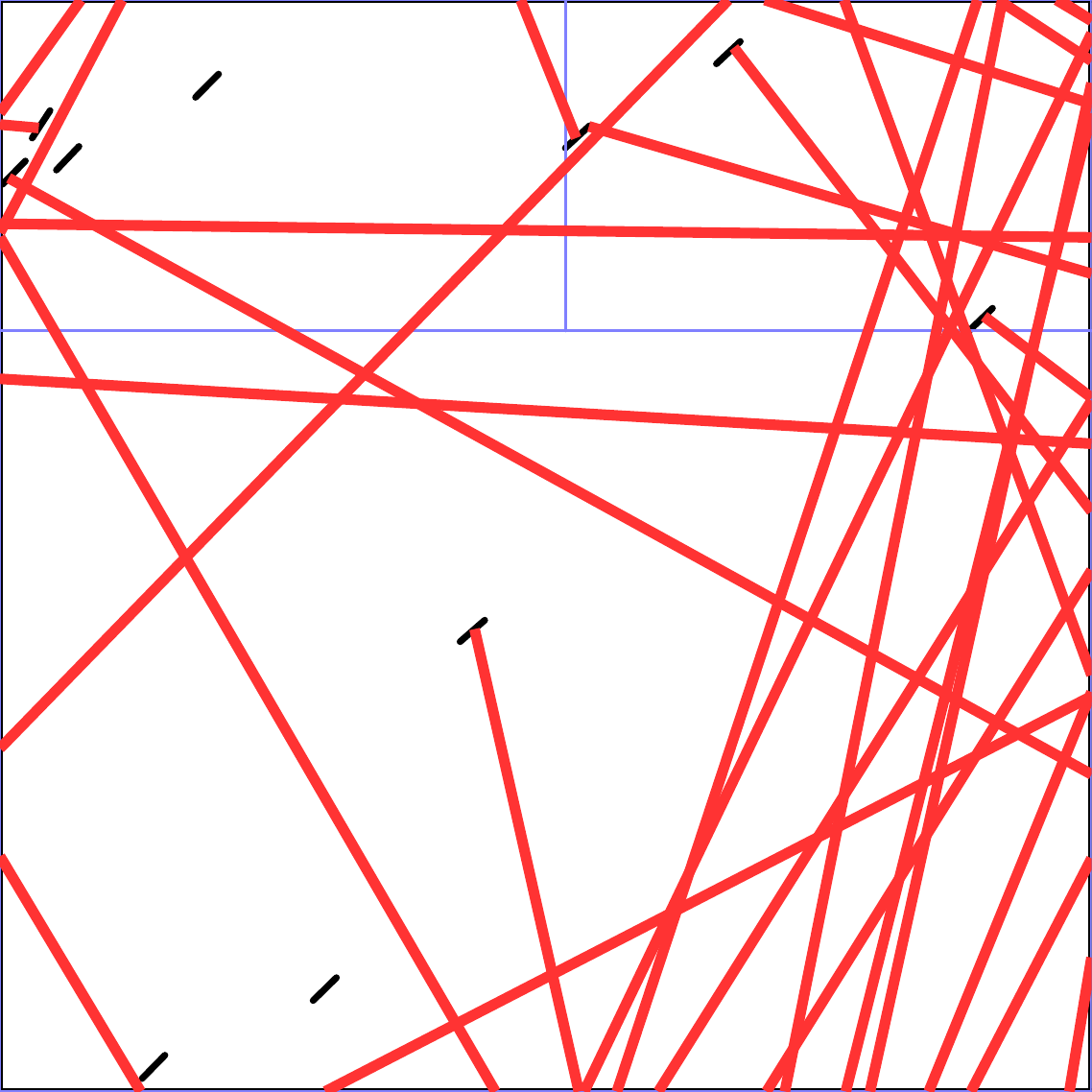}
    {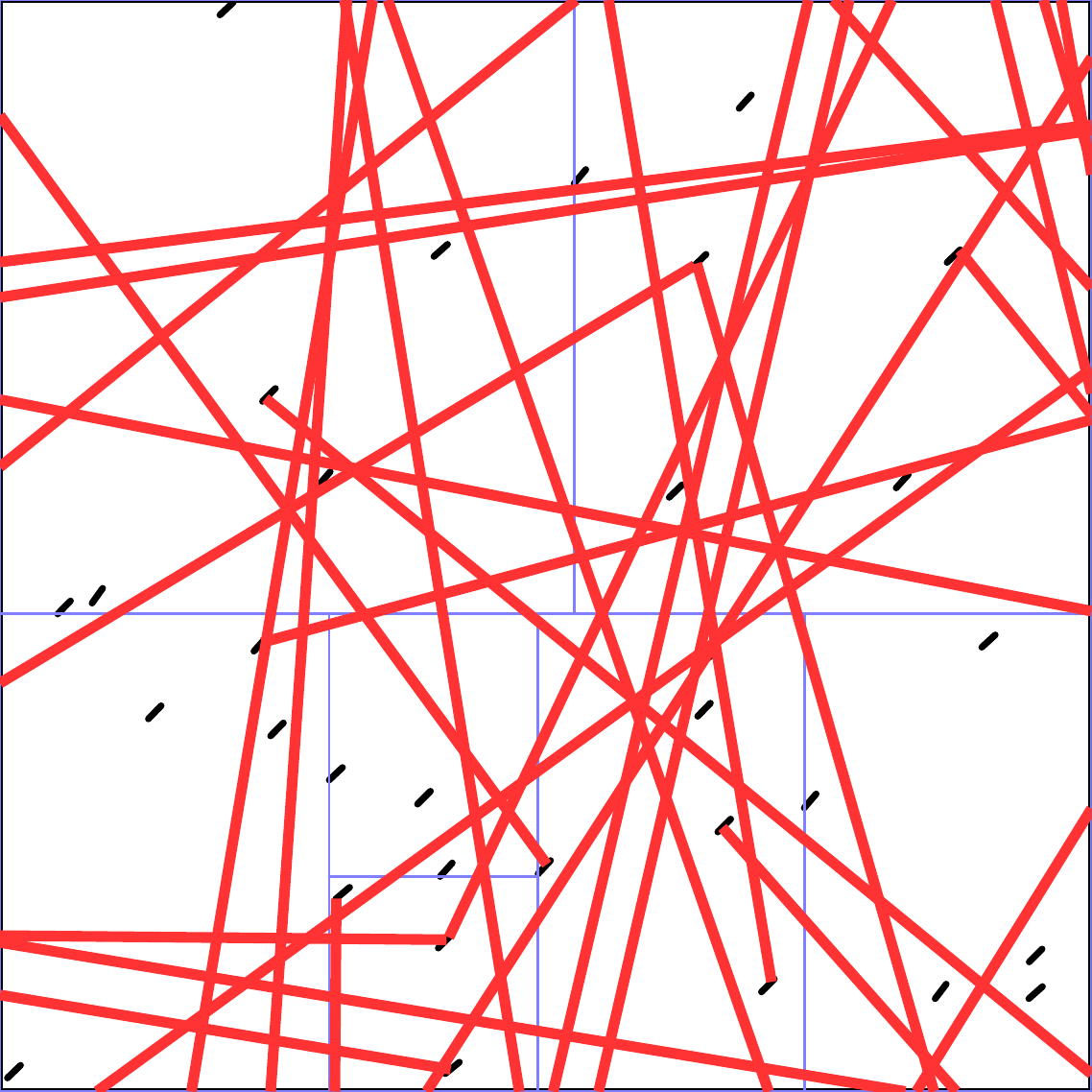}
    {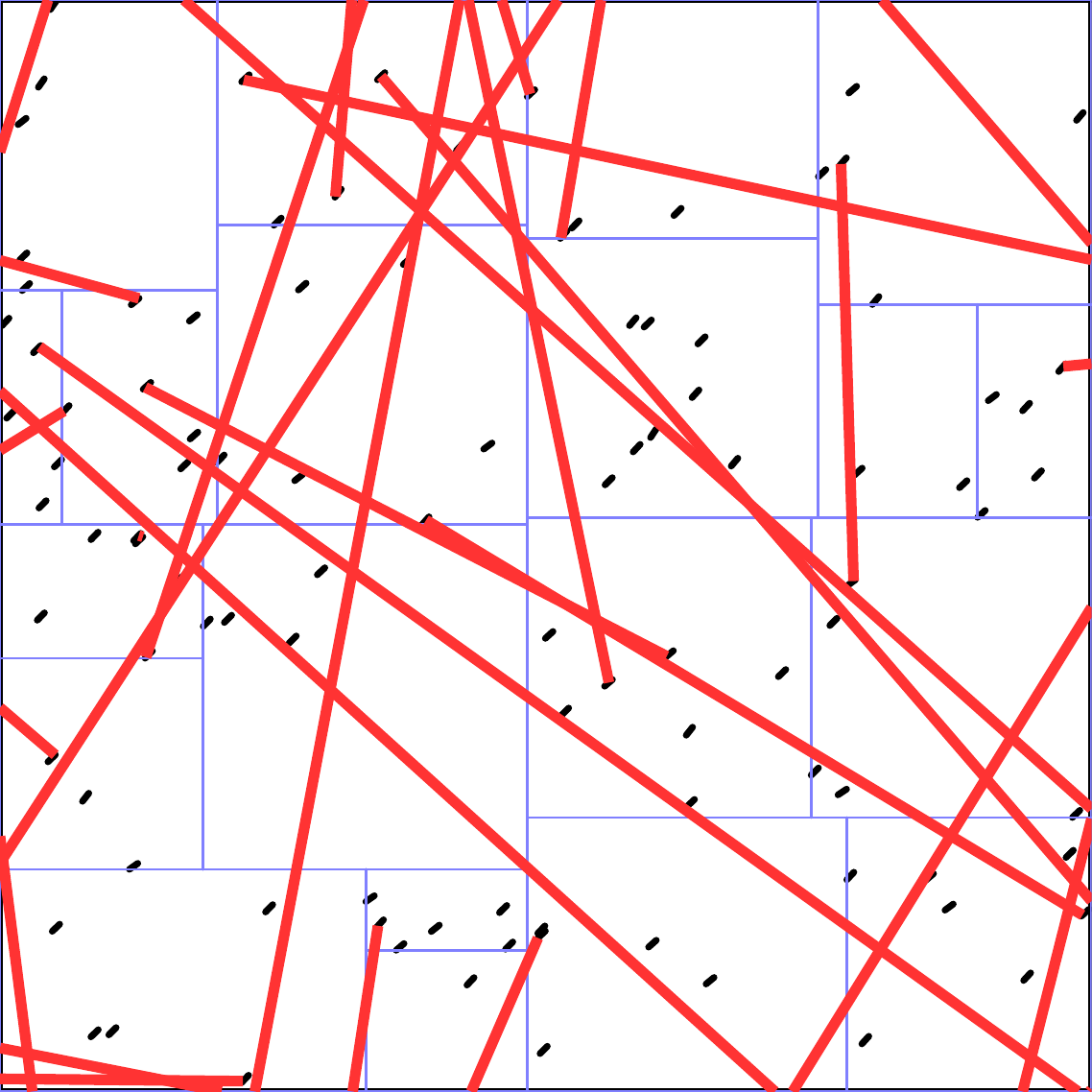}
    {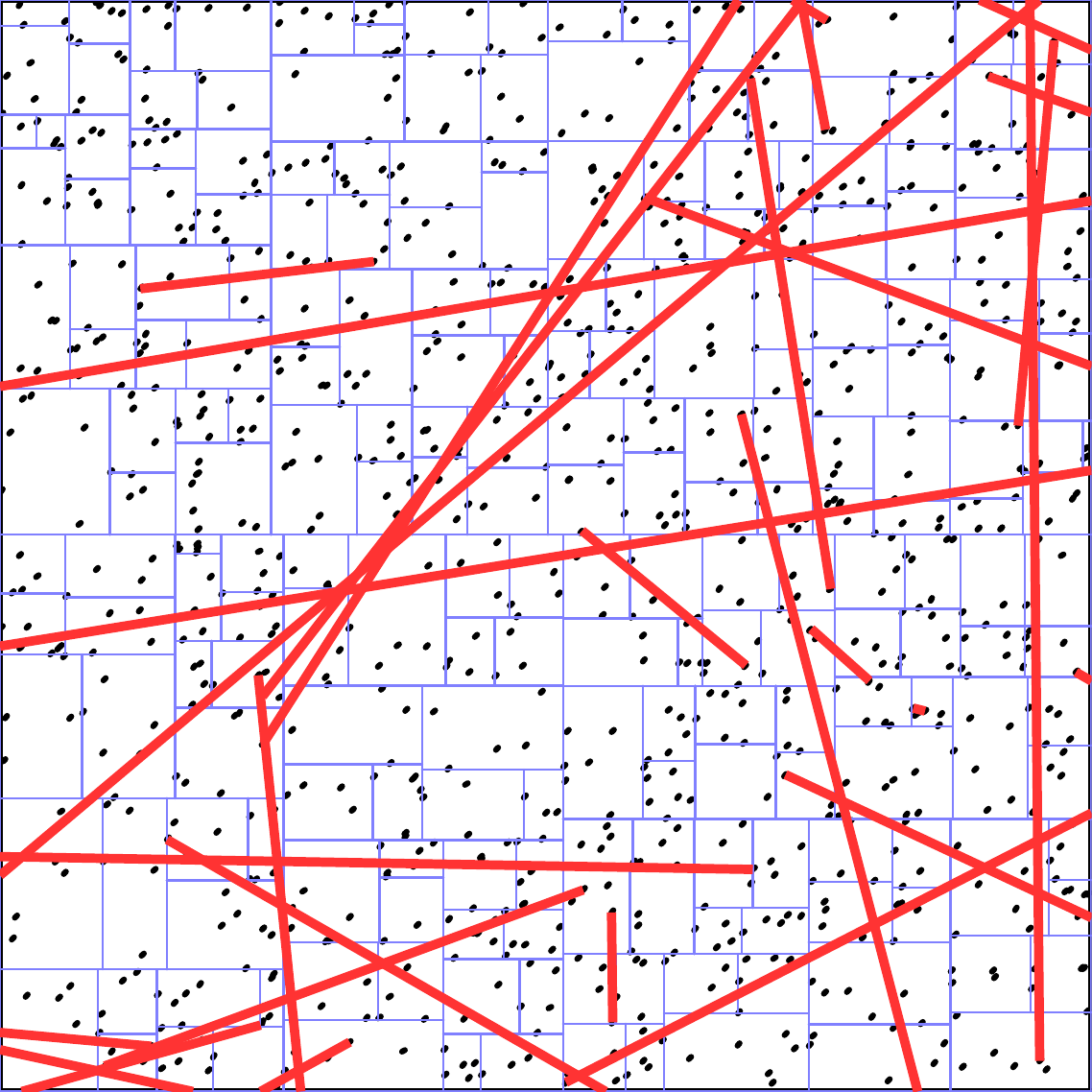}
\accelStructComparisonC{\linesAccelCompareSize}{\linesAccelCompareSize}
    {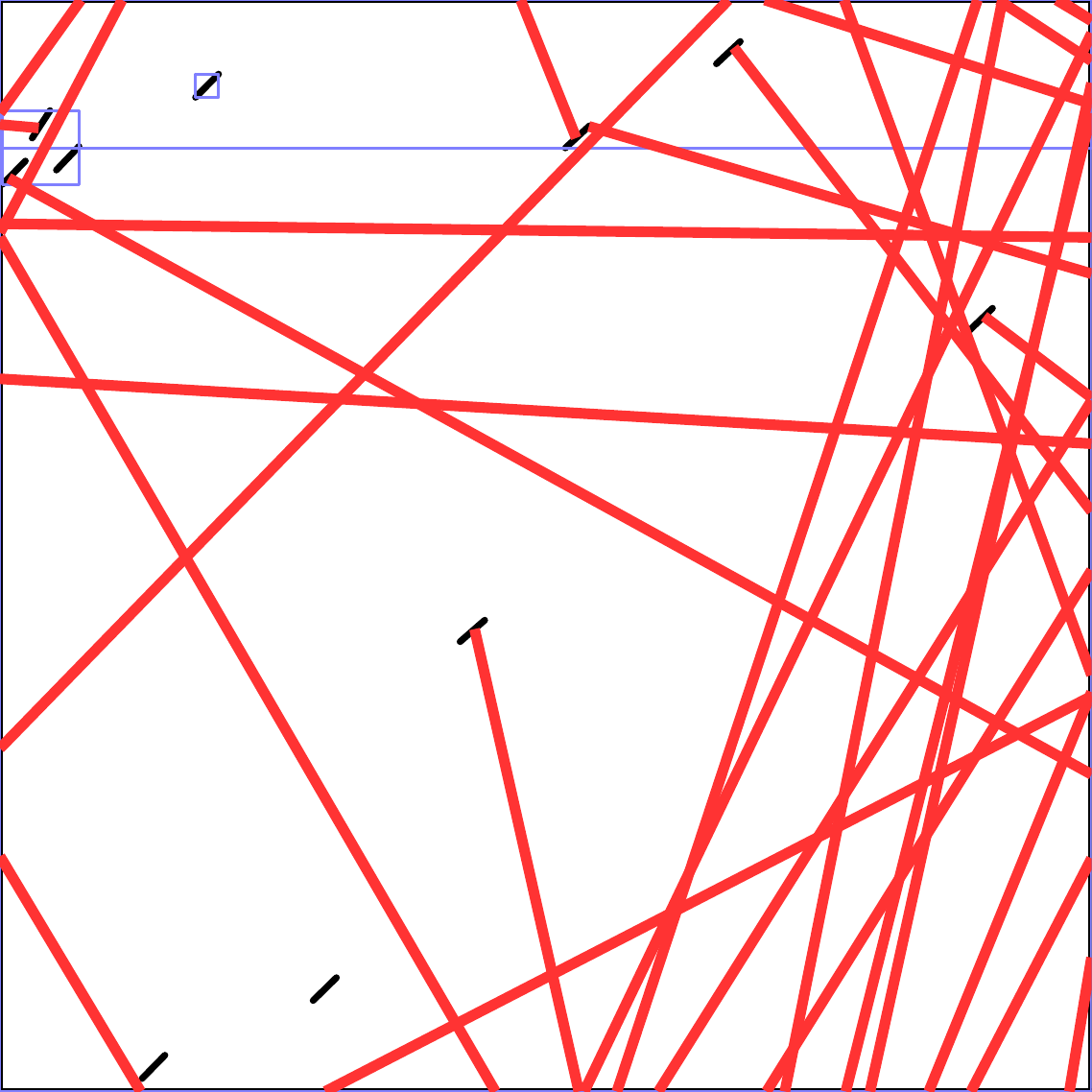}
    {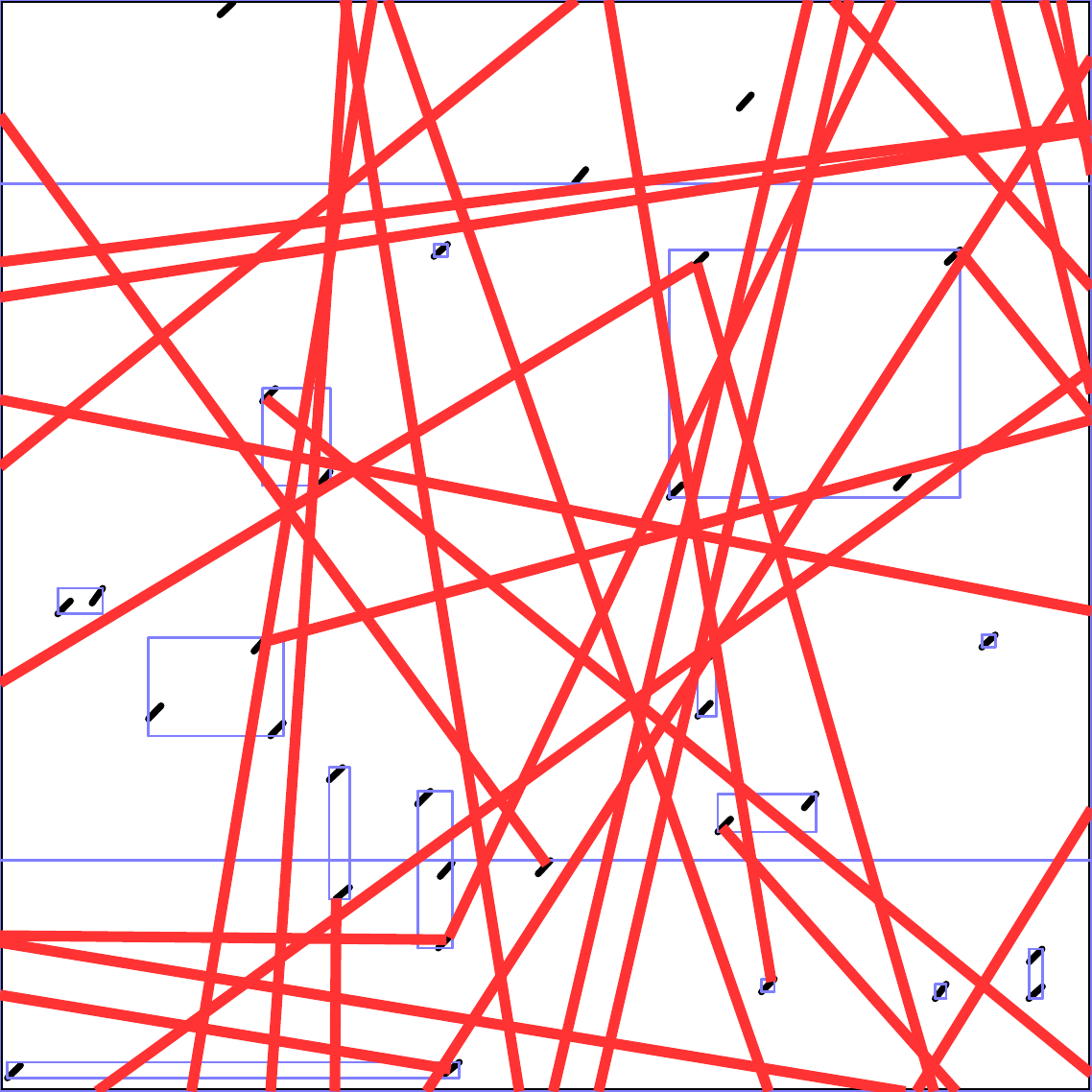}
    {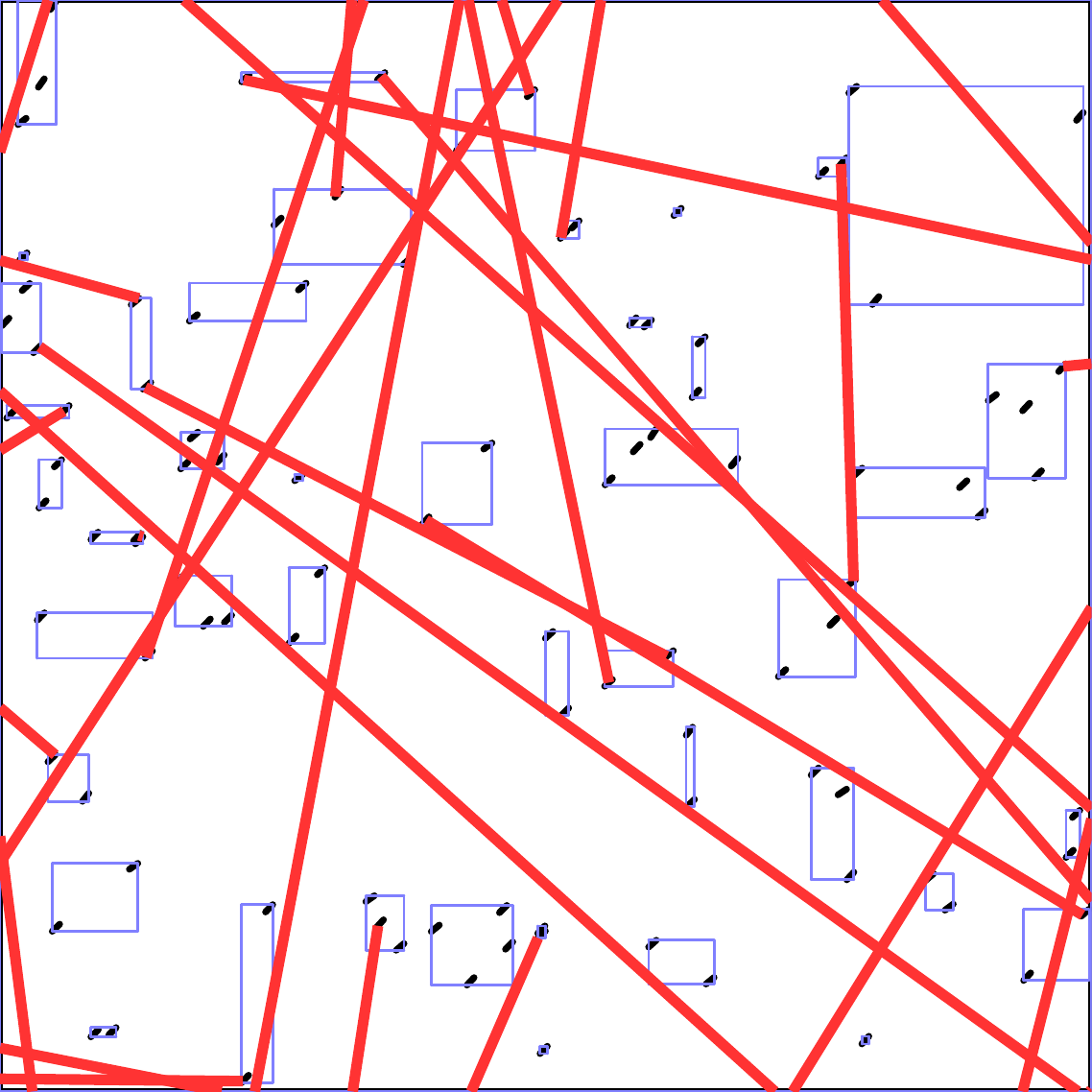}
    {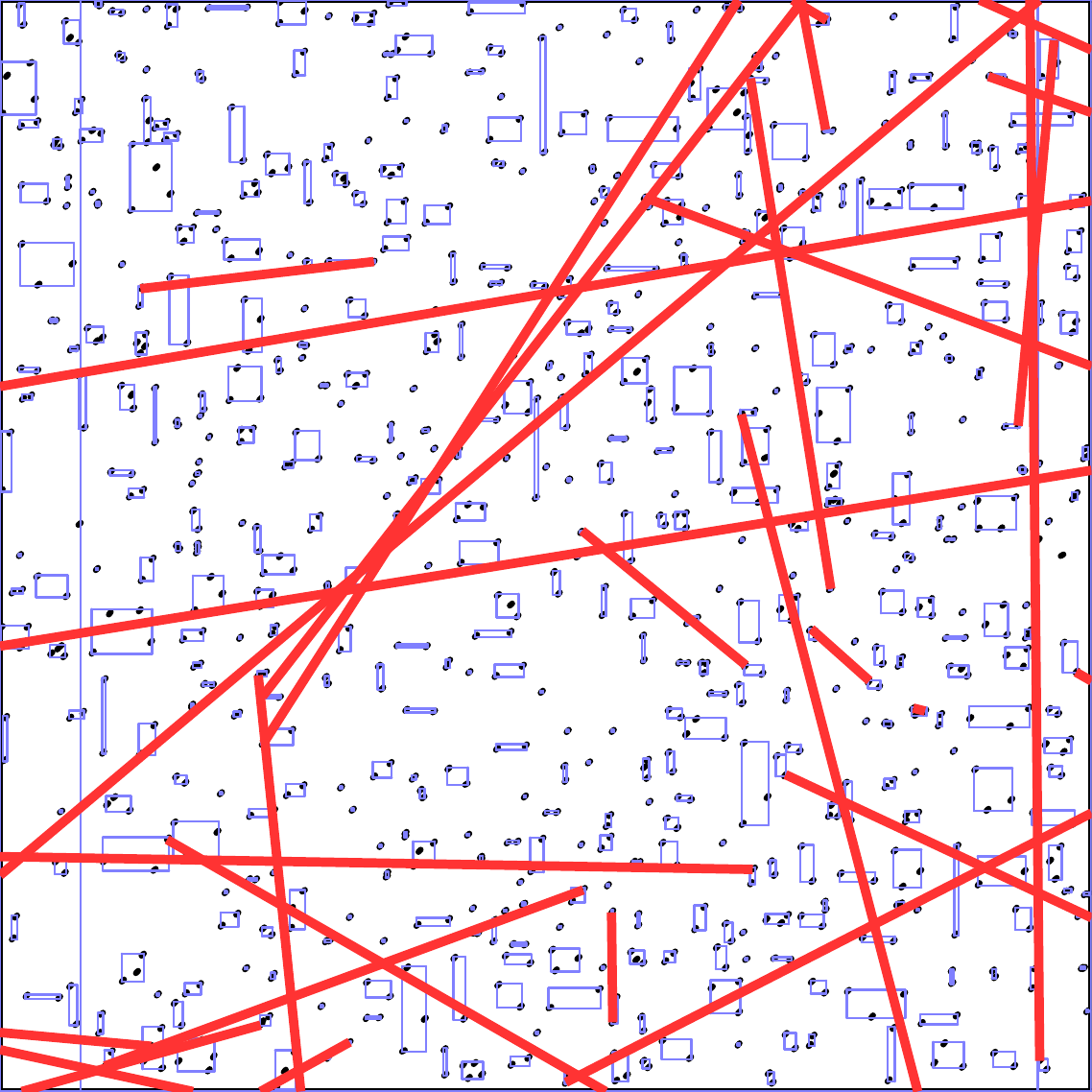}
\accelStructComparisonDlines

\accelStructComparisonA{\linesAccelCompareSize}{\linesAccelCompareSize}
    {Length factor 1 -- Vertical orientation}
    {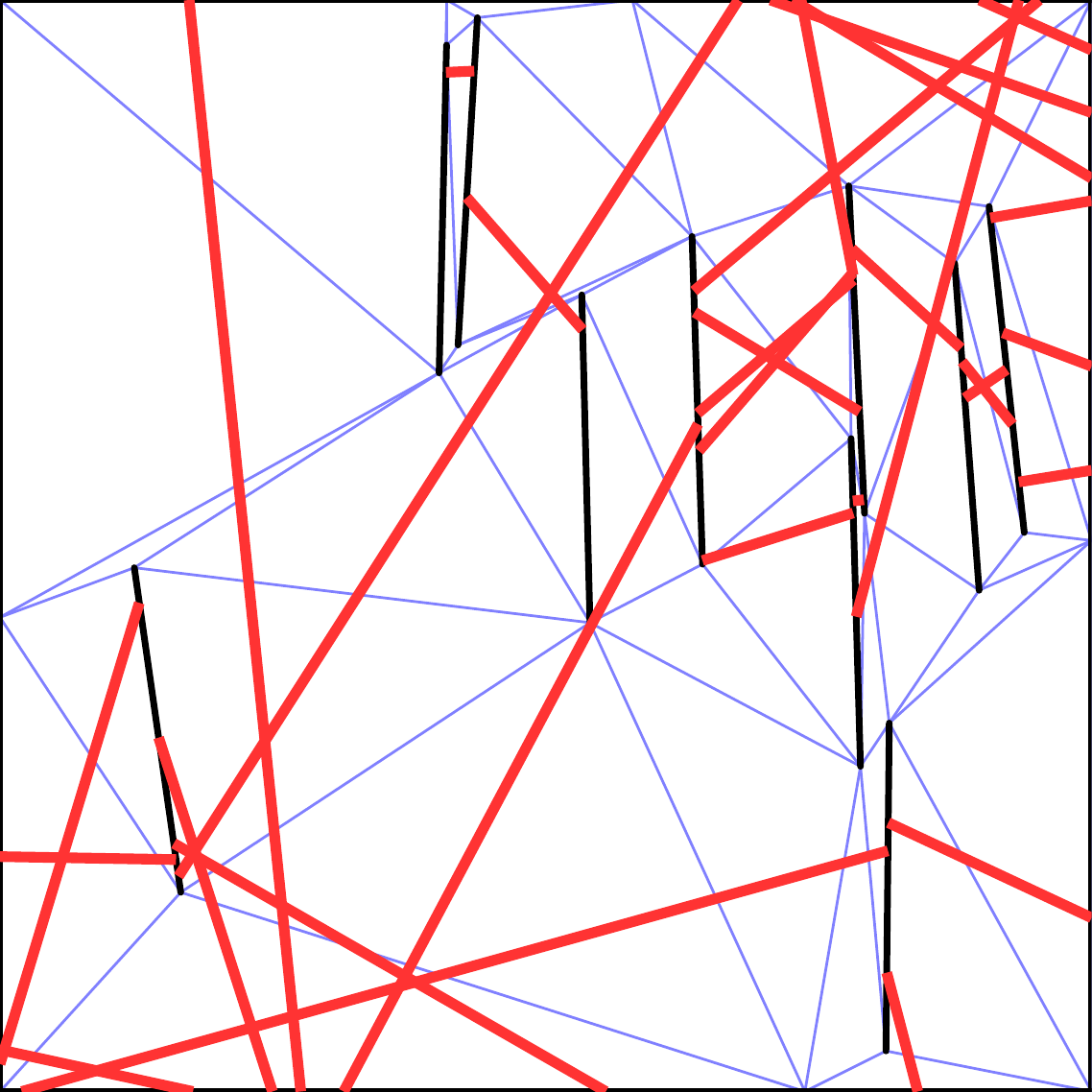}
    {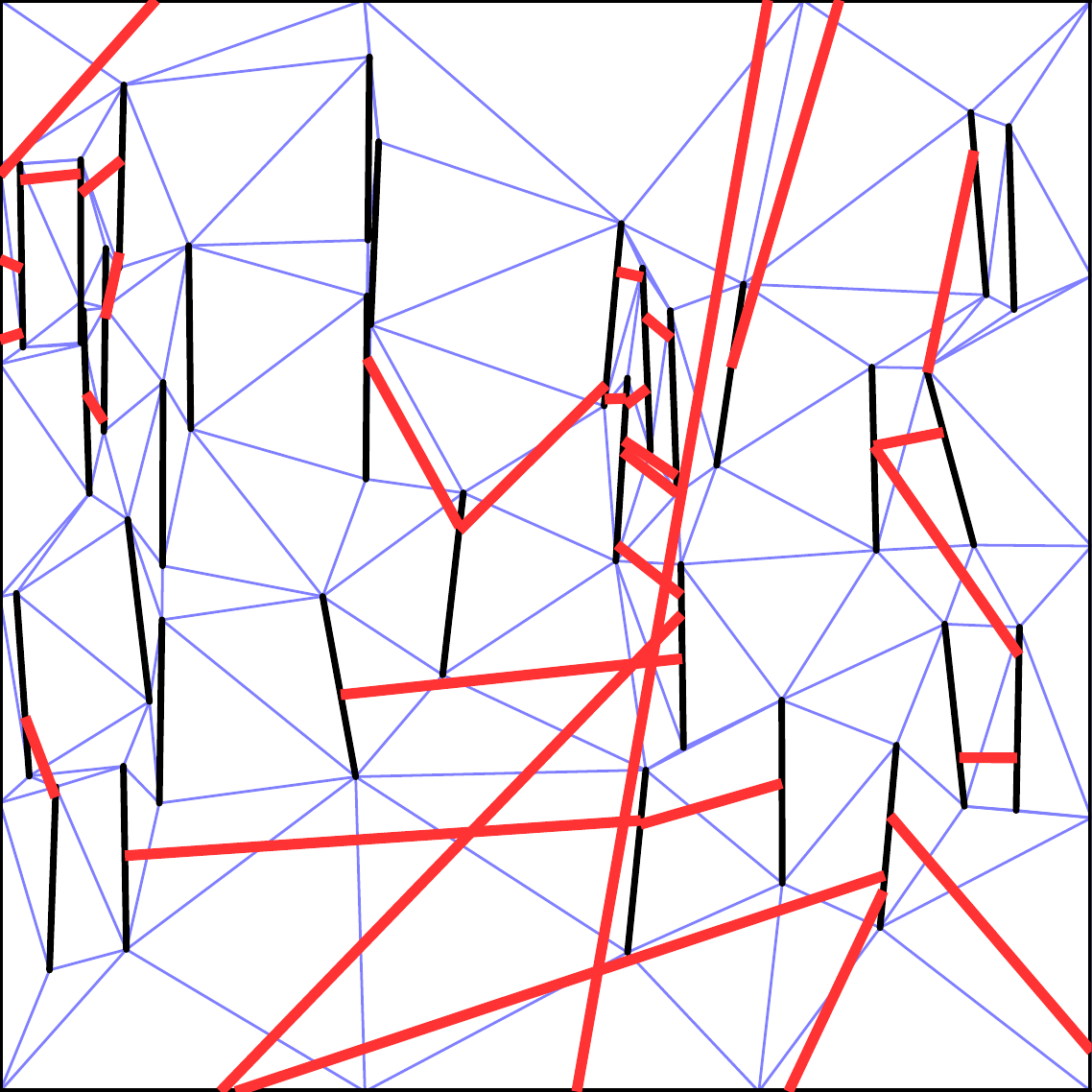}
    {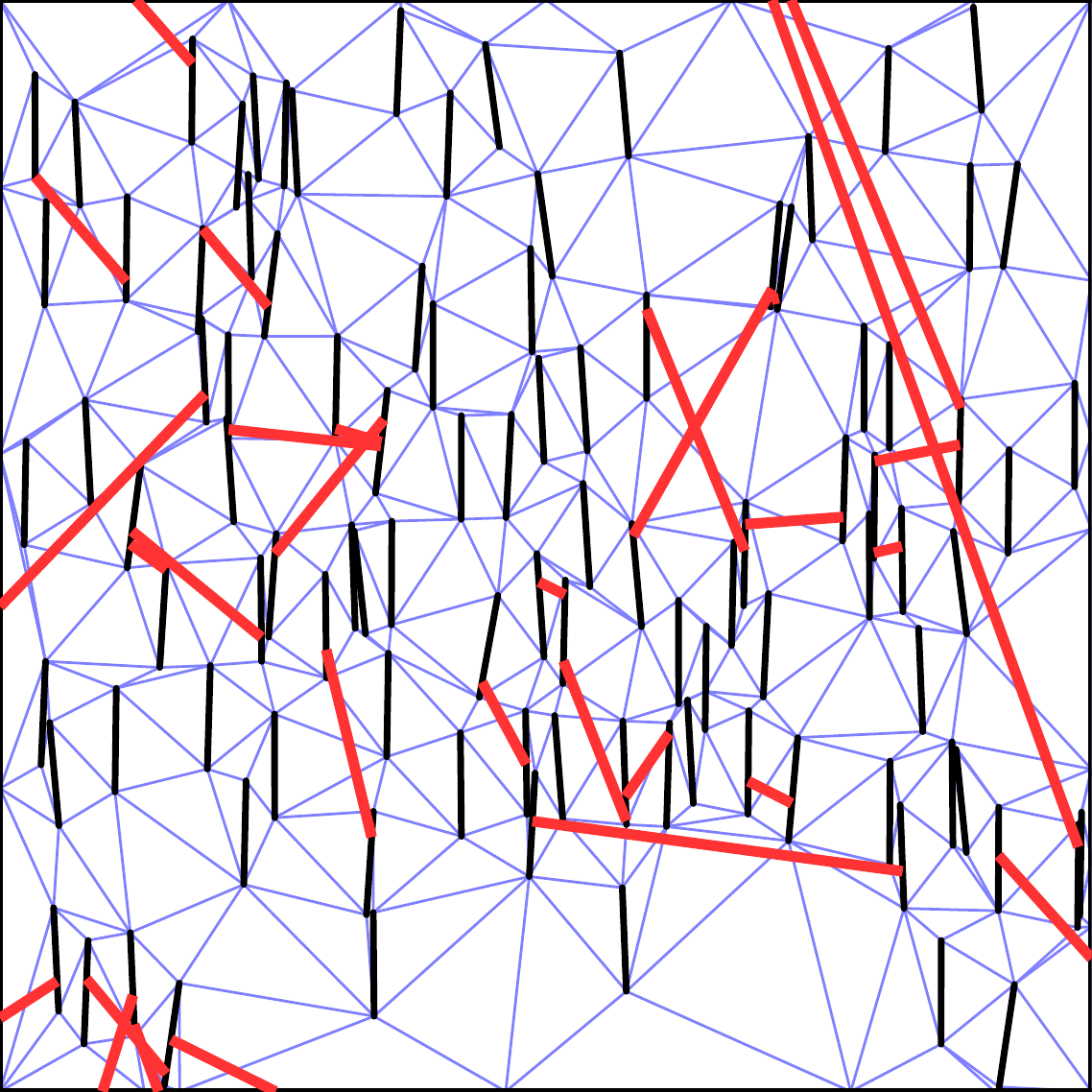}
    {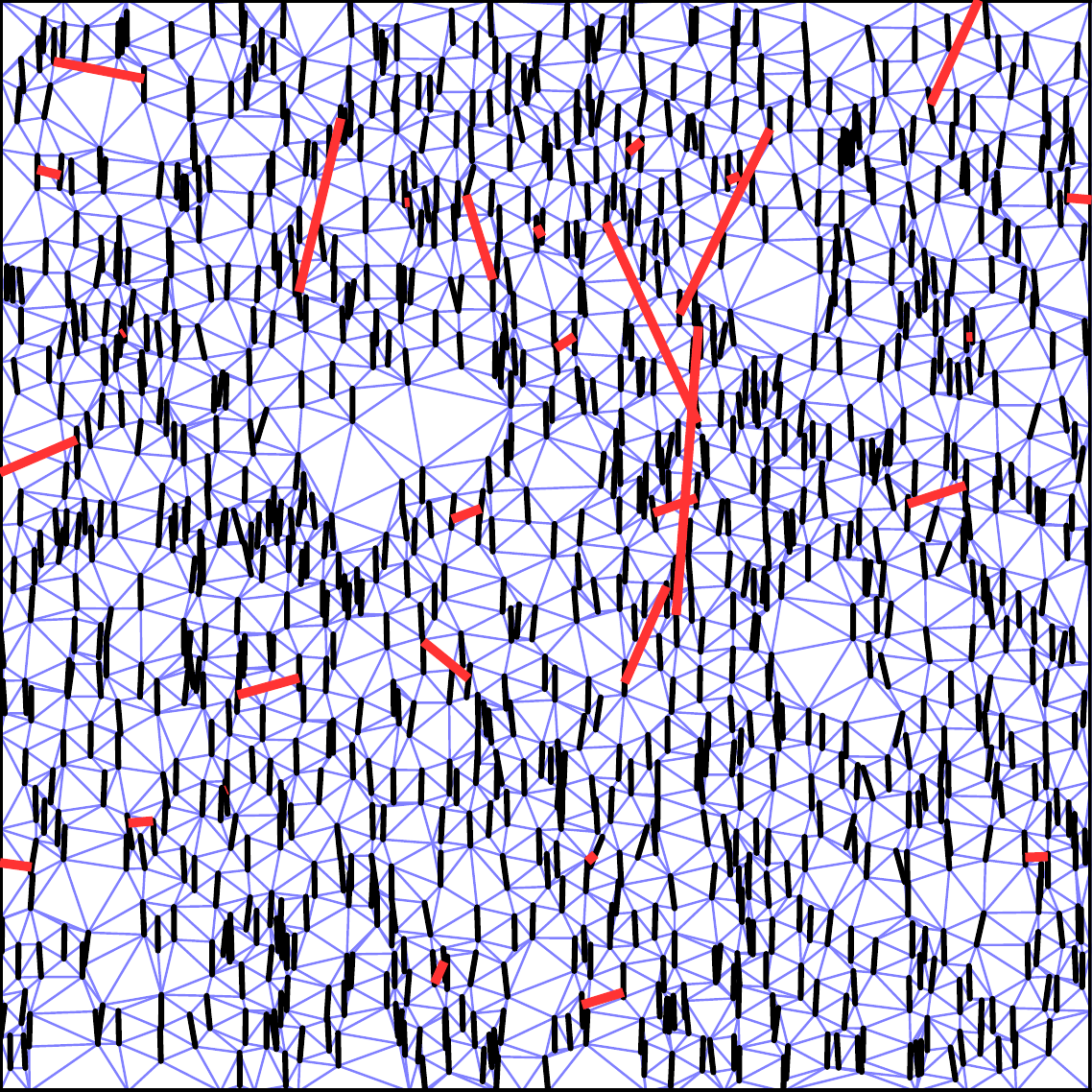}
\accelStructComparisonB{\linesAccelCompareSize}{\linesAccelCompareSize}
    {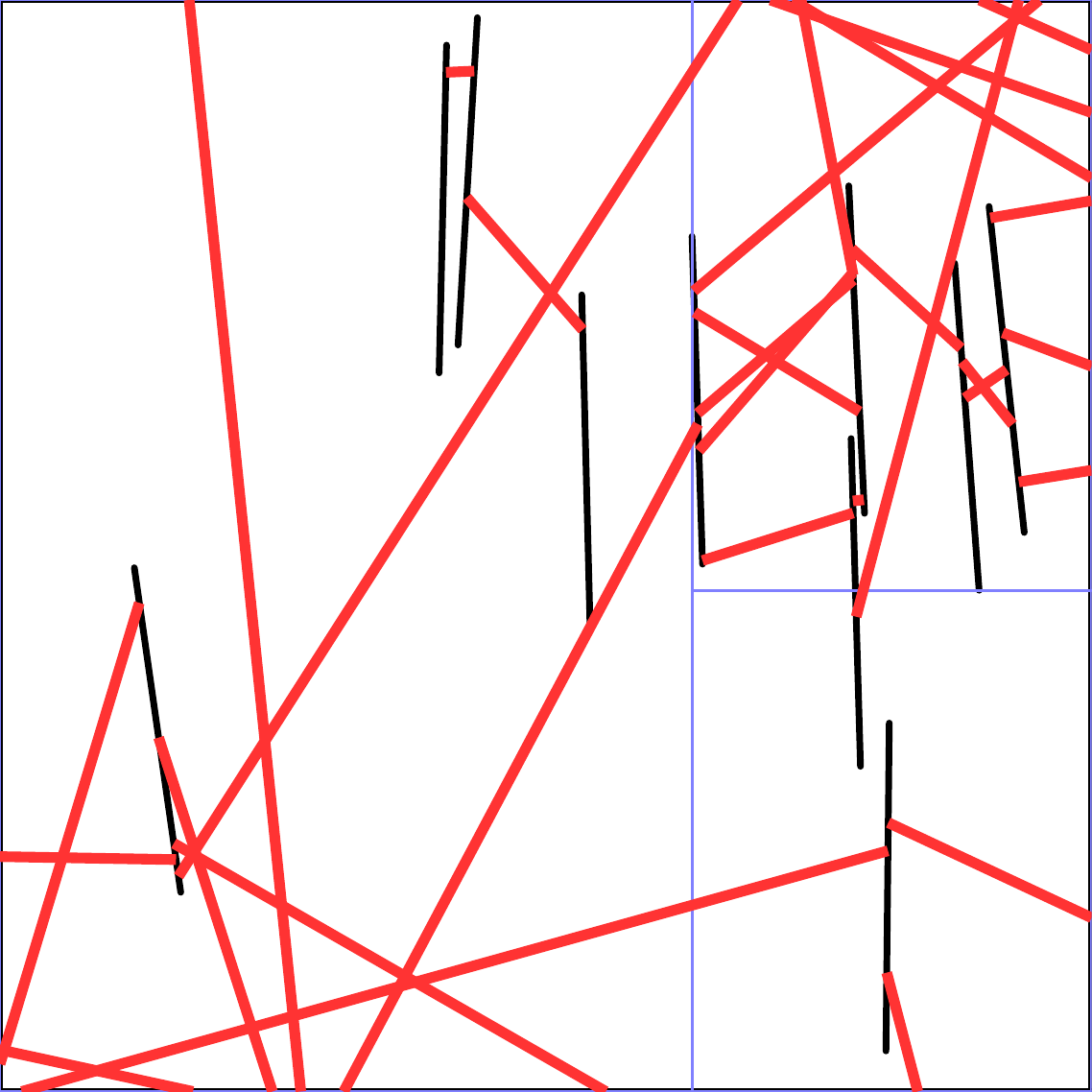}
    {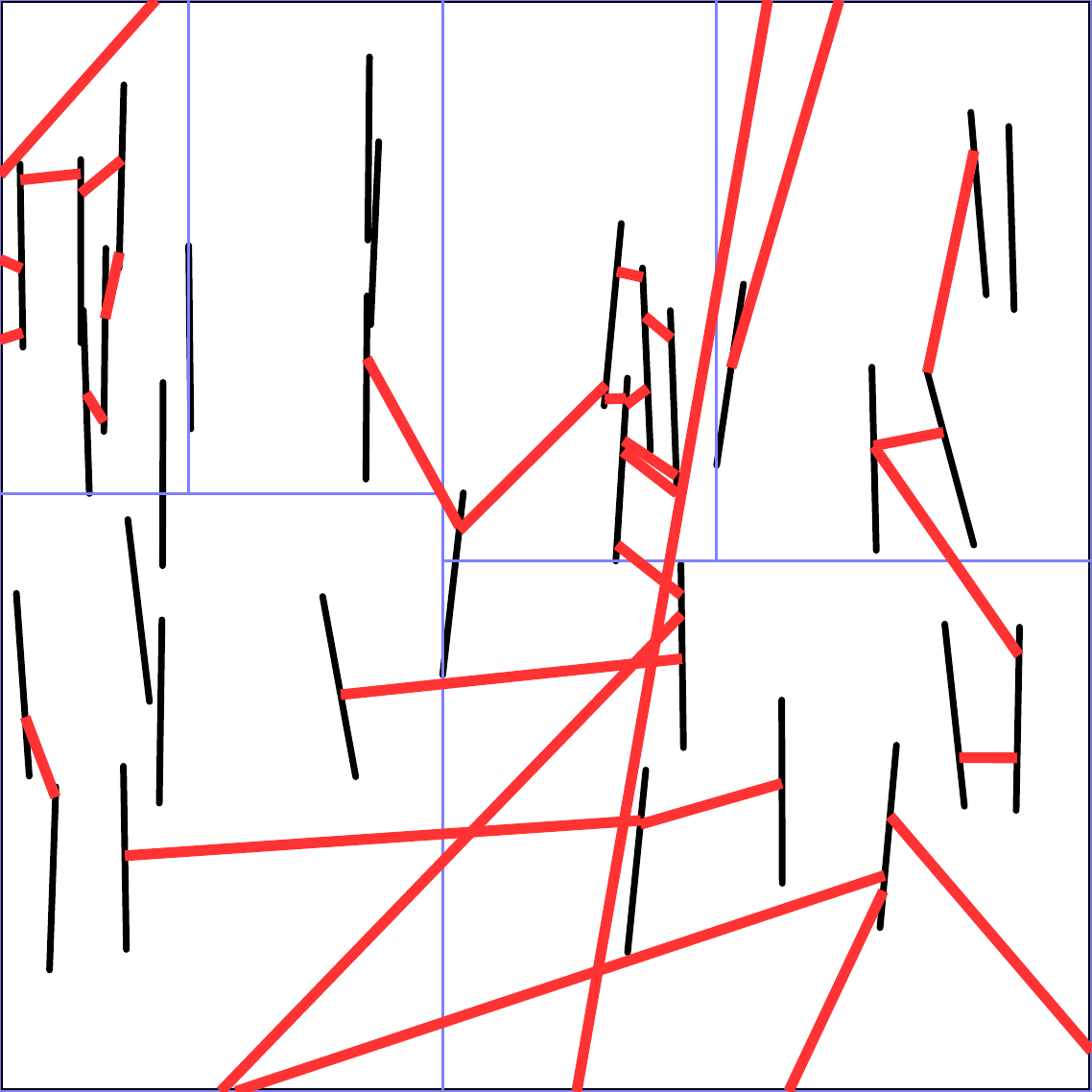}
    {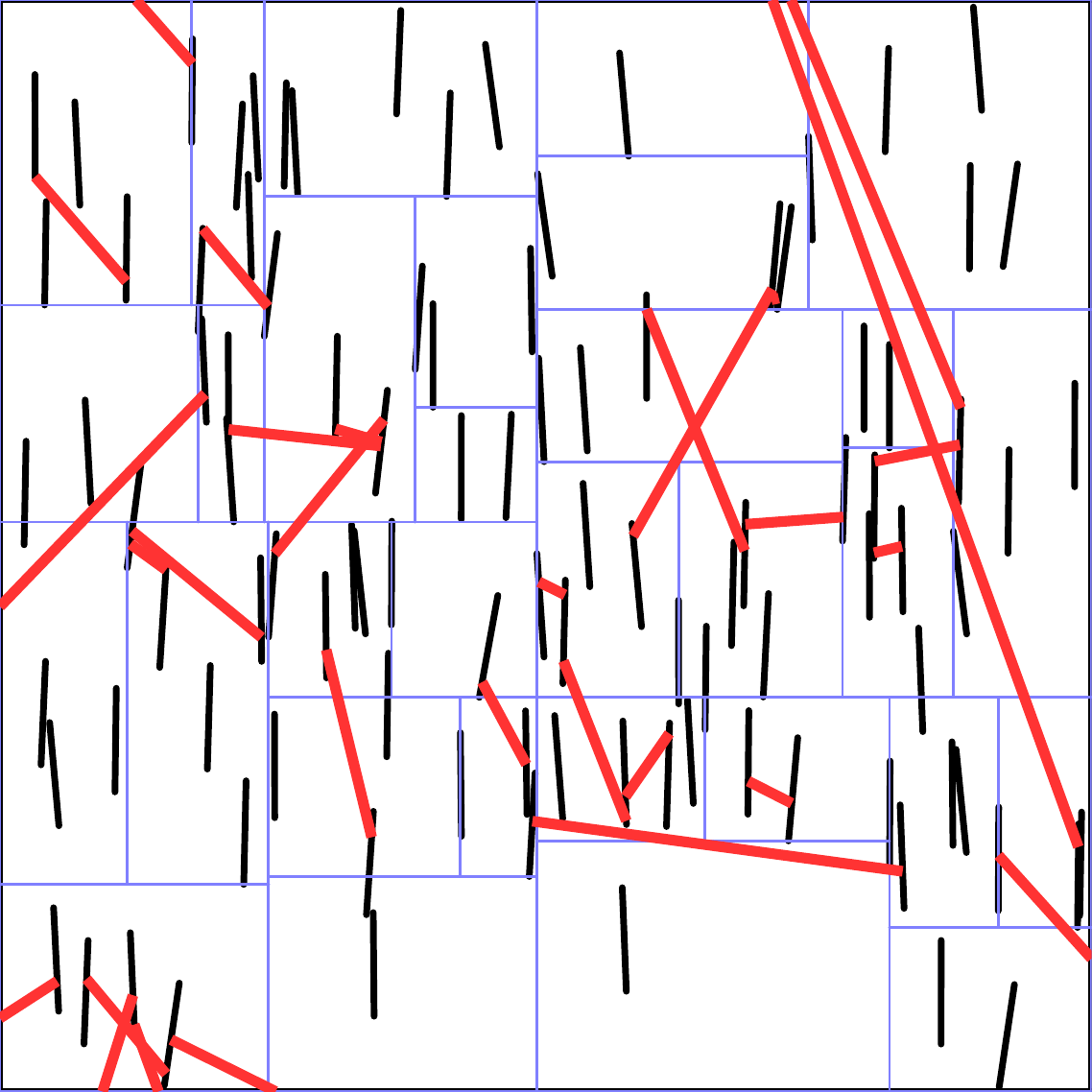}
    {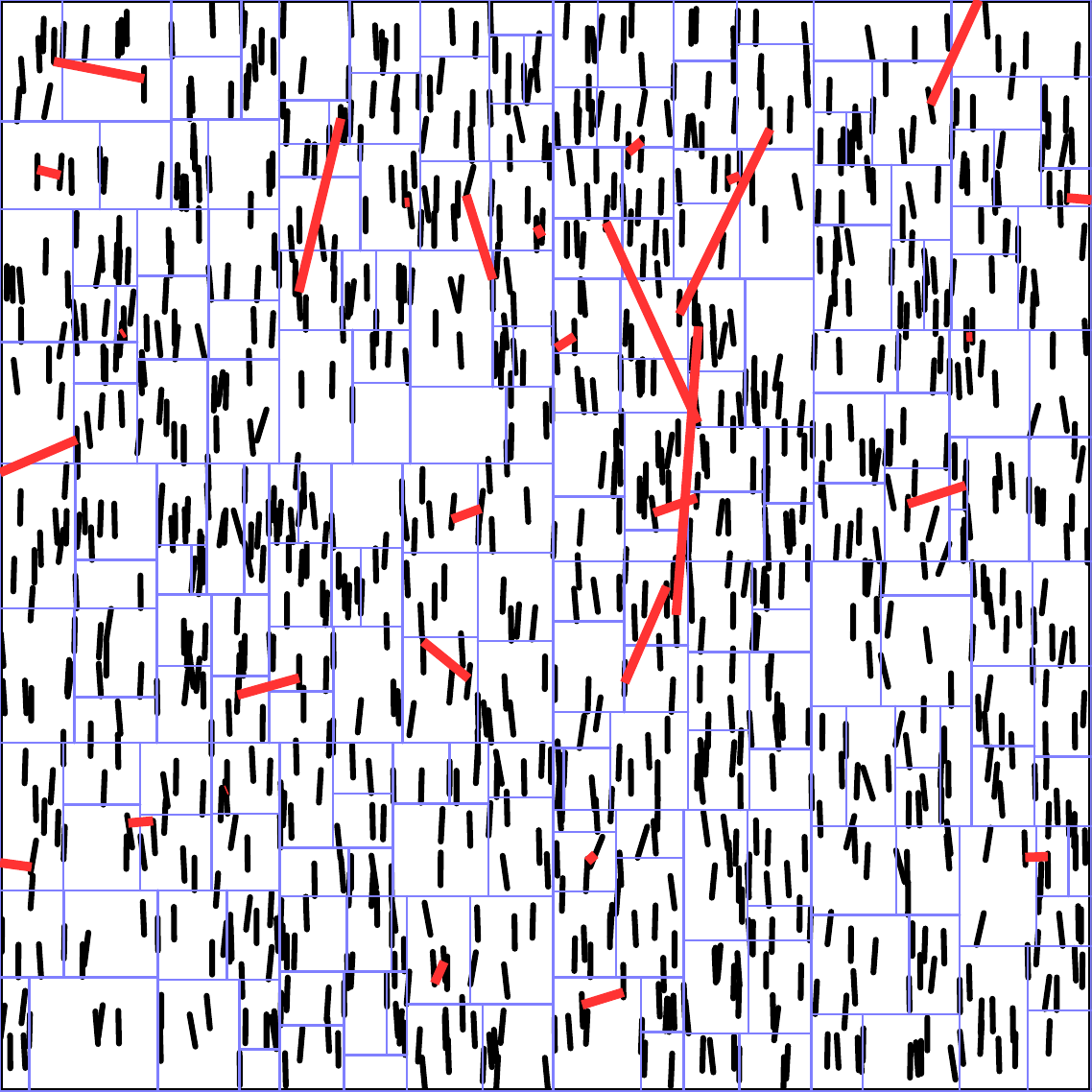}
\accelStructComparisonC{\linesAccelCompareSize}{\linesAccelCompareSize}
    {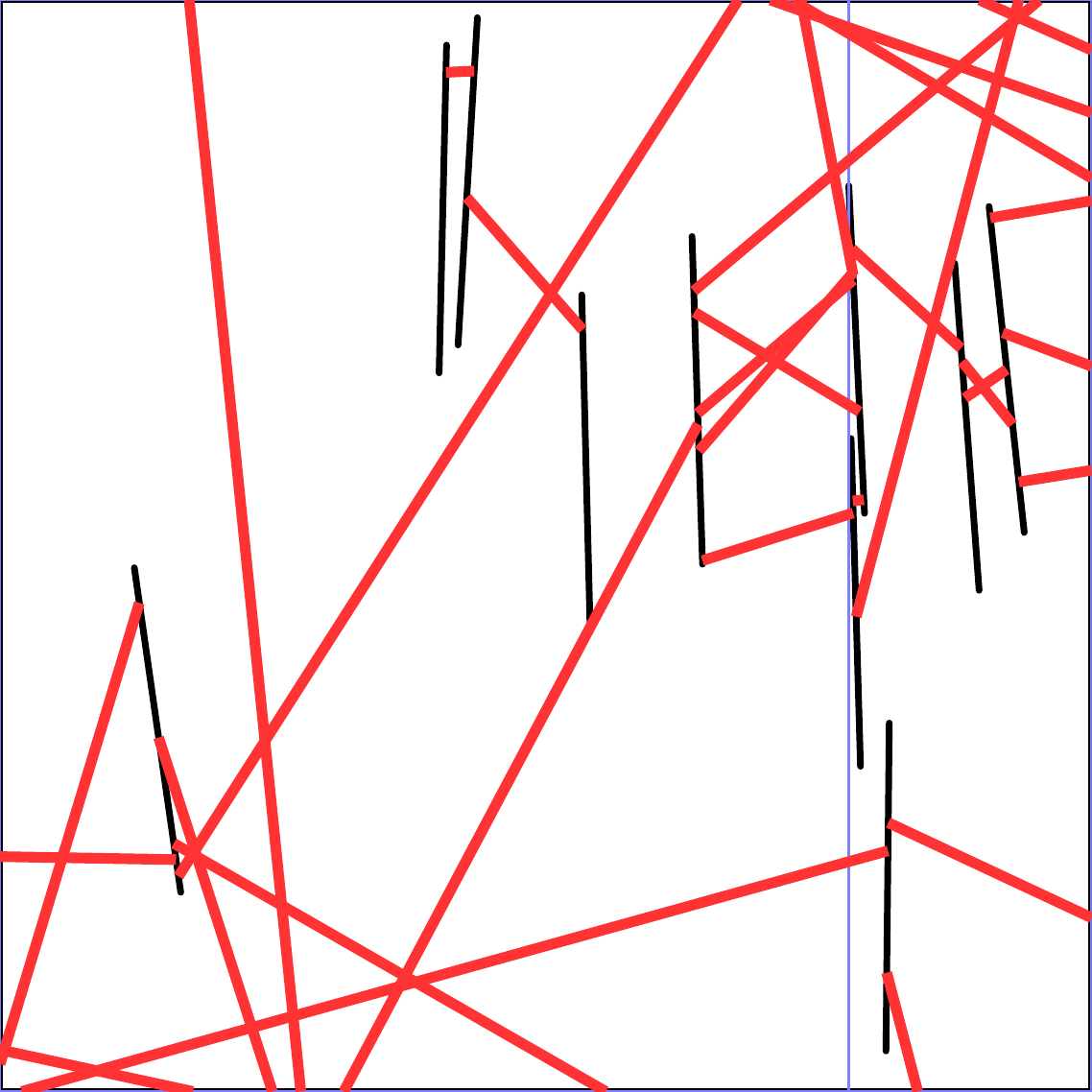}
    {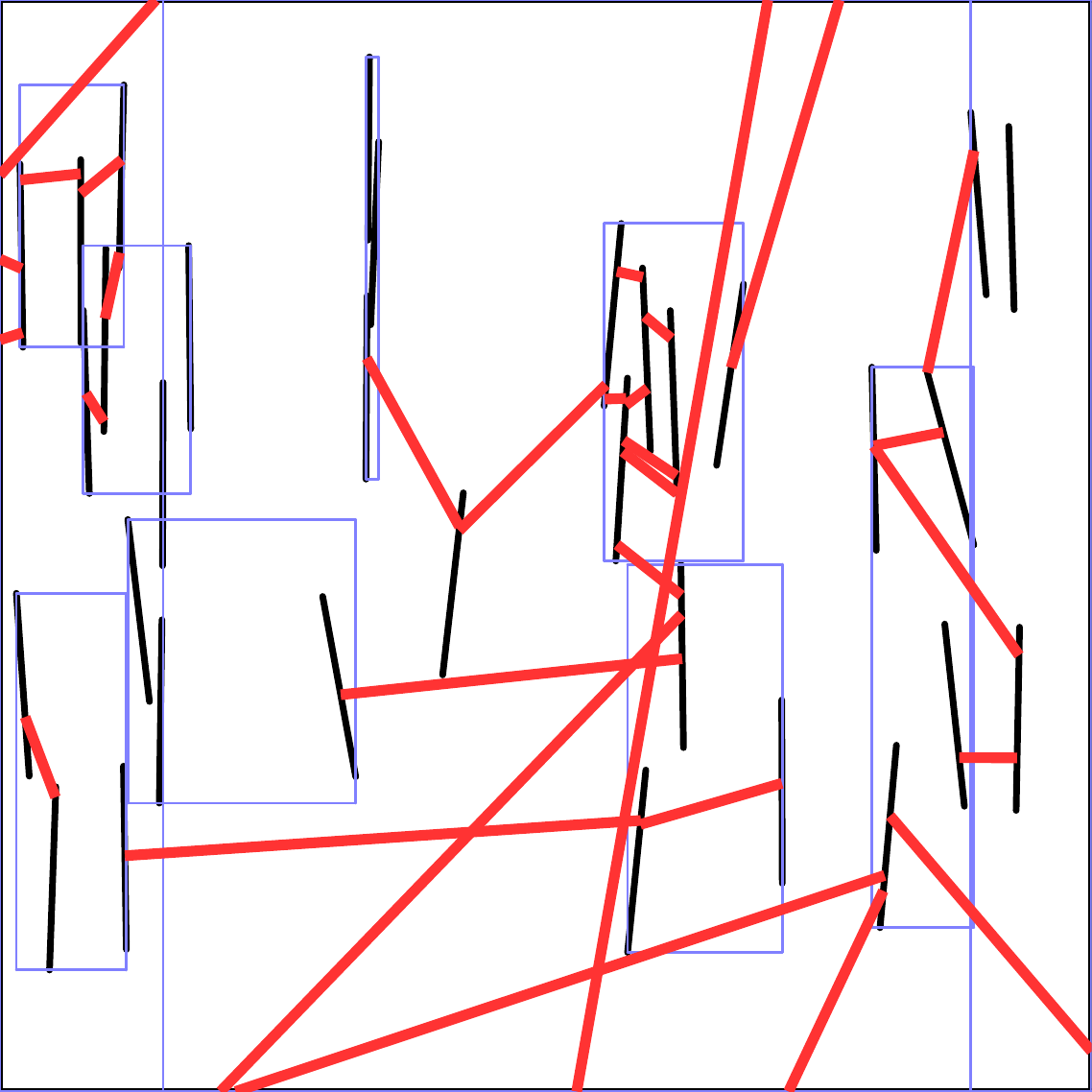}
    {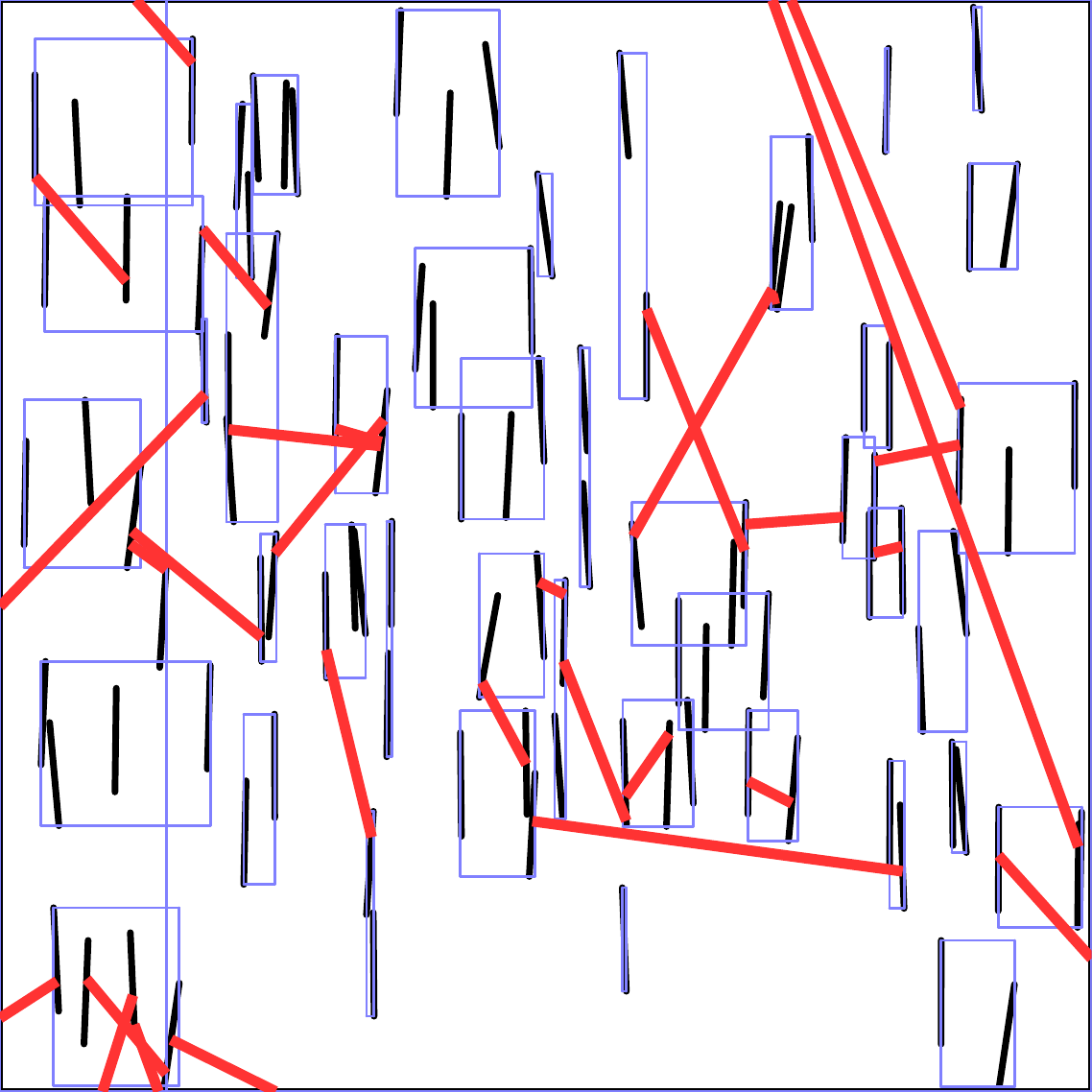}
    {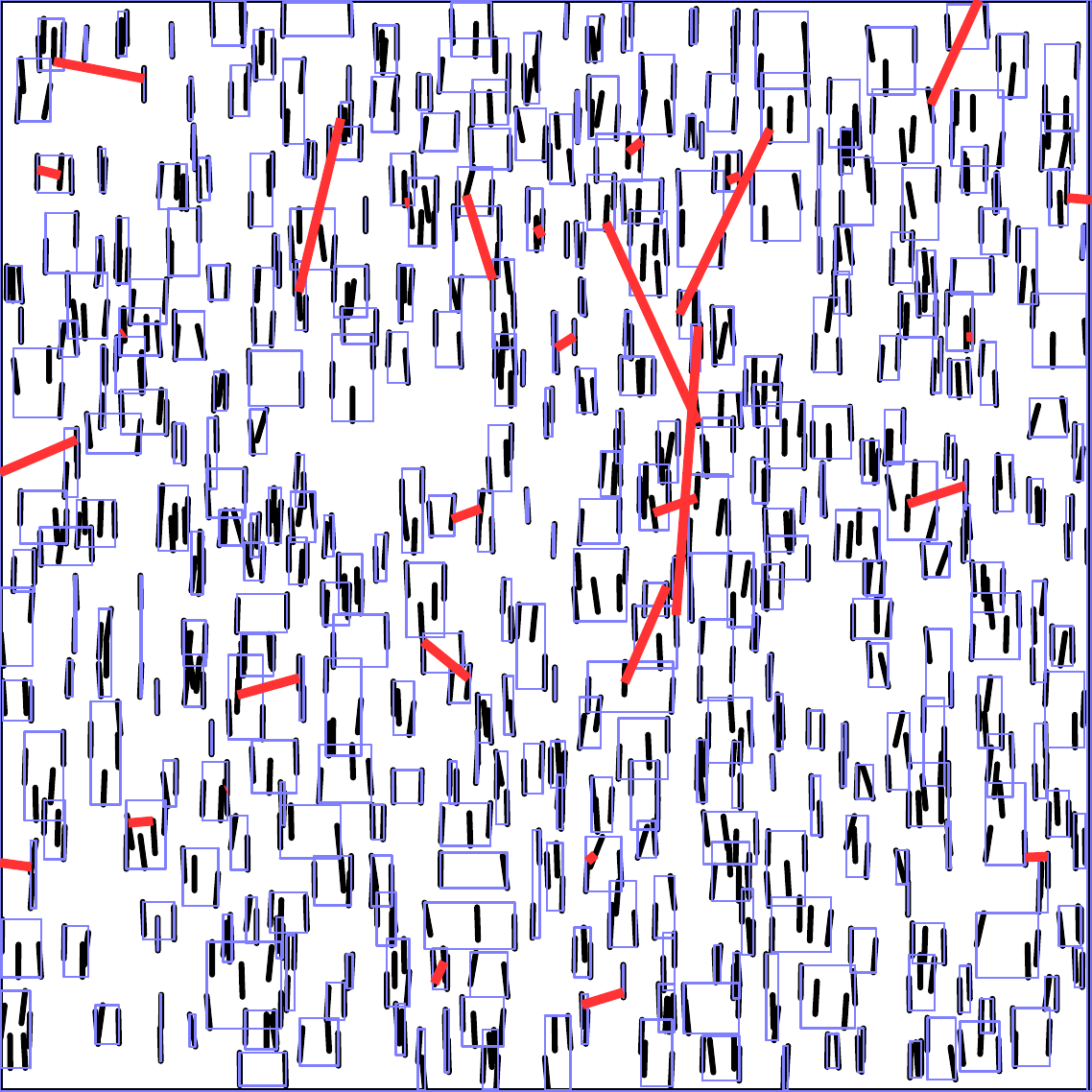}
\accelStructComparisonDlines

\accelStructComparisonA{\linesAccelCompareSize}{\linesAccelCompareSize}
    {Length factor 1 -- Uniform orientation}
    {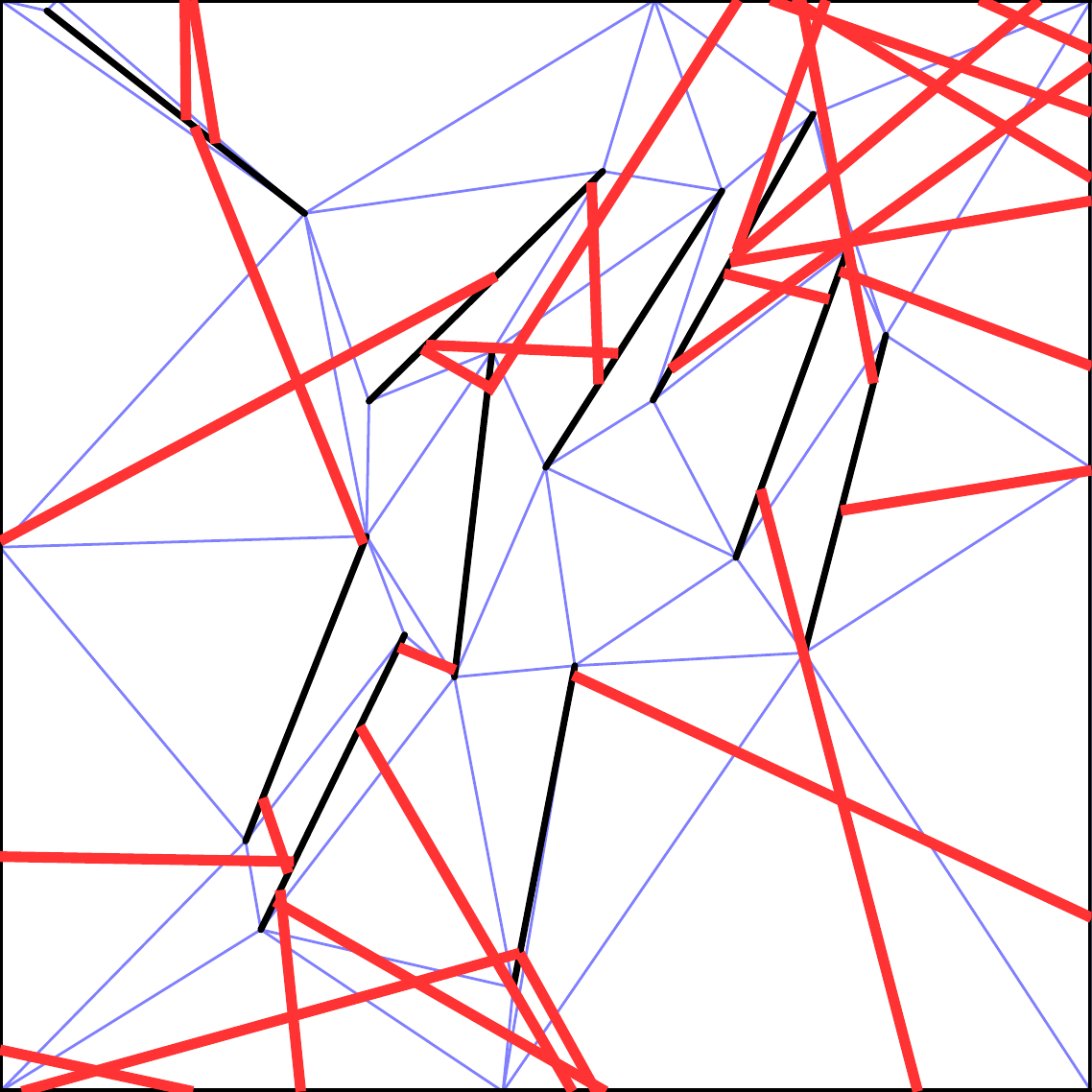}
    {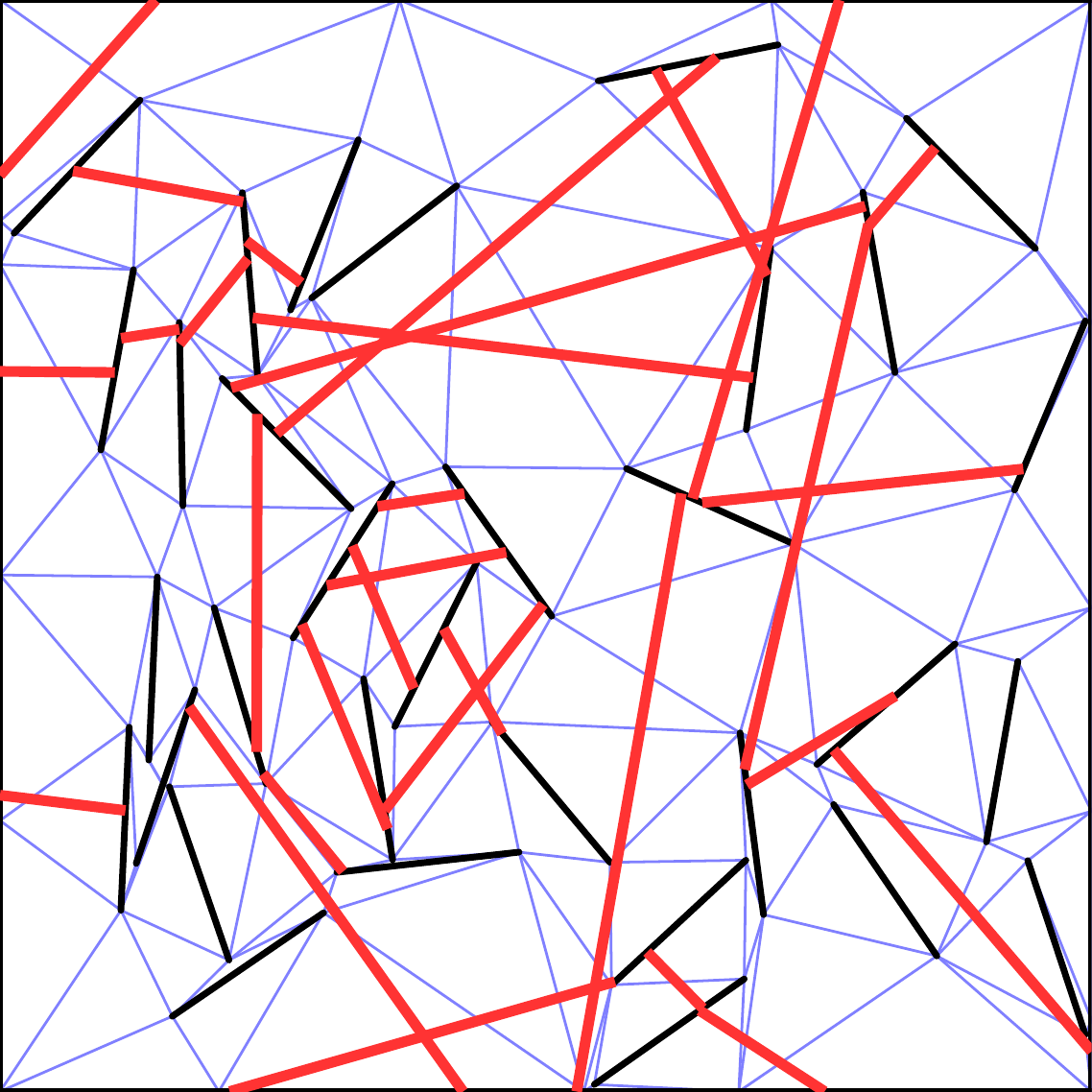}
    {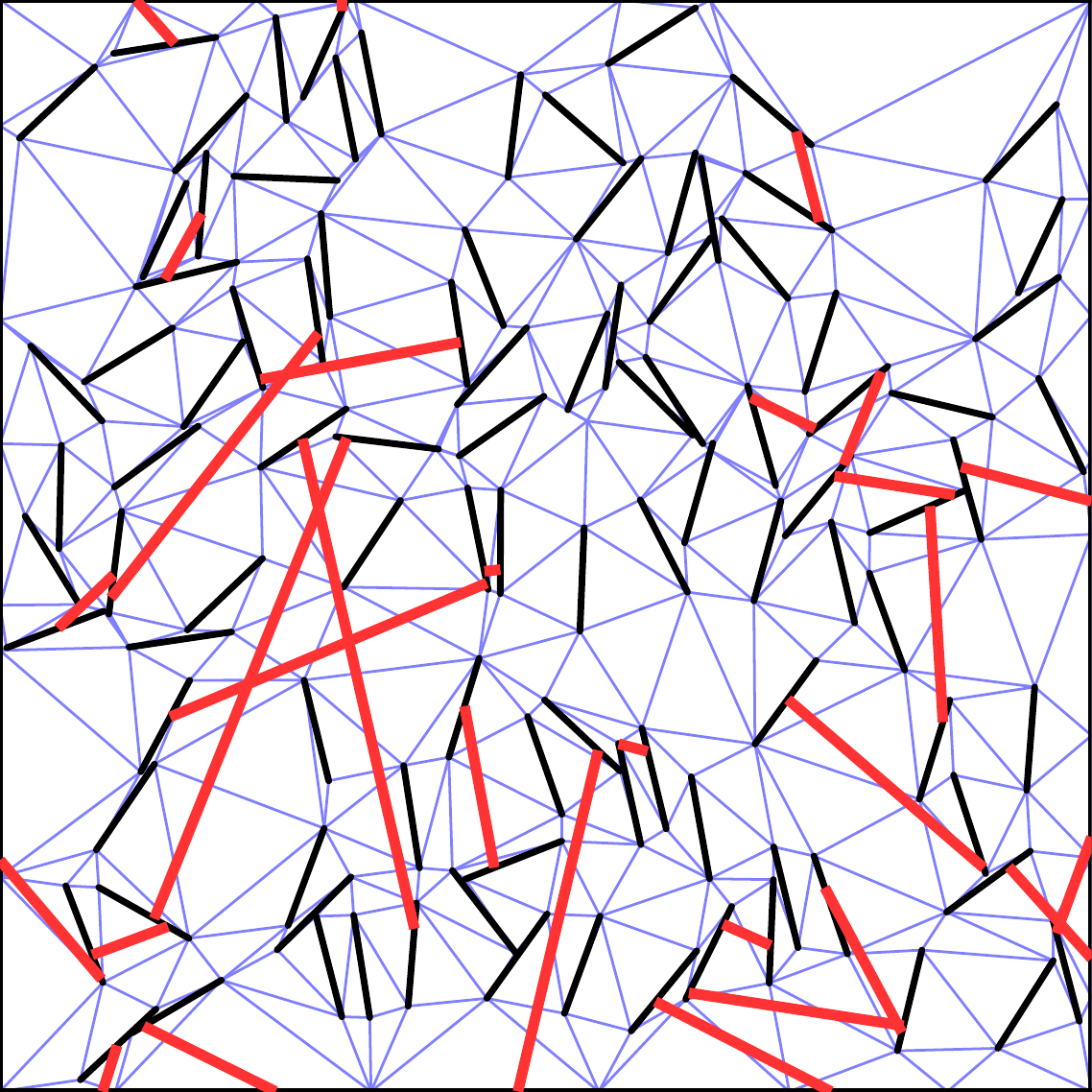}
    {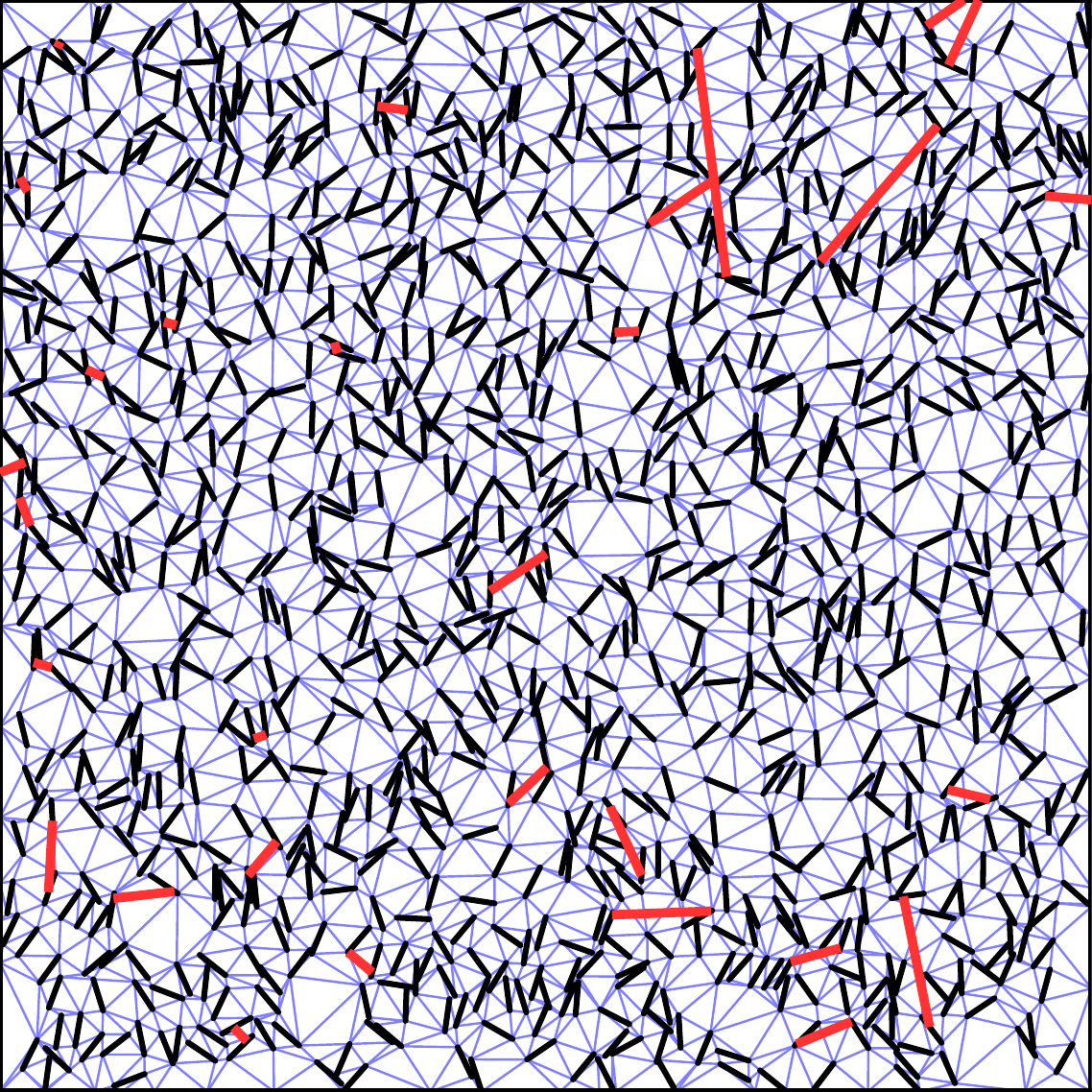}
\accelStructComparisonB{\linesAccelCompareSize}{\linesAccelCompareSize}
    {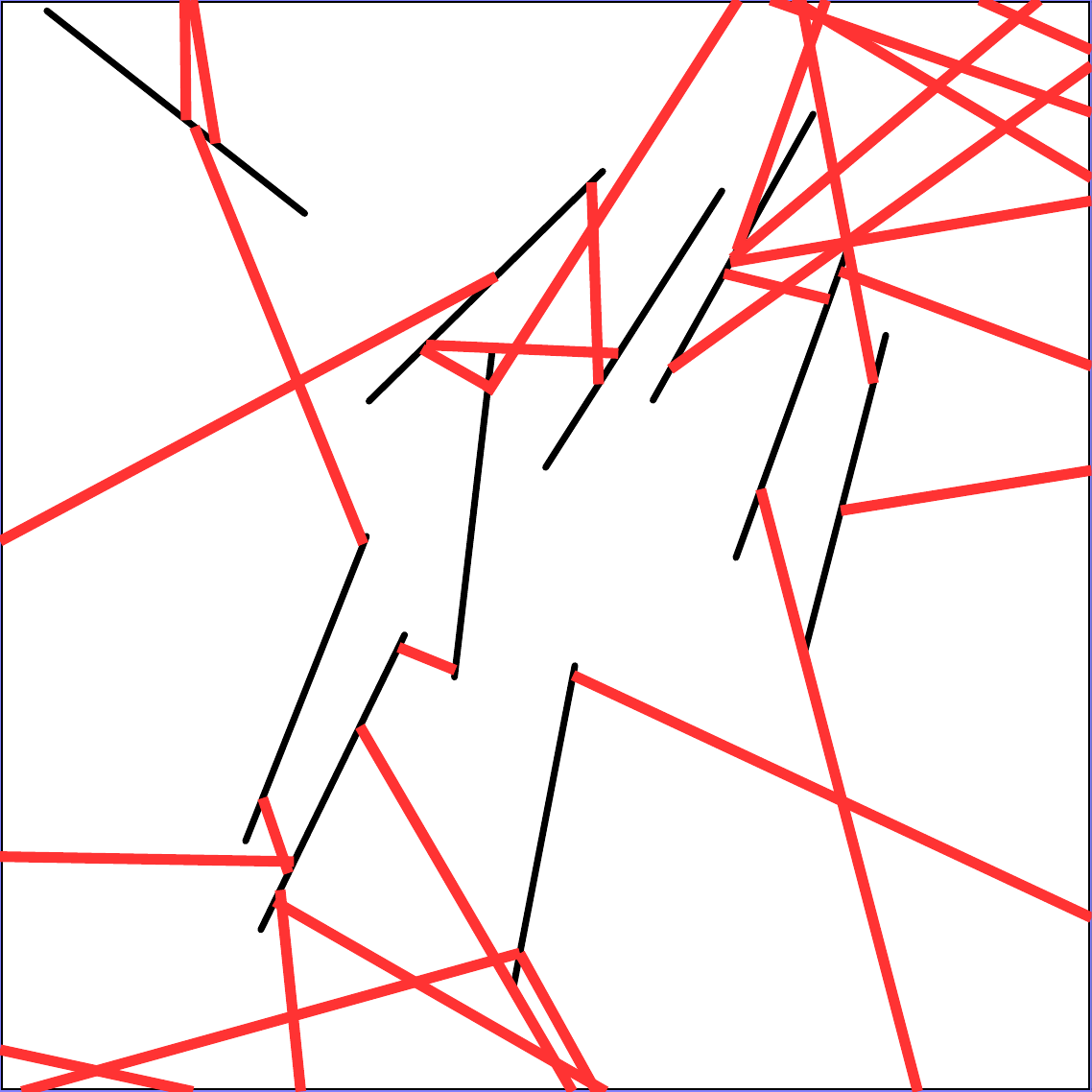}
    {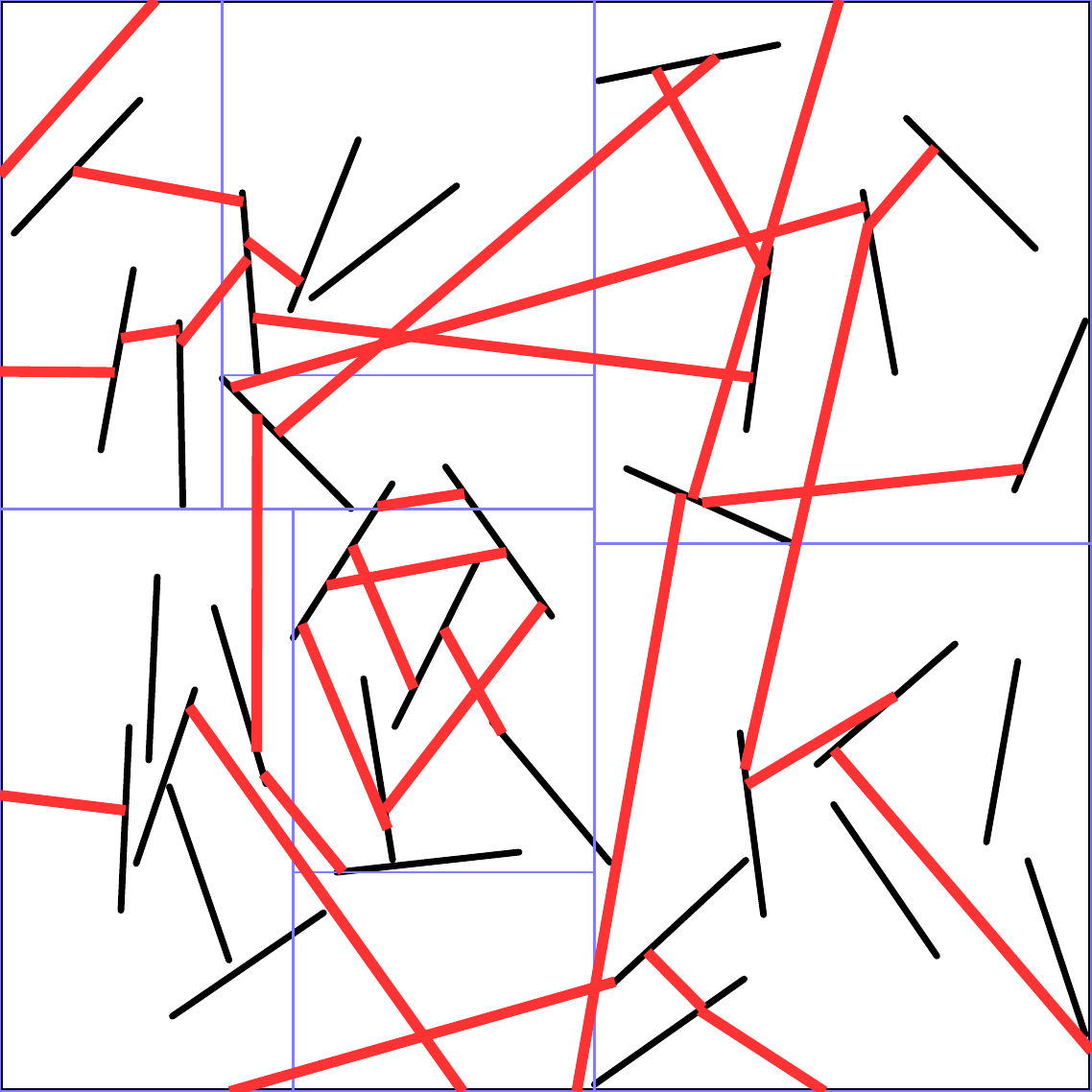}
    {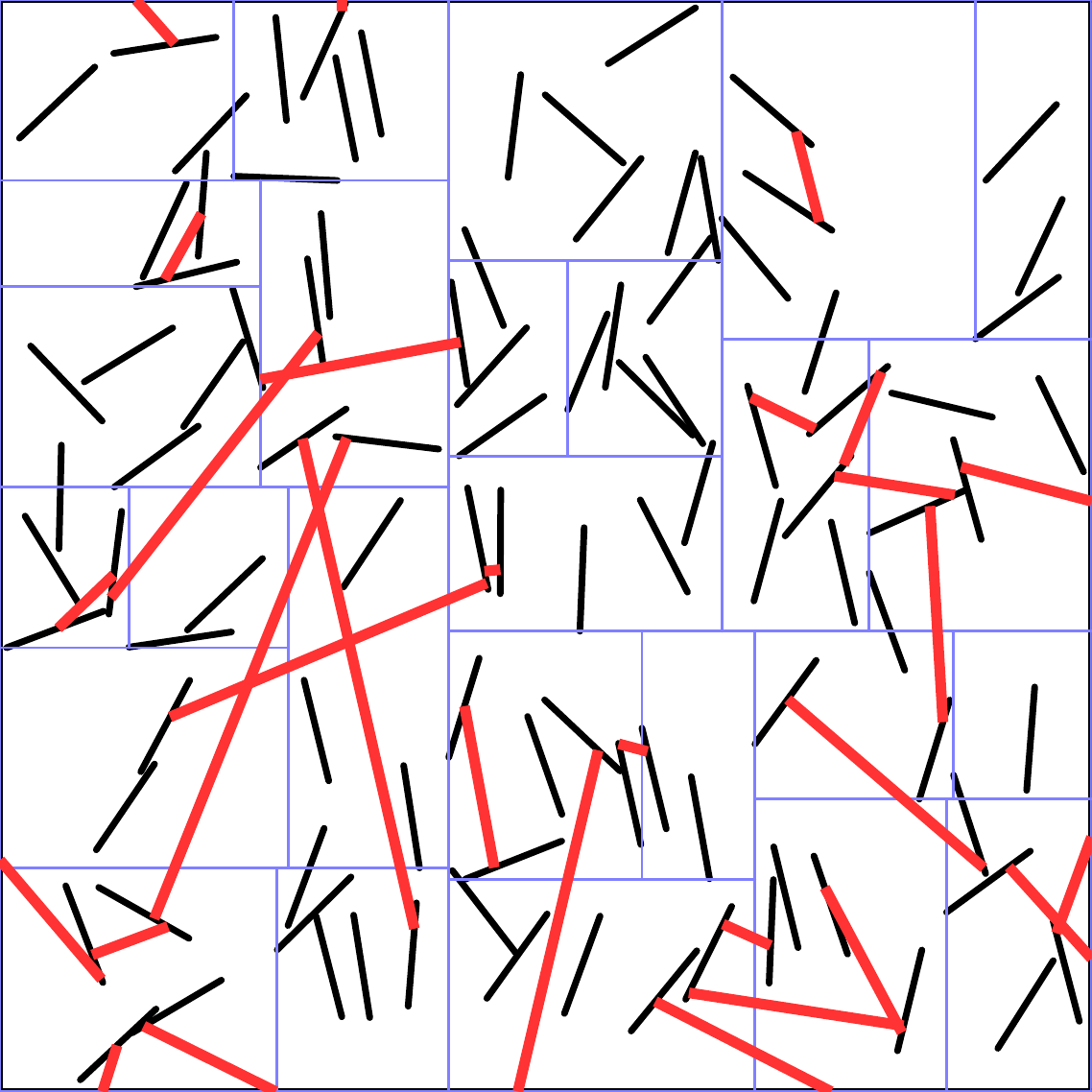}
    {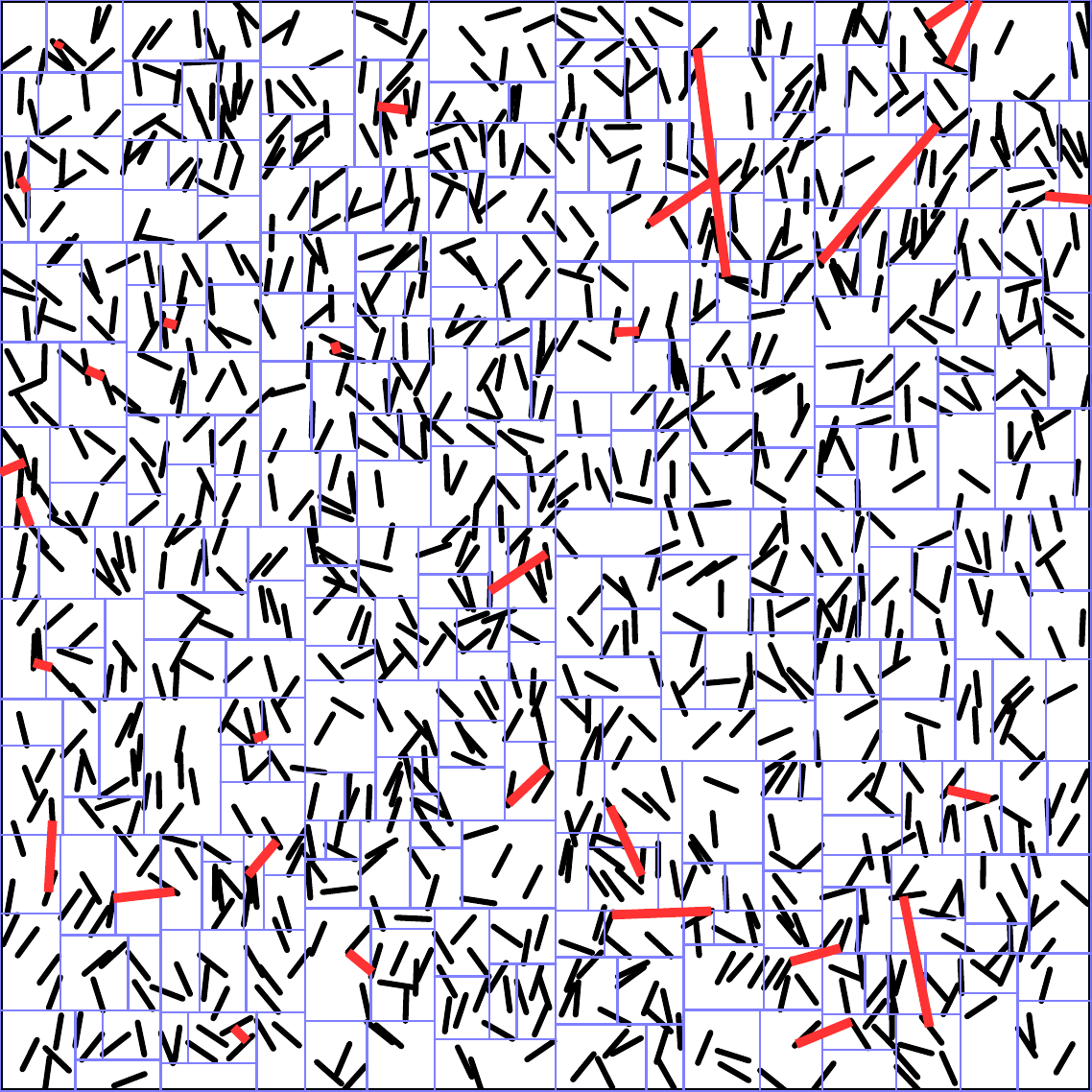}
\accelStructComparisonC{\linesAccelCompareSize}{\linesAccelCompareSize}
    {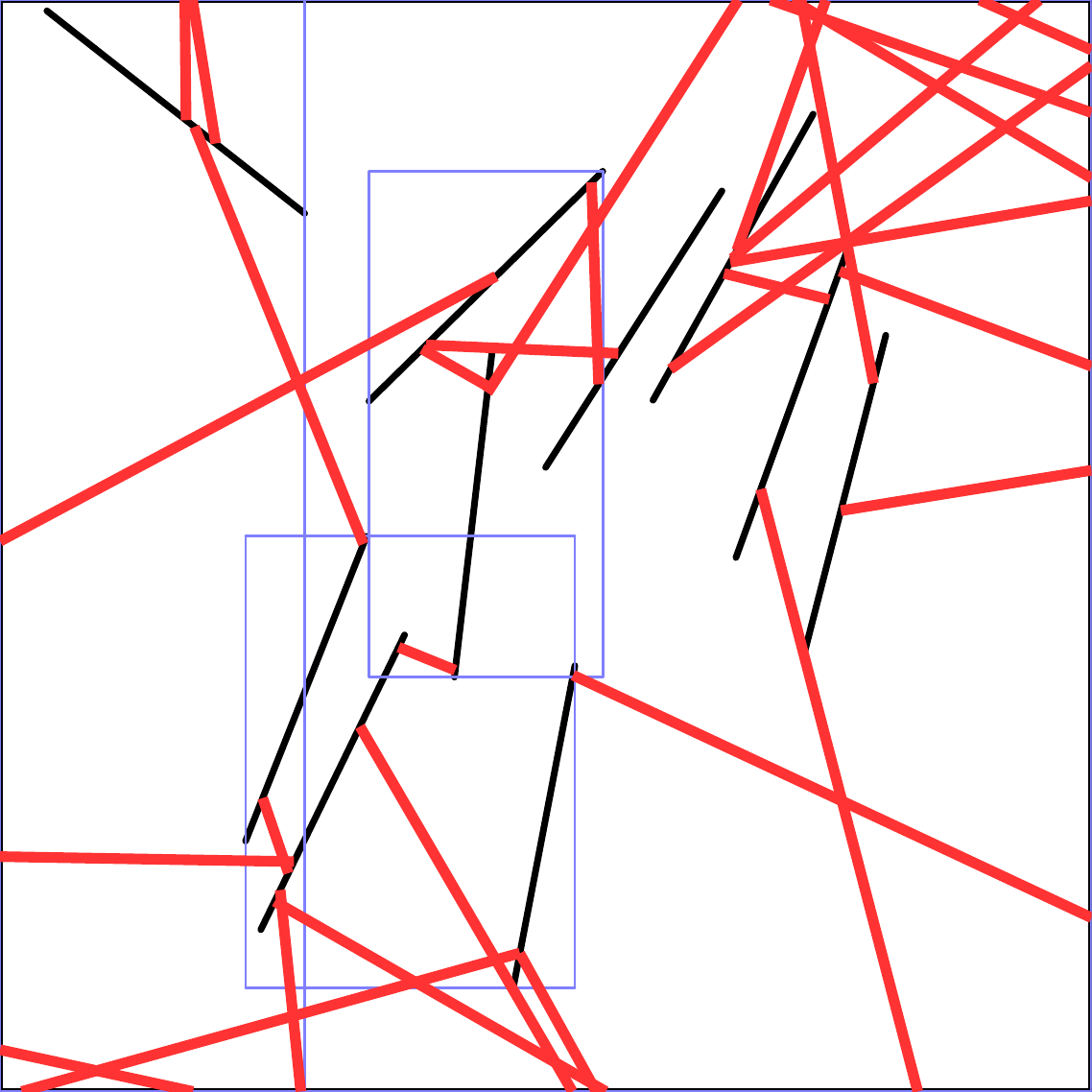}
    {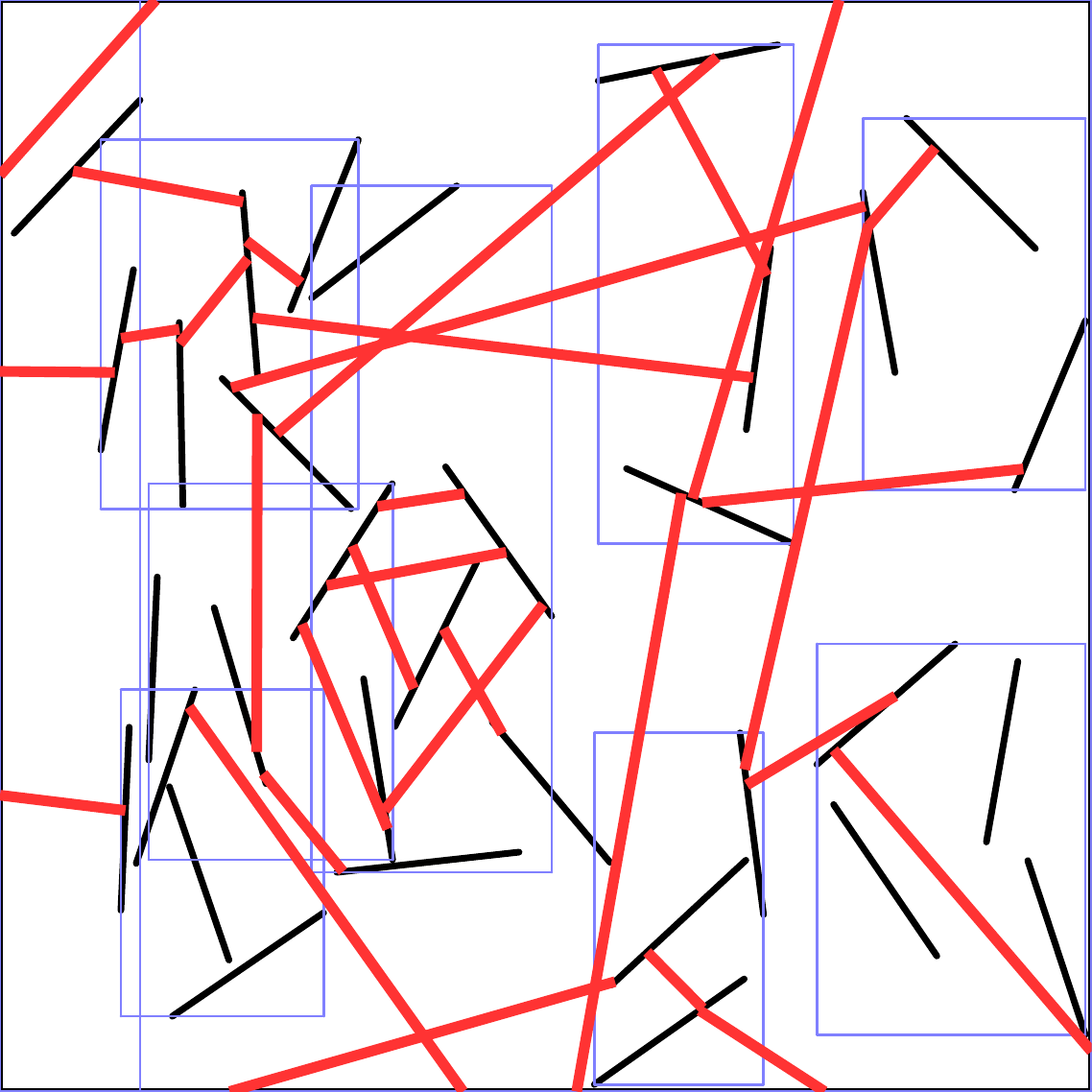}
    {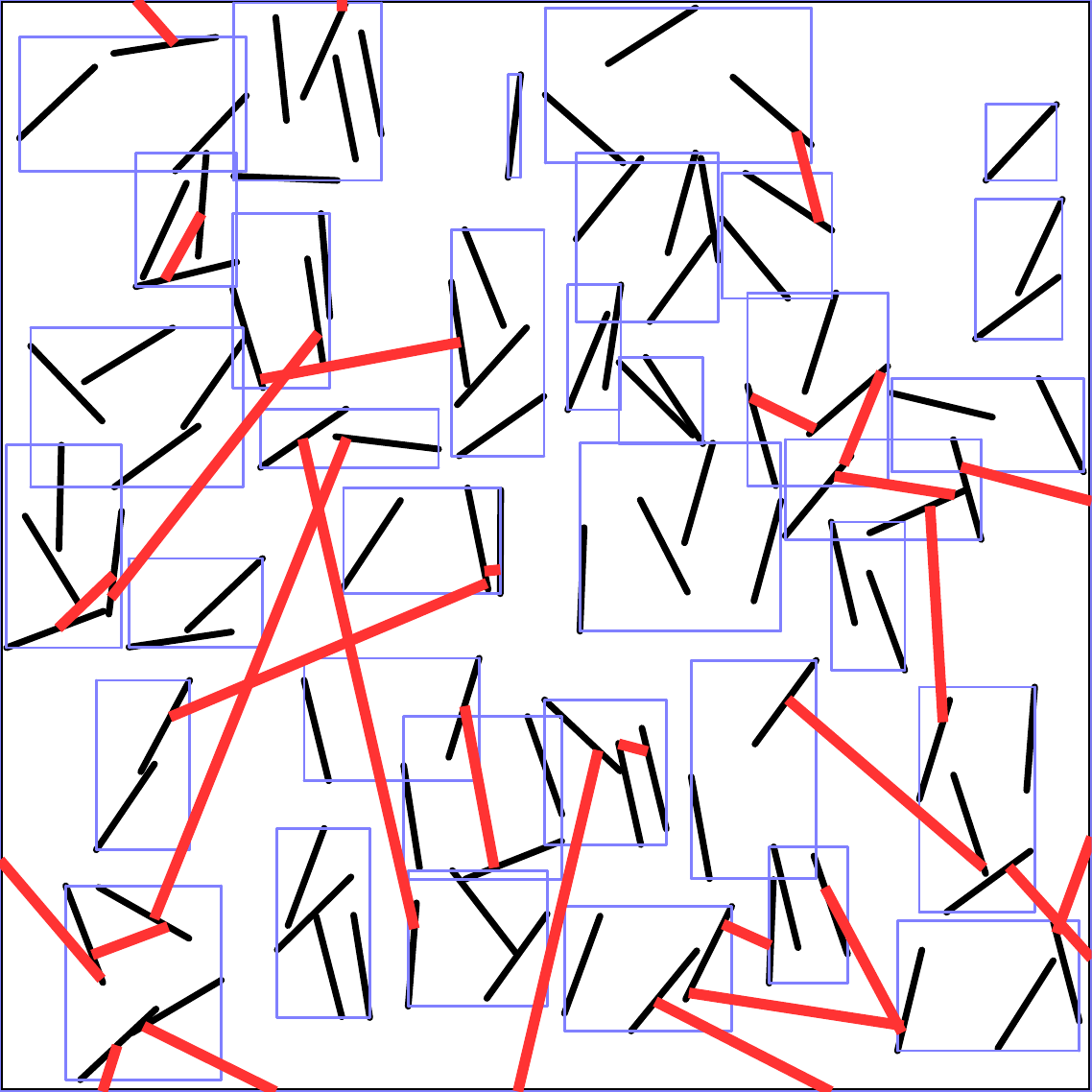}
    {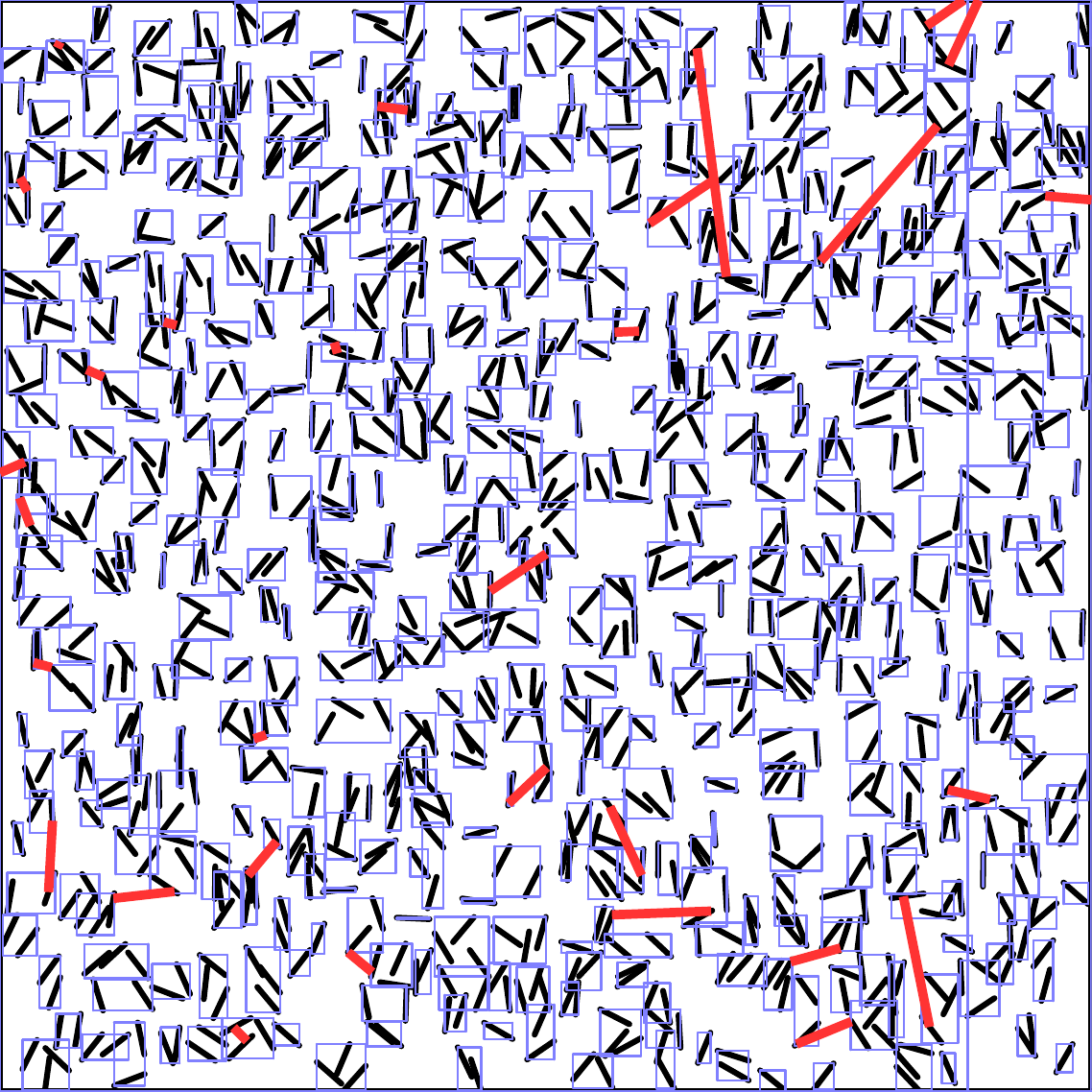}
\accelStructComparisonDlines

\accelStructComparisonA{\linesAccelCompareSize}{\linesAccelCompareSize}
    {Length factor 1 -- Diagonal orientation}
    {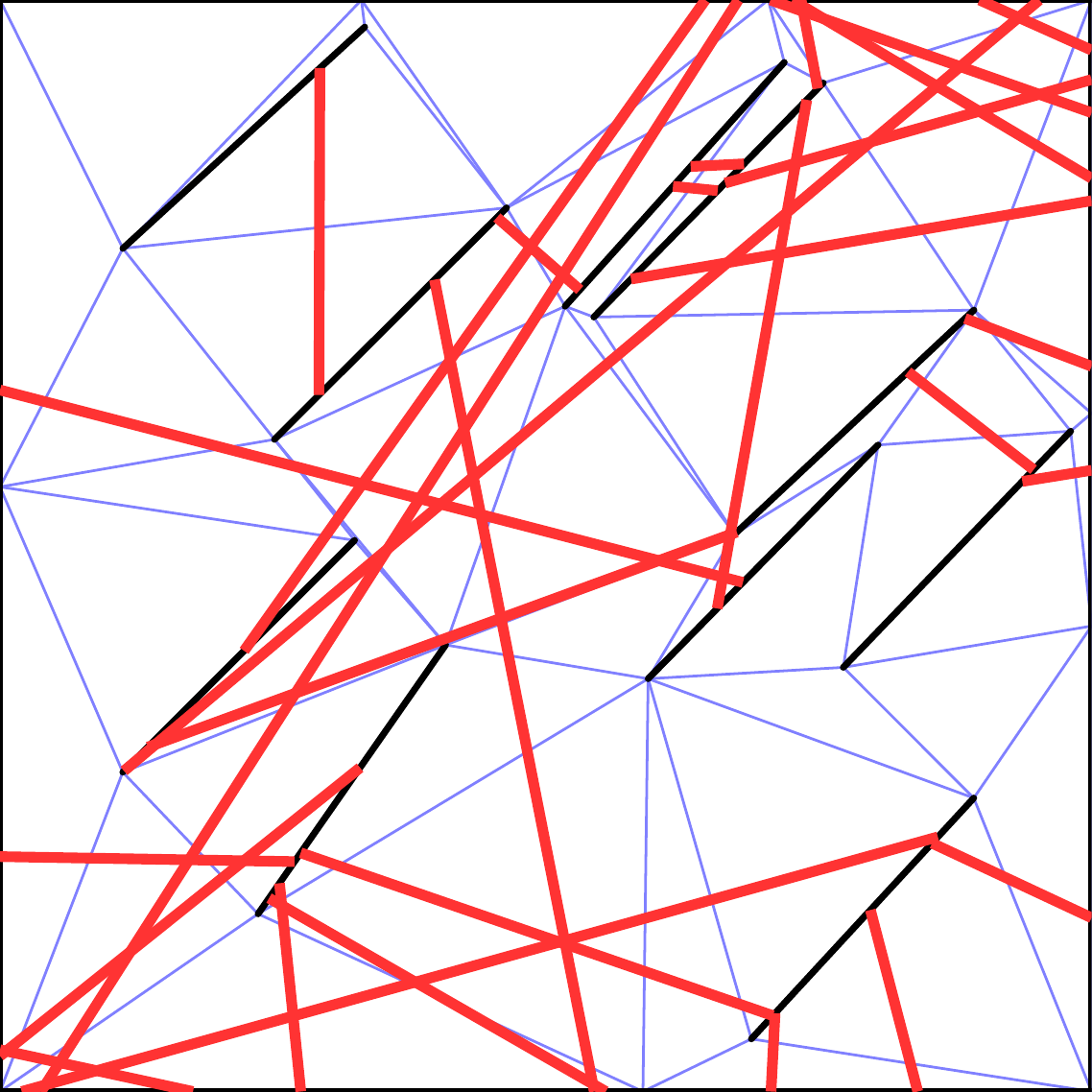}
    {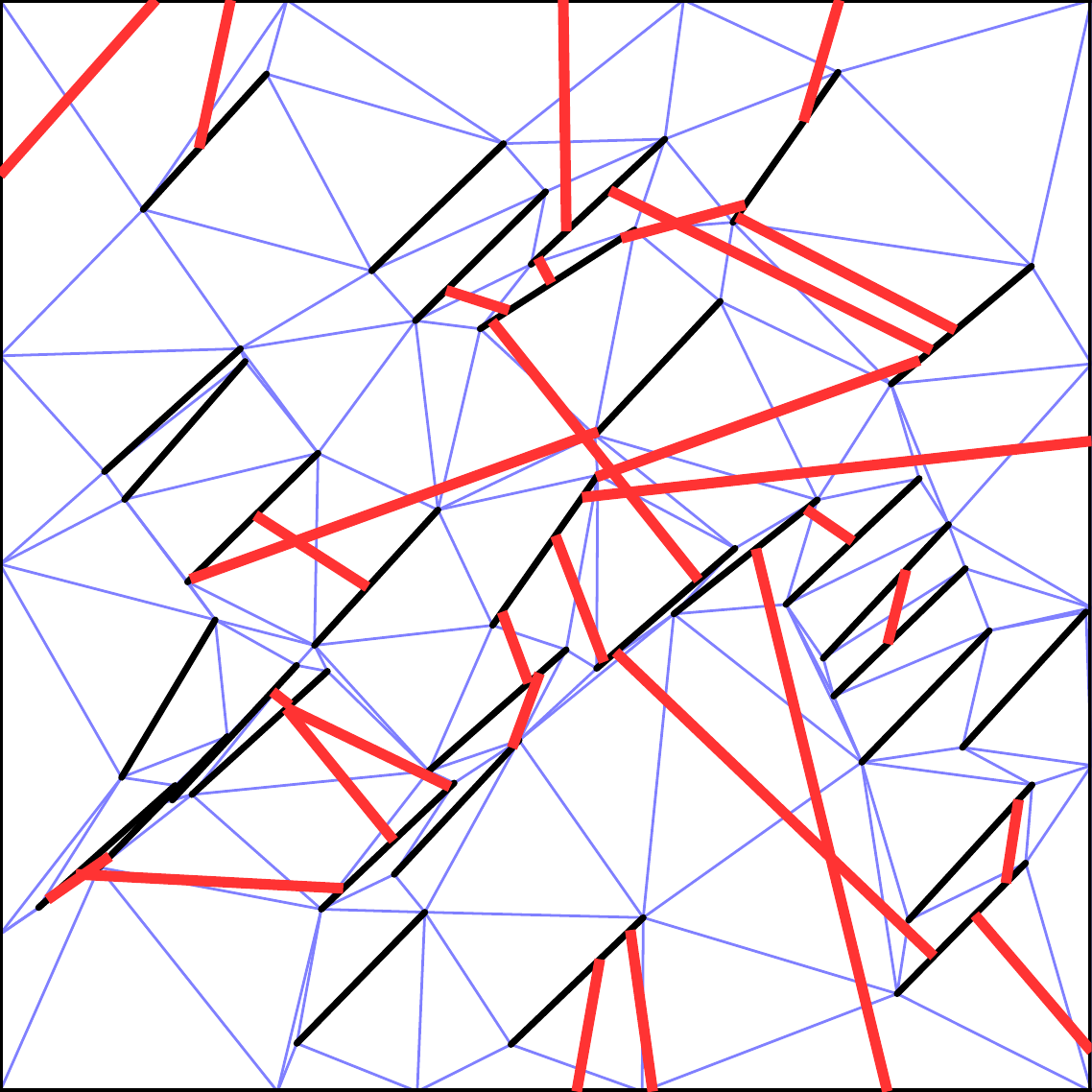}
    {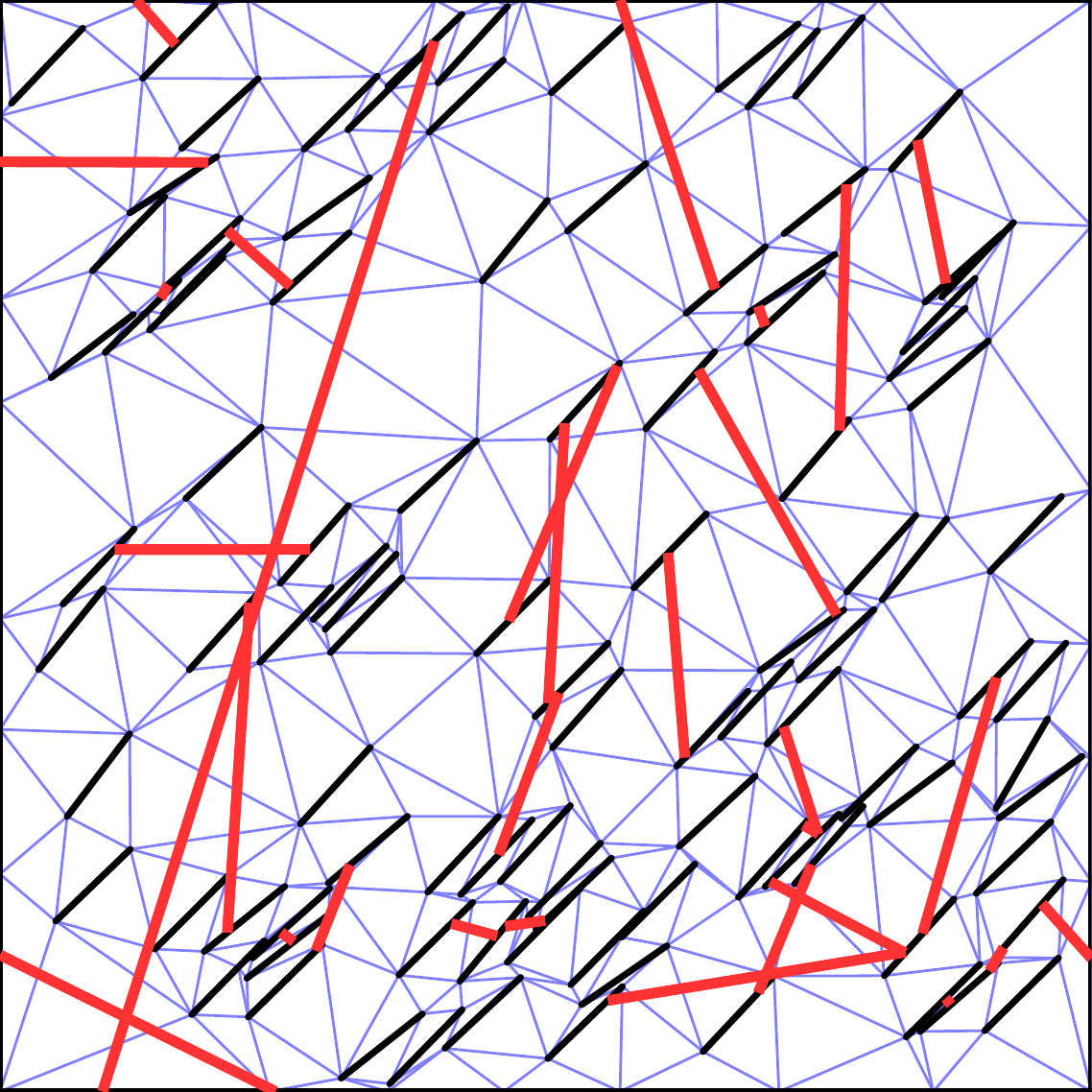}
    {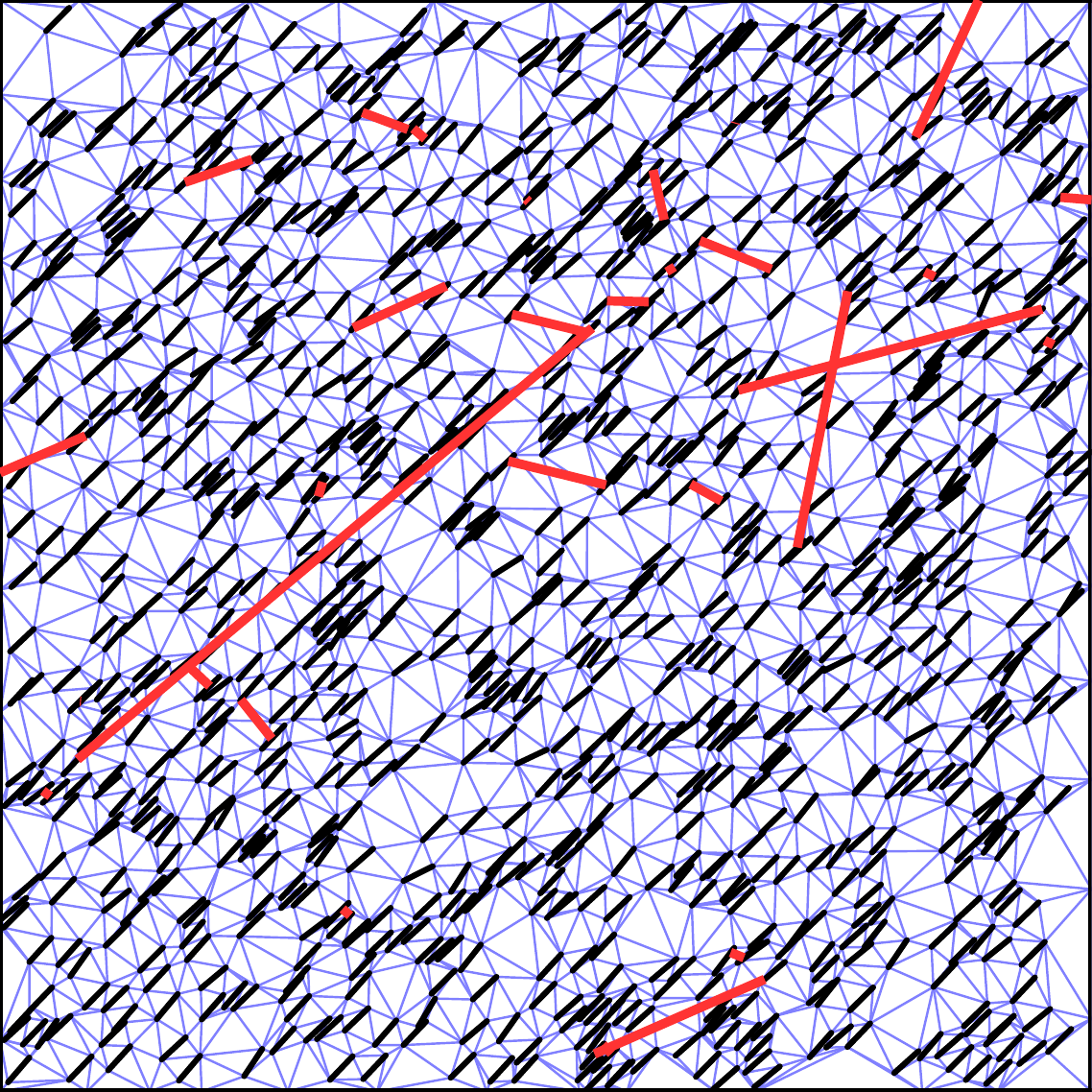}
\accelStructComparisonB{\linesAccelCompareSize}{\linesAccelCompareSize}
    {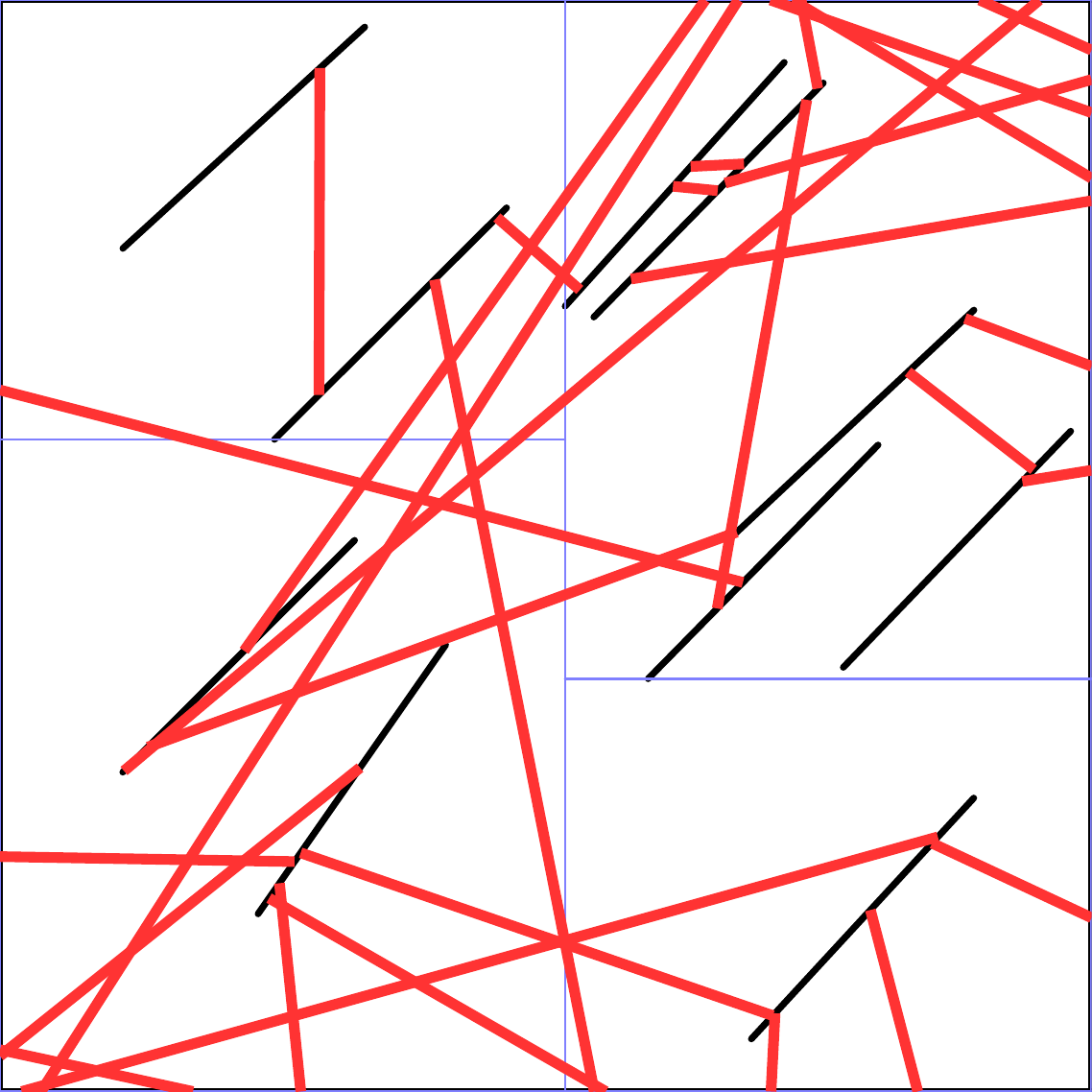}
    {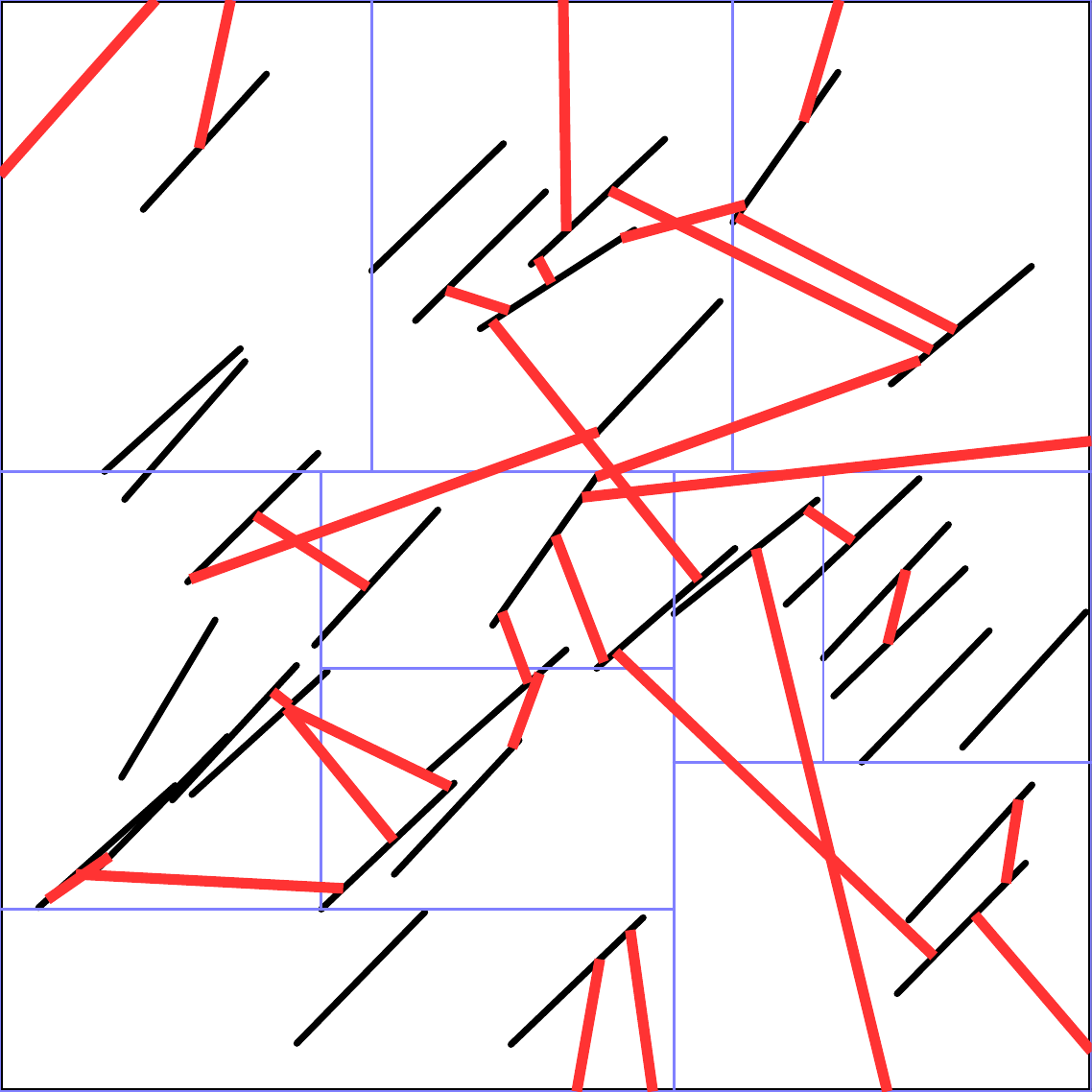}
    {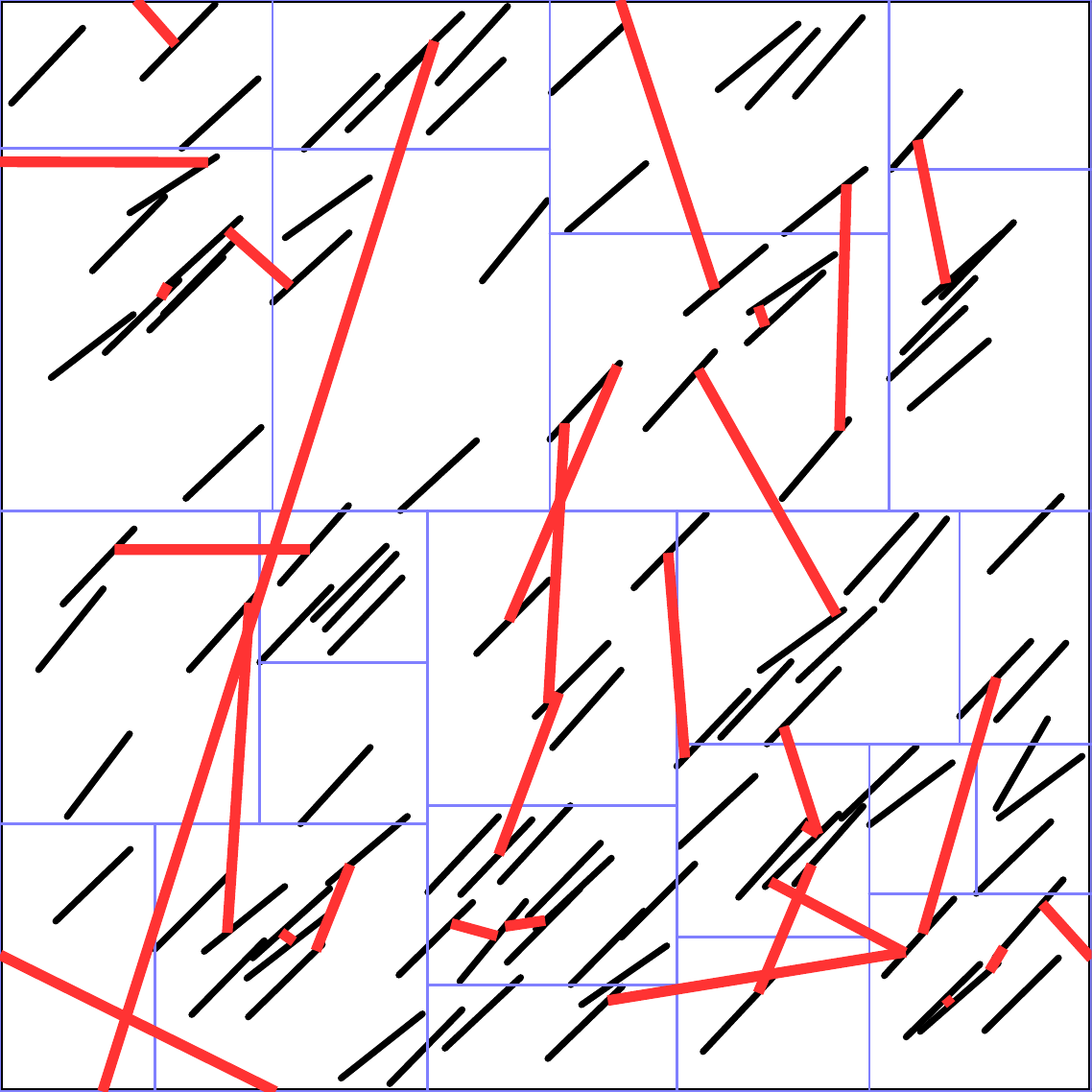}
    {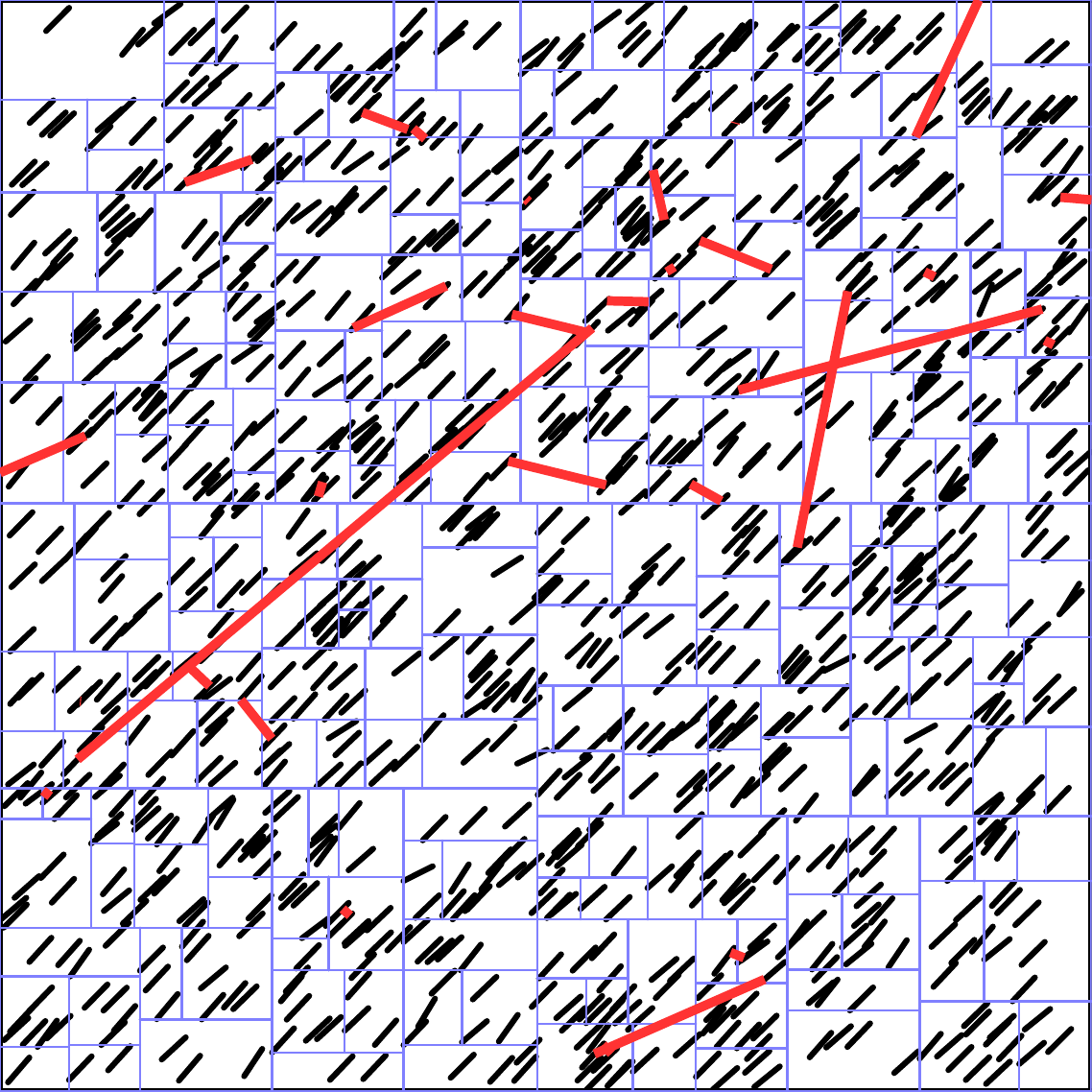}
\accelStructComparisonC{\linesAccelCompareSize}{\linesAccelCompareSize}
    {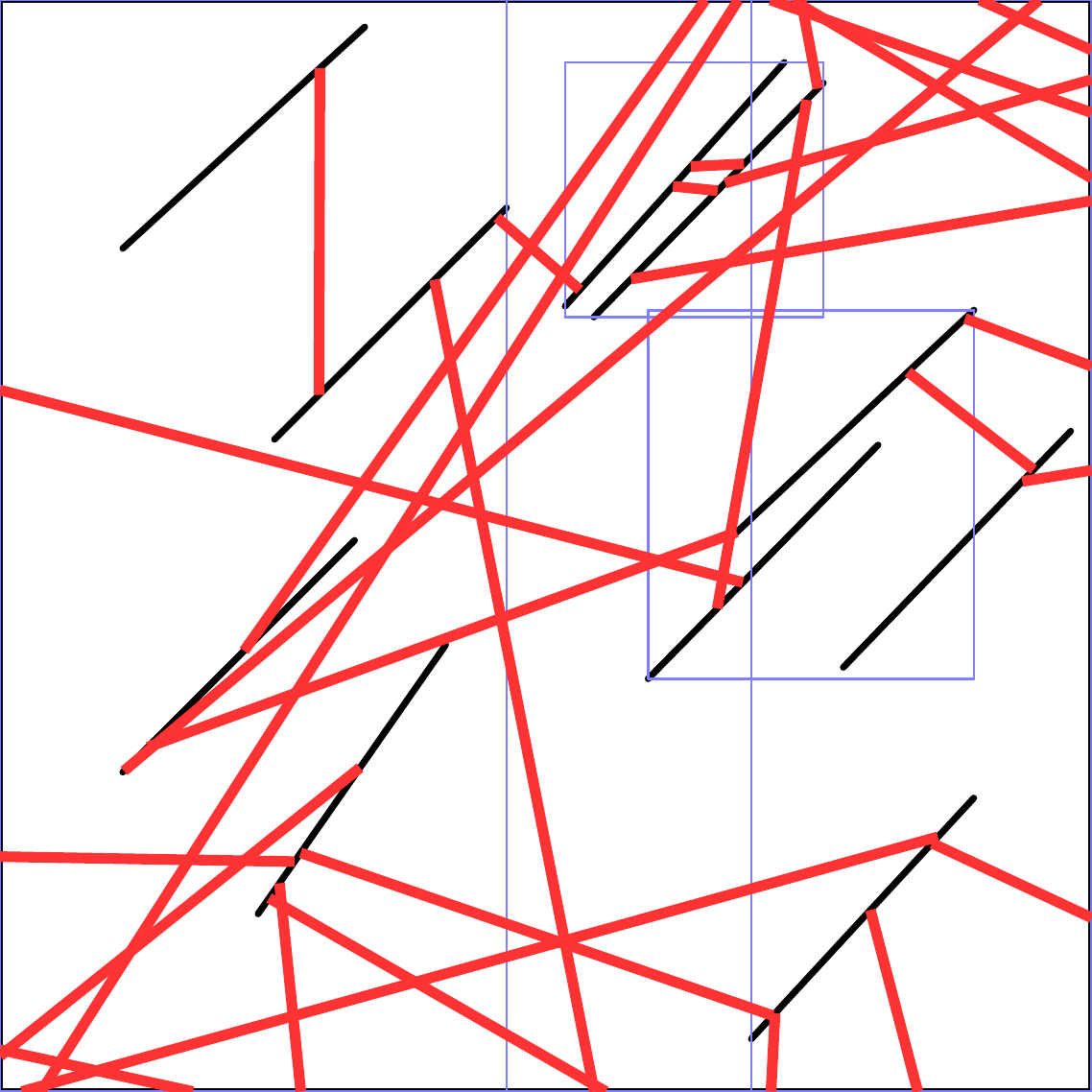}
    {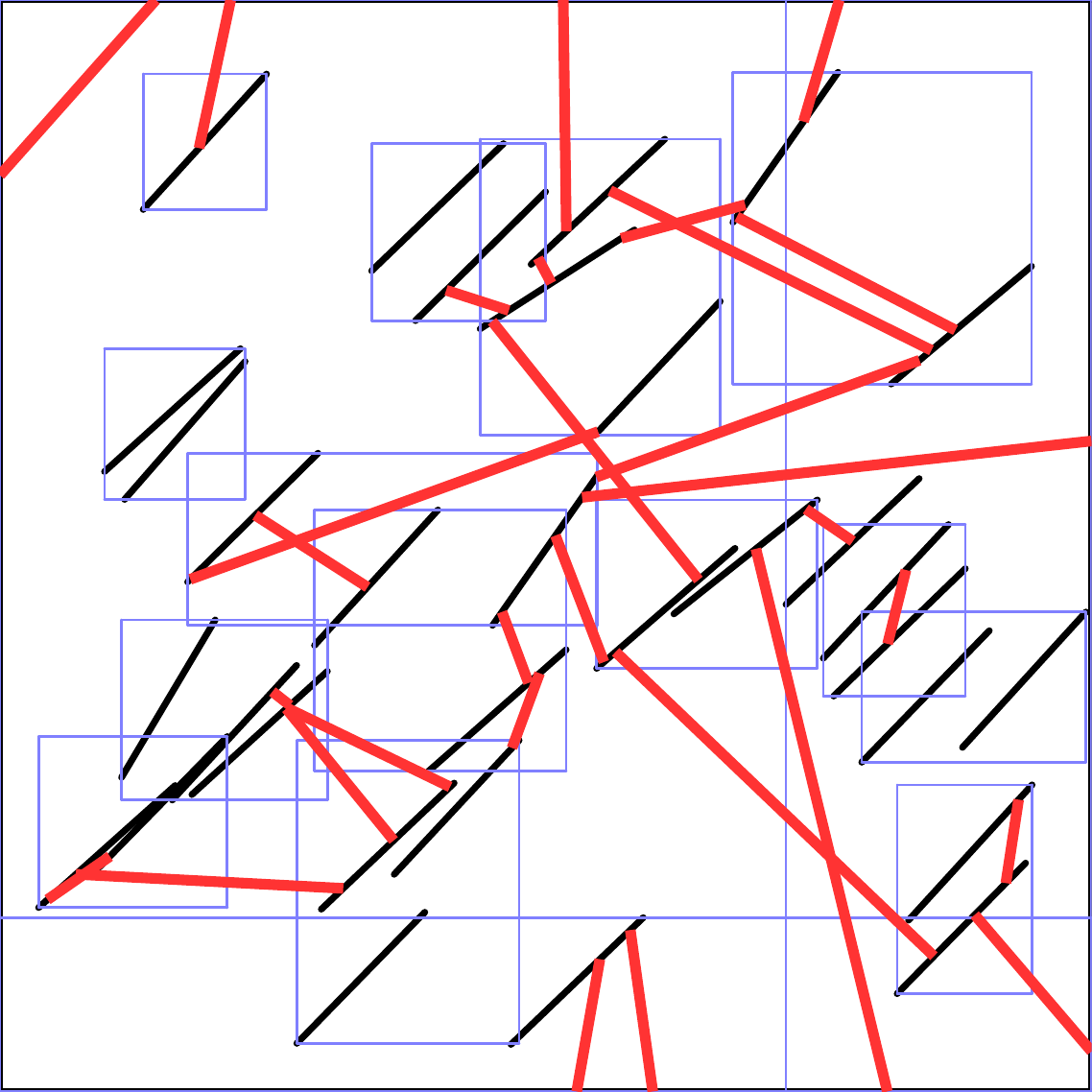}
    {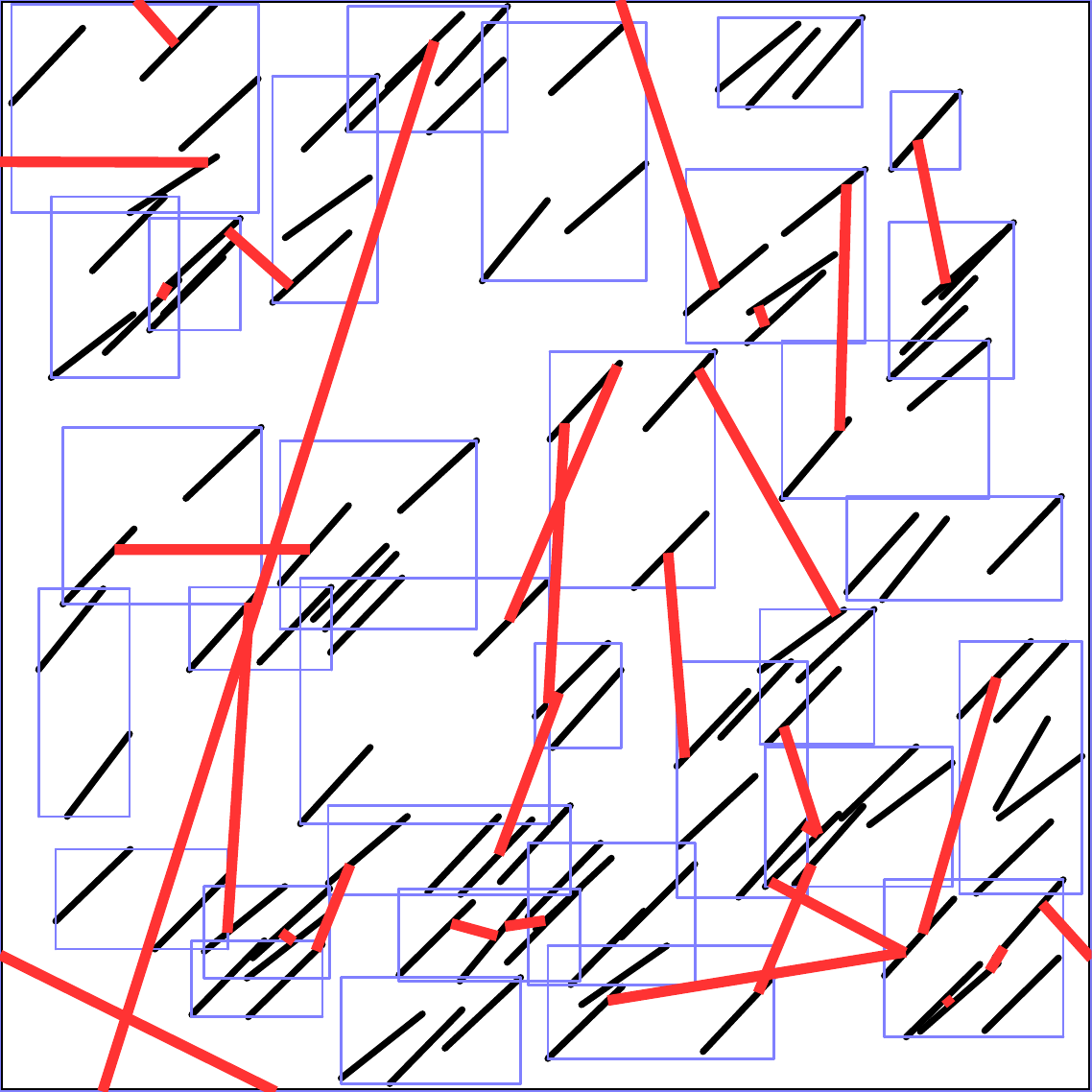}
    {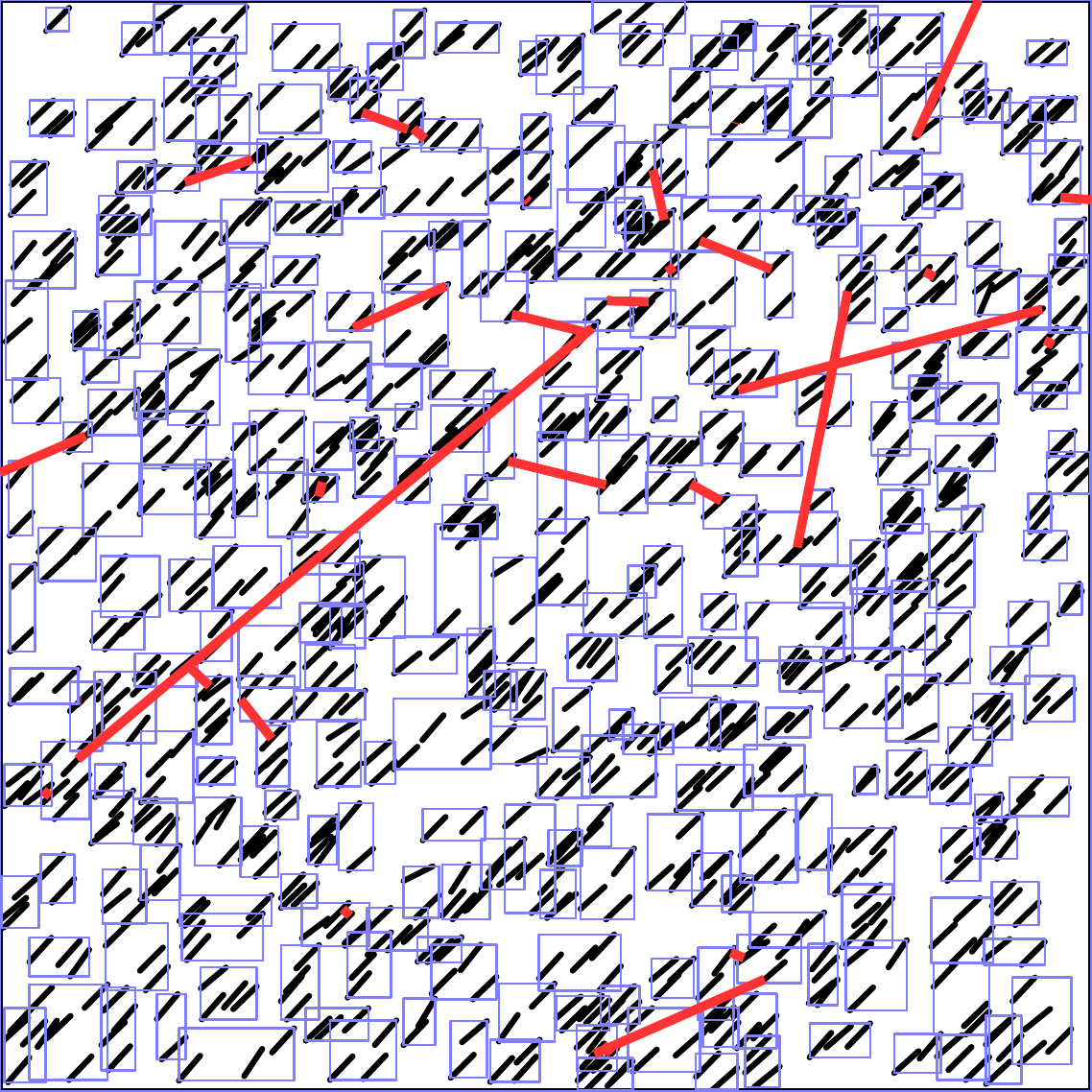}
\accelStructComparisonDlines

\accelStructComparisonA{\linesAccelCompareSize}{\linesAccelCompareSize}
    {Length factor 3 -- Vertical orientation}
    {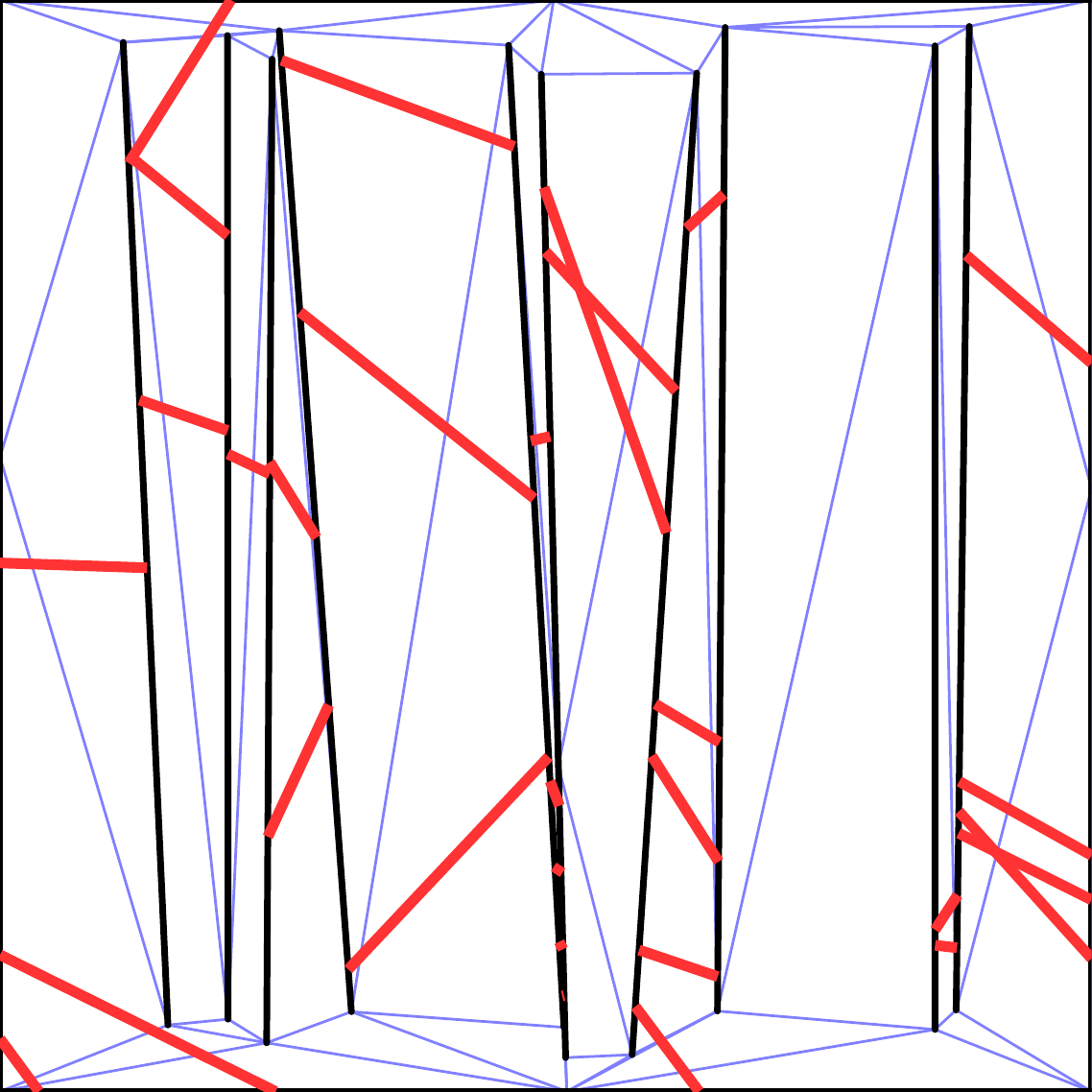}
    {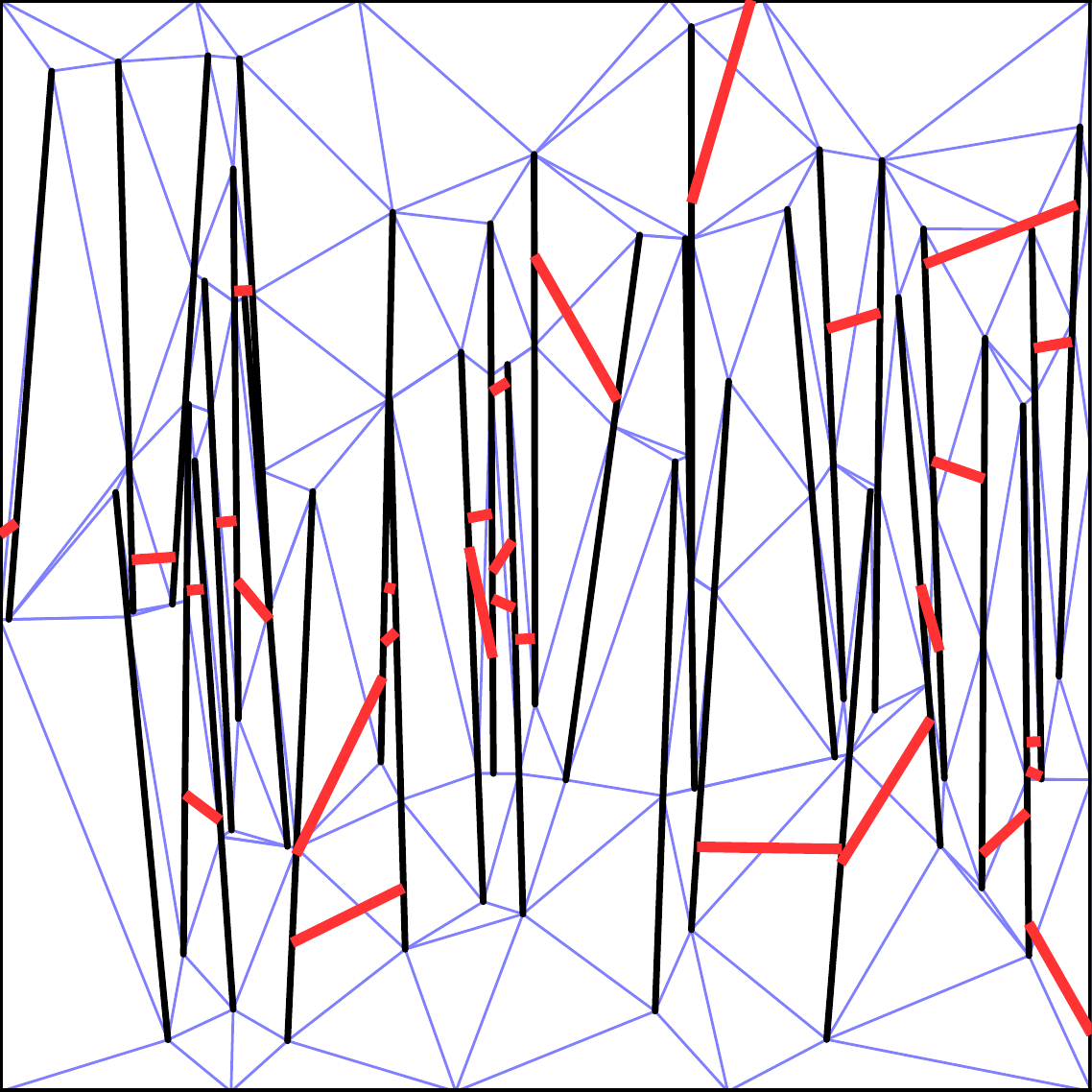}
    {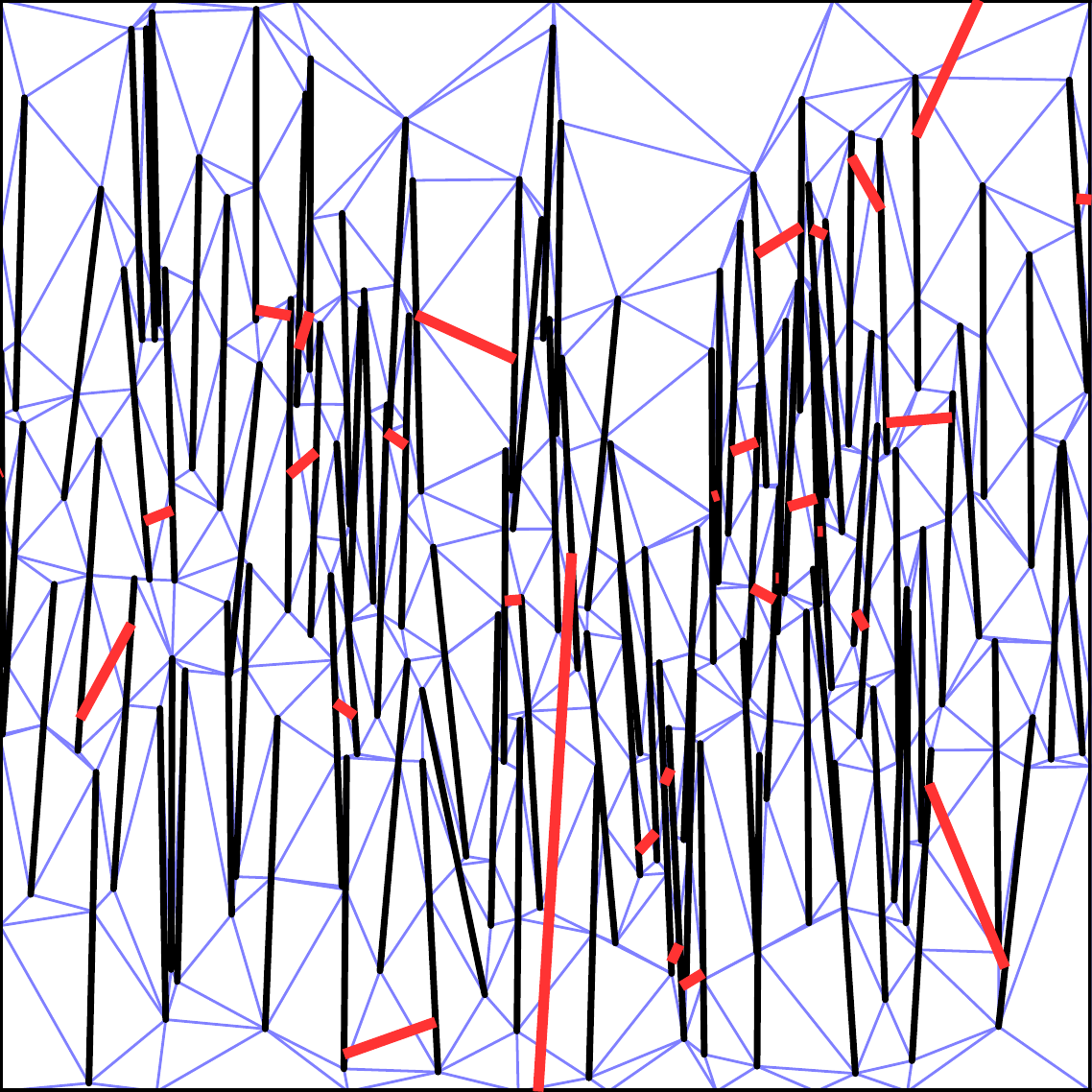}
    {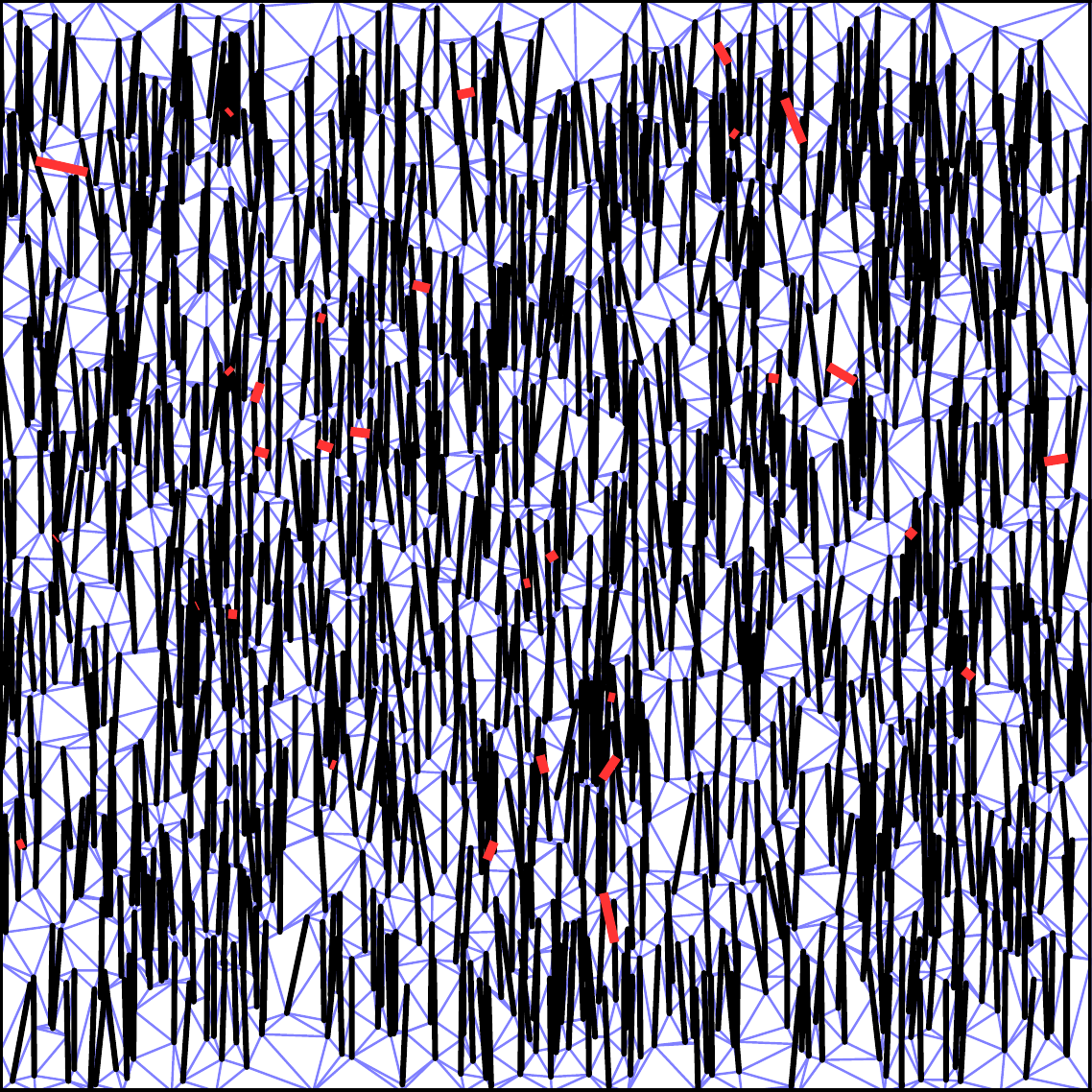}
\accelStructComparisonB{\linesAccelCompareSize}{\linesAccelCompareSize}
    {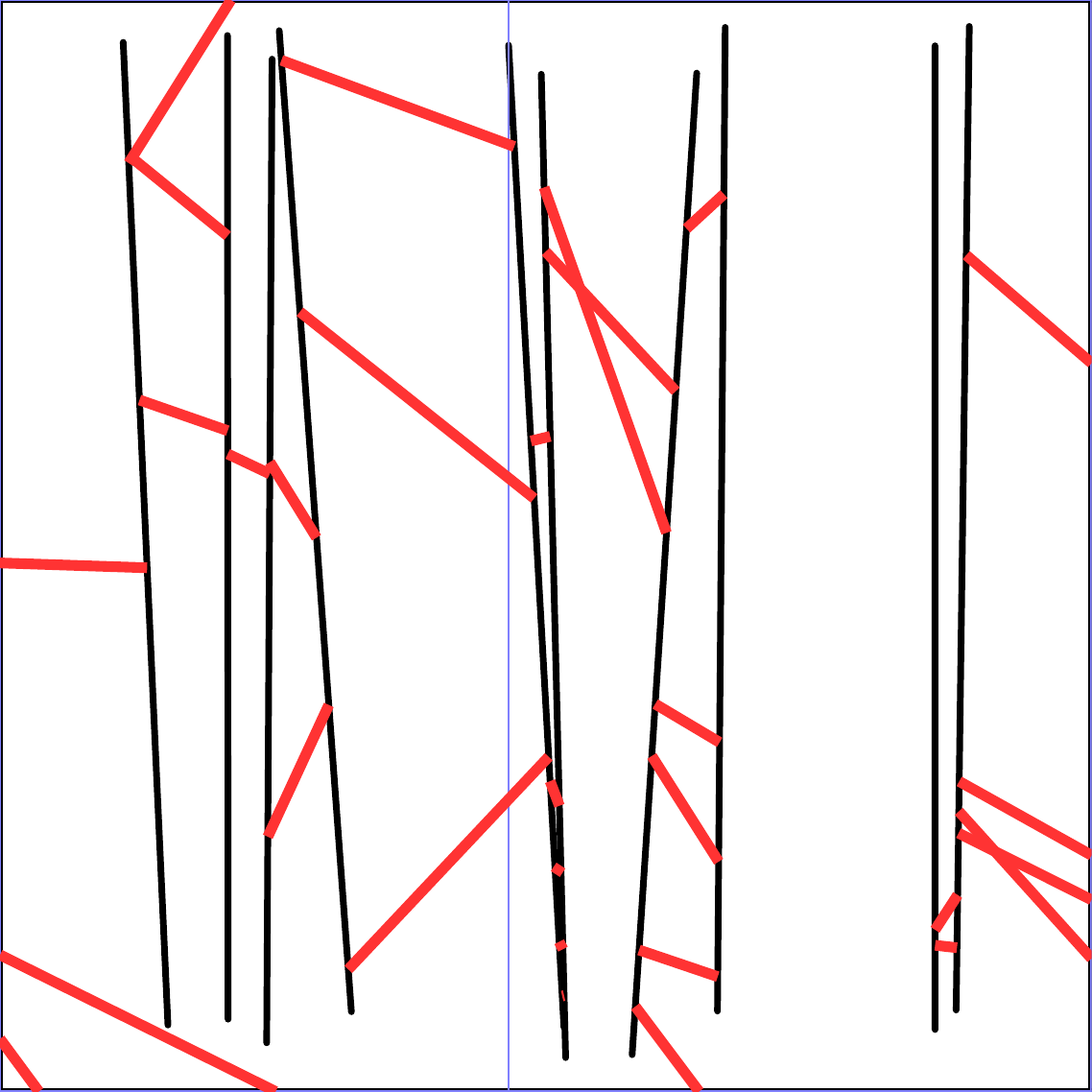}
    {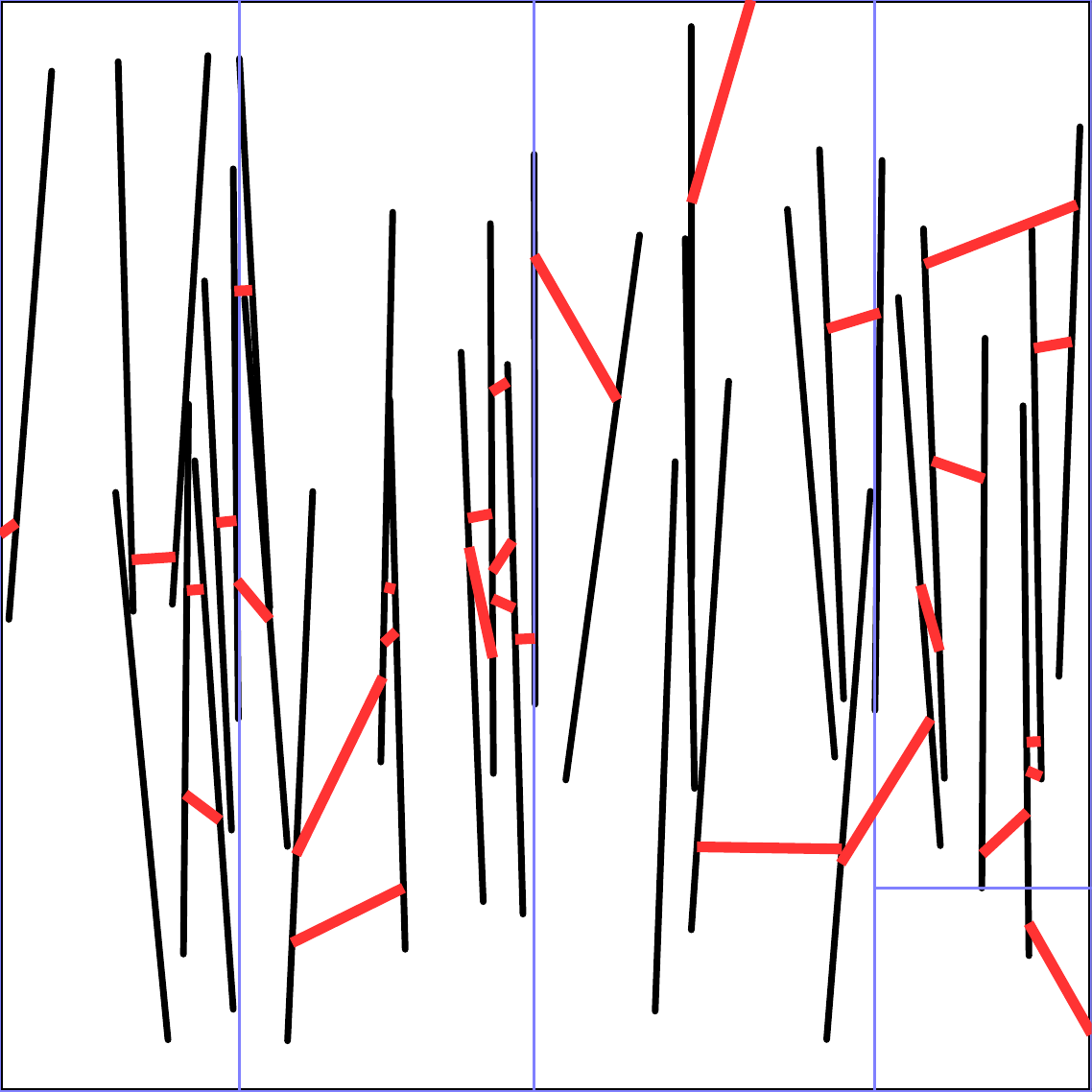}
    {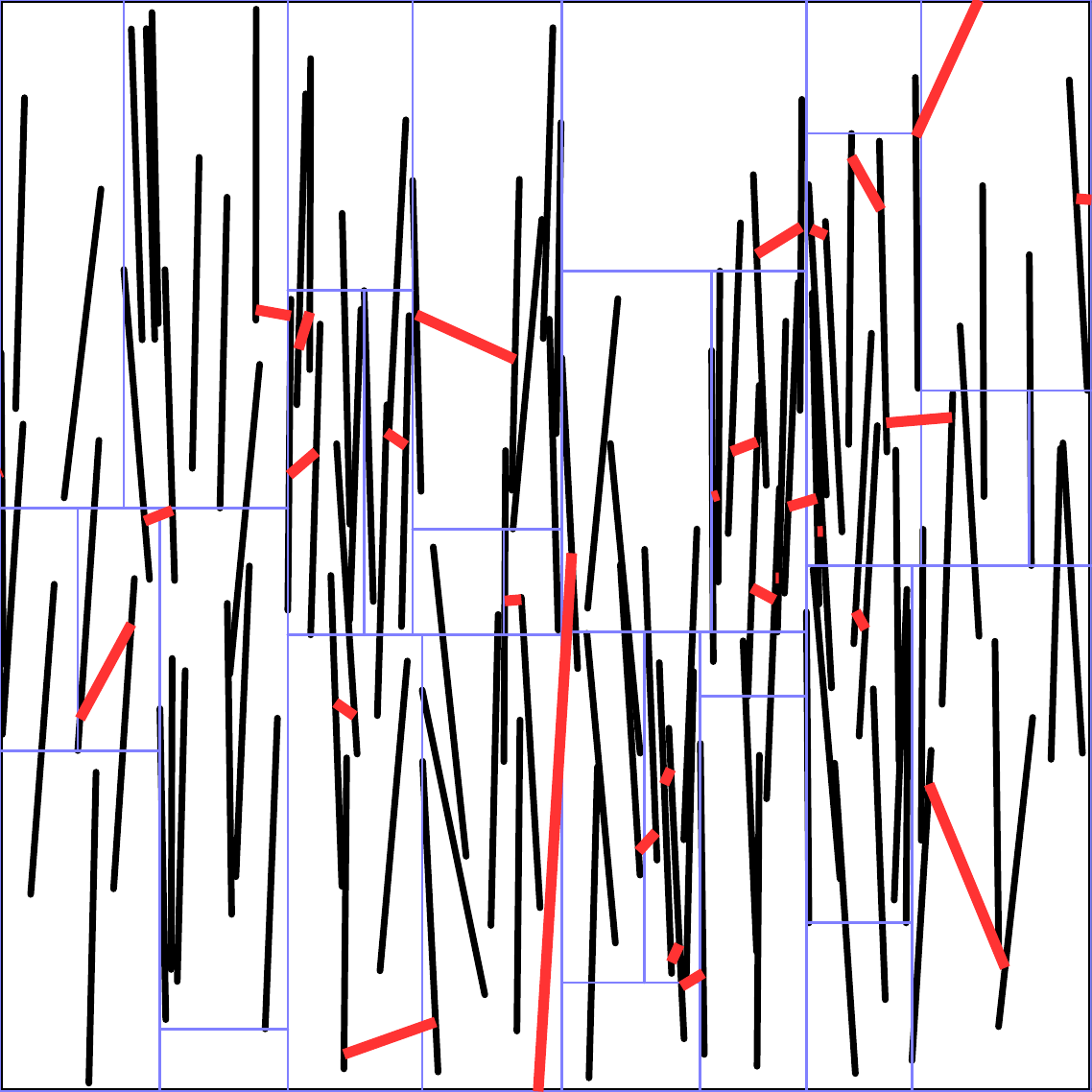}
    {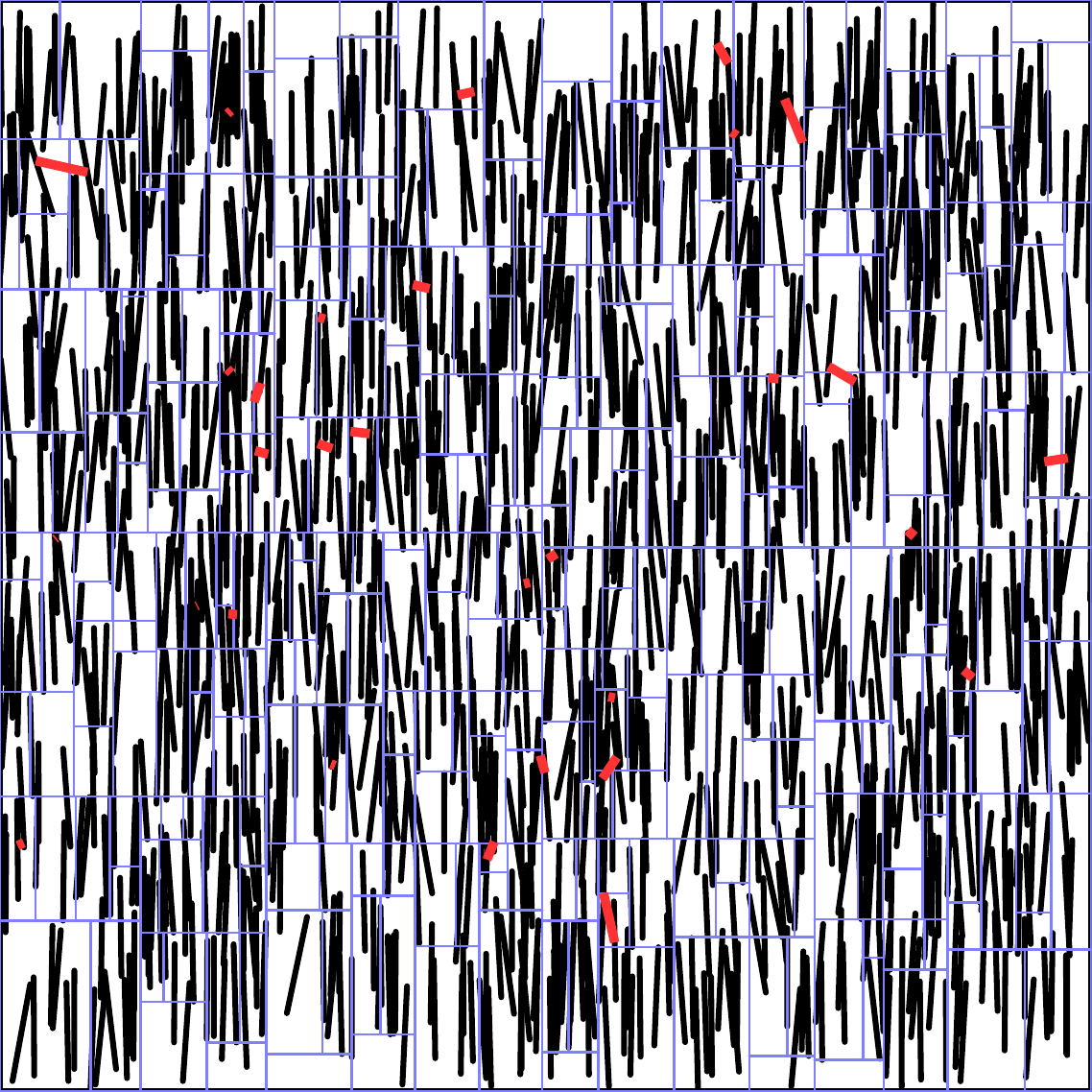}
\accelStructComparisonC{\linesAccelCompareSize}{\linesAccelCompareSize}
    {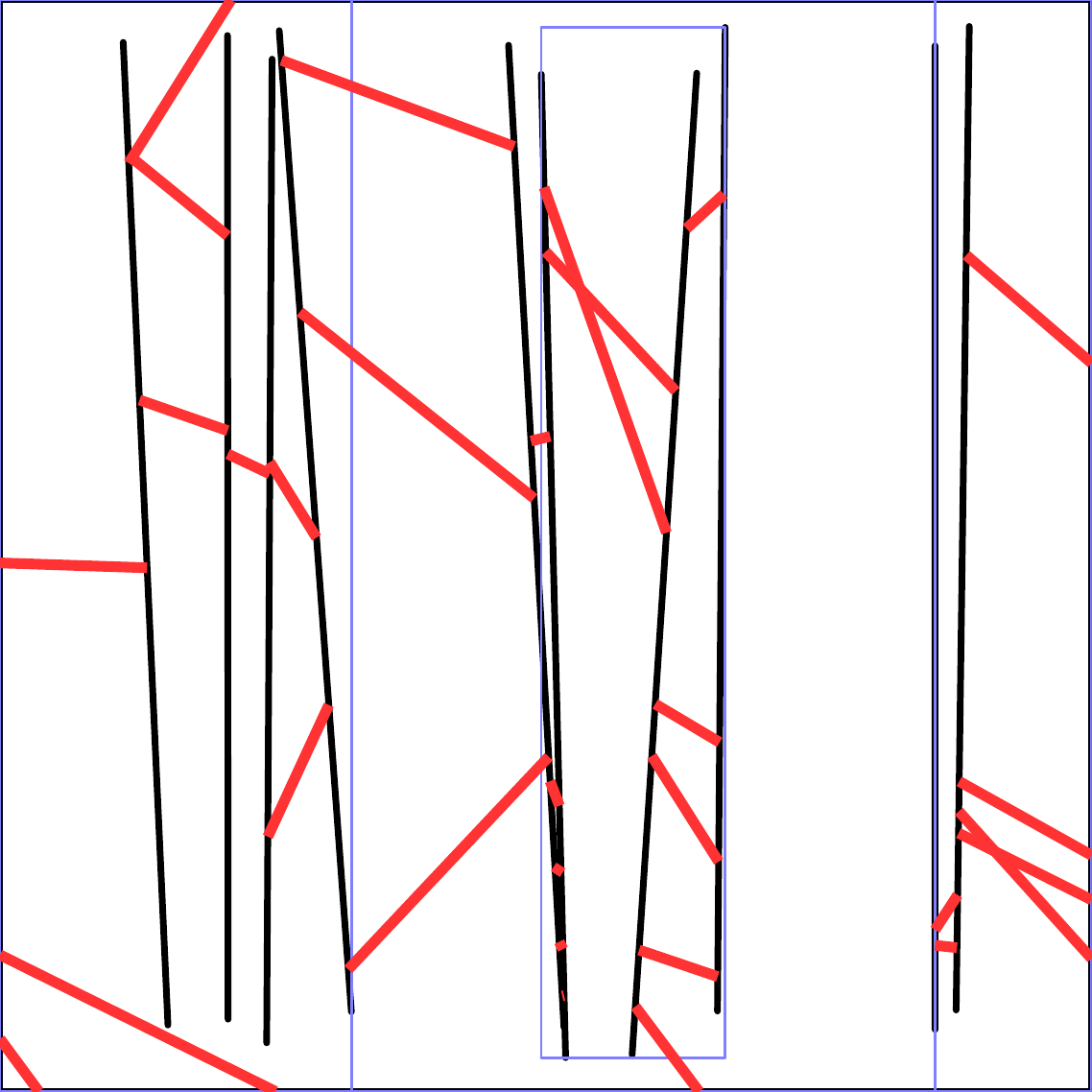}
    {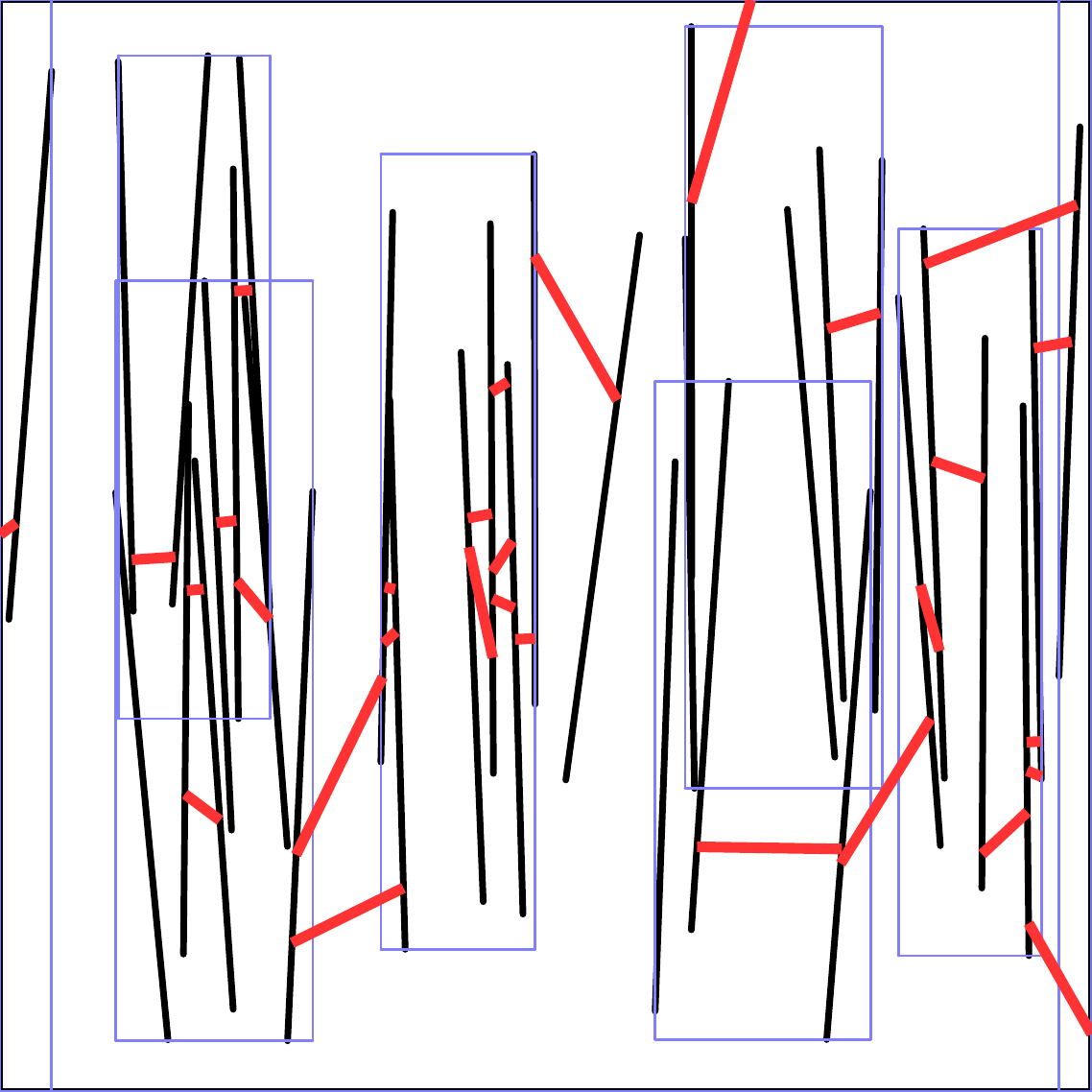}
    {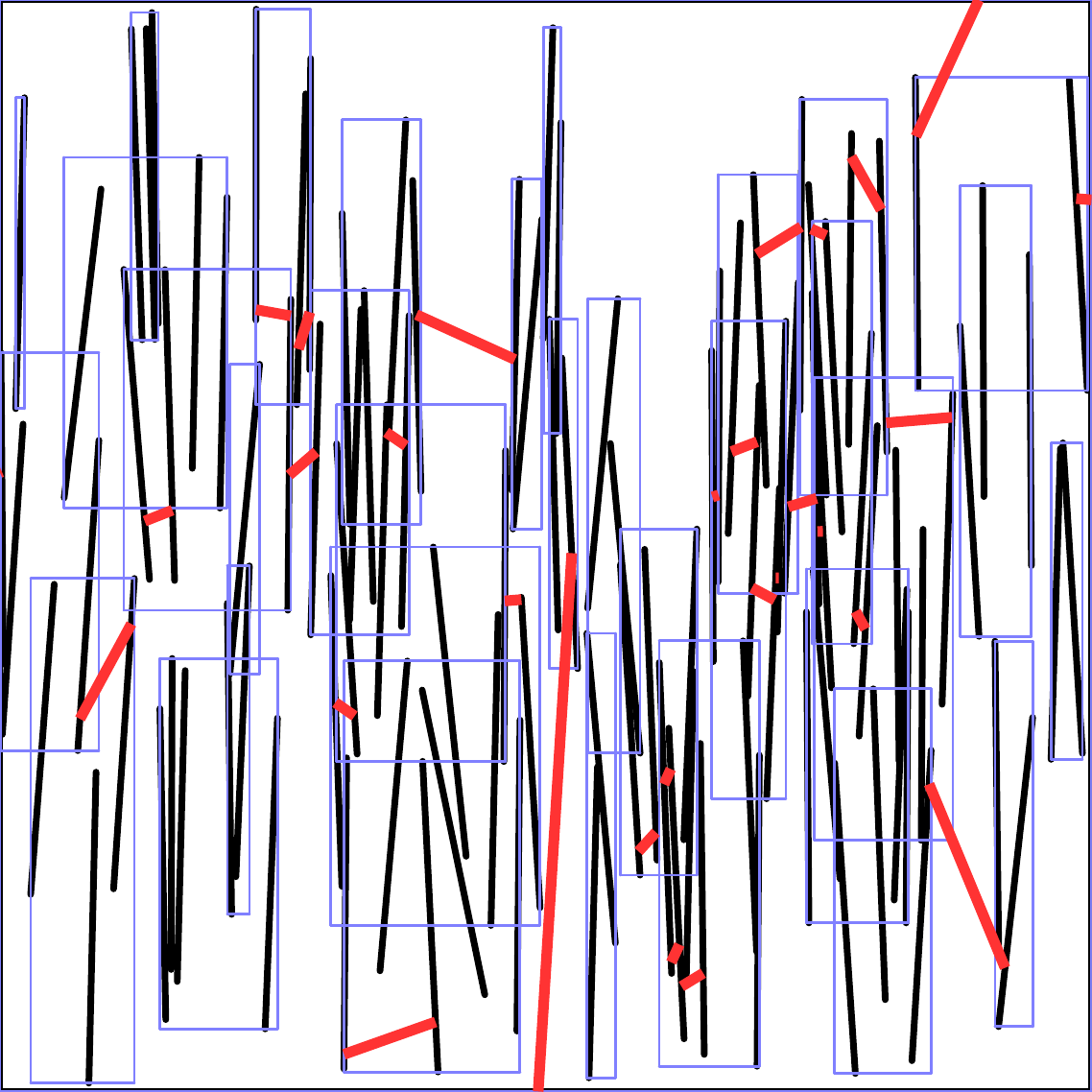}
    {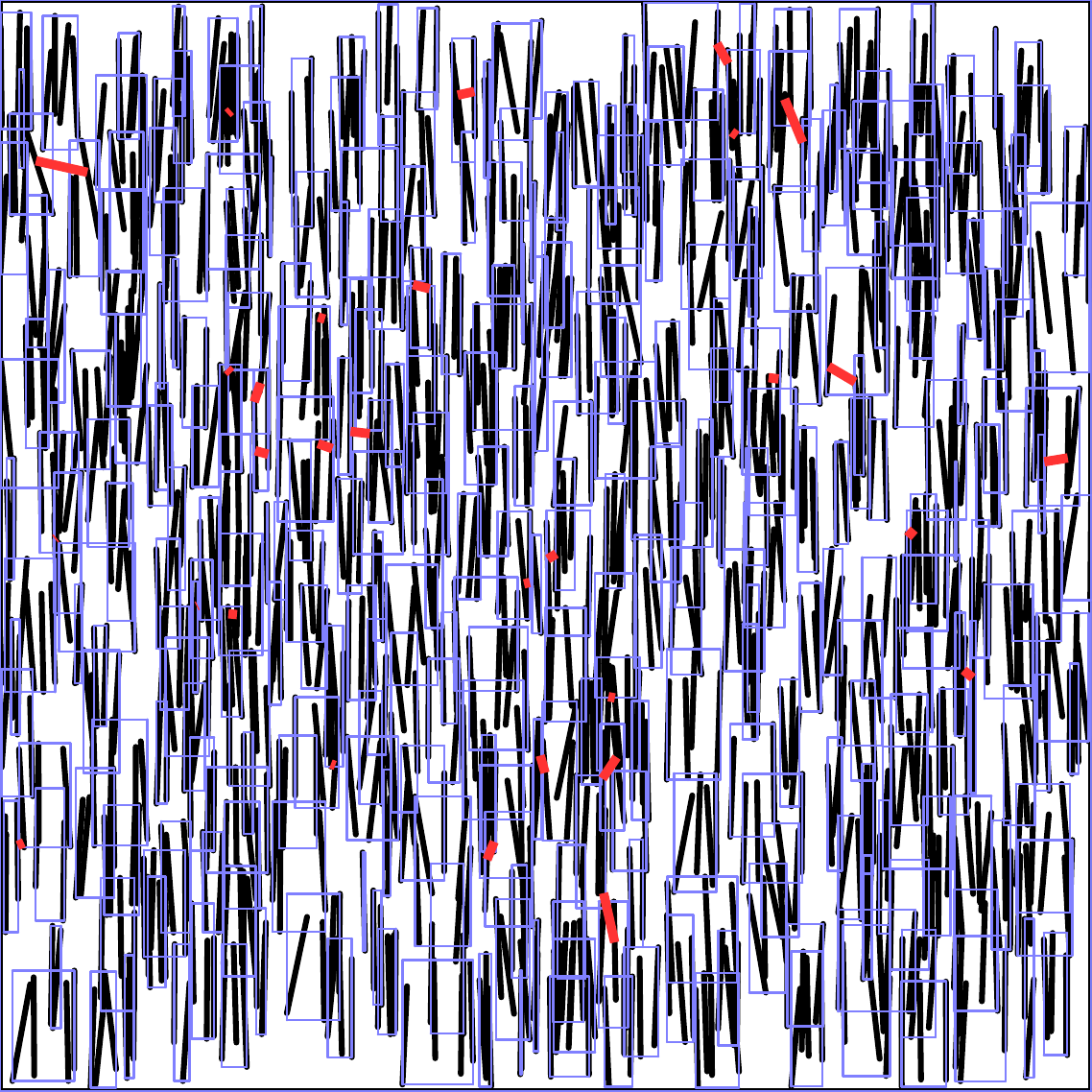}
\accelStructComparisonDlines

\accelStructComparisonA{\linesAccelCompareSize}{\linesAccelCompareSize}
    {Length factor 3 -- Uniform orientation}
    {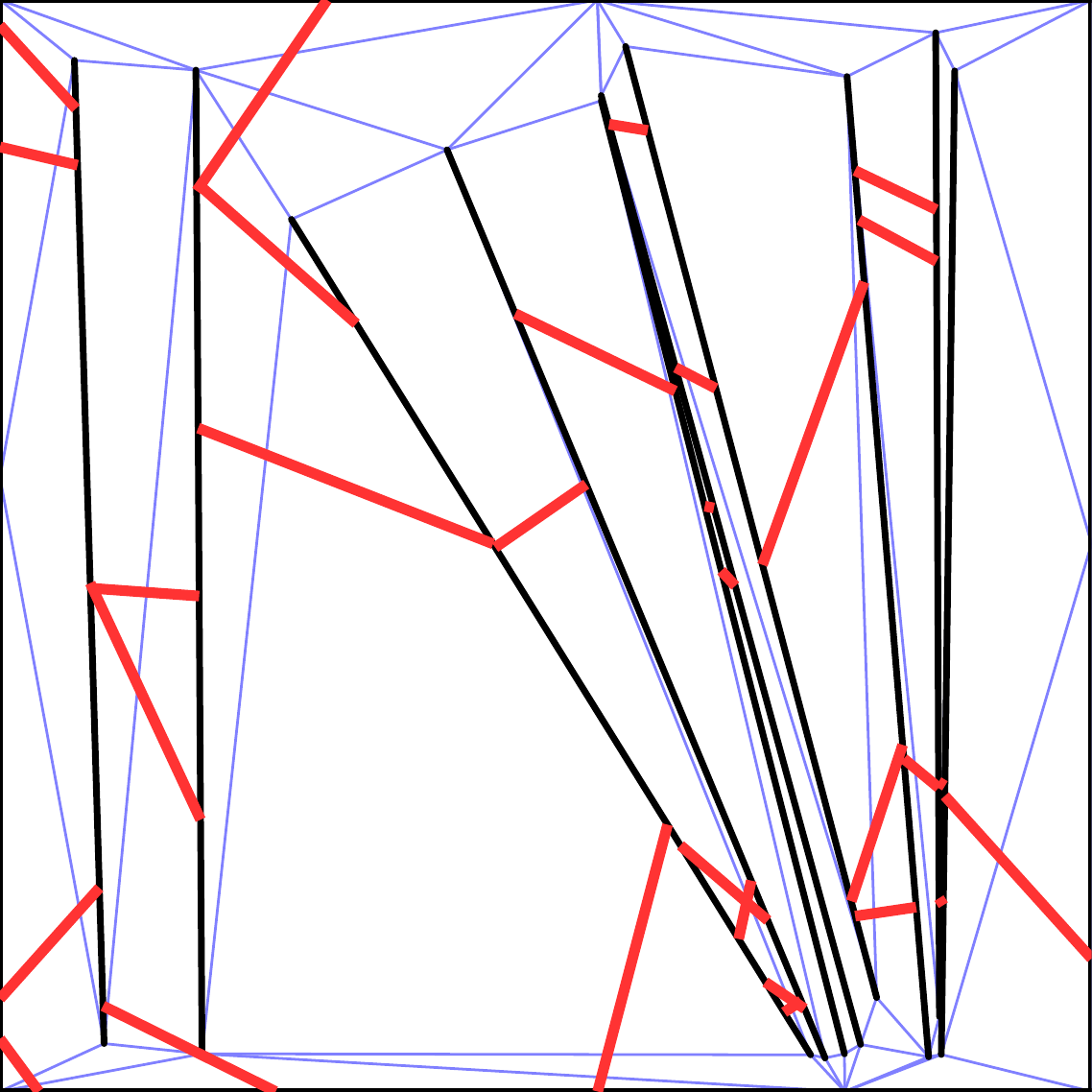}
    {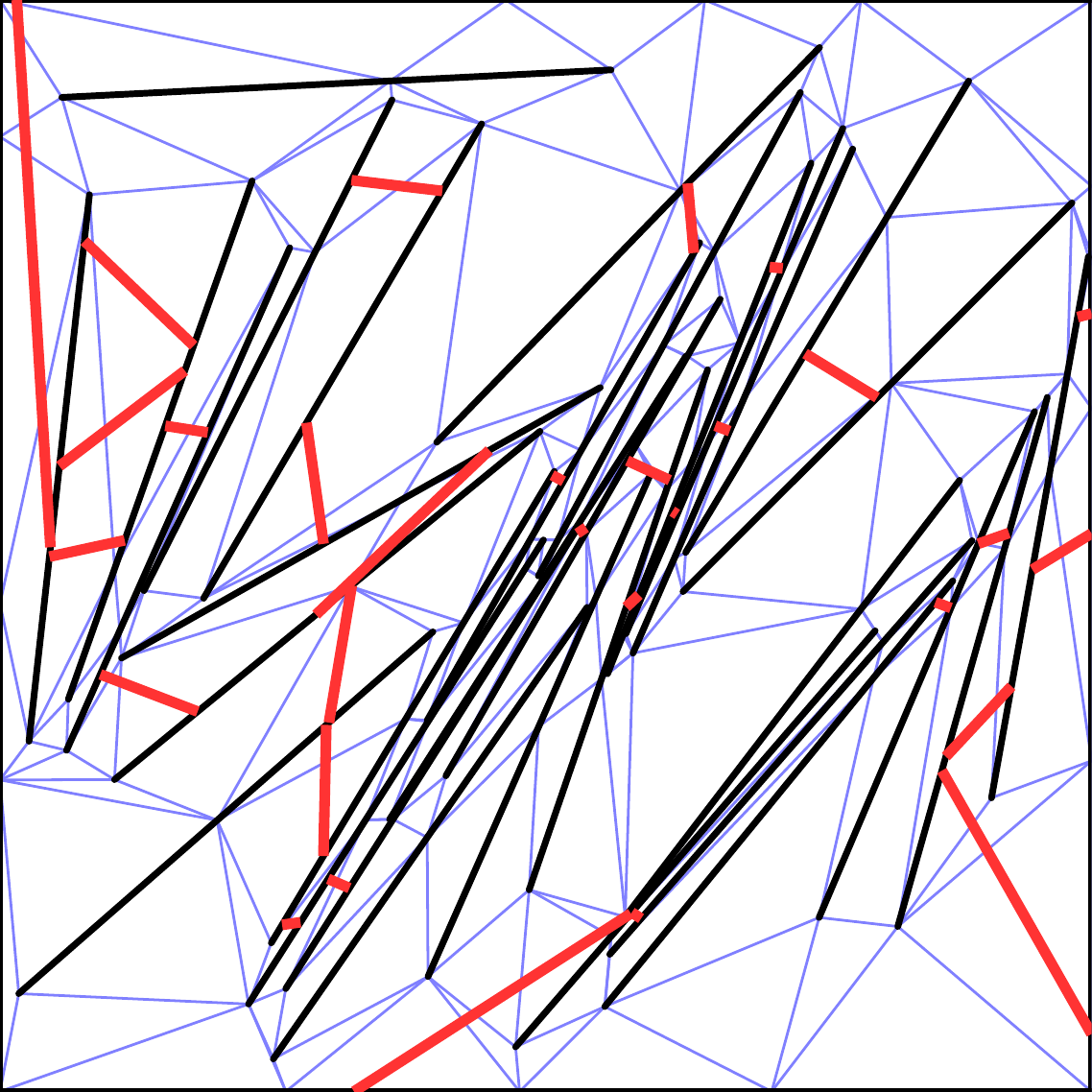}
    {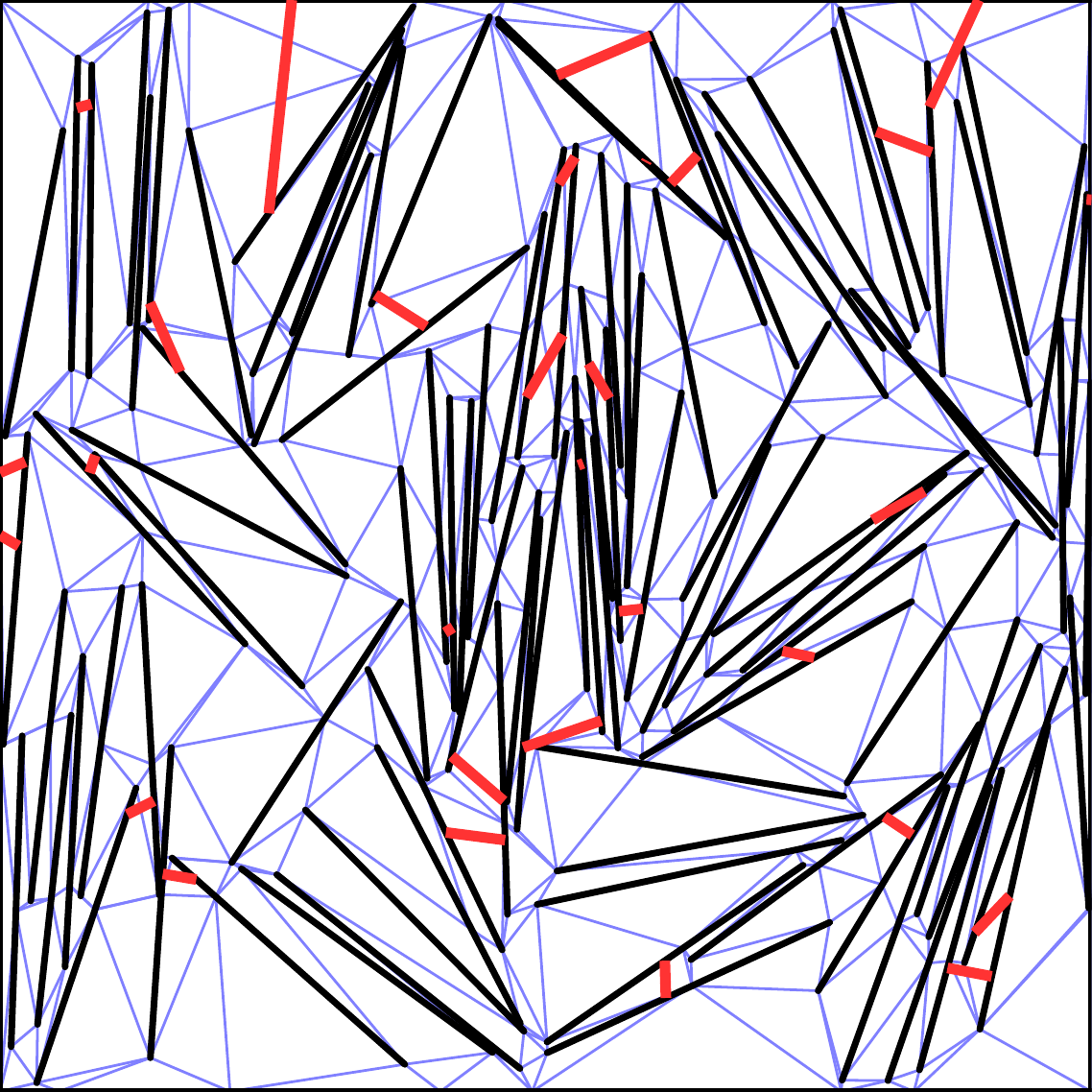}
    {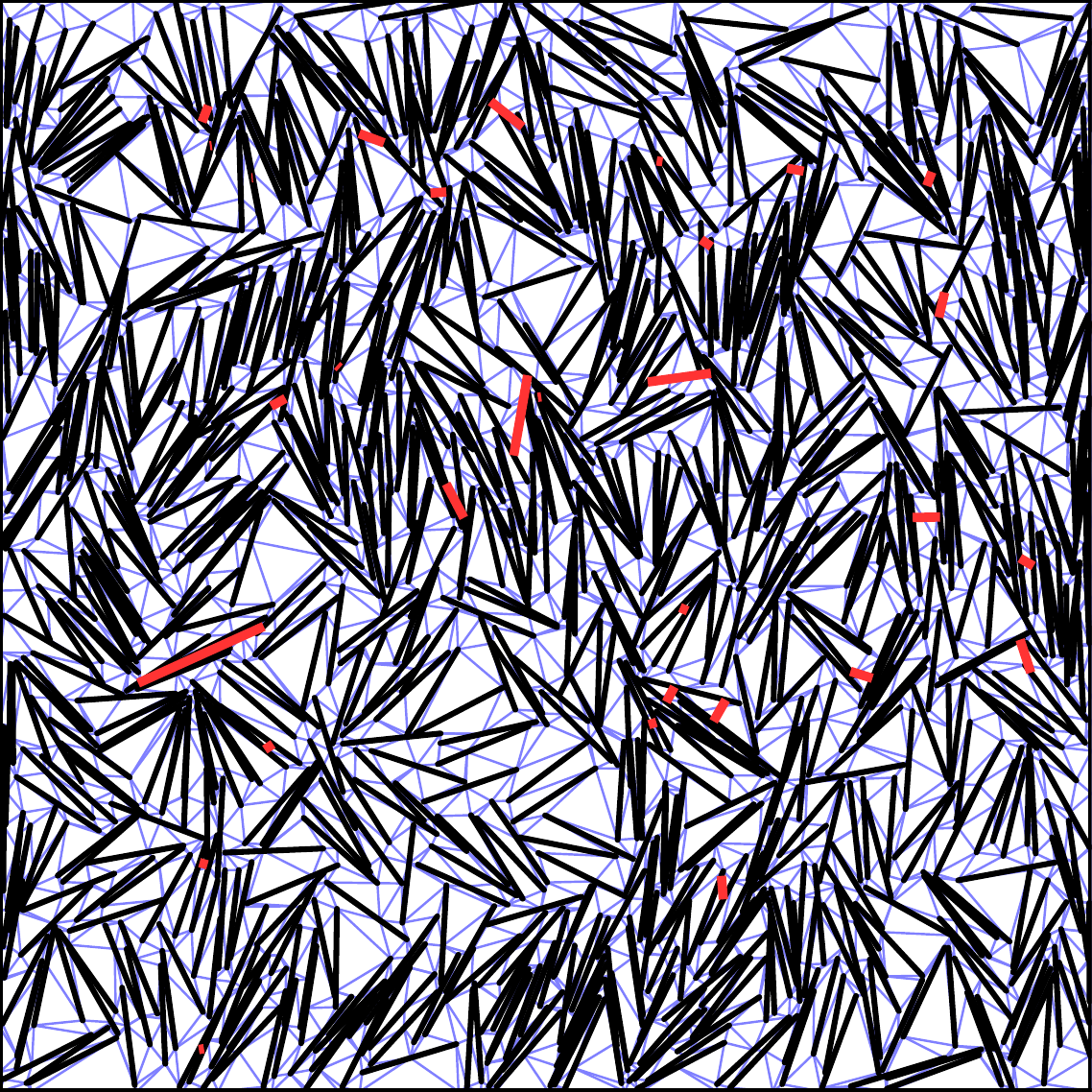}
\accelStructComparisonB{\linesAccelCompareSize}{\linesAccelCompareSize}
    {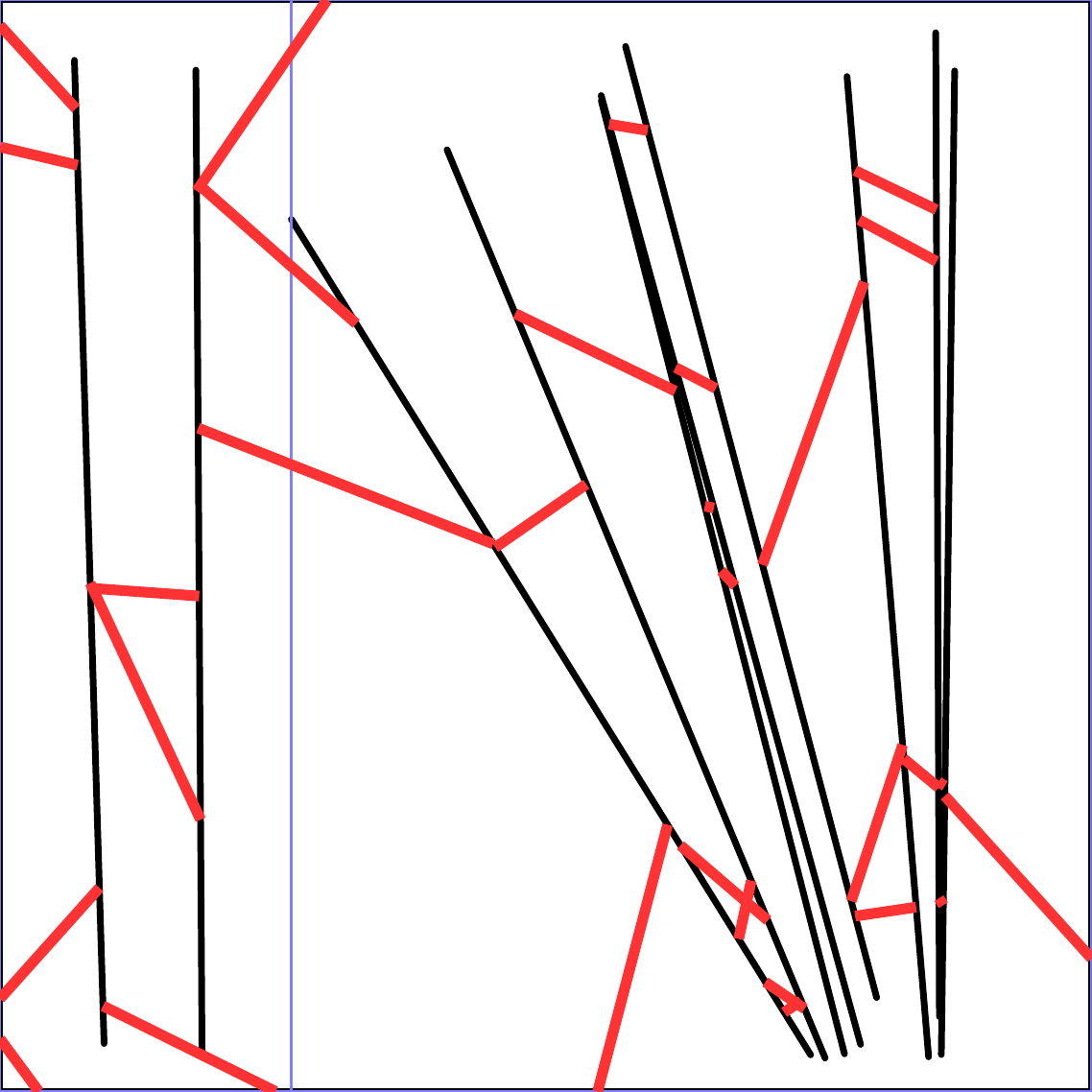}
    {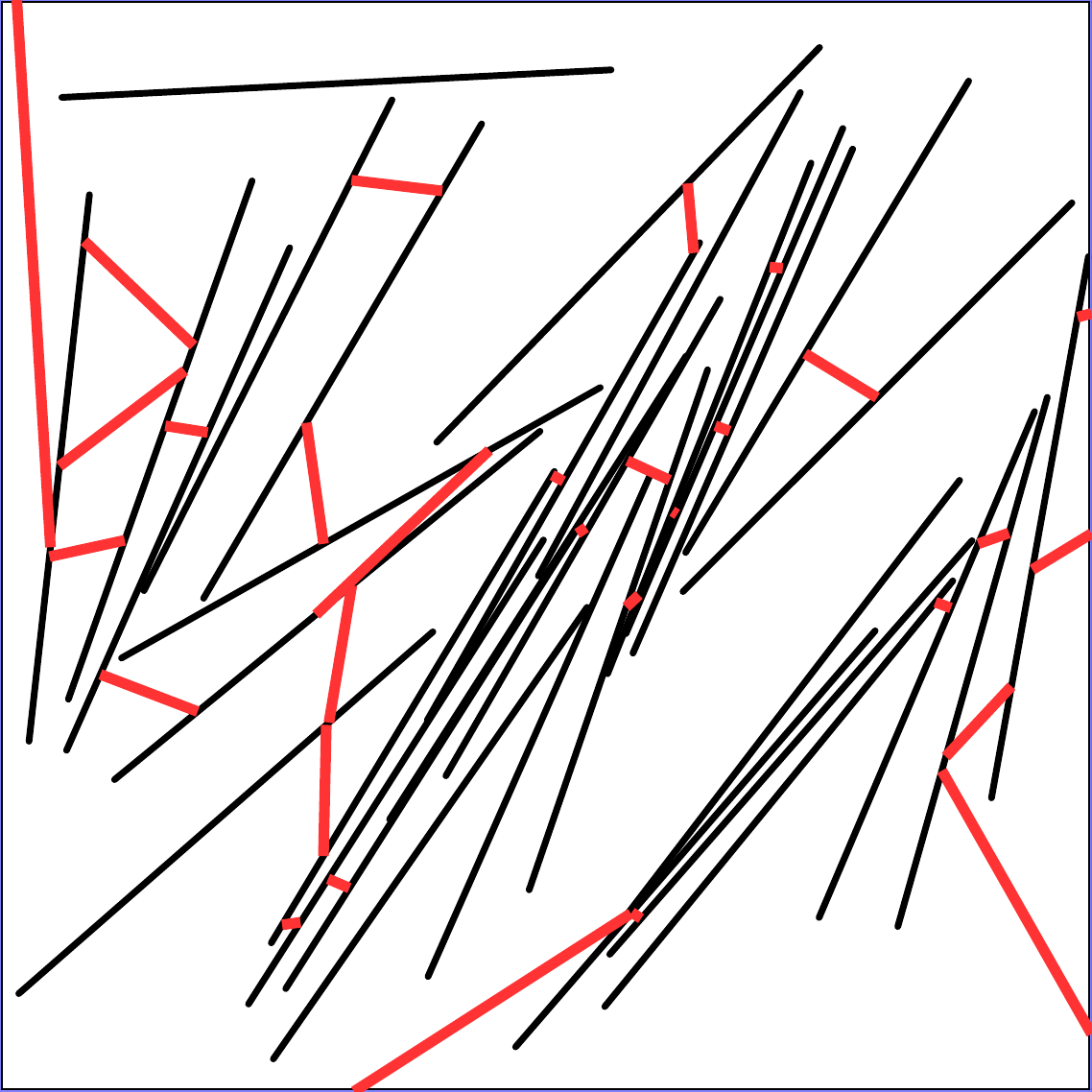}
    {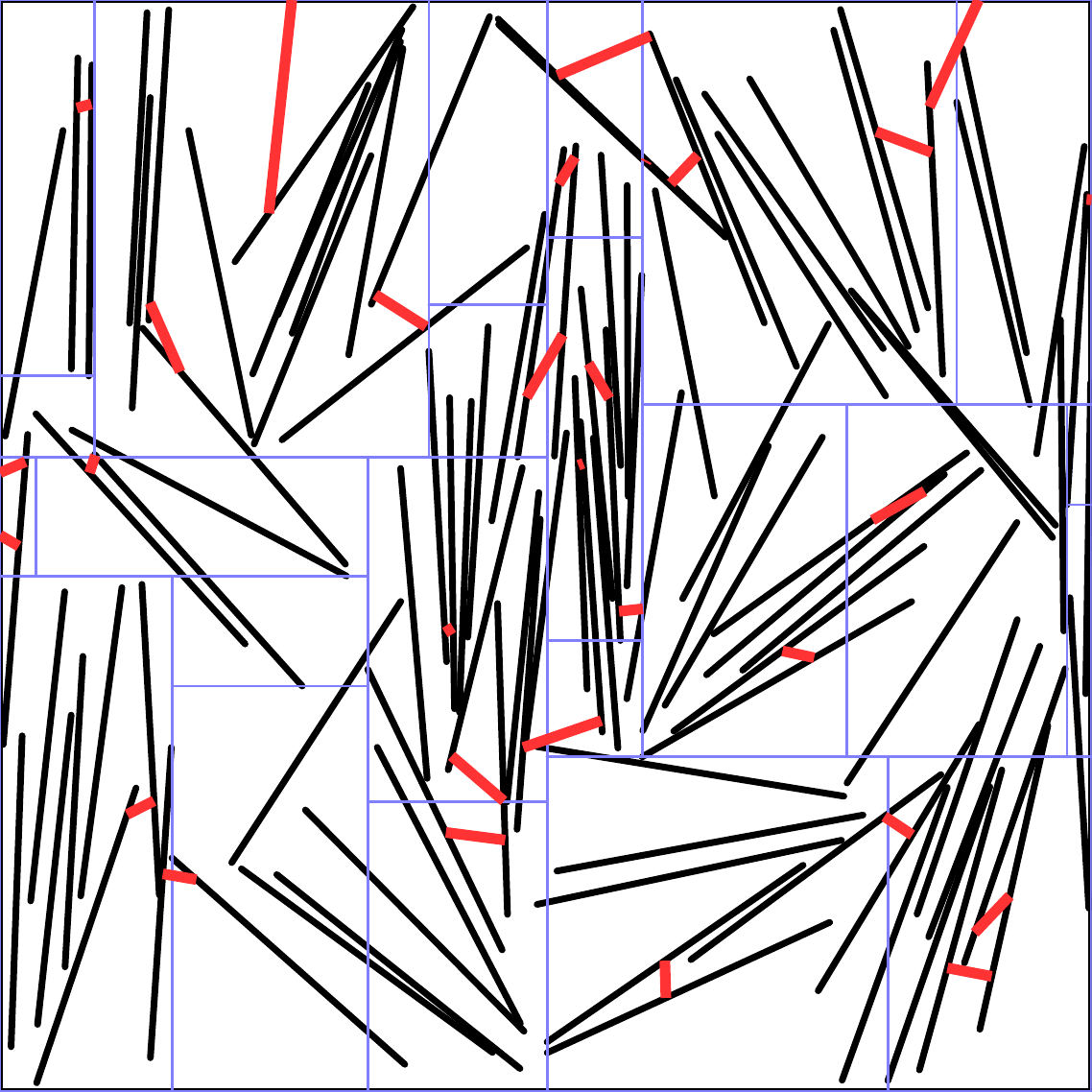}
    {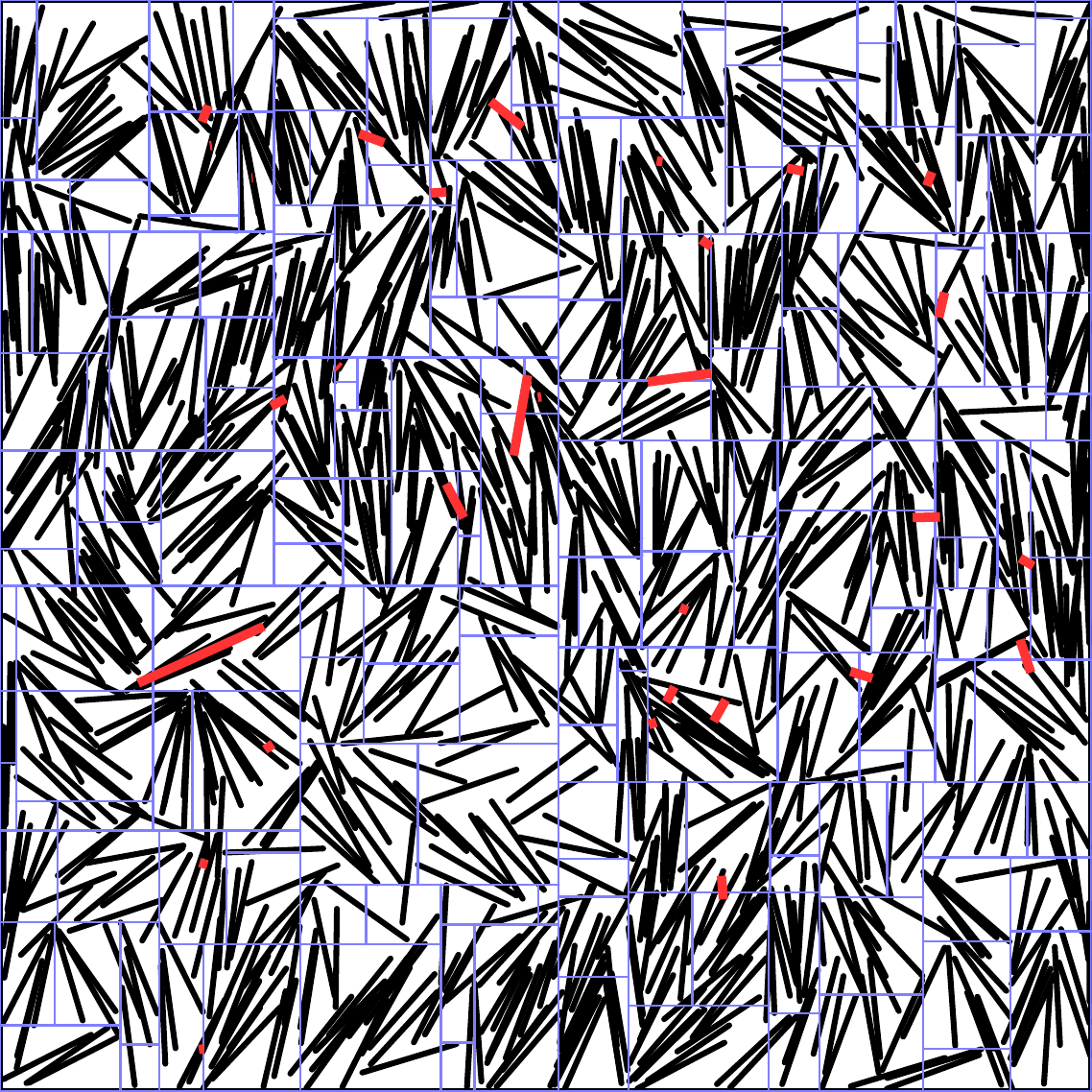}
\accelStructComparisonC{\linesAccelCompareSize}{\linesAccelCompareSize}
    {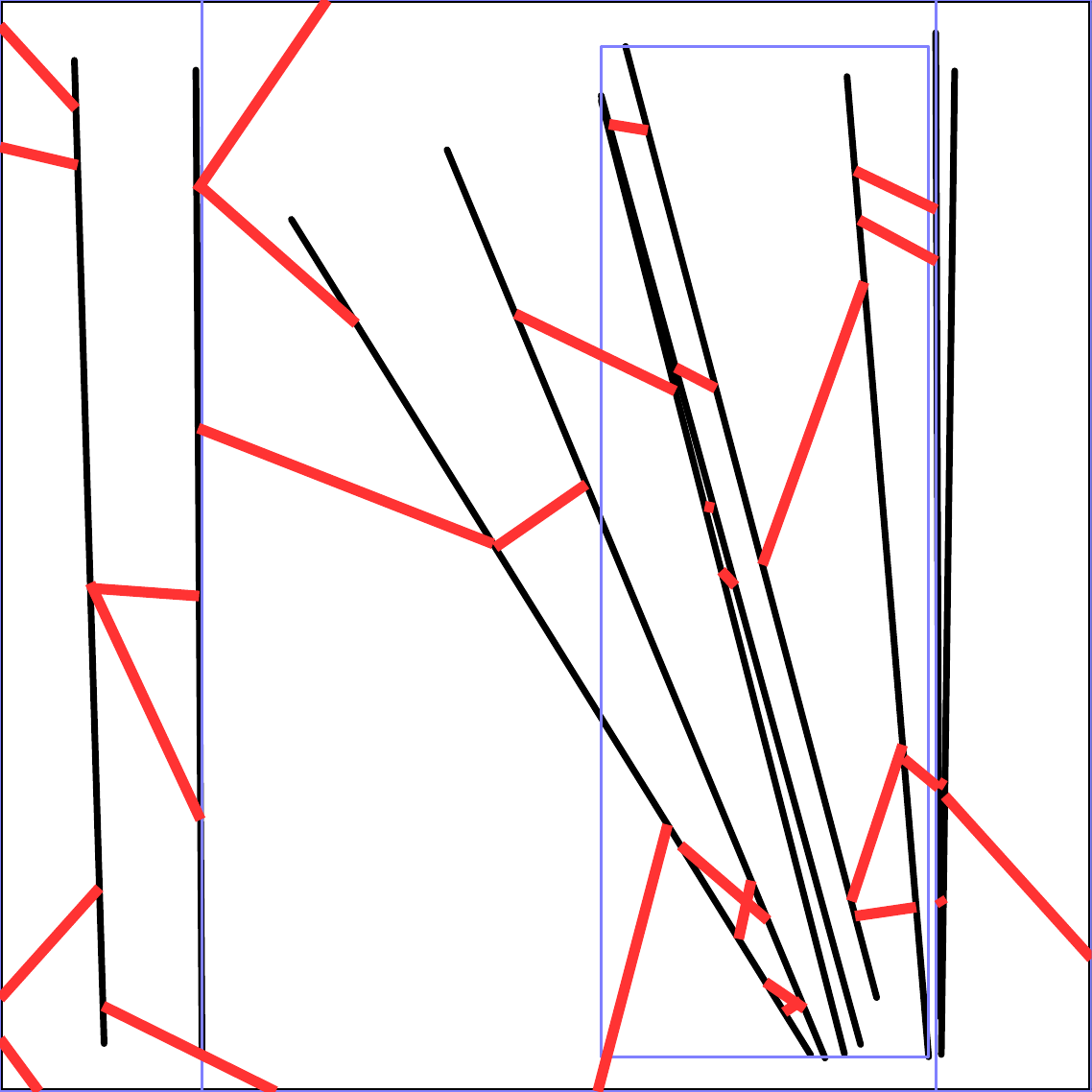}
    {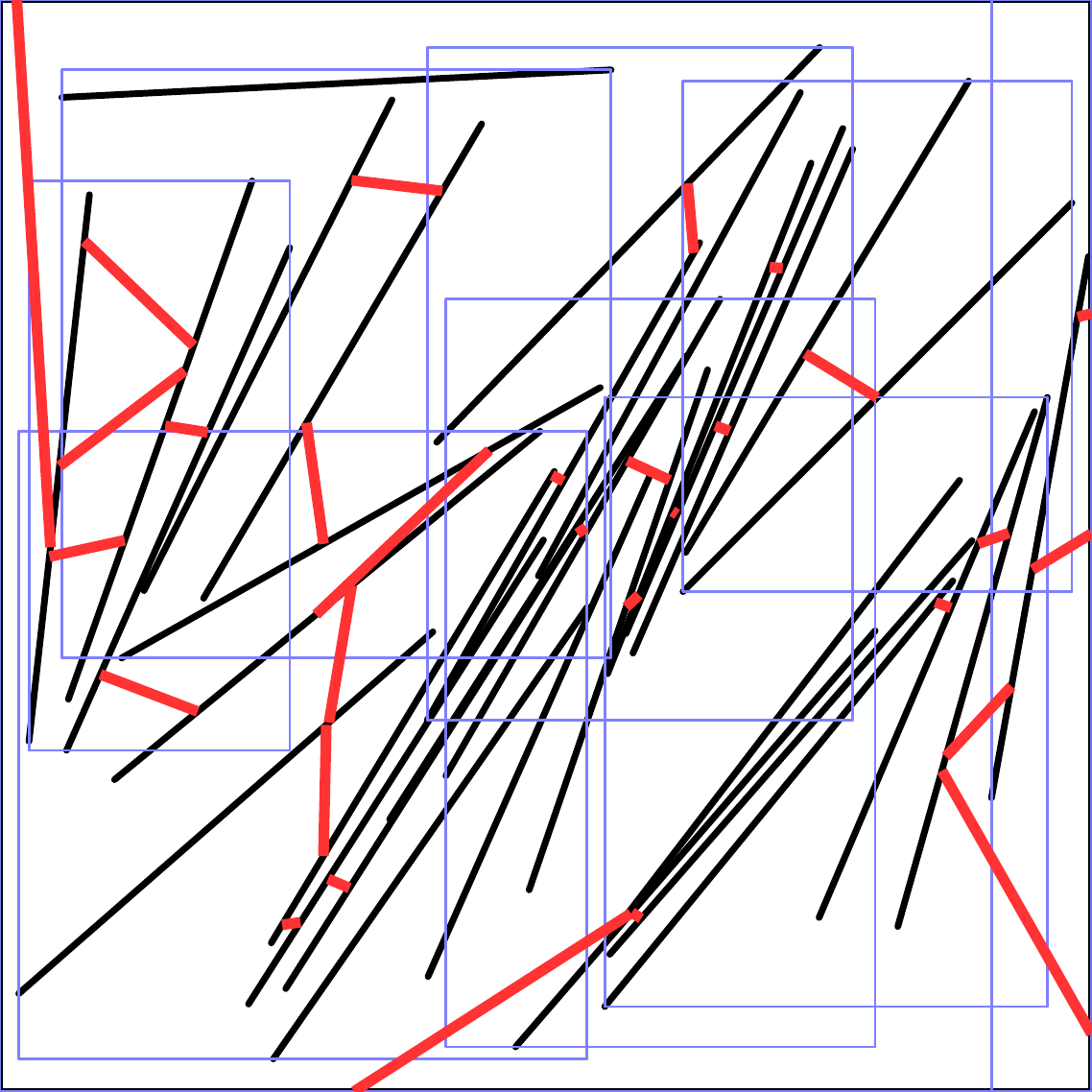}
    {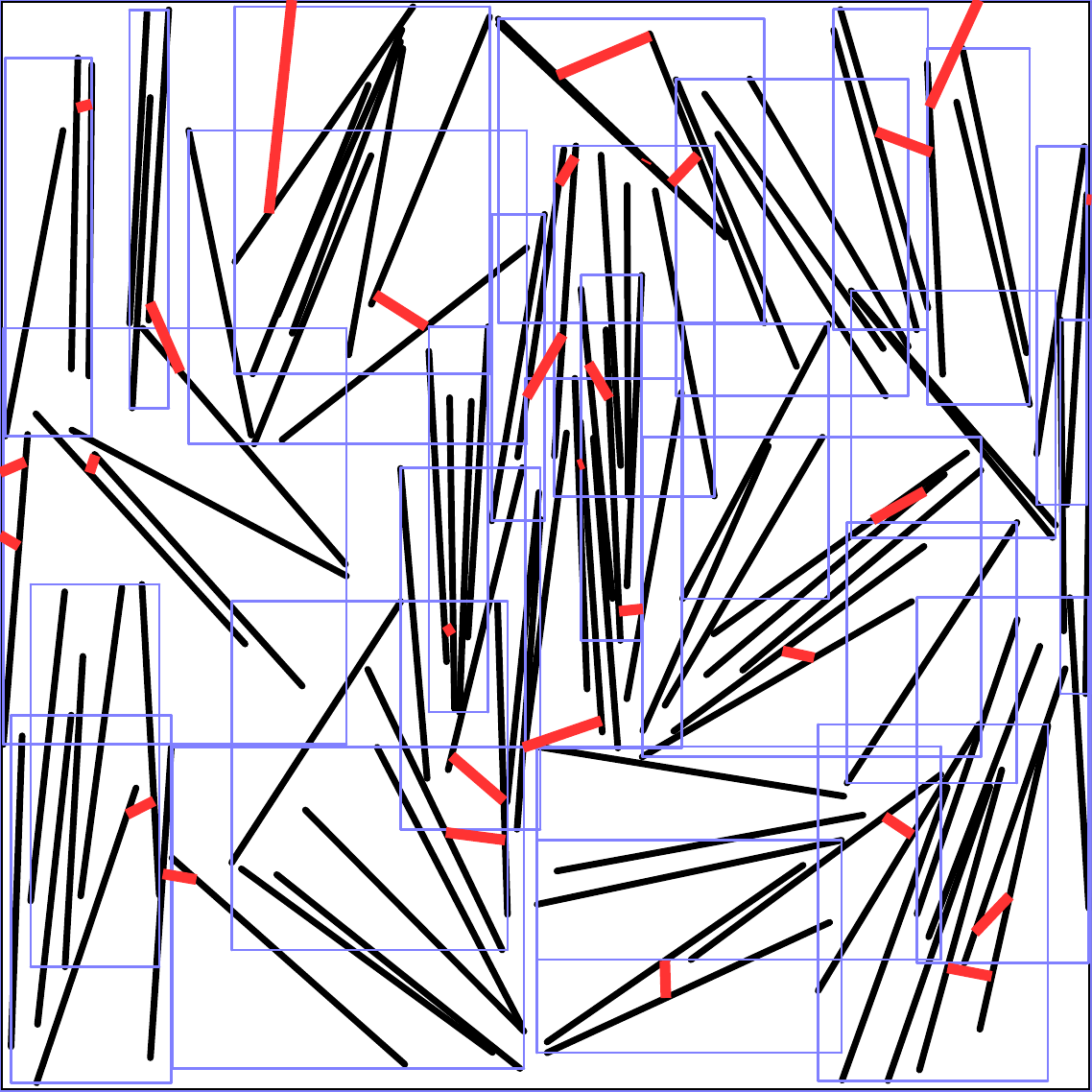}
    {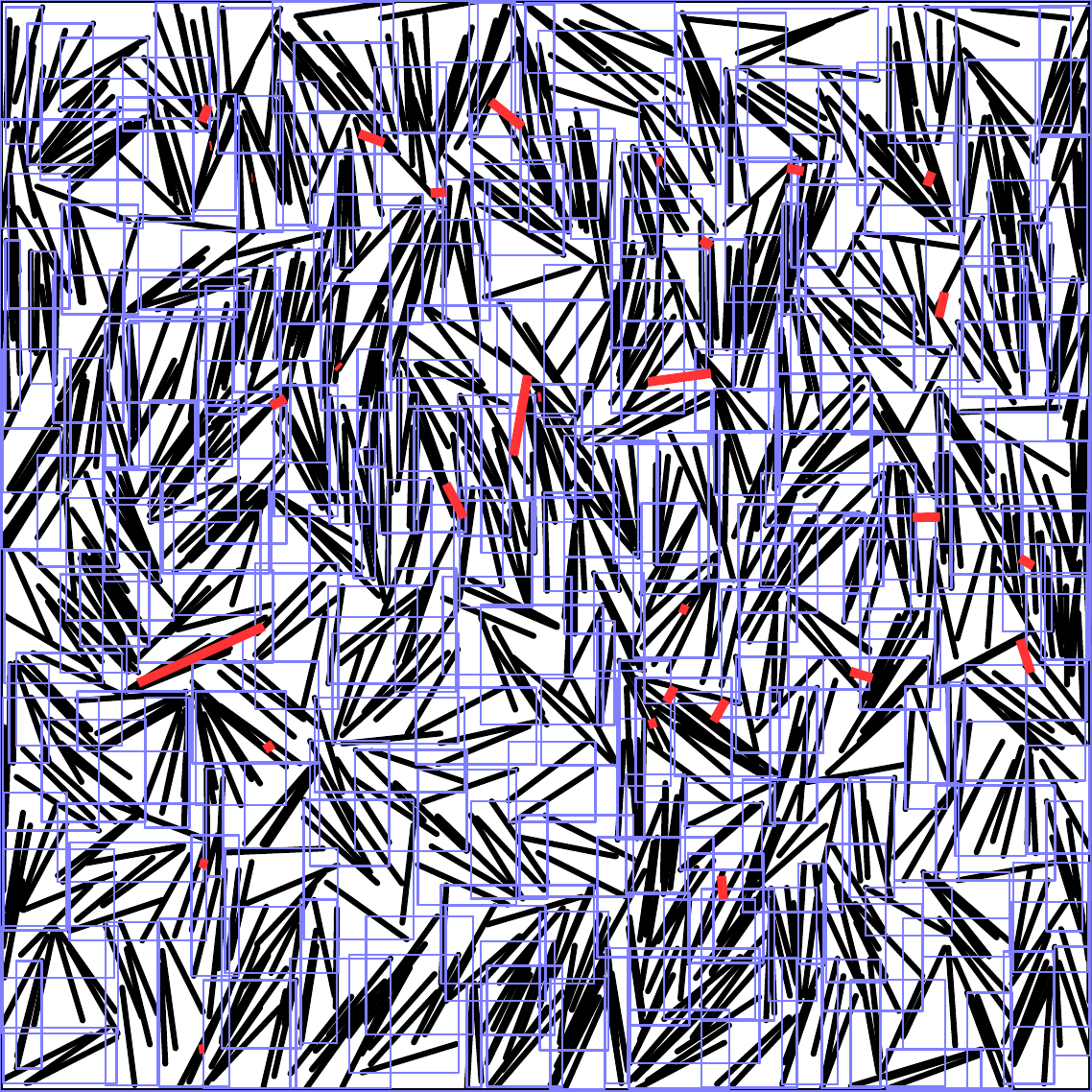}
\accelStructComparisonDlines

\accelStructComparisonA{\linesAccelCompareSize}{\linesAccelCompareSize}
    {Length factor 3 -- Diagonal orientation}
    {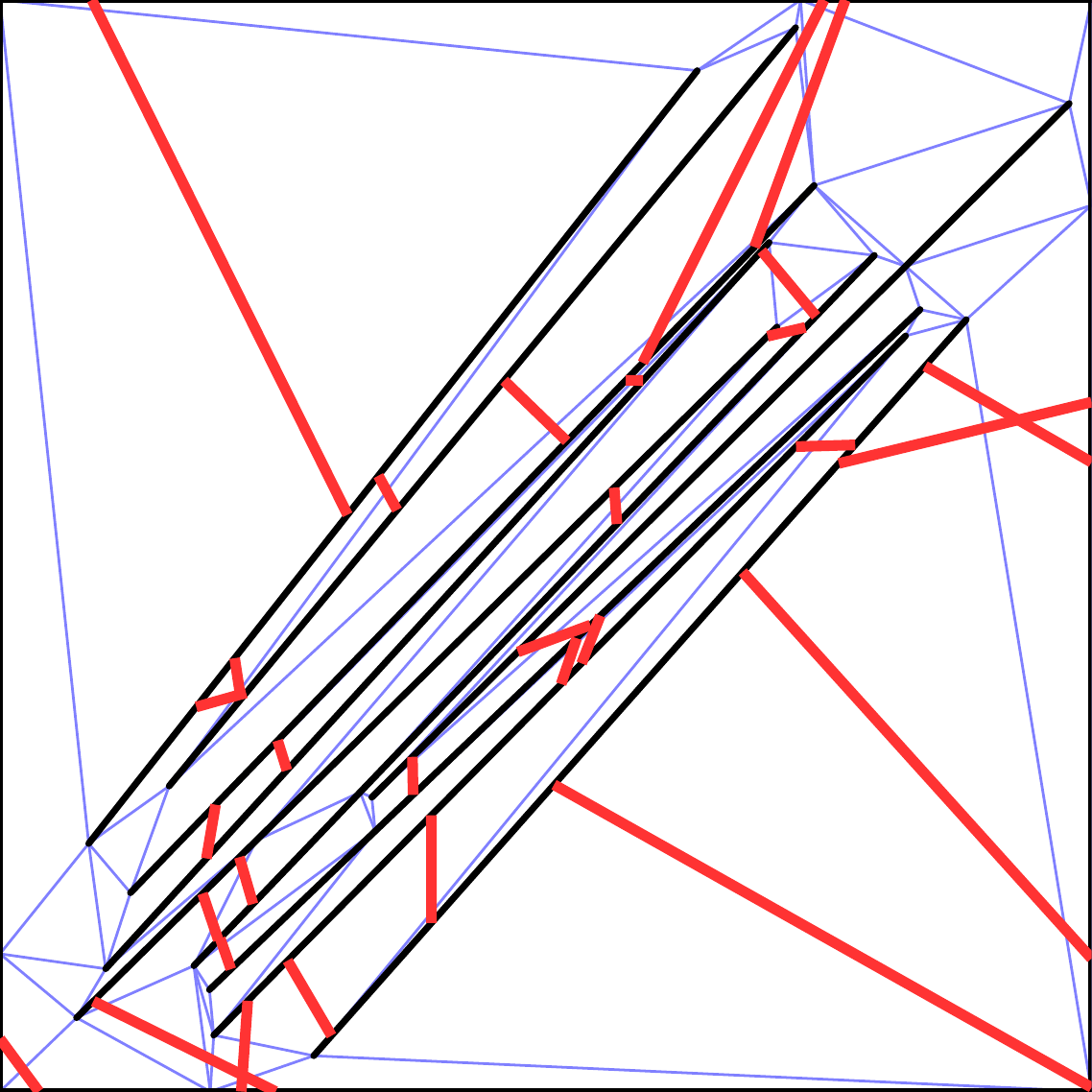}
    {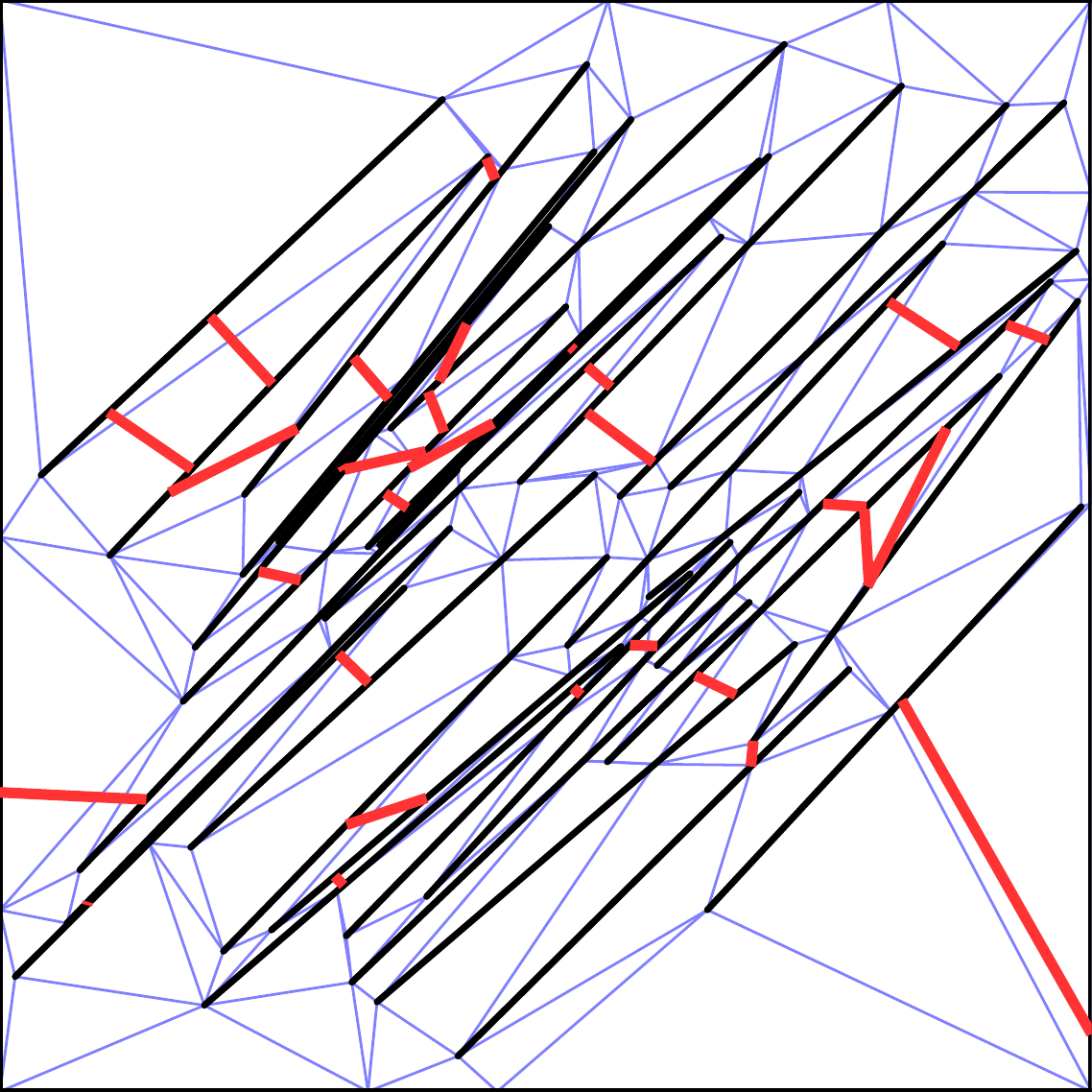}
    {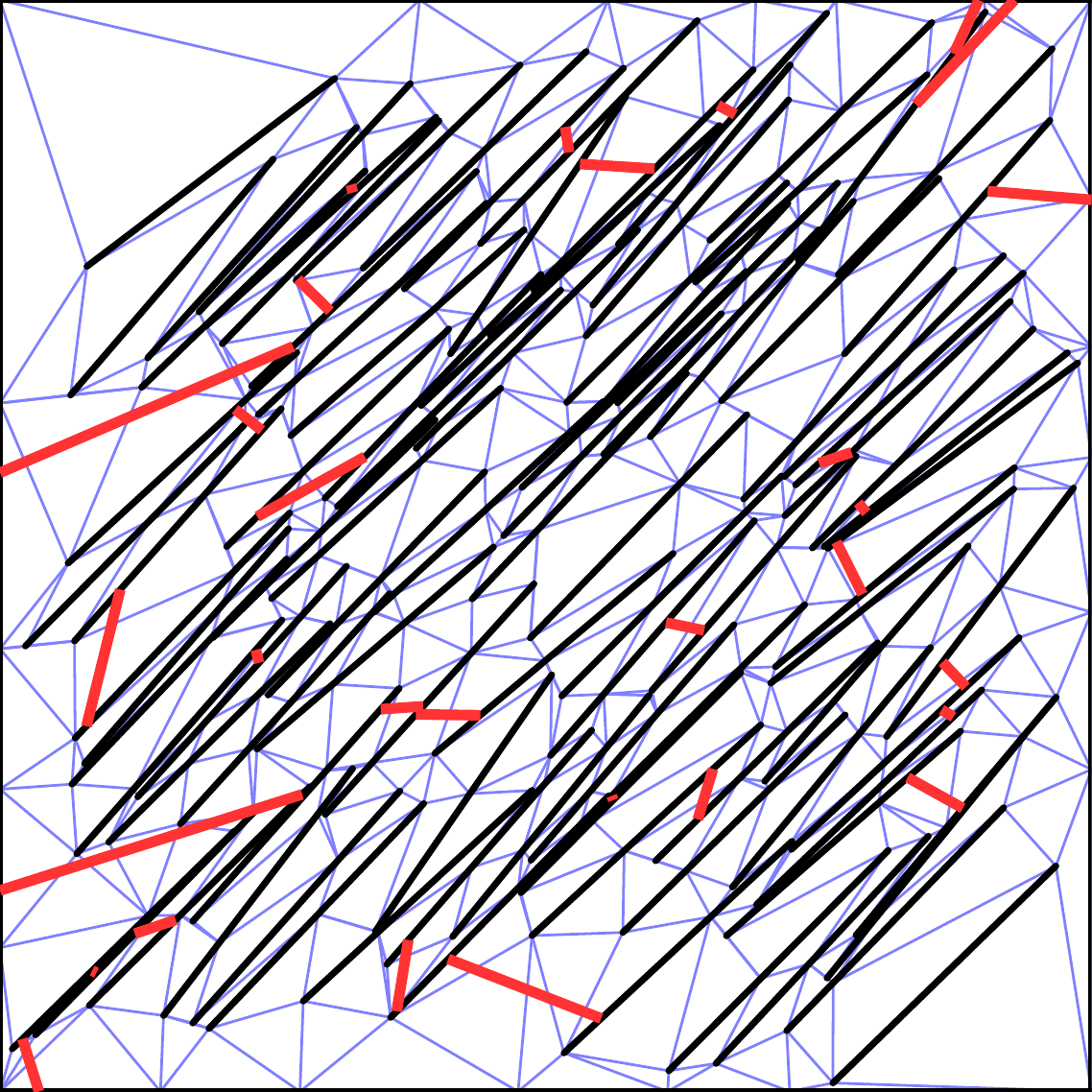}
    {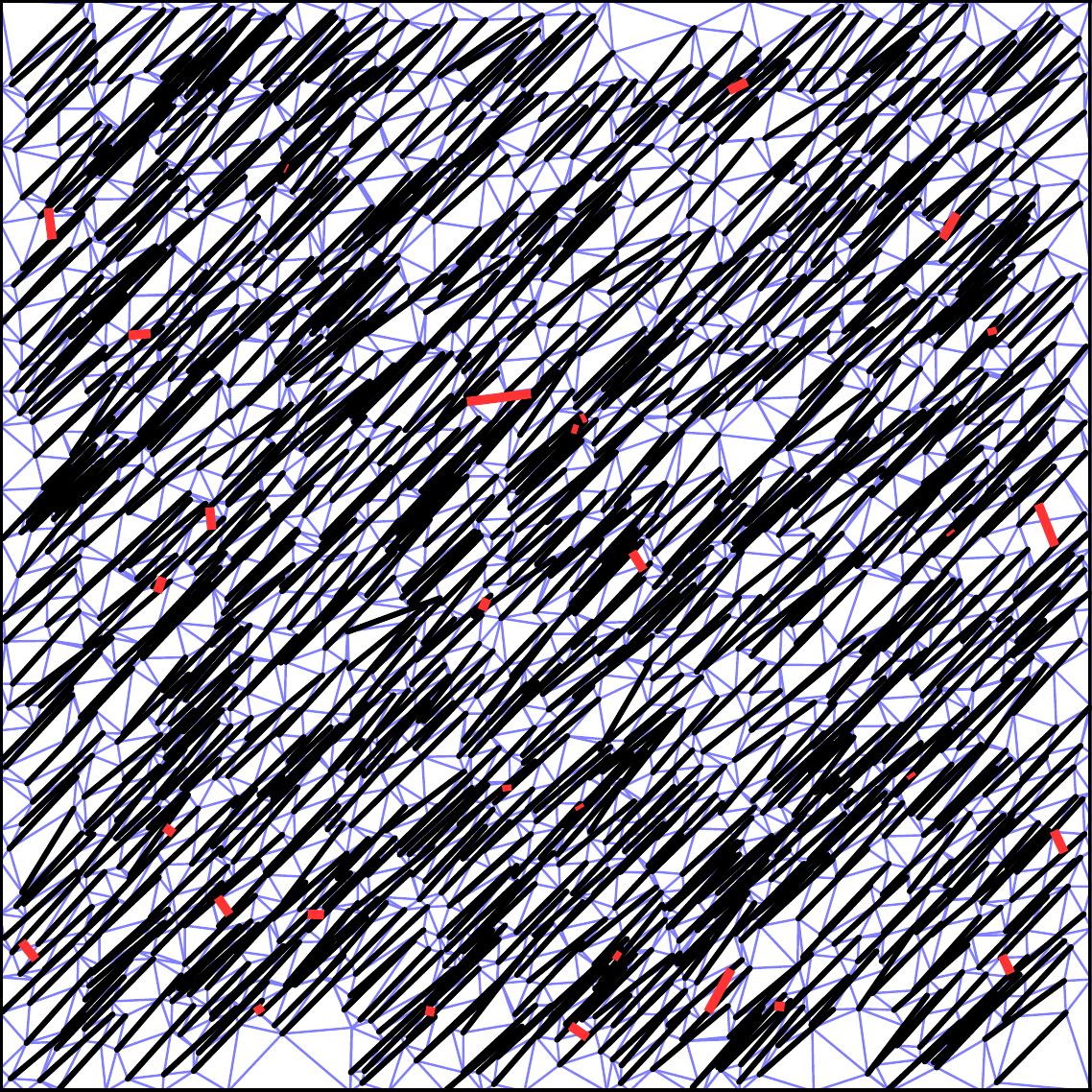}
\accelStructComparisonB{\linesAccelCompareSize}{\linesAccelCompareSize}
    {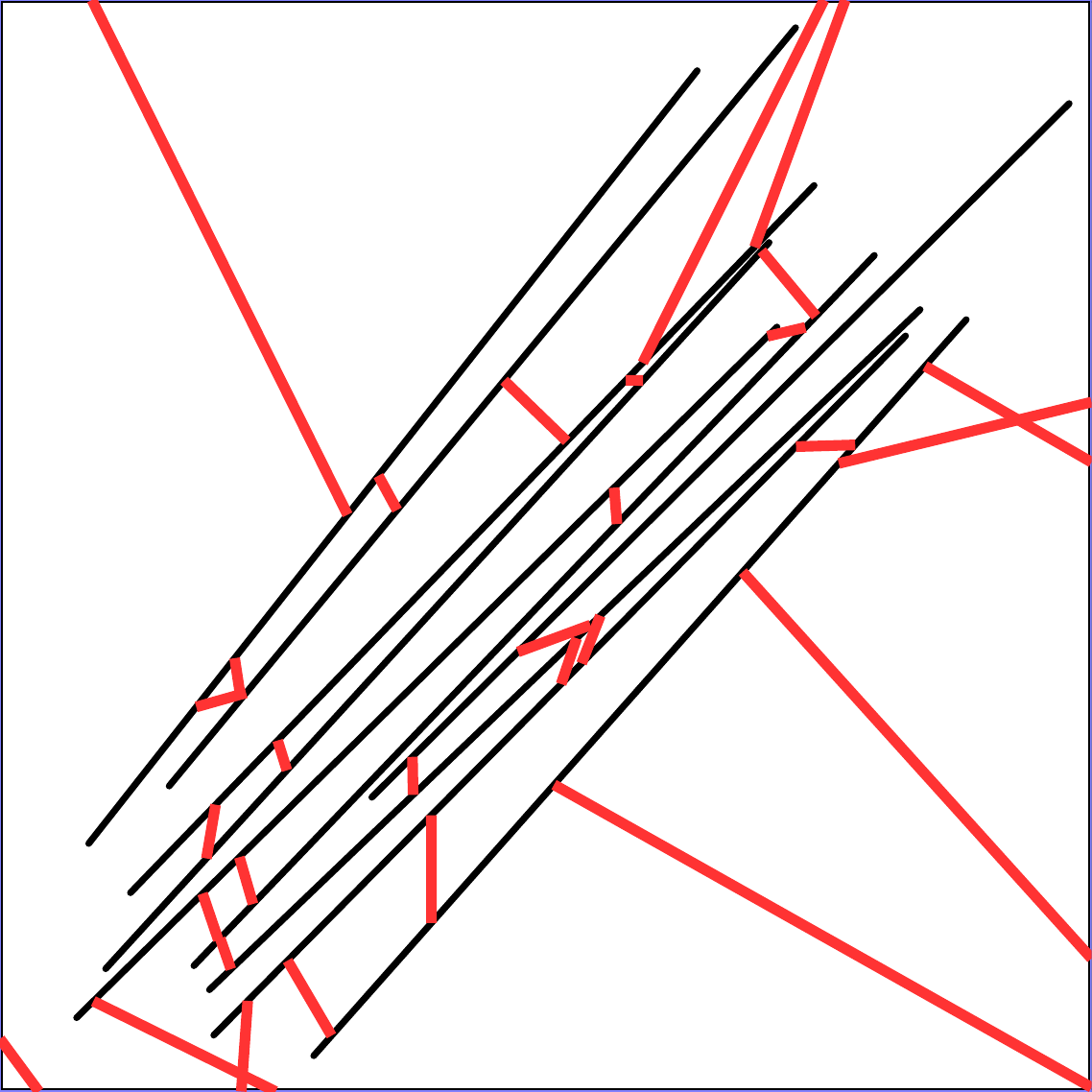}
    {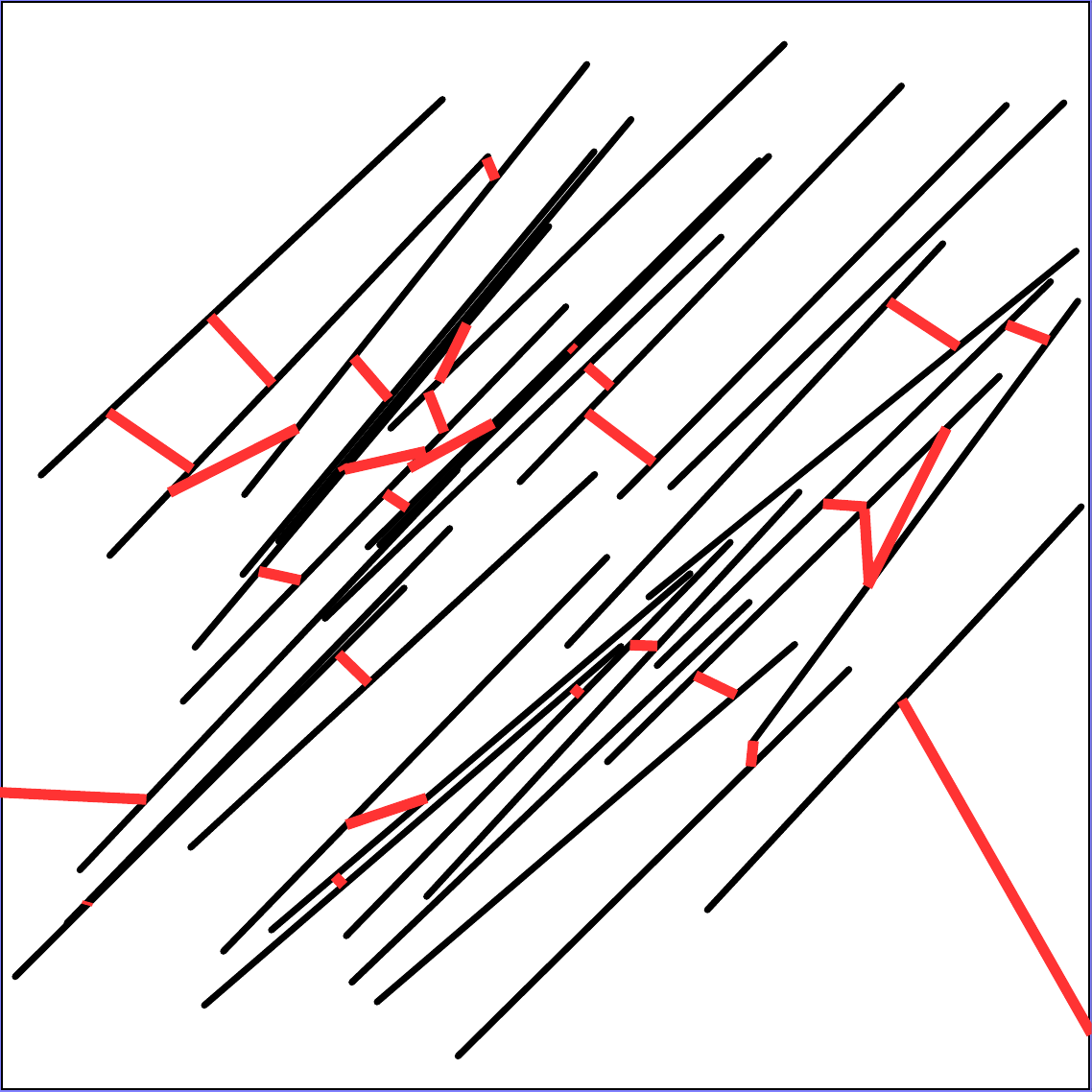}
    {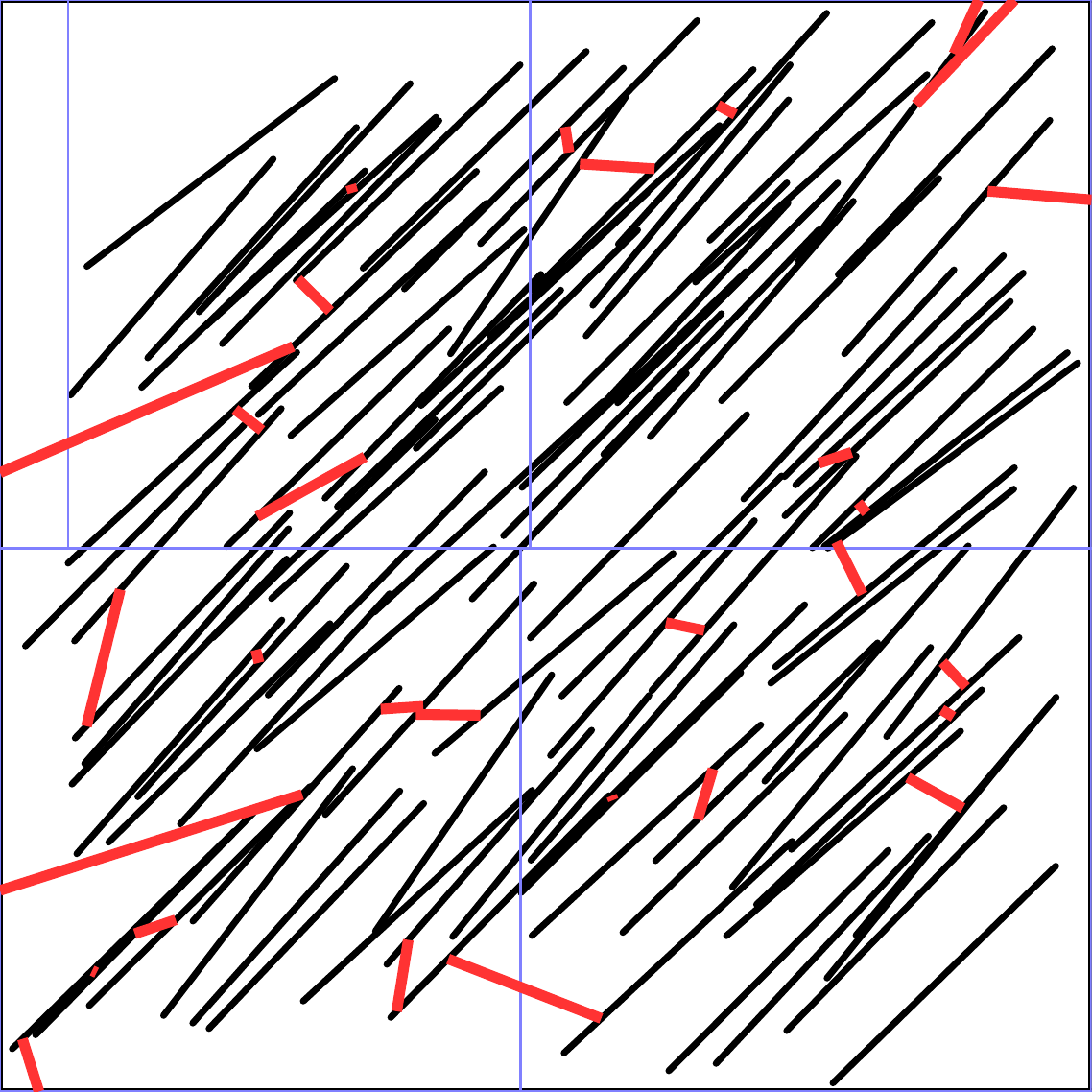}
    {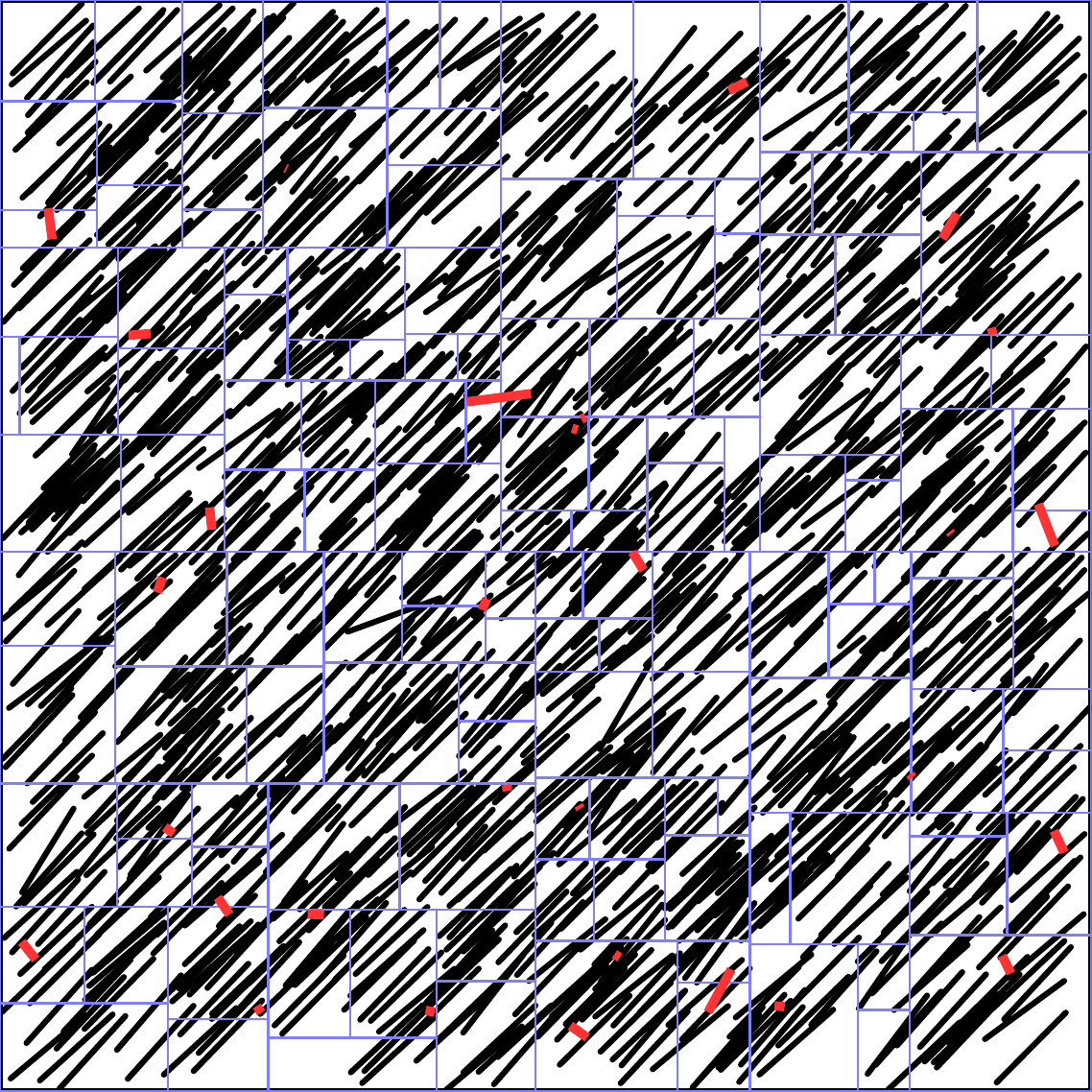}
\accelStructComparisonC{\linesAccelCompareSize}{\linesAccelCompareSize}
    {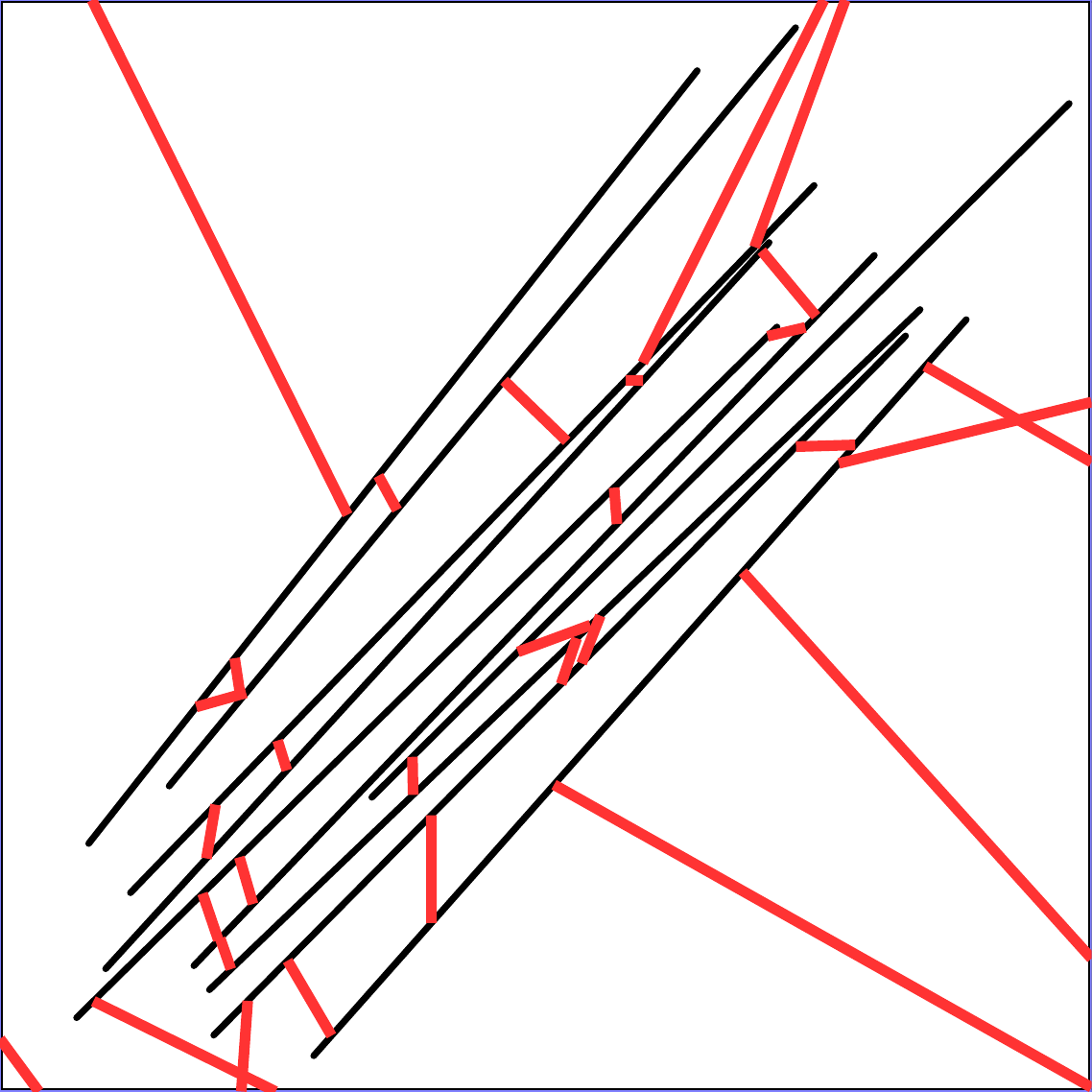}
    {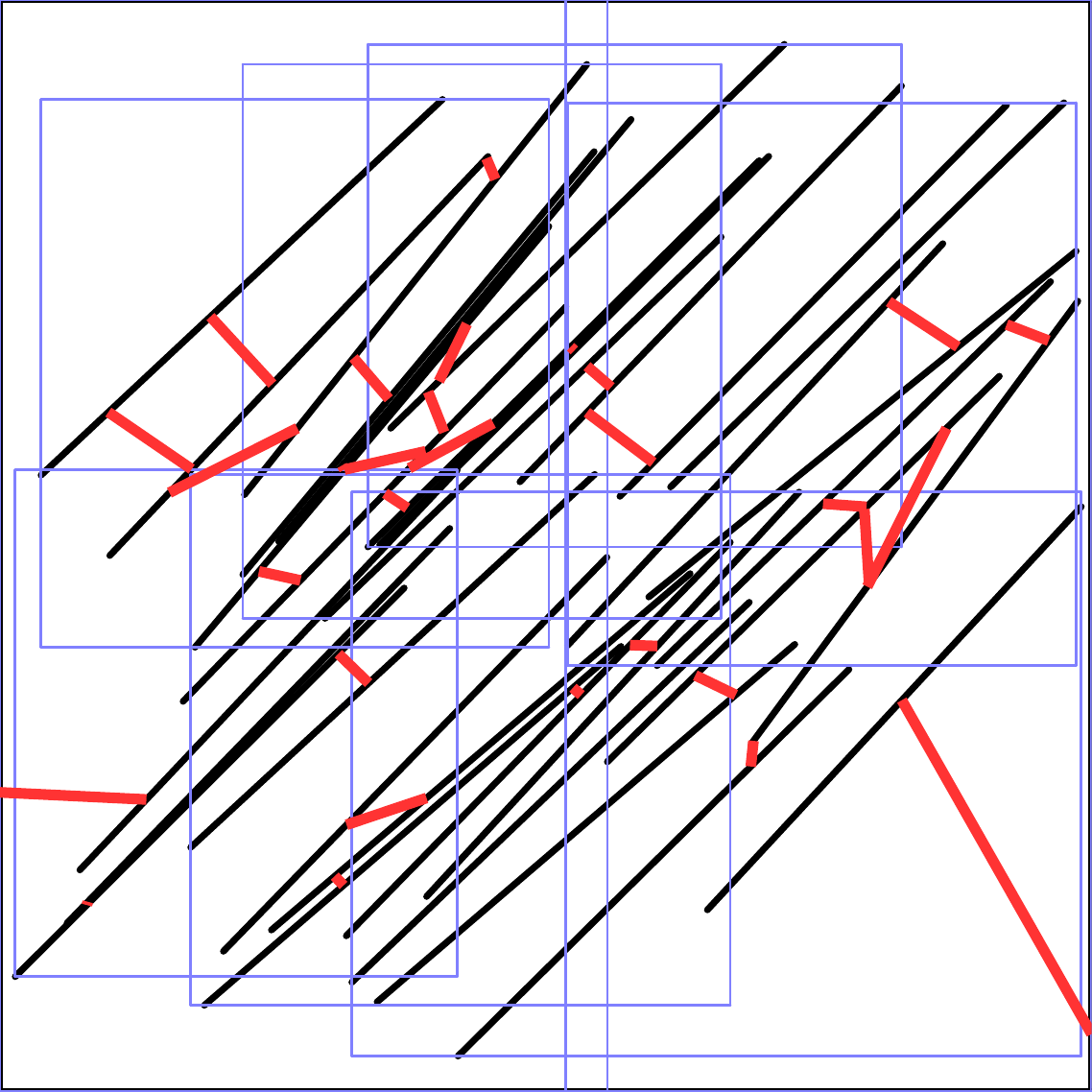}
    {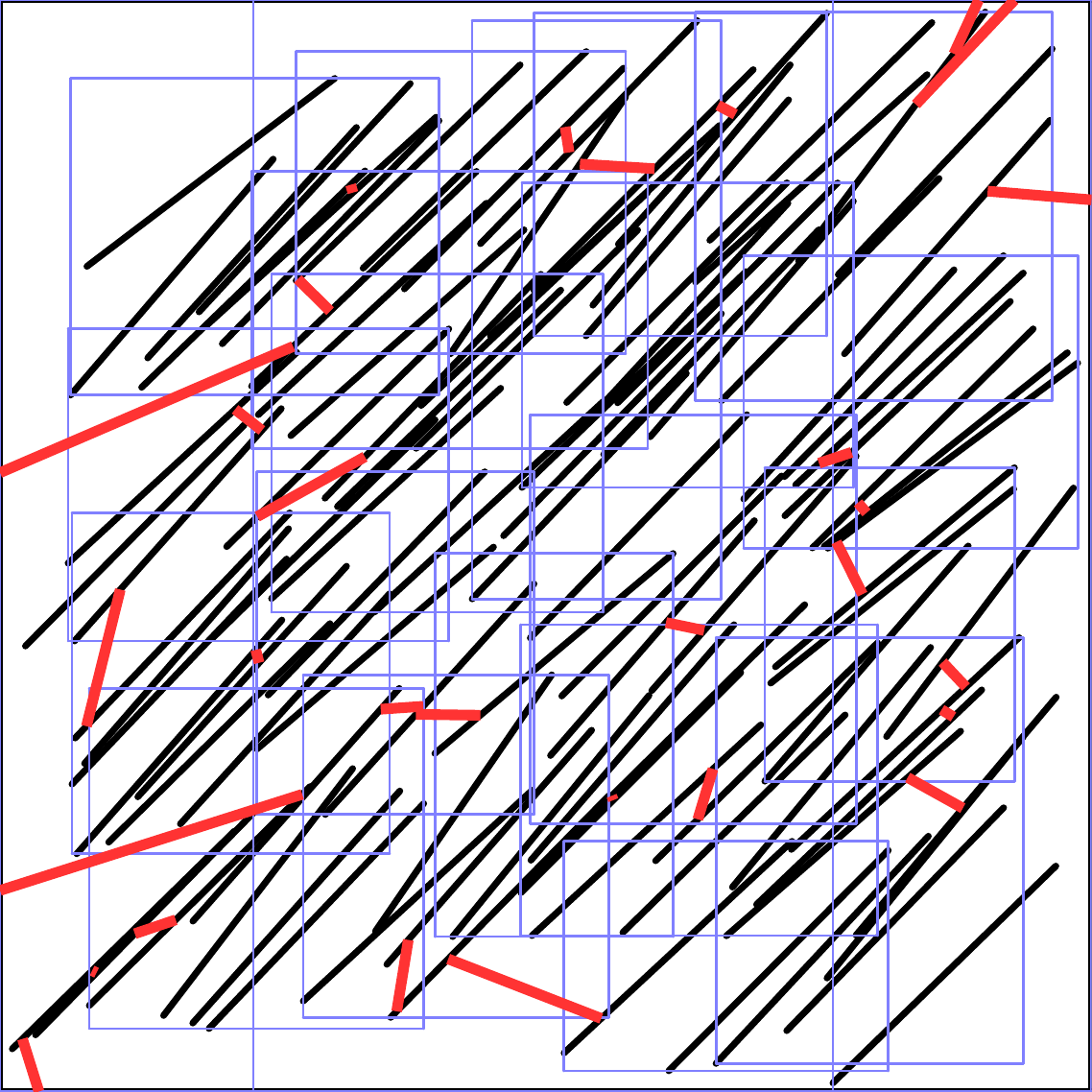}
    {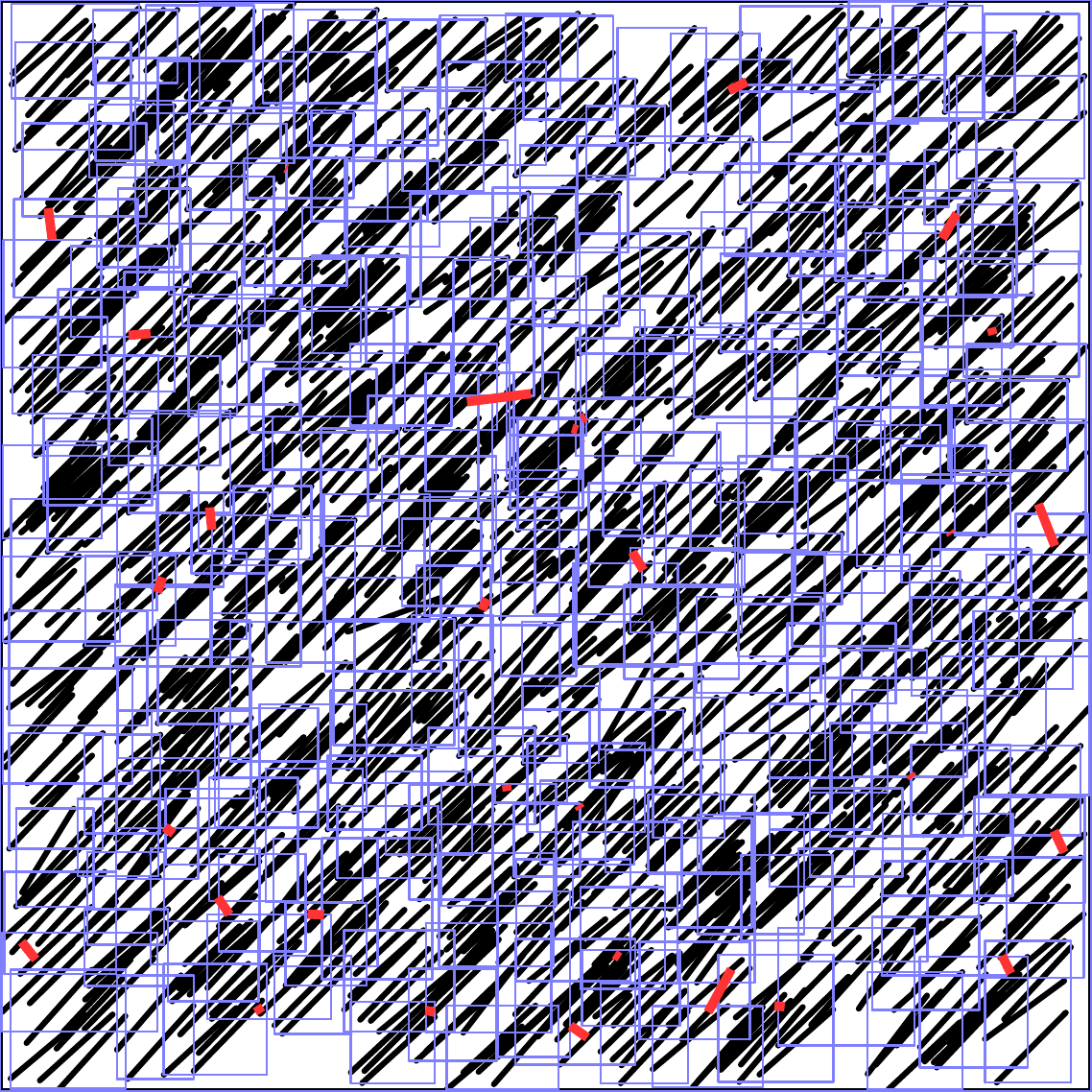}
\accelStructComparisonDlines

\accelStructComparisonA{\grassAccelCompareHeightFirst}{\grassAccelCompareHeight}
    {Grass scenes -- 3 segments/leaf}
    {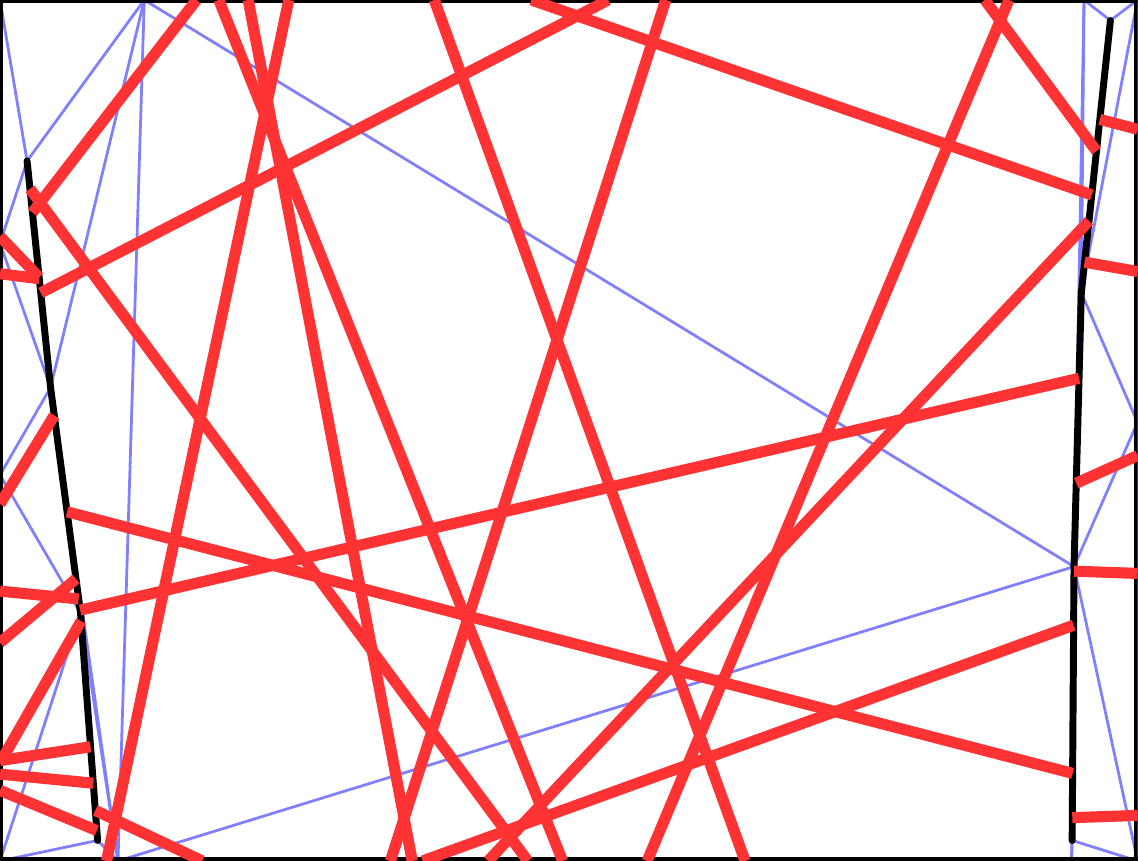}
    {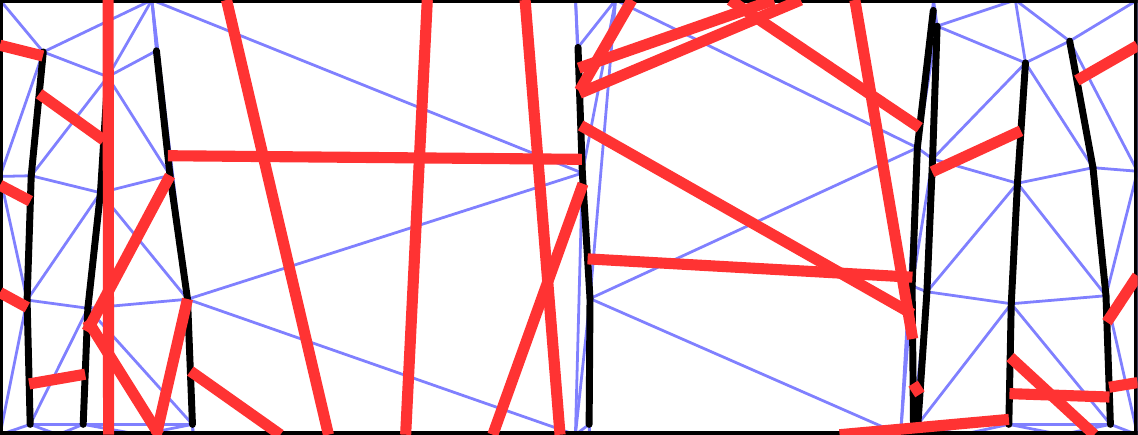}
    {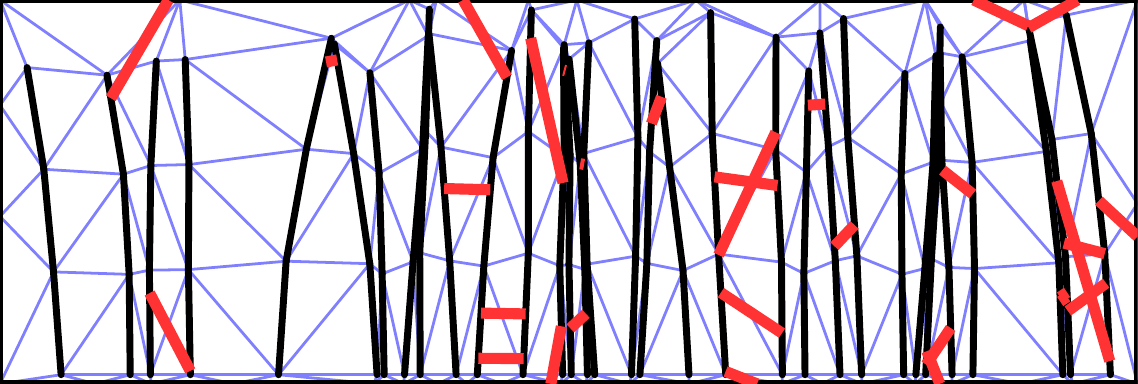}
    {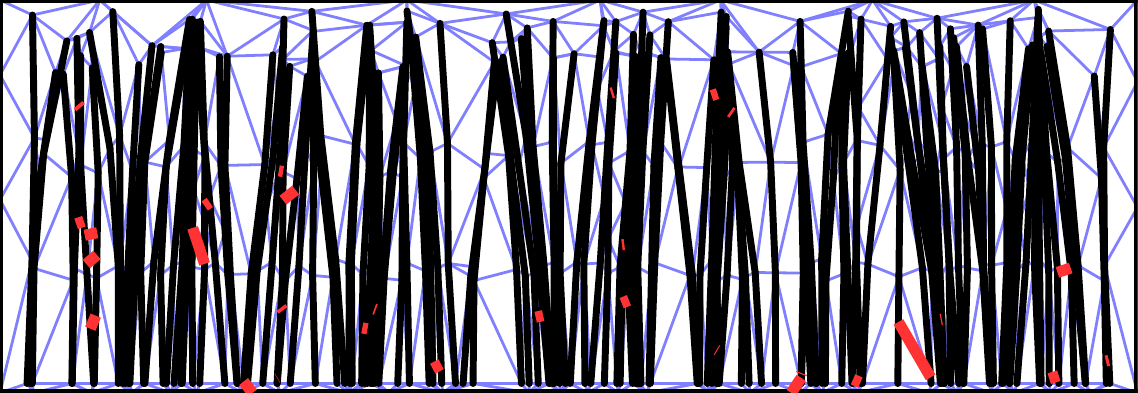}
\accelStructComparisonB{\grassAccelCompareHeightFirst}{\grassAccelCompareHeight}
    {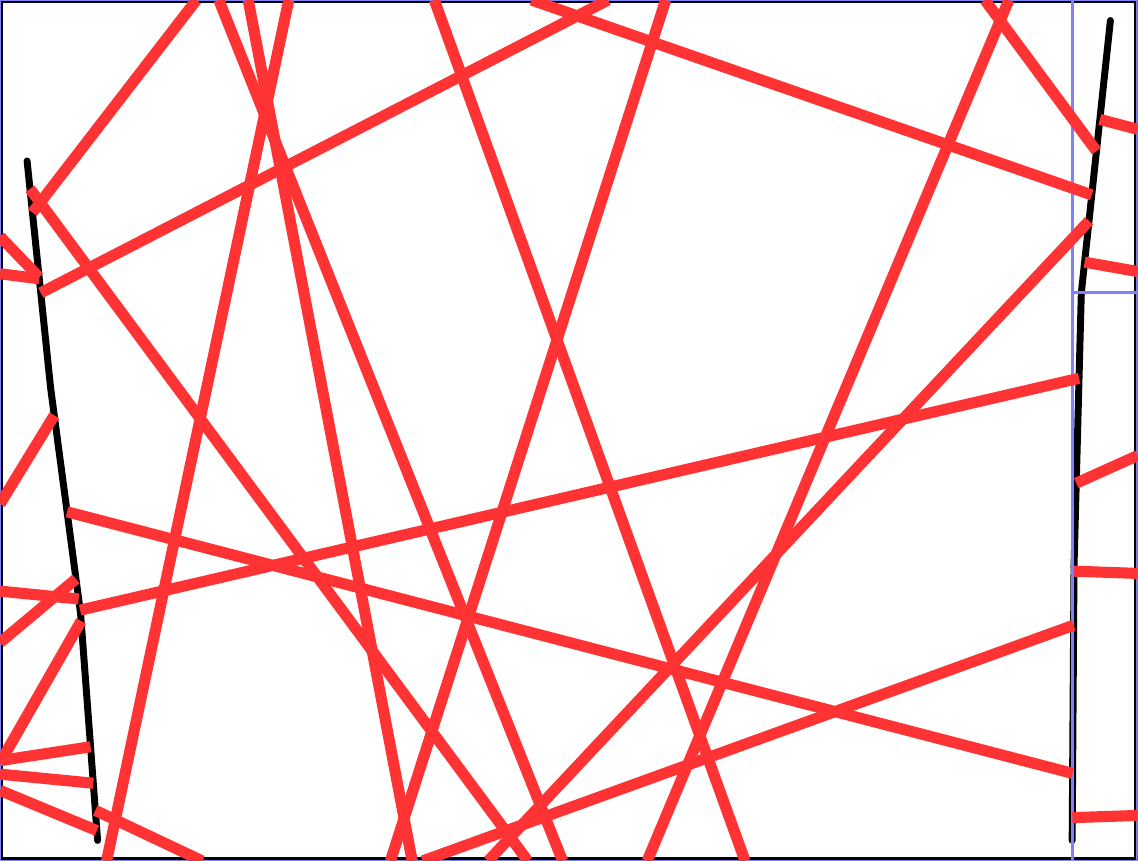}
    {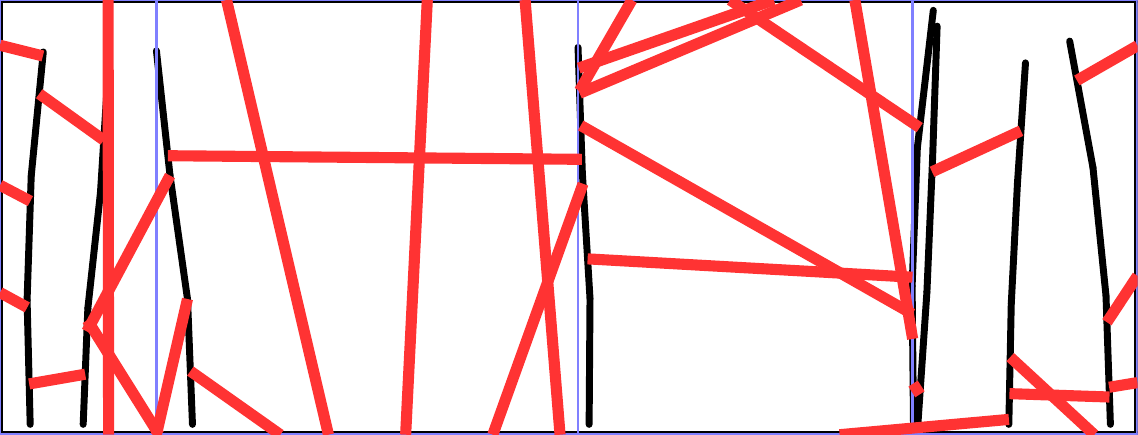}
    {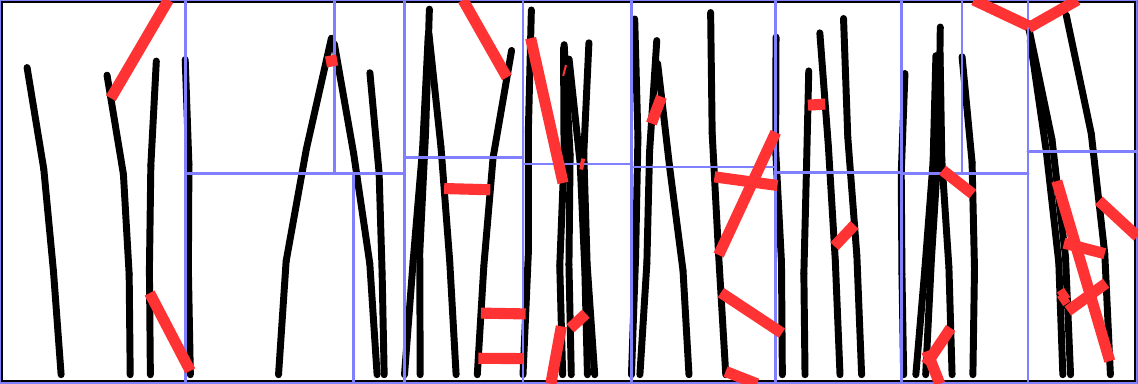}
    {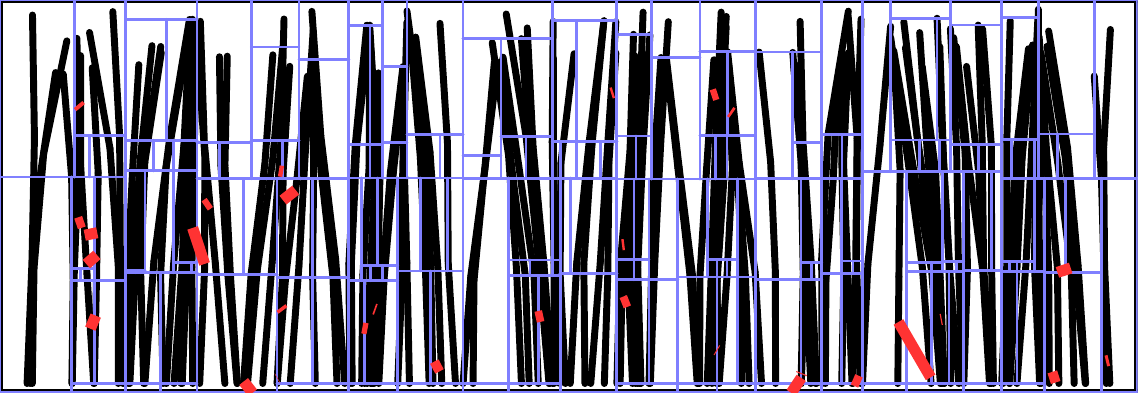}
\accelStructComparisonC{\grassAccelCompareHeightFirst}{\grassAccelCompareHeight}
    {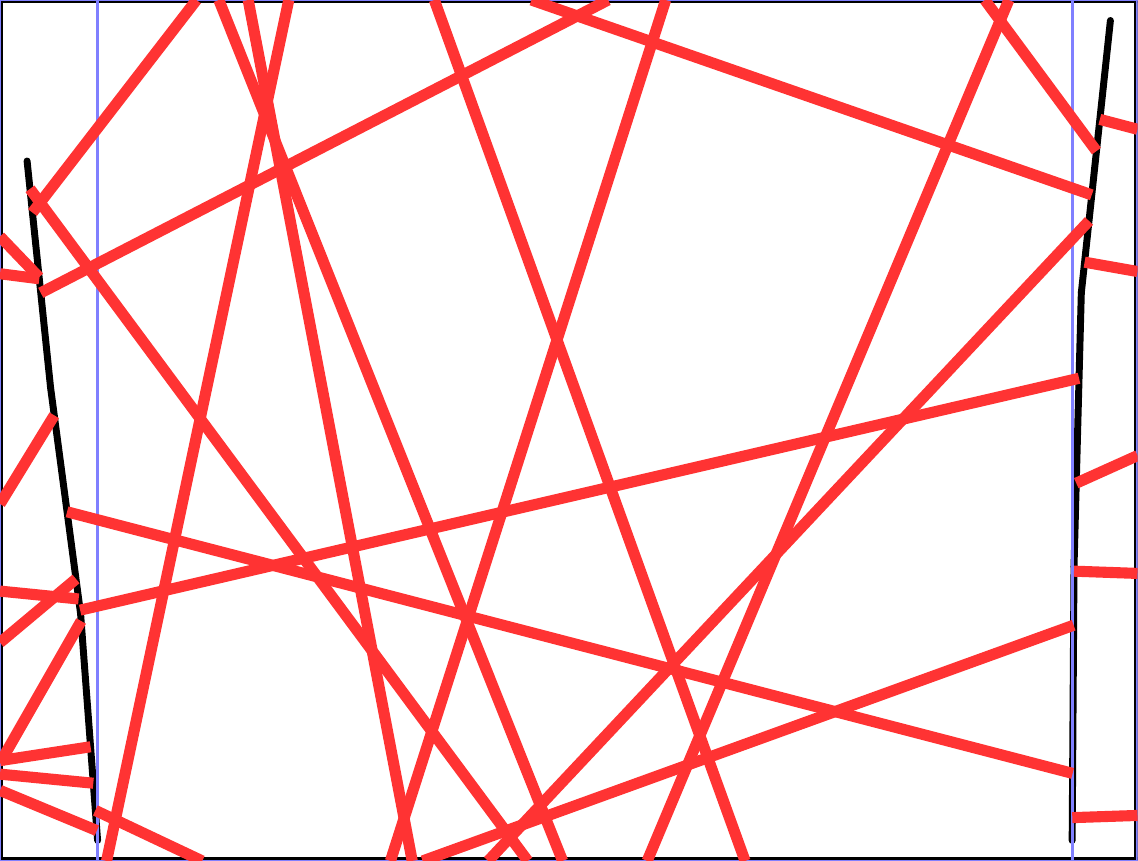}
    {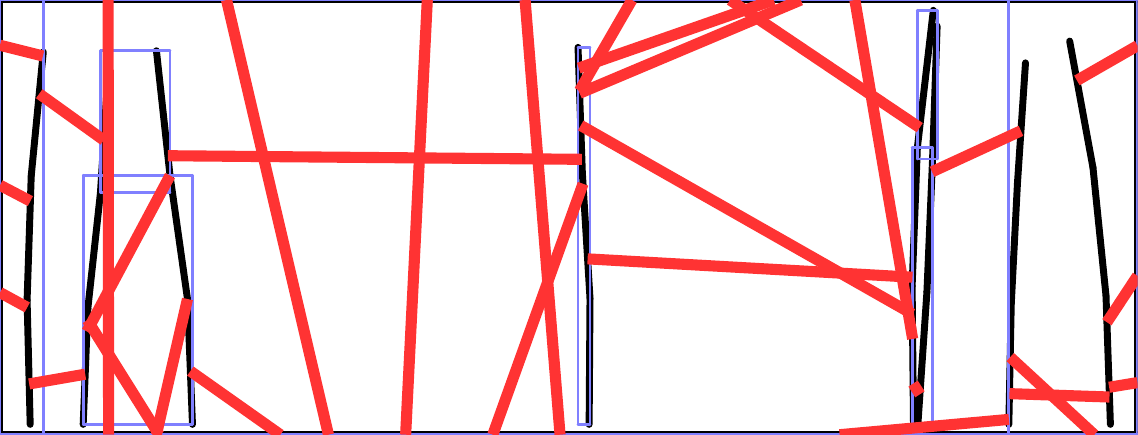}
    {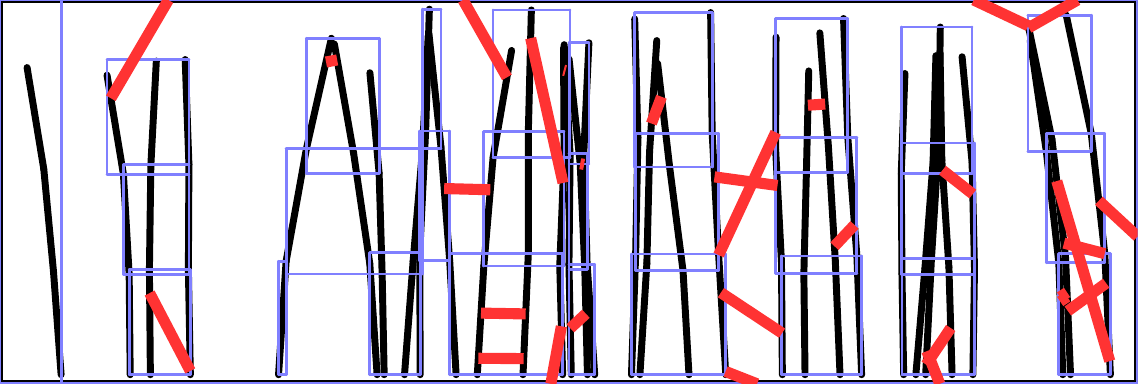}
    {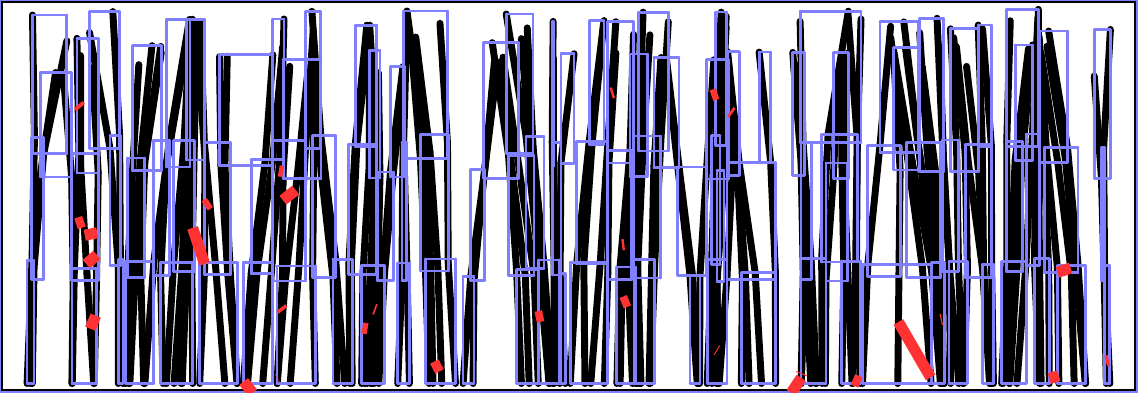}
\accelStructComparisonDgrass

\accelStructComparisonA{\grassAccelCompareHeightFirst}{\grassAccelCompareHeight}
    {Grass scenes -- 10 segments/leaf}
    {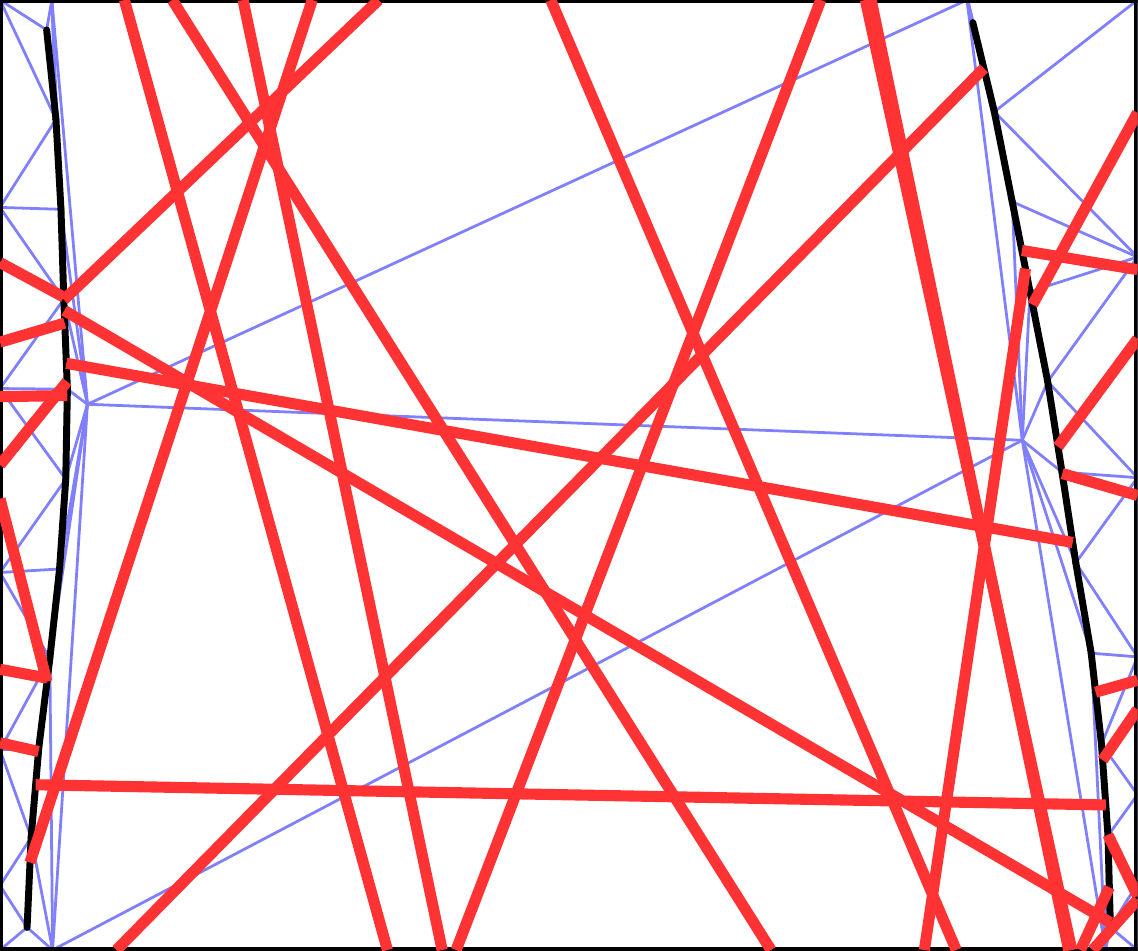}
    {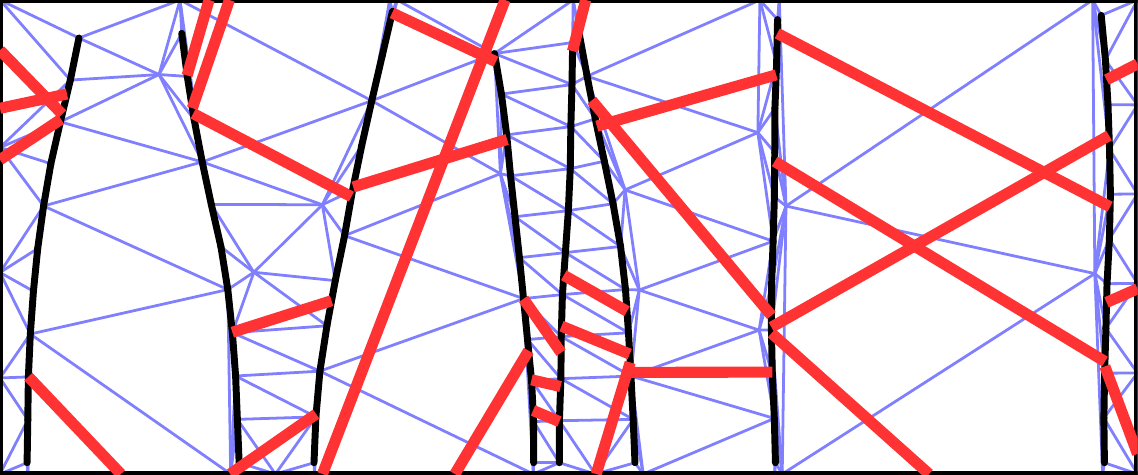}
    {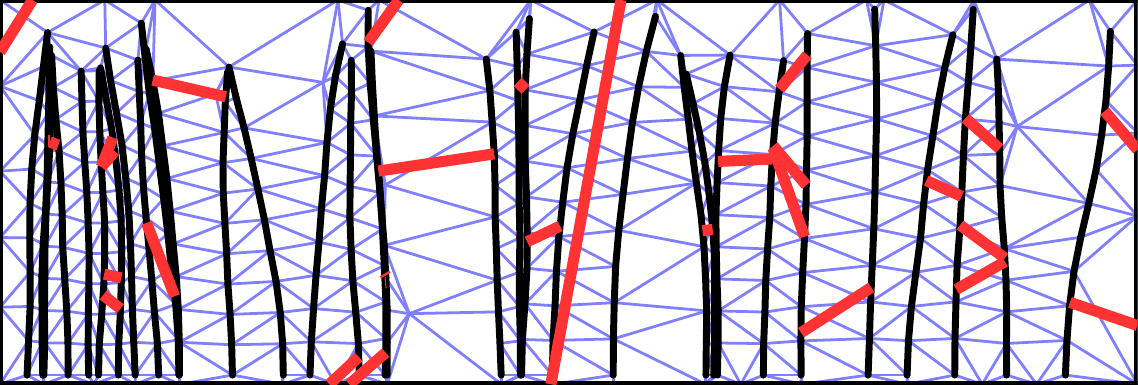}
    {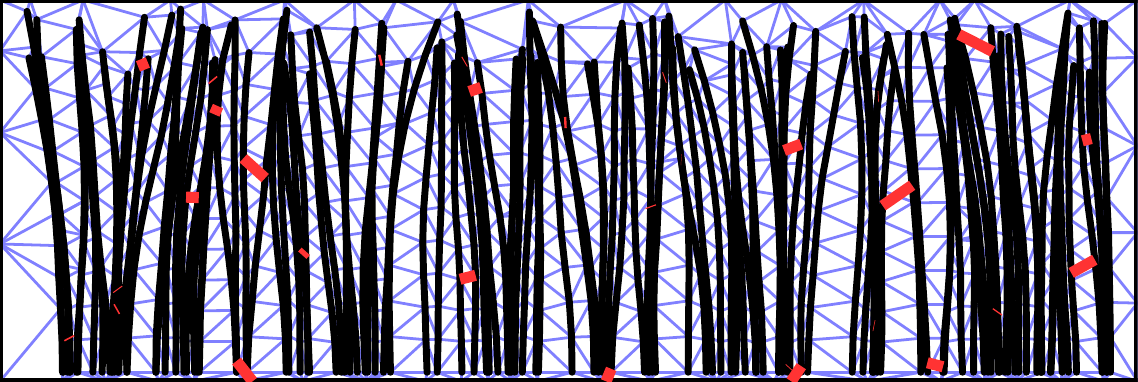}
\accelStructComparisonB{\grassAccelCompareHeightFirst}{\grassAccelCompareHeight}
    {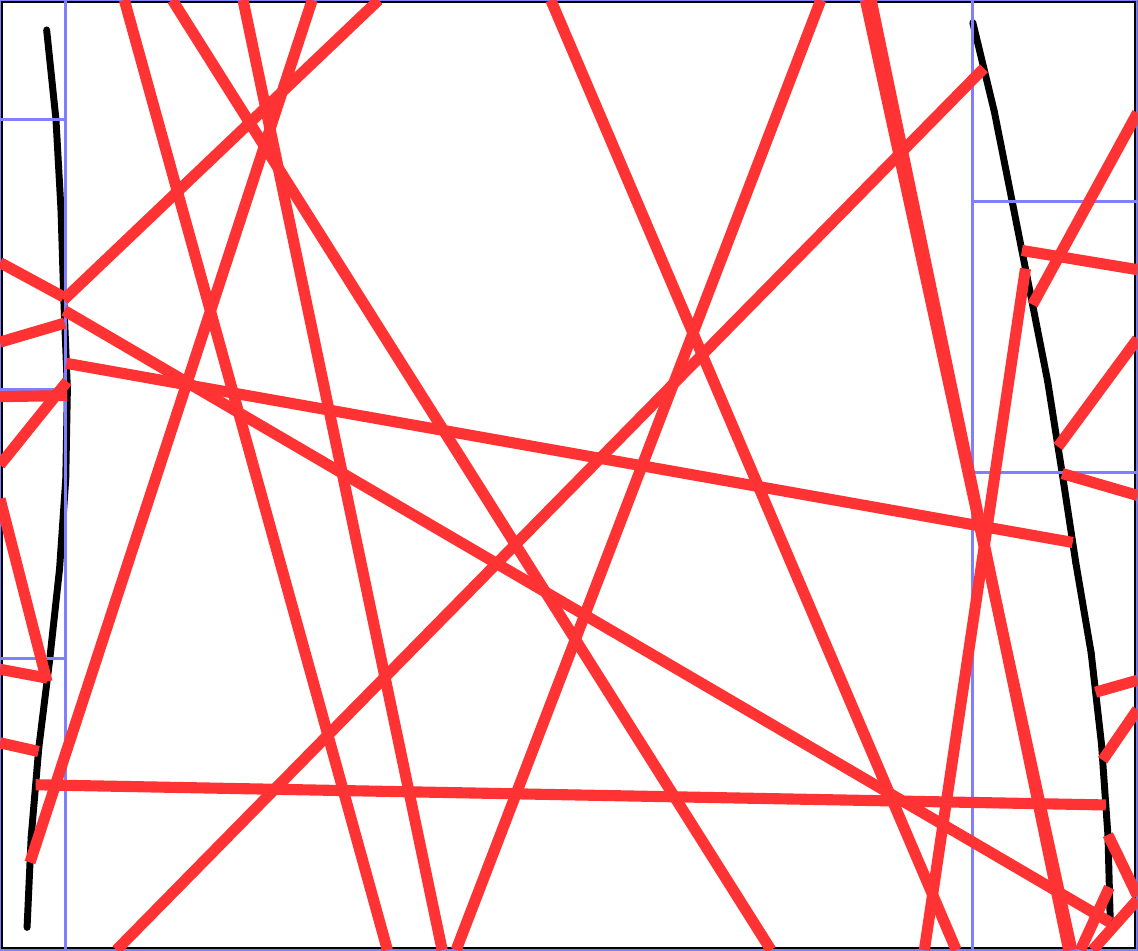}
    {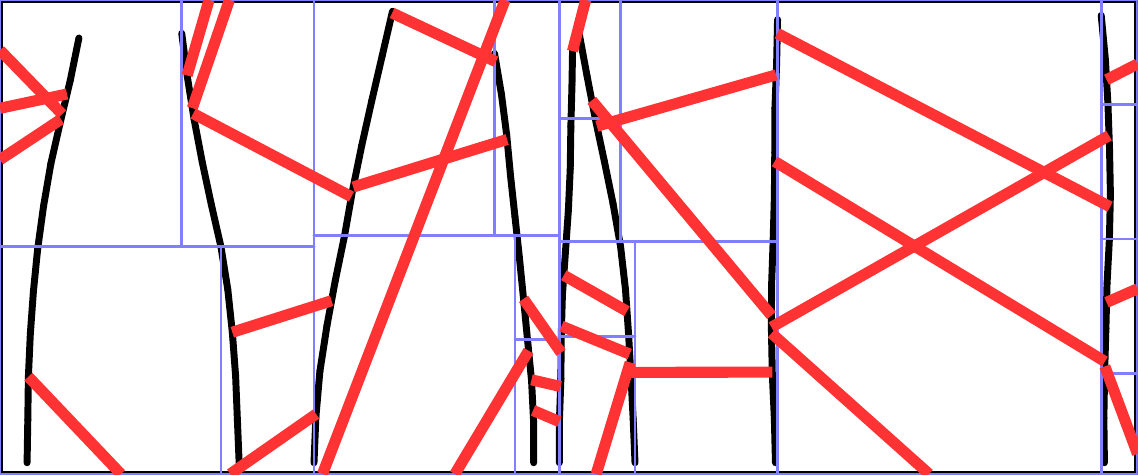}
    {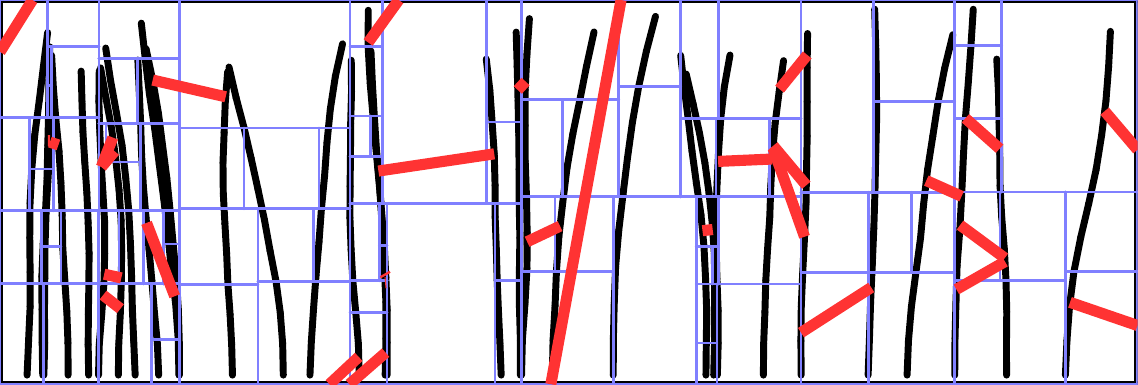}
    {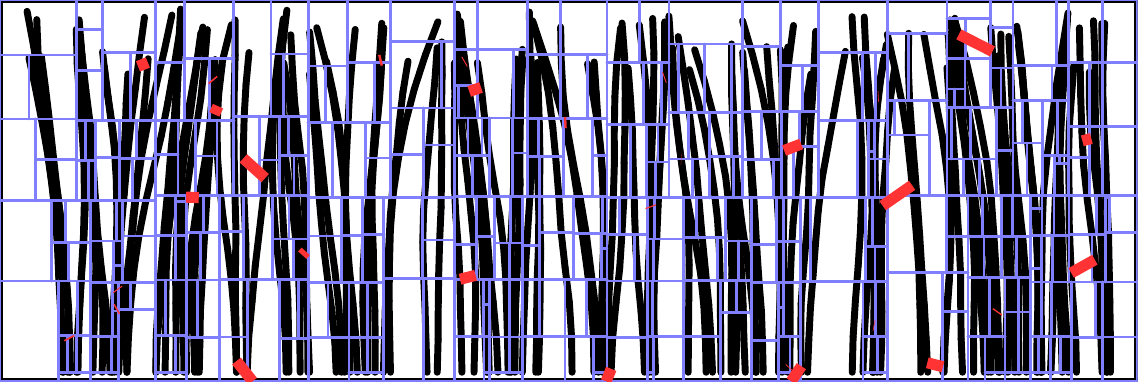}
\accelStructComparisonC{\grassAccelCompareHeightFirst}{\grassAccelCompareHeight}
    {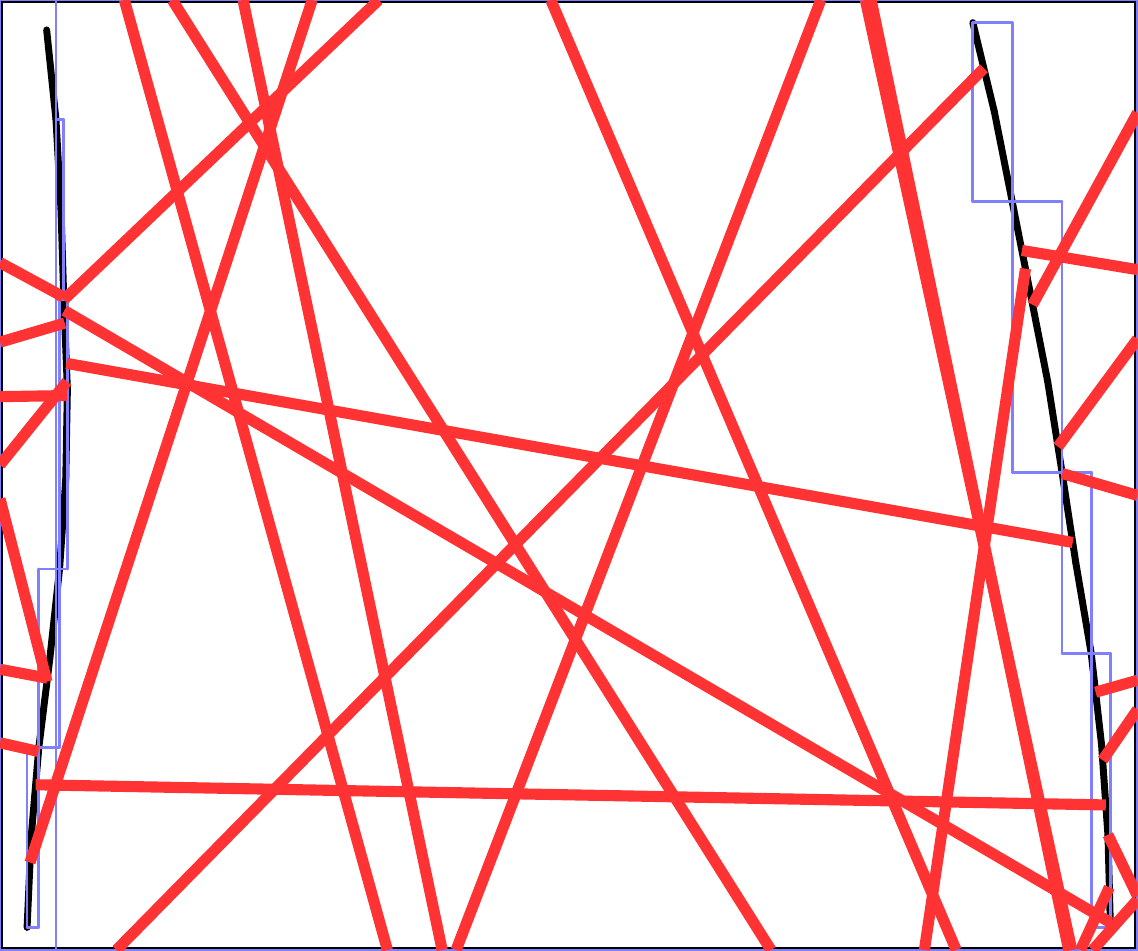}
    {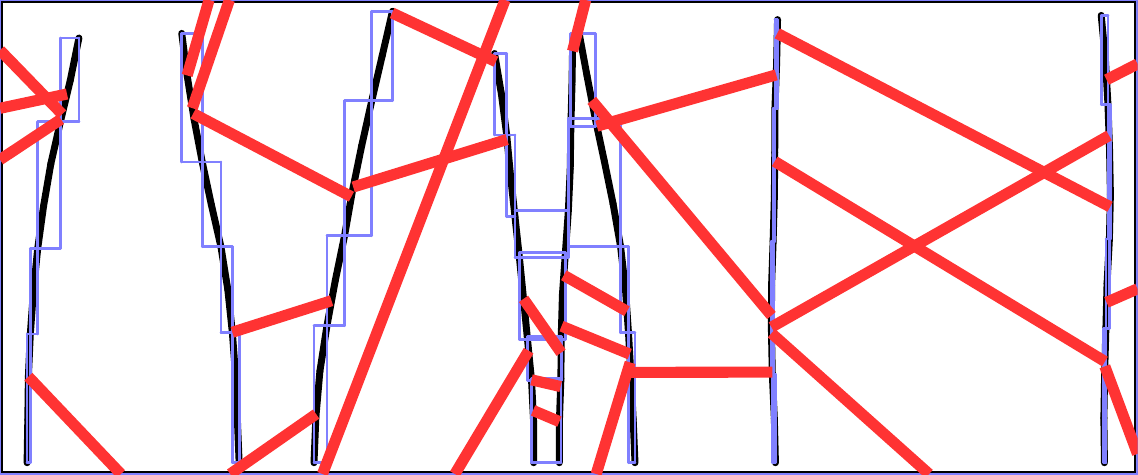}
    {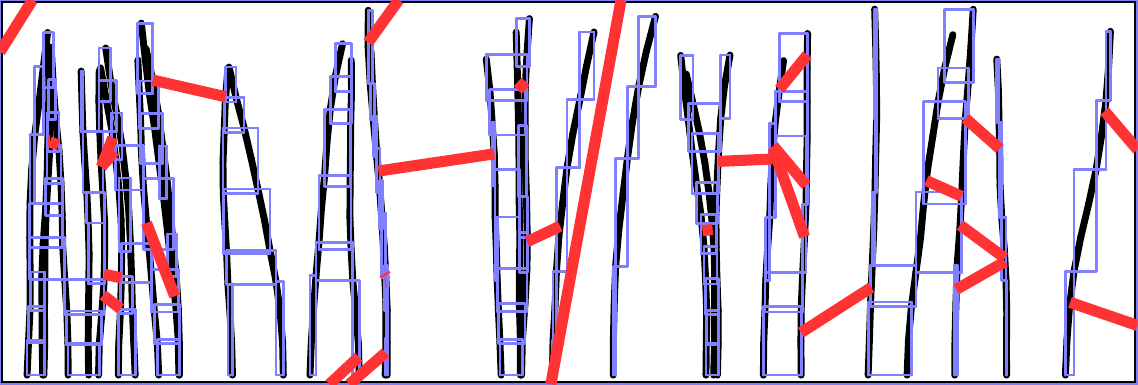}
    {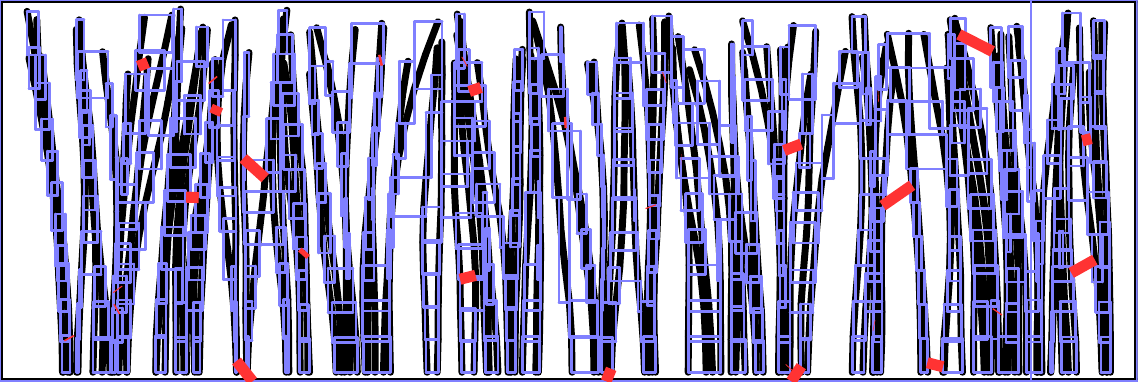}
\accelStructComparisonDgrass

\accelStructComparisonA{\hairAccelCompareHeight}{\hairAccelCompareHeight}
    {Hair scenes -- 5 segments/strand}
    {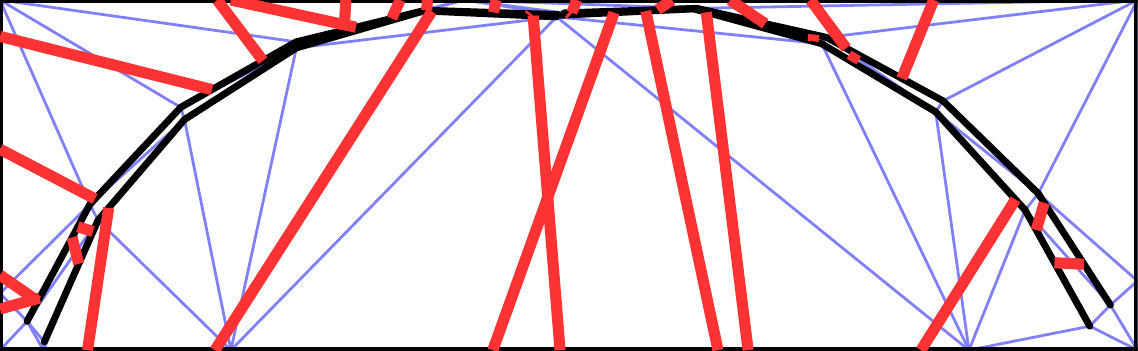}
    {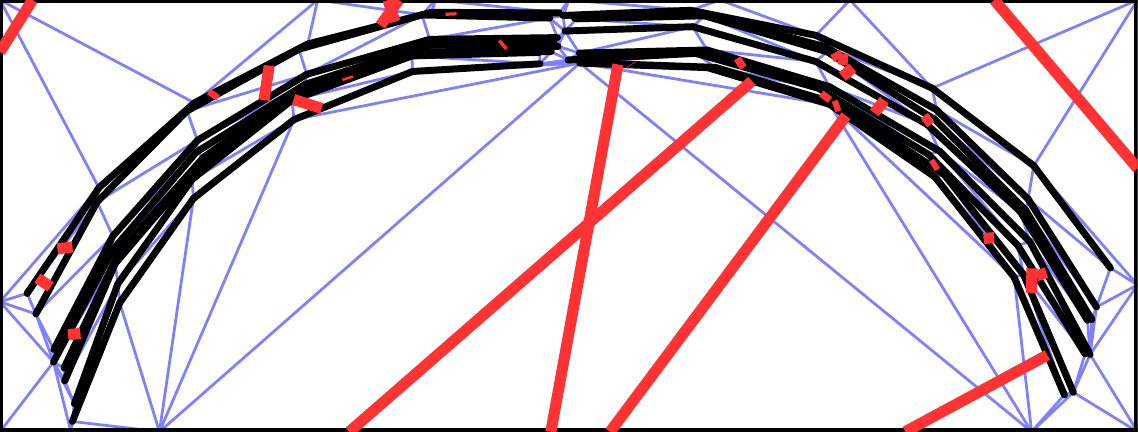}
    {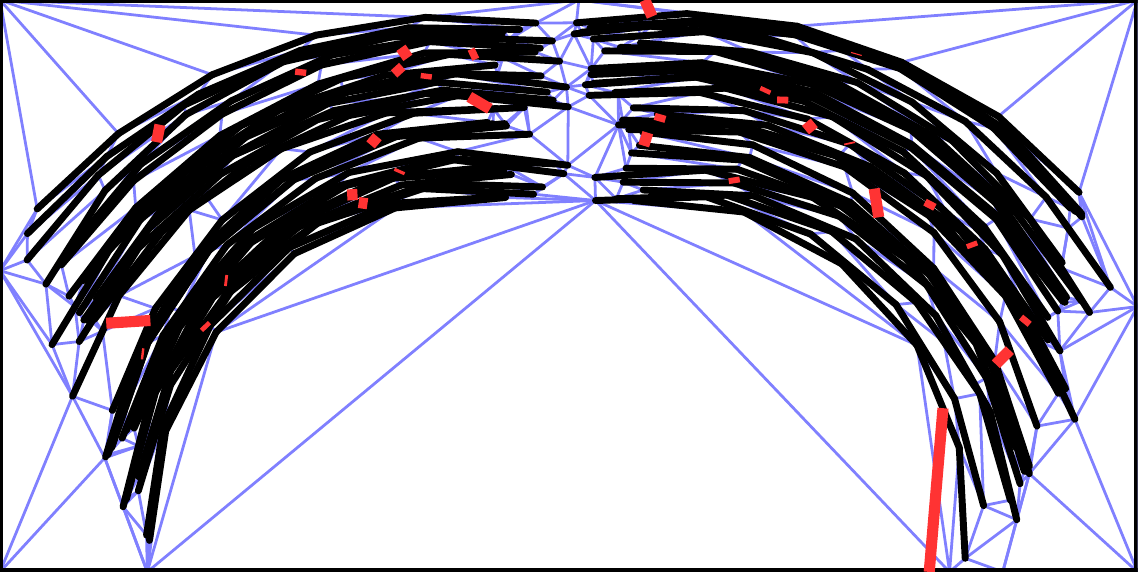}
    {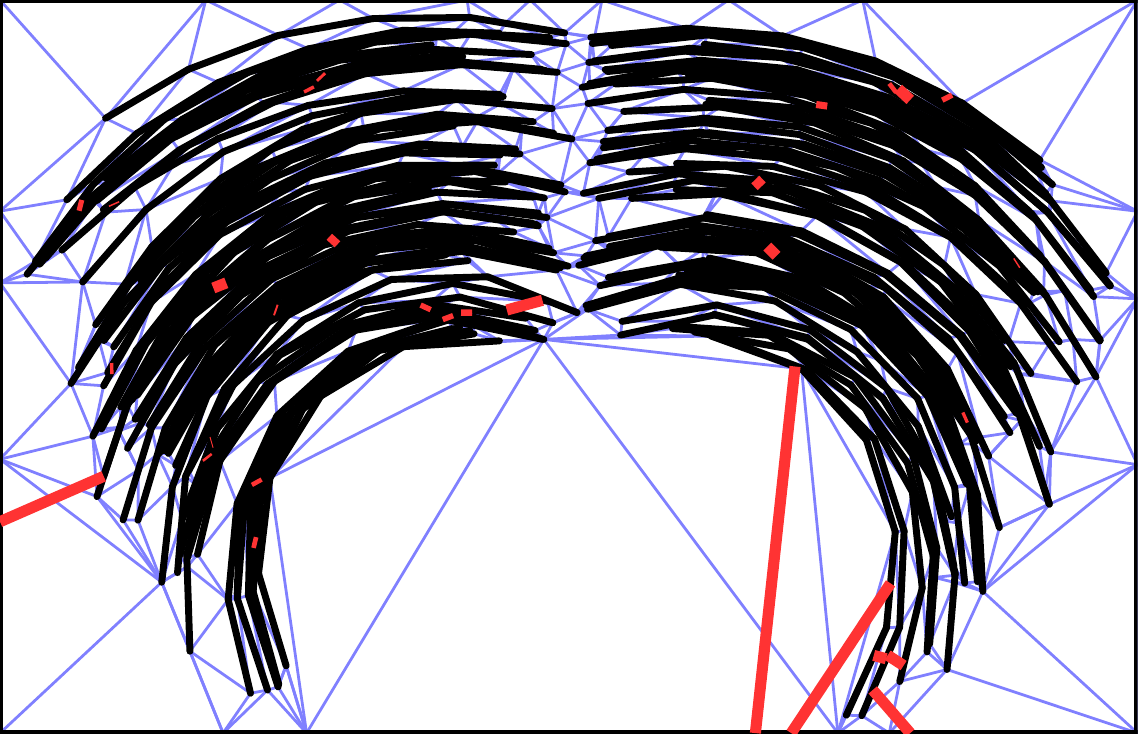}
\accelStructComparisonB{\hairAccelCompareHeight}{\hairAccelCompareHeight}
    {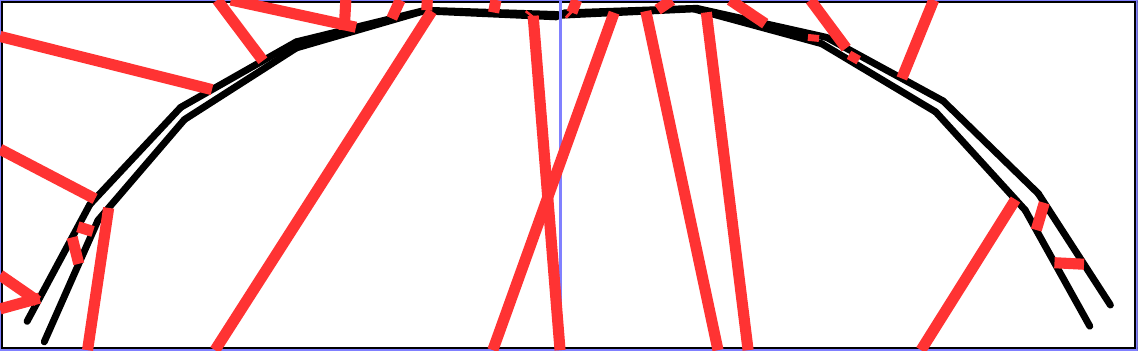}
    {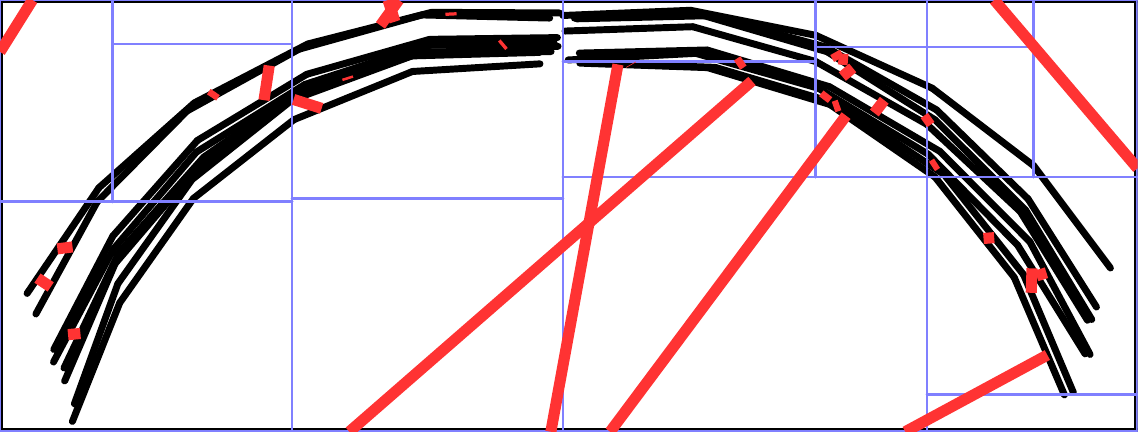}
    {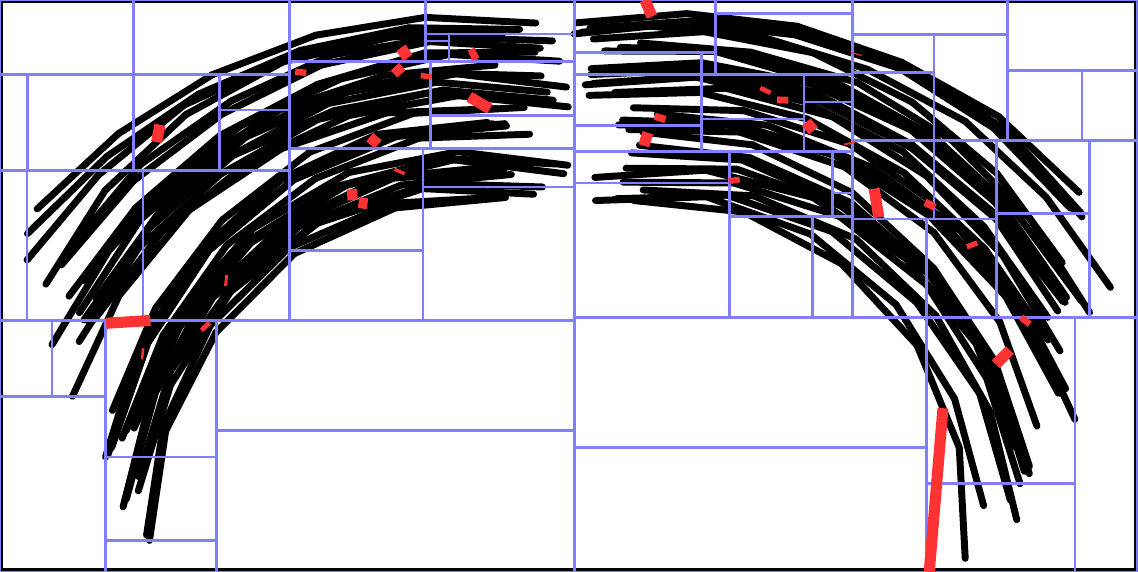}
    {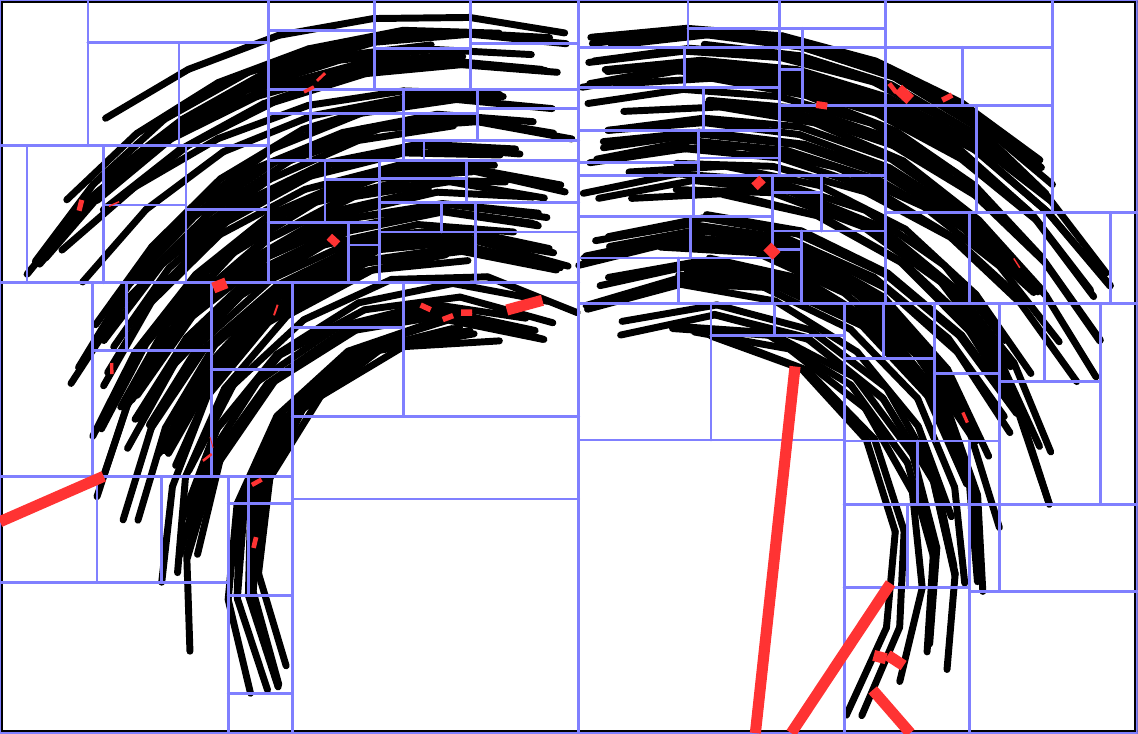}
\accelStructComparisonC{\hairAccelCompareHeight}{\hairAccelCompareHeight}
    {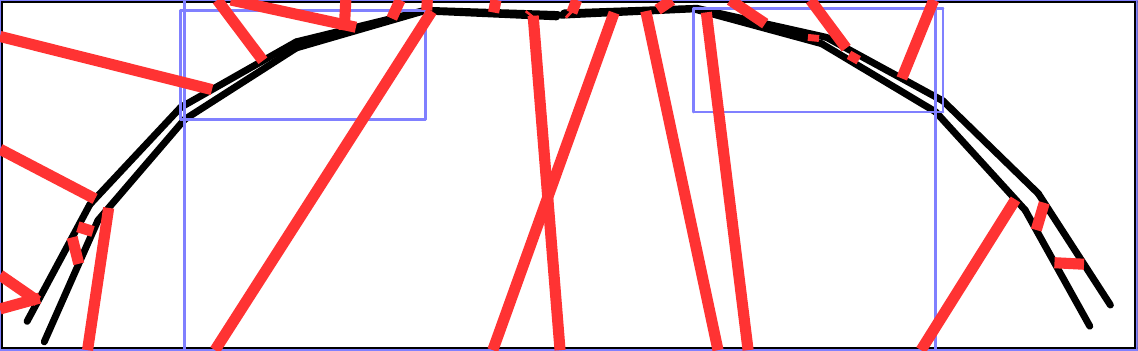}
    {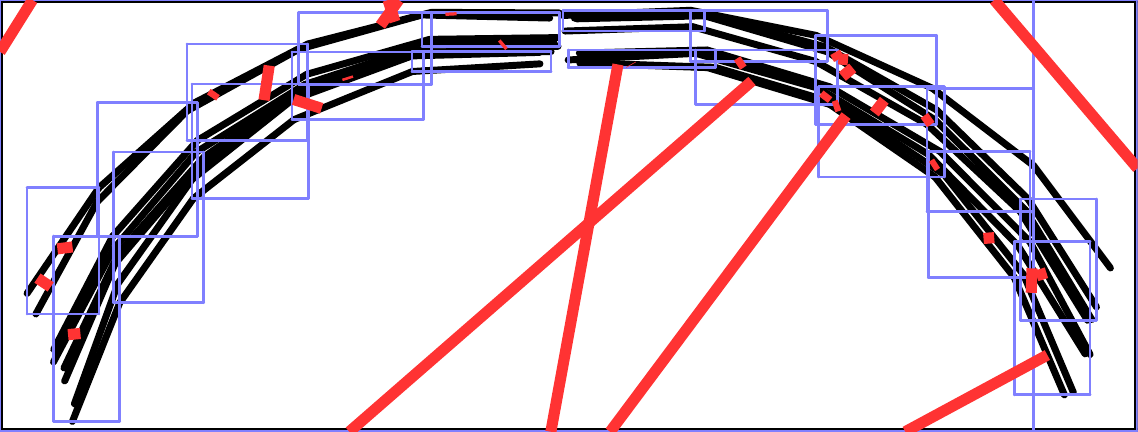}
    {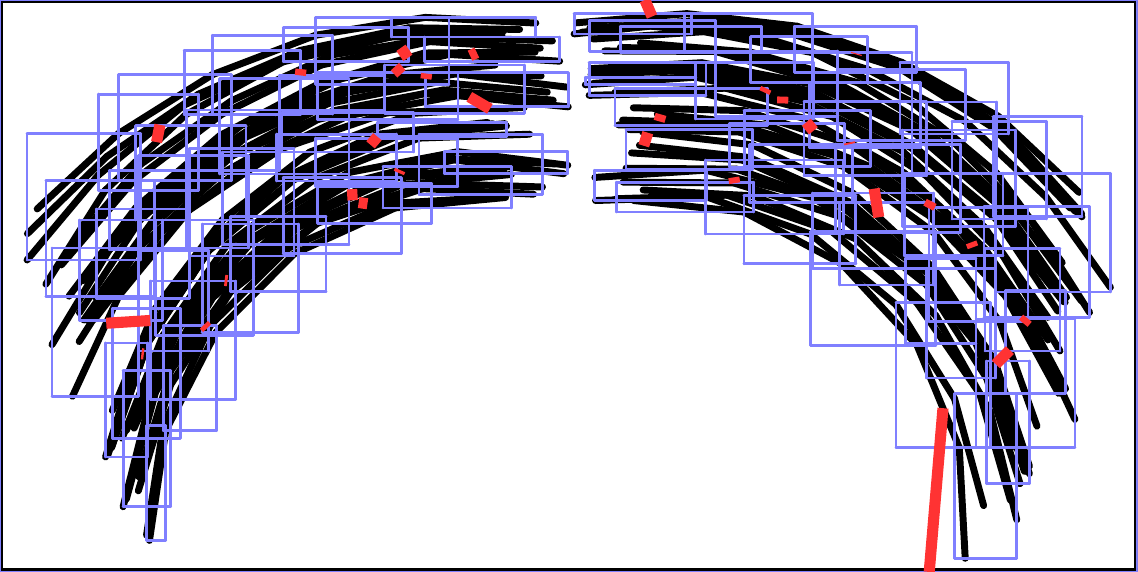}
    {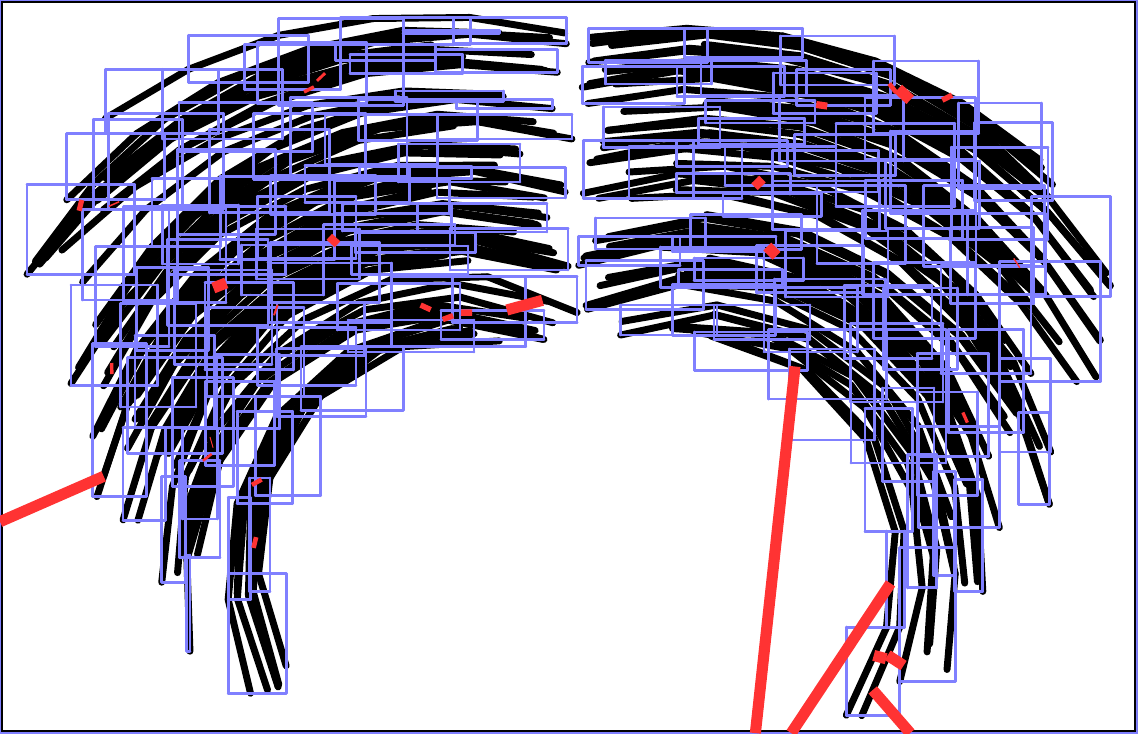}
\accelStructComparisonDhair

\accelStructComparisonA{\hairAccelCompareHeight}{\hairAccelCompareHeight}
    {Hair scenes -- 20 segments/strand}
    {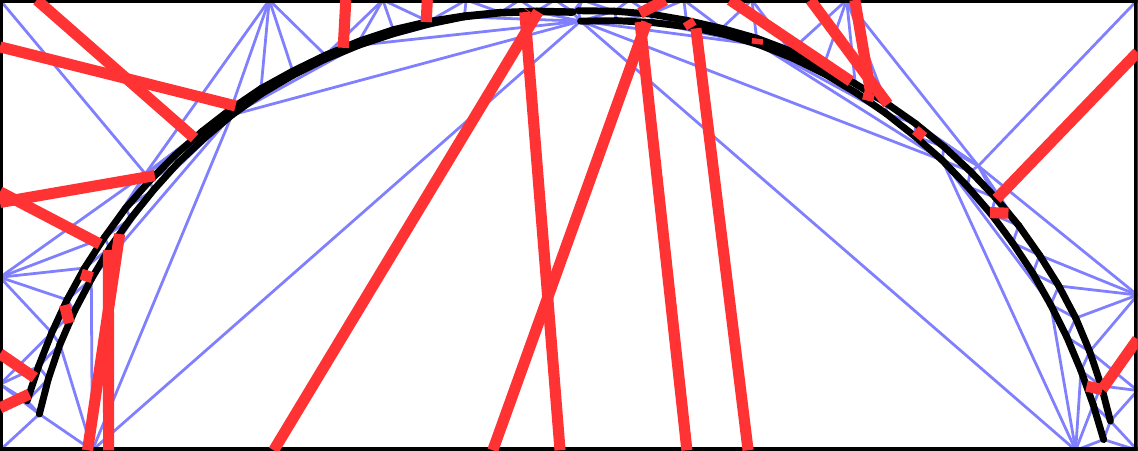}
    {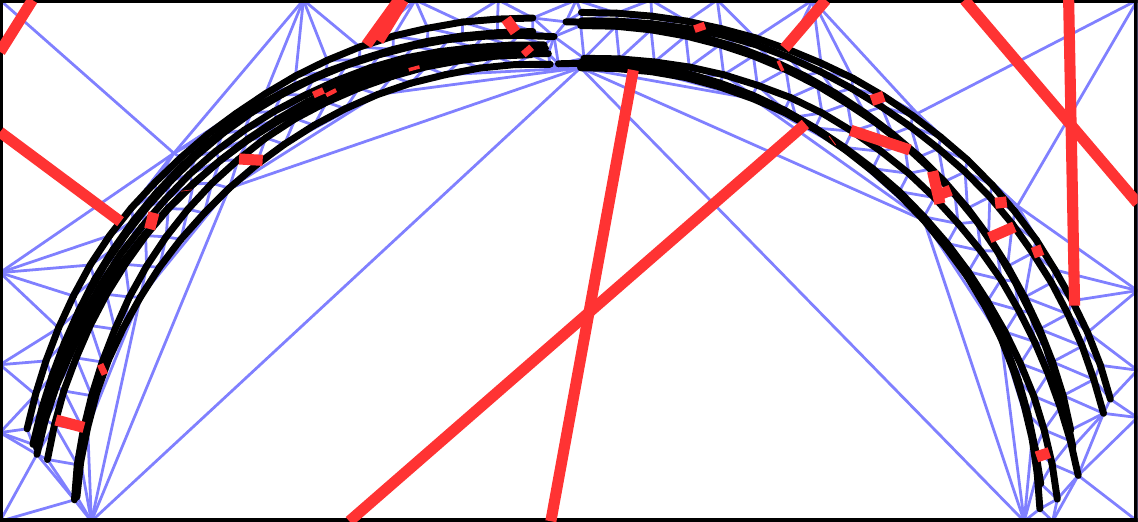}
    {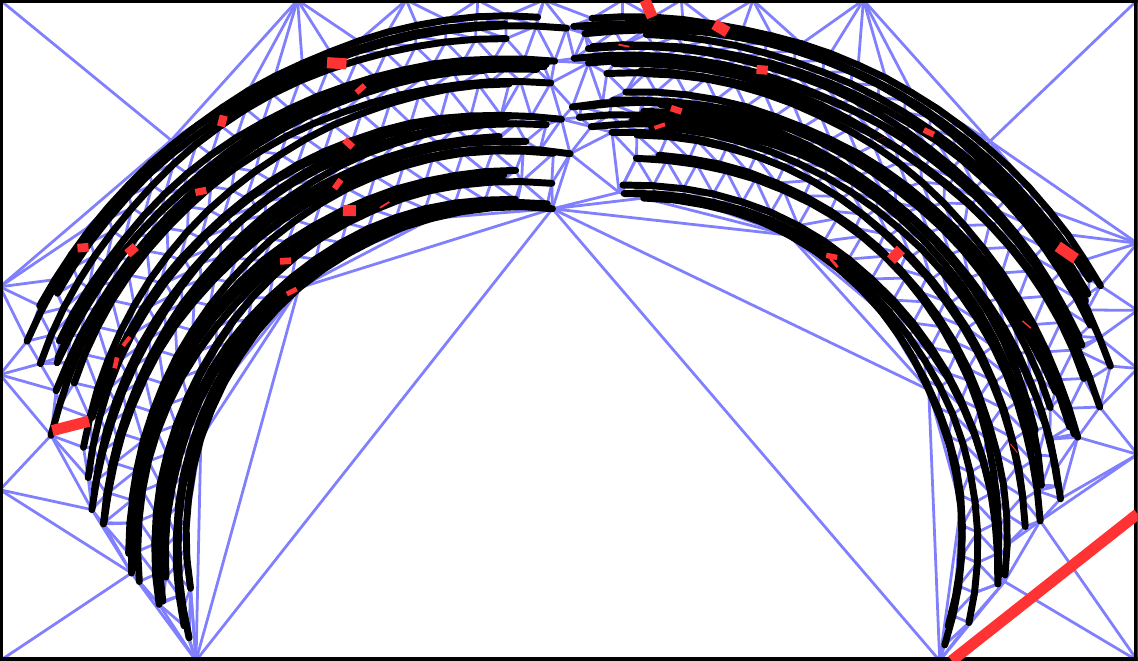}
    {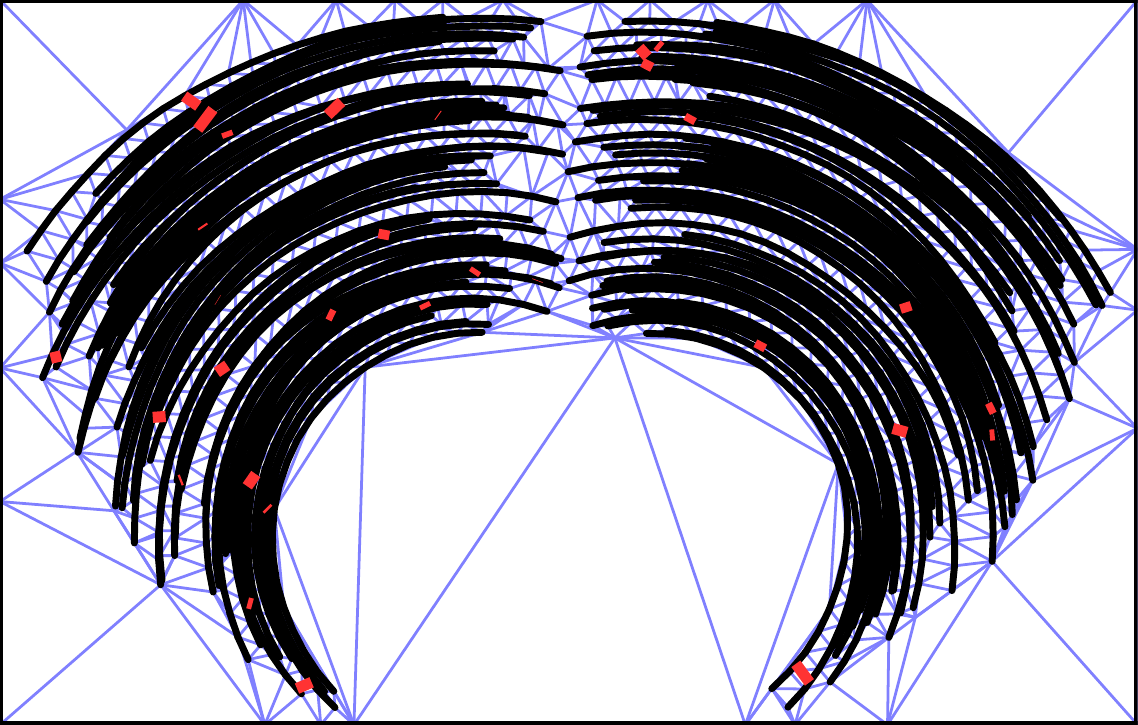}
\accelStructComparisonB{\hairAccelCompareHeight}{\hairAccelCompareHeight}
    {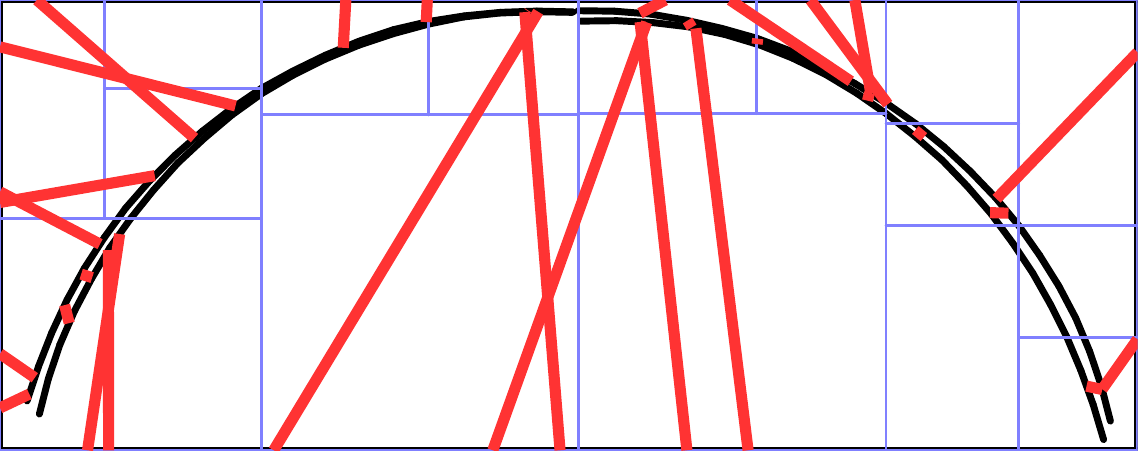}
    {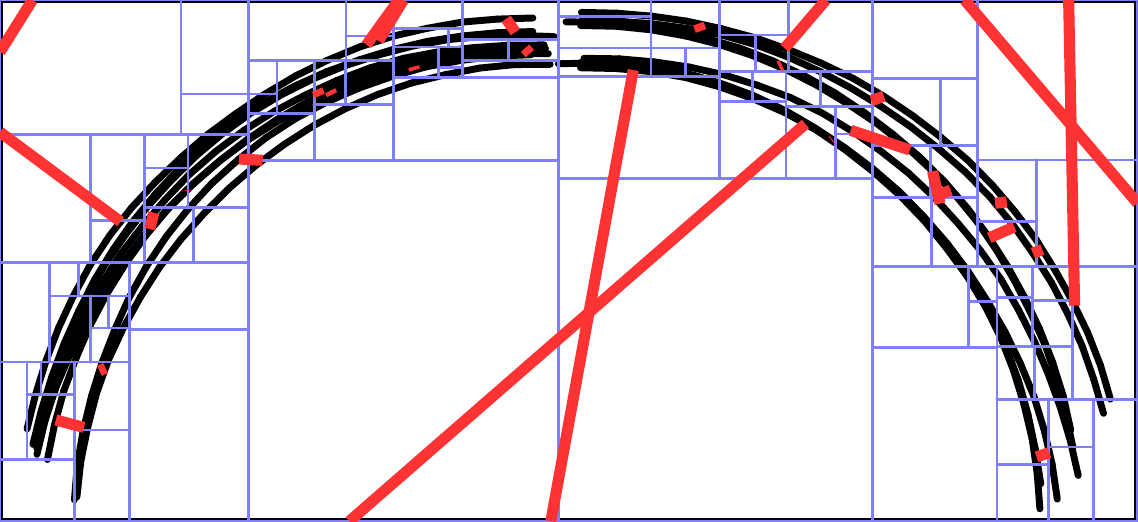}
    {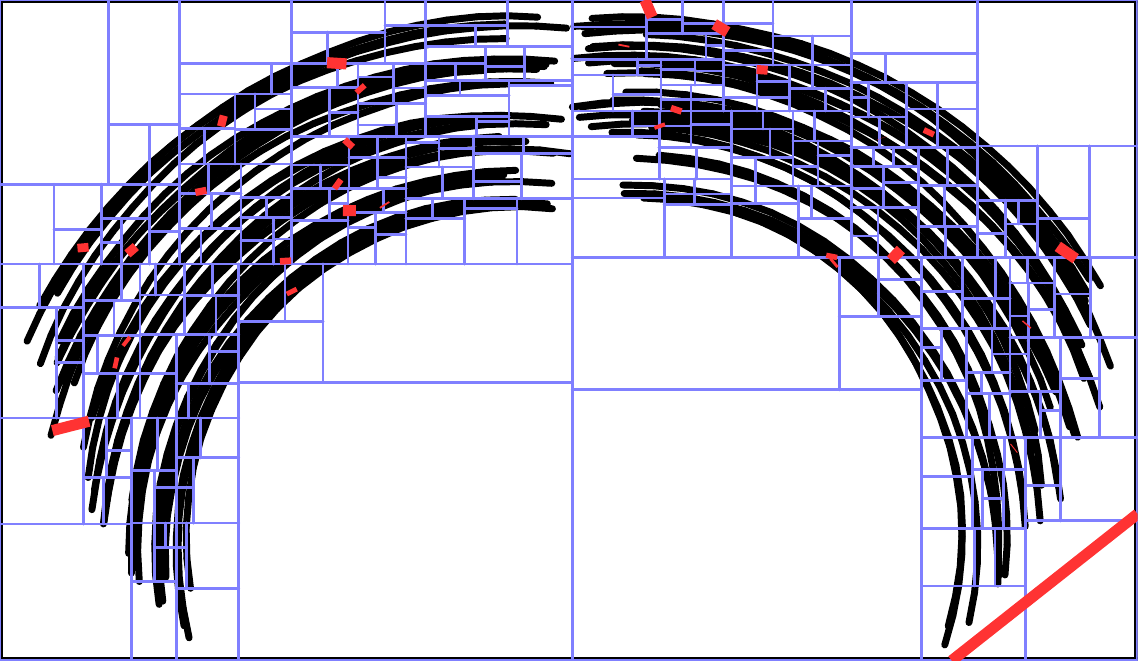}
    {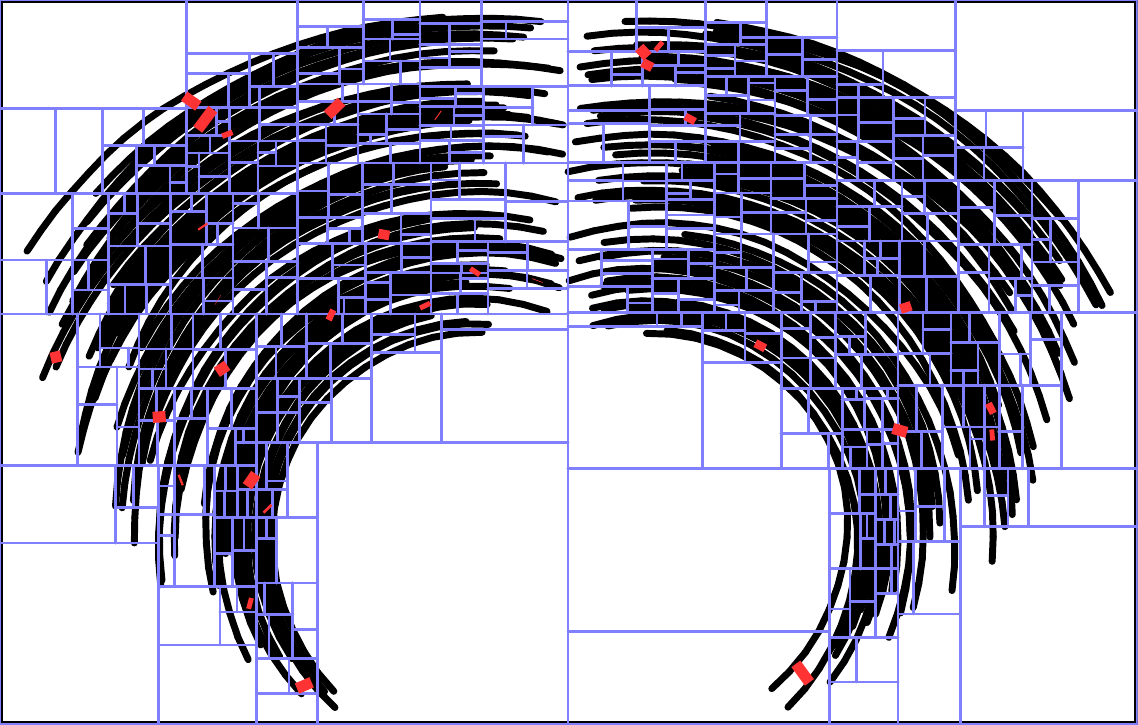}
\accelStructComparisonC{\hairAccelCompareHeight}{\hairAccelCompareHeight}
    {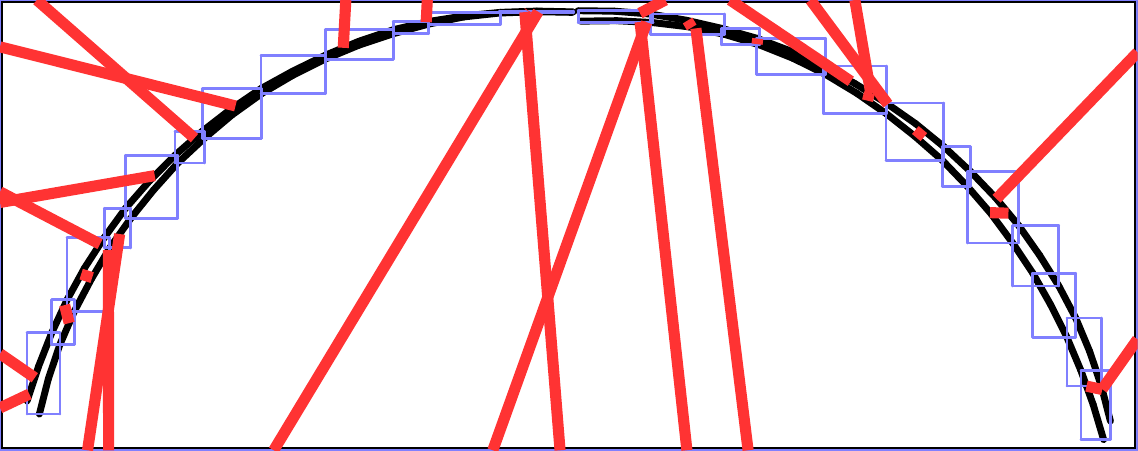}
    {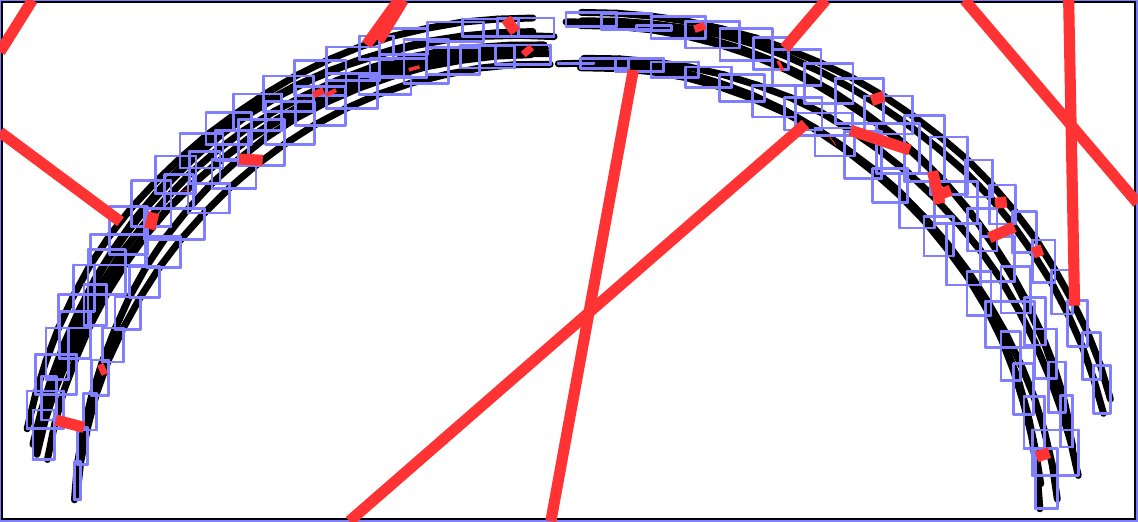}
    {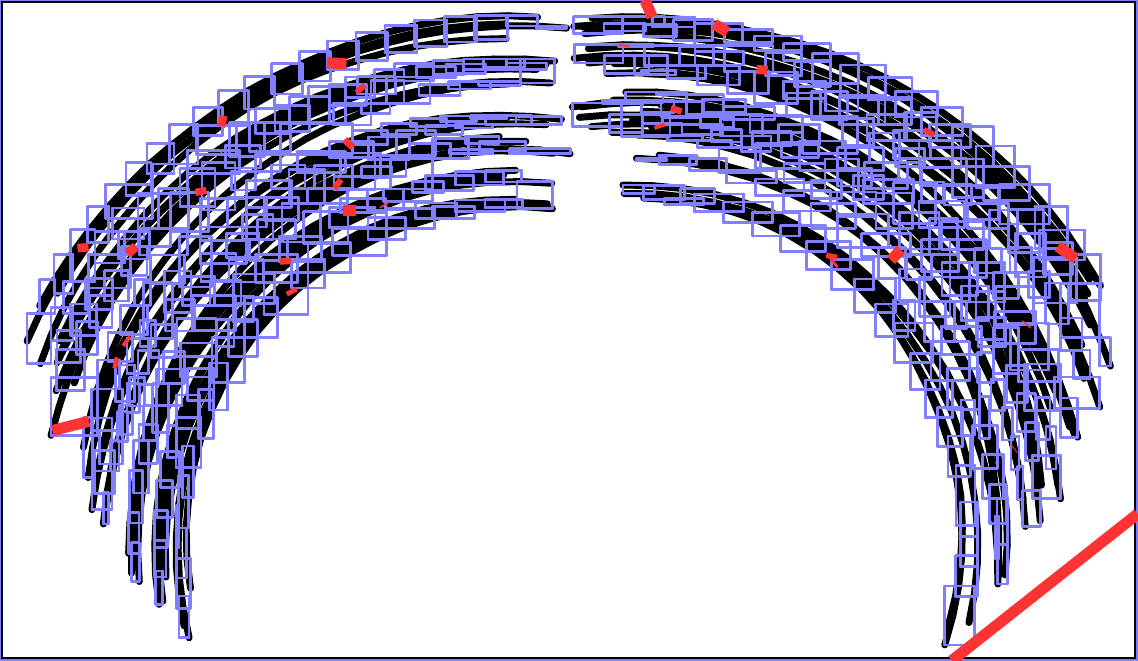}
    {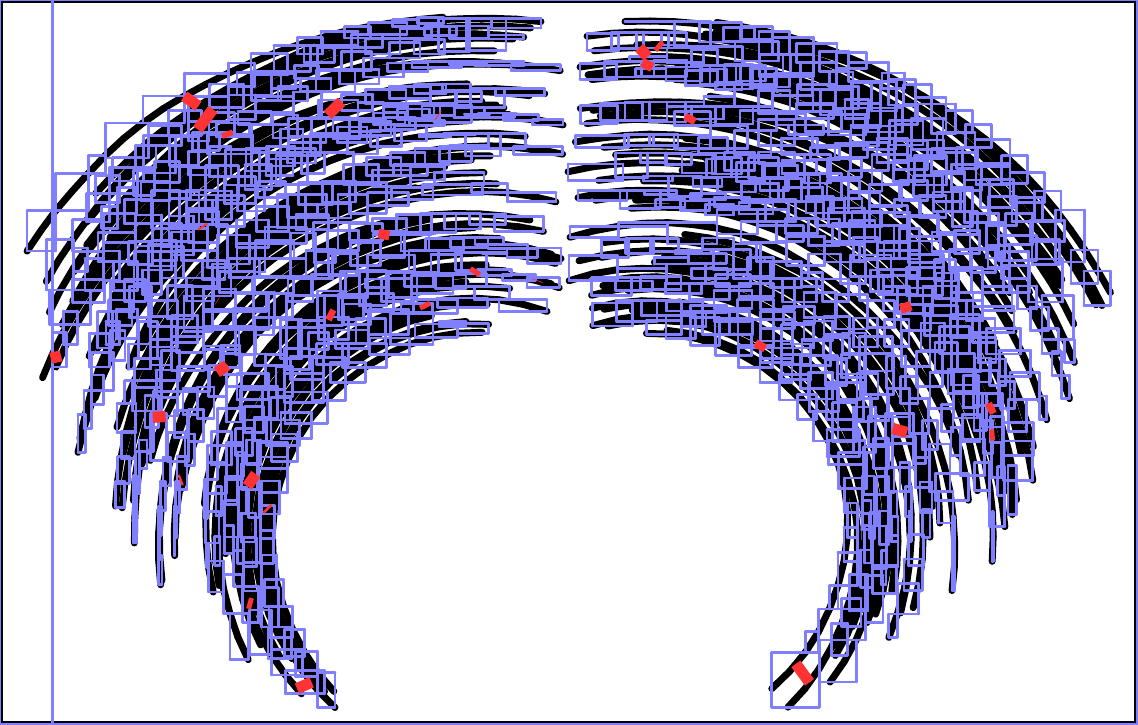}
\accelStructComparisonDhair

\newcommand{\svgStructComparison}[5]{
\begin{landscape}
\begin{figure}
\begin{center}
\begin{tabular}{c@{\ }c@{\ }c@{\ }c}
&
    {\small Optimized triangulation} &
    {\small Roped kd-tree} &
    {\small BVH leaves} \\
\rotatebox[origin=c]{90}{\small \textsc{#2}} &
    \includegraphics[width=#1,align=c]{plots/#3_randomRays.pdf} &
    \includegraphics[width=#1,align=c]{plots/#3_KdTree.pdf} &
    \includegraphics[width=#1,align=c]{plots/#3_BVH.pdf}
    \\
    \\[-4mm]
\rotatebox[origin=c]{90}{\small \textsc{#4}} &
    \includegraphics[width=#1,align=c]{plots/#5_randomRays.pdf} &
    \includegraphics[width=#1,align=c]{plots/#5_KdTree.pdf} &
    \includegraphics[width=#1,align=c]{plots/#5_BVH.pdf}
    \\ 
    \\
\end{tabular}
\end{center}
\end{figure}
\end{landscape}
}

\newcommand{\svgStructComparisonSingle}[3]{
\begin{landscape}
\begin{figure}
\begin{center}
\begin{tabular}{c@{\ }c@{\ }c@{\ }c}
&
    {\small Optimized triangulation} &
    {\small Roped kd-tree} &
    {\small BVH leaves} \\
\rotatebox[origin=c]{90}{\small \textsc{#2}} &
    \includegraphics[width=#1,align=c]{plots/#3_randomRays.pdf} &
    \includegraphics[width=#1,align=c]{plots/#3_KdTree.pdf} &
    \includegraphics[width=#1,align=c]{plots/#3_BVH.pdf}
    \\ 
    \\
\end{tabular}
\end{center}
\end{figure}
\end{landscape}
}

\svgStructComparison{75mm}
    {Louvre}{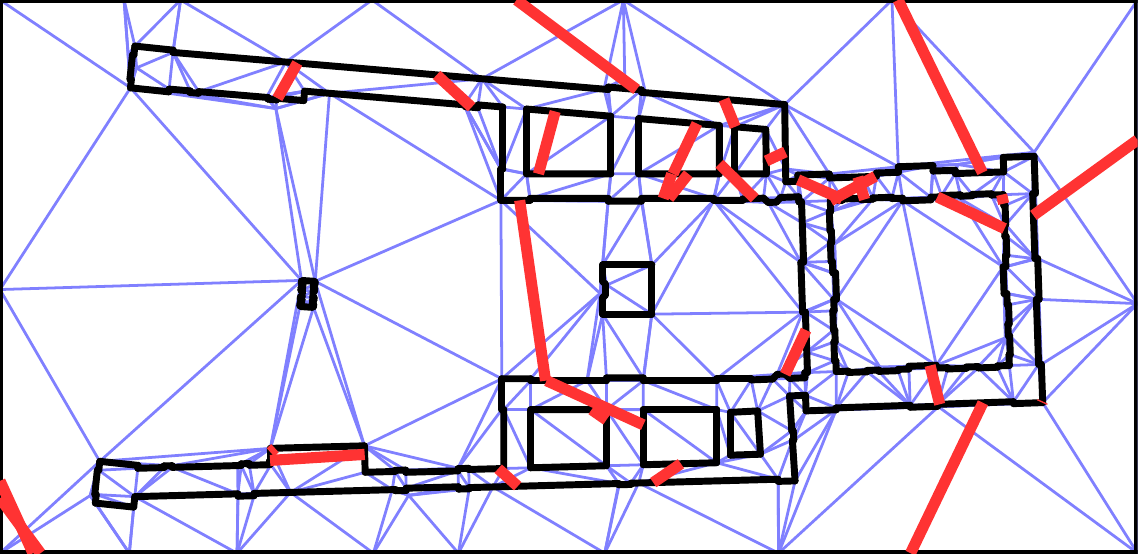}
    {Pont-l'Ev\^eque}{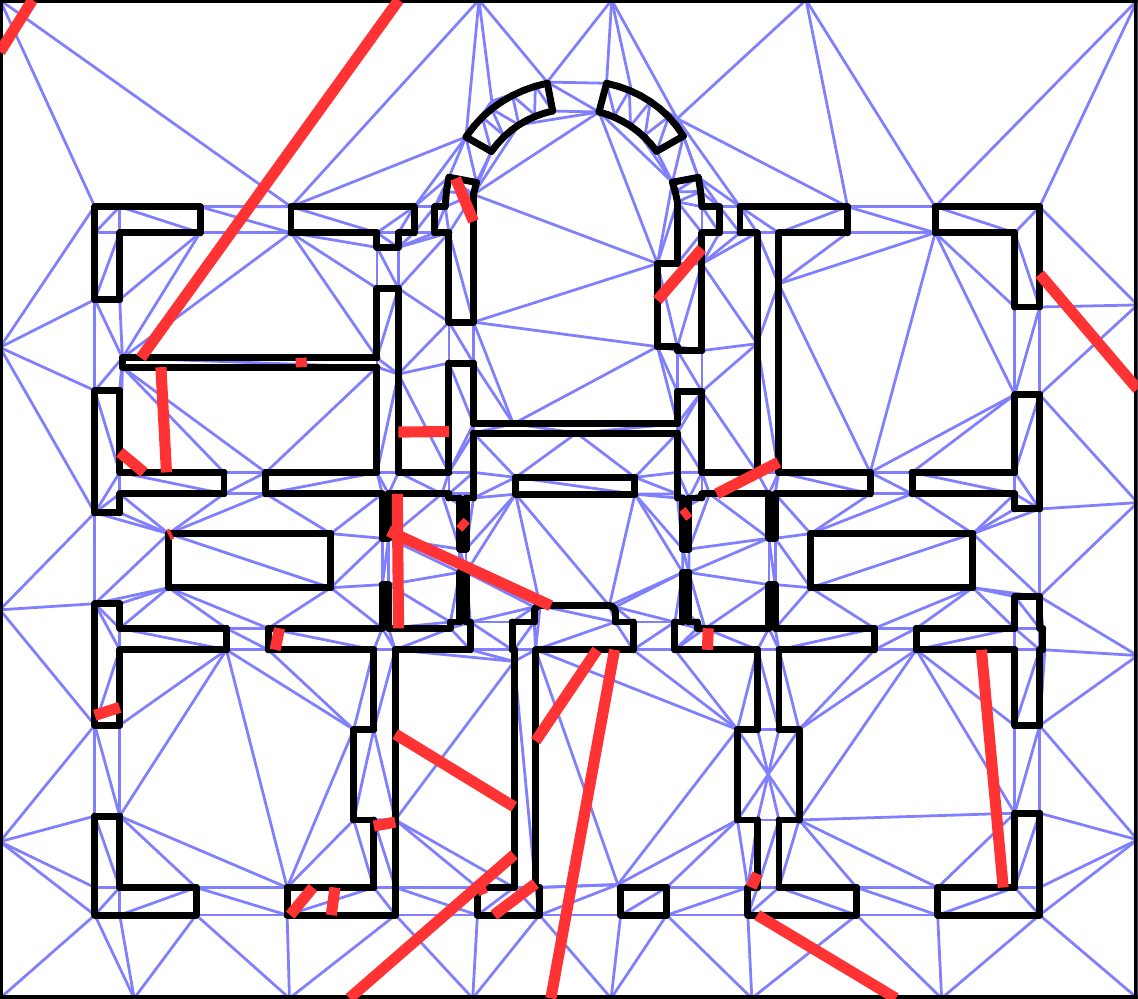}

\svgStructComparison{63mm}
    {White House}{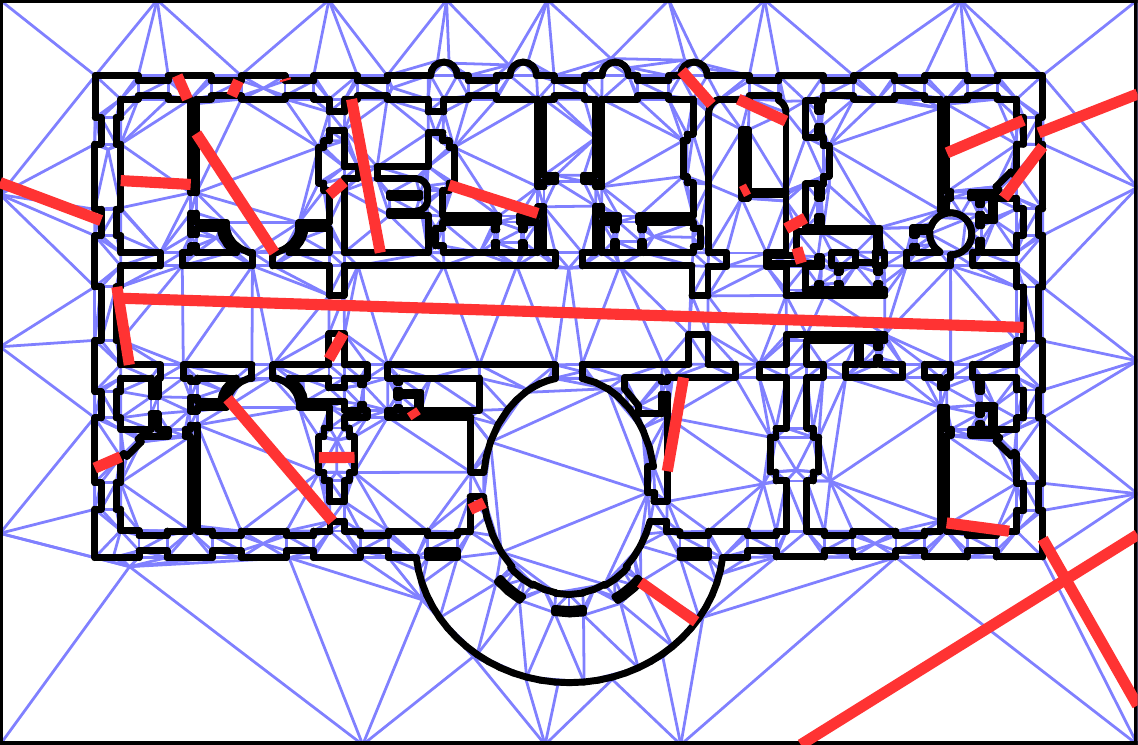}
    {Topkapı}{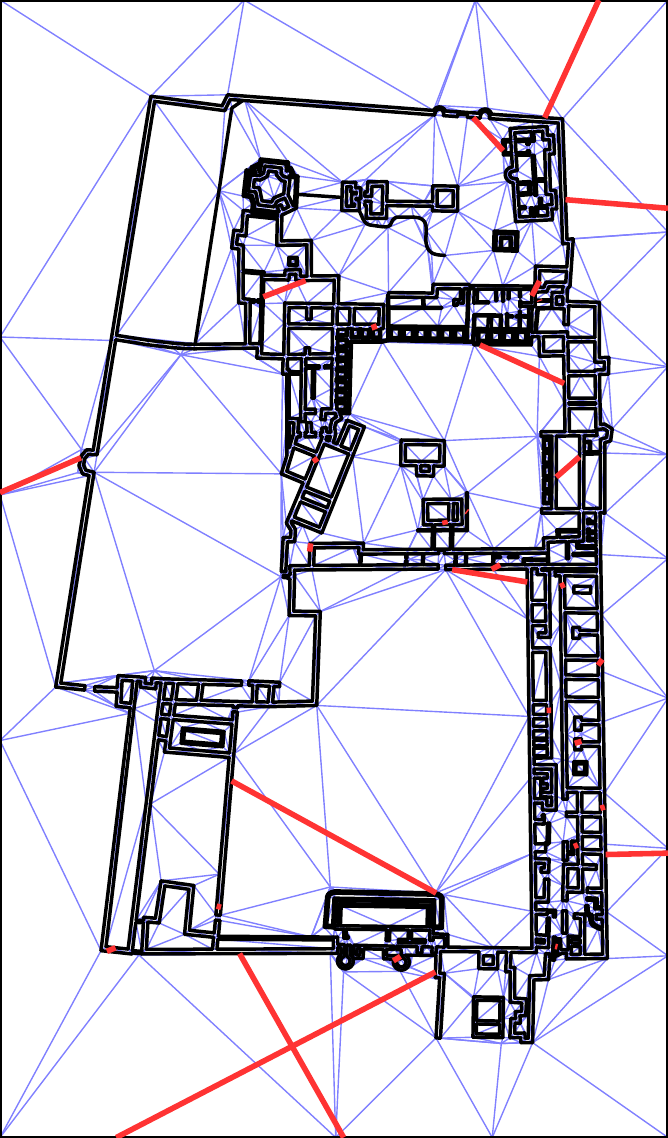}

\svgStructComparisonSingle{75mm}
    {Seville}{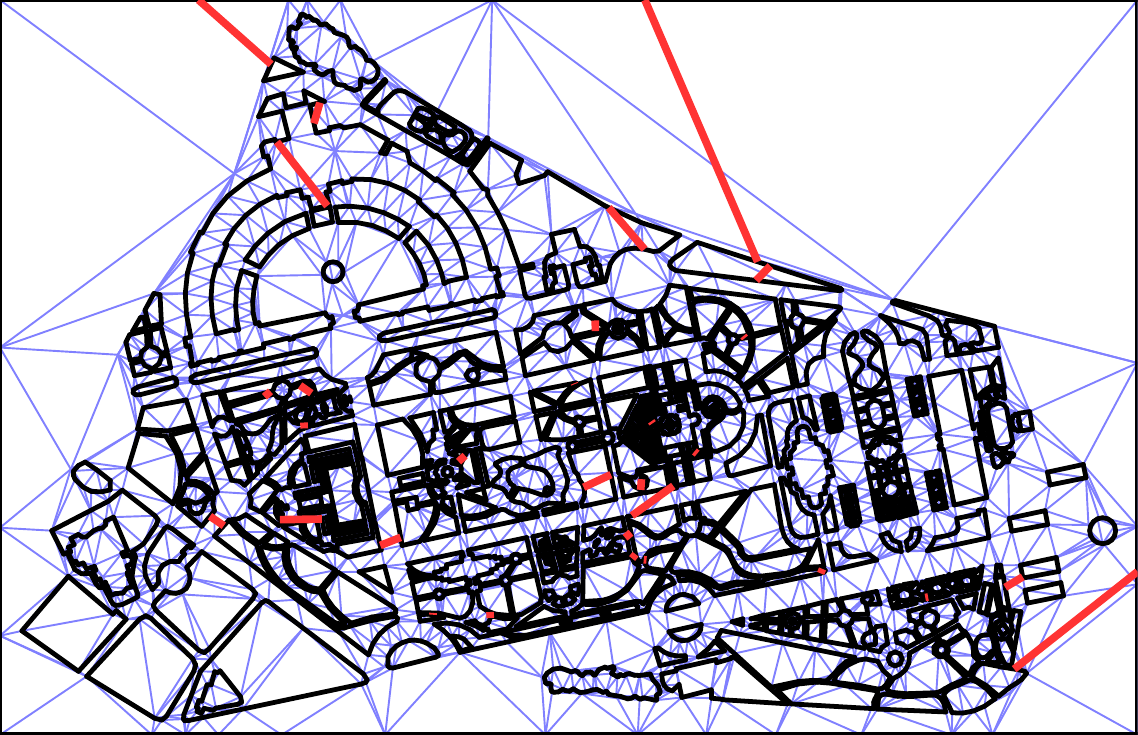}

\svgStructComparisonSingle{75mm}
    {Thessaloniki}{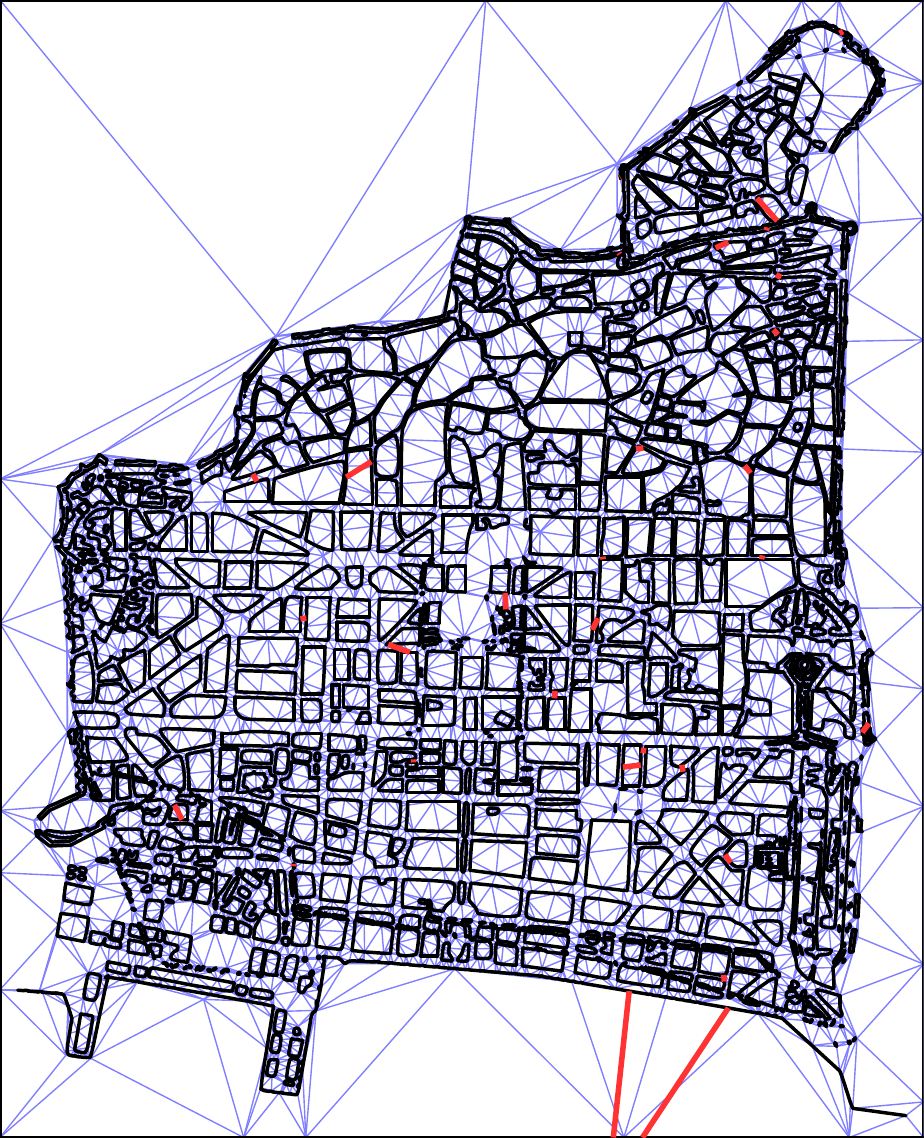}

\bibliographystyle{eg-alpha-doi}
\bibliography{techReport}

\end{document}